\documentclass{aastex63}

\usepackage{amsmath}
\usepackage{amssymb}
\usepackage{natbib}
\usepackage{multirow} 
\usepackage[mathletters]{ucs}
\usepackage[utf8x]{inputenc}
\hypersetup{linkcolor=magenta, citecolor=cyan, filecolor=yellow, urlcolor=blue}

\shorttitle{Standard GRB Spectral Models ``Misused"?}
\shortauthors{Li}

\begin{document}

\title{Standard GRB Spectral Models ``Misused"?}

\author[0000-0002-1343-3089]{Liang Li}
\affiliation{ICRANet, Piazza della Repubblica 10, I-65122 Pescara, Italy;liang.li@icranet.org}
\affiliation{INAF -- Osservatorio Astronomico d'Abruzzo, Via M. Maggini snc, I-64100, Teramo, Italy}
\affiliation{ICRA, Dipartimento di Fisica, Università di Roma ‘La Sapienza’, Piazzale Aldo Moro 5, I-00185 Roma, Italy}

\begin{abstract}

The standard model characterizing the gamma-ray burst (GRB) spectrum invokes a four-parameter empirical function, the so-called the BAND model. An alternative model named cutoff power law (COMP) implements a power law with an exponential cutoff. These functions achieve almost equally good fits on observed spectra, and are adopted in nearly all of the GRB literature. Here, we reanalyze the sample defined in Li.et al.,2021,ApJS,254,35 (39 bursts including 944 spectra). We classify the spectra by two methods: (1) checking their corner-corner plots of the posteriors to determine well-constrained $\beta$ (BAND-better) and unconstrained $\beta$ (COMP-better) categories; and (2) defining the four groups by difference of the deviance information criterion (DIC). We find inconsistent peaks of the parameter distributions between the BAND-better spectra ($\alpha=-0.64\pm0.28$ and $\rm log_{10}(E_{\rm p})=\rm log_{10}(191)\pm0.41$) and the COMP-better spectra ($\alpha=-0.96\pm0.33$ and $\rm log_{10}(E_{\rm p})=\rm log_{10}(249)\pm0.40$). With the statistically preferred model and vice versa the misused model defined based on DIC statistics, we also find that the fitted parameters obtained by the misused model (COMP) significantly deviate from those obtained by the statistically preferred model (BAND). This means that if a spectrum is statistically preferred, described as the BAND, applying COMP to derive the spectral parameters will prominently deviate from their intrinsic shape, therefore affecting the physical interpretation. Our analysis indicates that the better or statistically preferred model should be duly examined during GRB spectral analysis. In addition, the $\beta$ distribution exhibits a bimodal structure containing the BAND-better and COMP-better spectra, respectively, implying that BAND and COMP both may have physical origin. 

\end{abstract}


\section{Introduction} \label{sec:intro}

The standard approach to characterize the observational gamma-ray burst (GRB) spectral properties invokes a four-parameter empirical function known as the BAND model \citep{Band1993}. The photon number spectrum of BAND is defined as
\begin{eqnarray}
f_{\rm BAND}(E)=A \left\{ \begin{array}{ll}
(\frac{E}{E_{\rm piv}})^{\alpha} \rm exp (-\frac{{\it E}}{{\it E_{\rm 0}}}), & E < (\alpha-\beta)E_{\rm 0}  \\
\lbrack\frac{(\alpha-\beta)E_{\rm 0}}{E_{\rm piv}}\rbrack^{(\alpha-\beta)} \rm exp(\beta-\alpha)(\frac{{\it E}}{{\it E_{\rm piv}}})^{\beta}, & E\ge (\alpha-\beta)E_{0}\\
\end{array} \right.
\label{eq:Band} 
\end{eqnarray}
where $A$ is the normalization factor in units of ph cm$^{-2}$keV$^{-1}$s$^{-1}$, $E_{\rm piv}$ is the pivot energy always fixed at 100 keV, $E_{0}$ is the break energy correlated with the peak energy of $\nu F_{\nu}$ spectrum (assuming $\beta<-2$) by $E_{\rm p}=(2+\alpha)E_{\rm 0}$, $\alpha$ and $\beta$ are the low-energy and high-energy asymptotic power-law photon indices, respectively. The spectral indices ($\alpha$ and $\beta$) and the peak energy\footnote{The peak energy also represents the energy at which most of the energy of the selected spectrum (time-resolved analysis) or the entire burst (time-integrated analysis) is released.} ($E_{\rm p}$) are typically distributed around $\alpha$=-0.8 (below the break energy), $\beta$=-2.5 (above the break energy), and $E_{\rm p}$=210 keV, respectively.

An alternative empirical approach involves a simpler function called the cutoff power-law (COMP, aka the Comptonized model) model. This approach is valid when the power-law index $\beta$ is poorly constrained (having fairly large absolute values and large uncertainties; see, e.g. in Fig \ref{fig:conerCPL}). The COMP function is recovered from the BAND function as $\beta$ tends to $-\infty$. The COMP function is given by 
\begin{equation}
f_{\rm COMP}(E) =A \left(\frac{E}{E_{\rm piv}}\right)^{\alpha}\rm exp(-\frac{\it E}{\it E_{c}}),
\label{CPL}
\end{equation}
where the peak energy $E_{\rm p}$ of the $\nu F_\nu$ spectrum is related to the $E_{c}$ through $E_{\rm p}$=(2+$\alpha$)$E_{\rm c}$.

The physical origins of these empirical functions, however, have yet to be identified, although they have been the most widely used to fit GRB spectra. Neither BAND nor COMP functions correspond to an explicit emission mechanism. Whether these models are due to thermal or northermal emission is highly debated, depending on the slope values of their spectral parameters. Physically, the leading mechanisms for interpreting GRB prompt emission invoke either nonthermal photons originating from synchrotron emission (or inverse Compton scattering) \cite[e.g.,][]{Meszaros1994,Rees1994,Zhang2011b,Geng2018,Meng2018,Meng2019,Li2019b} or Comptonized quasi-thermal photons associated with photosphere emission \cite[e.g.,][]{Thompson1994,Peer2007,Ryde2010,Ruffini2013,Li2019a,Li2019c,Li2020,Xue2021}. The fast-cooling ($\alpha$=-3/2) and slow-cooling ($\alpha$=-2/3, so-called the line of death of synchrotron emission, \citealt{Preece1998}) synchrotron emission predicts two different values of $\alpha$, whereas photosphere models predict much harder values of $\alpha$ (e.g., above $\alpha$=-2/3). For instance, \cite{Acuner2020} argued that the spectra that prefer the photospheric model all have low-energy power-law indices $\alpha\sim >$-0.5, as long as the data has a high significance. Therefore, applying these empirical models to the GRB spectral analysis plays an important role in identifying the GRB radiation mechanism, and investigation of spectral parameters, and therefore, will shed light on our understanding of GRB physics \citep[e.g.,][and references therein]{Dai2006,Kaneko2006,Zhang2006,Gruber2014, Yu2016,Ruffini2018,Li2019a,Li2019c,Li2019b,Li2021a,Xue2021,Moradi2021}.

In general, we can apply either time-integrated or time-resolved spectral analysis to study the spectral properties of a GRB. Several spectral catalogs of GRBs exist in the literature based on either the time-integrated analysis \cite[e.g.,][]{Goldstein2012,Gruber2014} or the time-resolved analysis \cite[e.g.,][]{Yu2016,Li2021b}. The time-integrated spectrum represents the average spectral properties since the entire period of emission is treated as a single time bin. However, GRB prompt emission is well known to have strong spectral evolution \citep[e.g.,][and references therein]{Kaneko2006, Yu2019, Li2021b}, which requires the more detailed time-resolved spectral analysis \cite[treating the whole period of emission as multiple timing bins, and spectral analysis is therefore performed on each timing event individually, e.g.,][]{Yu2016,Yu2019,Li2021b}.

Several early GRB spectral catalogs make use of the frequentist approach \cite[e.g.,][]{Kaneko2006}. In recent years, a fully Bayesian analysis method has been increasingly developed. For example, time-resolved spectral catalogs based on such a fully Bayesian analysis method for single-pulse bursts \citep{Yu2019} and multi-pulse bursts \citep{Li2021b} have been created. In the Bayesian analysis, Bayesian inference is used to account for relevant prior information and the resulting posterior probability distributions of parameters are obtained by the Markov Chain Monte Carlo (MCMC) iterations.

Phenomenologically, the BAND-like spectrum with well-constrained model parameters is typically observed in the time-integrated spectral analysis, while the simpler COMP-like spectrum is commonly observed in the time-resolved spectral analysis. This is because time-resolved spectral properties typically do not have good high-energy photon statistics, and therefore, the high-energy spectral index $\beta$ for time-resolved spectra usually cannot be well evaluated due to the small number of photons available.

It is important to stress that the difference in fitting by BAND or COMP functions is not fully examined when performing the time-resolved analysis of large samples \citep[e.g.,][]{Kaneko2006, Goldstein2012, Yu2019, Li2021b}. Moreover, in some articles COMP is applied throughout without a comparison with other models since the COMP is usually preferred for the majority of the time-resolved spectra. We have doubts about the statistical conclusions and the physical implications generated from possible misused model. Therefore, we dedicate this article to examining the deviation of spectral fittings between BAND and COMP. We wish to answer the question: Is the impact on parameters significant if misusing a model? Do BAND and COMP both have physical backgrounds? Here we reanalyze the sample (39 bursts including 944 spectra) defined in \cite{Li2021b} to examine the spectral properties statistically of these two standard spectral models.

This paper is organized as follows. The methods are presented in \S \ref{sec:data}. The detailed results are summarized in \S \ref{sec:results}. The discussion and conclusion are presented in \S \ref{sec:discussion} and \S \ref{sec:conclusion}, respectively. The convention $Q=10^{x}Q_{x}$ is adopted in cgs units throughout the paper. The standard $\Lambda$ cold dark matter (CDM) cosmology with the parameters $H_{0}= 67.4$ ${\rm km s^{-1}}$ ${\rm Mpc^{-1}}$, $\Omega_{M}=0.315$, and $\Omega_{\Lambda}=0.685$ are adopted \citep{PlanckCollaboration2018}.

\section{Methodology} \label{sec:data}

\subsection{Sample Revisited} 

The Gamma-ray Burst Monitor (GBM; 8 KeV-40 MeV, \citealt{Meegan2009}), and the Large Area Telescope (LAT; 20 MeV-300 GeV, \citealt{Atwood2009}), are the two instruments on board \emph{Fermi} providing unprecedented spectral coverage for seven orders of magnitude in energy. \emph{Fermi}-GBM, together with \emph{Fermi}-LAT, has been triggered by more than 2000 bursts since its launch in 2008. Here, we revisit the sample defined in \cite{Li2021b}. The sample is collected from the \emph{Fermi}-GBM burst catalog published at HEASARC\footnote{\url{https://heasarc.gsfc.nasa.gov/W3Browse/fermi/fermigbrst.html}}, and it focuses on well-separated multi-pulse GBM-detected bursts. It consists of 39 bursts, 117 pulses, and 1228 spectra. There are two reasons that we included the sample in this task. First, all the spectra in the sample were selected to have a high statistical significance in order to allow us to perform a detailed time-resolved spectral analysis and ensure that the spectral fits are well determined, this is the key point. Second, the prompt-emission light curves of GRBs typically exhibit irregular, multi-pulse temporal profiles.

The sample selection in \cite{Li2021b} includes the following main steps: (1) The first is to visually inspect the light curves for each burst that was observed by {\it Fermi}-GBM during its first 11 yr of mission (more than 2000 bursts), and about 120 bursts that have well-separated multi-pulse features are roughly identified; (2) The second is to capture the variations of the Time-Tagged Events (TTE) light curve and divide the light curve into time segments by following the Bayesian blocks (BBlocks; \citealt{Scargle2013}) algorithm, and the significance ($S$; \citealt{Vianello2018}) for each time bin was also calculated; (3) The third is to select at least two pulses in each burst whose individual pulse light curve has at least four time bins with high significance ($S\geq 20$); \footnote{Although the BBlocks method can better capture the intrinsic variation of light curve \cite[e.g.,][]{Li2021b}, the time bins created by such a method usually have varied signal-to-noise ratios. This means that we cannot ensure that there are enough photons in each time bin in order to establish a reliably spectral fit. In order to ensure that the spectral fits can be well determined, a relatively high statistical significance for each selected time bin is needed. On the other hand, the threshold levels of statistical significance required by different spectral models may also be different. Practically, more complicated models (with more free parameters) require more signal photons in order to establish a reliable fit result. Therefore, a threshold level of $S$ $\geq$ 20 is typically used for the BAND model while $S \geq 15$ for the COMP model since the BAND model ($A$, $\alpha$, $\beta$, $E_{\rm p}$) has one more free parameter than the COMP model ($A$, $\alpha$, $E_{\rm c}$). The sample defined in \cite{Li2021b} adopted $S$ $\geq$ 20 to select the time bins, which is enough to study both two models.}, hence, the final sample was defined (39 bursts, 103 pulses, and 944 spectra); (4) The final goal is to obtain the best spectral parameters by adopting a fully Bayesian analysis using the MCMC method and performing both the BAND and COMP functions to fit all the spectra, respectively. For information on the data procedure, including the burst, detector, source, and background selections, light curve binning method, sample definition, and Bayesian and MCMC spectral fitting approaches; please refer to \cite{Li2021b,Li2019a,Li2021a} for more details.

\subsection{High energy power law, $\beta$, and the ``better'' models}\label{sec:betamodel}

In reality, for a given spectrum, in order to determine which one (BAND or COMP) is better, one needs to check whether a well-constrained $\beta$ can be determined. If $\beta$ is not well constrained in some cases, there are two possibilities. Firstly, lack of photons in the analysed bins (e.g., $S<$ 20), so that the spectral fit cannot be well determined. Secondly, the number of photons in the analyzed bins is sufficient (e.g., $S\geq$ 20), but the model that better characterizes the spectral shape is indeed the COMP. Our sample defined in \cite{Li2021b} with $S\geq$ 20 rules out the first possibility. We therefore inspected all the posteriors of the BAND spectra to check their $\beta$ indices. If a well constrained $\beta$ is clearly identified by a certain spectrum, the BAND model is considered as better, otherwise the COMP model is better. Under these criteria, all the spectra can be identified into two categories: 

\begin{itemize}

\item BAND-better spectra: All the spectra selected in this category are identified with a well-constrained $\beta$, indicating that the BAND model is indeed better. It contains 35\% of the total number of spectra. 
\item COMP-better spectra: All the spectra in this category are identified with an unconstrained $\beta$, implying that the COMP model is better. This is 65\% of the total number of spectra. 

\end{itemize}

\begin{figure*}
\includegraphics[angle=0,scale=0.6]{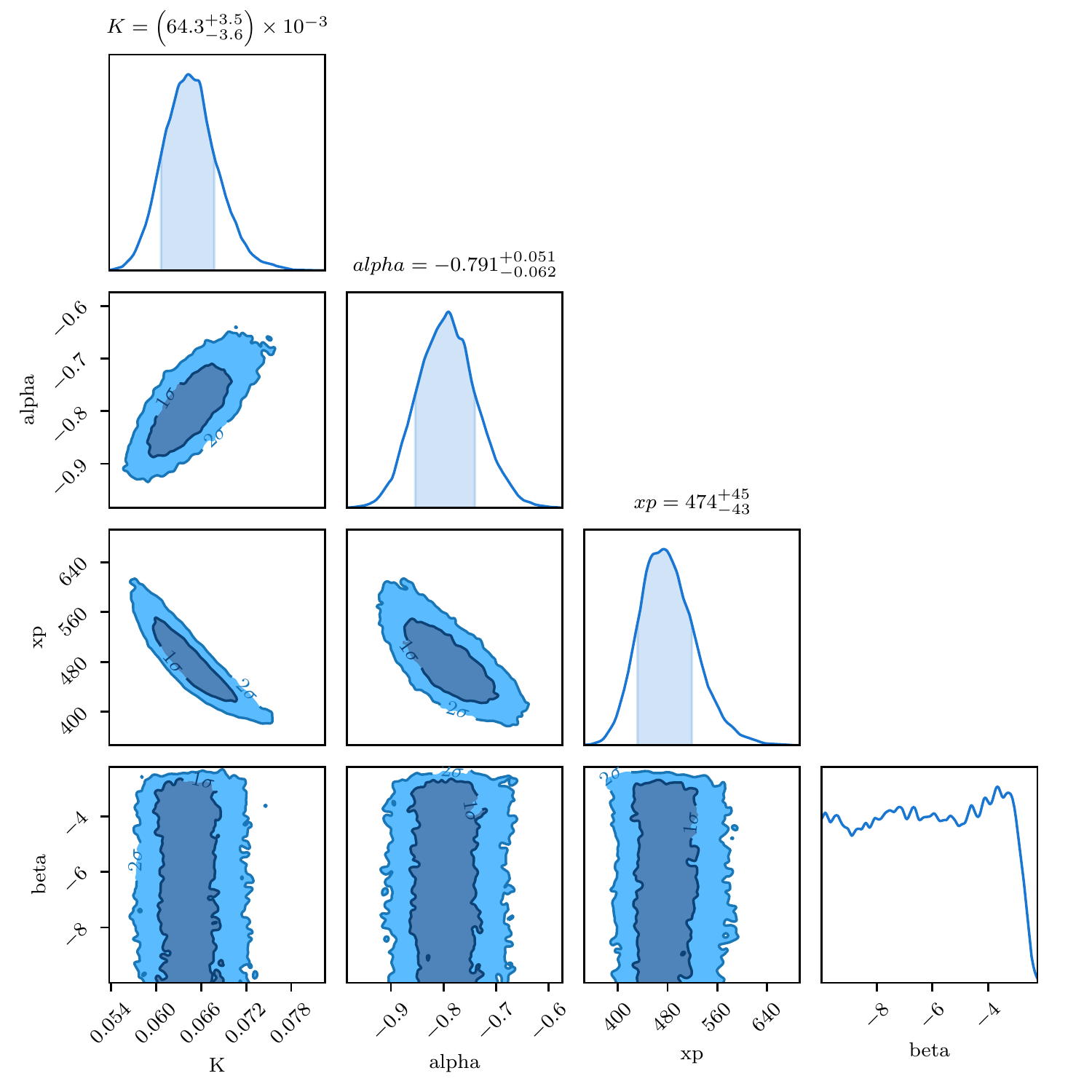}
\includegraphics[angle=0,scale=0.8]{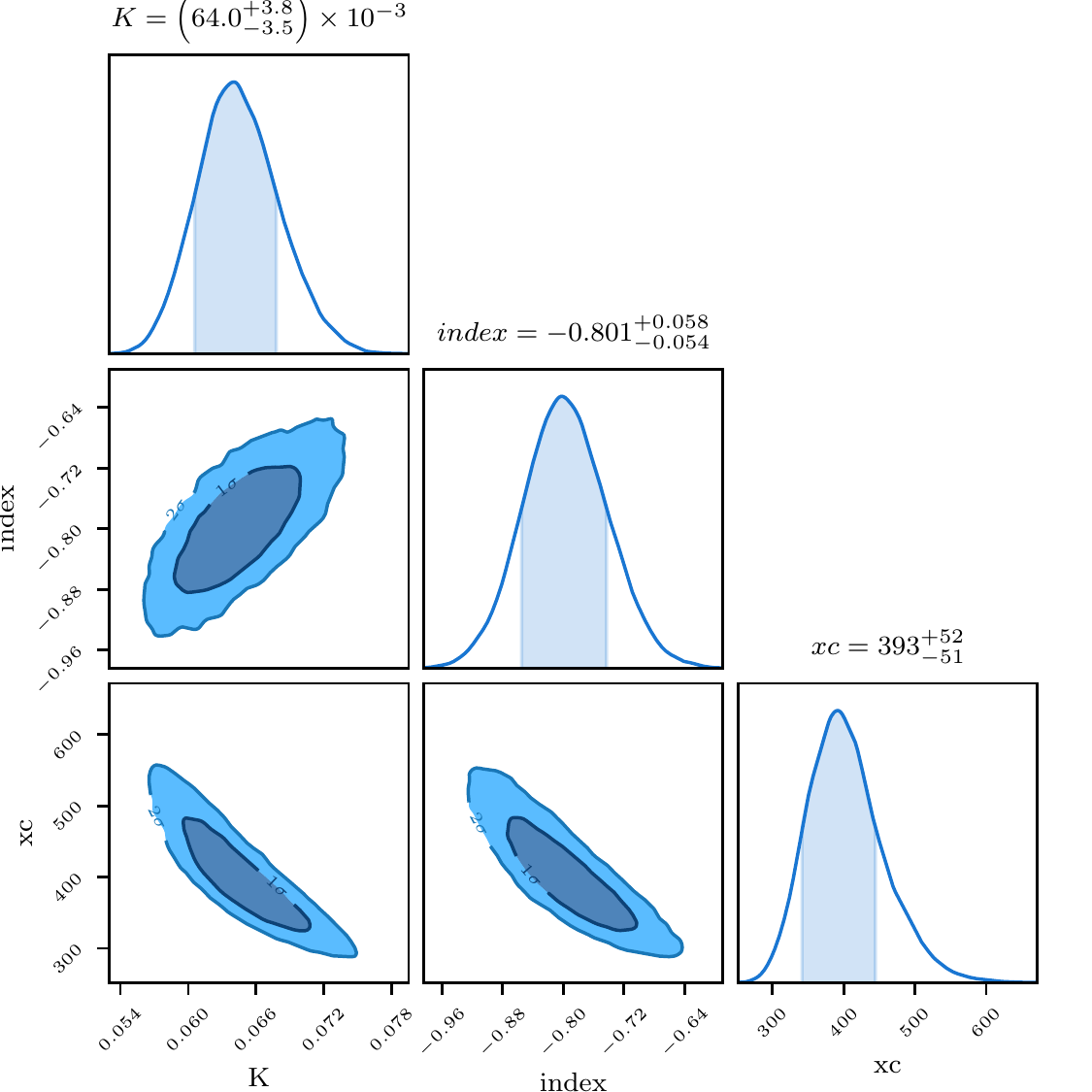}
\caption{Bayesian MCMC spectral fits to the data in one time bin (between 24.215 and 25.597 s) of GRB 171227 comparing the BAND with COMP models. The left panel shows the BAND fit to the data with a well-constrained $\beta$ while the right panel displays the COMP fit to the same data. The plots shows a COMP-preferred spectrum, with $\Delta \rm DIC$=1.7.}\label{fig:conerCPL}
\end{figure*}

\begin{figure*}
\includegraphics[angle=0,scale=0.6]{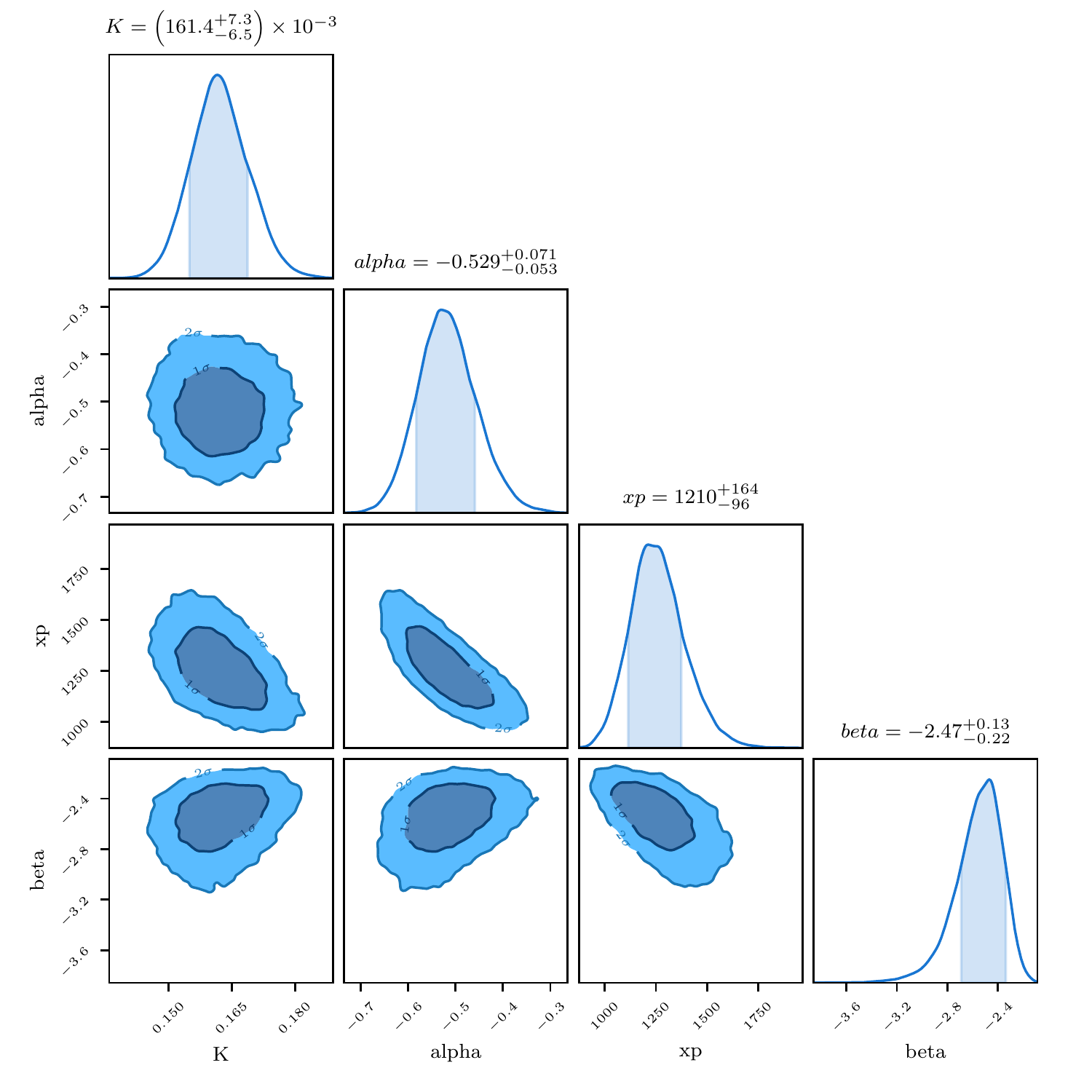}
\includegraphics[angle=0,scale=0.8]{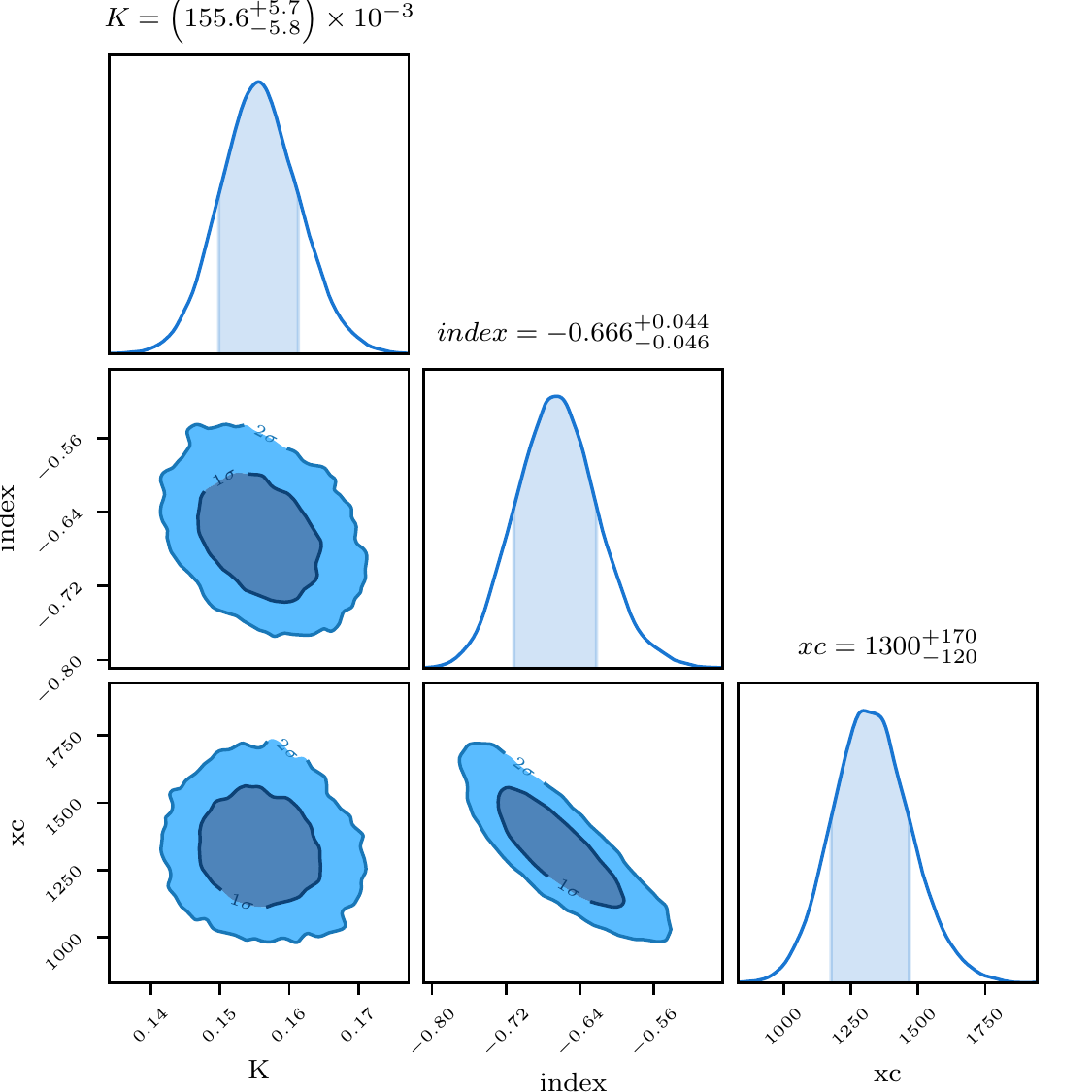}
\caption{Same as Figure \ref{fig:conerCPL} but for another time bin (between 17.648 and 17.820 s) of GRB 171227. The left panel shows the BAND fit to the data with a well-constrained $\beta$ while the right panel displays the COMP fit to the same data. The plots shows a BAND-preferred spectrum, with $\Delta \rm DIC$=-24.8.}\label{fig:conerBand}
\end{figure*}

In Figure \ref{fig:conerCPL}, the left panel shows two-dimensional corner-corner plots of the spectral parameters using the Bayesian MCMC method used to perform the BAND fit. The spectral data is obtained from one time bin (between 24.215 and 25.597 s) from GRB 171227, and an unconstrained $\beta$ is clearly identified from the posterior density map. While that of the COMP fit for the same spectral data is displayed in the right panel of Figure \ref{fig:conerCPL}. For comparison, we also present the same plots using another time bin (between 24.215 and 25.597 s) from GRB 171227 in Figure \ref{fig:conerBand}, where a well-constrained $\beta$ is clearly identified in the BAND fit.

In total, the fractions of the constrained-$\beta$ and unconstrained-$\beta$ spectra are 35\% and 65\%, respectively. This suggests that the majority of the spectra (two-thirds) can indeed be better fitted by the COMP, confirming the previous similar findings \citep{Yu2019,Li2021b}. 

\subsection{Statistically Preferred Models Determined by Information Criteria}\label{sec:group}

\begin{figure*}
\begin{center}
\includegraphics[angle=0,scale=0.75]{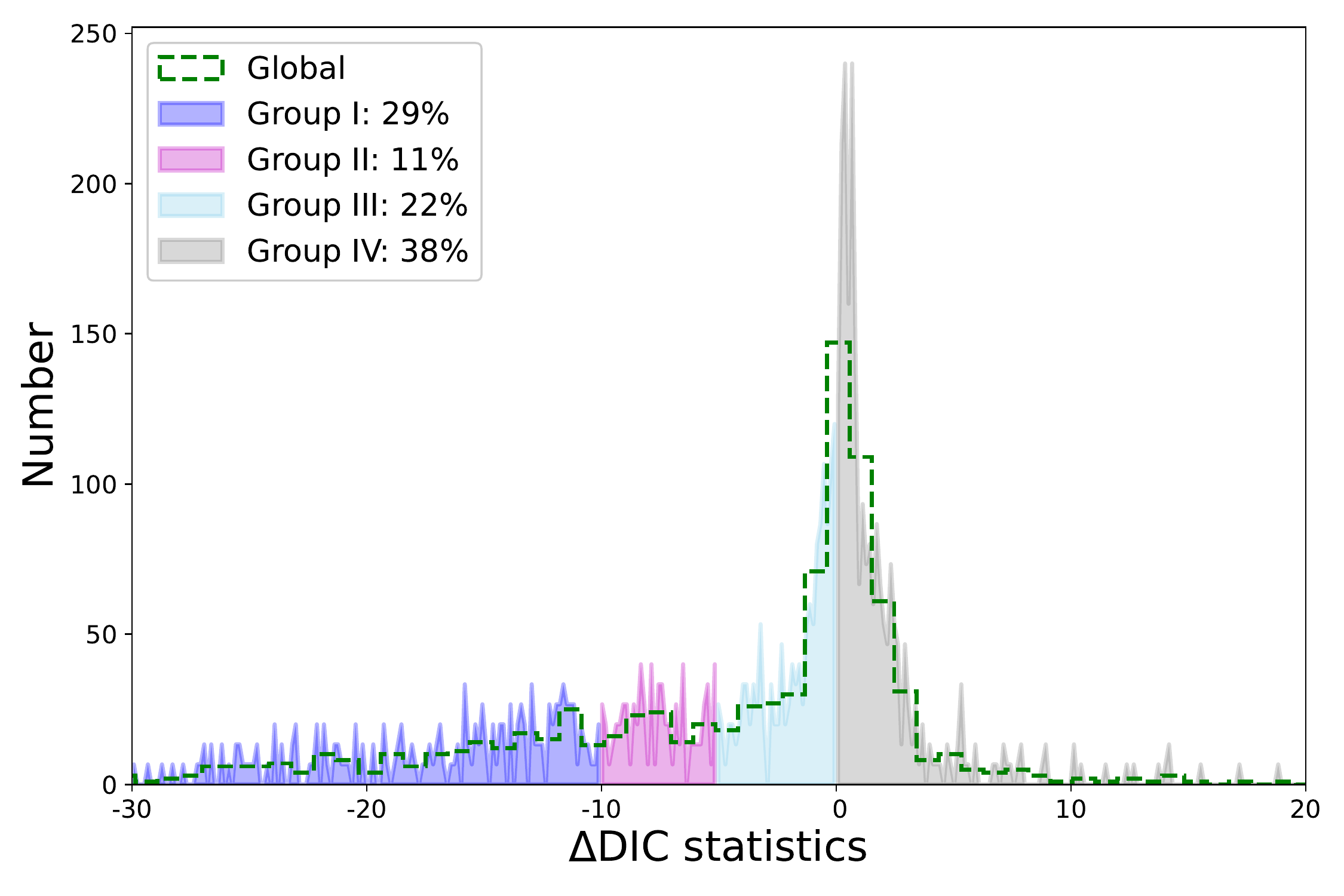}
\end{center}
\caption{Distribution of $\Delta \rm DIC$. The different groups are overlaid by different colors: Group I (blue), Group II (magenta), Group III (sky blue), and Group IV (gray). While the global distribution is shown by the green curve.}\label{fig:DIC}
\end{figure*}

In practice, a more common approach for model comparison is by using information criteria, such as Akaike Information Criteria (AIC), Bayesian Information Criteria (BIC), and the Deviance Information Criterion. The Bayesian analysis and MCMC method are fully applied in this work, the deviance information criterion (DIC, \citealt{Spiegelhalter2002, Moreno2013}) is computed to compare models, it is defined as DIC=-2log[$p$(data$\mid\hat{\theta}$)]+2$p_{\rm DIC}$, where $\hat{\theta}$ is the posterior mean of the parameters, and $p_{\rm DIC}$ is a term to penalize the more complex model for overfitting \citep{Gelman2014}. The values of the difference between the BAND's and the COMP's, defined as $\Delta$DIC = DIC$_{\rm BAND}$-DIC$_{\rm COMP}$, can be used to indicate the preferred one. 
For each individual spectrum, a negative DIC value indicates that the observational data favors a Band-like spectrum. All of the spectra can be separated into the following groups using different threshold levels based on DIC statistics for Bayesian models \citep[e.g.,][]{Gelman2014,Pooley2018}, in which the BAND-preferred or COMP-preferred spectra could also be determined and gathered.
\begin{itemize}
\item Group I: $\Delta \rm DIC<=$-10. BAND model is statistically preferred. This group contains 29\% of the spectra, as shown in Figure \ref{fig:DIC}.
\item Group II: -10$<\Delta \rm DIC<=$-5. BAND model is still statistically preferred, but is not as strong as Group I. This group contains 11\% of the spectra (Figure \ref{fig:DIC}). 
\item Group III: -5$<\Delta \rm DIC<=$0. COMP model is statistically preferred. This group contains 22\% of the spectra (Figure \ref{fig:DIC}). 
\item Group IV: $\Delta \rm DIC>$0. COMP model is statistically preferred, and is stronger than Group III. This group contains 38\% of the spectra (Figure \ref{fig:DIC}). 
\end{itemize}

Based on the BAND-better and COMP-better spectra defined in \S \ref{sec:betamodel}, we then check the fractions of the spectra with or without a constrained $\beta$ for each DIC-defined group. Groupwisely, the corresponding fractions are [93\%, 7\%], [58\%, 42\%], [5\%, 95\%], and [1\%, 99\%] for Group I, Group II, Group III, and Group IV, respectively. In Group I, we find that for almost all the spectra (up to $\sim$93\%) a well-constrained $\beta$ can be clearly identified. However, there are very few spectra showing a well-constrained $\beta$ in both Group III and Group IV. These results suggest that Band-like spectra dominate Group I while COMP-like spectra dominate Group III and Group IV. Interestingly, we also find that in Group-II, both Band-like (58\%) and COMP-like (42\%) spectra, are almost identical. The results also suggest that our two methods of classifying GRBs are consistent, but one needs to consider a relatively large DIC value (e.g. -5 is good and -10 is perfect, these values are in agreement with previous works \citep{Acuner2020,Li2021b}.

\section{Results} \label{sec:results}

Before we move forward, a few remarks need to be made here. Firstly, we focus on the two most widely used models for GRB spectra (BAND and COMP), and we miss several other models (e.g., power-law model, smooth broken power-law model, and the BETA model) that were used in the previous catalogs (e.g., \citealt{Kaneko2006}). In some cases, these models should be able to fit the spectra better than the BAND or COMP that we used in this task. For instance, if a break's energy lies outside the detector passband, or the source photon signal beyond the break energy is weak enough so that the break energy cannot be well determined. In such cases, the simpler power-law model (see one recent work, \citealt {Tang2021}, for instance) is superior to the other, more complicated models (having more parameters). As such, the models we used would be the better ones that characterize GRB intrinsic spectra, rather than the best ones. Secondly, our analysis is based on a sample of well-separated multi-pulse GBM-detected bursts and a criterion of statistical significance $S>20$ was used to select the bright spectra for each individual burst. These may be causing some bias in our analysis results. Lastly, compared to previous catalogs (e.g., \citealt{Kaneko2006,Yu2016}) that used the $\chi^{2}$ method to statistically compare the models and determine the best-fit model for each individual spectrum, our analysis is based on a fully Bayesian analysis approach using the MCMC method and we used the information criteria to compare the models. Unlike the $\chi^{2}$ method involving a different degree of freedom in different models and resulting in the comparison not being straightforwardly performed, the information criteria (e.g., DIC statistics) that we used in this task may easily offer a straightforward comparison among different models because penalty factors for overfitting of more complex models have also been taken into account. Based on such a Bayesian analysis and MCMC spectral fit method, we may also be able to select the better models more straightforward by inspecting their posterior distributions from MCMC sampling, as compared to some previous studies that invoke a more complicated selection method to determine their {\it good} class of parameters \citep[e.g.,][]{Kaneko2006,Yu2016}.

\subsection{Statistically Preferred Model Misused}\label{sec:model-misused}

In this task, our primary interest is to assess the effect of misuse of the models on fitting results. For instance, for a given spectrum that can be better (or statistically preferred) described by the BAND model, are the spectral parameters obtained from the simpler COMP model still consistent with that from the BAND model? Vice versa. To better address this question, we define (1) the BAND-to-BAND case (Column 6 in Table \ref{table:distribution_CPL_Band}): a statistically preferred BAND model is used for a given Category (defined in \S \ref{sec:betamodel}) or Group (defined in \S \ref{sec:group}); (2) the COMP-to-COMP case: a statistically preferred COMP model is used. Alternatively, if a statistically preferred model is not being used, in contrast, the model used is a statistically undesirable one. This may involve the better model being misused. We therefore also define (3) the BAND-to-COMP case: a statistically preferred model is BAND but COMP is misused, this invokes the case of underfitting; (4) the COMP-to-BAND" case: a statistically preferred model is COMP but BAND is misused, this invokes the case of overfitting.

\begin{figure*}
\includegraphics[angle=0,scale=0.45]{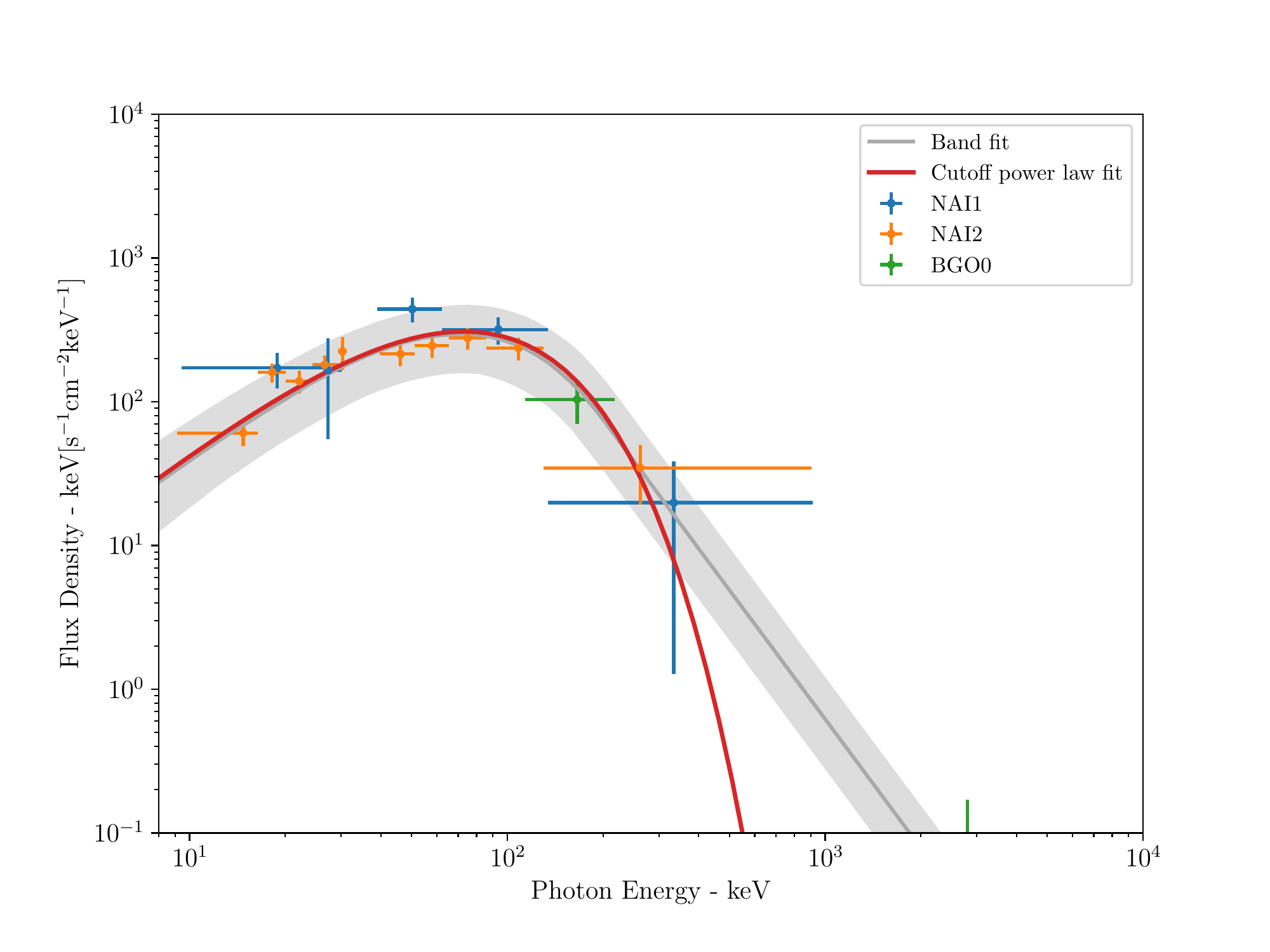}
\includegraphics[angle=0,scale=0.45]{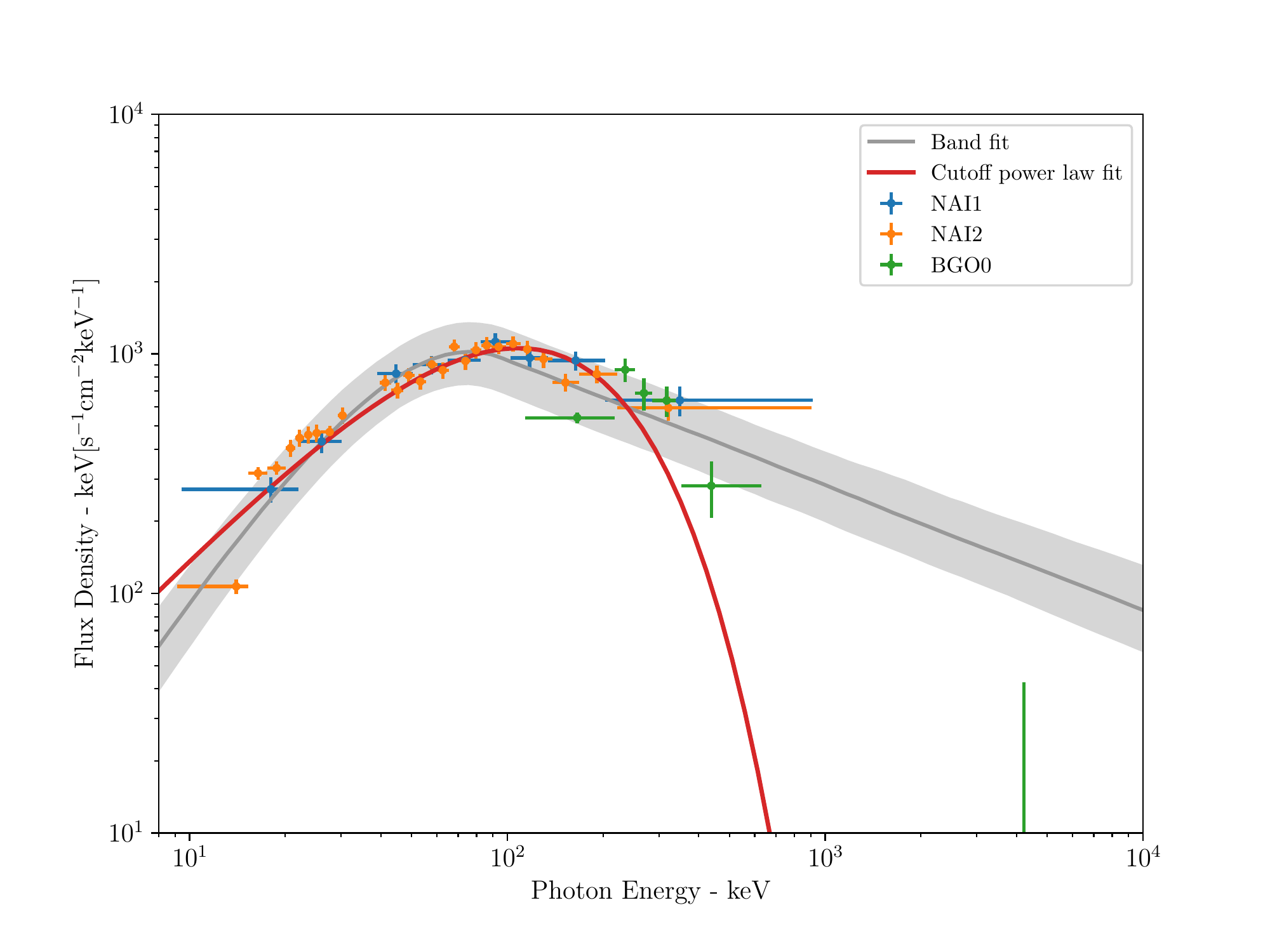}
\caption{Comparison of the fitting of the same spectral data using BAND and COMP models. Left panel: COMP is the statistically preferred model for the time interval between 86.338 and 86.877 s of GRB 120728. Right panel: BAND is the statistically preferred model for the time interval between 69.274 and 71.015 s of GRB 120728}.\label{fig:spectrum}
\end{figure*}

We then investigate these misused cases using the following two typical examples, as shown in Figure \ref{fig:spectrum}. In the left panel of Figure \ref{fig:spectrum}, we present the spectral fit to a time bin (between 86.338 and 86.877 s in GRB 120728) using both BAND and COMP models. This spectrum can be statistically-preferred fitted by the COMP model, confirmed by the DIC statistics with a value of $\Delta$DIC=1.2. The spectral parameters obtained by the COMP fit are $\alpha=-0.26^{+0.23}_{-0.23}$, and $E_{\rm p}=74^{+16}_{-16}$ and obtained by the BAND fit are $\alpha=-0.17^{+0.25}_{-0.27}$, $\beta=-6.46^{+2.38}_{-2.38}$, and $E_{\rm p}=71^{+6}_{-5}$. These results suggest that the fitted spectral parameters ($\alpha$ and $E_{\rm p}$) between the COMP-to-BAND case all seem to agree.  
In the right panel in Figure \ref{fig:spectrum}, we present the spectral fit to another time bin (between 69.274 and 71.015 s in GRB 120728). The BAND model is the statistically preferred one that describes the spectral shape, which is also confirmed by the DIC statistics with a value of $\Delta$DIC=-123.8. For comparison, we also fit the spectral data using the COMP model. For the BAND model: $\alpha=0.09^{+0.12}_{-0.12}$, $\beta=-2.52^{+0.06}_{-0.06}$, and $E_{\rm p}=76^{+4}_{-4}$. For the COMP model: $\alpha=-0.61^{+0.05}_{-0.05}$, and $E_{\rm p}=109^{+7}_{-7}$. We find a remarkable discrepancy in the spectral parameters between the BAND-to-COMP case (the COMP model with softer $\alpha$ values and higher $E_{\rm p}$, as compared with the BAND model). This is due to compensating for the lack of a high-energy spectral component in the model resulting in the underfitting.

We also test the misuse of models by fitting the simulated spectra, which are generated by the GBM Data Tools\footnote{\url{https://fermi.gsfc.nasa.gov/ssc/data/analysis/rmfit/gbm_data_tools/gdt-docs/}}. For the source spectra, we take the functional modeling of the BAND and set the initial model parameters of $A$=0.03, $\alpha$=-0.55, $E_{\rm p}$=500, and $\beta$=-2.5, and those of the COMP with $A$=0.03, $\alpha$=-0.55, and $E_{\rm c}$=500. For the background spectrum, we generate it using a phenomenological method that first to fit the background of GRB 210518A\footnote{This burst is randomly selected, and its background is adopted, not the burst signal.}  then to simulate the background spectrum from the fitted parameters. The response matrix is taken from the first 10 seconds of GRB 210518A.

\begin{figure*}
\includegraphics[angle=0,scale=0.45]{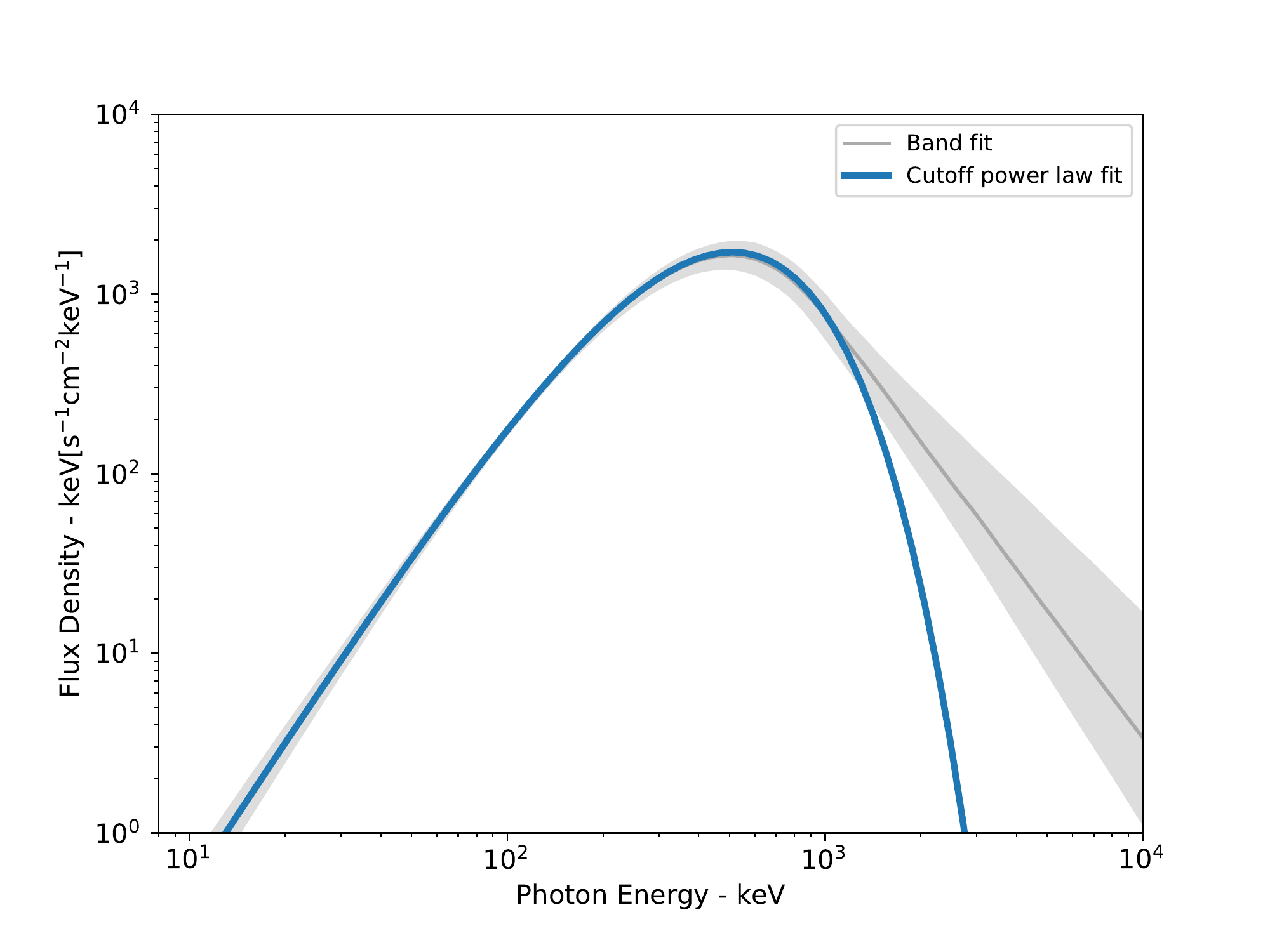}
\includegraphics[angle=0,scale=0.45]{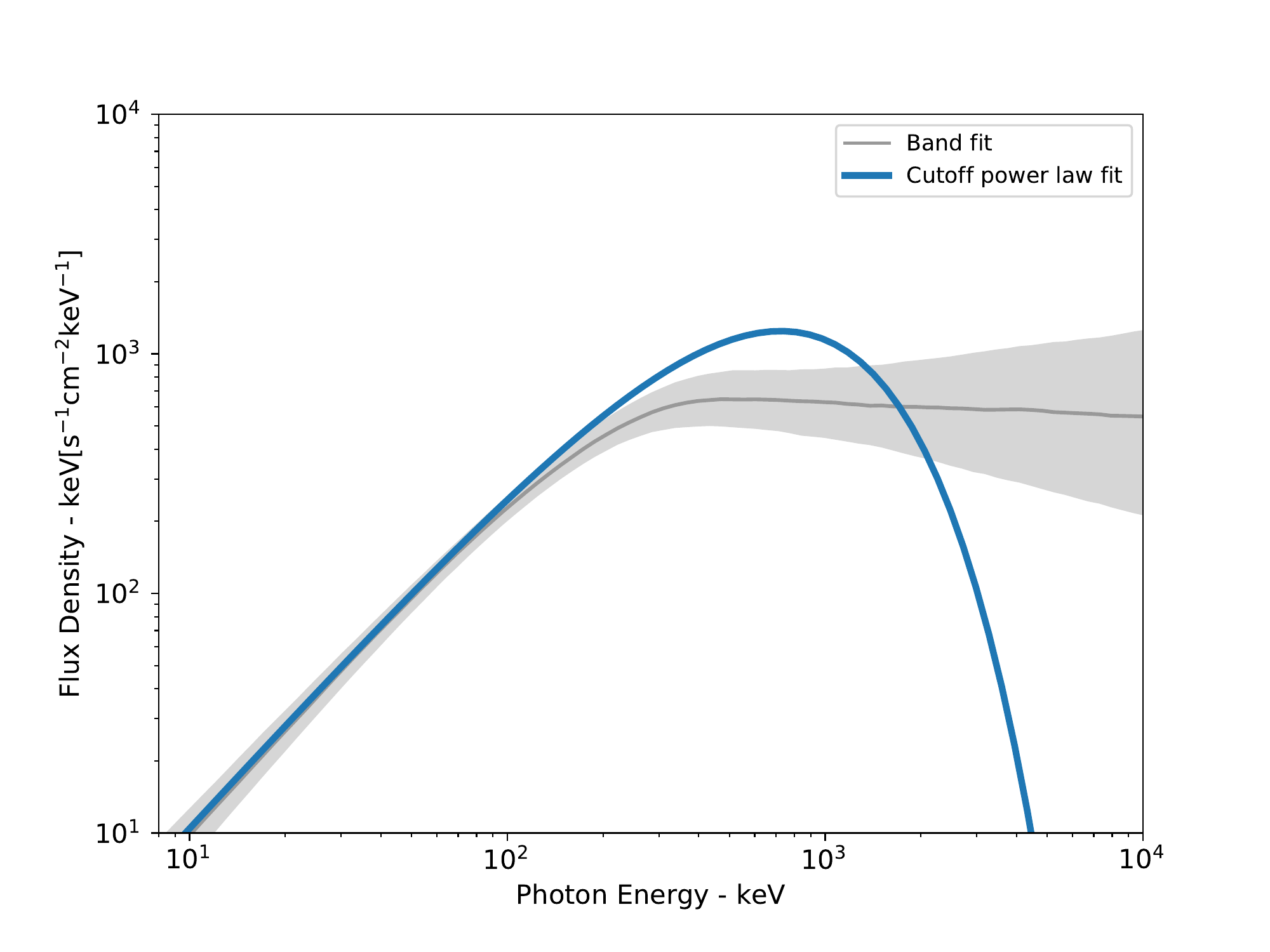}
\caption{Same as Figure \ref{fig:spectrum} but for the simulated spectra. The left panel is the fittings for the simulated COMP-like spectrum while the right panel is the fittings for the simulated BAND-like spectrum.}\label{fig:SimulatedSpectrum}
\end{figure*}

The simulated COMP-like spectrum fitted using BAND and COMP gives rise to a differential value of $\Delta$DIC=0.4 (see the left panel in Figure \ref{fig:SimulatedSpectrum}). A small difference in DIC statistics suggests that adding a high-energy component to such a COMP-like spectrum does not significantly improve the fit. Whereas the simulated BAND-like spectrum results in large DIC statistics ($\Delta$DIC=-38.6) improvements (see the right panel of Figure \ref{fig:SimulatedSpectrum}), indicating a statistically significant high-energy power-law component. These simulated results are consistent with the findings using true spectral fittings as described above. 

\subsection{Comparisons of Parameter Distribution for $\beta$-Statistic-Based Categories}

\begin{figure*}
\includegraphics[angle=0,scale=0.4]{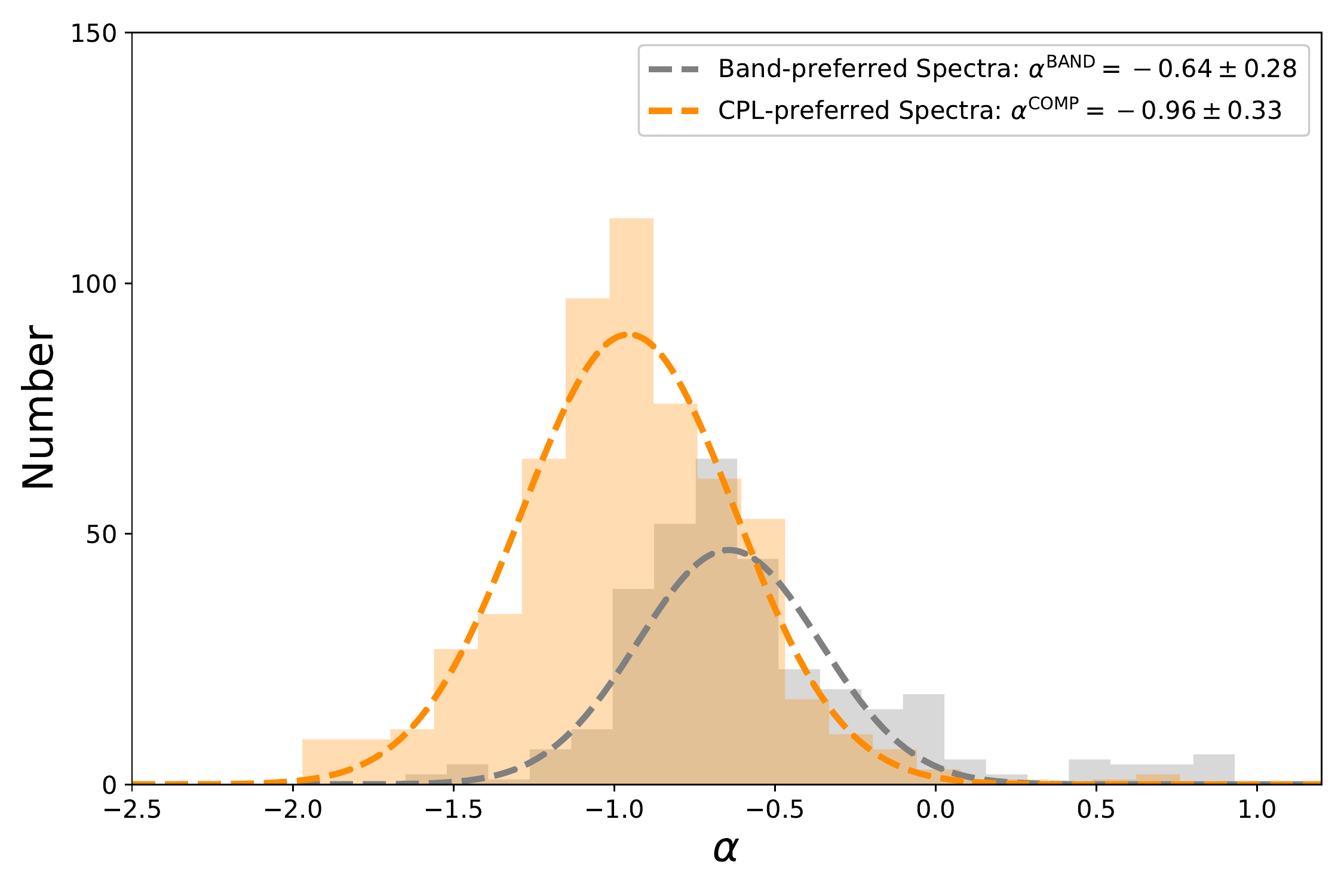}
\includegraphics[angle=0,scale=0.4]{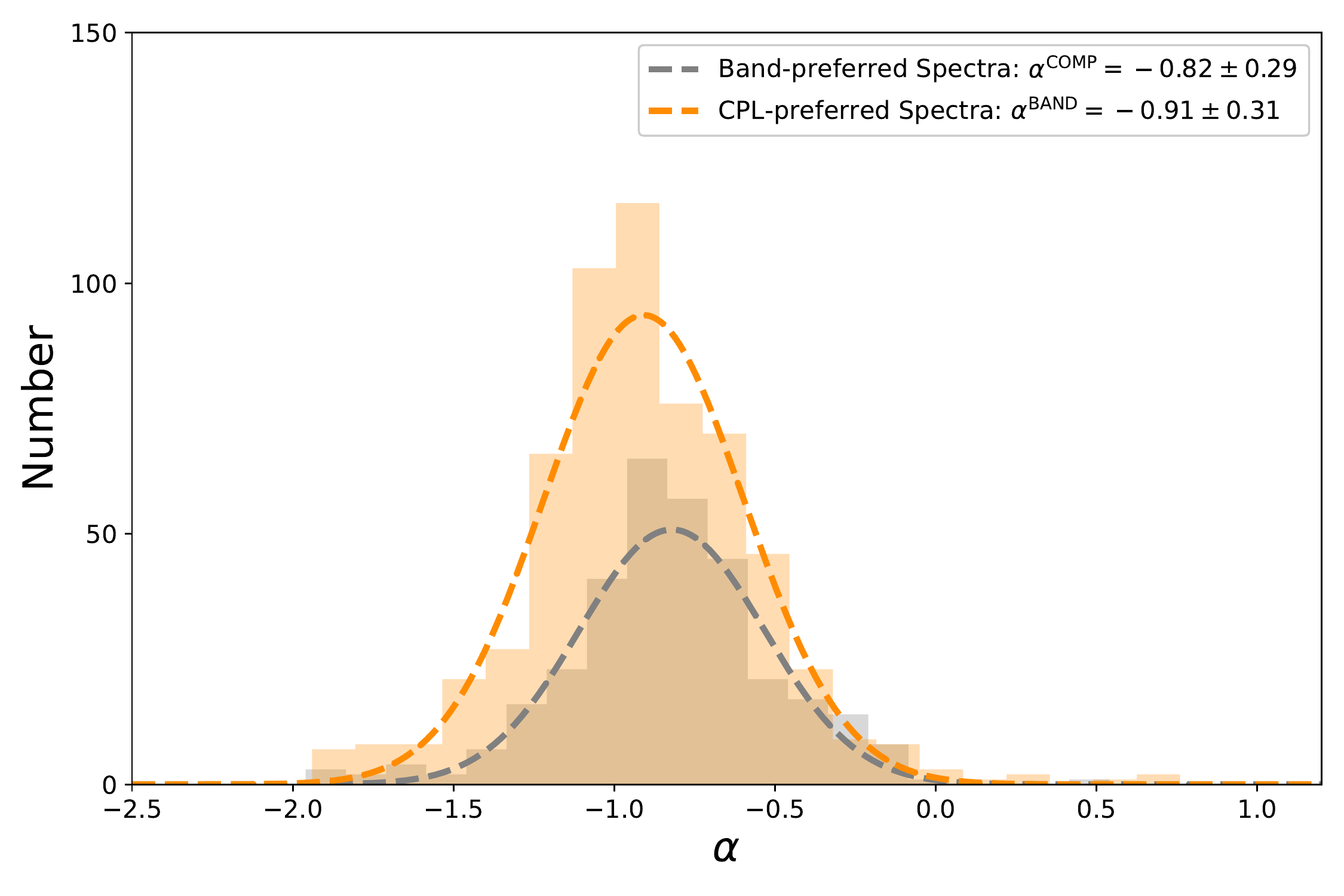}
\includegraphics[angle=0,scale=0.4]{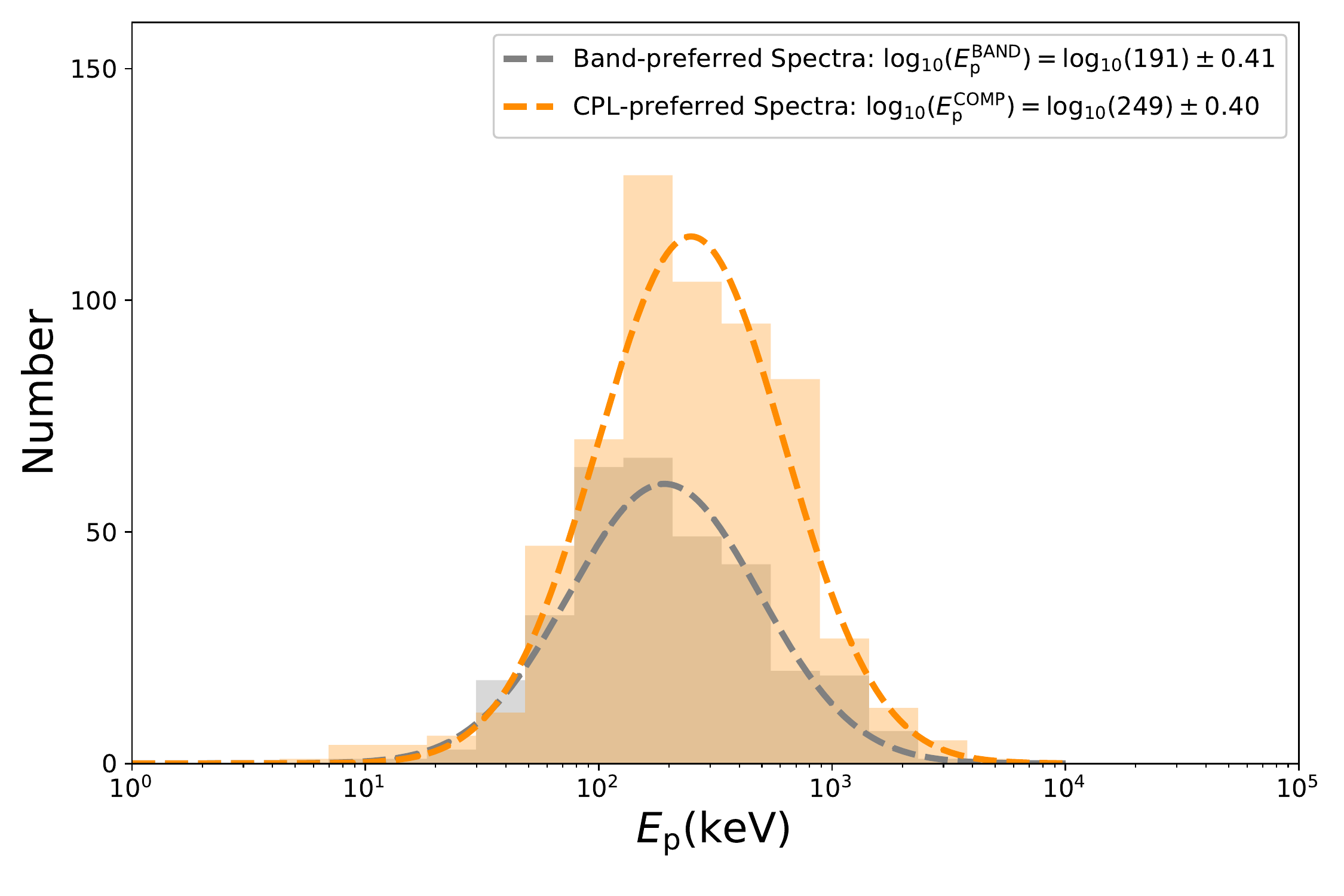}
\includegraphics[angle=0,scale=0.4]{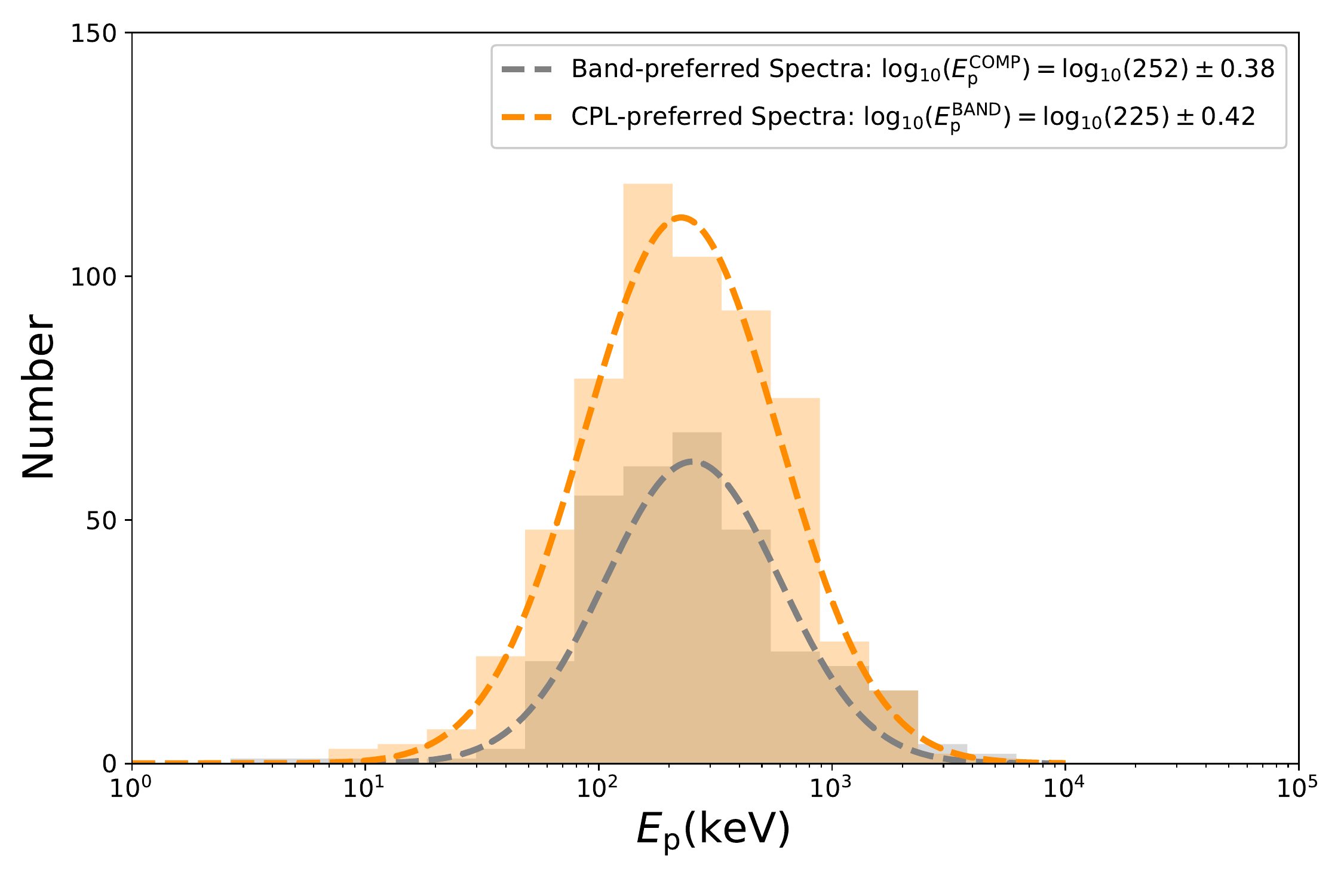}
\includegraphics[angle=0,scale=0.4]{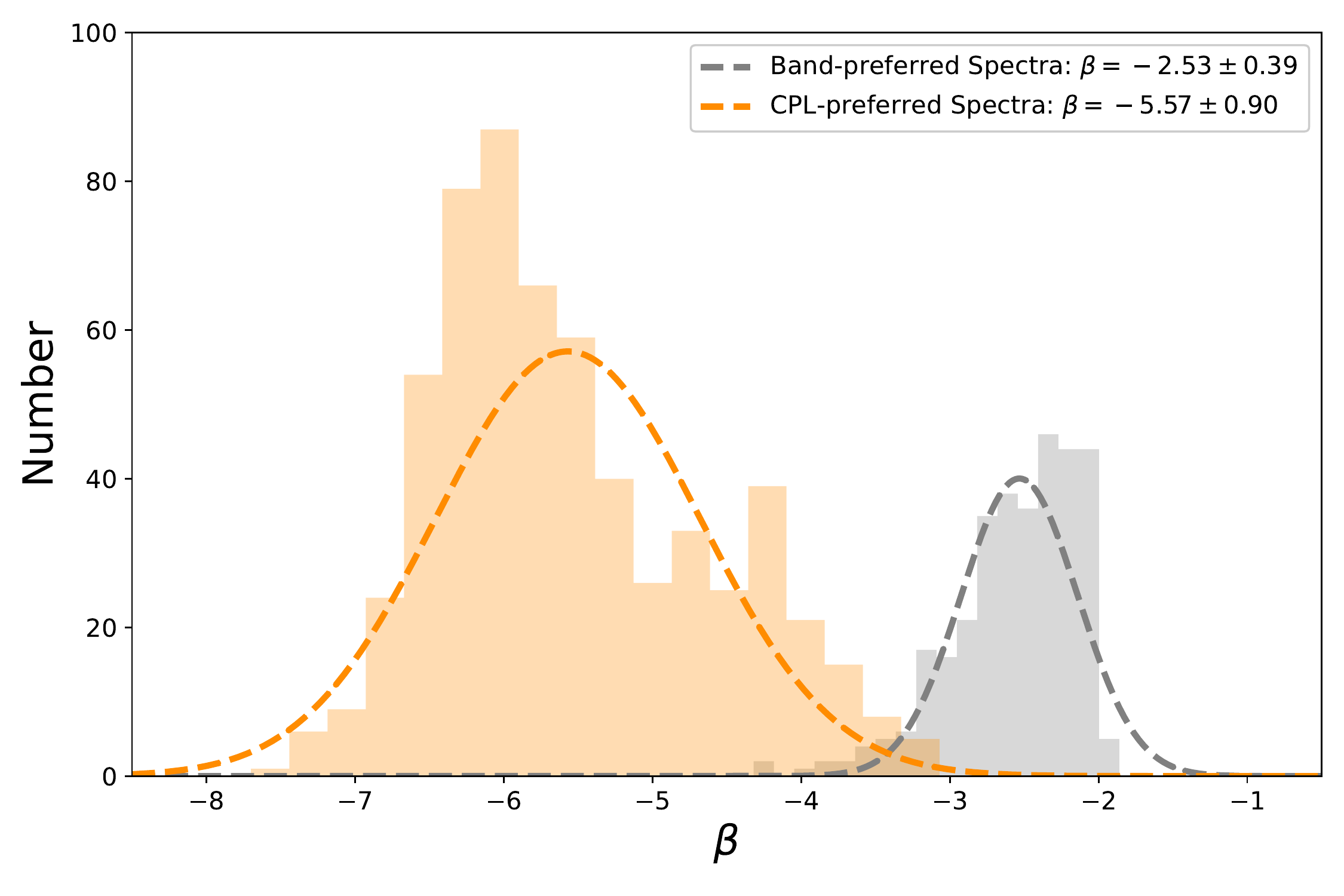}
\caption{Distributions of $\alpha$ (upper panels),  $E_{\rm p}$ (middle panels), and $\beta$ (lower panel). All are based on statistical significance $S \geq 20$ (944 spectra). The spectra separated by the better models via to check their posterior plots. For each spectral parameter, the left panel shows the model preferred cases while the right panel displays the model misused cases. The BAND-preferred spectra is indicated by gray color while those of the COMP-preferred spectra is indicated by orange color.}\label{fig:distributionbeta}
\end{figure*}

With the categories defined in \S \ref{sec:betamodel}, we present the distributions of the spectral parameters that are used to compare the BAND-preferred spectra and COMP-preferred spectra in Figure \ref{fig:distributionbeta}. The average values and the corresponding standard deviation obtained from the best Gaussian fits for parameter distributions are summarized in Table \ref{table:distribution_CPL_Band}, including $\alpha$ (BAND and COMP), $\beta$ (BAND only), and $E_{\rm p}$ (BAND and COMP).

\begin{figure*}
\includegraphics[angle=0,scale=0.45]{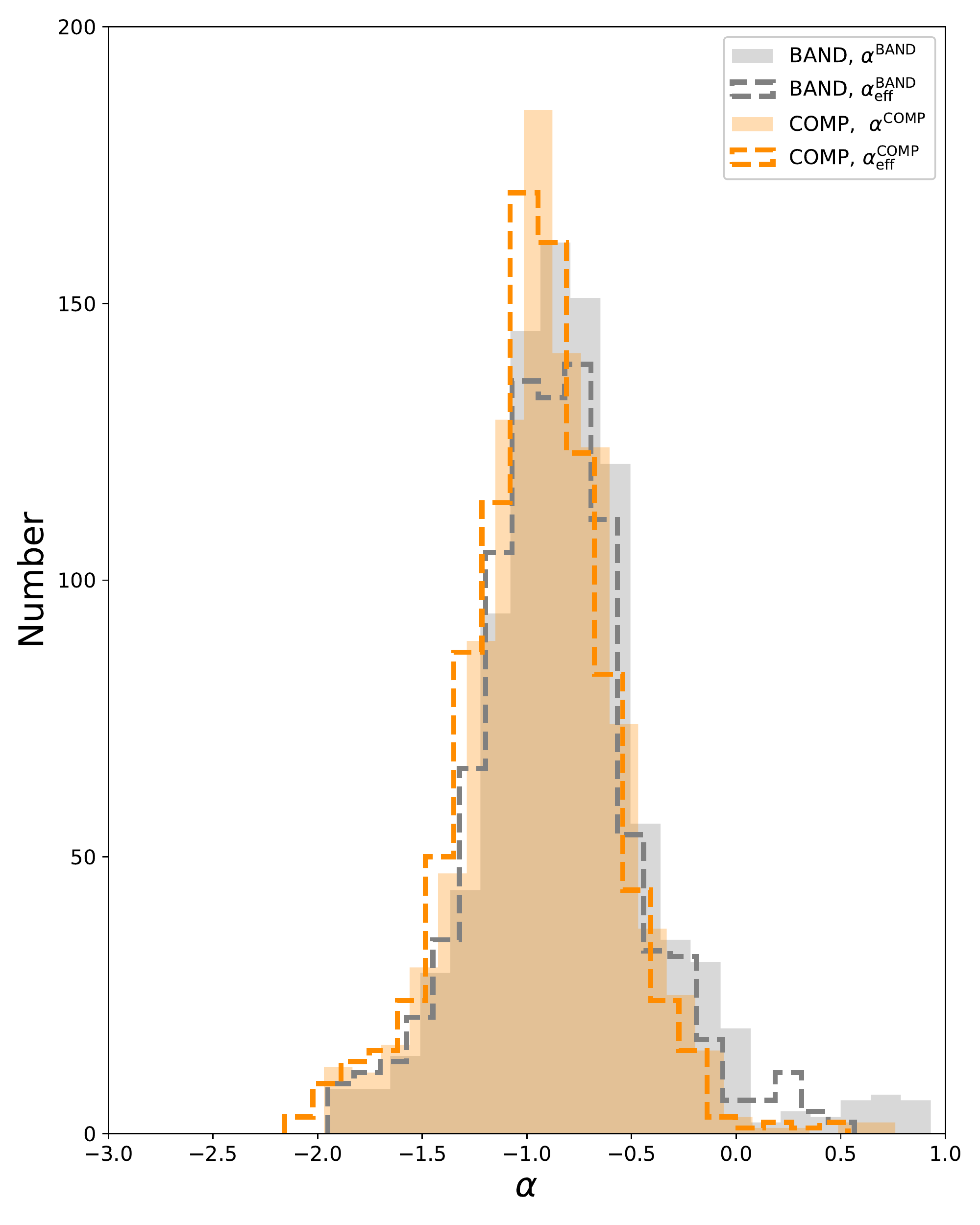}
\includegraphics[angle=0,scale=0.45]{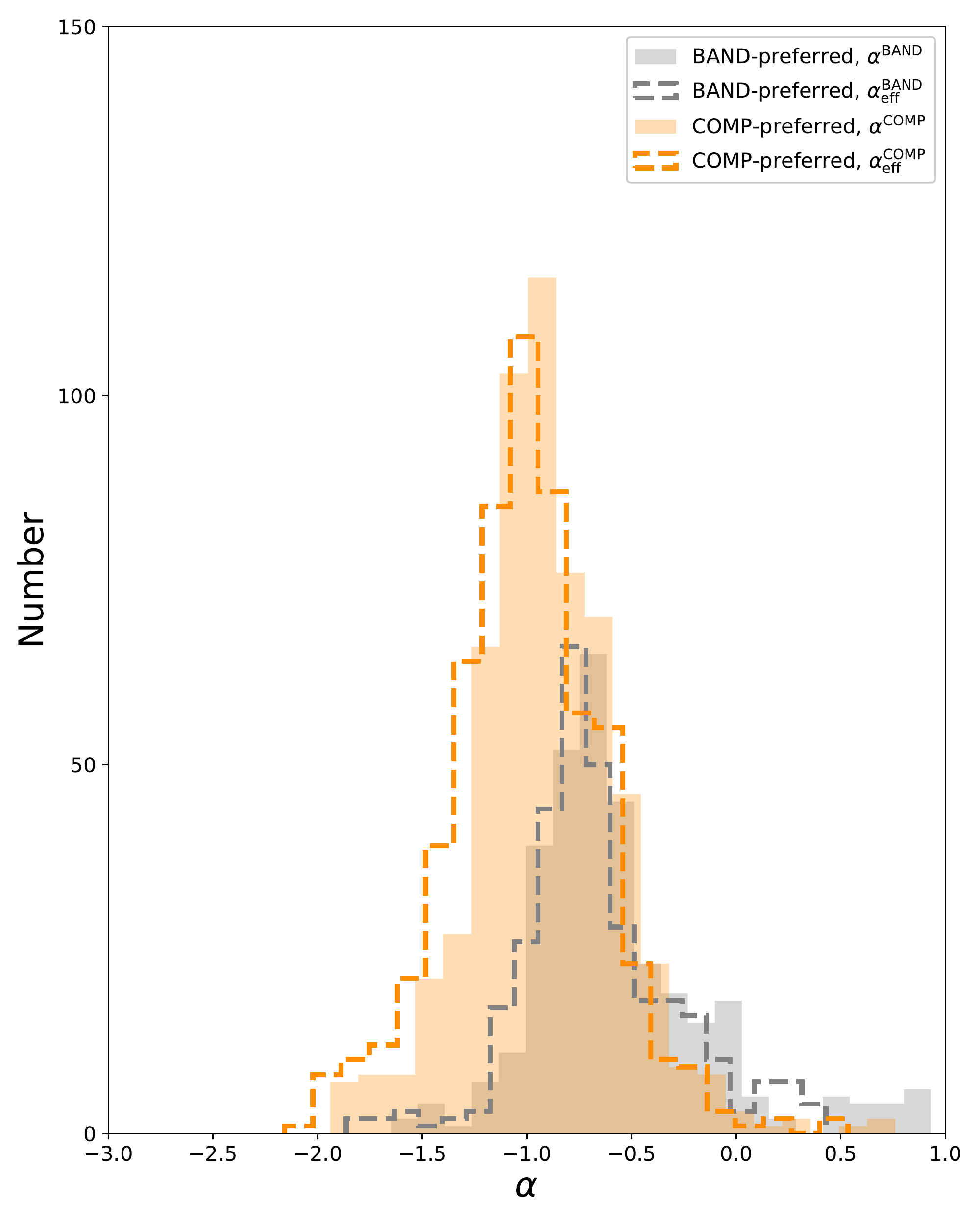}
\caption{Distributions of fitted values of $\alpha$ obtained from the BAND and COMP as compared to their effective values $\alpha_{\rm eff}$, computed at 8 keV using the Equation (2) in \cite{Preece1998}. Left panel: for the global spectra. Right panel: for the $\beta$-based spectra.}\label{fig:alphaeff}
\end{figure*}

Before comparing the fitted parameters of BAND and COMP with different categories and groups, we caution that the low-energy spectral indices $\alpha$ obtained from BAND and COMP are asymptotic values rather than actual slopes, and therefore cannot be directly compared. In order to minimize the discrepancy, an effective $\alpha_{\rm eff}$, computed at 25 keV (the BATSE\footnote{The Burst and Transient Source Experiment (BATSE) on board the Compton Gamma-Ray Observatory (CGRO).} detector lower limit), was introduced by \cite{Preece1998}. In the GBM observations, the lower limit of the detector is at 8 keV, which is much smaller than the BATSE, and the difference between the asymptotic values and the actual slopes can be negligible (Figure \ref{fig:alphaeff}). The fit values of $\alpha$, therefore, can be directly used for our further analyses.

For $\alpha$ distribution (the right panel of Figure \ref{fig:distributionbeta}), we find the BAND-better spectra and COMP-better spectra showing inconsistent peaks. The best fit gives $\alpha^{\rm BAND}=-0.64\pm0.28$ for the BAND-better spectra and $\alpha^{\rm COMP}=-0.96\pm0.33$ for the COMP-better spectra, with a difference between $\alpha$ of the BAND fits and the COMP fits of $\Delta \alpha$=0.32, where $\Delta \alpha$ is defined as $\Delta \alpha=\alpha^{\rm BAND}-\alpha^{\rm COMP}$.  

While if the BAND-better spectra are misused by the COMP fit, one has $\alpha^{\rm COMP}=-0.82\pm0.29$, with a value of $\Delta \alpha$ of 0.18. Likewise, if the COMP-better spectra are misused by the BAND fit, one has $\alpha^{\rm BAND}=-0.91\pm0.31$, with a value of $\Delta \alpha$ of 0.05. Based on these results, several interesting results can be drawn: (1) a significantly statistical difference of the spectral parameters between BAND-better spectra and COMP-better spectra is found; (2) The deviation (the COMP model with higher $E_{\rm p}$ and softer $\alpha$ indices than the BAND model) between the BAND-to-COMP case is much more significant than the COMP-to-BAND case. Similar results can also be found in the $E_{\rm p}$ distribution (see the middle panels in Figure \ref{fig:distributionbeta} and Column 9 in Table \ref{table:distribution_CPL_Band}).

We use the Kolmogorov–Smirnov (K-S) test to assess whether the distributions change between the two distinct categories. The chance probability, $P$, determined by the K-S test, leads to a value of $P_{\rm K-S}(\alpha^{\rm BAND}, \alpha^{\rm COMP})<10^{-4}$ for the $\alpha$ distributions and of $P_{\rm K-S}(E_{\rm p}^{\rm BAND}, E_{\rm p}^{\rm COMP})=1.06\times10^{-4}$ for the $E_{\rm p}$ distributions between the BAND-better spectra and the COMP-better spectra, indicating that these distributions are indeed different from one another.

With separated well-constrained and unconstrained $\beta$ categories, the $\beta$ distributions show a single peak for each category, with the best fits giving $\beta=-2.53\pm0.39$ for well-constrained $\beta$ categories and $\beta=-5.57\pm0.90$ for unconstrained $\beta$ category, respectively (the lower panel in Figure \ref{fig:distributionbeta} and Column 8 in Table \ref{table:distribution_CPL_Band}).
 
\subsection{Comparisons of Parameter Distribution for DIC-statistic-based Groups}\label{sec:dis}

\begin{figure*}
\includegraphics[angle=0,scale=0.30]{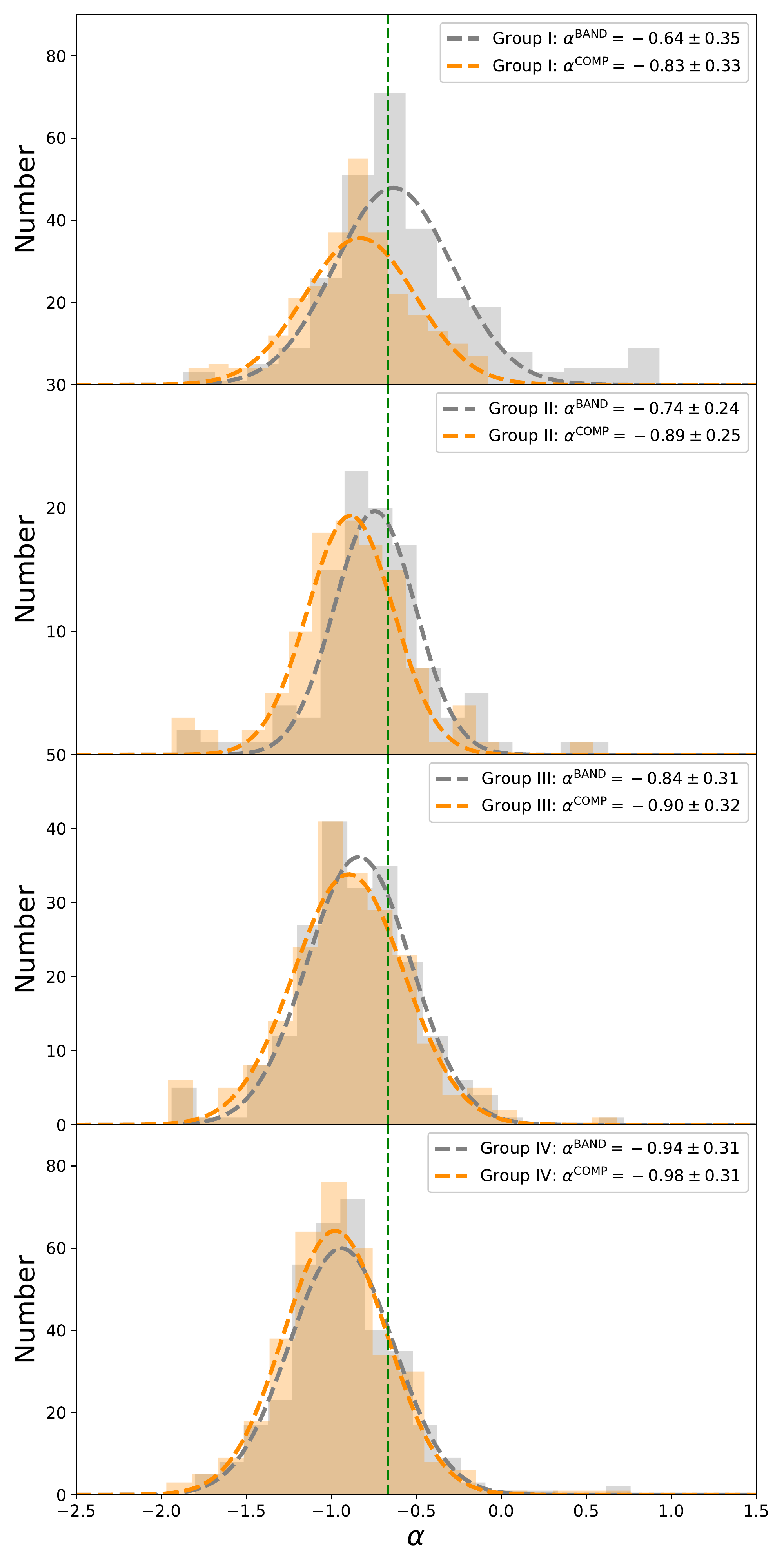}
\includegraphics[angle=0,scale=0.30]{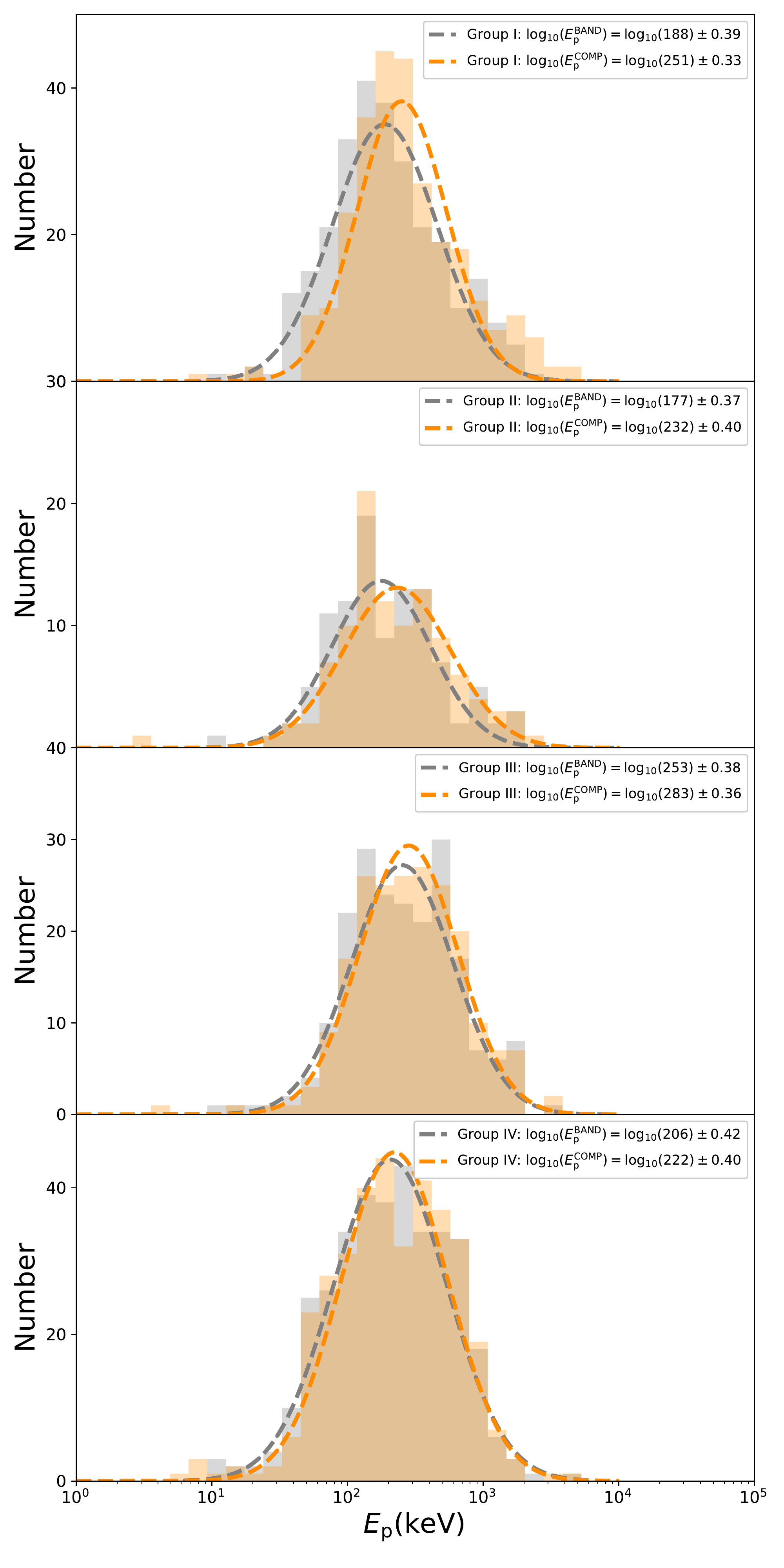}
\includegraphics[angle=0,scale=0.30]{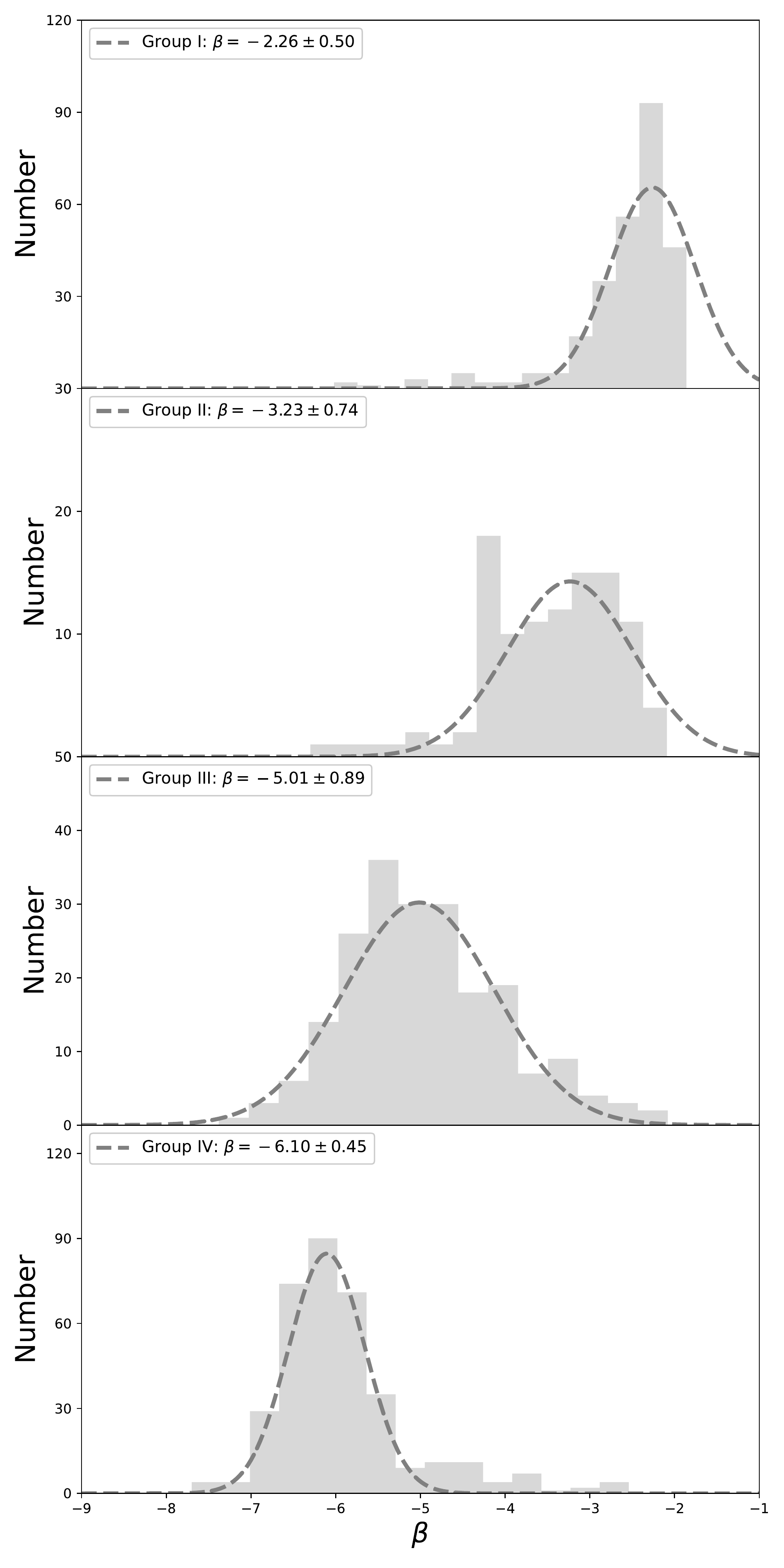}
\caption{Same as Figure \ref{fig:distributionbeta} but for the spectra grouped based on DIC statistic. The green line indicated the line-of-death for the synchrotron emission ($\alpha$=-2/3).}\label{fig:distributionDIC}
\end{figure*}

We present the distributions of $\Delta$DIC in Figure \ref{fig:DIC}. We find that Group I, Group II, Group III, and Group IV can account for 29\%, 11\%, 22\%, and 38\% of the total number of the spectra, respectively. Based on these DIC-statistic-based groups, we then present the parameter distributions ($\alpha$, $E_{\rm p}$, and $\beta$) by comparing BAND and COMP (Figure \ref{fig:distributionDIC}) group-wisely. The parameter distributions obtained from the BAND model with the best Gaussian fit are shown by grey lines and those from the COMP model are shown by orange lines.

For $\alpha$ distribution (see the left panel in Figure \ref{fig:distributionDIC} and Column 7 in Table \ref{table:distribution_CPL_Band}), we find that $\alpha$ indices obtained from BAND are significantly harder than those obtained from COMP in each group\footnote{Following the traditional classification, the ``hard" spectra are denoted as large values of both $\alpha$ and $E_{\rm p}$, while the ``soft" spectra are denoted as low values of $\alpha$ and $E_{\rm p}$.}. More interestingly, such a statistically significant difference in parameters tends to be weaker during the transition from Group I (minimum-$\Delta$DIC) to Group IV (maximum-$\Delta$DIC). 

Physically, we could diagnose the underlying physical mechanism through the distributions of $\alpha$ indices. This is because different theoretical models predict different distributions of $\alpha$. The photosphere emission models usually associate with harder $\alpha$ indices while the synchrotron emission models typically relate to softer $\alpha$ indices. As pointed out by some previous works \citep[e.g.,][]{Preece1998, Acuner2020}, the low-energy index $\alpha$ is a good estimator for which model is preferred by the data. For example, the synchrotron emission explains the spectral indices with a limit, known as the ``line-of-death", $\alpha=-\frac{2}{3}$ \citep{Preece1998}. \cite{Acuner2020} argued that the spectra that prefer the photospheric model all have low-energy power-law indices $\alpha\gtrsim$ −0.5. In Figure \ref{fig:distributionDIC}, the line-of-death of the synchrotron emission is indicated by the green lines for each group. As a result, we find that the fraction of the spectra with $\alpha$ beyond the synchrotron limit obtained from the BAND model is apparently greater than those obtained from the COMP model. Moreover, these fractions decrease for subsequent DIC-based groups. Groupwise, the corresponding fractions [BAND, COMP] are [49\%, 25\%], [34\%, 25\%], [30\%, 25\%], and [19\%, 17\%] for Group I, Group II, Group III, and Group IV, respectively. We also notice that the distribution of $\alpha$ has a smooth and well-defined Gaussian shape of at the ``line-of-death", challenging the existence of the ``line-of-death".

For $E_{\rm p}$ distributions (the middle panel of Figure \ref{fig:distributionDIC} and Column 9 in Table \ref{table:distribution_CPL_Band}), unlike $\alpha$ distribution, we do not find a strong trend among the groups. Interestingly, the statistical significant difference in parameters tends to be weaker for subsequent DIC-based groups, resembling to the finding in the $\alpha$ distribution.

We present the groupwise $\beta$ distributions in the right panel of Figure \ref{fig:distributionDIC}. Using the same data, \cite{Li2021b} found a similar bimodal distribution based on the better model determined for each individual pulse, with the harder peak at $\sim$ -2.3 and the softer peak at $\sim$ -6.1. The results indicate that the BAND-better and COMP-better spectra should be mixed to compose the distributions, and the harder peak should be contributed by the BAND-better spectra while the softer peak should be contributed by the COMP-better spectra.

Separated by the DIC statistics, we group the spectra into four groups as defined in \S \ref{sec:group}. Interestingly, we find that all of the groups show a single peak, but the peak is clearly shifted from Group I to Group IV with a hard-to-soft trend. The hardest peak is at $\sim$ -2.3 found in the Group I (minimum-$\Delta$DIC) (Figure \ref{fig:distributionDIC}). This value is the same as the harder peak of the bimodal distribution found in the pulse-wise categories \citep{Li2021b}, implying that the peak is dominated by the BAND-like spectra. Likewise, the softest peak is at $\sim$ -6.1 (Figure \ref{fig:distributionDIC}), which is found in Group IV (maximum-$\Delta$DIC). This value is the same as the softer peak of the bimodal distribution found in \cite{Li2021b}, suggesting the peak is dominated by the COMP-like spectra. However, the peak (Figure \ref{fig:distributionDIC}) for Group II is $\beta=-3.23\pm0.74$ while that for Group III is $\beta=-5.01\pm0.89$, suggesting a mix of Band-like and COMP-like spectra. 

\subsection{Comparisons of Parameter Relations}

In GRB physics, the study of parameter correlations is an open question, and it plays an important role in understanding the underlying physical processes and radiation mechanisms\cite[e.g.,][and references therein]{Amati2002,Geng2013,Srinivasaragavan2020,Li2021b}.

\begin{figure*}
\includegraphics[angle=0,scale=0.30]{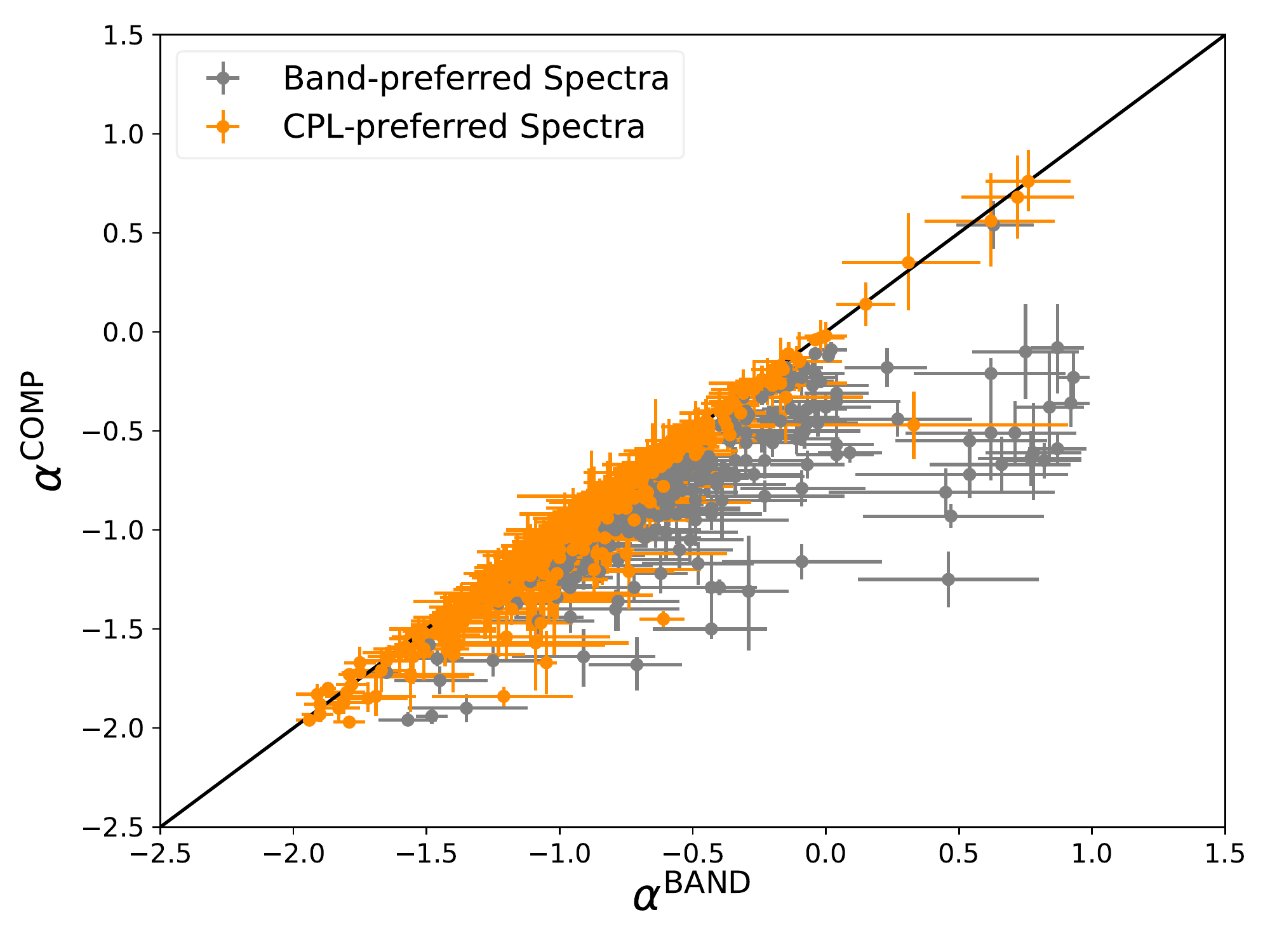}
\includegraphics[angle=0,scale=0.30]{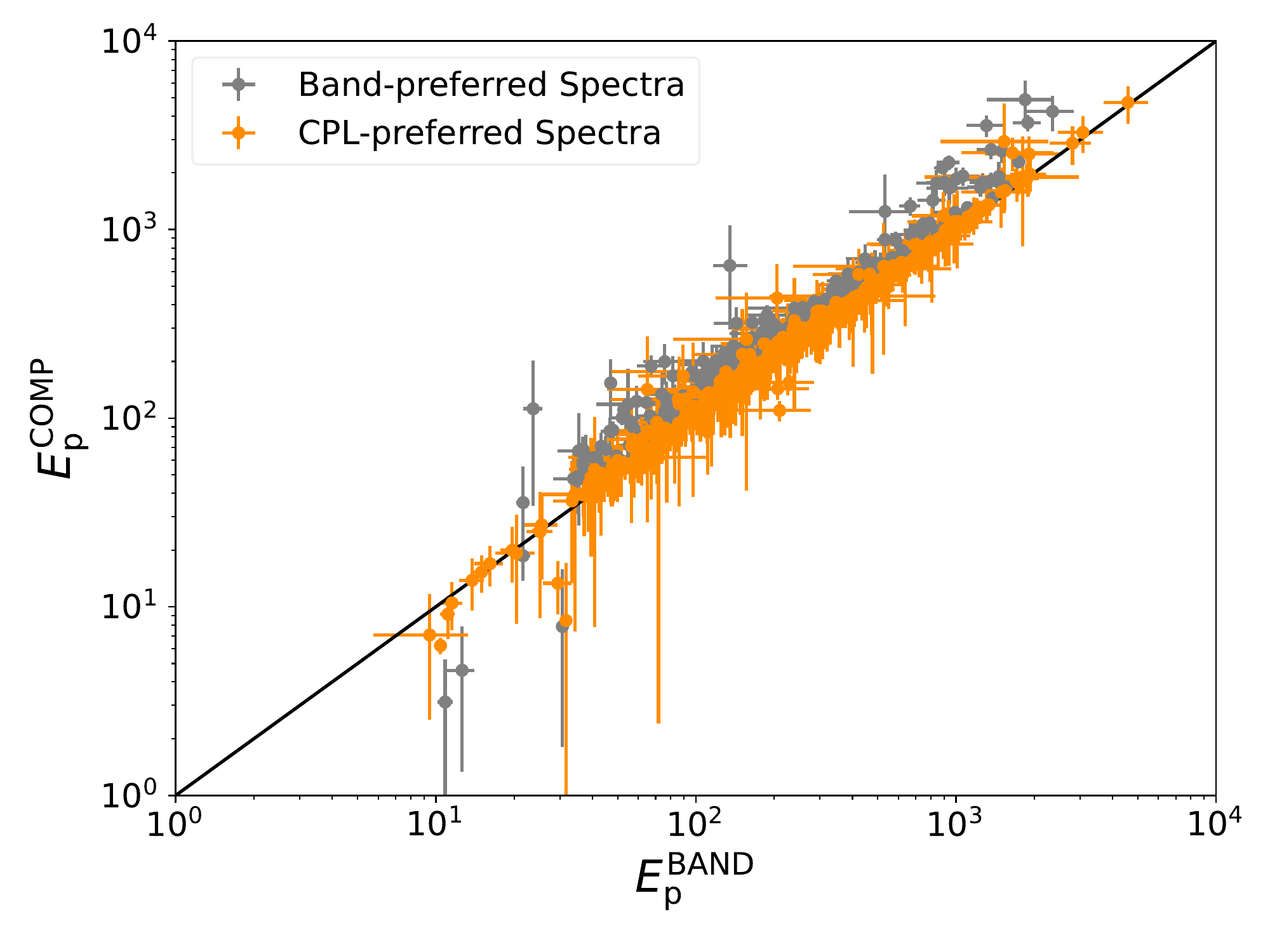}
\includegraphics[angle=0,scale=0.30]{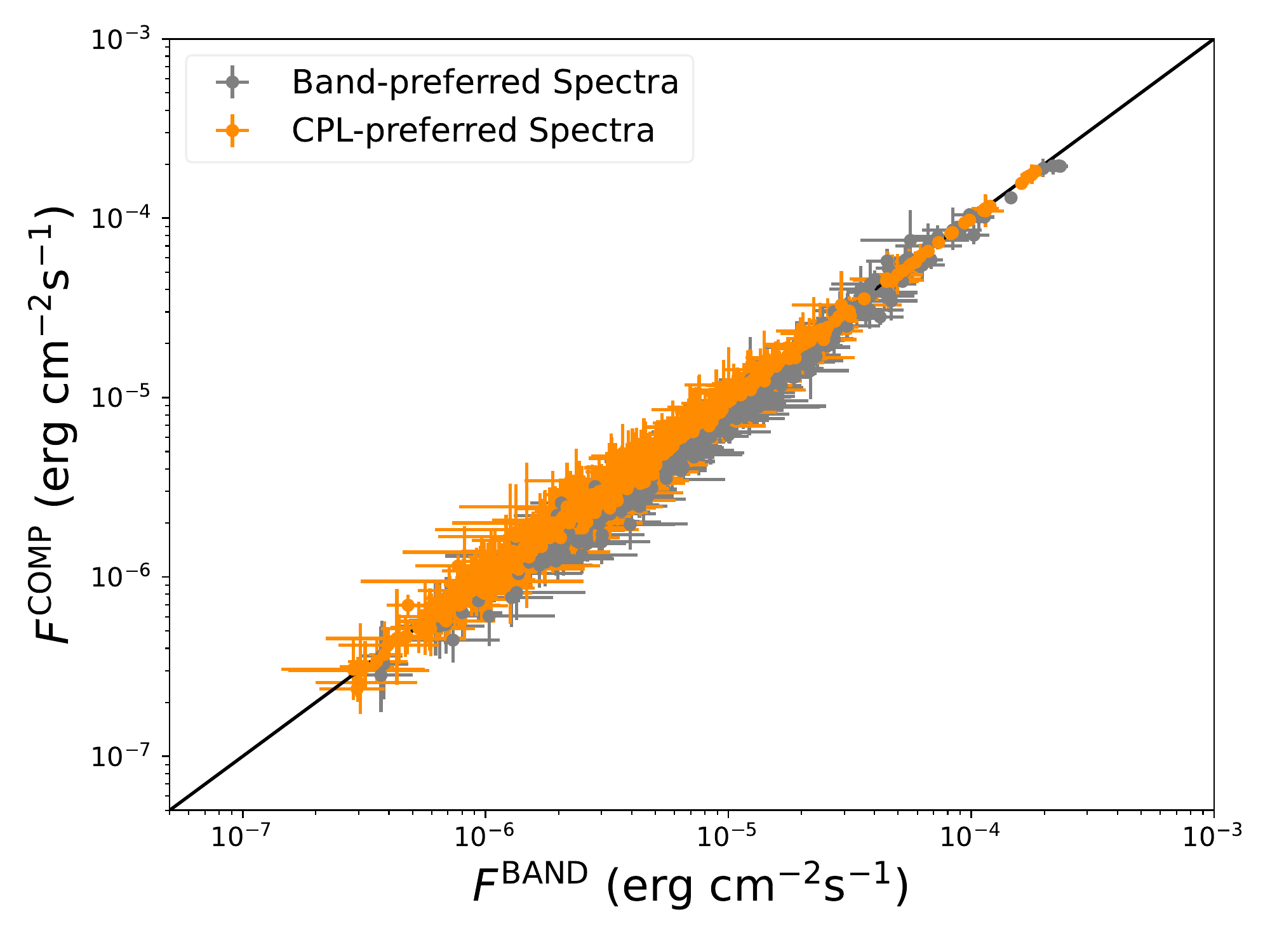}
\includegraphics[angle=0,scale=0.30]{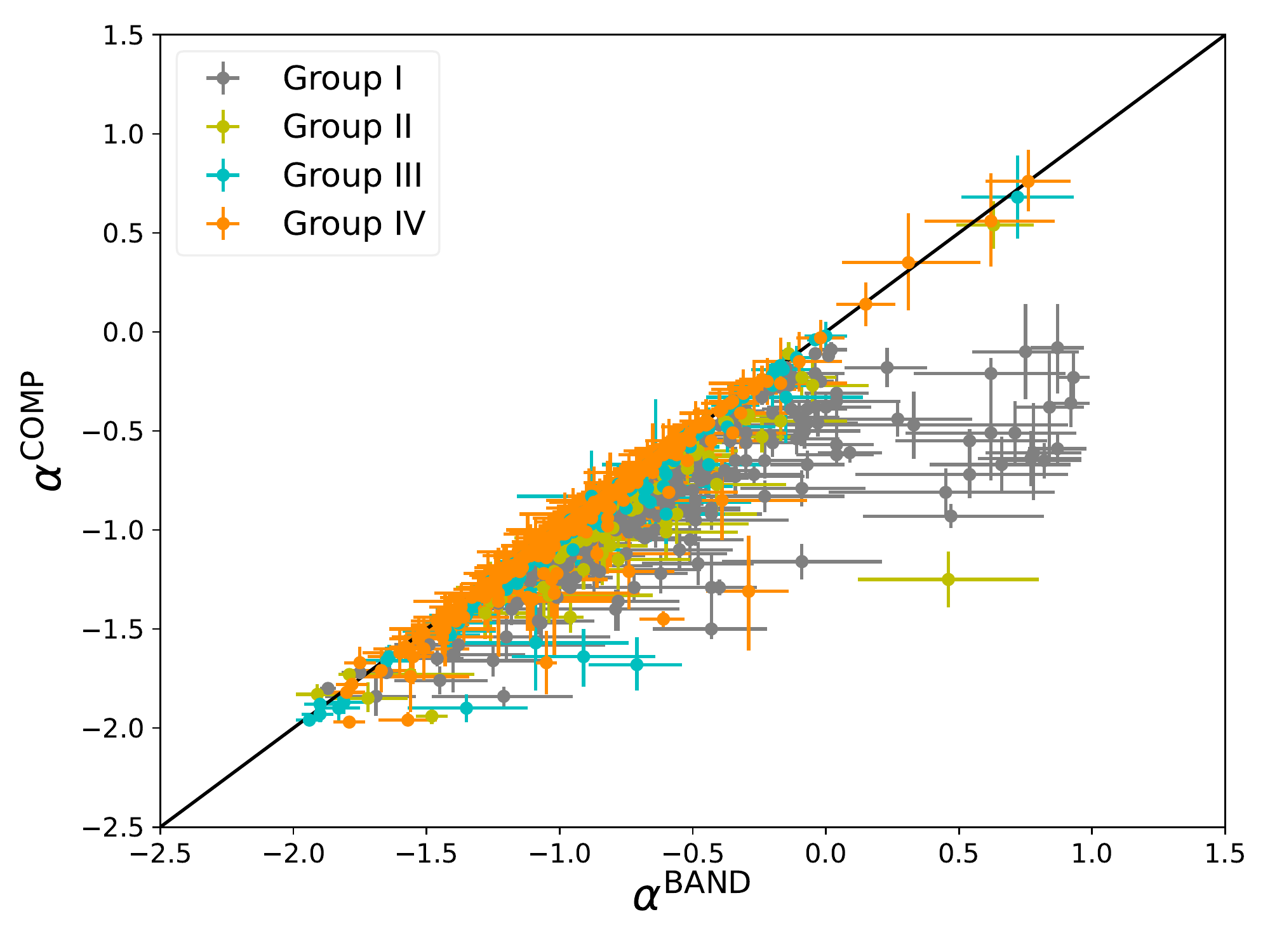}
\includegraphics[angle=0,scale=0.30]{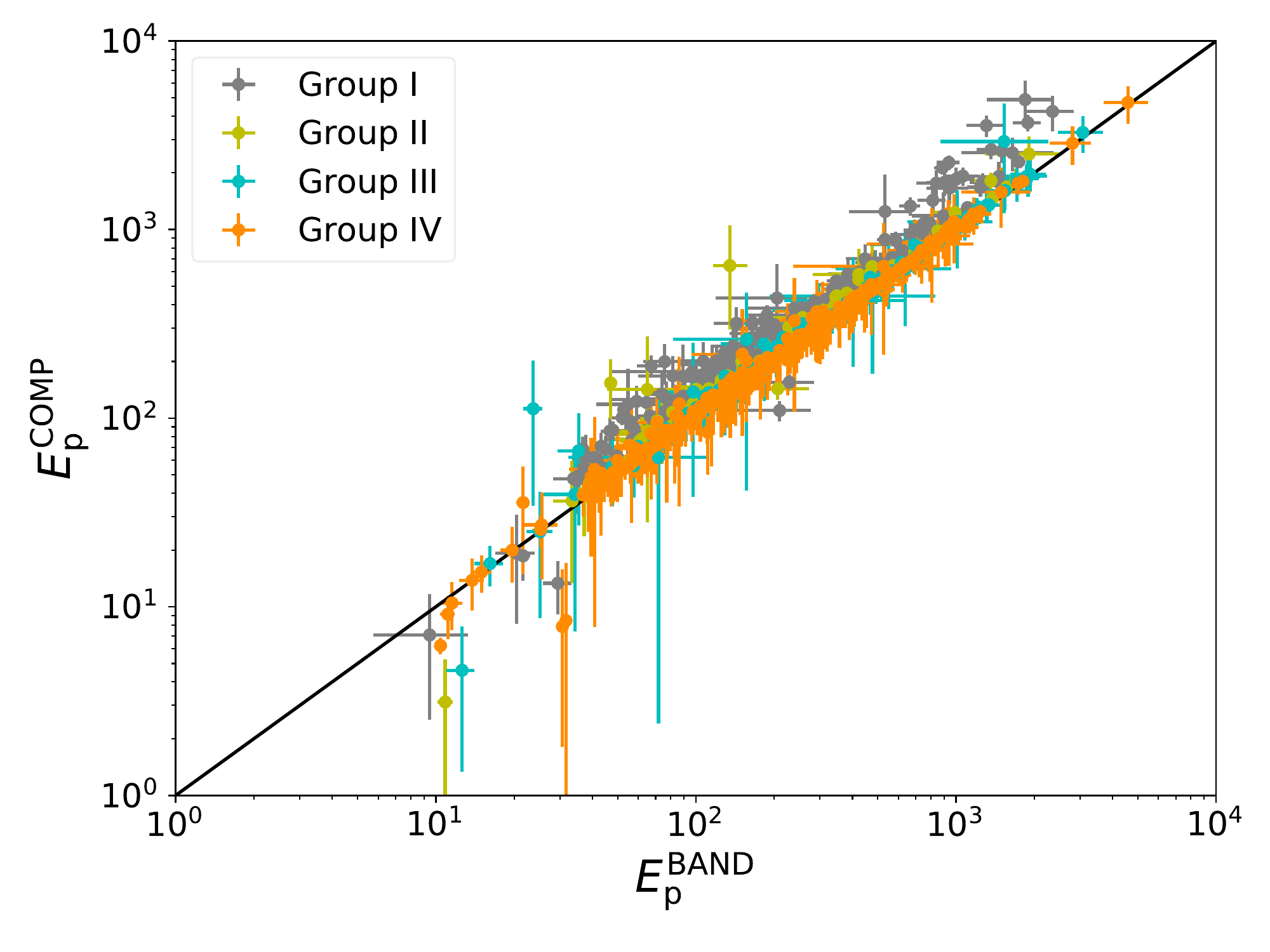}
\includegraphics[angle=0,scale=0.30]{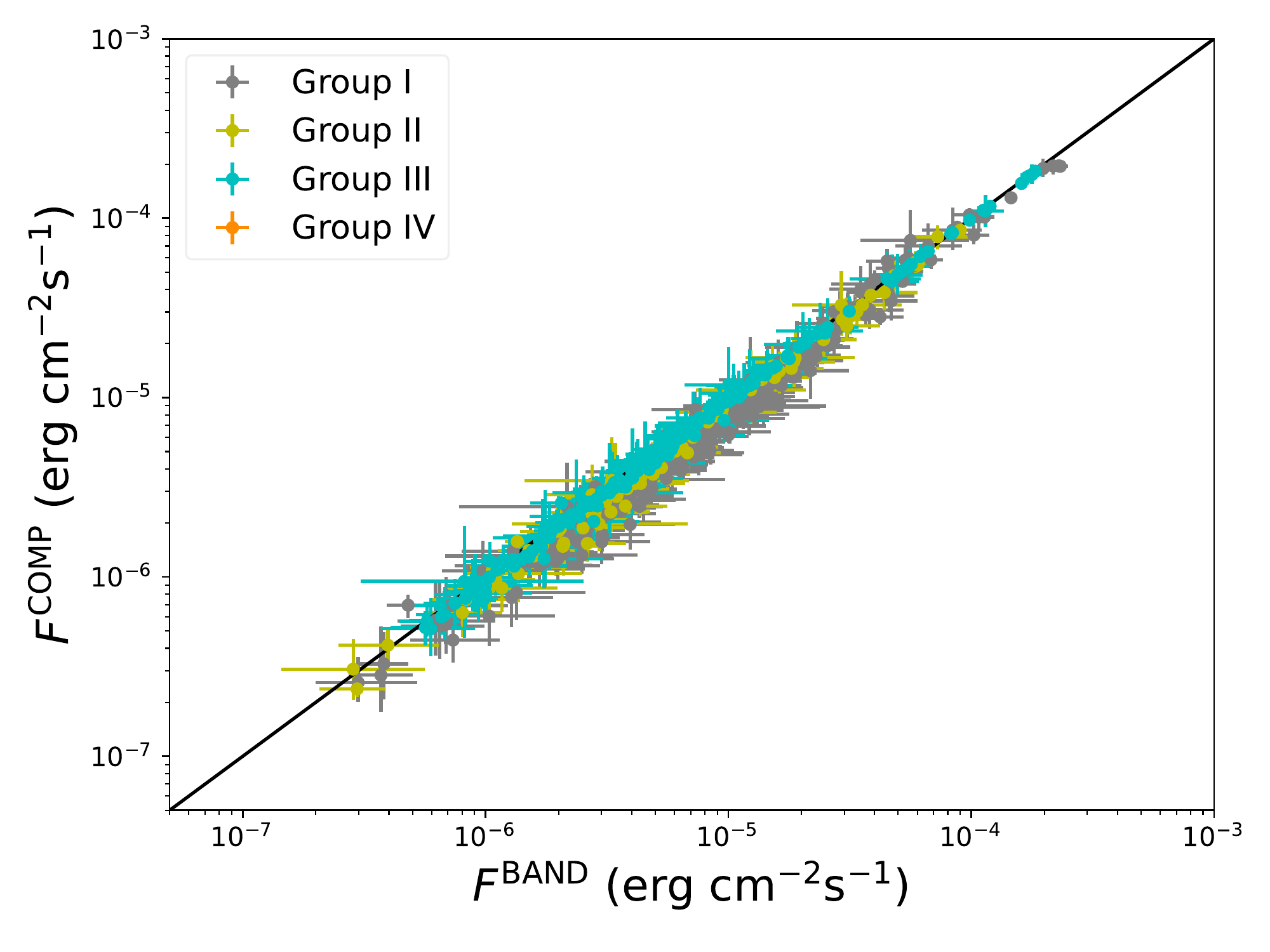}
\caption{Comparison of the same spectral parameters between BAND and COMP: the $\alpha^{\rm BAND}$-$\alpha^{\rm COMP}$ (left panel), $E^{\rm BAND}_{\rm p}$-$E^{\rm COMP}_{\rm p}$ (middle panel), and $F^{\rm BAND}$-$F^{\rm COMP}$ (right panel) plottings. Upper panels: for the $\beta$-based categories. Lower panels: for the DIC-based groups.}\label{fig:relation1}
\end{figure*}

We first compare the same spectral parameters between two models by plotting $\alpha^{\rm COMP}$--$\alpha^{\rm BAND}$ (the left panels of Figure \ref{fig:relation1}), $E^{\rm COMP}_{\rm p}$--$E^{\rm BAND}_{\rm p}$ (the middle panels of Figure \ref{fig:relation1}), and $F^{\rm COMP}_{\rm p}$--$F^{\rm BAND}_{\rm p}$ (the right panels of Figure \ref{fig:relation1}). We find that the $\alpha$ indices obtained from the BAND model are systematically harder than the ones obtained from the COMP model, particularly in the BAND-better spectra. However, this trend is weaker in the BAND-wise spectra as compared to the COMP-wise spectra, which is consistent with the finding based on parameter distributions as discussed in \S \ref{sec:dis}. Similar results are also found in the $E^{\rm COMP}_{\rm p}$--$E^{\rm BAND}_{\rm p}$ plot. An interesting result is found in the $F^{\rm COMP}_{\rm p}$--$F^{\rm BAND}_{\rm p}$ plot, where the energy flux obtained from BAND and COMP is similar, crossing different categories and groups. 

\begin{figure*}
\includegraphics[angle=0,scale=0.30]{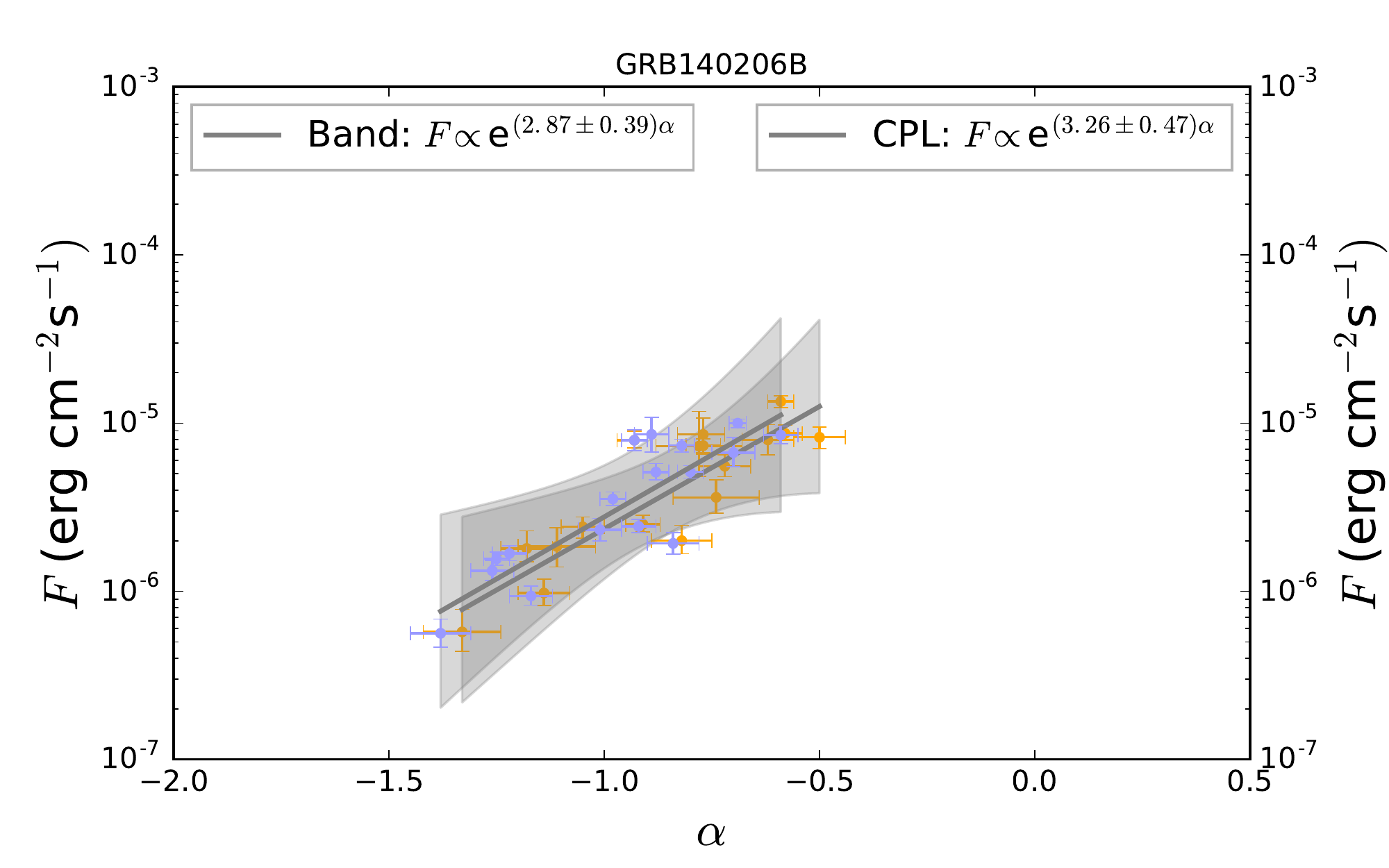}
\includegraphics[angle=0,scale=0.30]{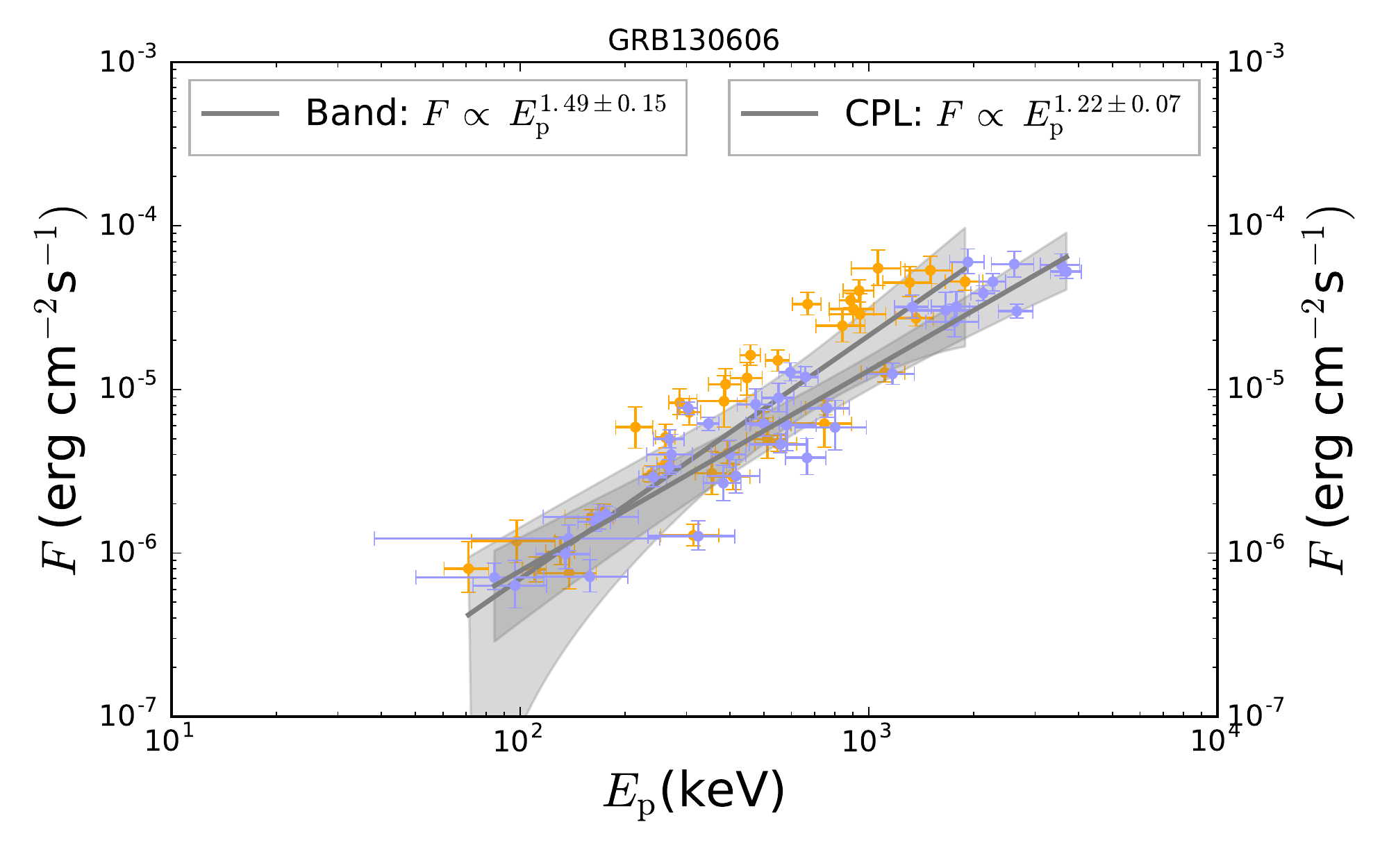}
\includegraphics[angle=0,scale=0.30]{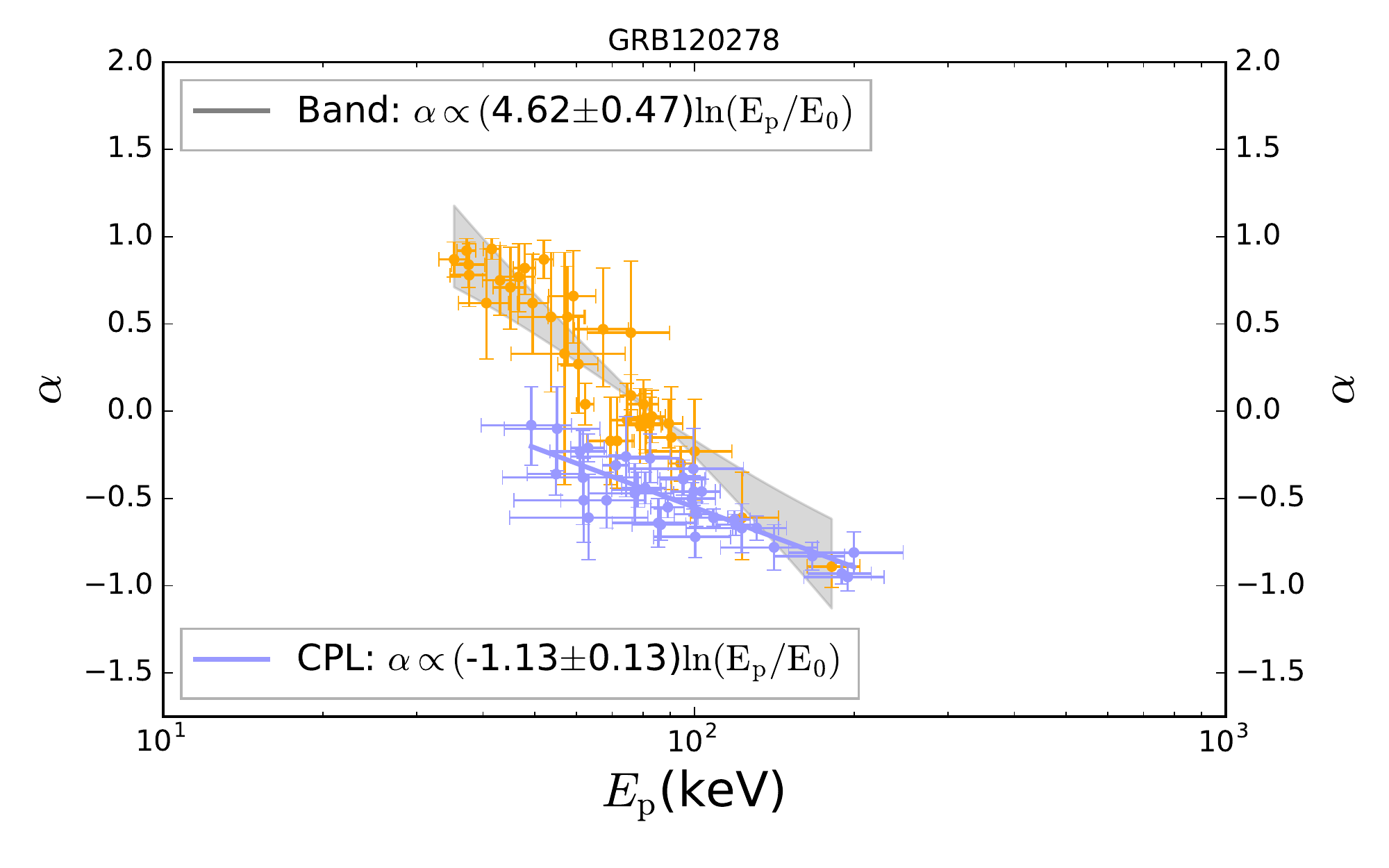}
\caption{The parameter relation of the $F$-$\alpha$ (left panel), $F$-$E_{\rm p}$ (middle panel), and $\alpha$-$E_{\rm p}$ (right panel), as well as the best-fit relations with the 2$\sigma$ error region.}\label{fig:relation2}
\end{figure*}

It is even more interesting to see how these parameter relations are affected by the misused models. Based on the categories defined in \S \ref{sec:betamodel}, we therefore investigate the following pair parameter relations comparing BAND with COMP: ($\rm{log}$$F$, $\alpha$), ($\rm{log}$$F$, $\rm{log}$$E_{\rm p}$), ($\alpha$, $\rm{log}$$E_{\rm p}$). For each individual parameter relation, in order to ensure that the majority of spectra are BAND-like, we select three typical bursts (GRB 140206B for the $F-\alpha$ relation; GRB 130306B for the $F-E_{\rm p}$ relation; and GRB 120827 for the $\alpha-E_{\rm p}$ relation), where the vast majority of spectra in these bursts satisfy $\Delta \rm DIC<$-10 (seven out of 10 from GRB 140206B, 13 out of 16 from GRB 130306B, and 30 out of 36 from GRB 120827). We use the following function to fit the data: $F = F_0 \, e^{k_1 \alpha}$ for the $F-\alpha$ plot; $F = F_0 \, E_{\rm p}^{k_2}$ for the $F-E_{\rm p}$ plot; and $\alpha = k_3 \ln (E_{\rm p}/E_0) + \alpha_0$ for the $\alpha-E_{\rm p}$ plane.  The time-resolved $F$-$\alpha$ relation \citep[e.g.,][]{Ryde2019,Li2021b}, $F-E_{\rm p}$ relation \citep{Golenetskii1983}, and $\alpha-E_{\rm p}$ relation \citep[e.g.,][]{Li2021b} are presented in the left, middle, and right panels in Figure \ref{fig:relation2}, respectively. The results of our linear regression analysis for these parameter relations comparing BAND with COMP are reported in Table \ref{tab:linear}. We find that $k^{\rm Band}_{1} \sim 2.87 \pm 0.39$ is significantly shallower than $k^{\rm CPL}_{1} \sim 3.26 \pm 0.47$, whereas $k^{\rm Band}_{2} \sim 1.49 \pm 0.15$ is clearly steeper than $k^{\rm CPL}_{2} \sim 1.22 \pm 0.07$, and likewise, $k^{\rm Band}_{3} \sim -4.62 \pm 0.47$ is apparently steeper than $k^{\rm CPL}_{3} \sim -1.14 \pm 0.13$.

\section{Discussion} \label{sec:discussion}

BAND function and COMP are preferred respectively by a given group of GRBs, so a question is raised as to whether these two empirical functions have different physical origins? Or, is COMP just an approximation of BAND as demonstrated in Figure \ref{fig:spectrum} when $\beta << 0$? We may find some clues in Figure \ref{fig:distributionbeta}, of which the histogram of $\beta$ does not form the shape of the unimodal distribution; instead, it displays a clear bimodal structure which peaks at $\beta\simeq-2.5$ and $\beta\simeq-6$ respectively. Moreover, the two modes are separated at $\beta\simeq-3.5$, and each one is almost independently contributed by COMP-preferred spectra or BAND-preferred spectra. This separation is hardly due to that COMP is preferred by noisy data, because first the separation is distinct, and second all the spectra have $S>20$ which is high enough to ensure data quality\footnote{We also tested the spectra with $S>50$, the bimodal structure of $\beta$ distribution clearly exists as well.}. Figure \ref{fig:distributionbeta} also exhibits the histograms of $\alpha$ and $E_{\rm p}$, for which the distributions of COMP and BAND preferred GRBs differ from each other, though not as distinguishable as the distribution of $\beta$. This statistical result infers that BAND and COMP may have different physical origins. One may propose that there exist two different mechanisms of prompt emission. One produces spectra consists of many power laws, for e.g., synchrotron emission of charged particles accelerated by kinetic shock waves \cite[e.g.,][]{Sari1998}; the other produces spectra of a power law with an exponential tail, for e.g., the convolution of blackbody spectra in photospheres, the cut of the highest temperature corresponds to the exponential tail \cite[e.g.,][]{Ryde2010}. 

\section{Conclusions} \label{sec:conclusion}

In this paper, we have revisited the catalog of the time-resolved spectrum of the multi-pulse {\it Fermi}-GBM bursts defined in \cite{Li2021b}. We used two methods to determine the better (or statistically preferred) spectra between two standard empirical spectral functions: BAND and COMP. Firstly, we grouped the spectra into the well-constrained $\beta$ (BAND-better) and unconstrained $\beta$ (COMP-better) categories by checking their two-dimensional corner-corner plots of the posteriors for each Bayesian MCMC spectral fit. Secondly, we also separated the spectra into four groups based on the DIC statistics: Group I with $\Delta \rm DIC<$-10 strongly suggests that BAND spectra are statistically preferred; Group II with -10$<\Delta \rm DIC<$-5 indicates that BAND spectra are still statistically preferred, but not as strong as Group I; Group III with -5$<\Delta \rm DIC<$0, indicating COMP spectra are statistically preferred; and Group IV with $\Delta \rm DIC>$0 significantly indicates that the COMP spectra are statistically preferred. 

With these categories and groups defended, we therefore compared the spectral properties obtained by both BAND and COMP functions, including their spectral distributions, spectral relations, and spectral evolution. 

In the categories defined by identifying well-constrained $\beta$ and unconstrained $\beta$, we found inconsistent peaks of the parameter distributions (both $\alpha$ and $E_{\rm p}$) showing between the BAND-better and COMP-better spectra. 

These results were also independently confirmed by an analysis based on the DIC statistics. Moreover, such a statistical difference in parameters tends to be weaker when transitioning from Group I (minimum-$\Delta$DIC) to Group IV (maximum-$\Delta$DIC). The BAND-$\beta$ distributions show a single peak for all the DIC-statistics-based groups, and the peaks obtained from the Group I and Group IV samples are the same as the harder and softer peaks found in \cite{Li2021b}, suggesting that these peaks are more likely dominated by the BAND-like spectra and COMP-like spectra, respectively.

We also discussed the effect of the misused model on the results for each category and group. We found that the apparent deviation from the parameters is found between the BAND-to-COMP cases, while the parameters between the COMP-to-BAND cases all seem to agree.

As a self-consistency test, we also compare the same spectral parameters between the BAND-better and COMP-better categories and among the DIC-statistic-based groups by investigating the $\alpha^{\rm COMP}$--$\alpha^{\rm BAND}$, $E^{\rm COMP}_{\rm p}$--$E^{\rm BAND}_{\rm p}$, and $F^{\rm COMP}_{\rm p}$--$F^{\rm BAND}_{\rm p}$ plotting. The greater dispersion for data points is still found between the BAND-to-COMP cases. We further investigated the $F-\alpha$, $F-E_{\rm p}$, and $\alpha-E_{\rm p}$ relations for such the misused case using three example cases. The obtained power-law index (slope) between the misused model (COMP) and the better model (BAND) are significantly different. The index ($k^{\rm COMP}$) derived from the ``misused" model is shallower ($F-E_{\rm p}$ relation and $\alpha-E_{\rm p}$ relation) and steeper ($F$-$\alpha$ relation) than that ($k^{\rm BAND}$) derived from the better model.

We also discussed the bimodal distribution of $\beta$, which indicates that BAND and COMP may have different physical origins.

In conclusion, our analysis suggests that the choice between BAND and COMP spectral model for the GRB spectral analysis should be made with caution. The fit from the misused model deviates from the real spectral shape and then may lead to incorrect physical interpretation.

\acknowledgments

I thank the anonymous referee for valuable comments and suggestions. I also thank Felix Ryde, M.G. Dainotti, Gregory Vereshchagin, Remo Ruffini, and ICRANet members for many discussions on GRB physics and phenomena. I particularly thank to Yu Wang for many useful discussions that greatly improved the paper. This research has made use of the High Energy Astrophysics Science Archive Research Center (HEASARC) Online Service at the NASA/Goddard Space Flight Center (GSFC).

\bibliography{Myreferences.bib}

\clearpage
\begin{deluxetable}{cccccccccc}
\tablewidth{0pt}
\tabletypesize{\tiny}
\tablecaption{Results of the Average and Deviation Values of the Parameter Distribution}\label{tab:Gauss}
\tablehead{
\colhead{Spectra}
&\colhead{Classified by}
&\colhead{Model}
&\colhead{Model}
&\colhead{Model}
&\colhead{Model}
&\colhead{Spectra}
&\colhead{$\alpha$}
&\colhead{$\beta$}
&\colhead{$E_{\rm p}$}\\
(Categorized)&&(Preferred)&(Used)&(Identified)&(Defined)&(Number)&&&(keV)
}
\colnumbers
\startdata
\hline
Overall&\nodata&...&BAND&\nodata&\nodata&944&-0.82$\pm$0.34&\nodata&$\rm log_{10}$(213)$\pm$0.41\\
Overall&\nodata&...&COMP&\nodata&\nodata&944&-0.91$\pm$0.33&\nodata&$\rm log_{10}$(239)$\pm$0.40\\
\hline
BAND preferred&Well-constrained $\beta$&BAND&BAND&Better&BAND-to-BAND&327&-0.64$\pm$0.28&-2.53$\pm$0.39&$\rm log_{10}$(191)$\pm$0.41\\
COMP preferred&Unconstrained $\beta$&COMP&COMP&Better&COMP-to-COMP&617&-0.96$\pm$0.33&...&$\rm log_{10}$(249)$\pm$0.40\\
BAND preferred&Well-constrained $\beta$&BAND&COMP&Misused&BAND-to-COMP&327&-0.82$\pm$0.29&...&$\rm log_{10}$(252)$\pm$0.38\\
COMP preferred&Unconstrained $\beta$&COMP&BAND&Misused&COMP-to-BAND&617&-0.91$\pm$0.31&-5.57$\pm$0.90&$\rm log_{10}$(225)$\pm$0.42\\
\hline
Group I&$\Delta \rm DIC<=$-10&BAND&BAND&Statistically preferred&BAND-to-BAND&272&-0.64$\pm$0.35&-2.26$\pm$0.50&$\rm log_{10}$(188)$\pm$0.39\\
Group I&$\Delta \rm DIC<=$-10&BAND&COMP&Misused&BAND-to-COMP&272&-0.83$\pm$0.33&\nodata&$\rm log_{10}$(251)$\pm$0.33\\
Group II&-10$<\Delta \rm DIC<=$-5&BAND&BAND&Statistically preferred&BAND-to-BAND&106&-0.74$\pm$0.24&-3.23$\pm$0.74&$\rm log_{10}$(177)$\pm$0.37\\
Group II&-10$<\Delta \rm DIC<=$-5&BAND&COMP&Misused&BAND-to-COMP&106&-0.89$\pm$0.25&\nodata&$\rm log_{10}$(232)$\pm$0.40\\
Group III&-5$<\Delta \rm DIC<=$0&COMP&BAND&Misused&COMP-to-BAND&208&-0.84$\pm$0.31&-5.01$\pm$0.89&$\rm log_{10}$(253)$\pm$0.38\\
Group III&-5$<\Delta \rm DIC<=$0&COMP&COMP&Statistically preferred&COMP-to-COMP&208&-0.90$\pm$0.32&\nodata&$\rm log_{10}$(283)$\pm$0.36\\
Group IV&$\Delta \rm DIC>$0&COMP&BAND&Misused&COMP-to-BAND&358&-0.94$\pm$0.31&-6.10$\pm$0.45&$\rm log_{10}$(206)$\pm$0.42\\
Group IV&$\Delta \rm DIC>$0&COMP&COMP&Statistically preferred&COMP-to-COMP&358&-0.98$\pm$0.31&\nodata&$\rm log_{10}$(222)$\pm$0.40\\
\hline
\enddata 
\vspace{1mm}
\tablecomments{Column (1) lists the spectral categories and groups, Column (2) lists our criteria to select better  or statistically preferred spectra, Columns (3)-(6) list the preferred model, the used model, and the identified model (better, or statistically preferred, or misused), and the defined model for each category and group, Column (7) lists the number of the spectra, Columns (8)-(10) list the average and its deviation (1$\sigma$) for the spectral parameters. Note that the global properties of these spectral parameters are displayed in the top panel of this table (see also in Table A1 of \citealt{Li2021b})}.
\end{deluxetable}\label{table:distribution_CPL_Band}

\clearpage
\begin{deluxetable*}{ccccccccc}
\tablewidth{0pt}
\tabletypesize{\scriptsize}
\tablecaption{Results of our Linear Regression Analysis for Parameter Relations}\label{tab:linear}
\tablehead{
\colhead{Relation}&
\colhead{Model}&
\colhead{Expression}&
\colhead{$N$}&
\colhead{$R$}&
\colhead{$p$}&
}
\startdata
$F$-$\alpha$&BAND&$F$/(erg cm$^{-2}$s$^{-1}$)=(9.61$\pm$0.42)e$^{(2.87\pm0.39)\alpha}$&10&0.86&$<$10$^{-4}$\\
$F$-$\alpha$&COMP&$F$/(erg cm$^{-2}$s$^{-1}$)=(9.40$\pm$0.49)e$^{(3.26\pm0.47)\alpha}$&10&0.82&$<$10$^{-4}$\\
$F$-$E_{\rm p}$&BAND&$F$/(erg cm$^{-2}$s$^{-1}$)=(9.23$\pm$1.06)e-6$\times$($E_{\rm p}$/keV)$^{(1.49\pm0.15)}$&16&0.88&$<$10$^{-4}$\\
$F$-$E_{\rm p}$&COMP&$F$/(erg cm$^{-2}$s$^{-1}$)=(1.19$\pm$0.08)e-5$\times$($E_{\rm p}$/keV)$^{(1.22\pm0.07)}$&16&0.92&$<$10$^{-4}$\\
$\alpha$-$E_{\rm p}$&BAND&$\alpha$=(-4.62$\pm$0.47)ln($E_{\rm p}$/$E_{0}$)+(4.90$\pm$0.47)$\times$($E_{\rm p}$/keV)&36&-0.91&$<$10$^{-4}$\\
$\alpha$-$E_{\rm p}$&COMP&$\alpha$=(-1.14$\pm$0.13)ln($E_{\rm p}$/$E_{0}$)+(1.72$\pm$0.26)$\times$($E_{\rm p}$/keV)&36&-0.79&$<$10$^{-4}$\\
\enddata
\end{deluxetable*}

\clearpage

\clearpage

\clearpage

\clearpage

\clearpage

\clearpage

\clearpage

\clearpage

\clearpage

\clearpage

\clearpage
\vspace{35mm}
\appendix
\setcounter{figure}{0}    
\setcounter{section}{0}
\setcounter{table}{0}
\renewcommand{\thesection}{A\arabic{section}}
\renewcommand{\thefigure}{A\arabic{figure}}
\renewcommand{\thetable}{A\arabic{table}}
\renewcommand{\theequation}{A\arabic{equation}}

\begin{figure*}
\includegraphics[angle=0,scale=0.3]{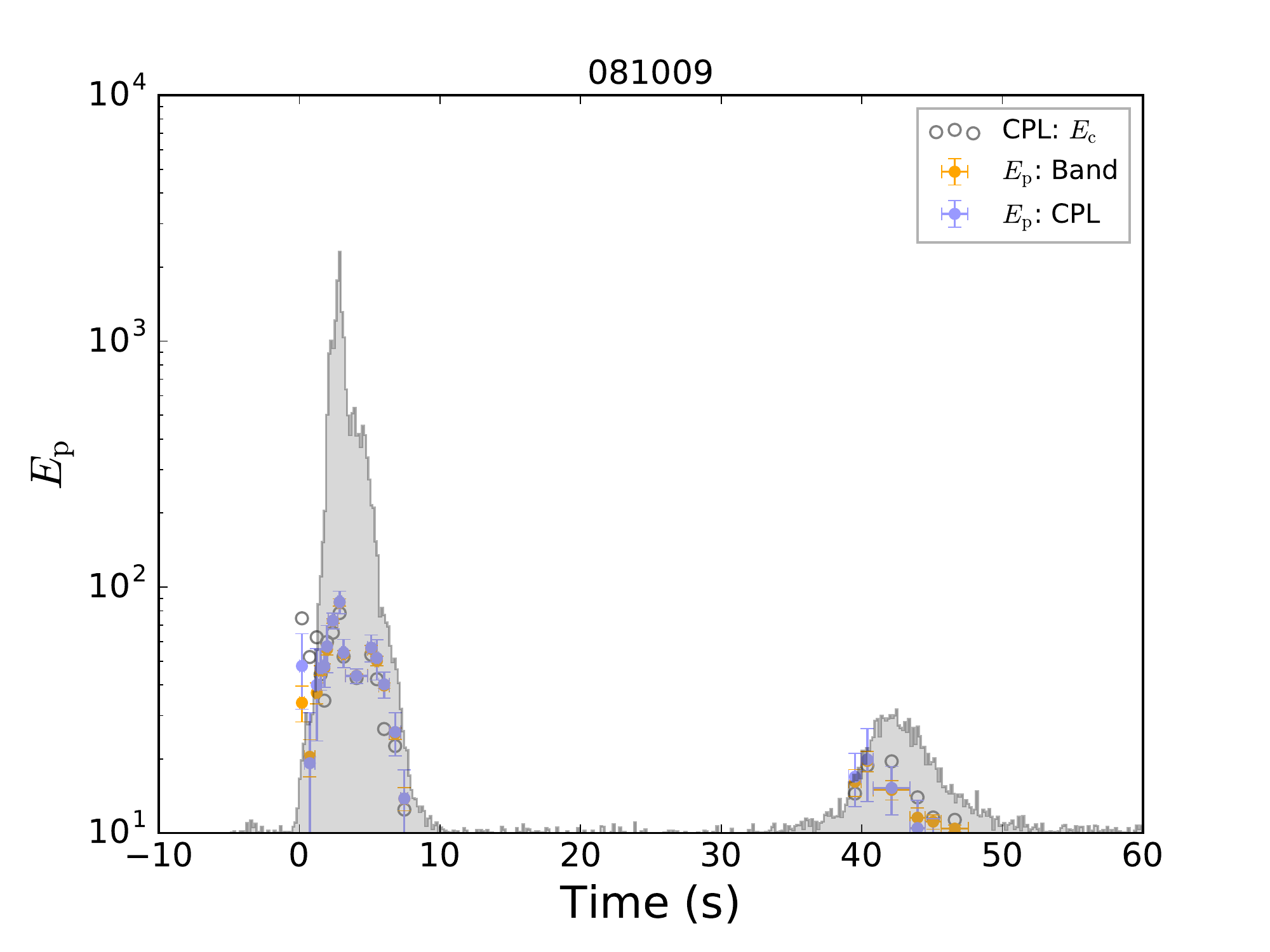}
\includegraphics[angle=0,scale=0.3]{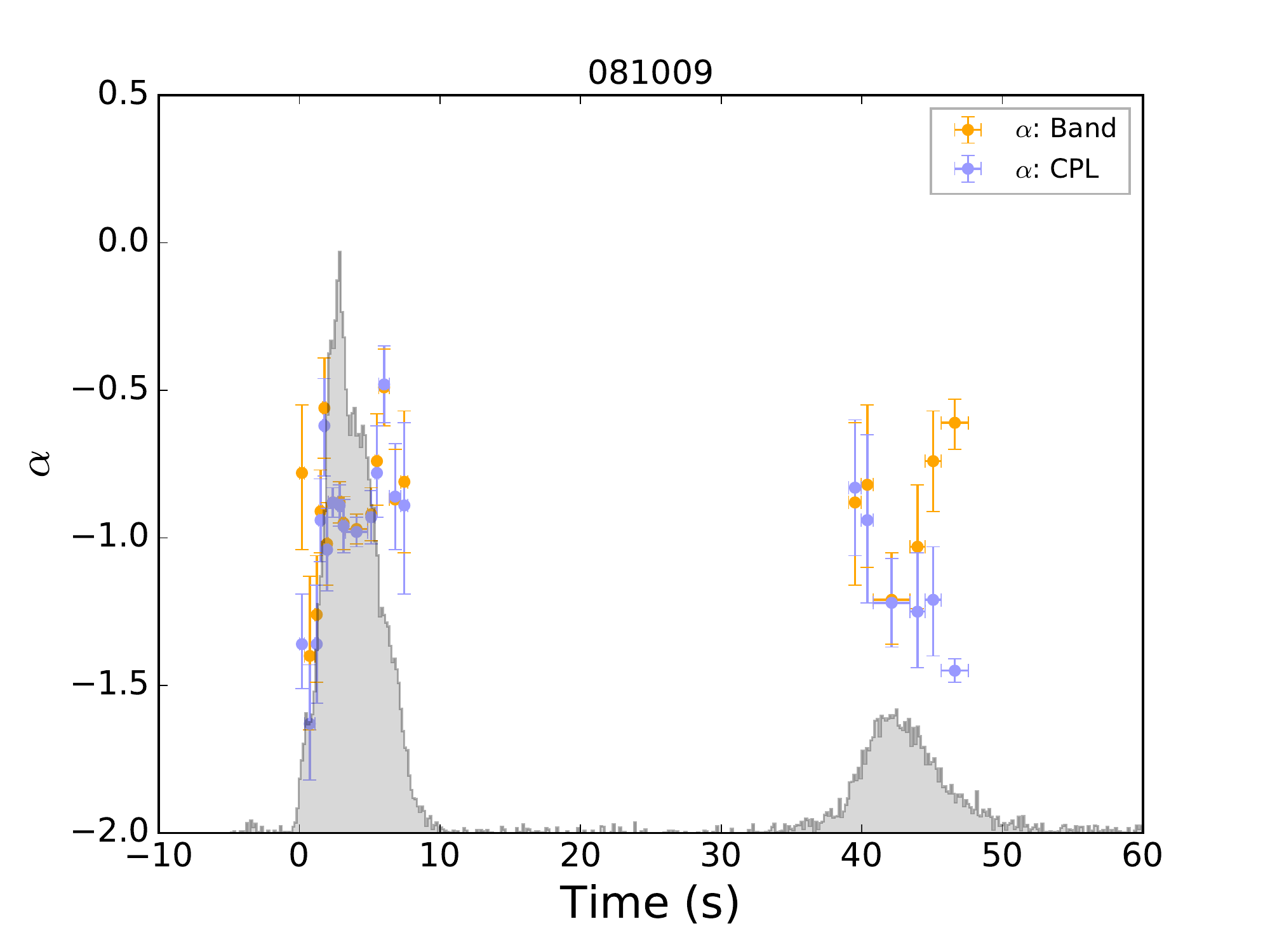}
\includegraphics[angle=0,scale=0.3]{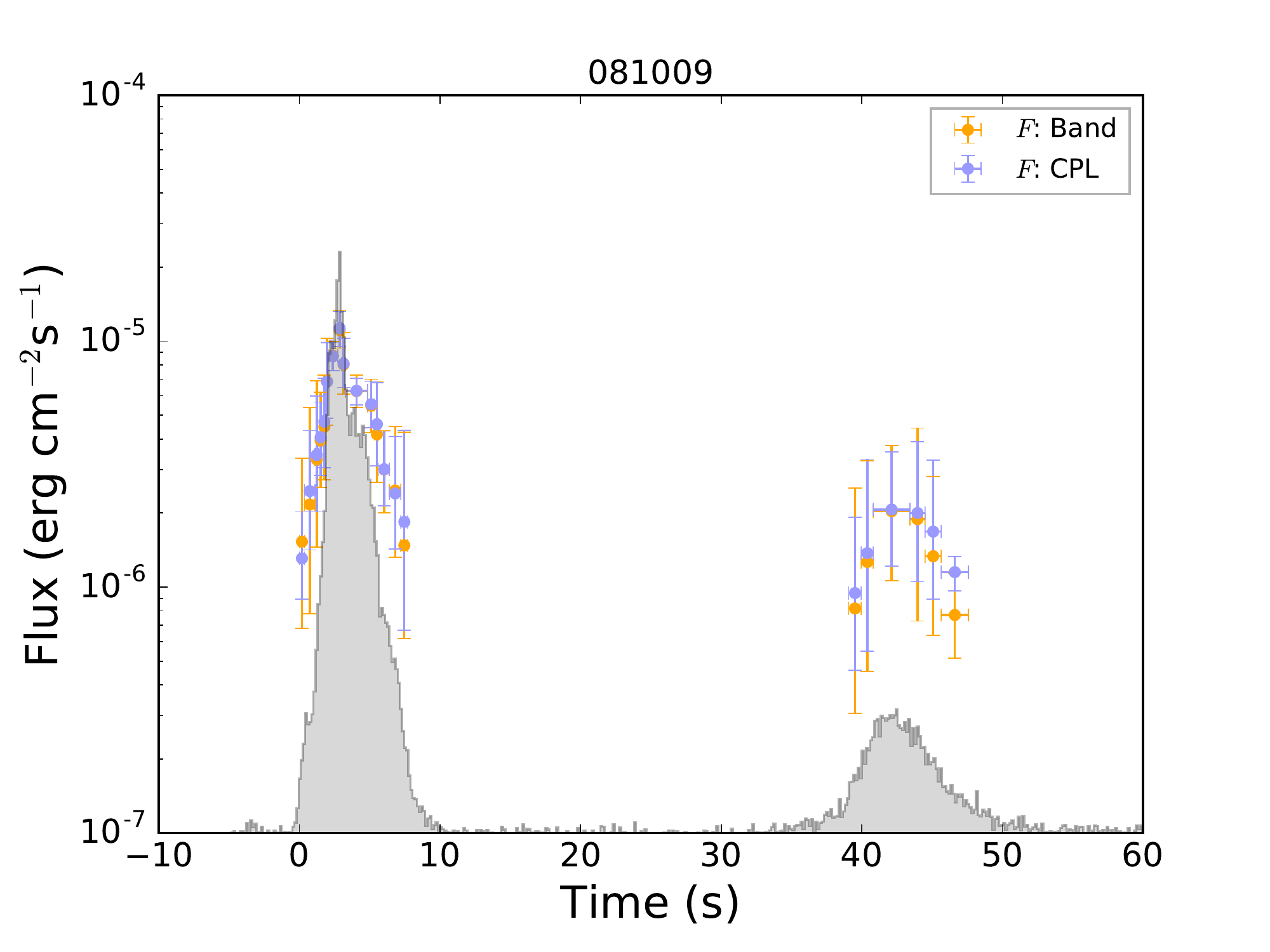}
\includegraphics[angle=0,scale=0.3]{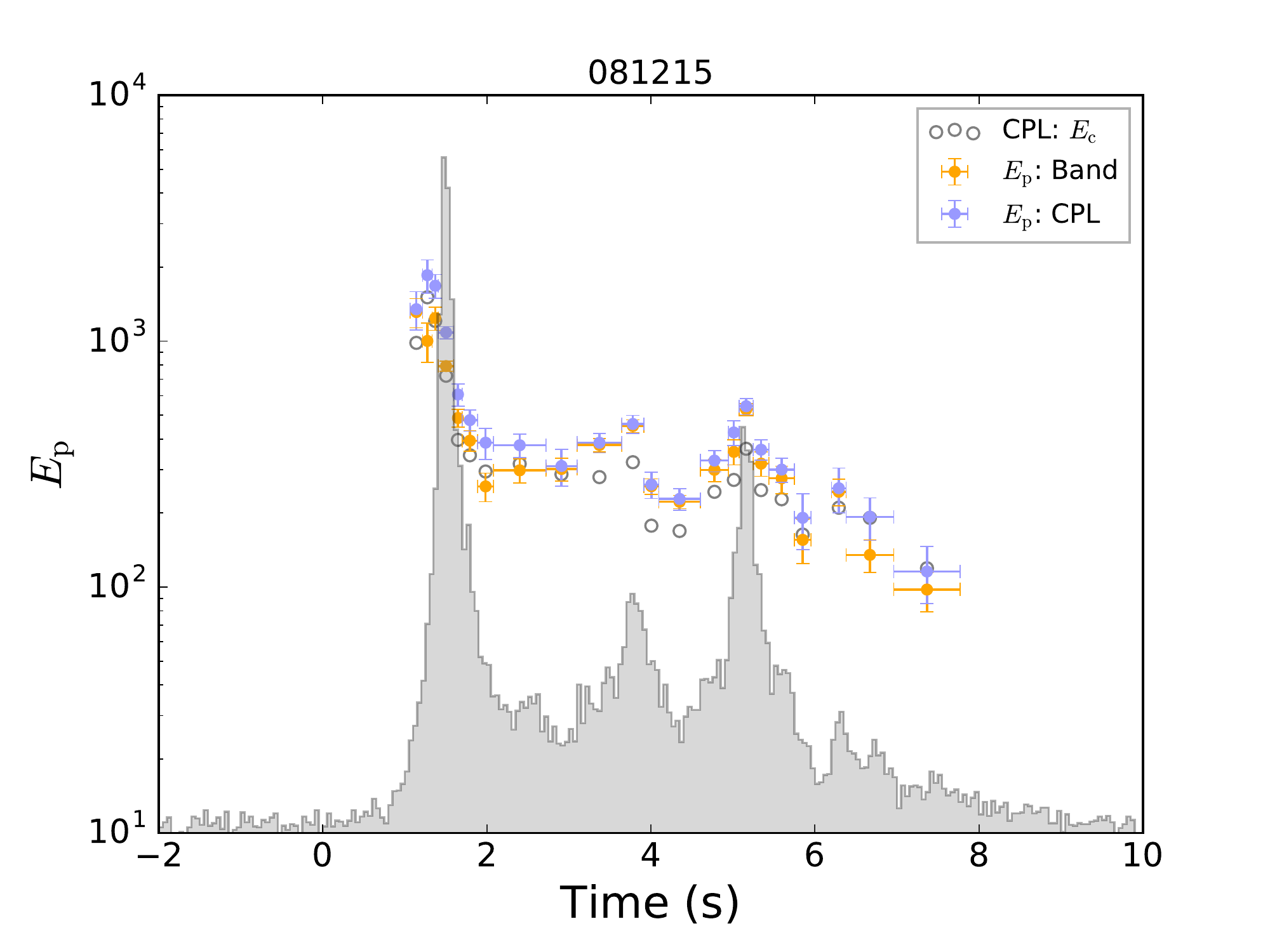}
\includegraphics[angle=0,scale=0.3]{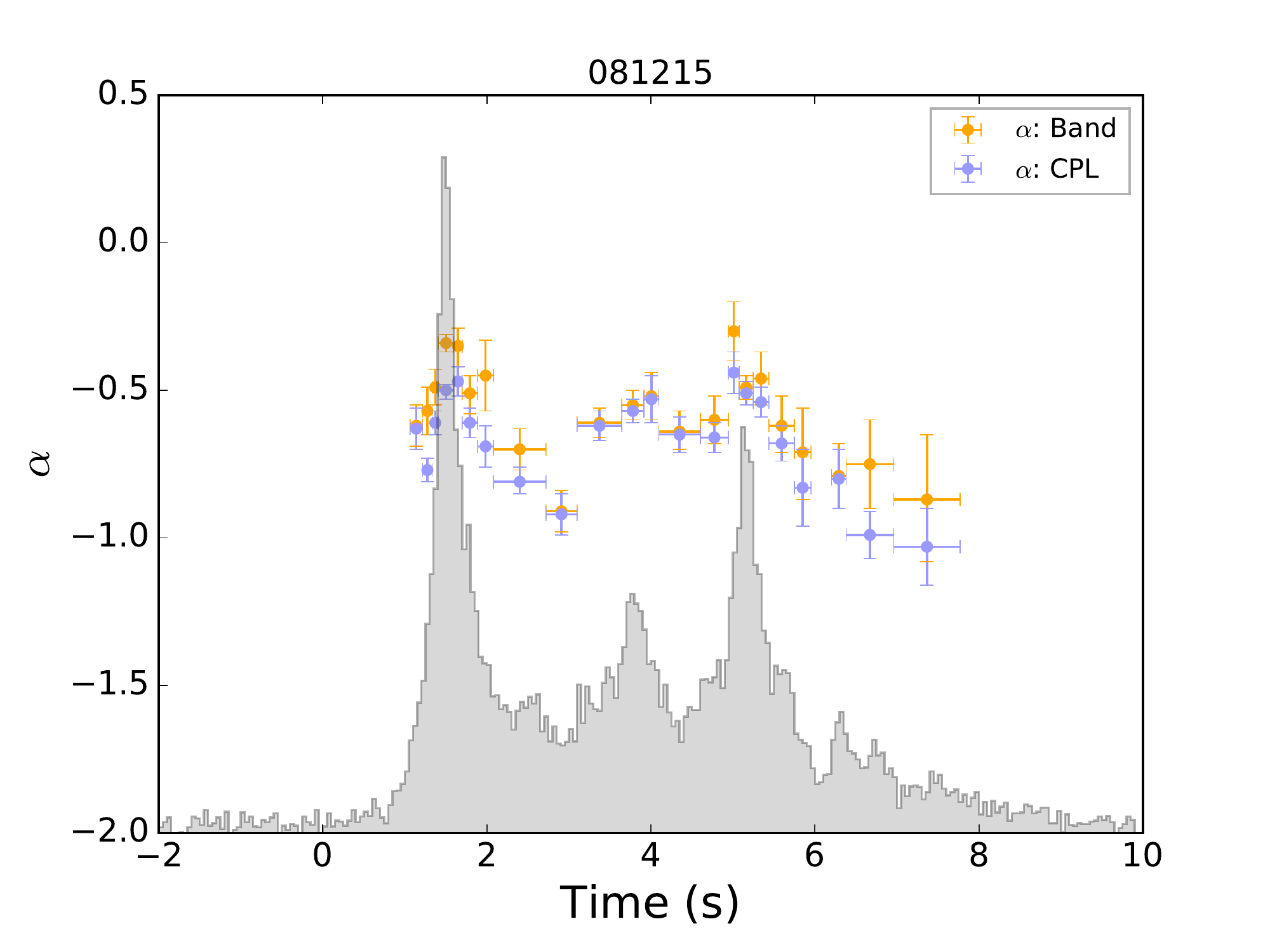}
\includegraphics[angle=0,scale=0.3]{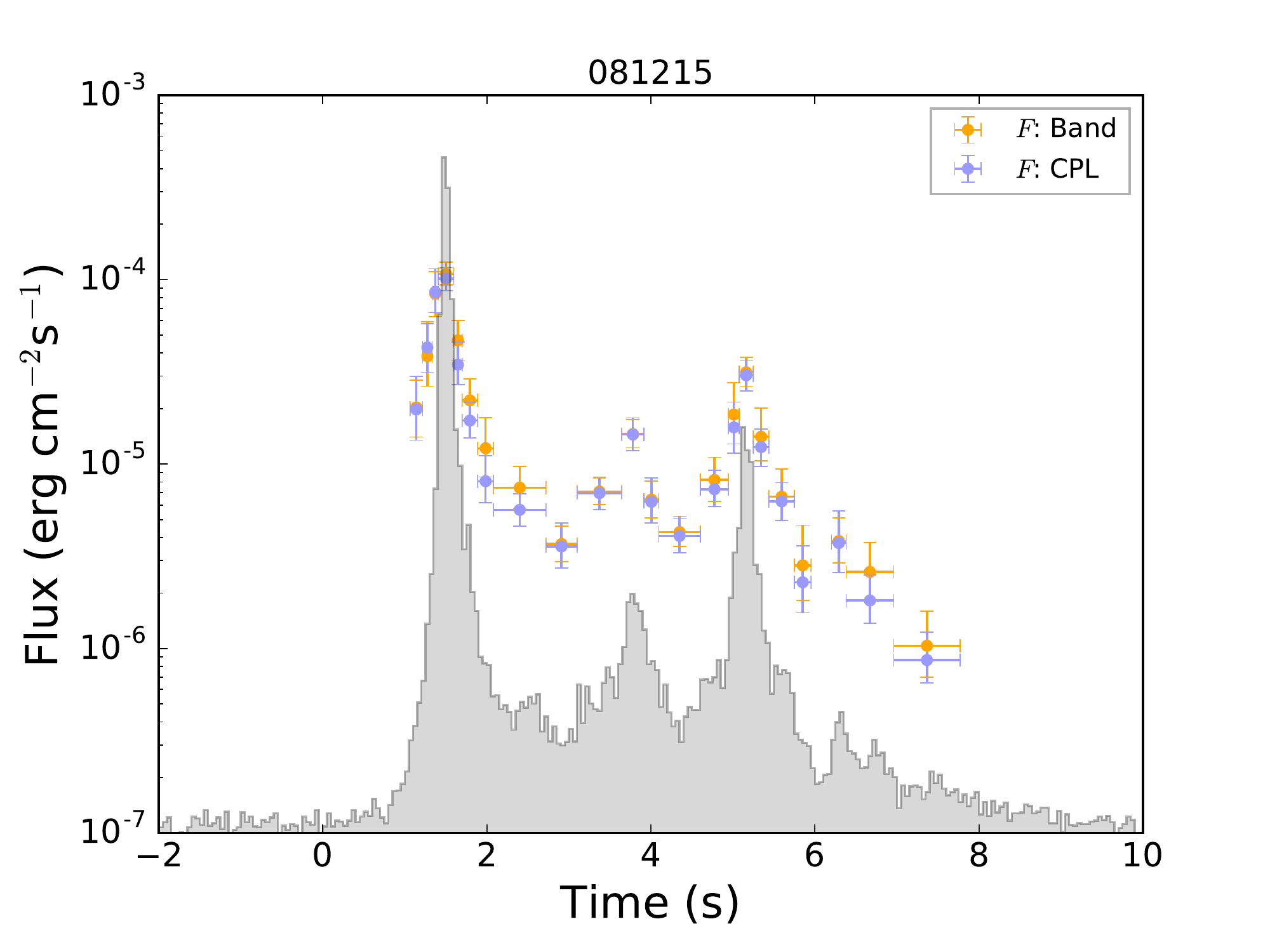}
\includegraphics[angle=0,scale=0.3]{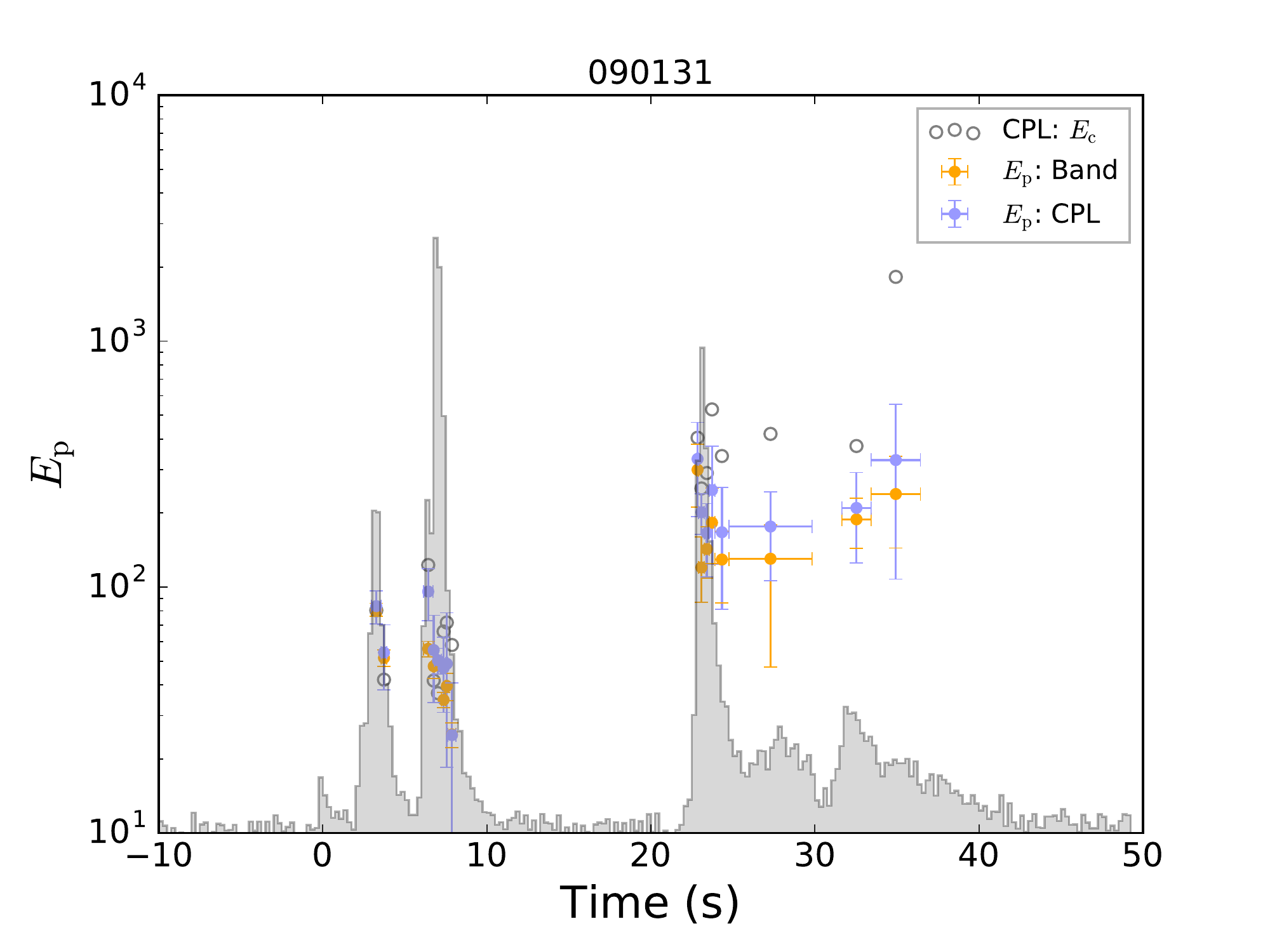}
\includegraphics[angle=0,scale=0.3]{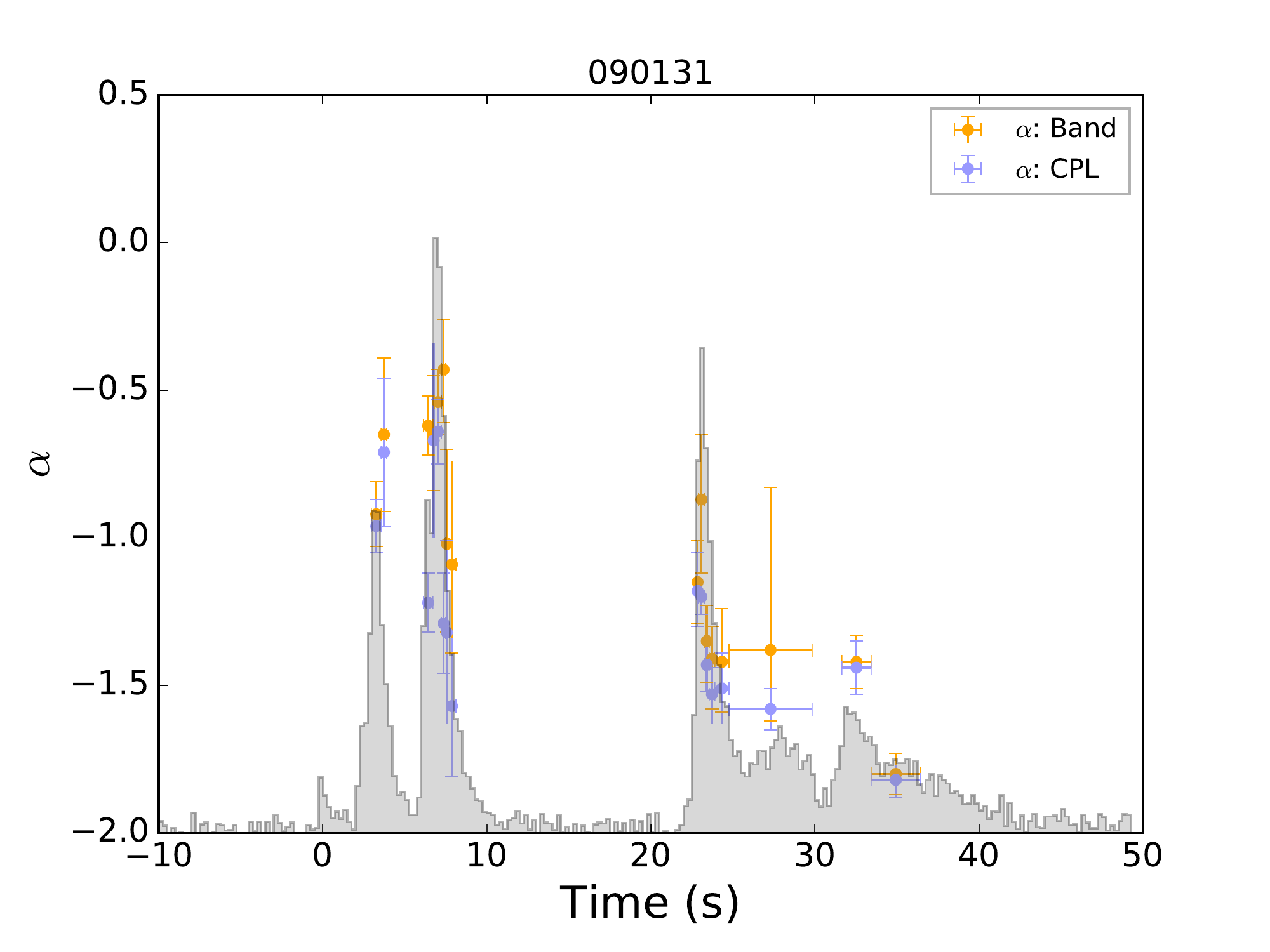}
\includegraphics[angle=0,scale=0.3]{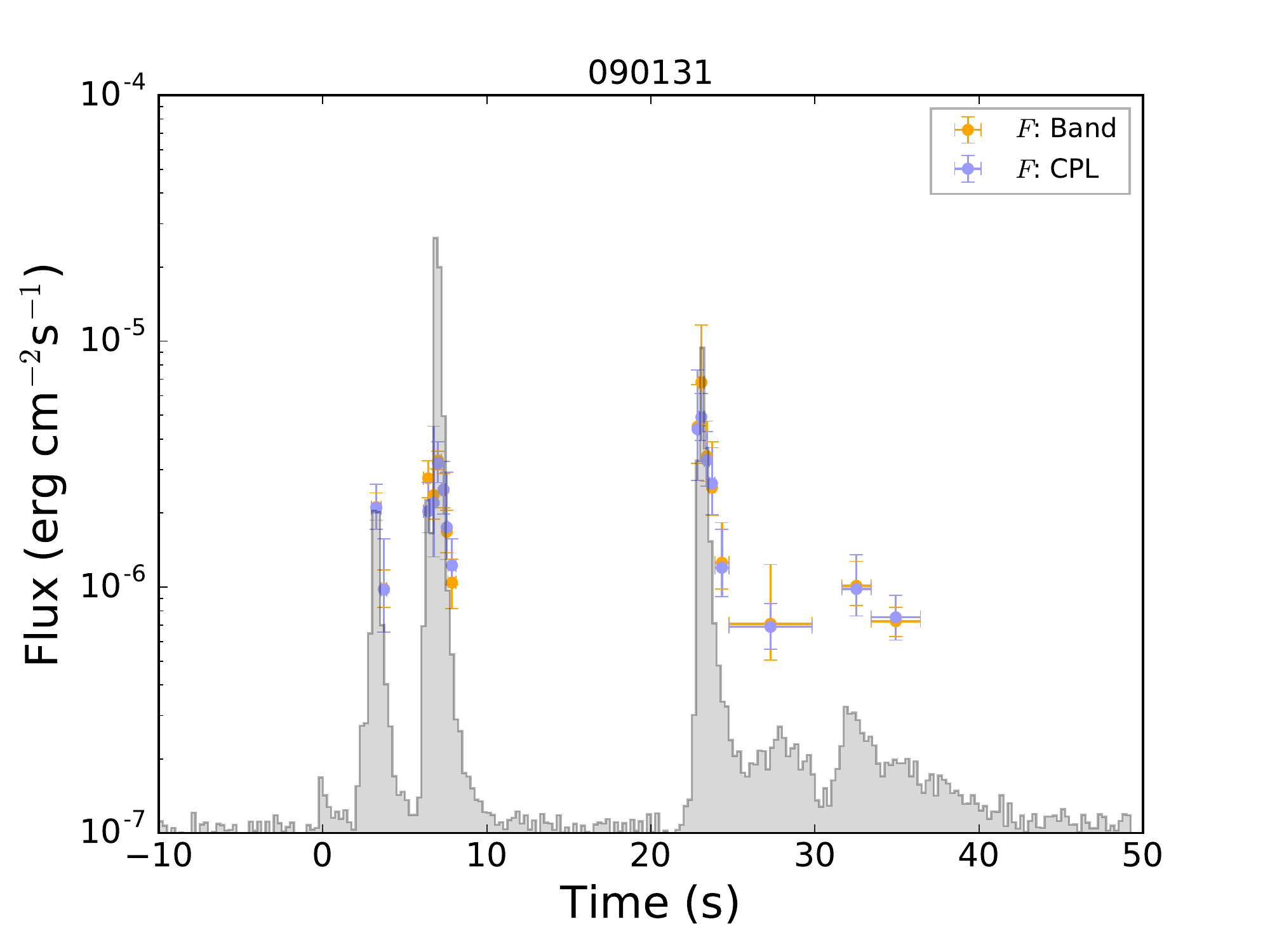}
\includegraphics[angle=0,scale=0.3]{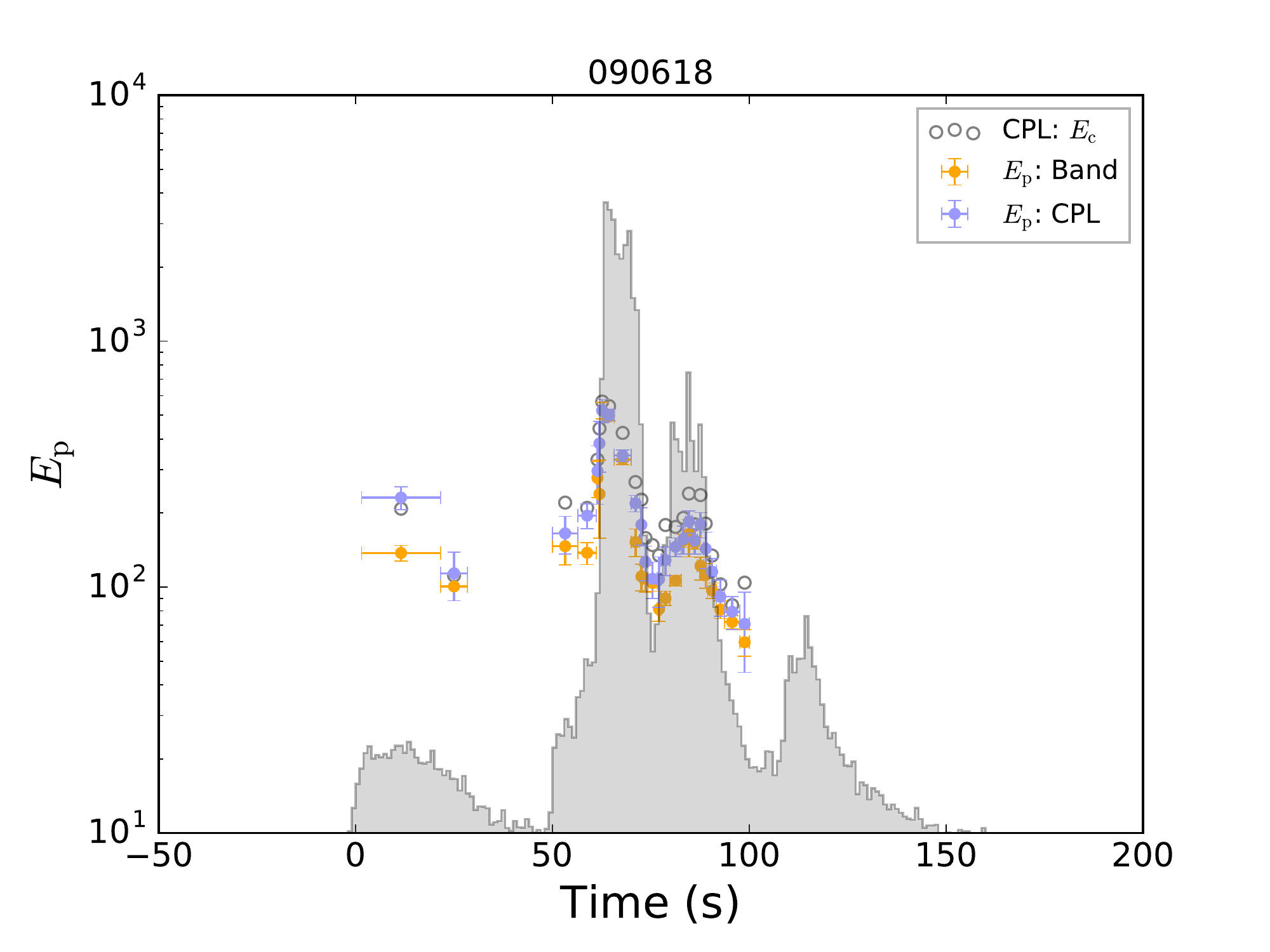}
\includegraphics[angle=0,scale=0.3]{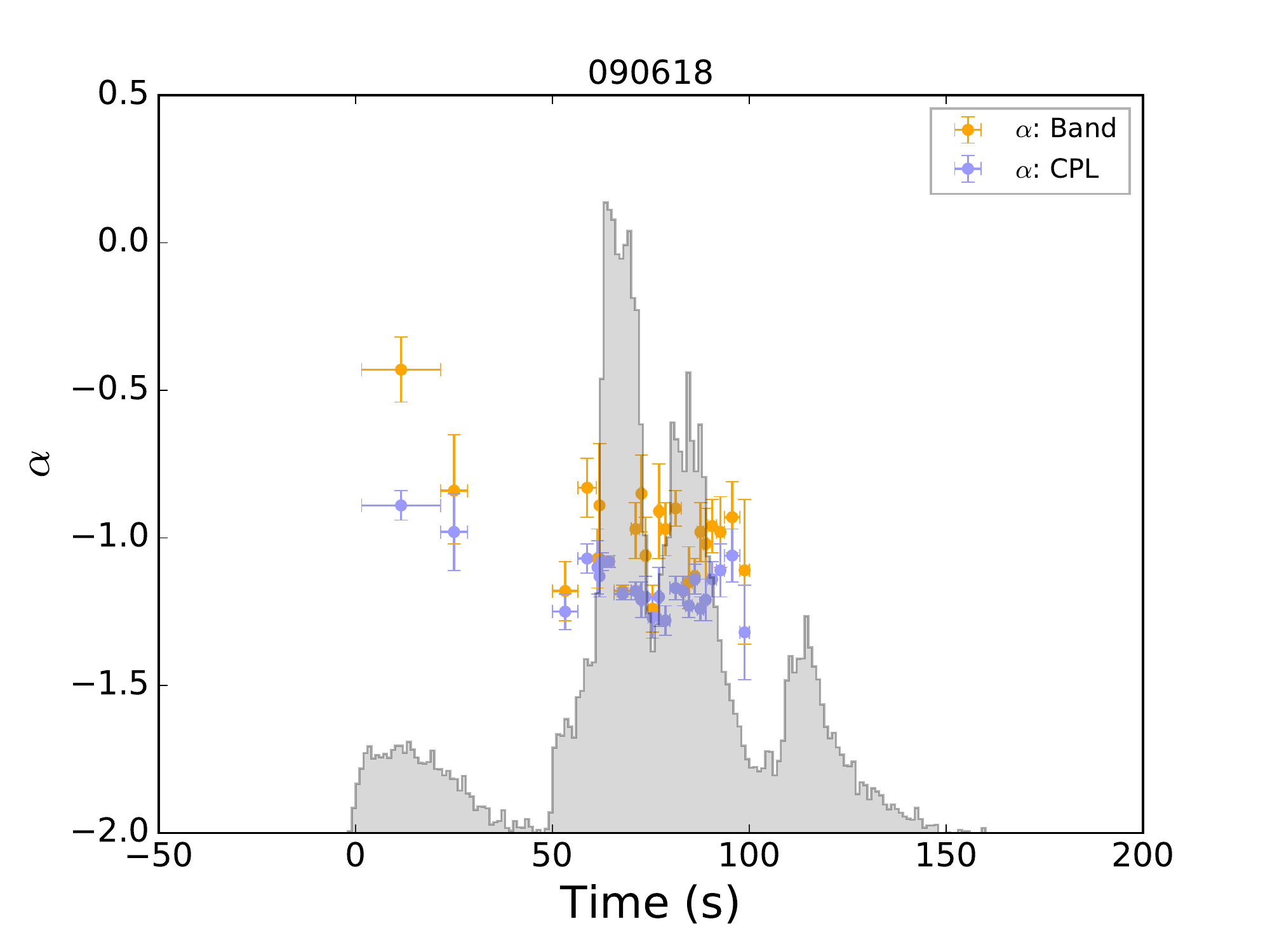}
\includegraphics[angle=0,scale=0.3]{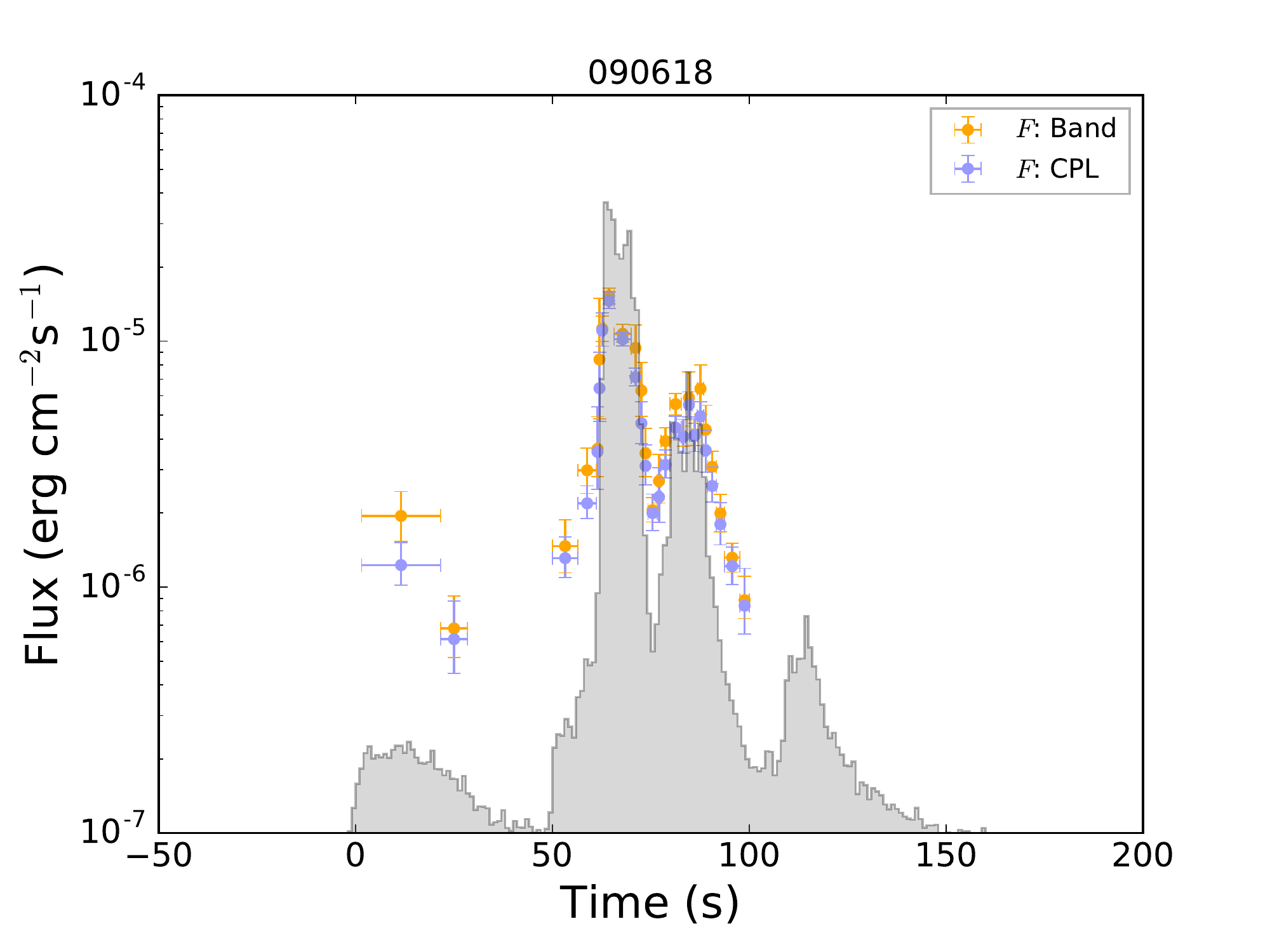}
\includegraphics[angle=0,scale=0.3]{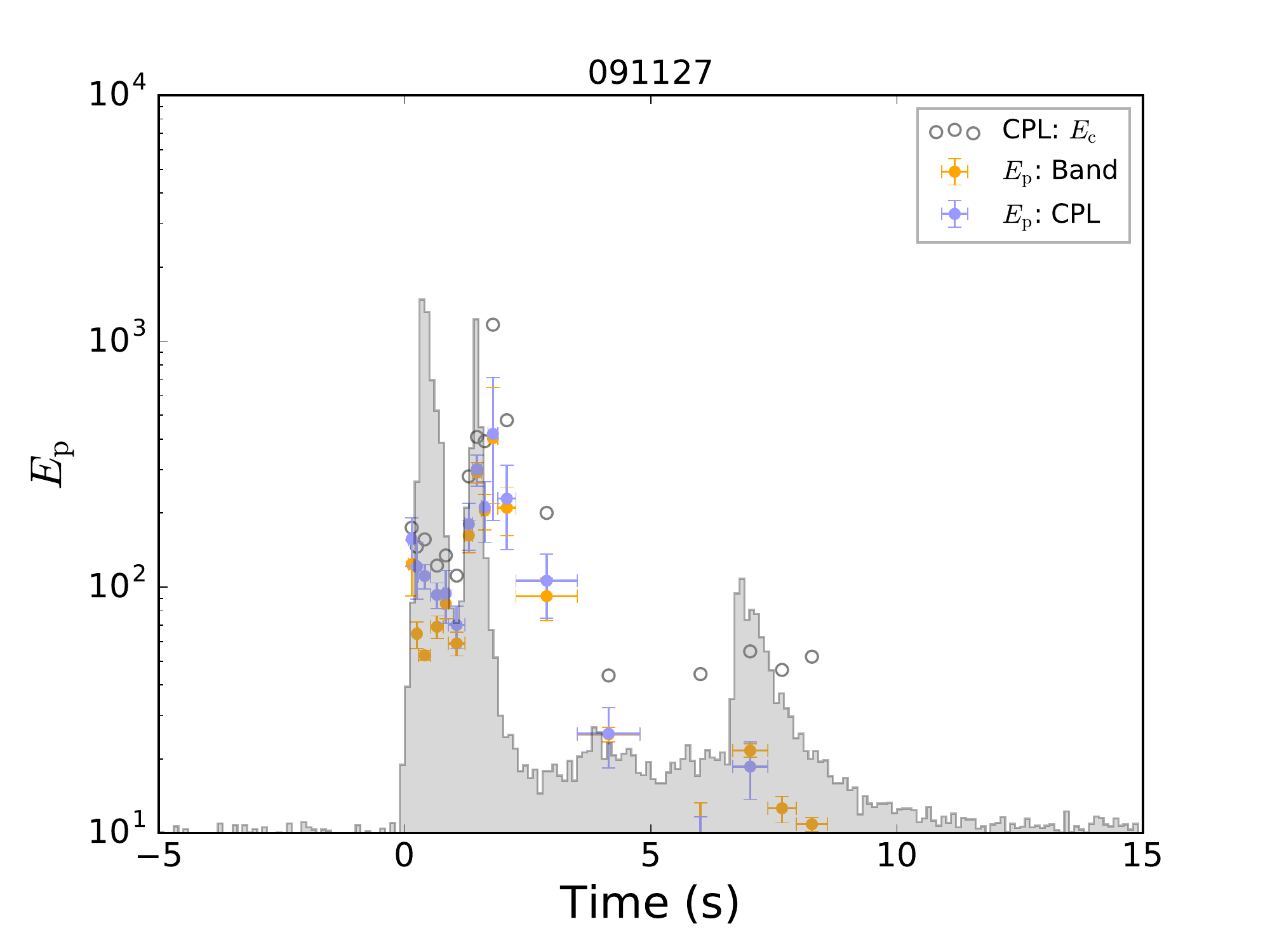}
\includegraphics[angle=0,scale=0.3]{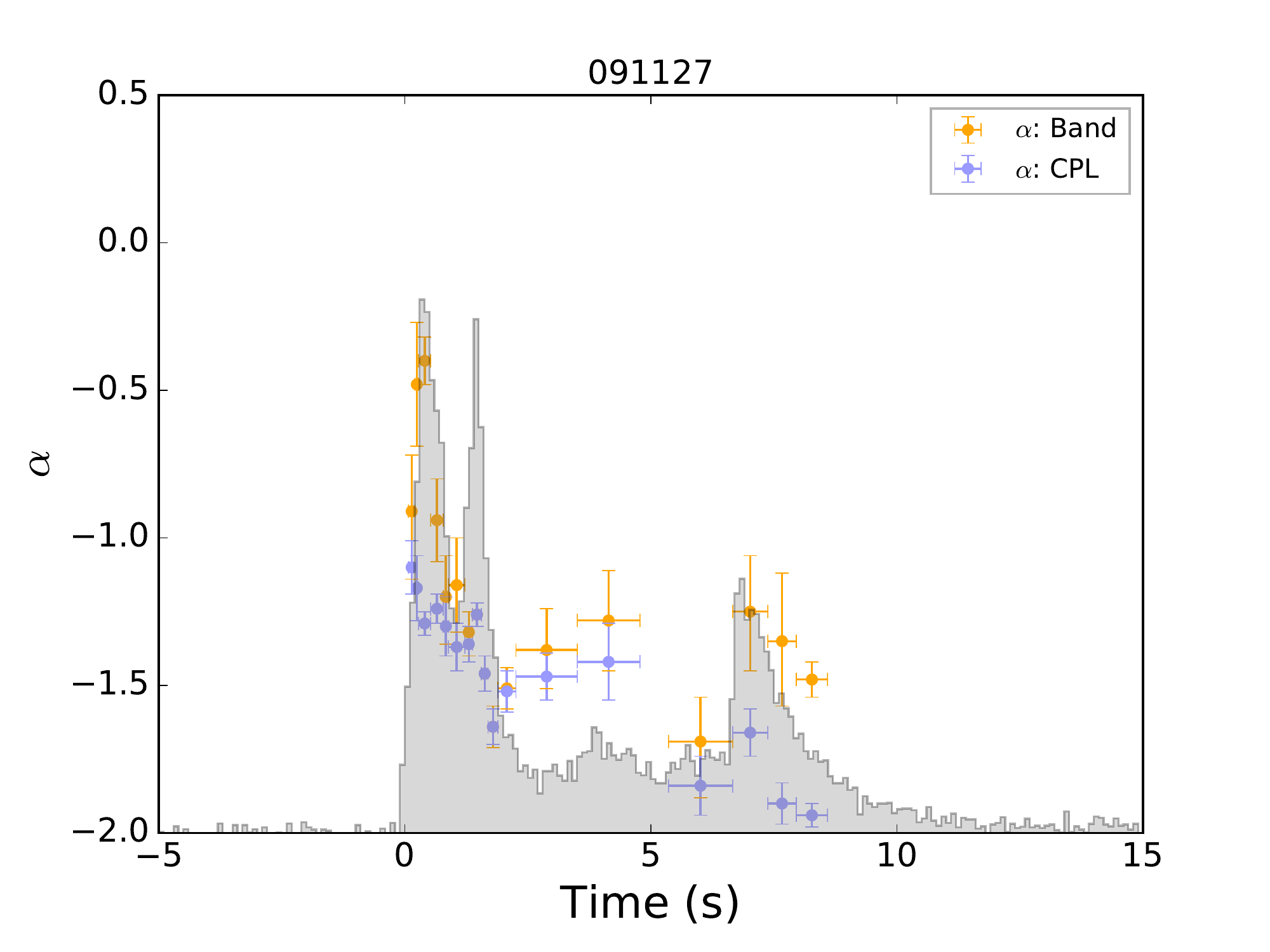}
\includegraphics[angle=0,scale=0.3]{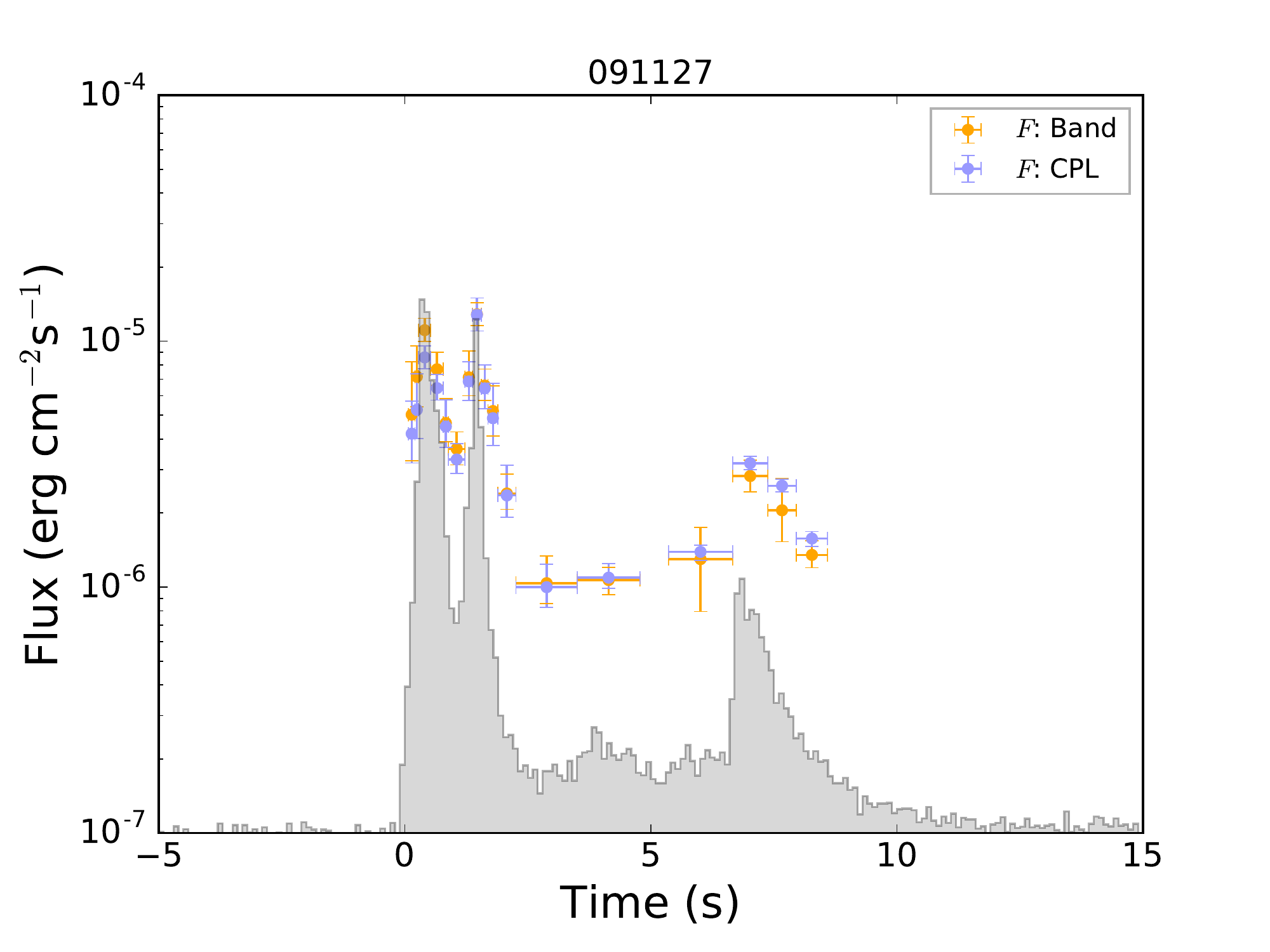}
\caption{Temporal evolution of $E_{\rm p}$, $\alpha$ index and energy flux $F$. Data points with solid pink and orange colors indicate Band and CPL respectively. Count rate lightcurves are overlaid in gray. All data points correspond to a statistical significance $S \geq 20$.}\label{fig:evolution}
\end{figure*}

\begin{figure*}
\includegraphics[angle=0,scale=0.3]{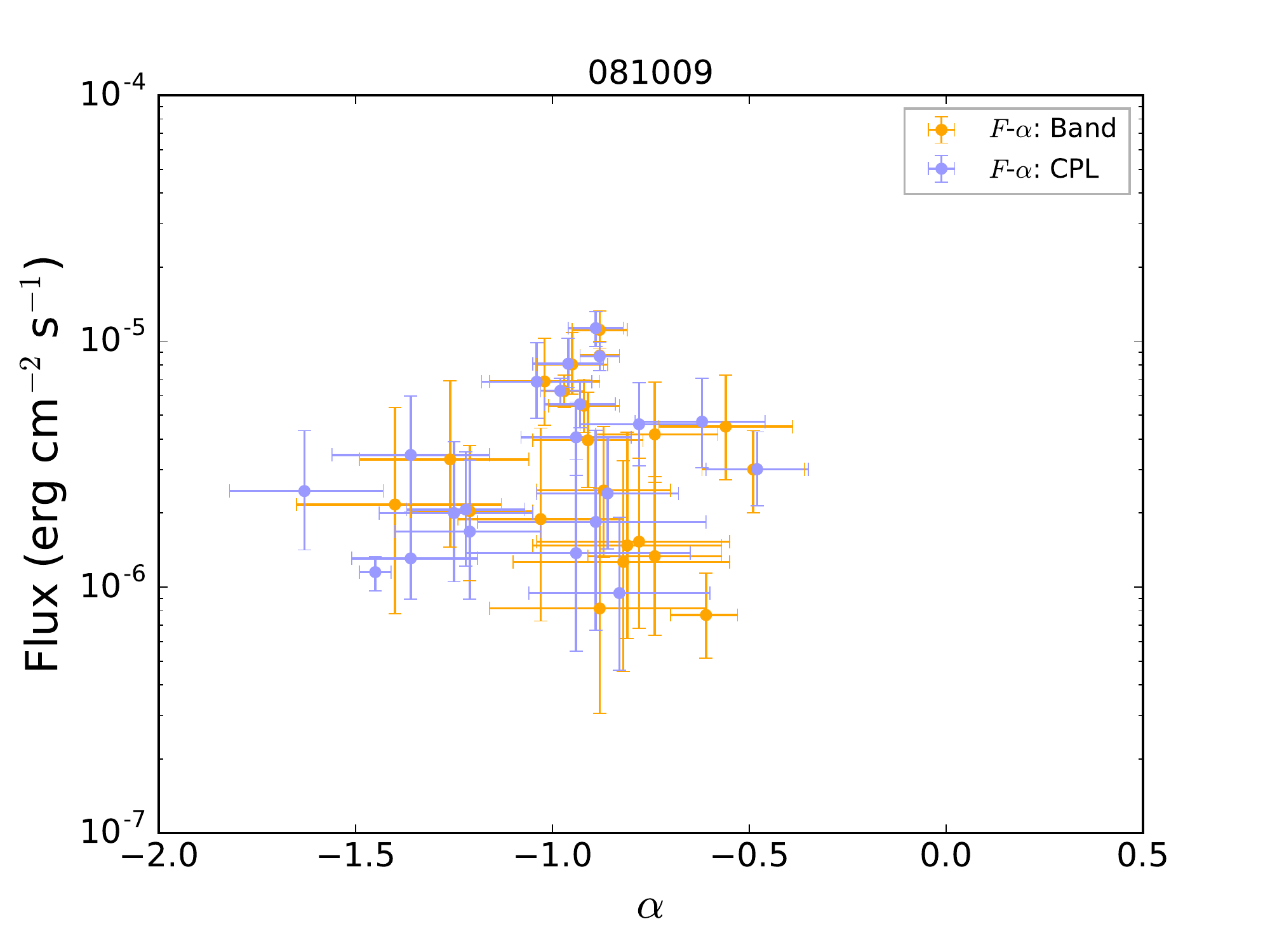}
\includegraphics[angle=0,scale=0.3]{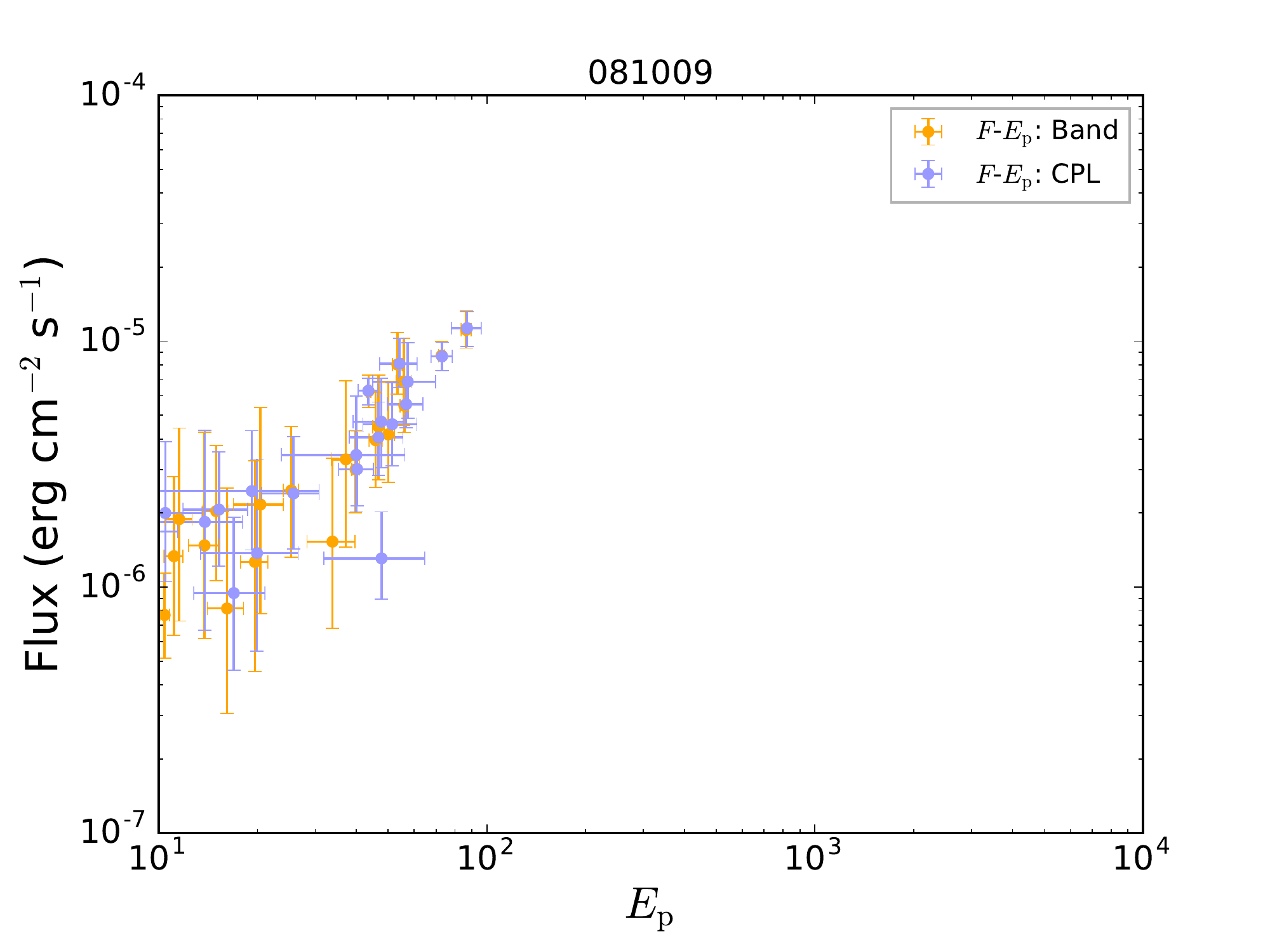}
\includegraphics[angle=0,scale=0.3]{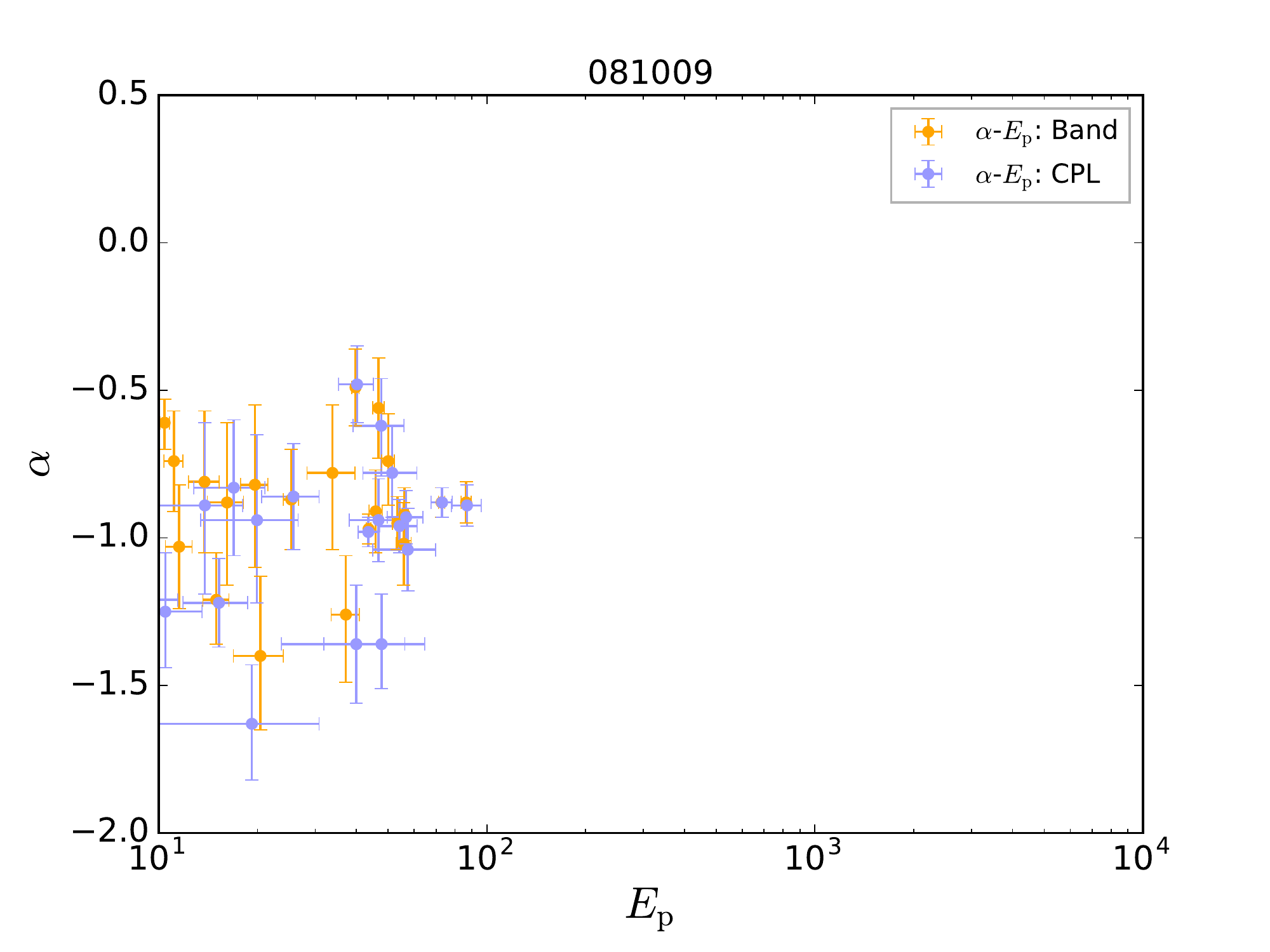}
\includegraphics[angle=0,scale=0.3]{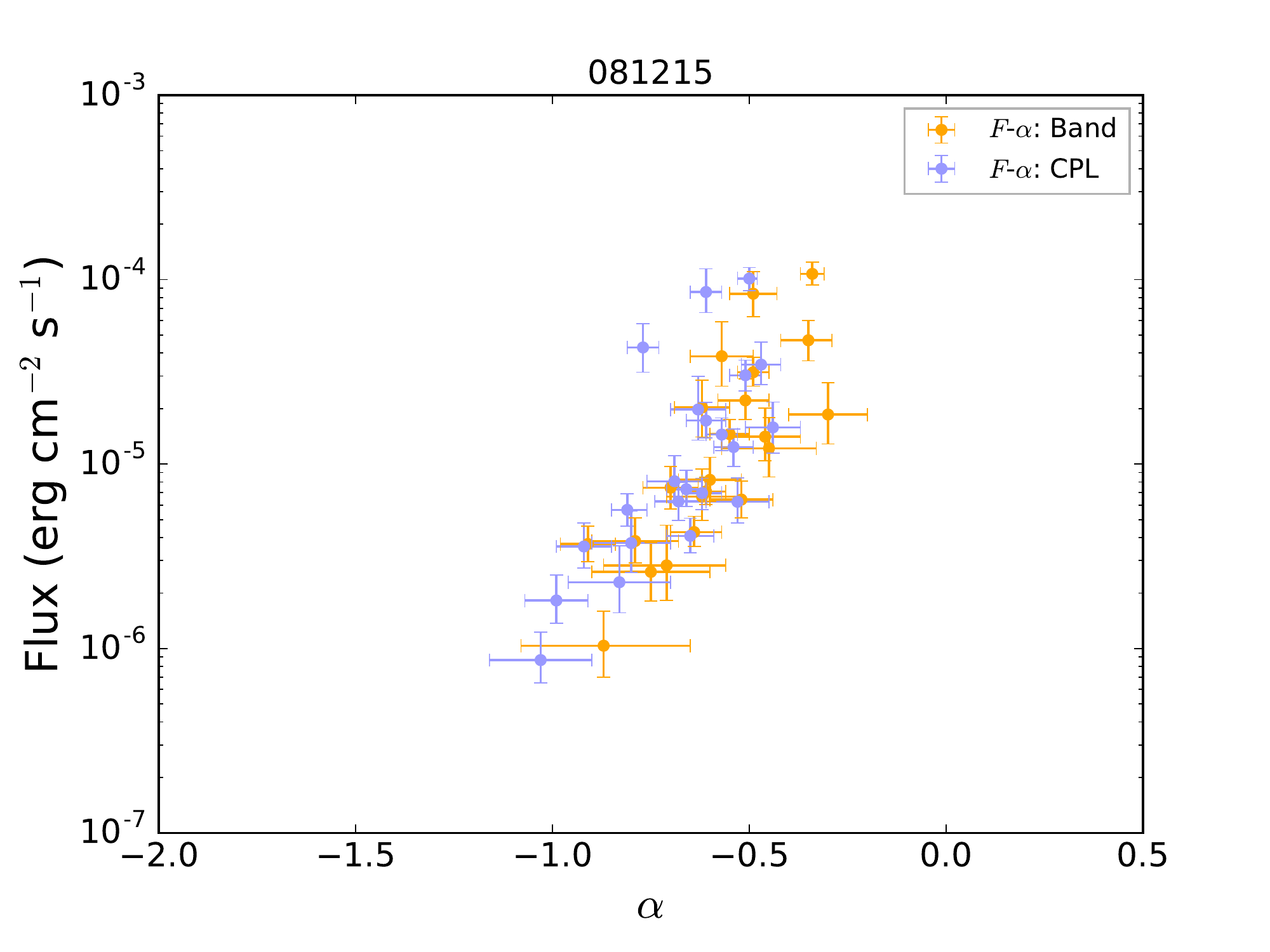}
\includegraphics[angle=0,scale=0.3]{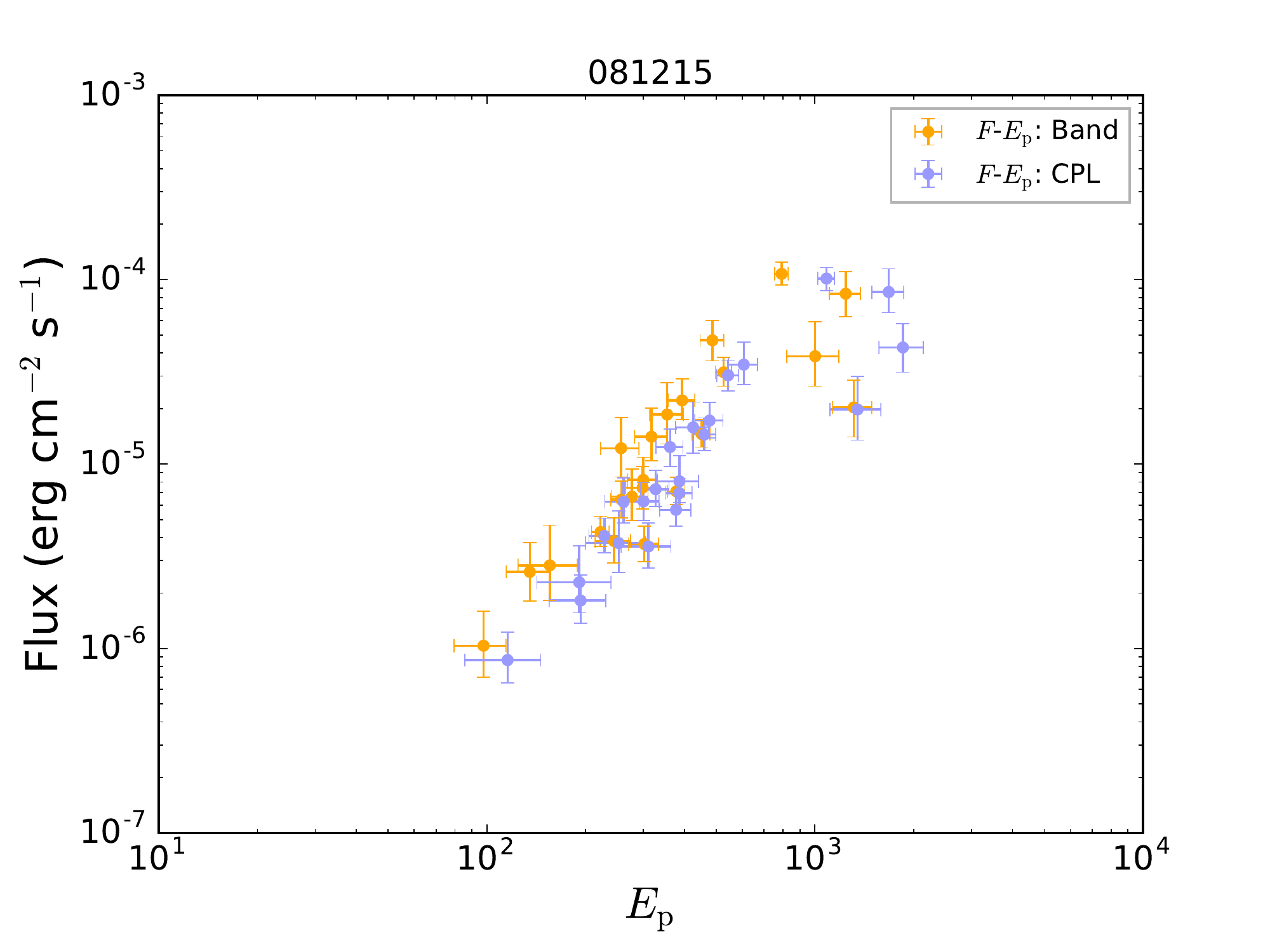}
\includegraphics[angle=0,scale=0.3]{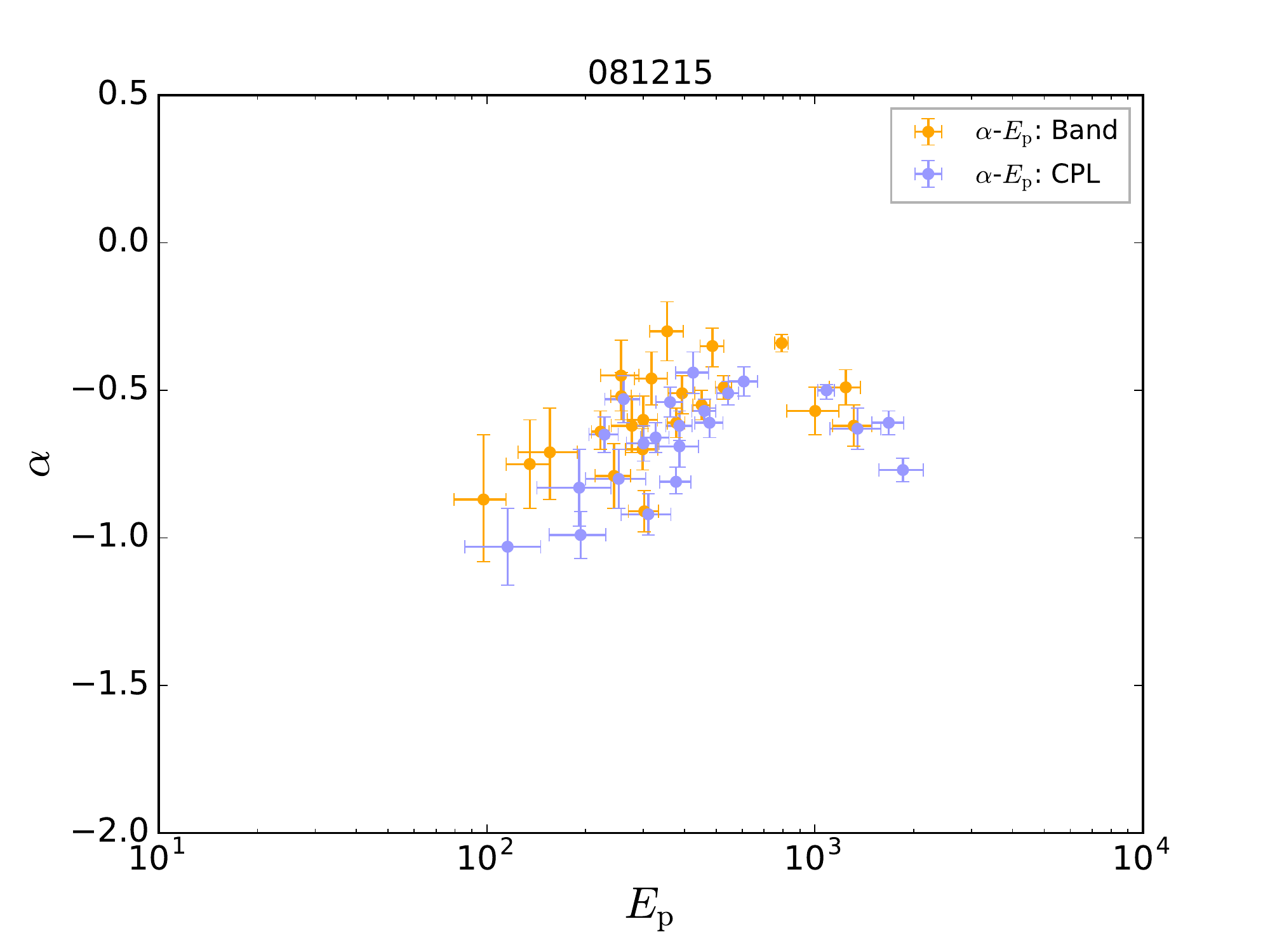}
\includegraphics[angle=0,scale=0.3]{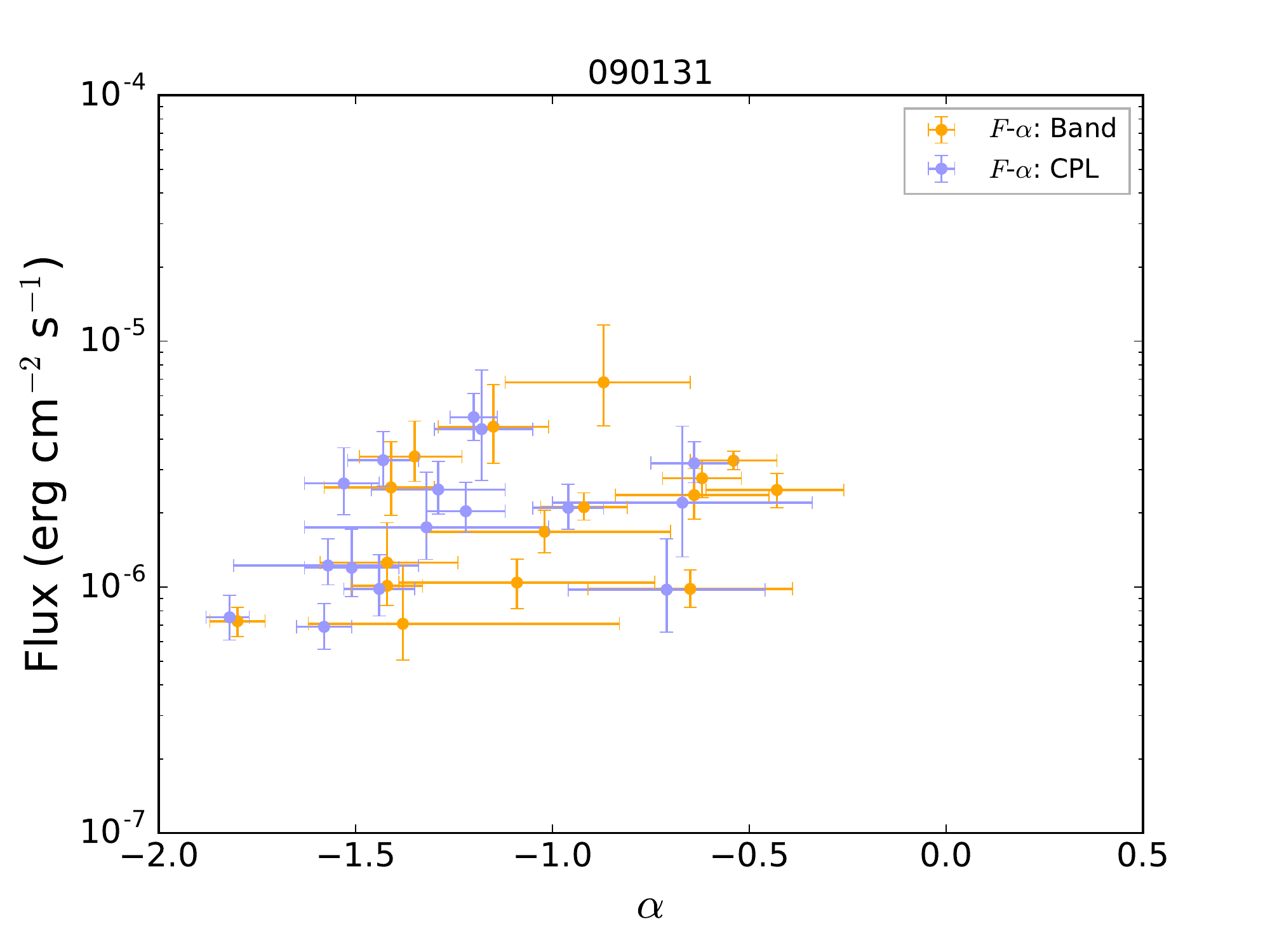}
\includegraphics[angle=0,scale=0.3]{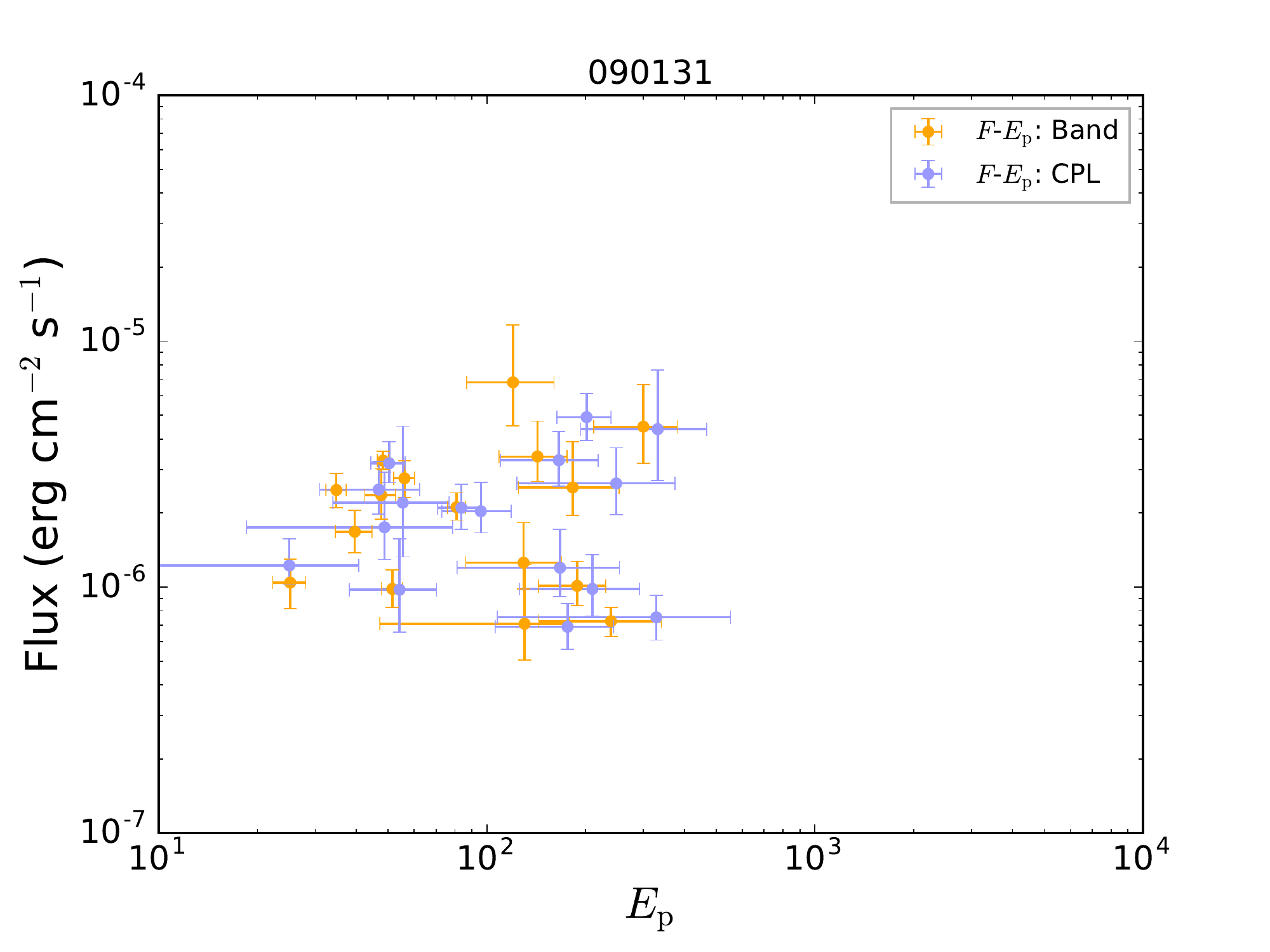}
\includegraphics[angle=0,scale=0.3]{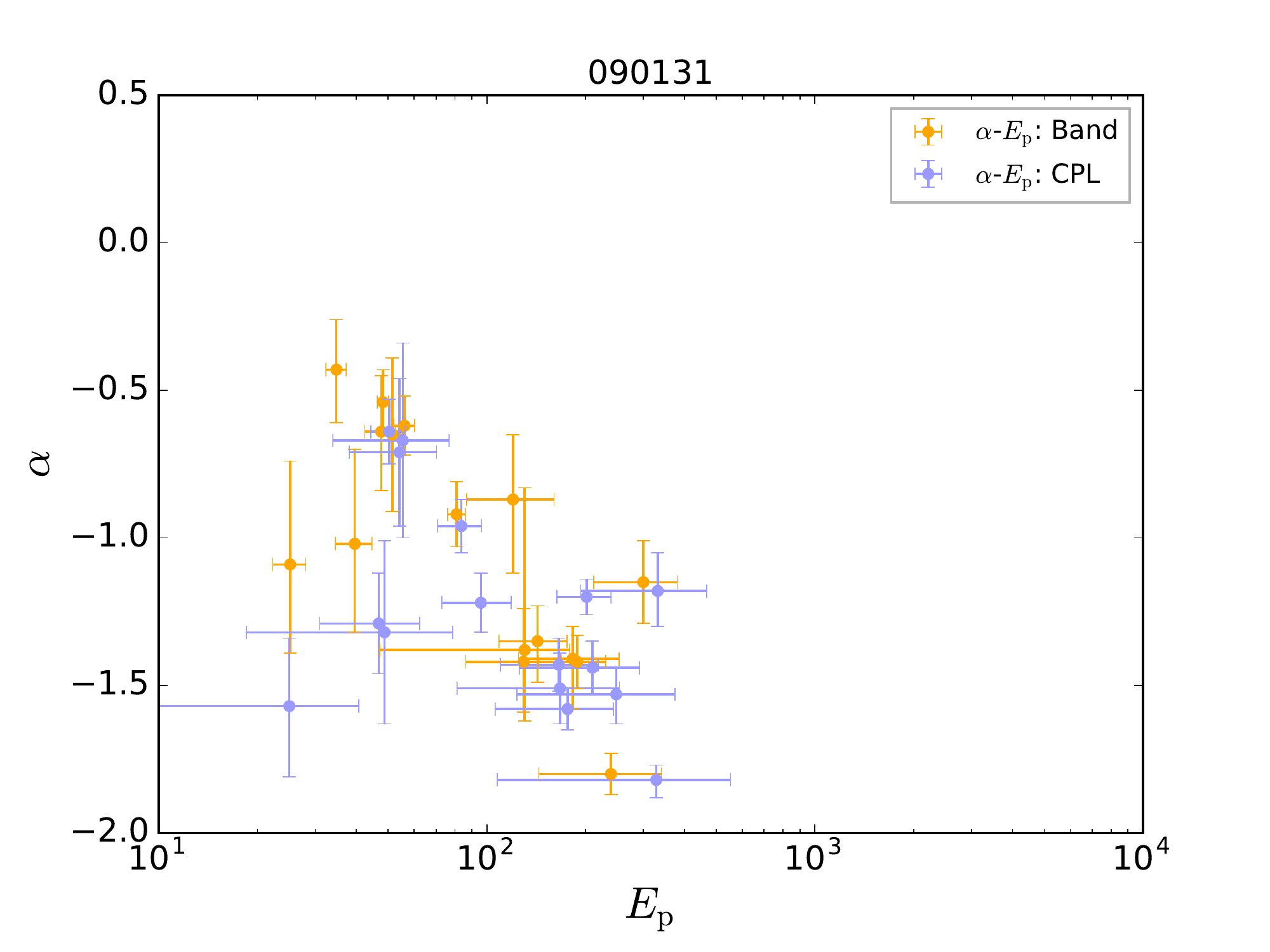}
\includegraphics[angle=0,scale=0.3]{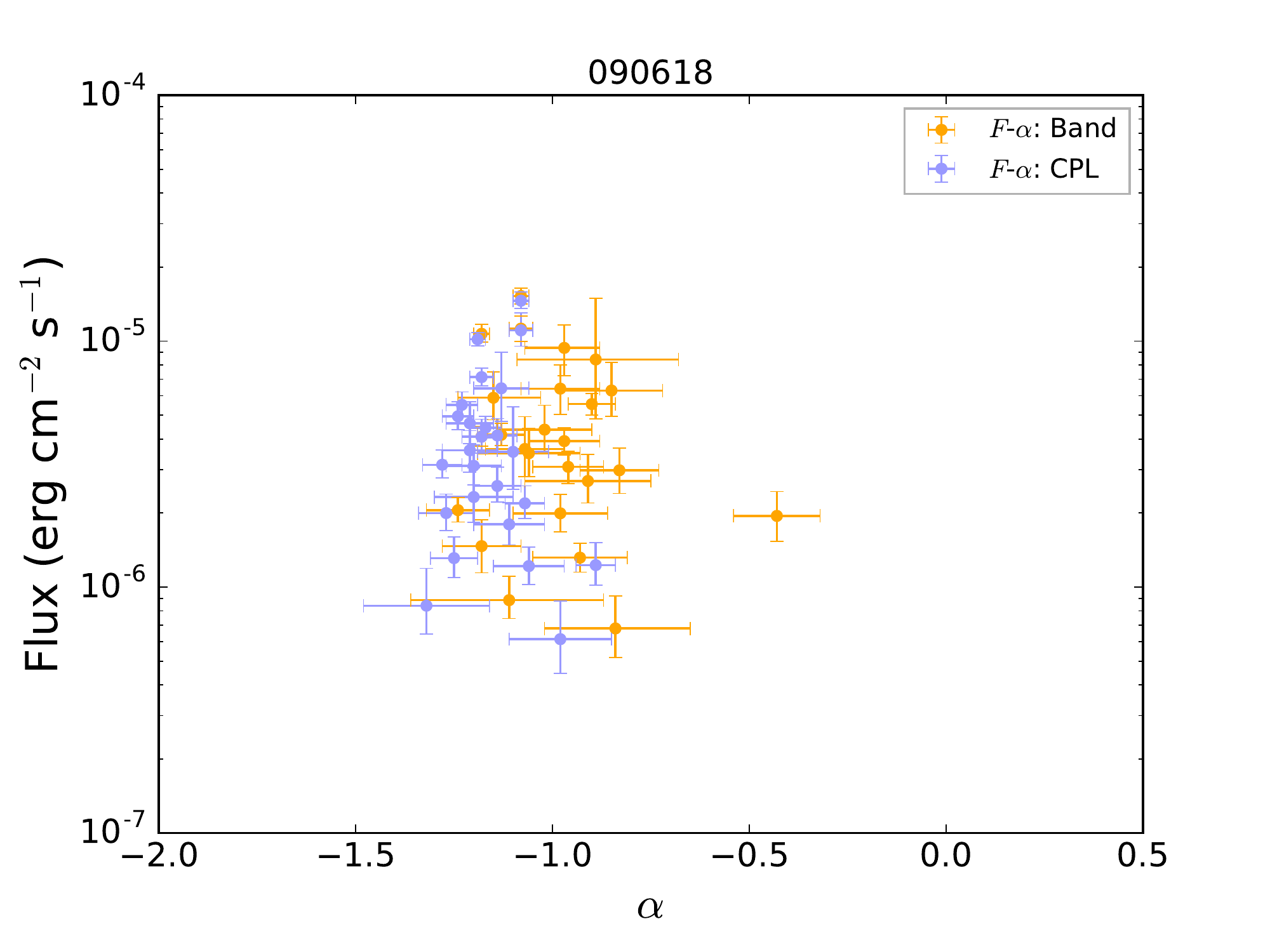}
\includegraphics[angle=0,scale=0.3]{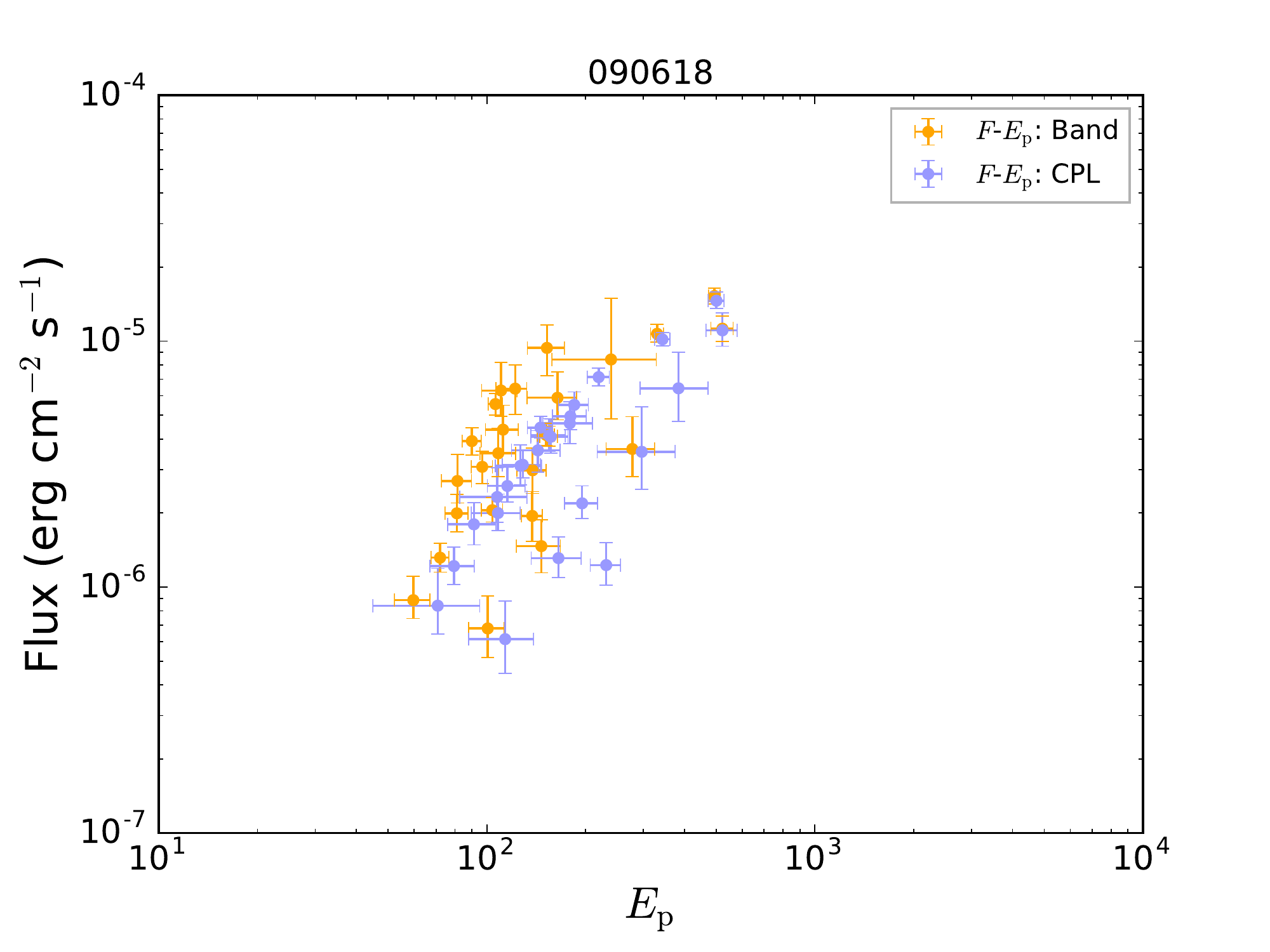}
\includegraphics[angle=0,scale=0.3]{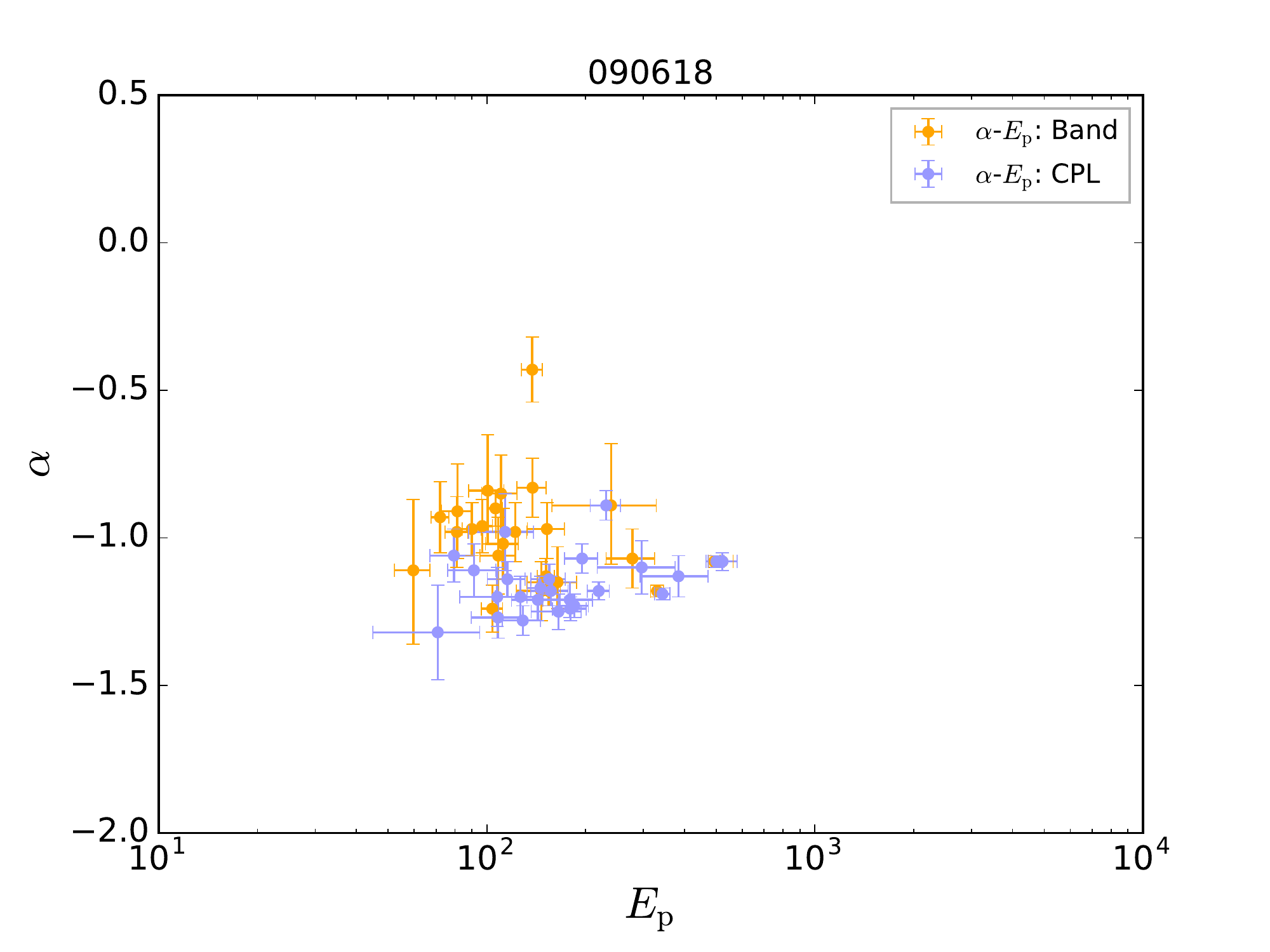}
\includegraphics[angle=0,scale=0.3]{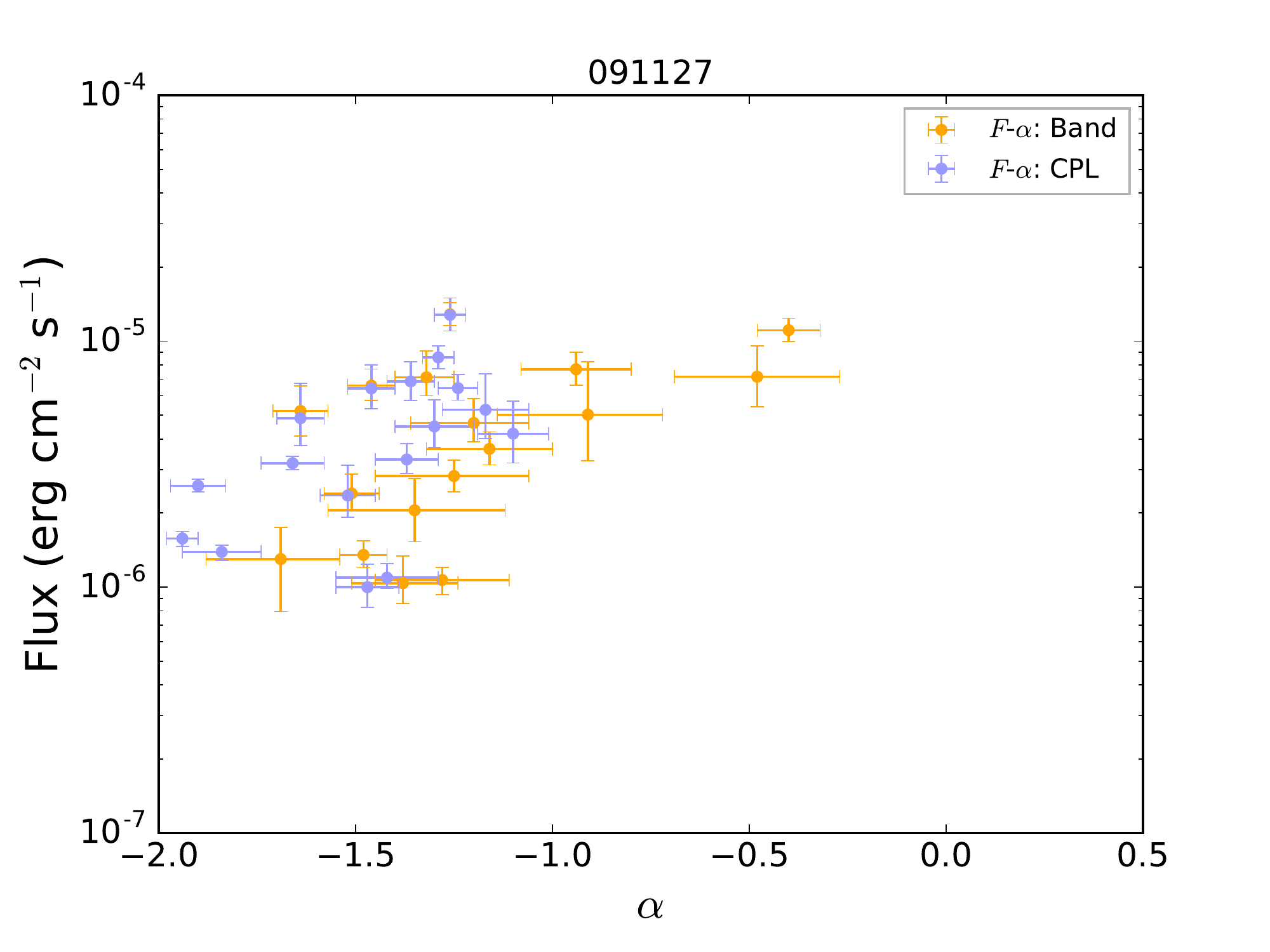}
\includegraphics[angle=0,scale=0.3]{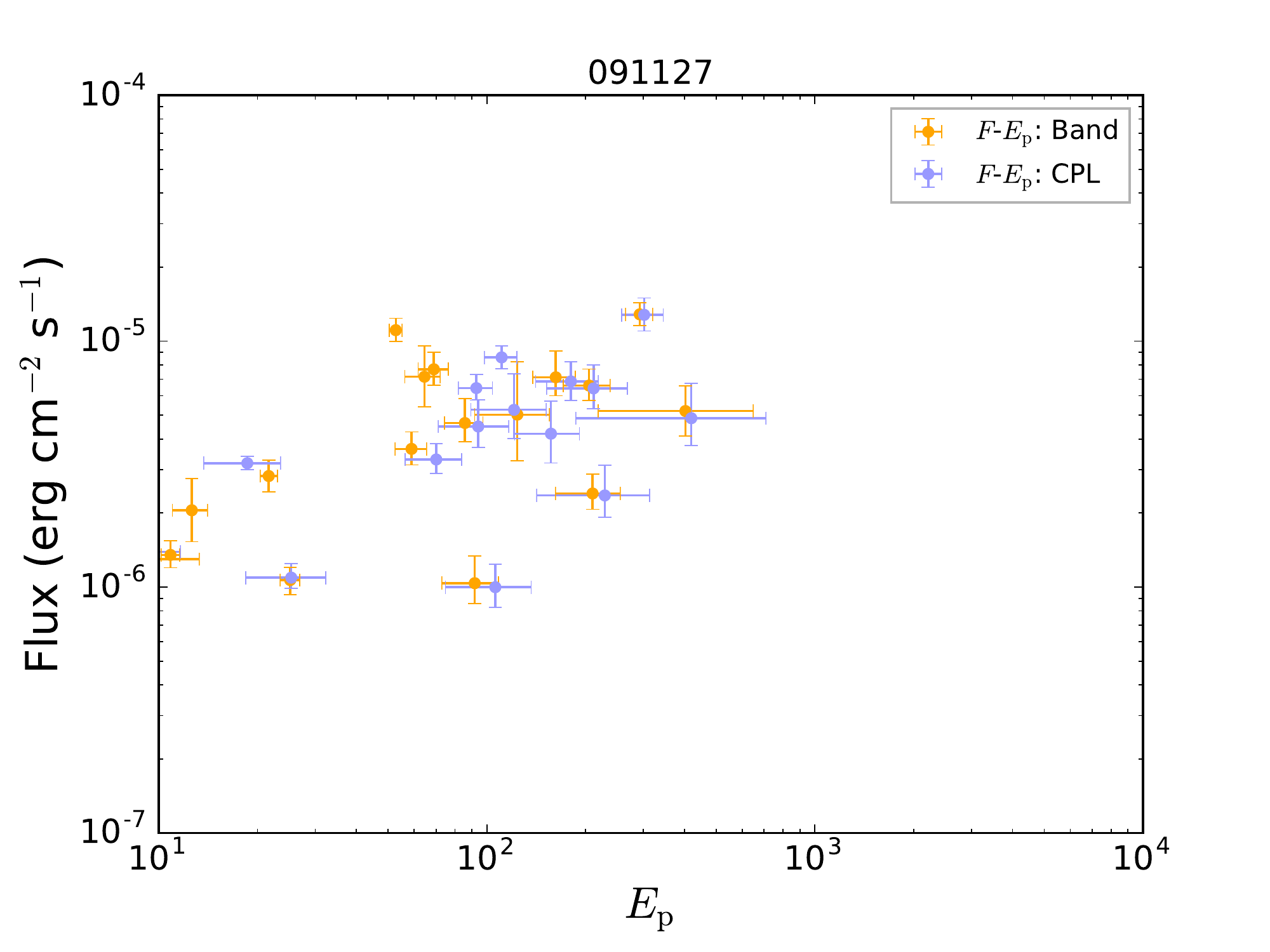}
\includegraphics[angle=0,scale=0.3]{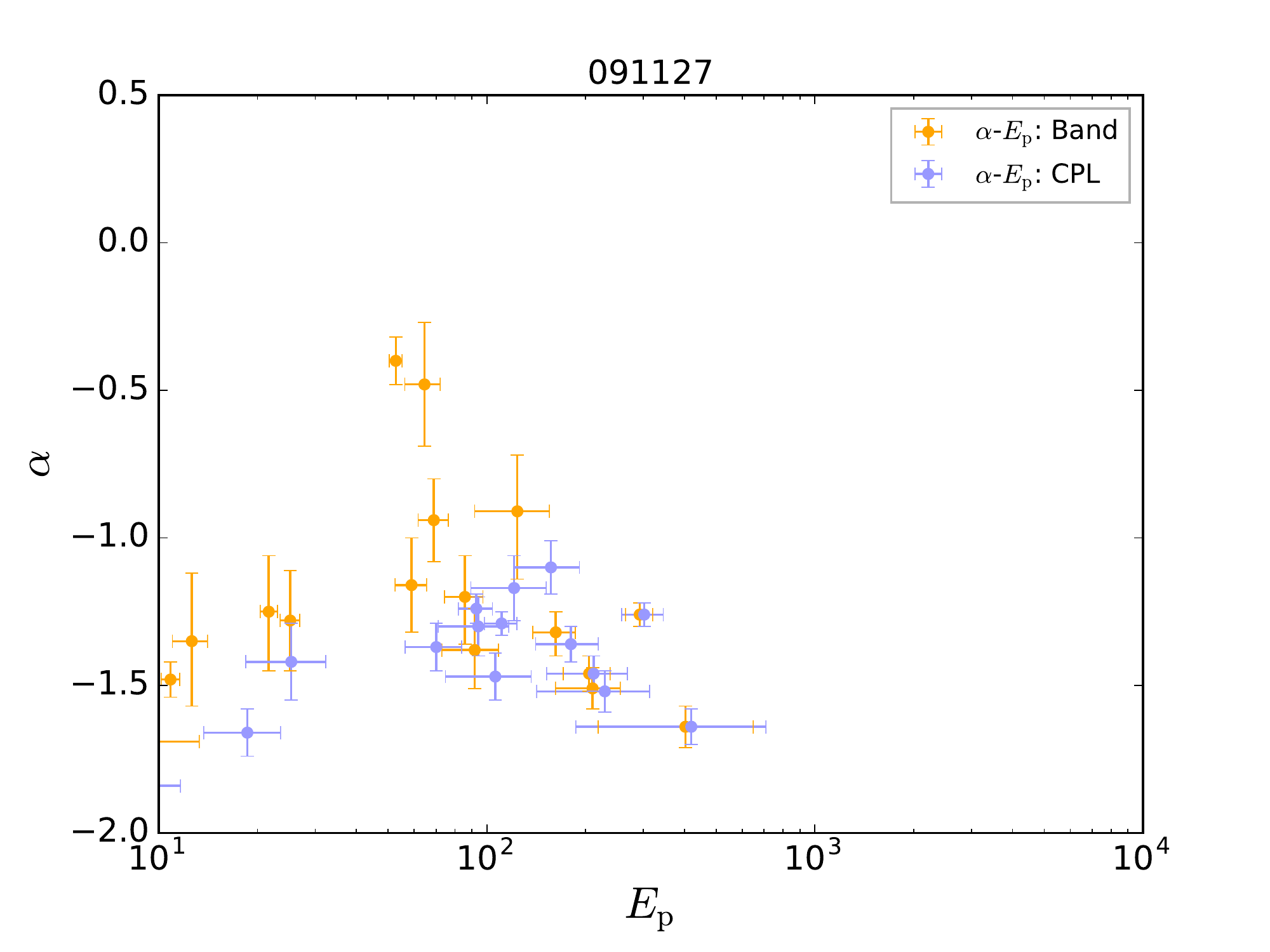}
\caption{The $F$-$\alpha$, $F$-$E_{\rm p}$, and $\alpha$-$E_{\rm p}$ relations. Data points with solid pink and orange colors indicate Band and CPL, respectively. All data points correspond to a statistical significance $S \geq 20$.}\label{fig:relation3}
\end{figure*}

In this Appendix, we also present the spectral parameter evolution and relations (BAND versus COMP) for each individual burst in Figures \ref{fig:evolution}-\ref{fig:relation3}.

\begin{figure*}
\includegraphics[angle=0,scale=0.3]{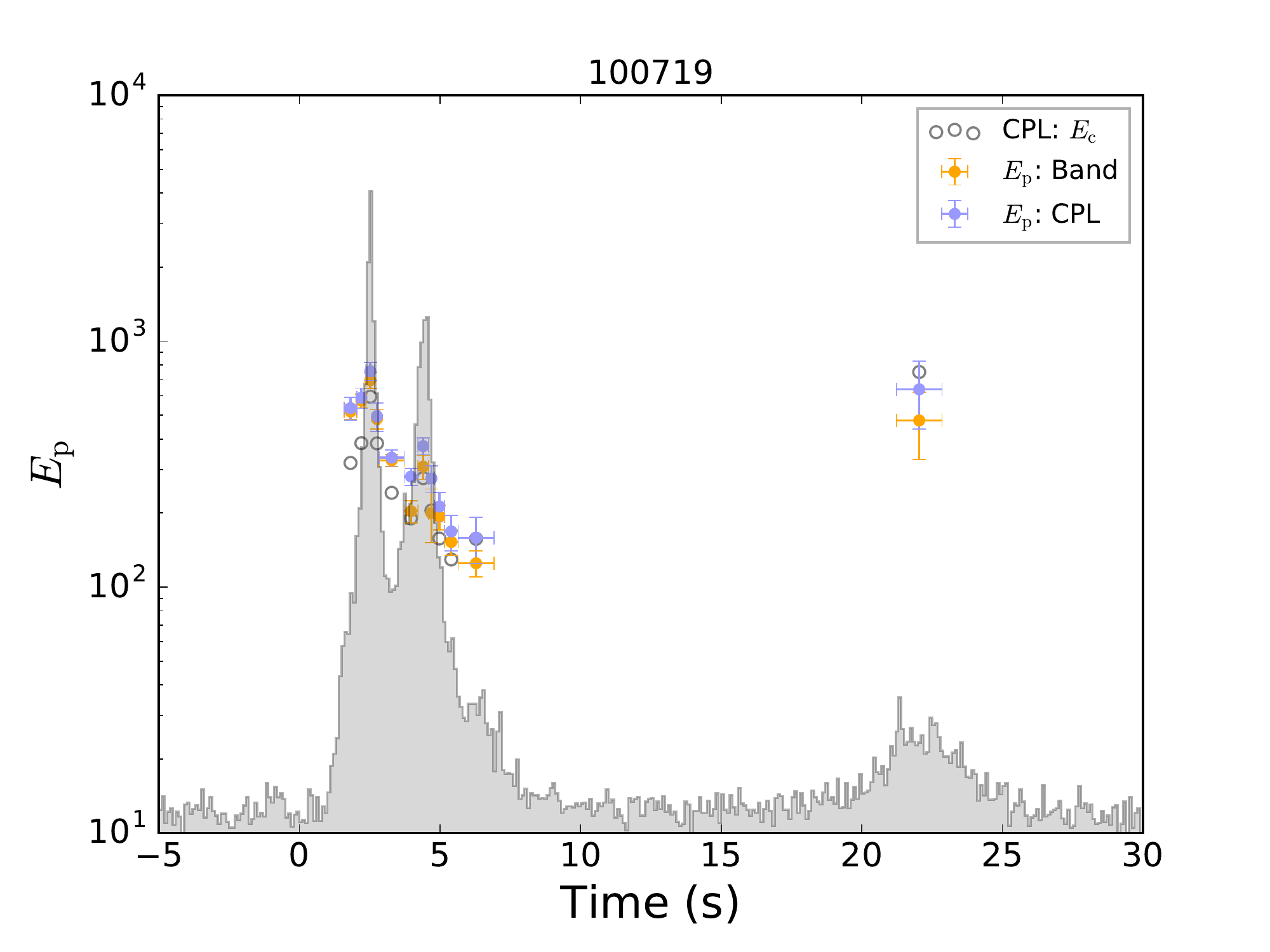}
\includegraphics[angle=0,scale=0.3]{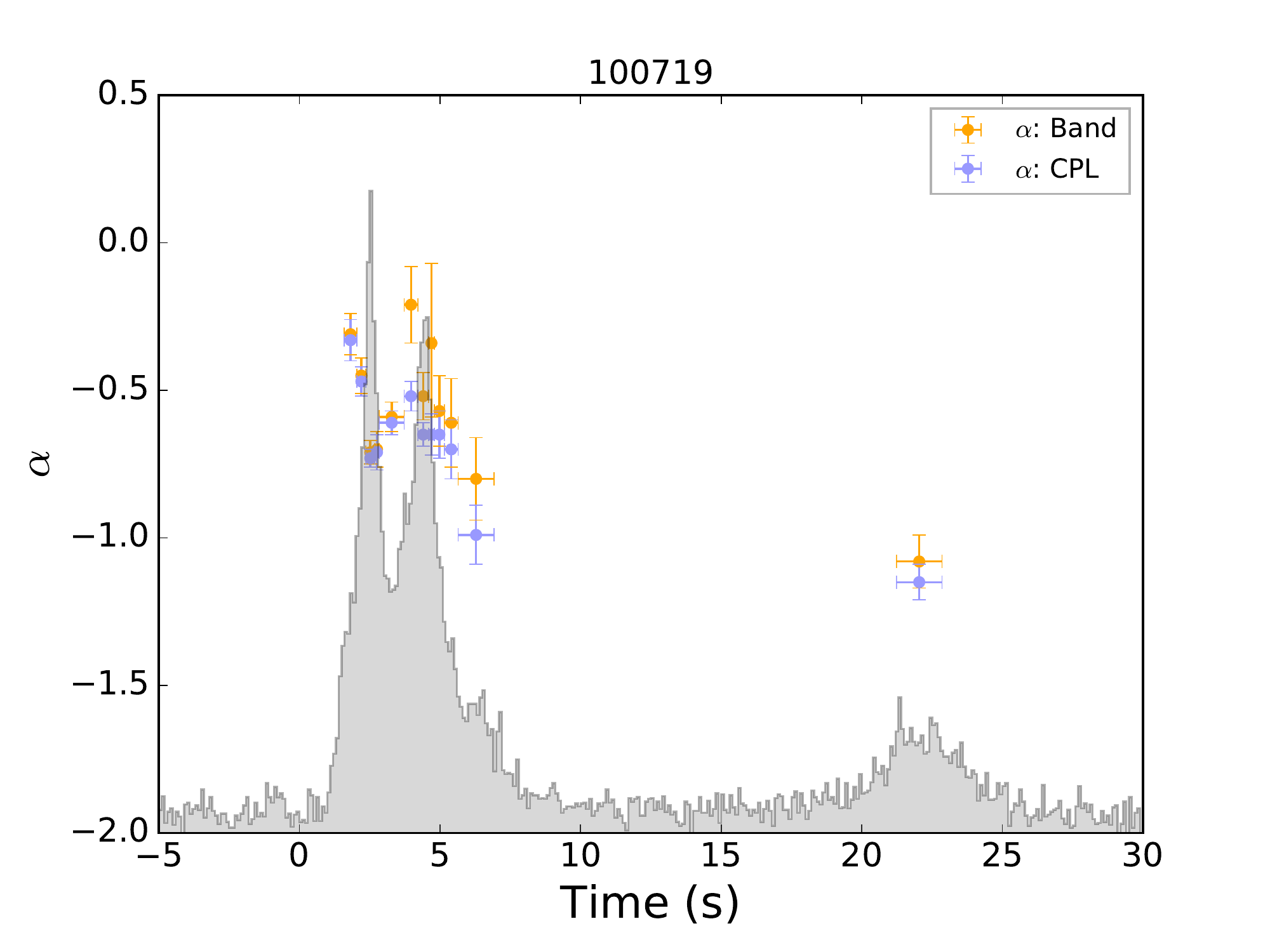}
\includegraphics[angle=0,scale=0.3]{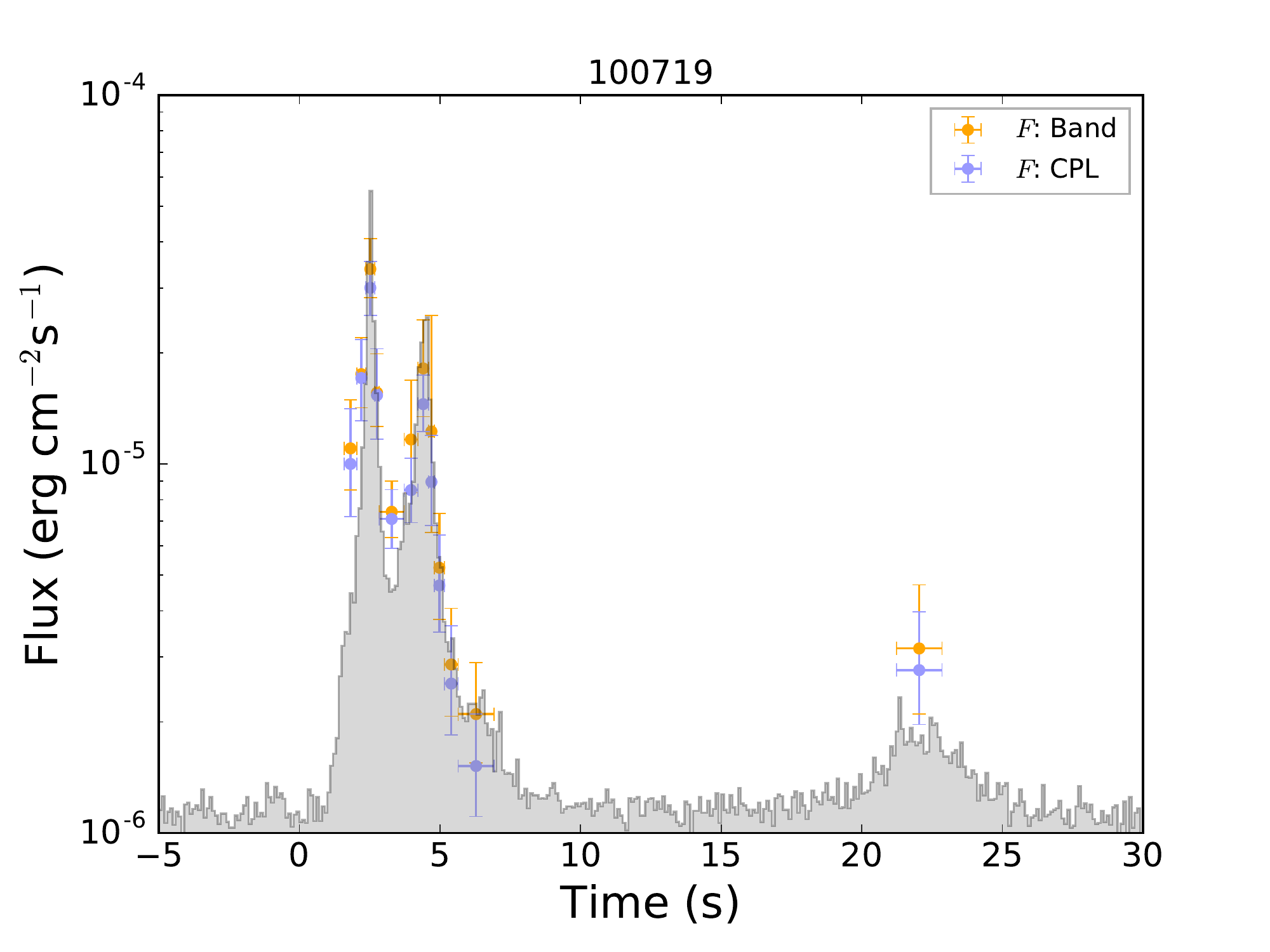}
\includegraphics[angle=0,scale=0.3]{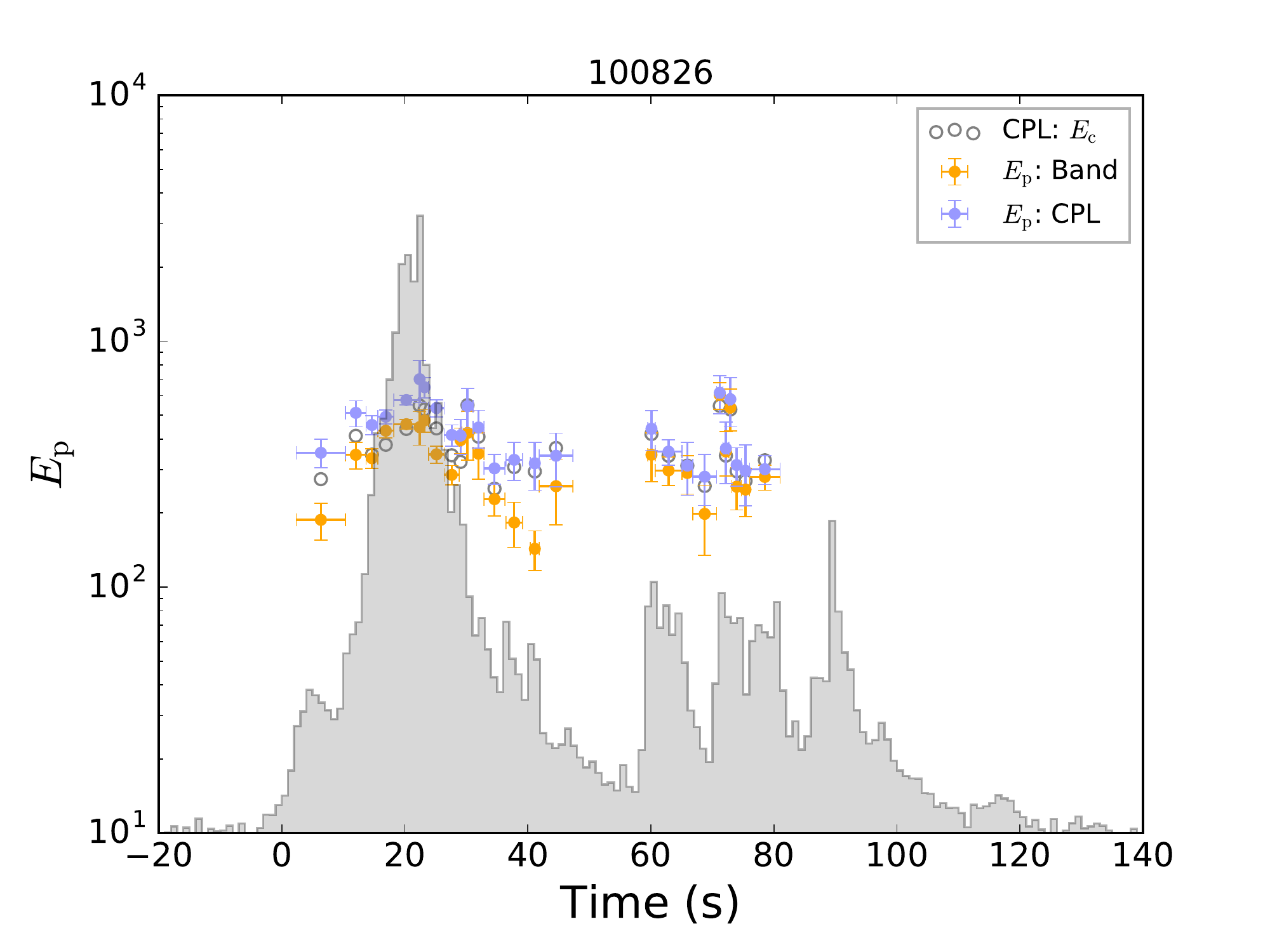}
\includegraphics[angle=0,scale=0.3]{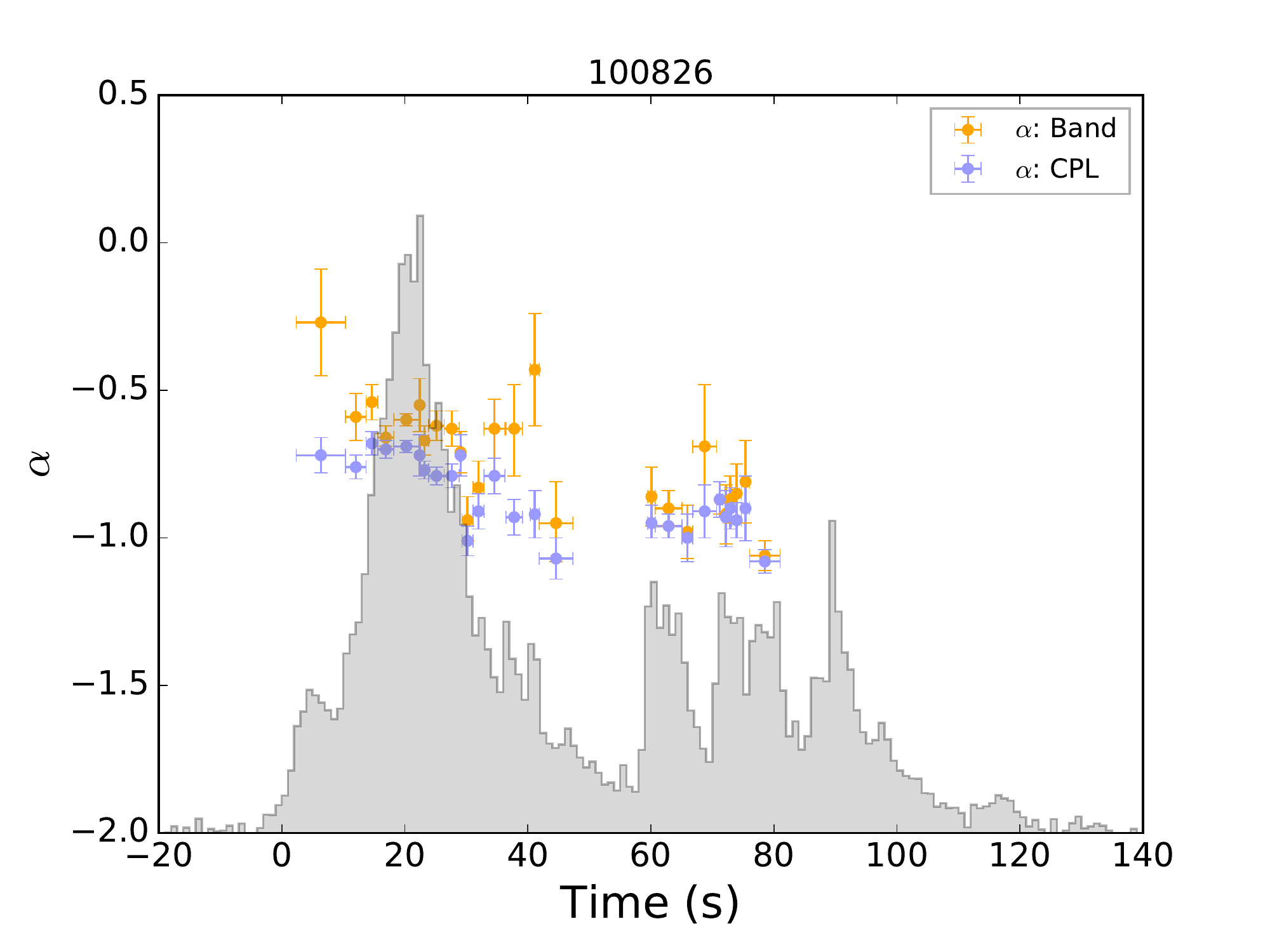}
\includegraphics[angle=0,scale=0.3]{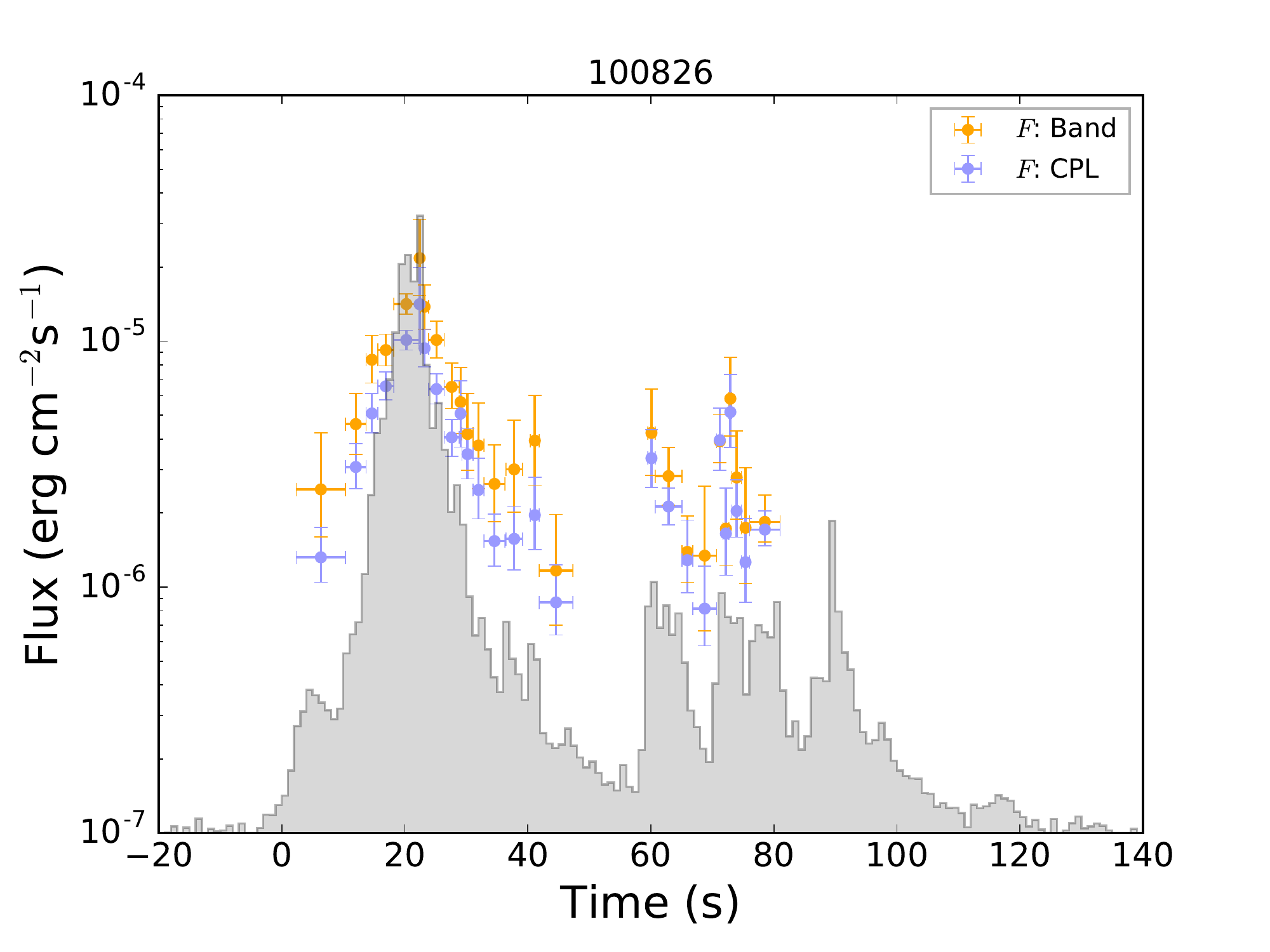}
\includegraphics[angle=0,scale=0.3]{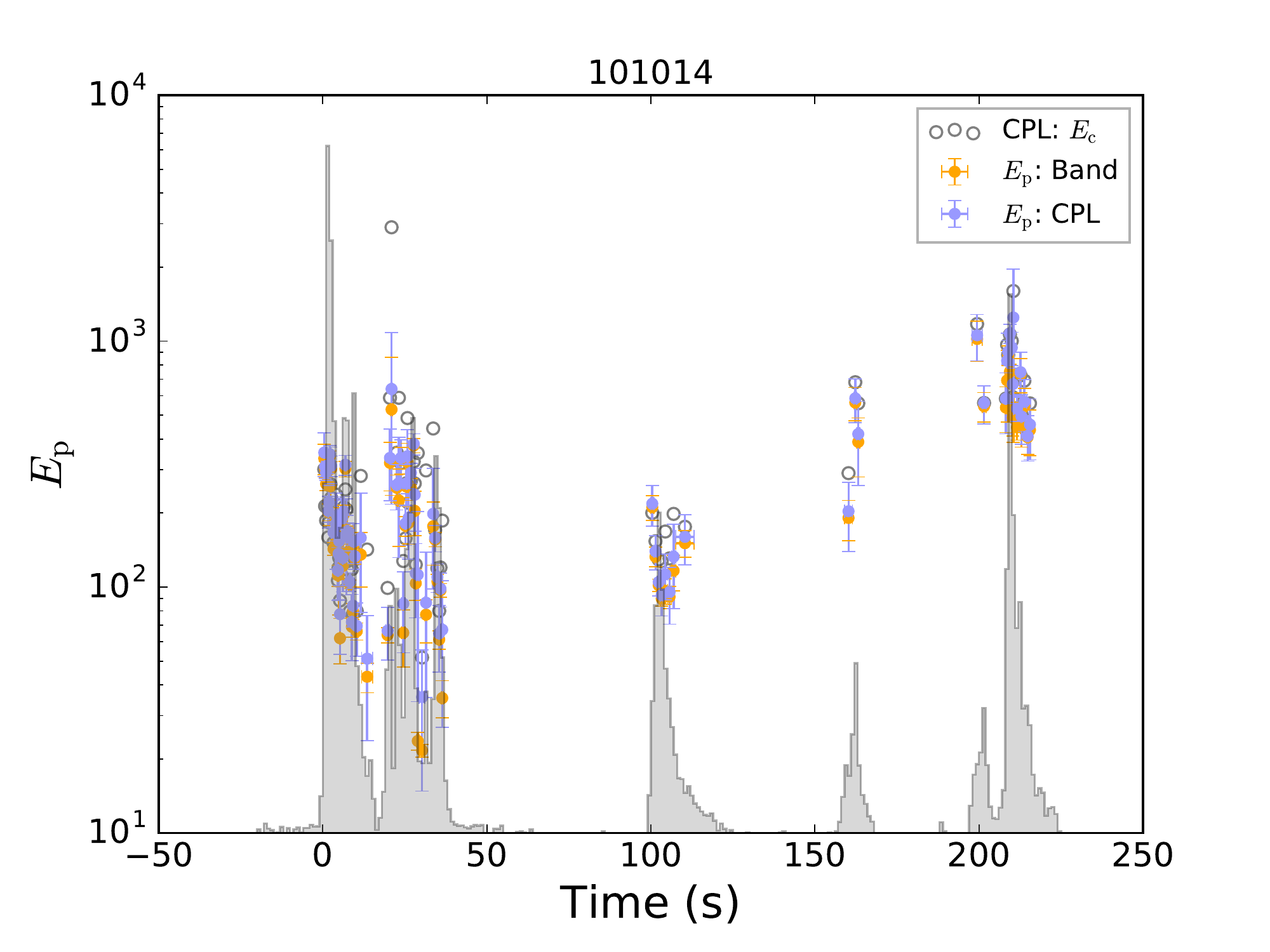}
\includegraphics[angle=0,scale=0.3]{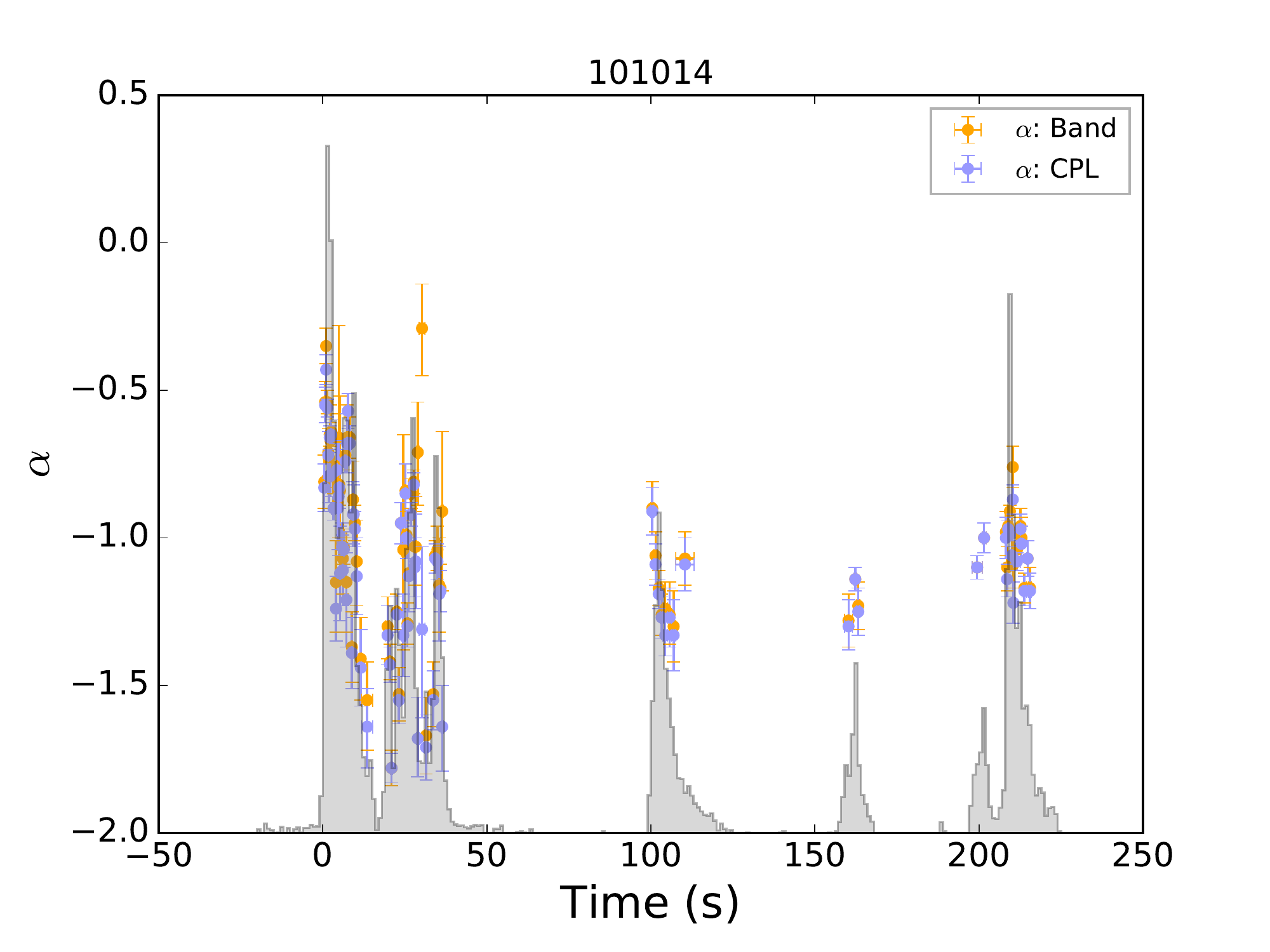}
\includegraphics[angle=0,scale=0.3]{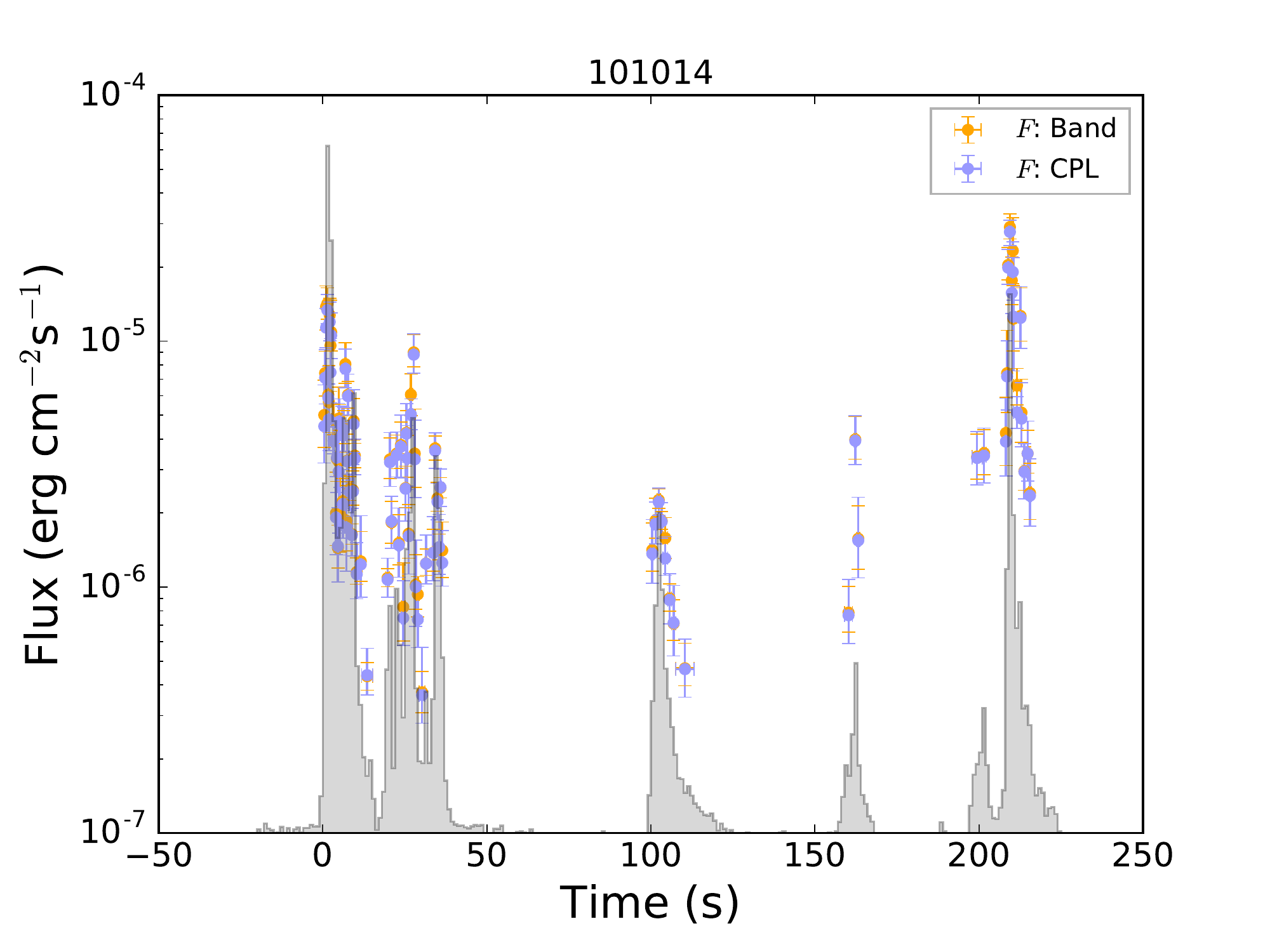}
\includegraphics[angle=0,scale=0.3]{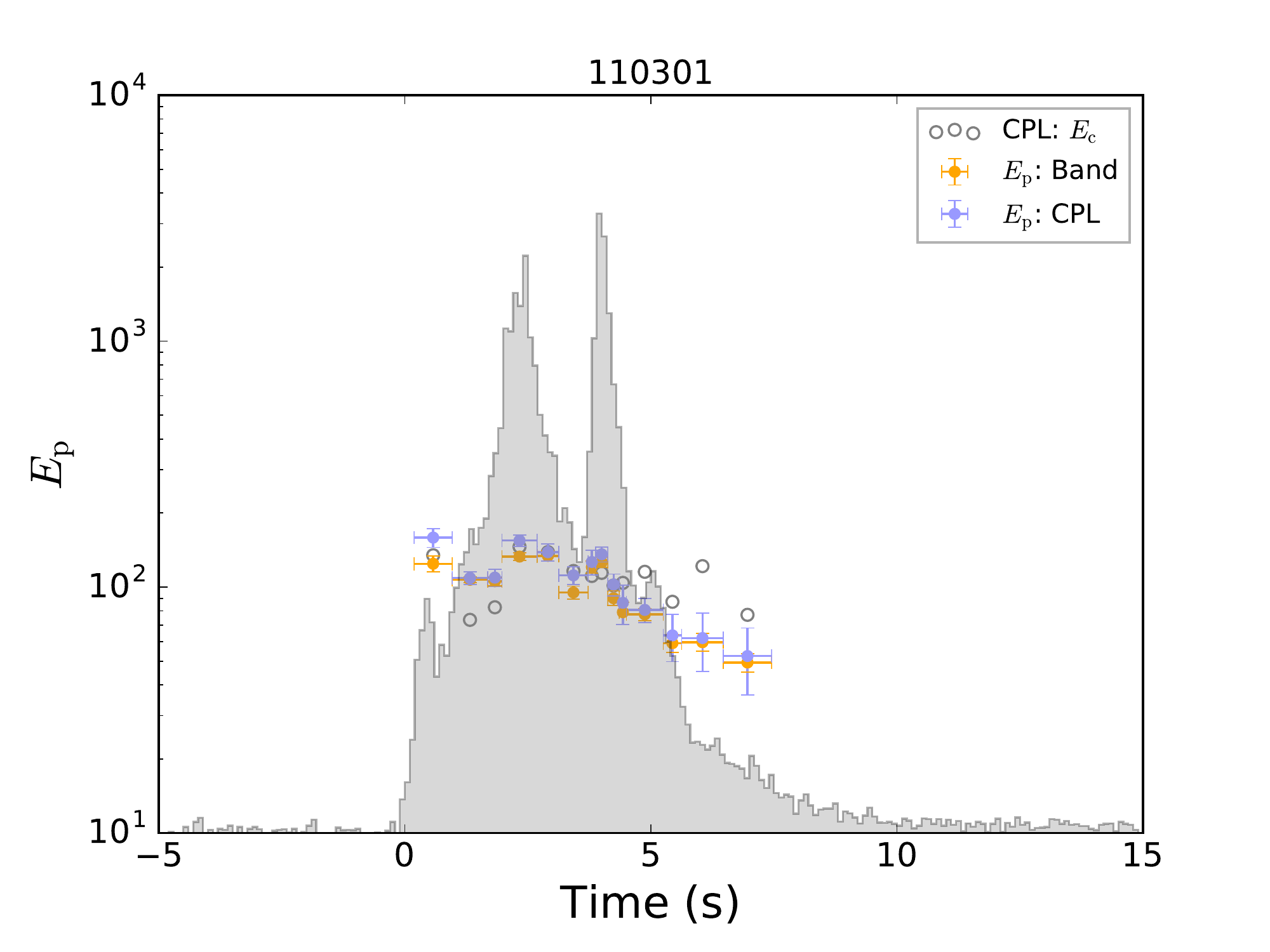}
\includegraphics[angle=0,scale=0.3]{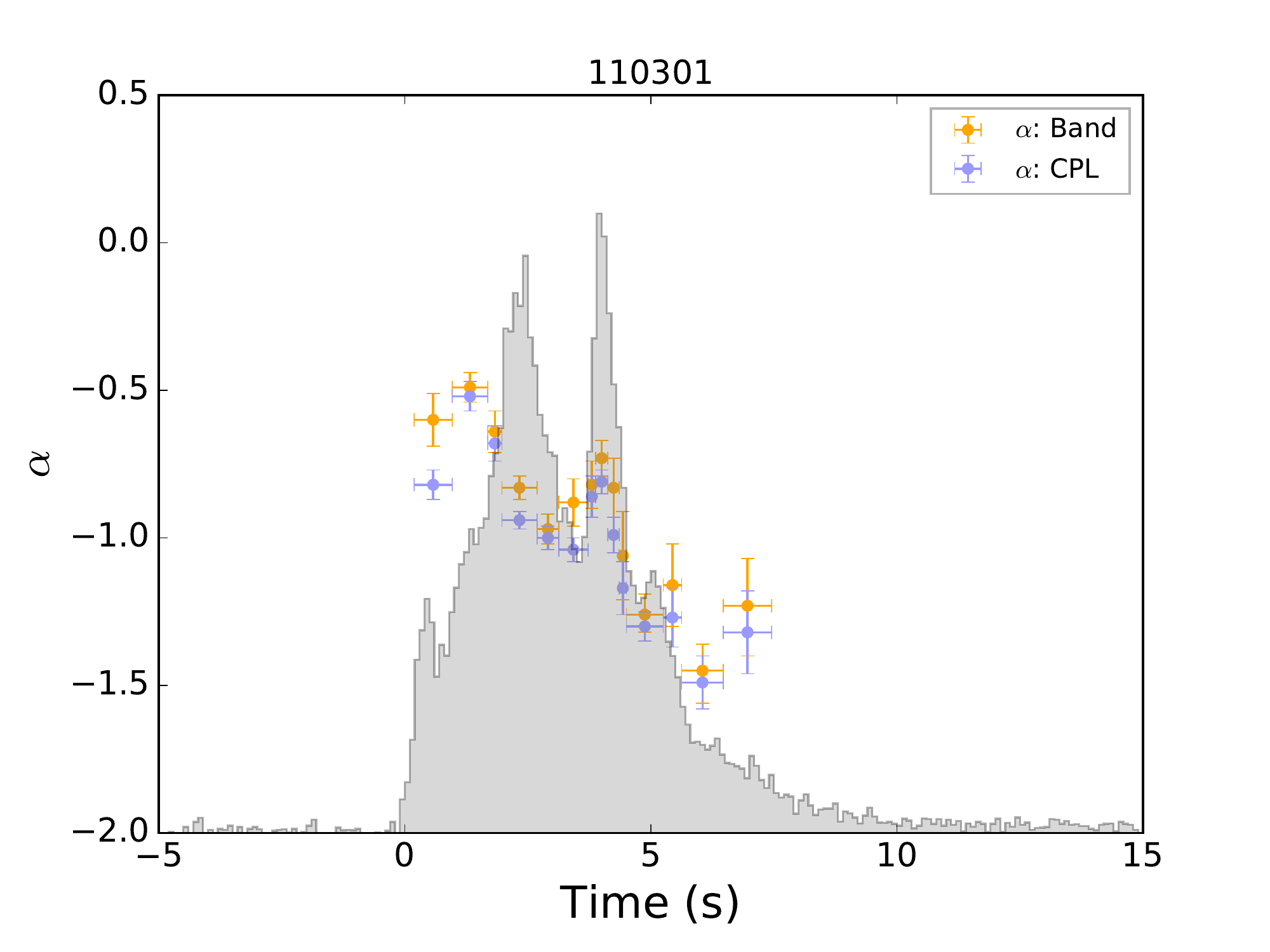}
\includegraphics[angle=0,scale=0.3]{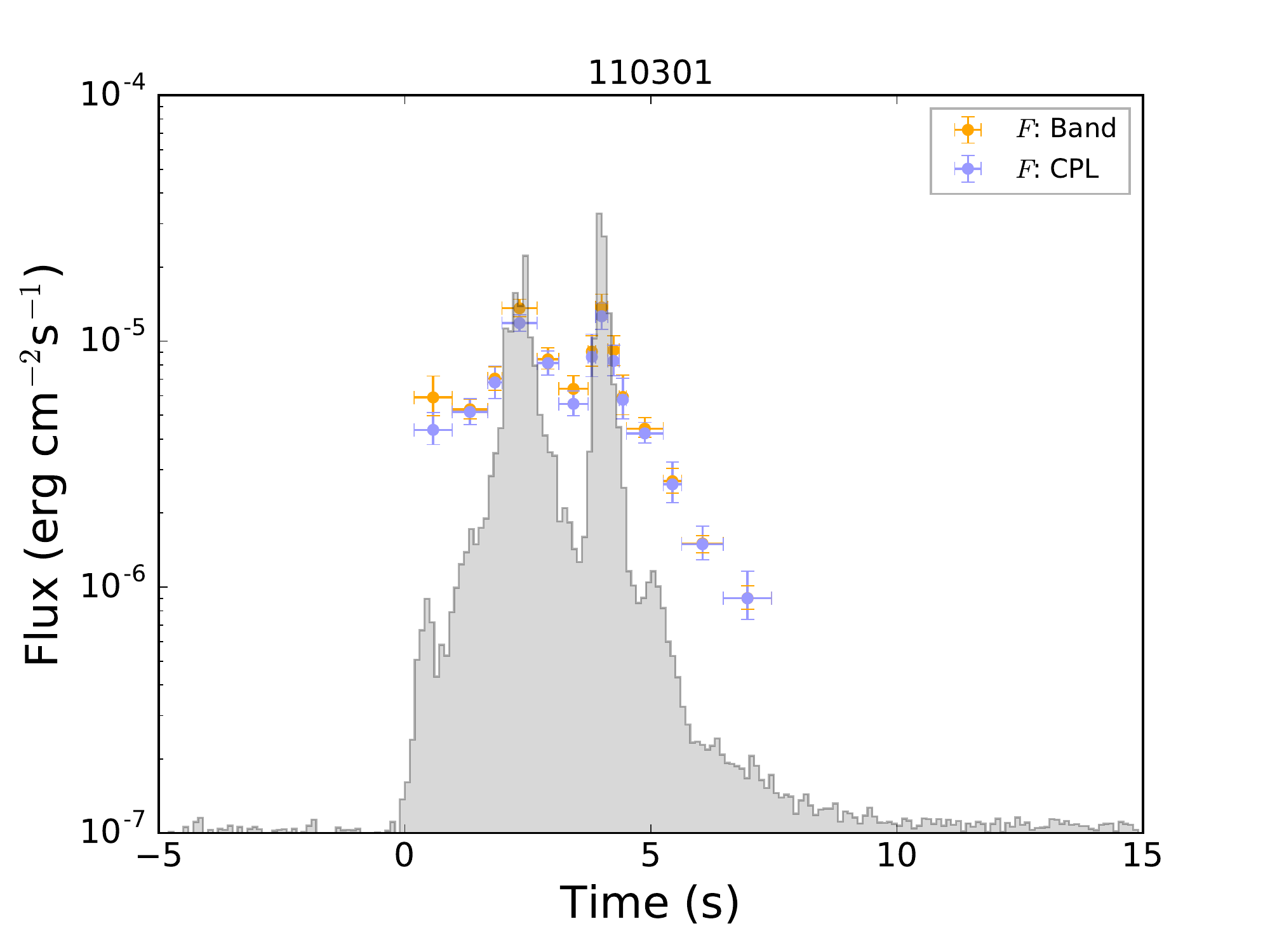}
\includegraphics[angle=0,scale=0.3]{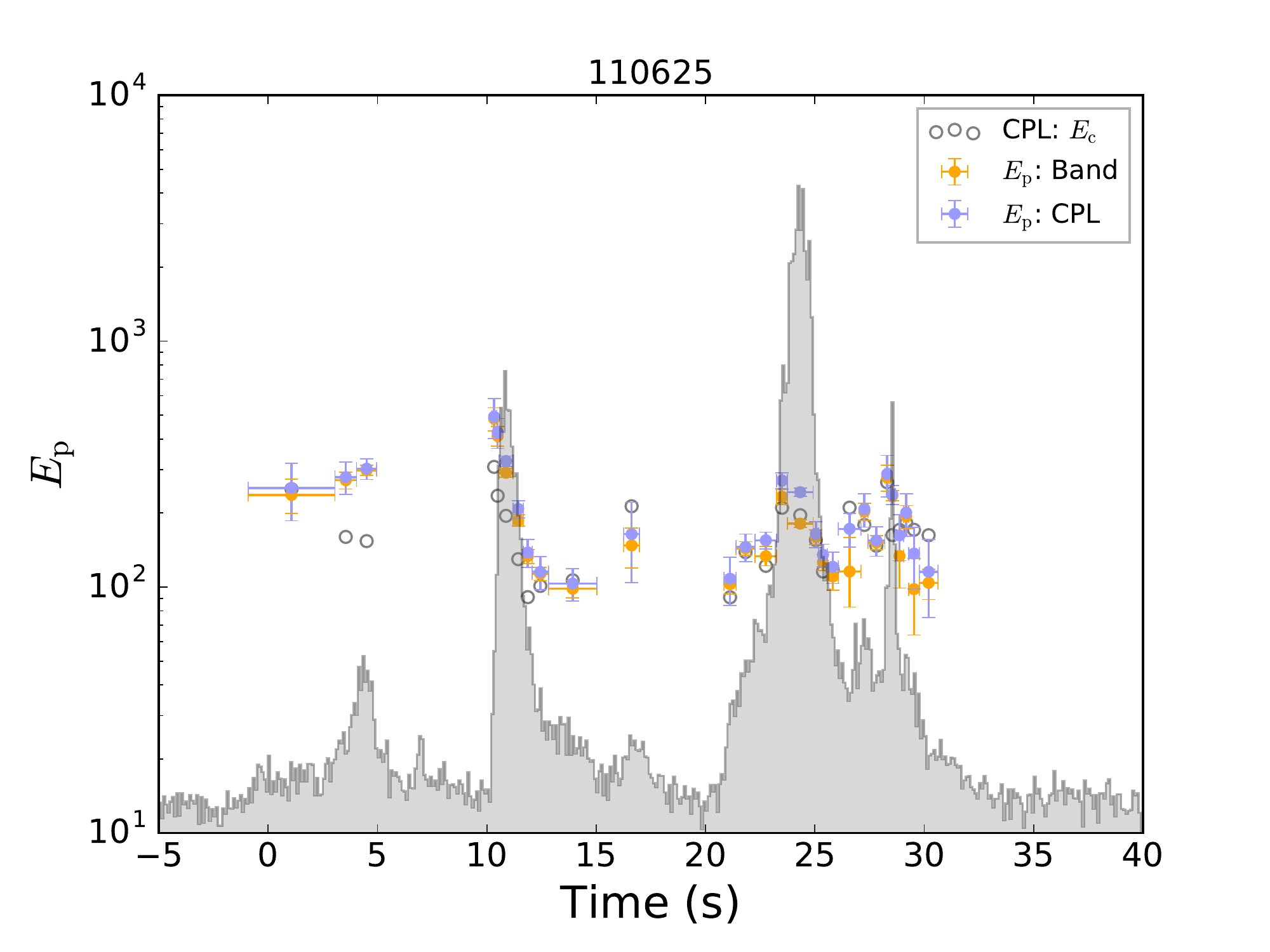}
\includegraphics[angle=0,scale=0.3]{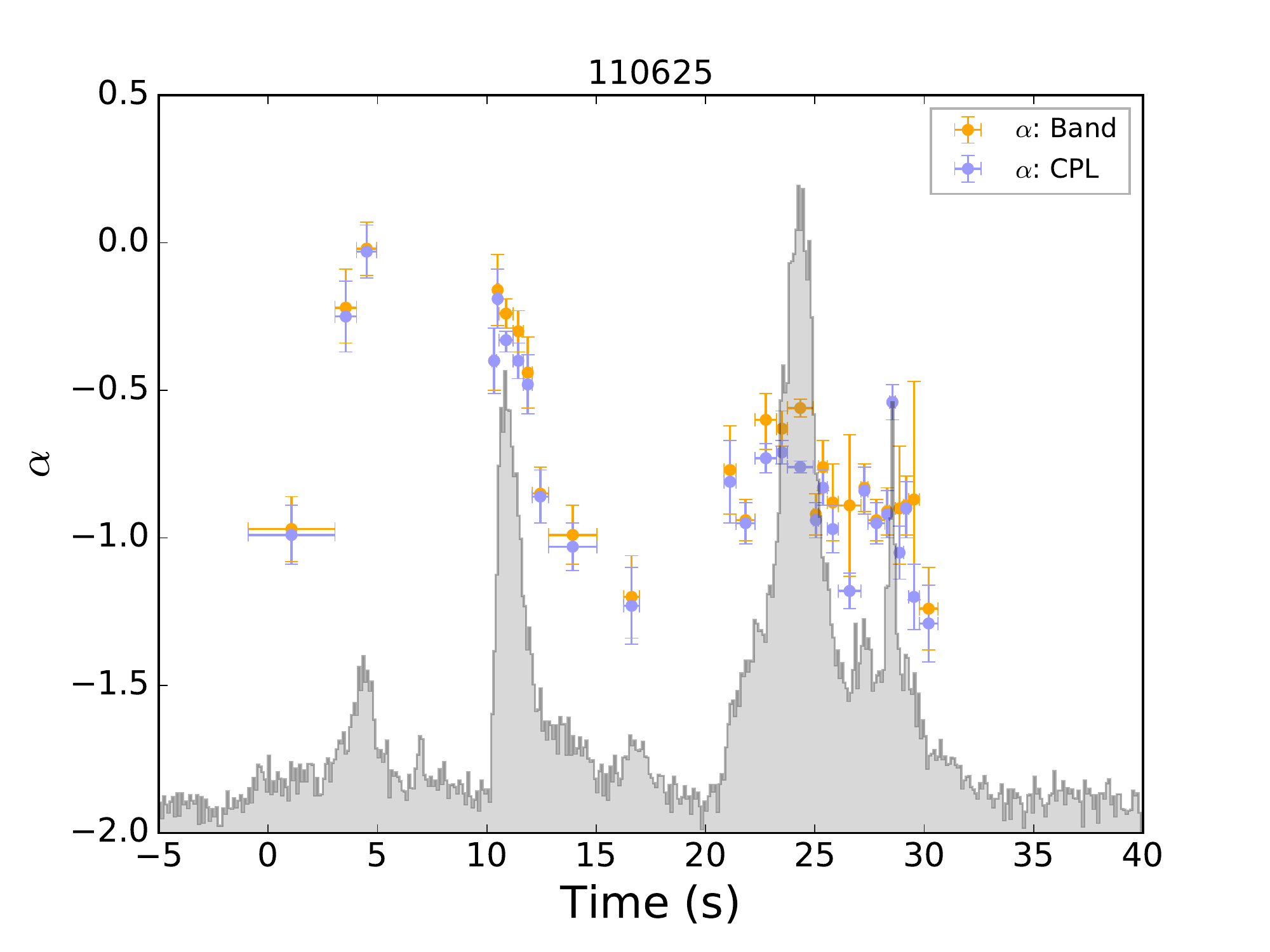}
\includegraphics[angle=0,scale=0.3]{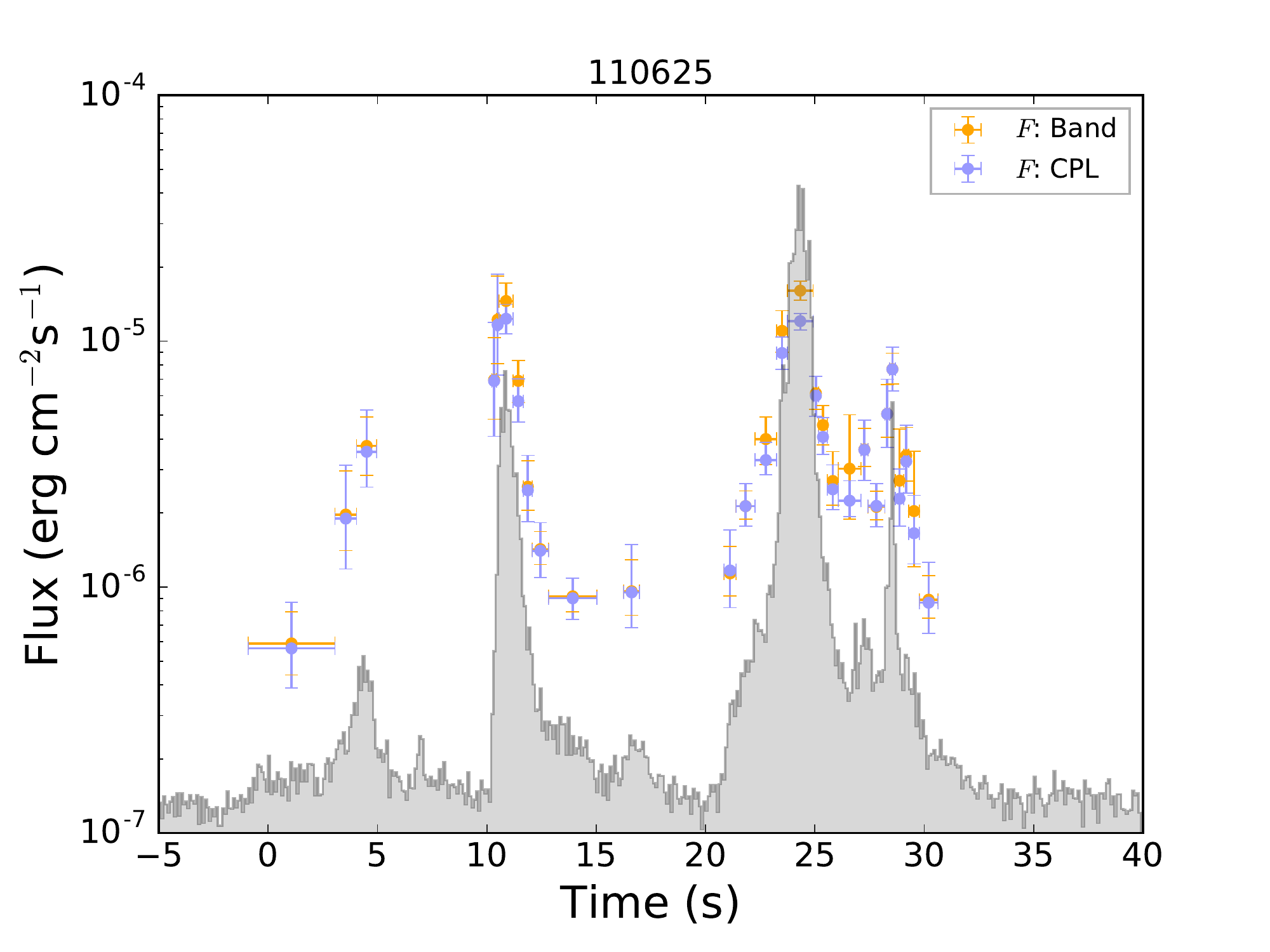}
\center{Fig. \ref{fig:evolution}--- Continued}
\end{figure*}
\begin{figure*}
\includegraphics[angle=0,scale=0.3]{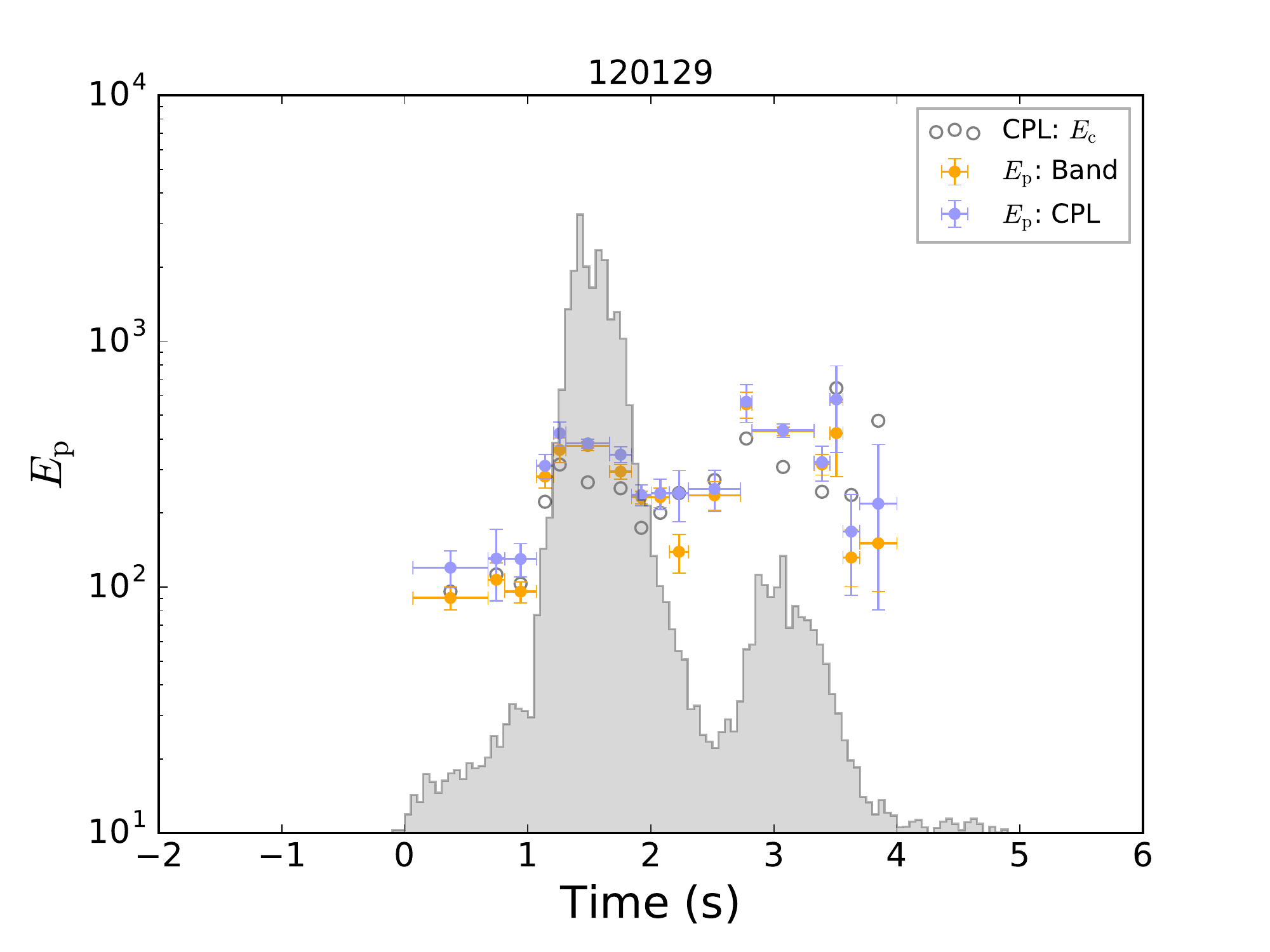}
\includegraphics[angle=0,scale=0.3]{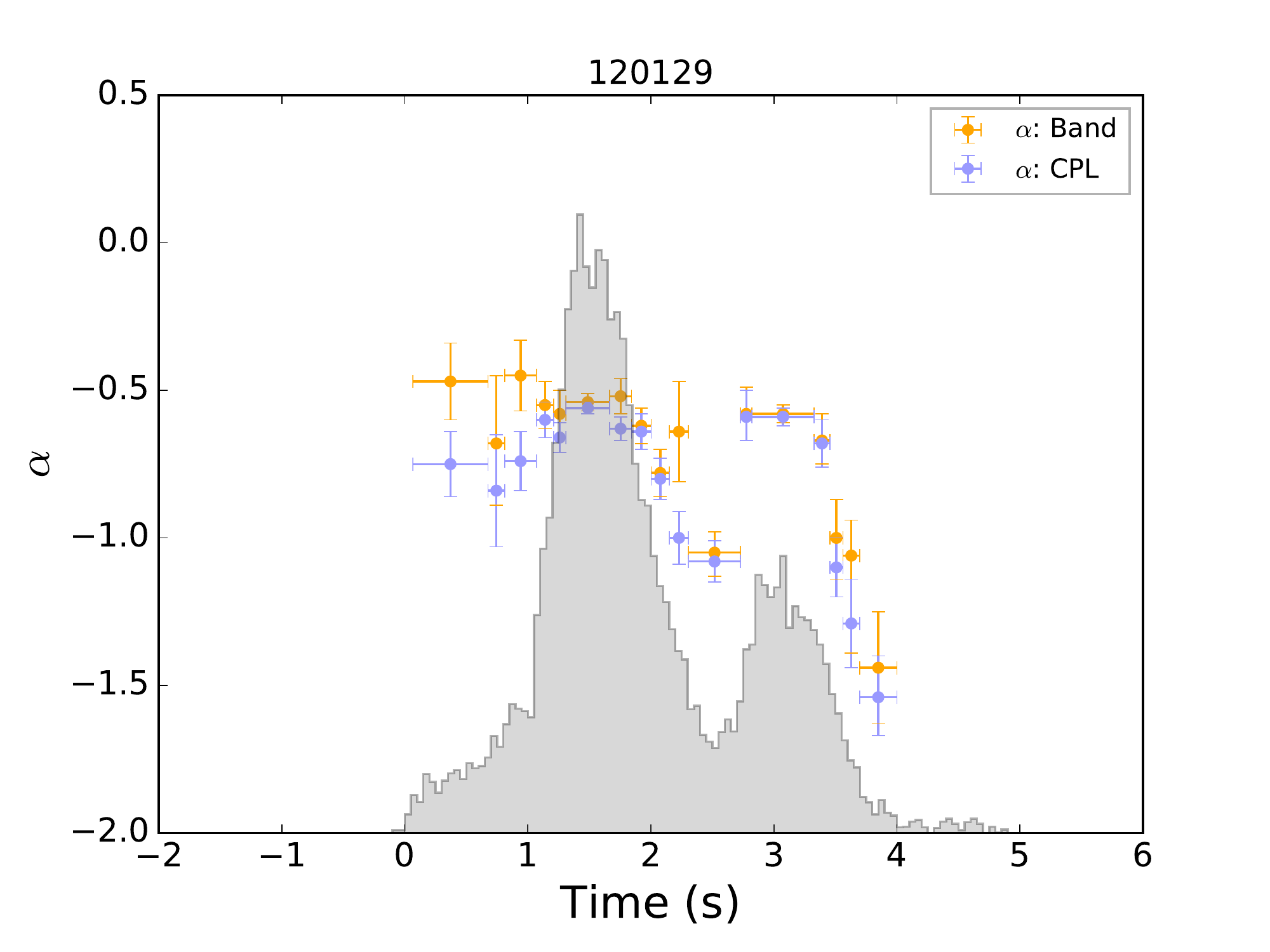}
\includegraphics[angle=0,scale=0.3]{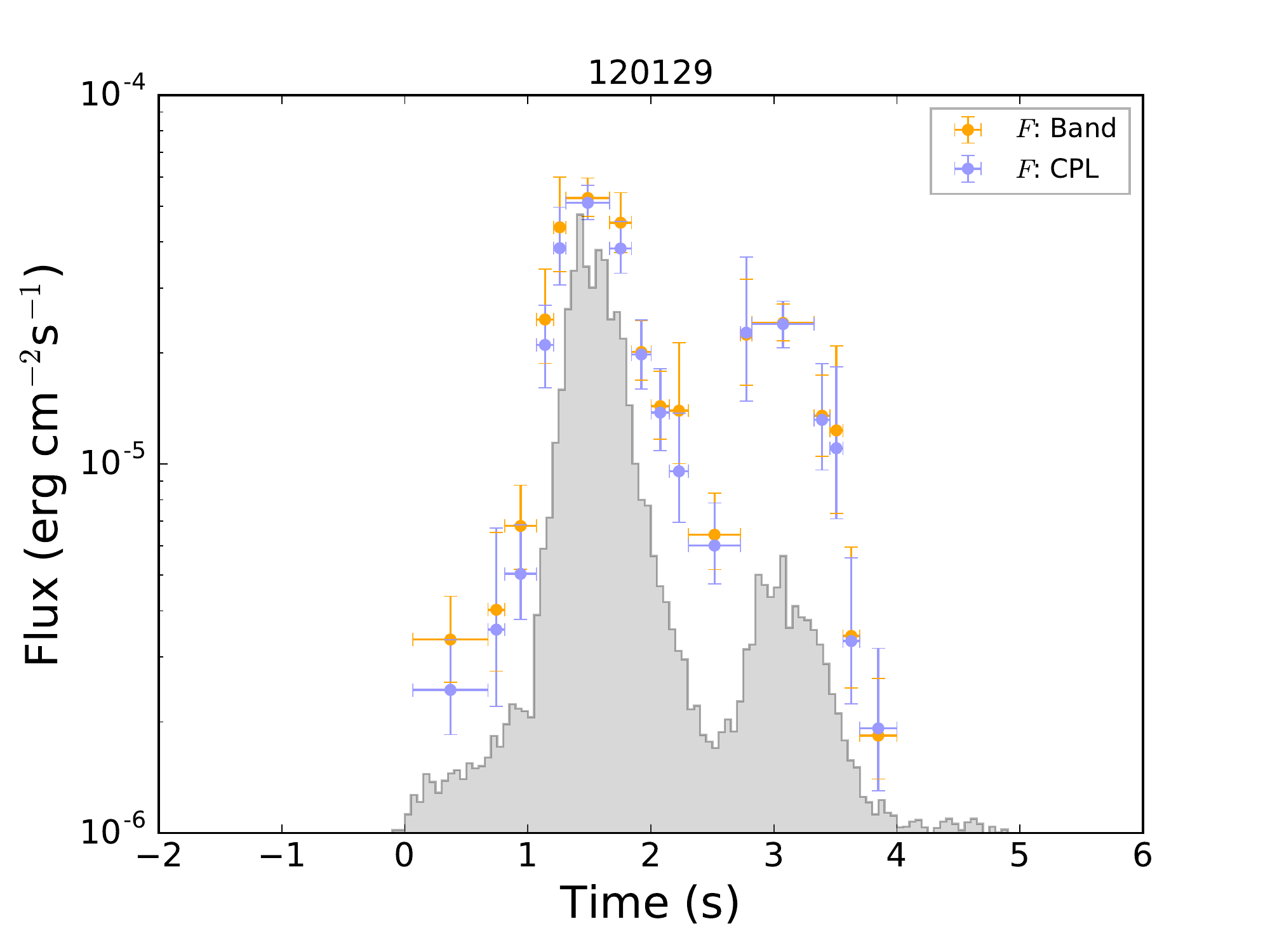}
\includegraphics[angle=0,scale=0.3]{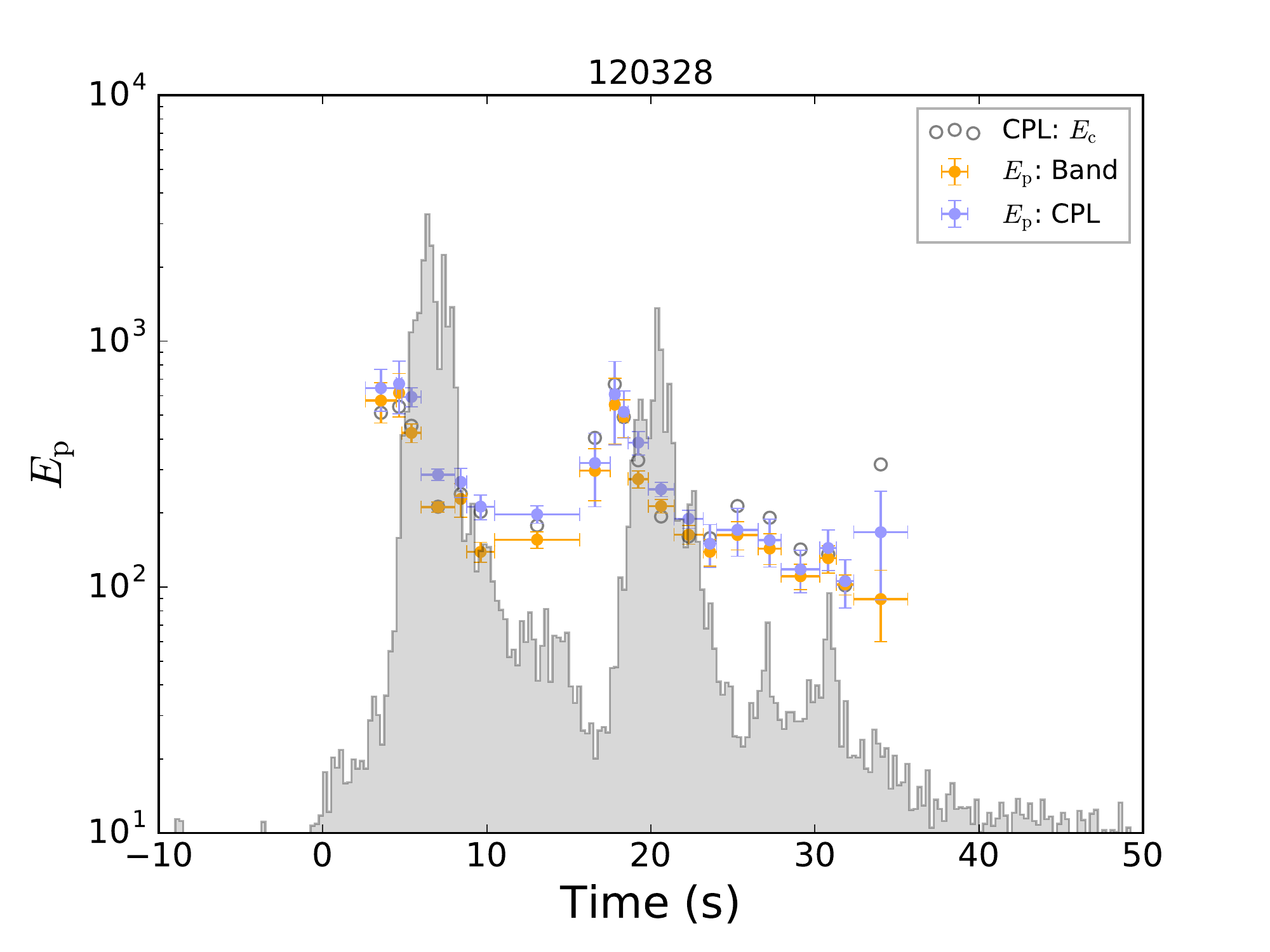}
\includegraphics[angle=0,scale=0.3]{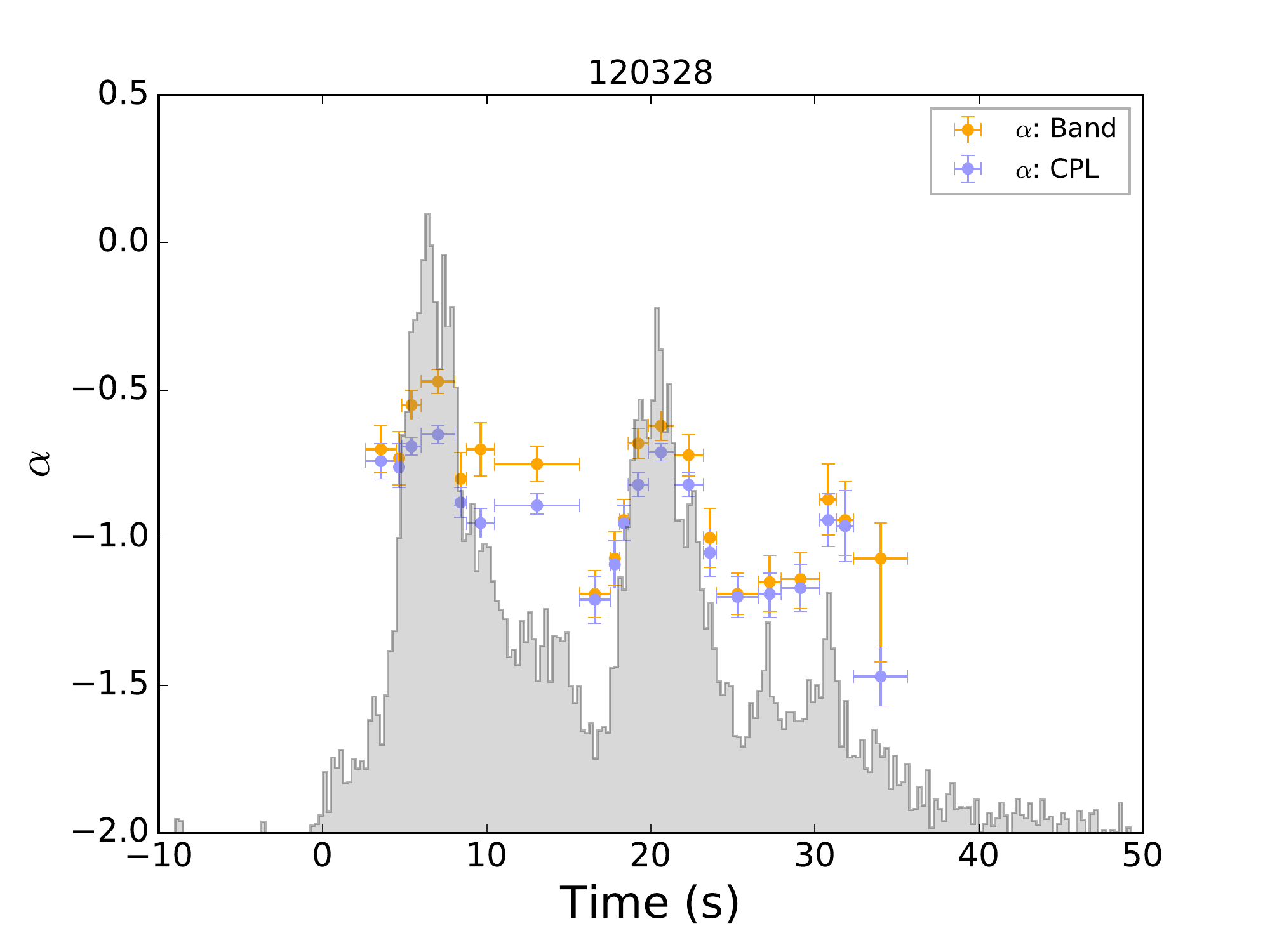}
\includegraphics[angle=0,scale=0.3]{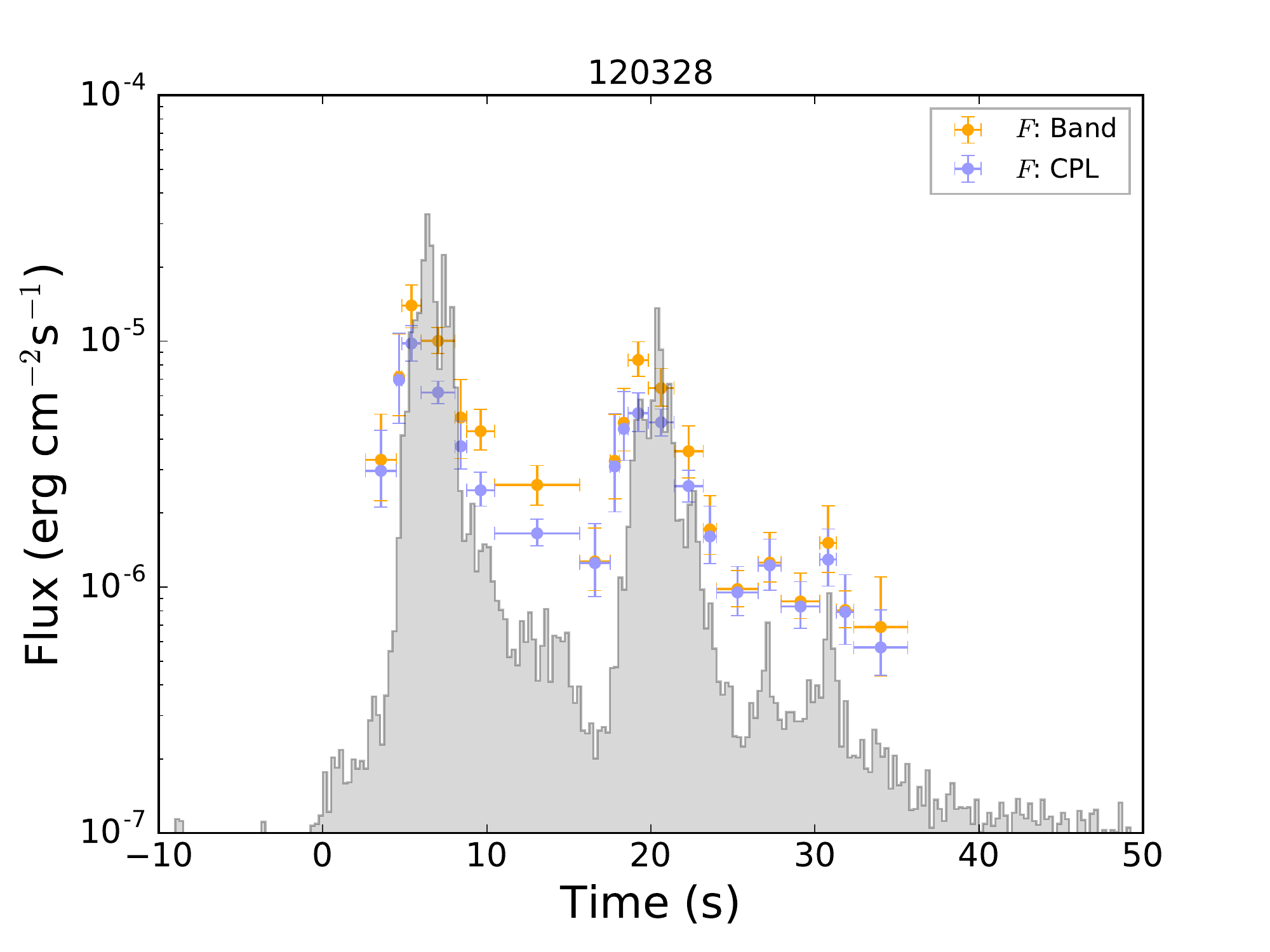}
\includegraphics[angle=0,scale=0.3]{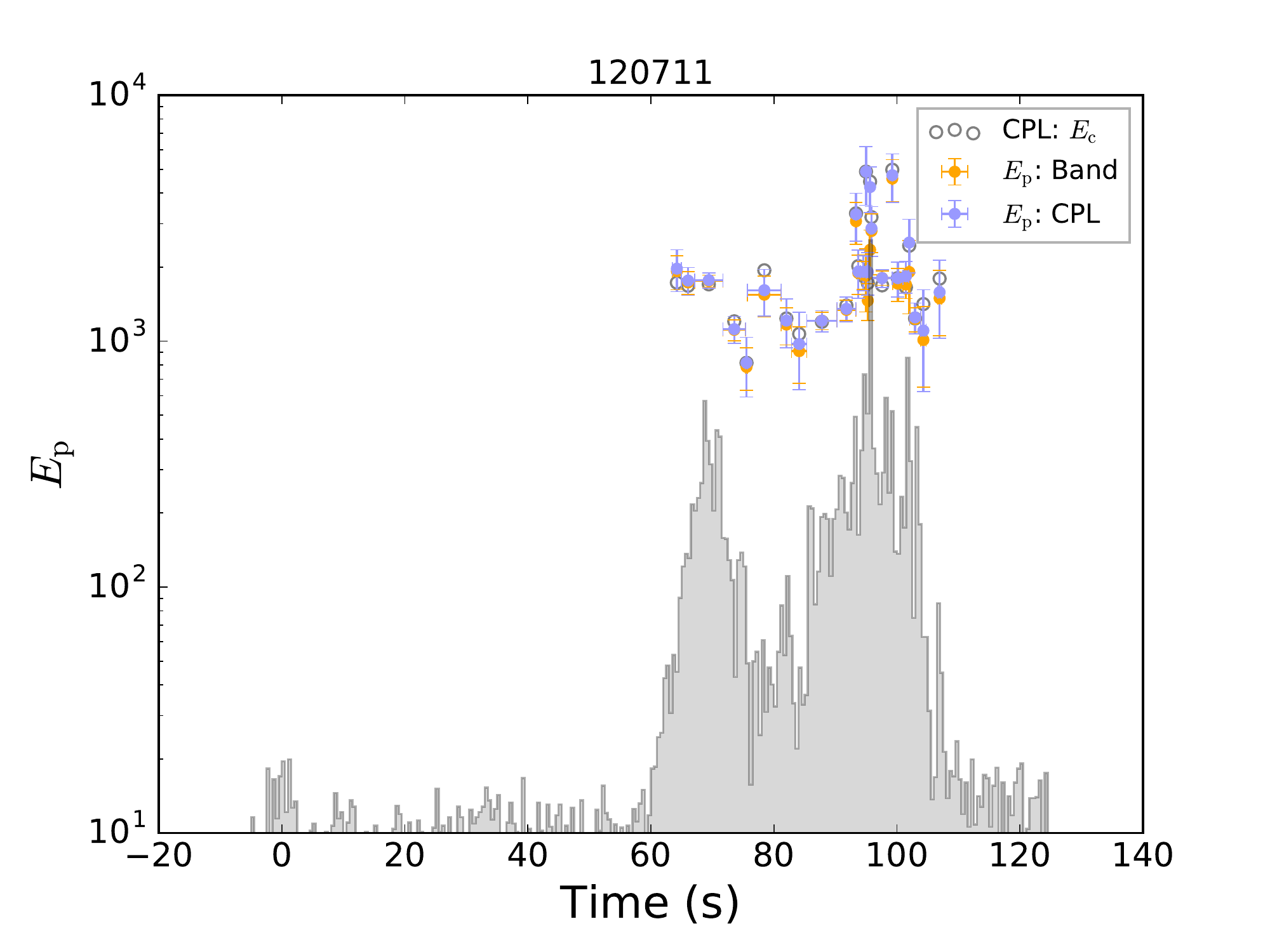}
\includegraphics[angle=0,scale=0.3]{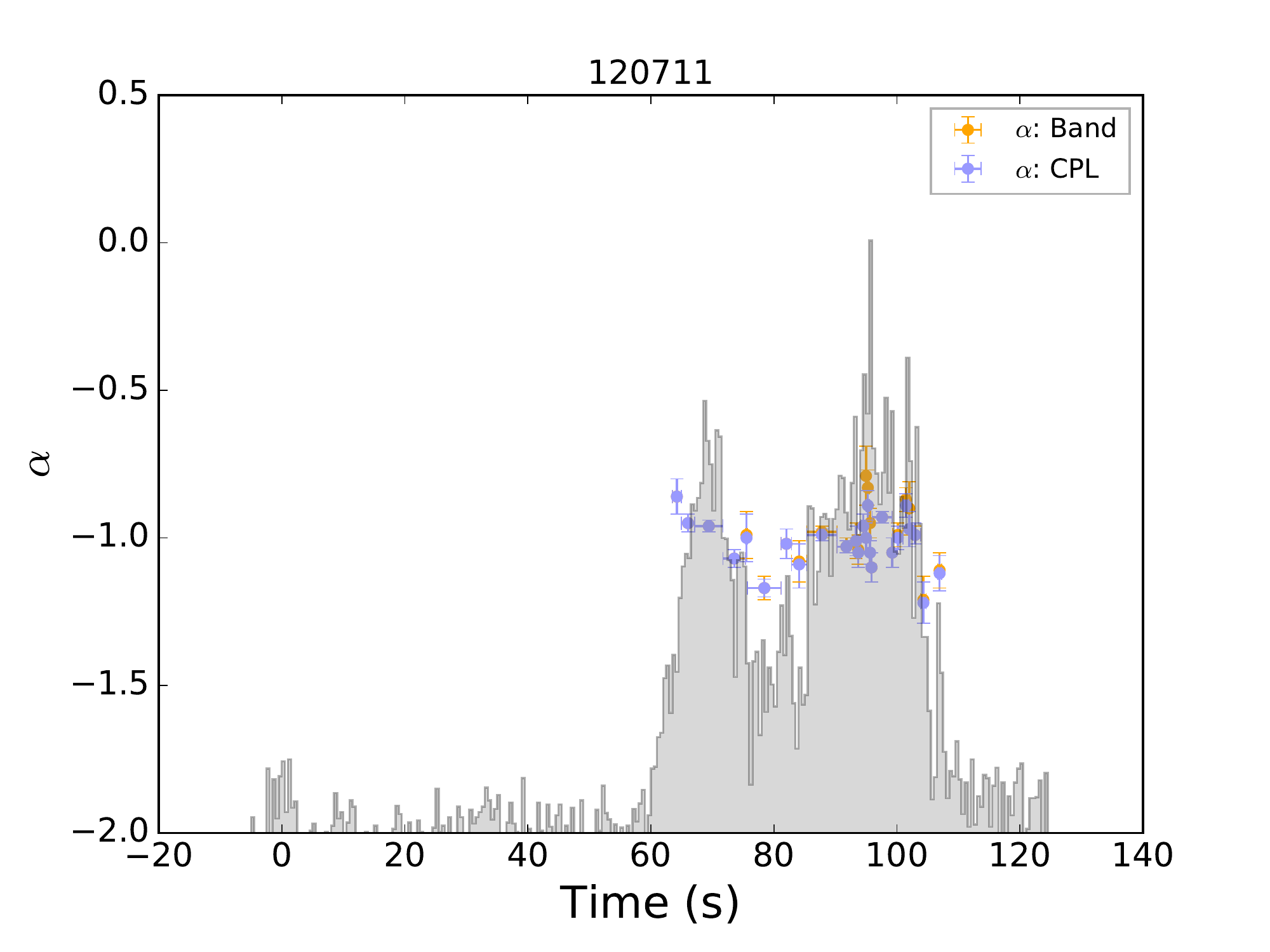}
\includegraphics[angle=0,scale=0.3]{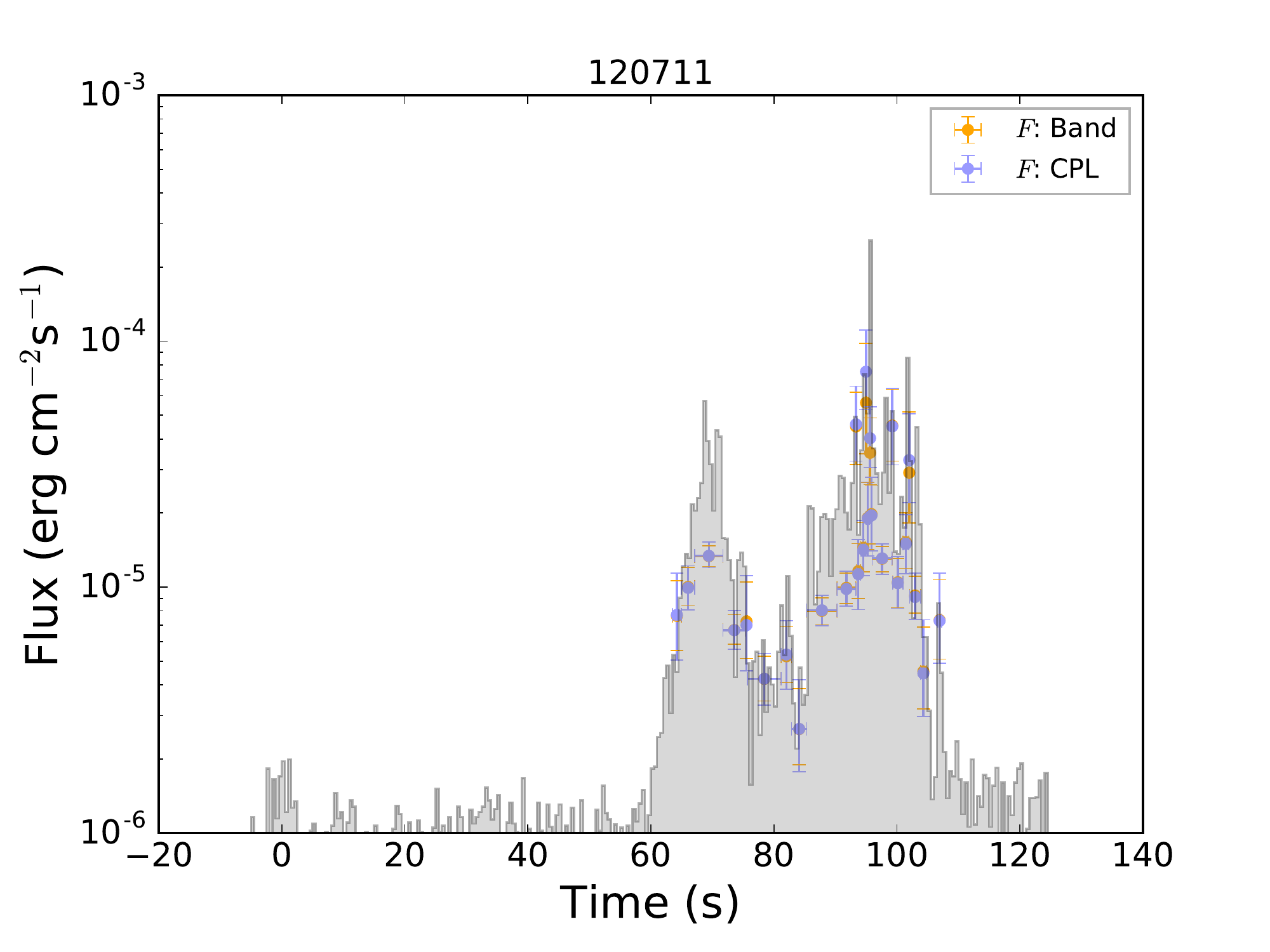}
\includegraphics[angle=0,scale=0.3]{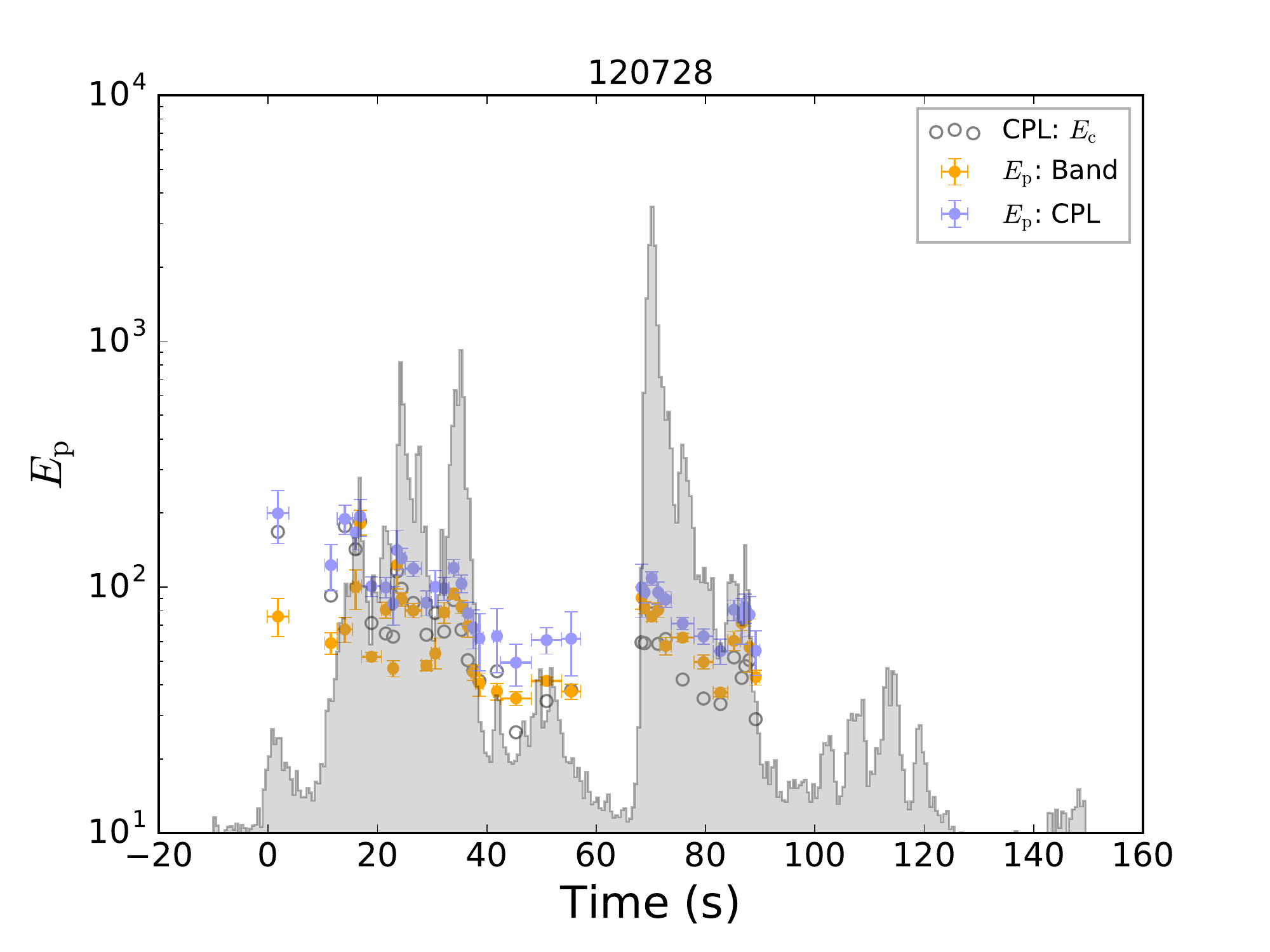}
\includegraphics[angle=0,scale=0.3]{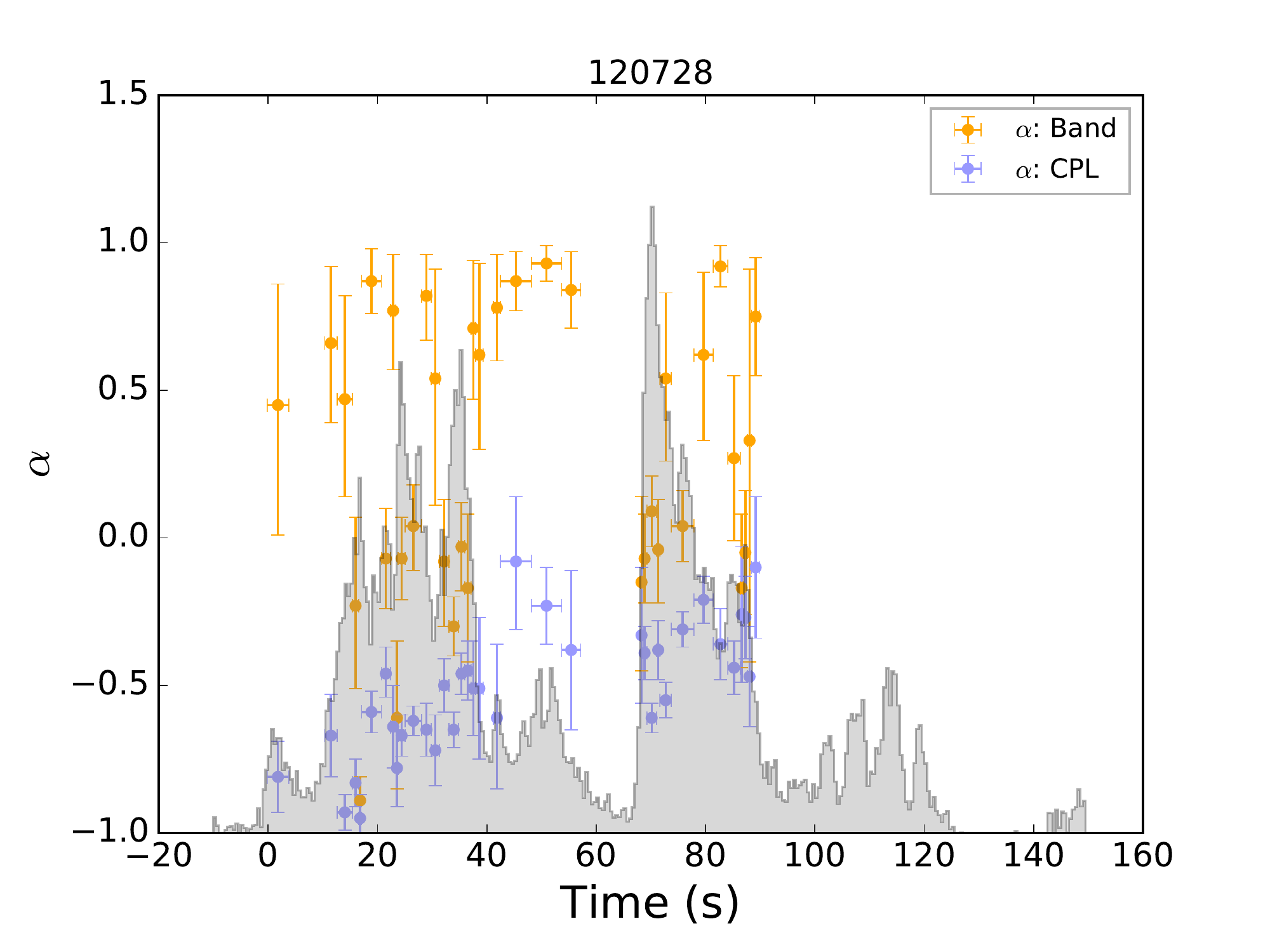}
\includegraphics[angle=0,scale=0.3]{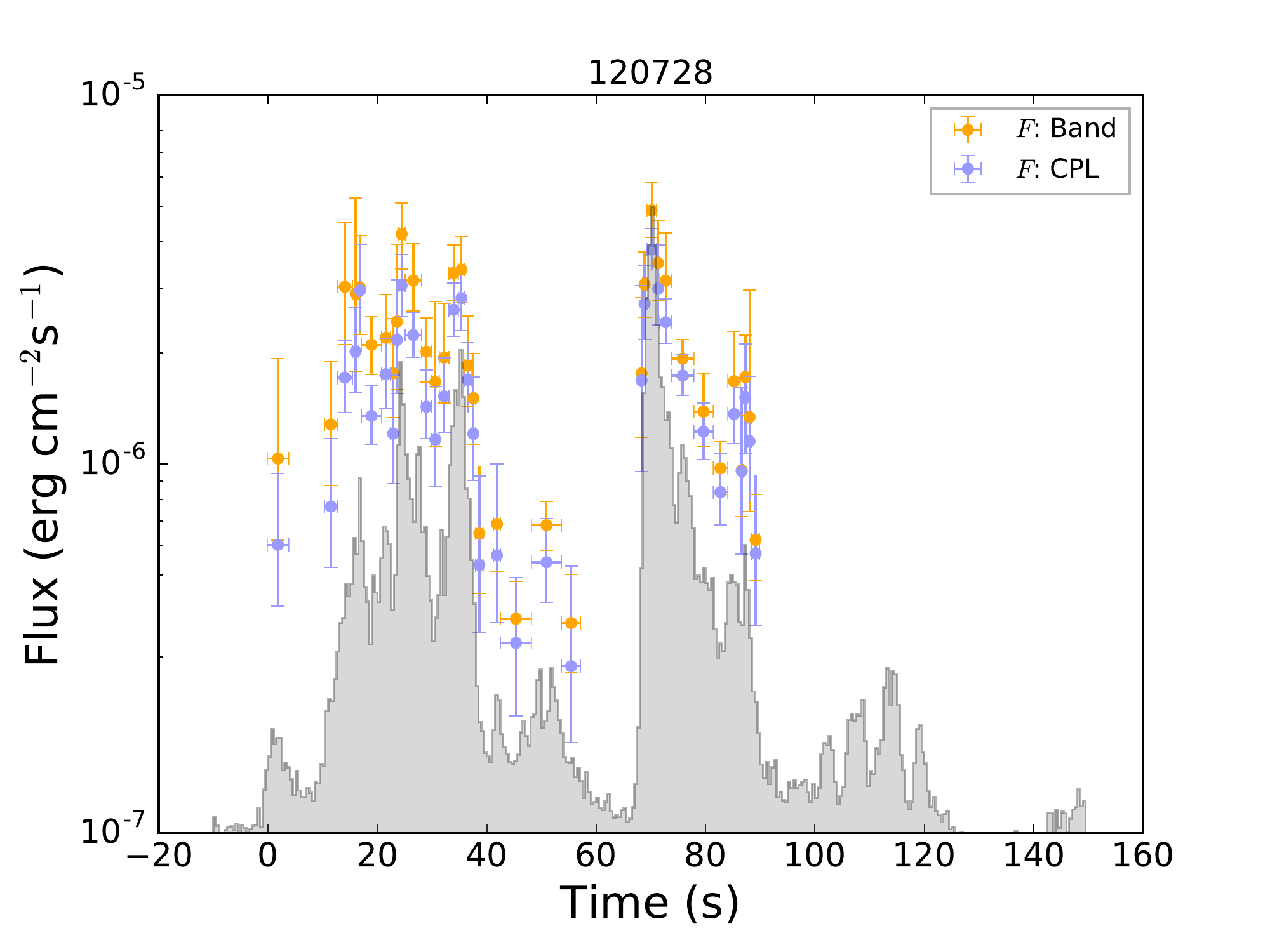}
\includegraphics[angle=0,scale=0.3]{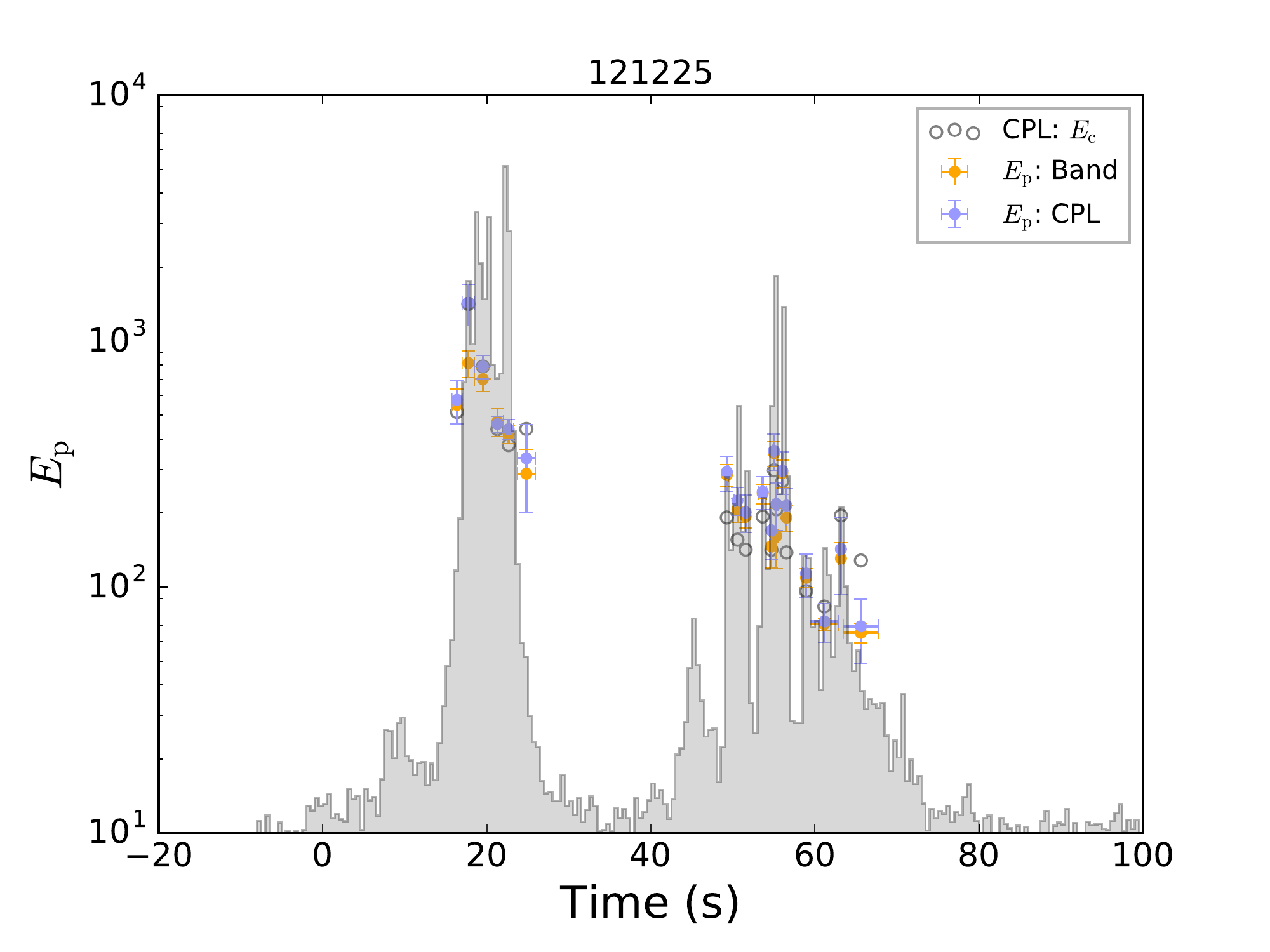}
\includegraphics[angle=0,scale=0.3]{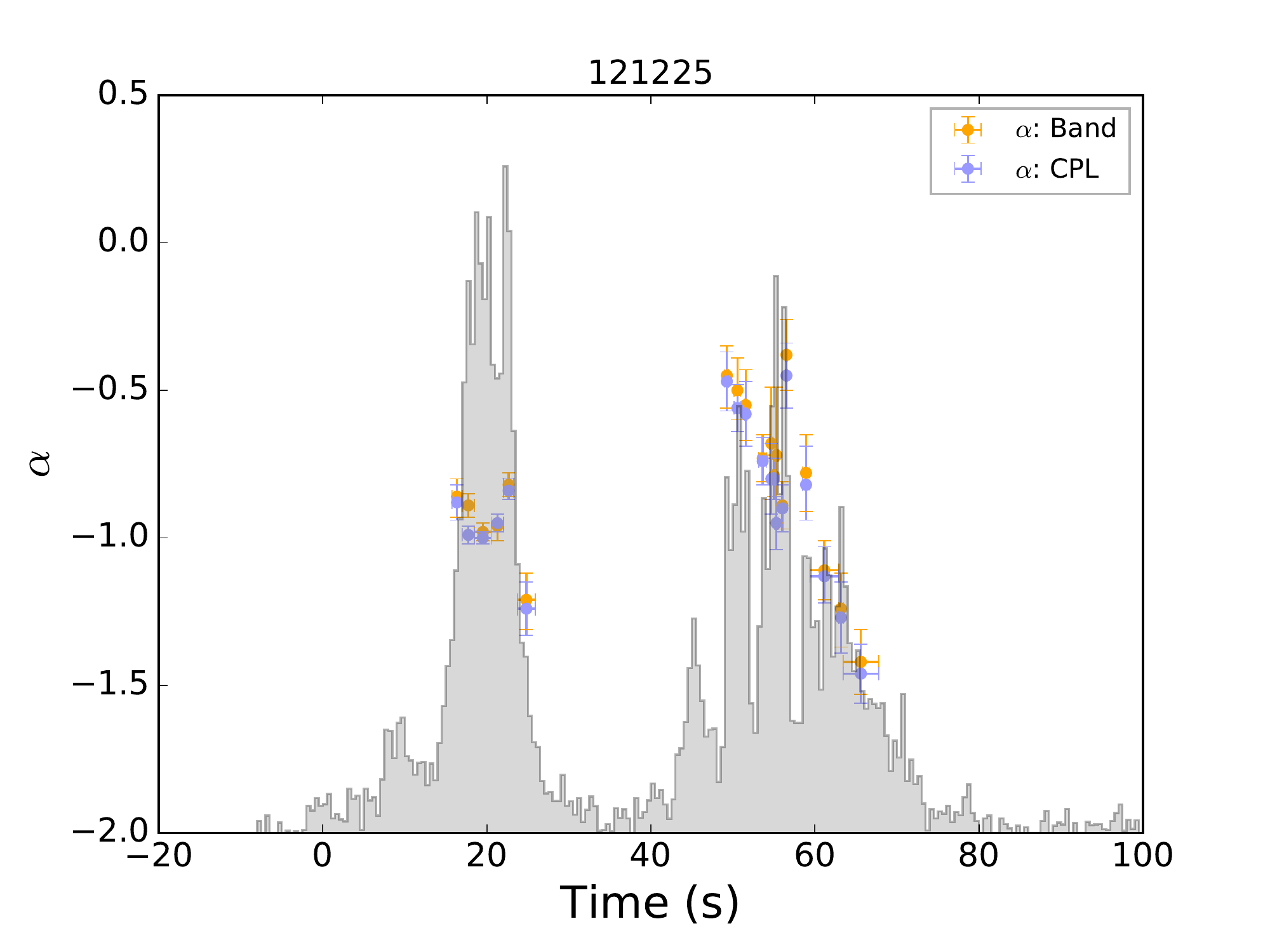}
\includegraphics[angle=0,scale=0.3]{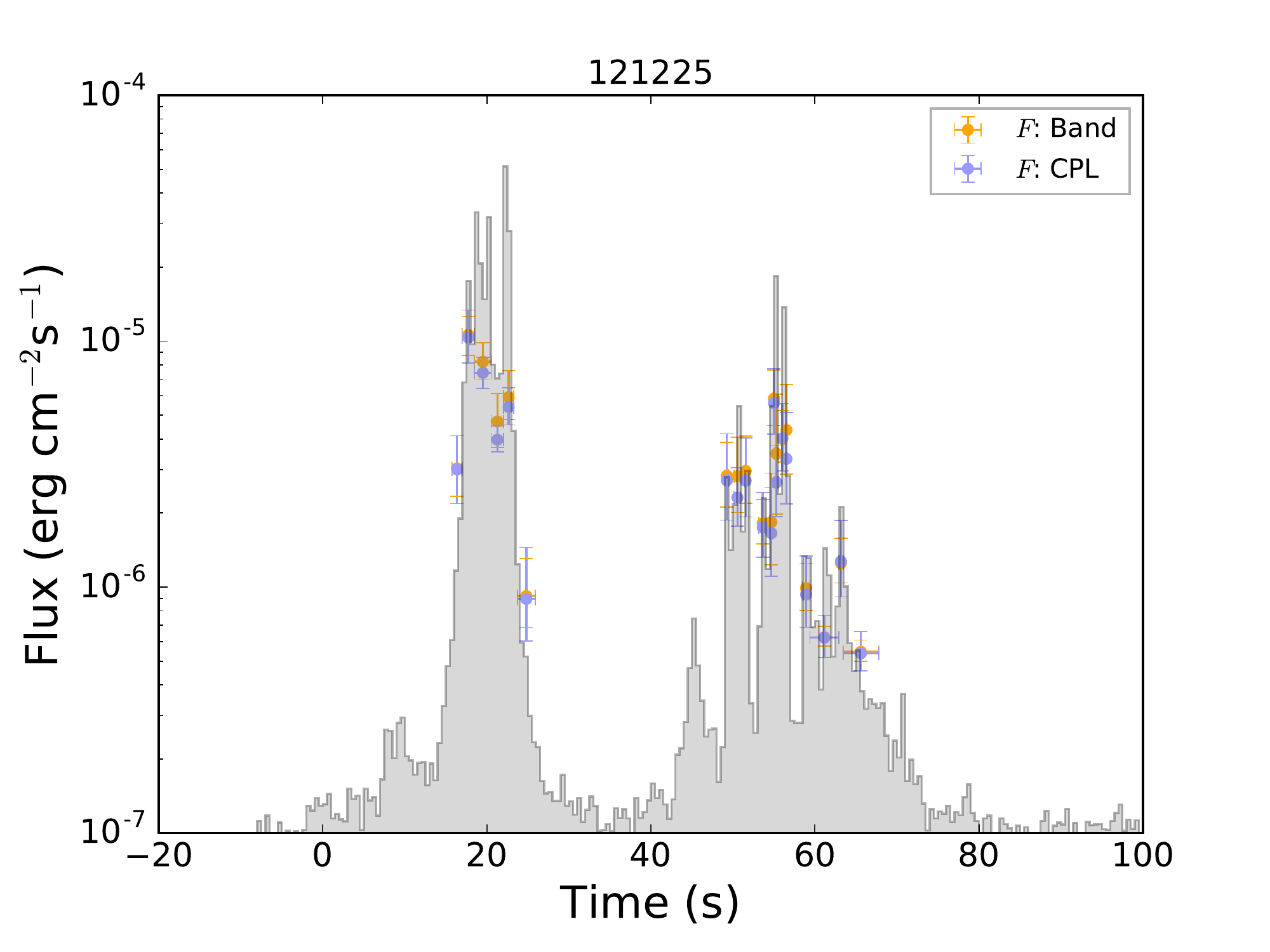}
\center{Fig. \ref{fig:evolution}--- Continued}
\end{figure*}
\begin{figure*}
\includegraphics[angle=0,scale=0.3]{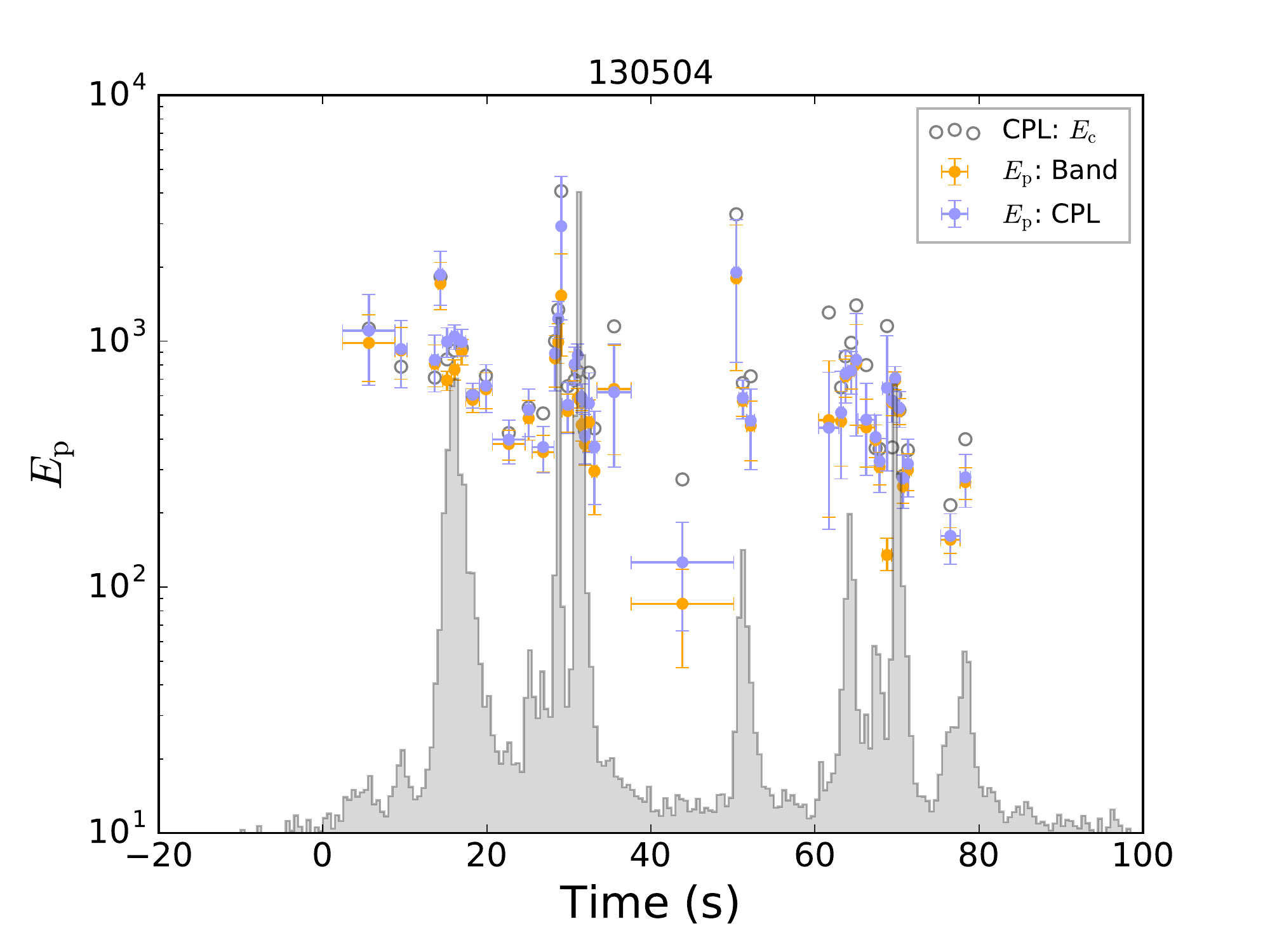}
\includegraphics[angle=0,scale=0.3]{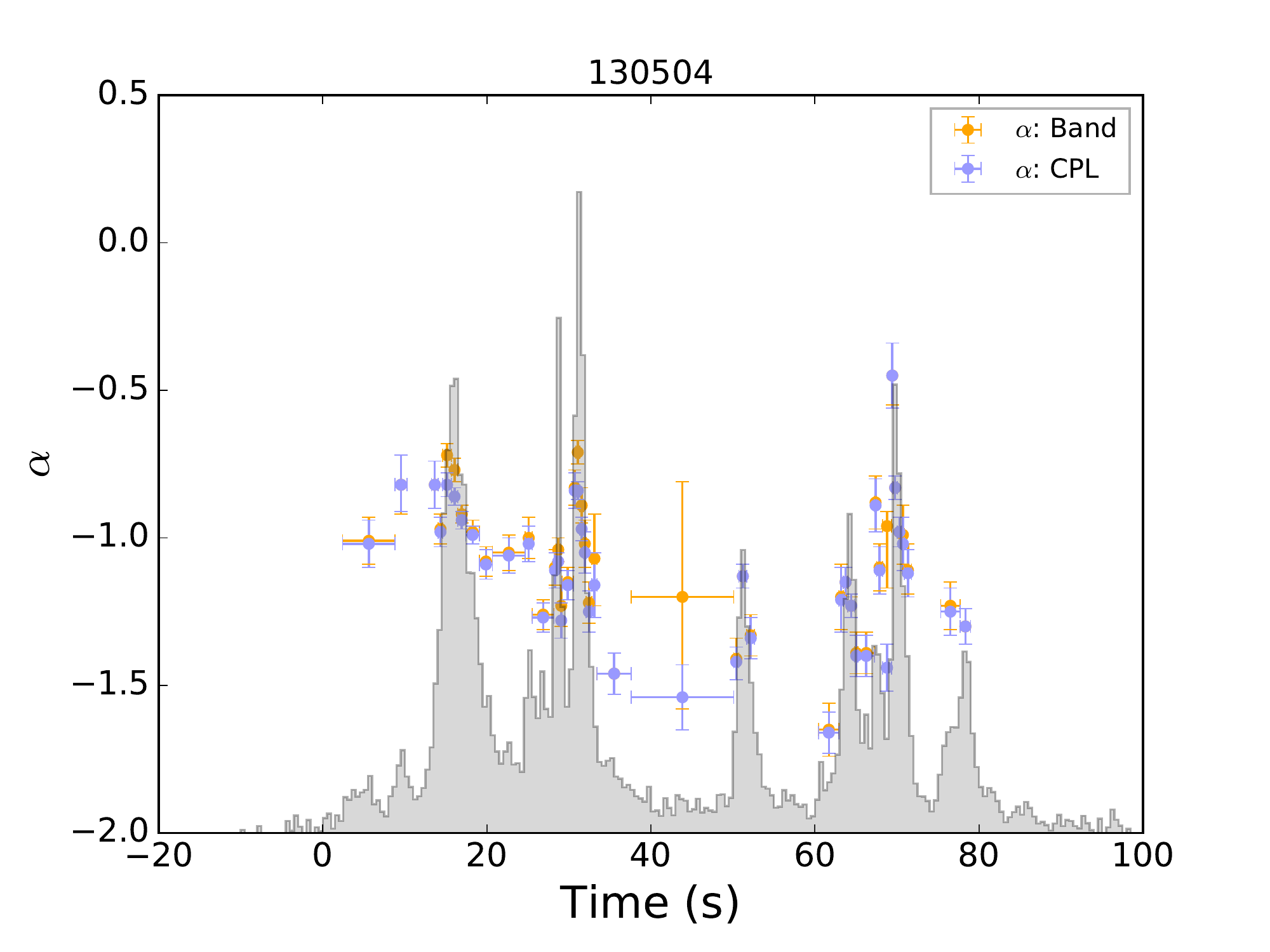}
\includegraphics[angle=0,scale=0.3]{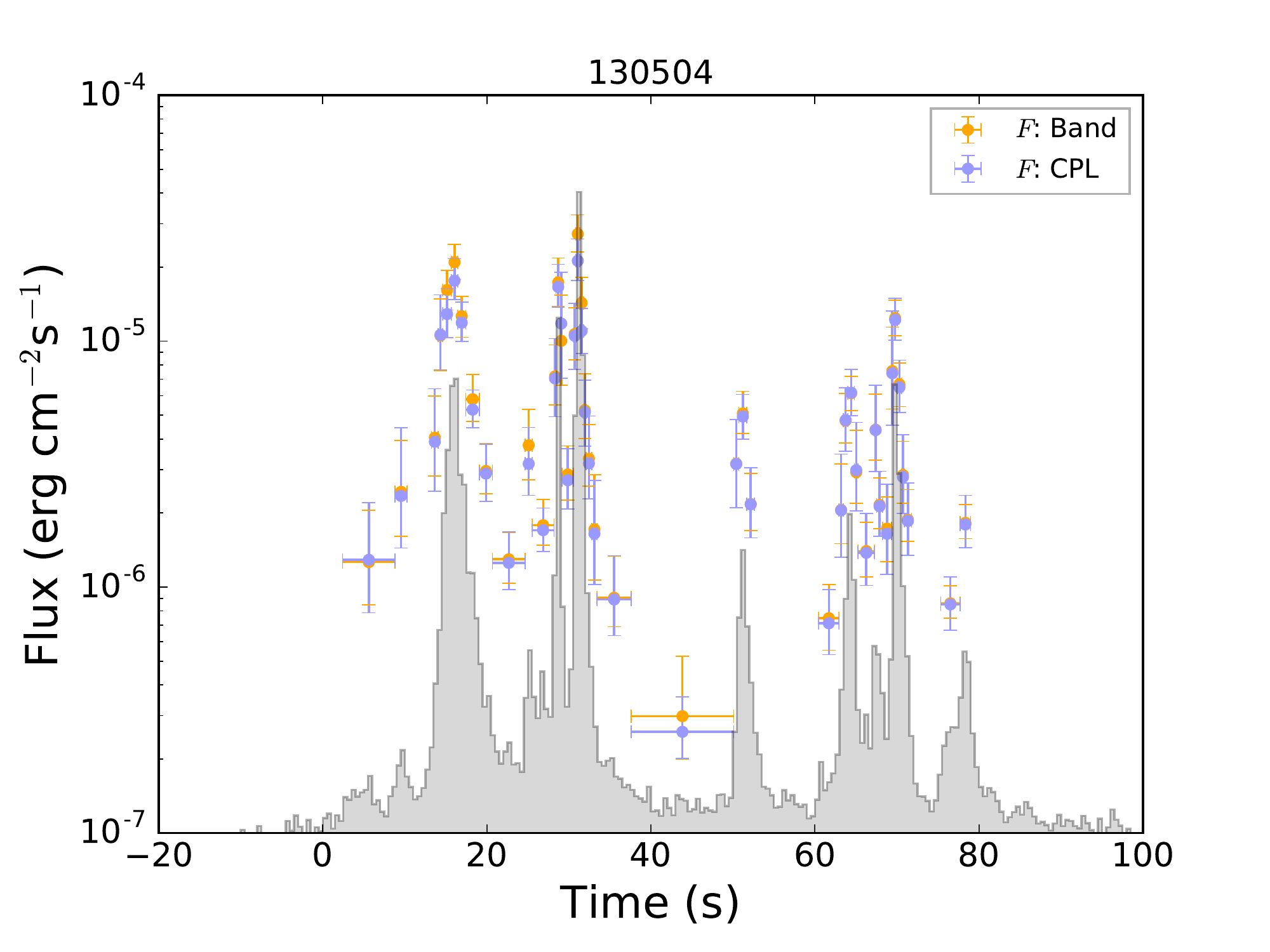}
\includegraphics[angle=0,scale=0.3]{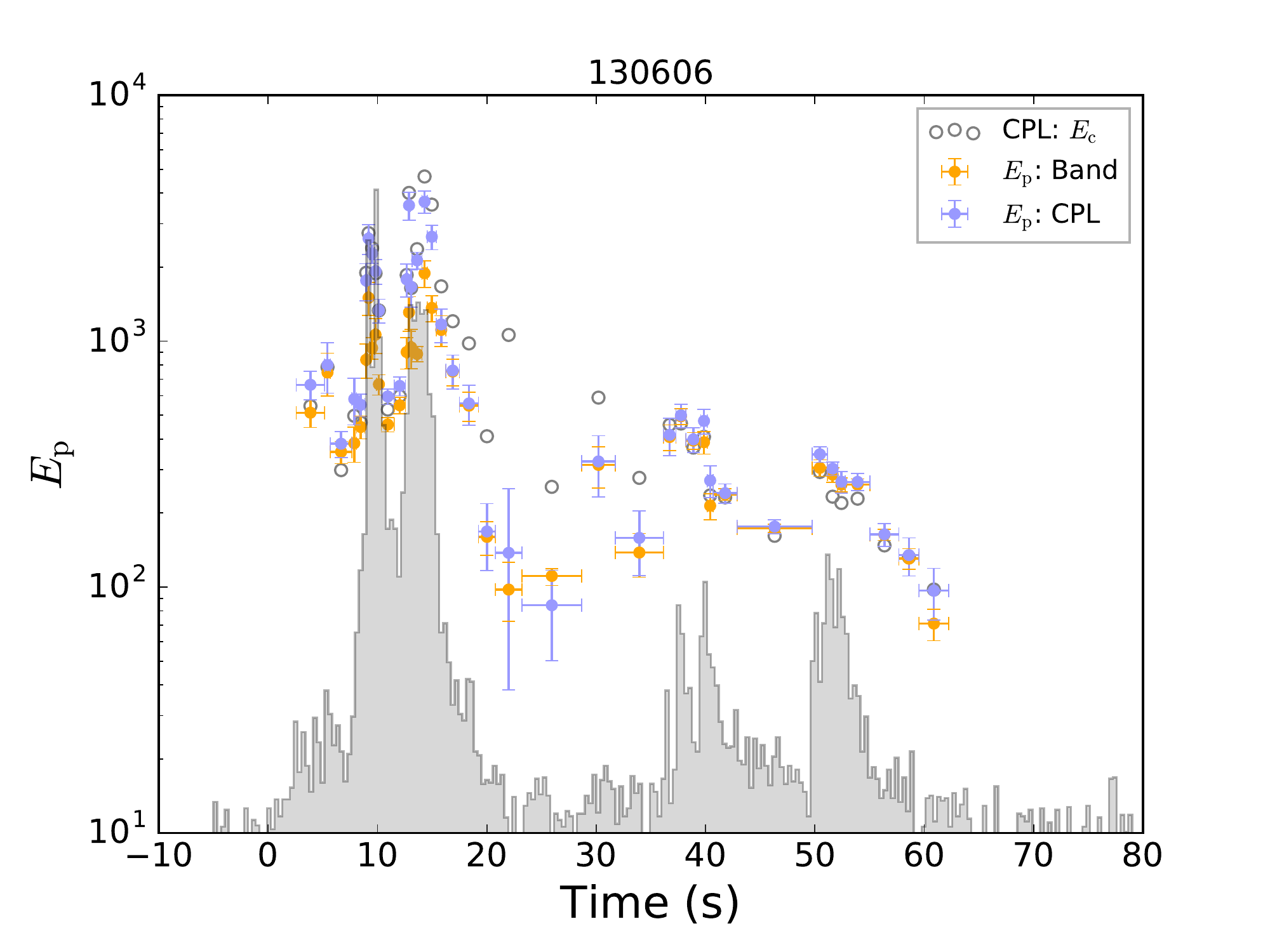}
\includegraphics[angle=0,scale=0.3]{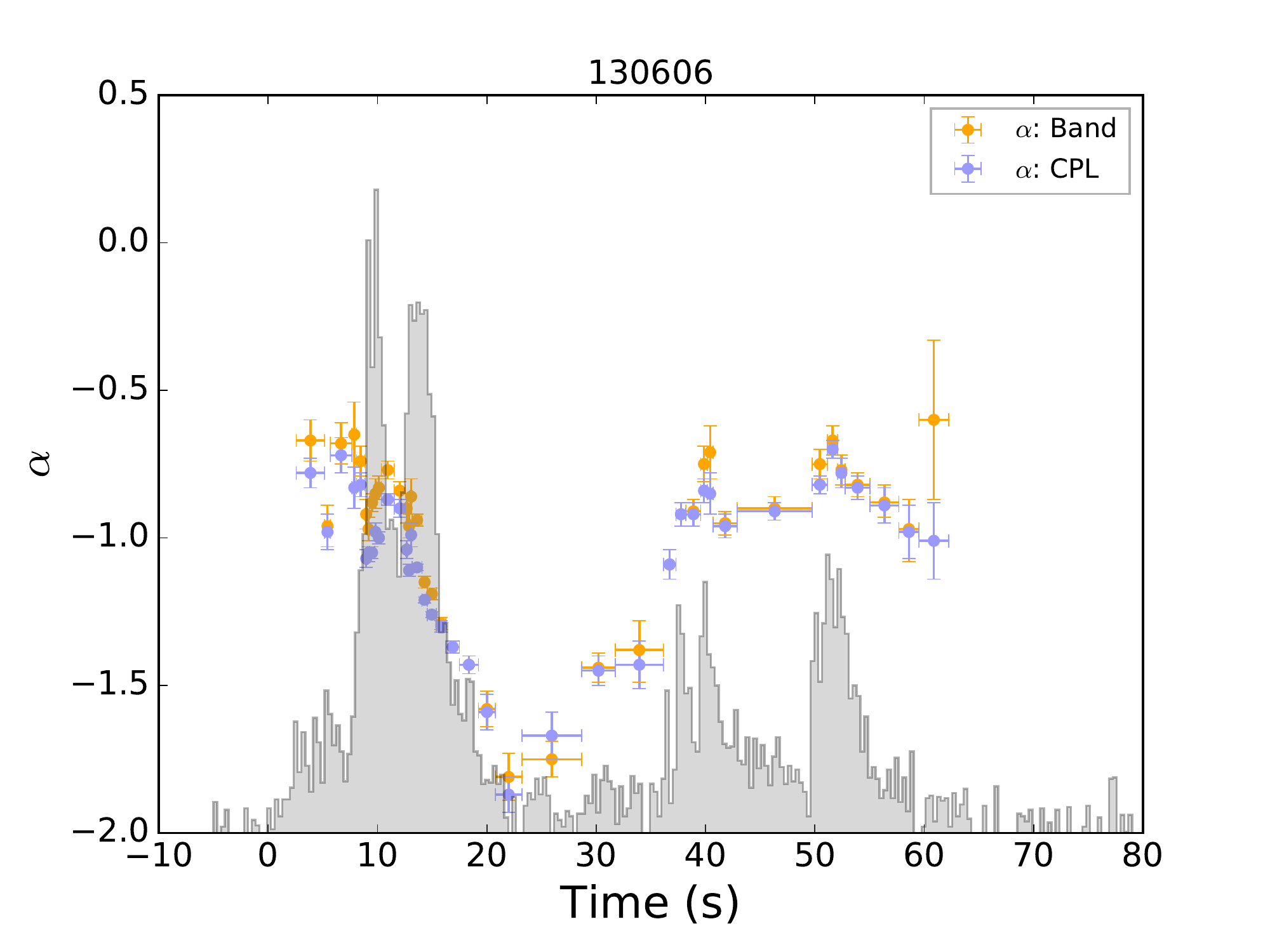}
\includegraphics[angle=0,scale=0.3]{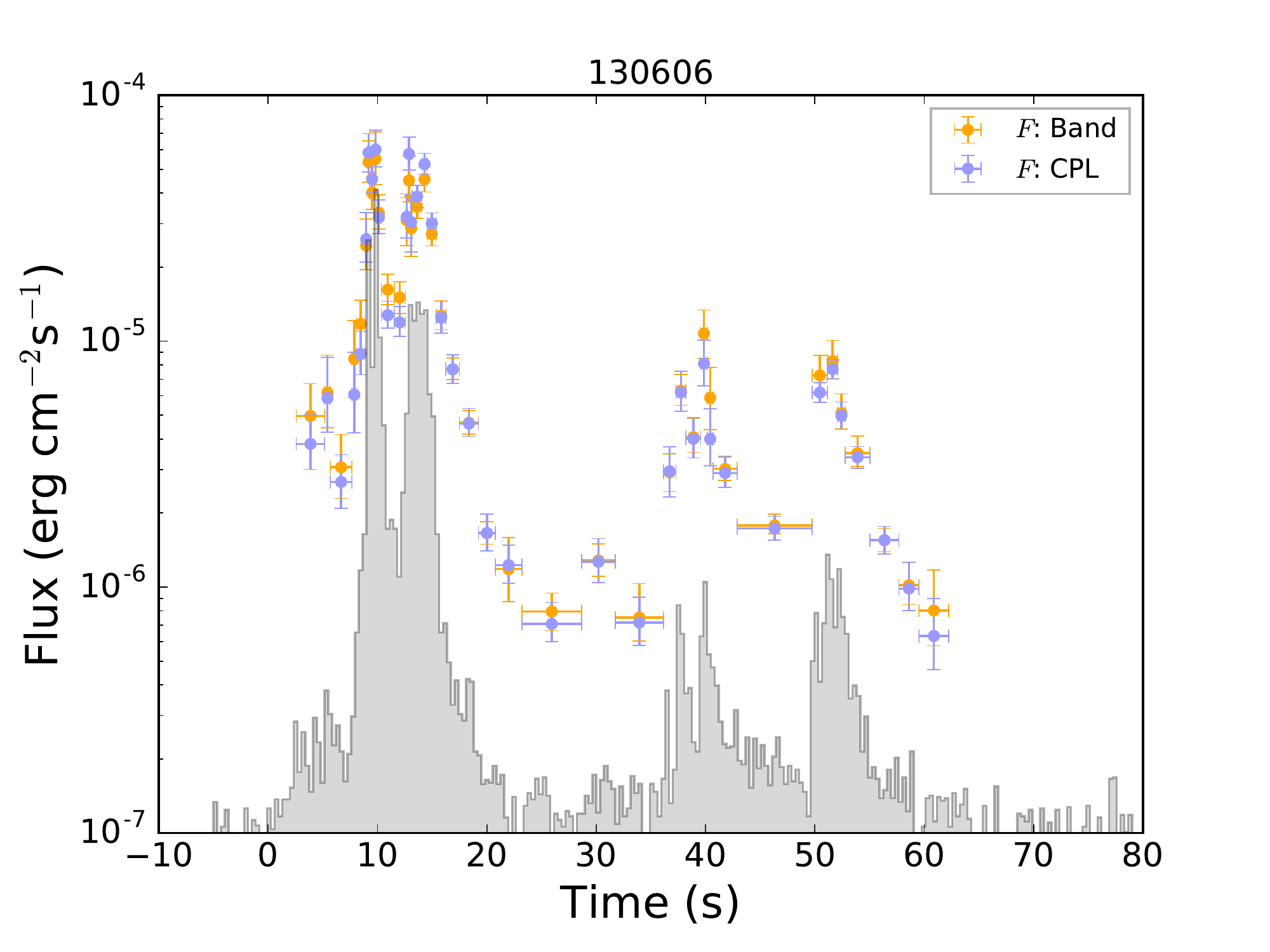}
\includegraphics[angle=0,scale=0.3]{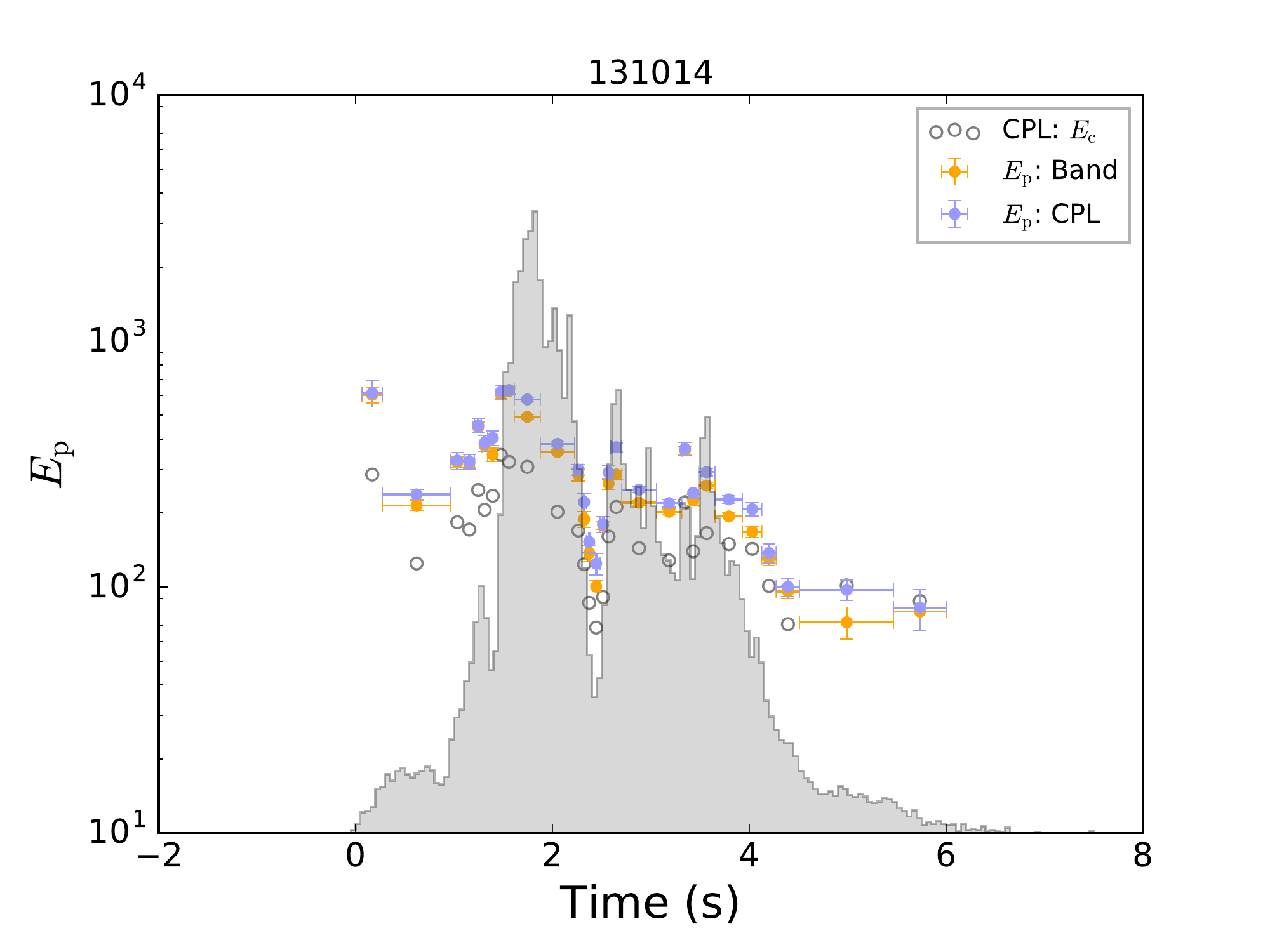}
\includegraphics[angle=0,scale=0.3]{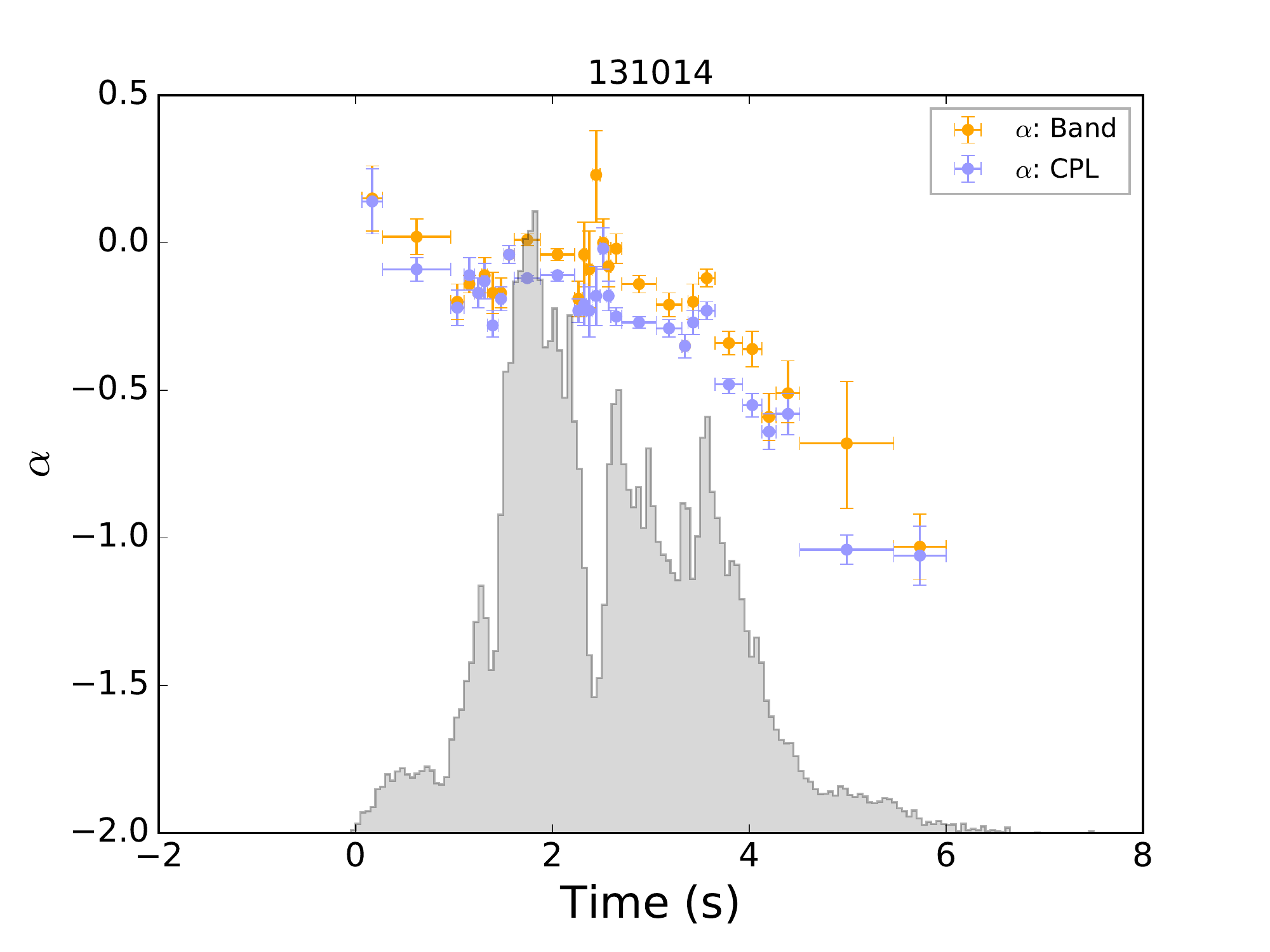}
\includegraphics[angle=0,scale=0.3]{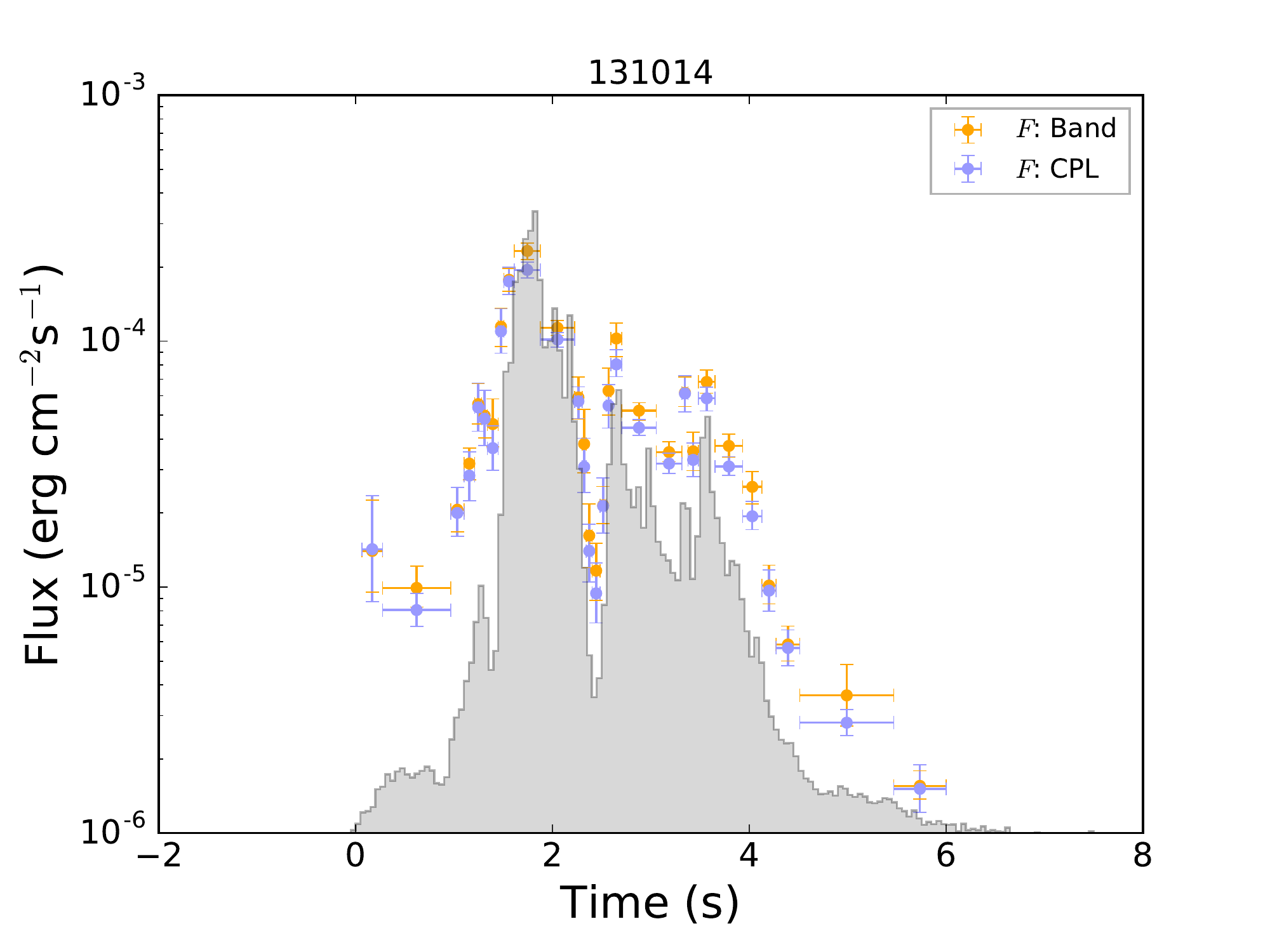}
\includegraphics[angle=0,scale=0.3]{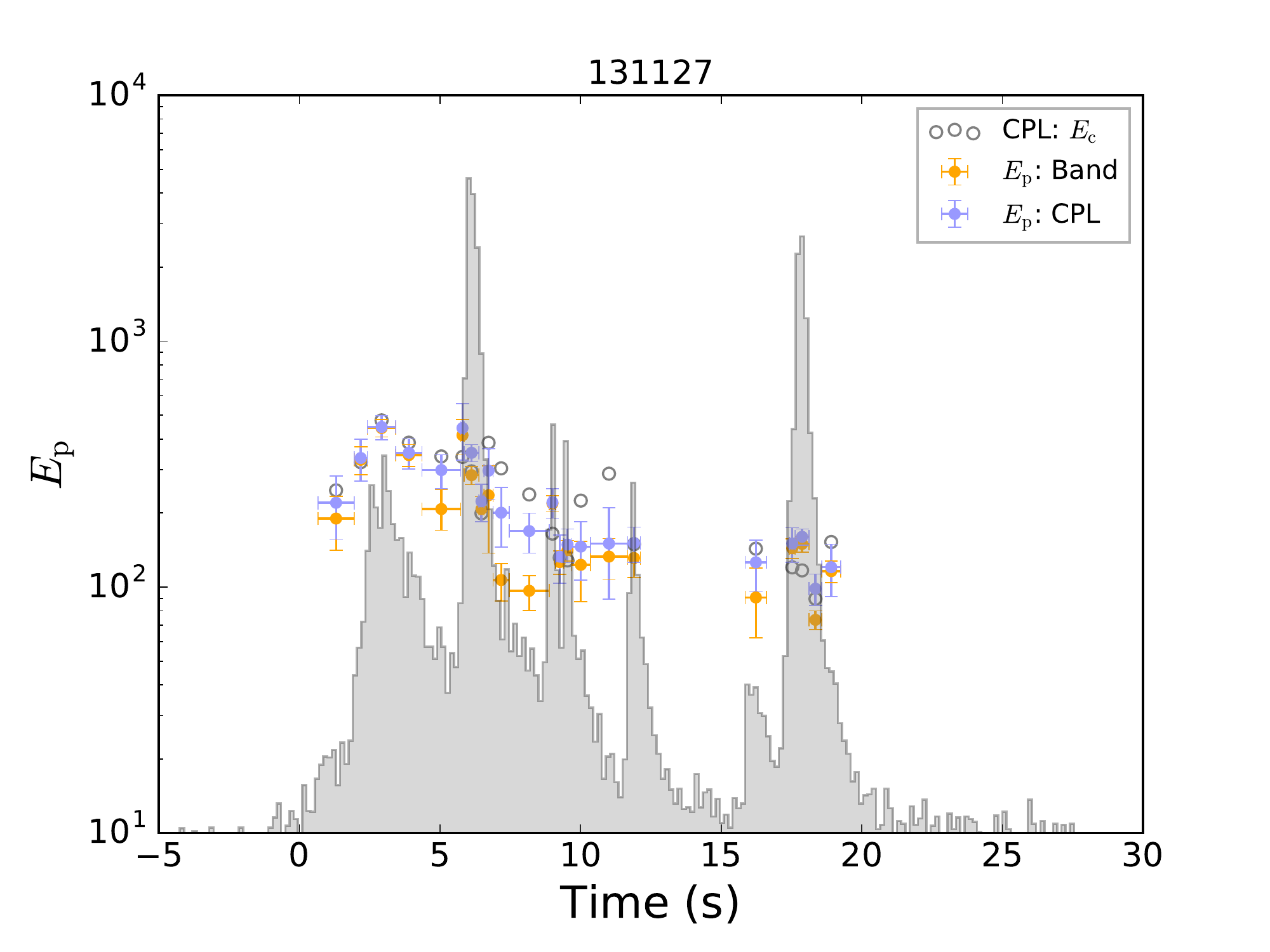}
\includegraphics[angle=0,scale=0.3]{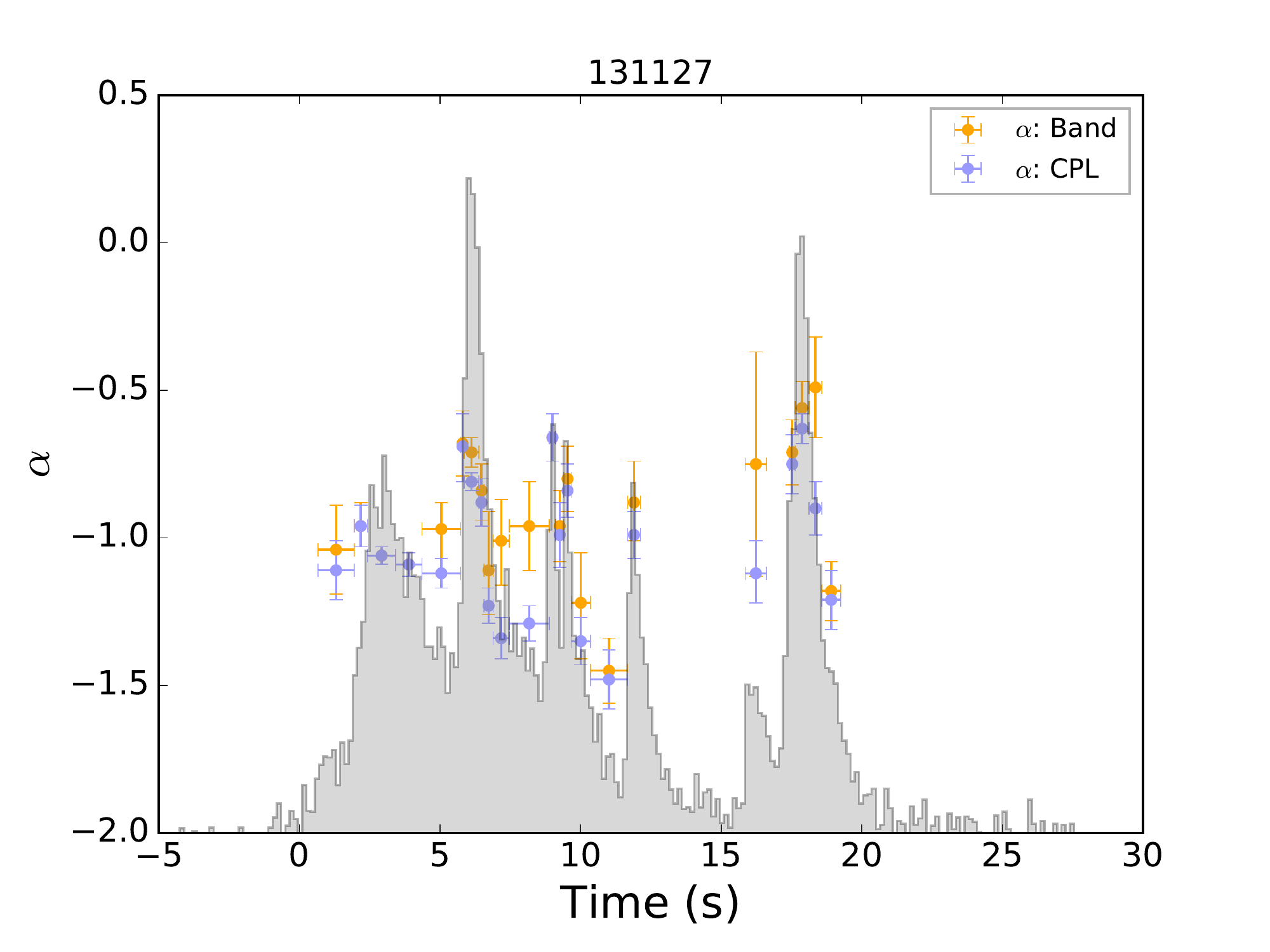}
\includegraphics[angle=0,scale=0.3]{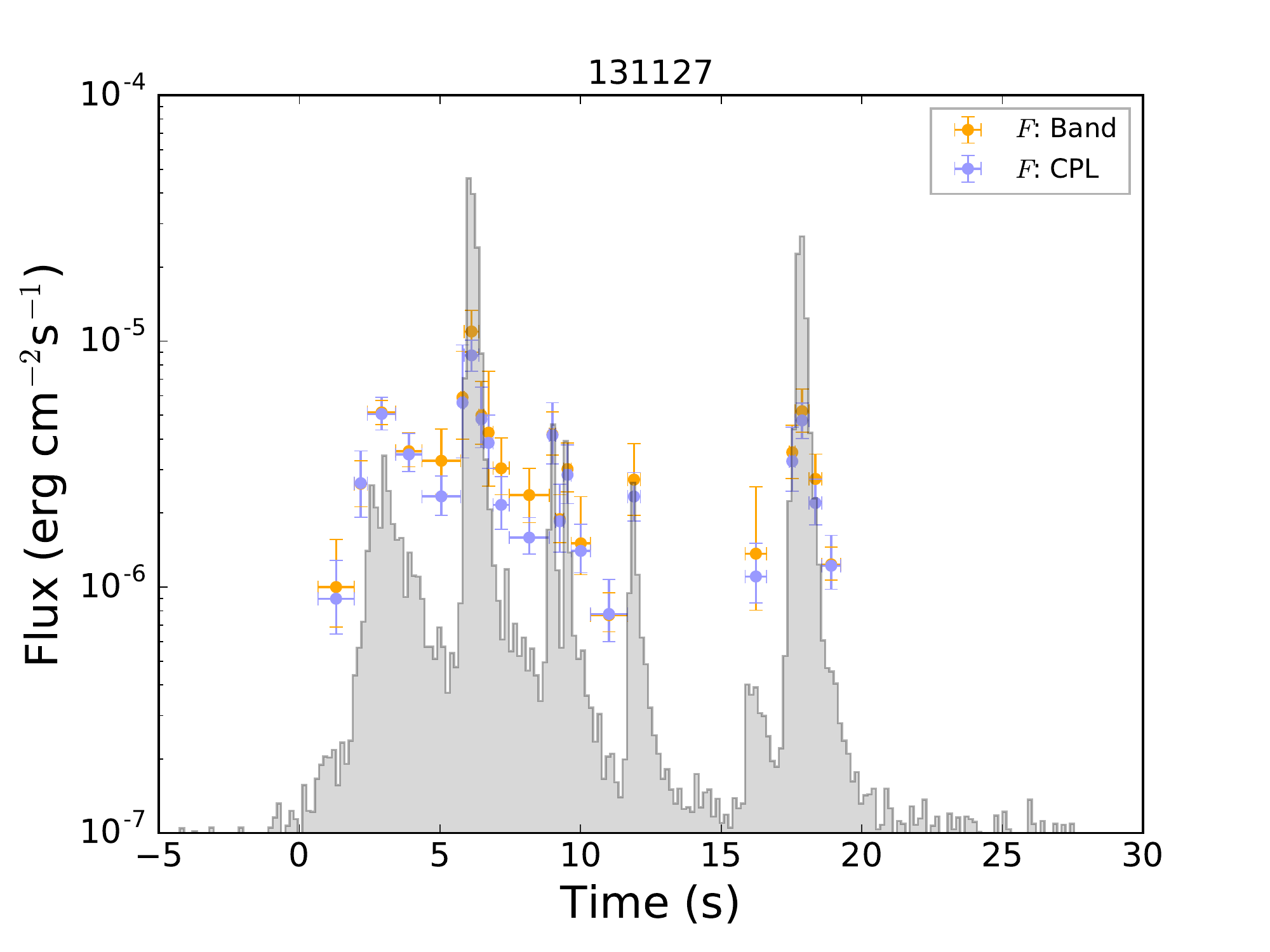}
\includegraphics[angle=0,scale=0.3]{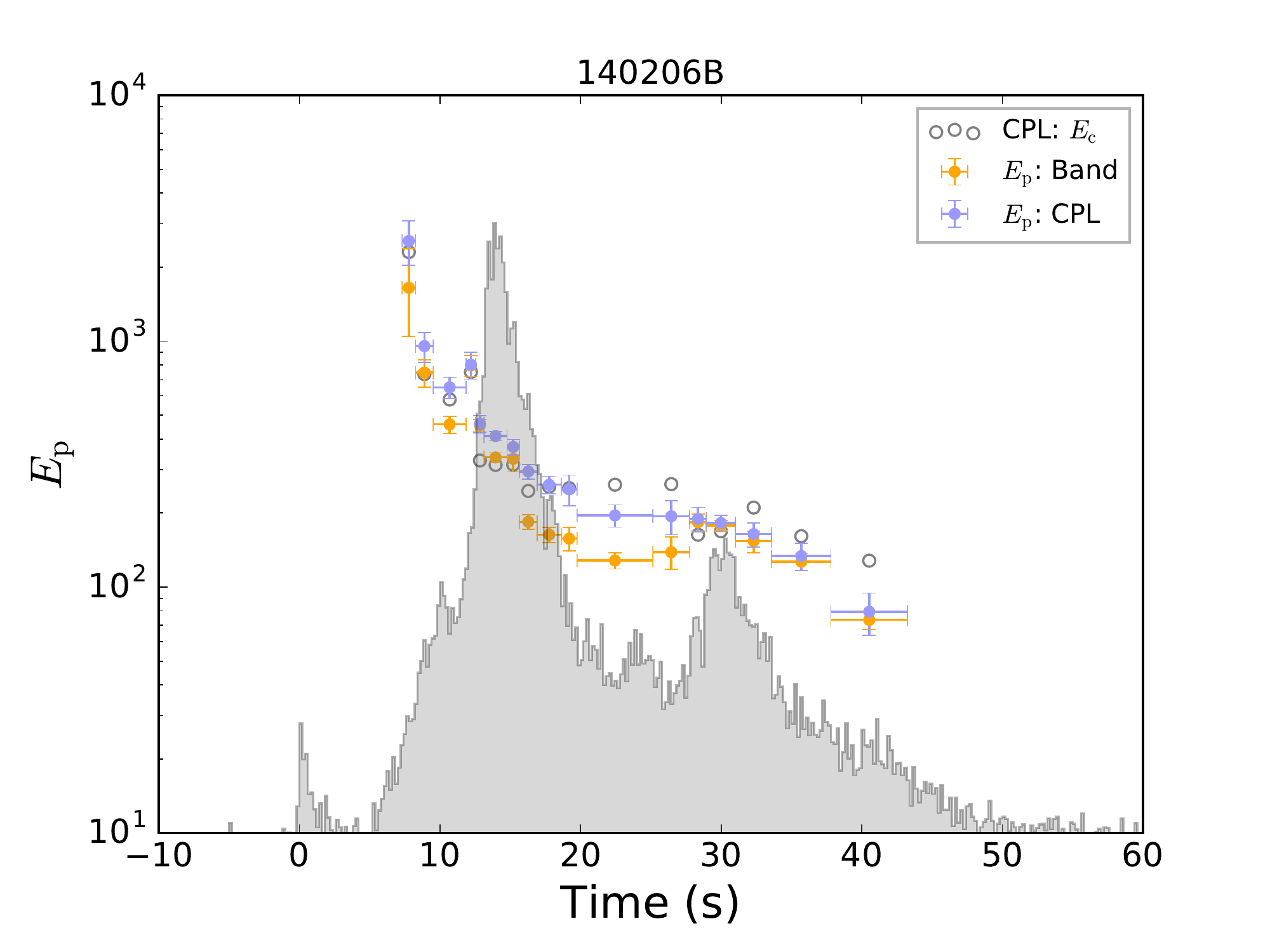}
\includegraphics[angle=0,scale=0.3]{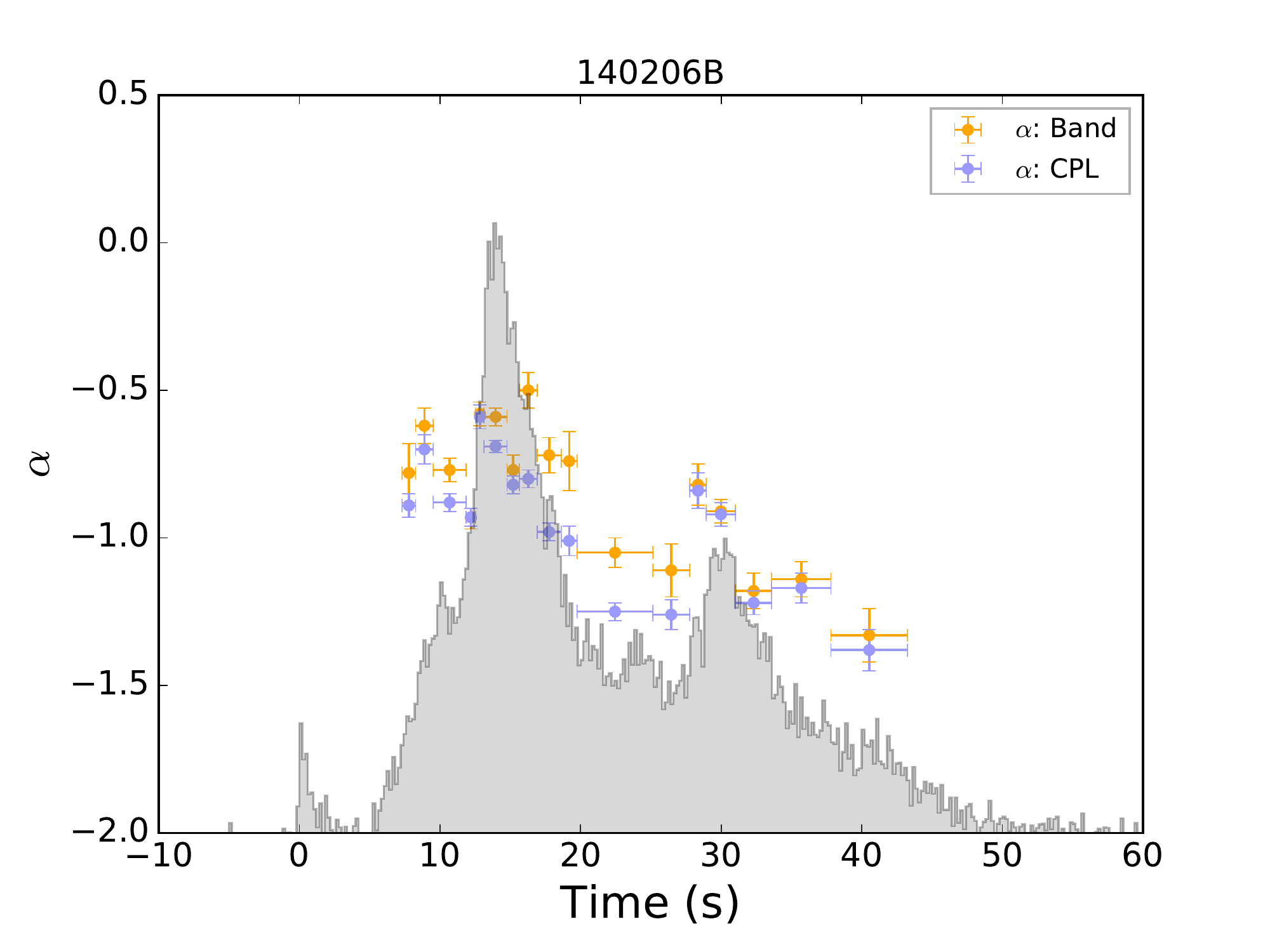}
\includegraphics[angle=0,scale=0.3]{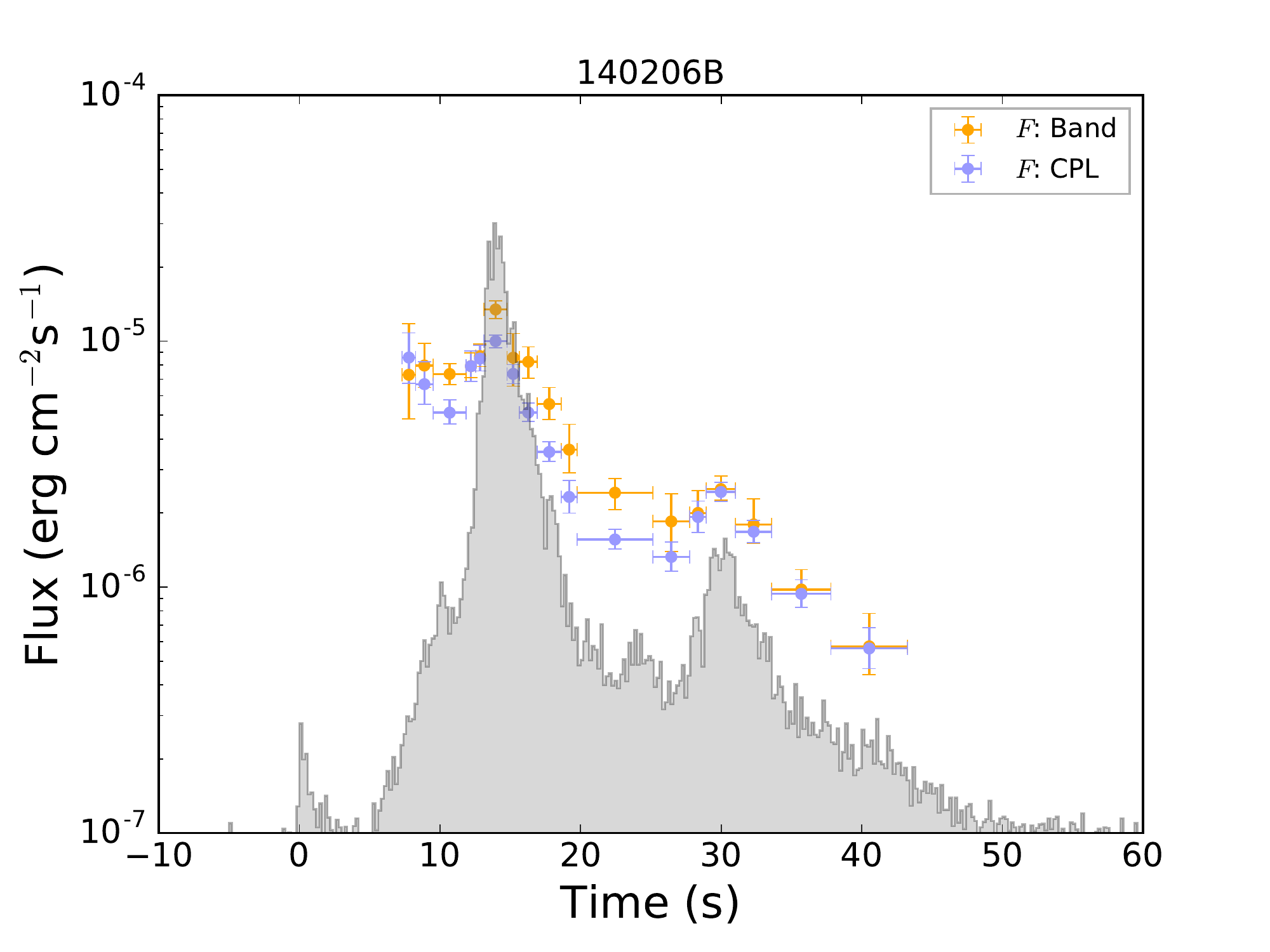}
\center{Fig. \ref{fig:evolution}--- Continued}
\end{figure*}
\begin{figure*}
\includegraphics[angle=0,scale=0.3]{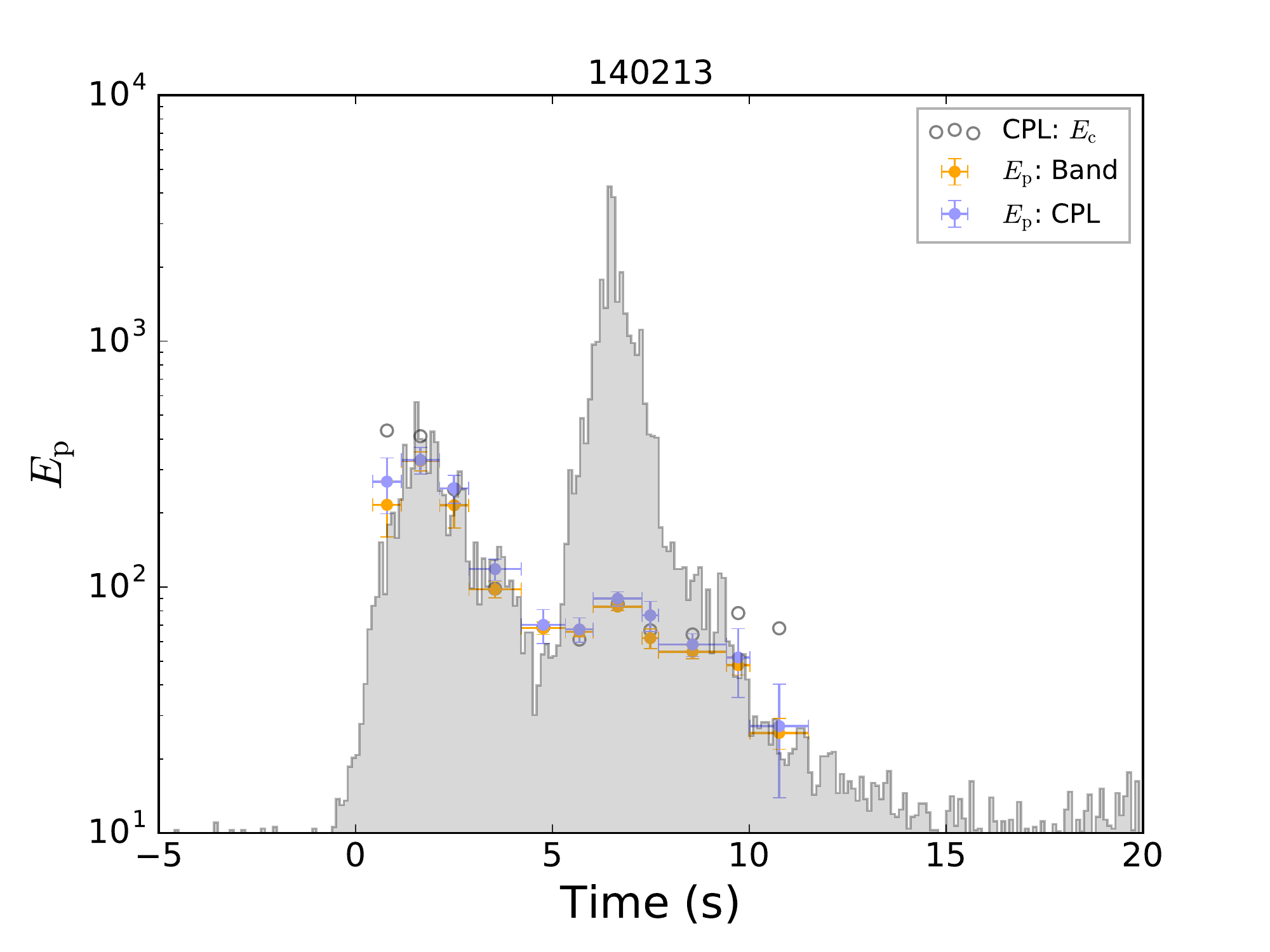}
\includegraphics[angle=0,scale=0.3]{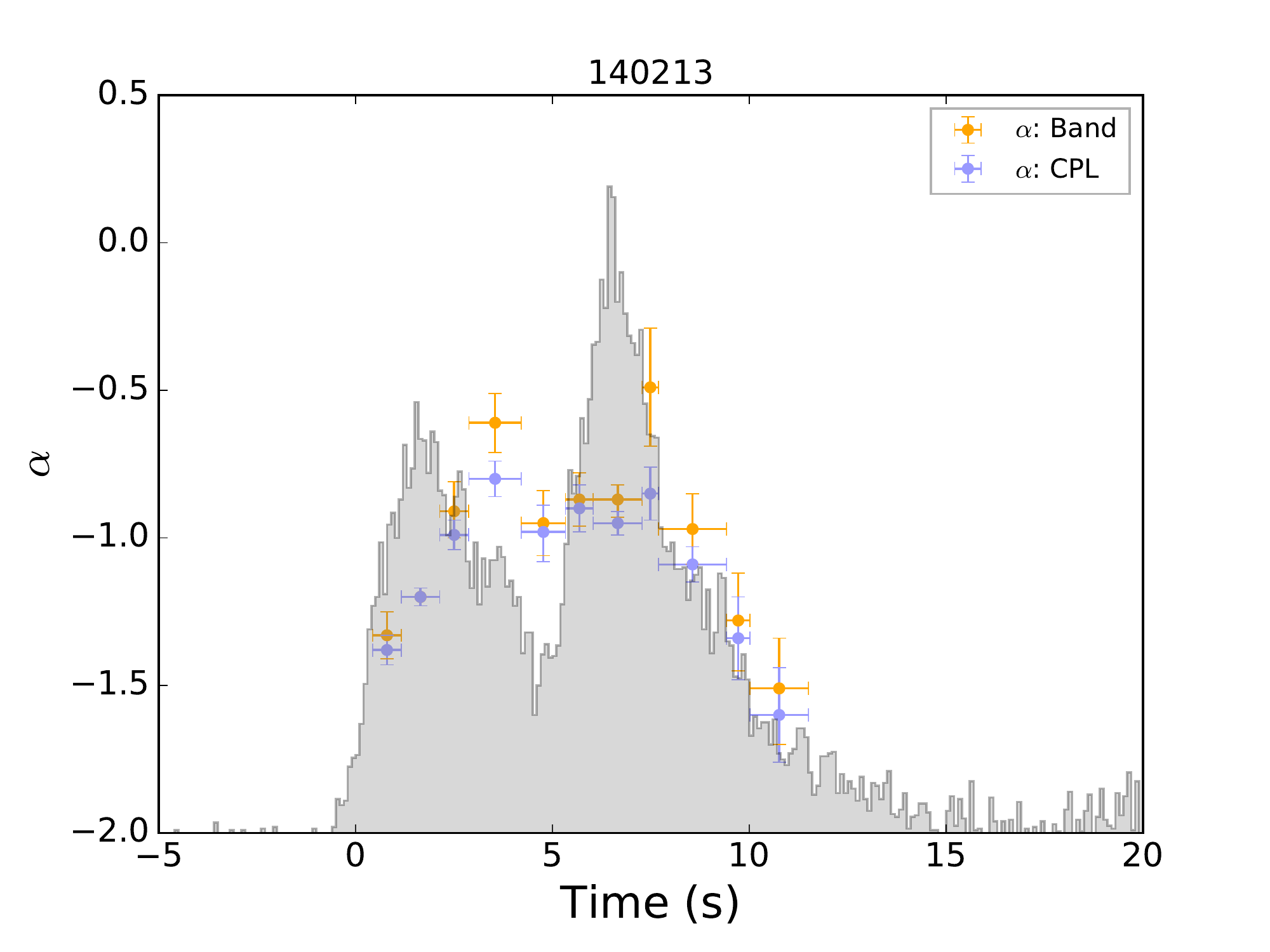}
\includegraphics[angle=0,scale=0.3]{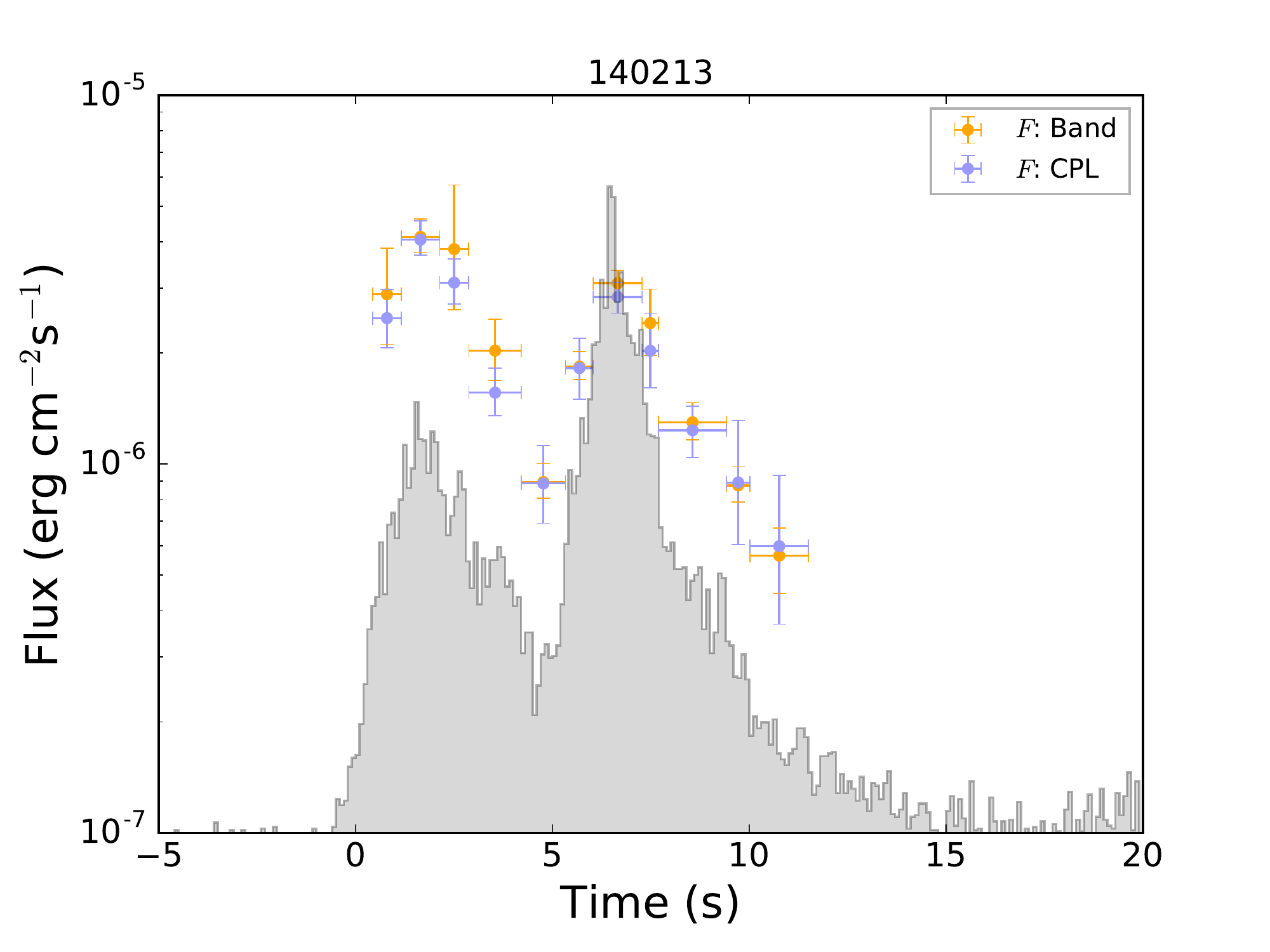}
\includegraphics[angle=0,scale=0.3]{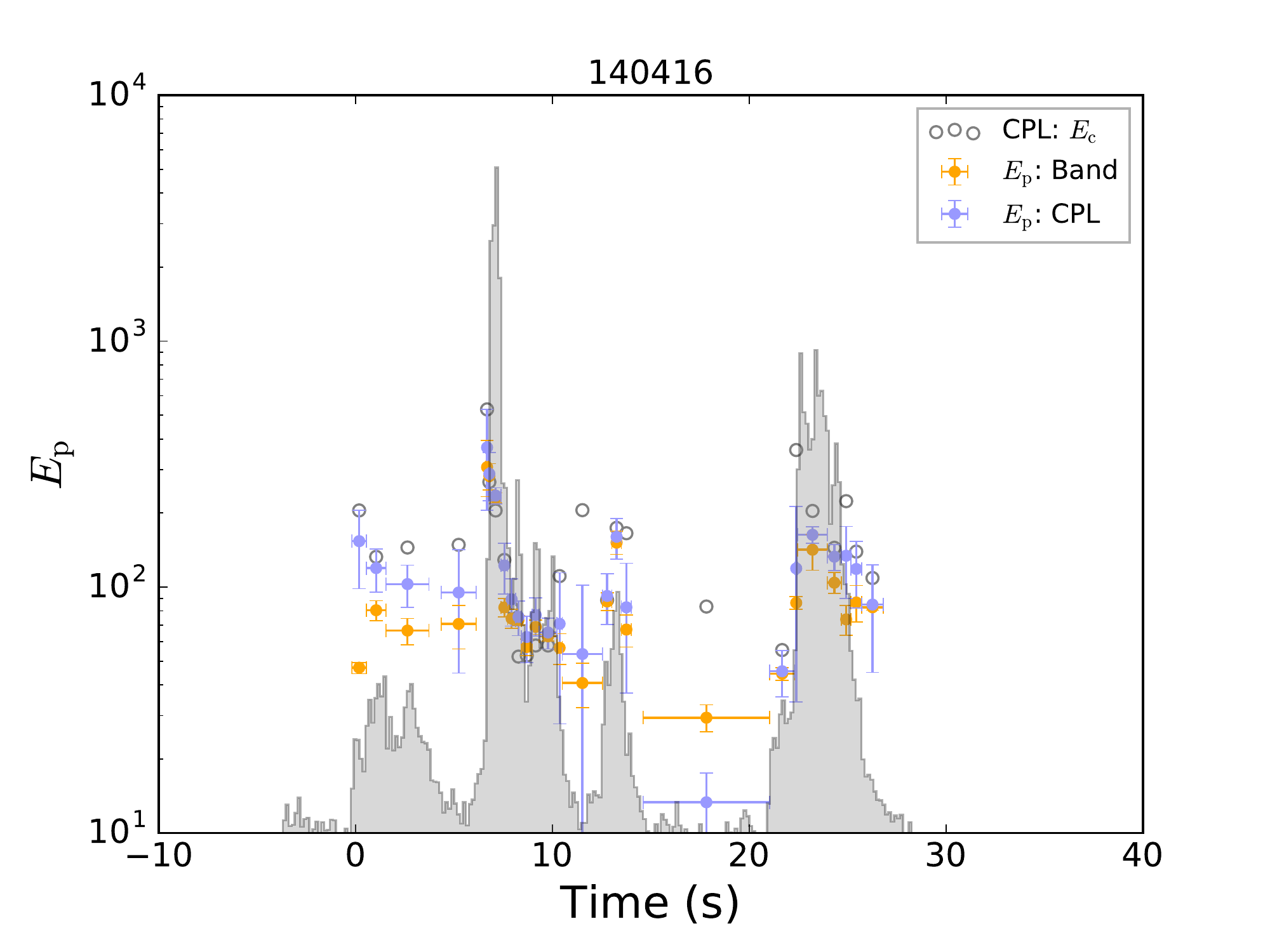}
\includegraphics[angle=0,scale=0.3]{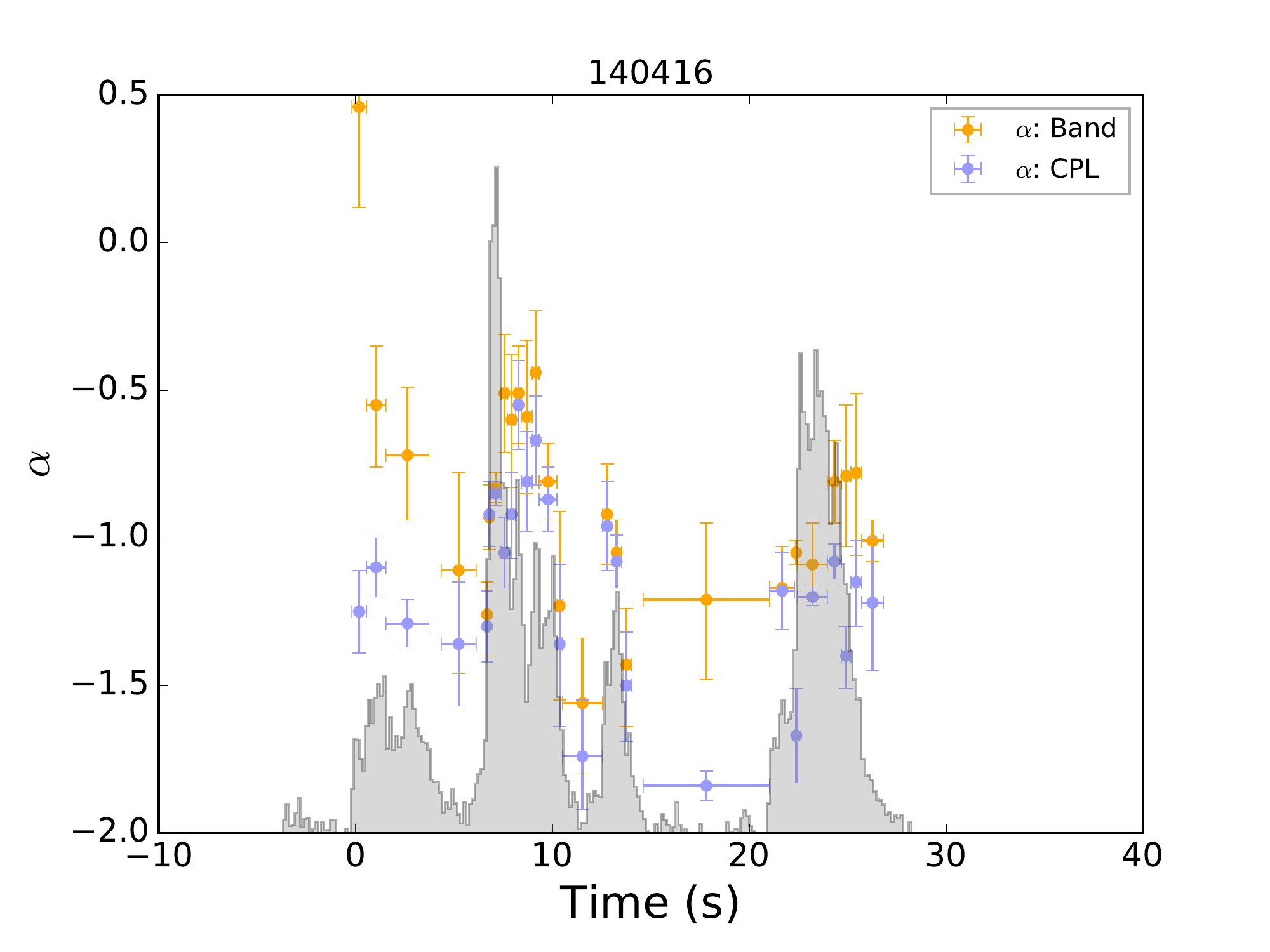}
\includegraphics[angle=0,scale=0.3]{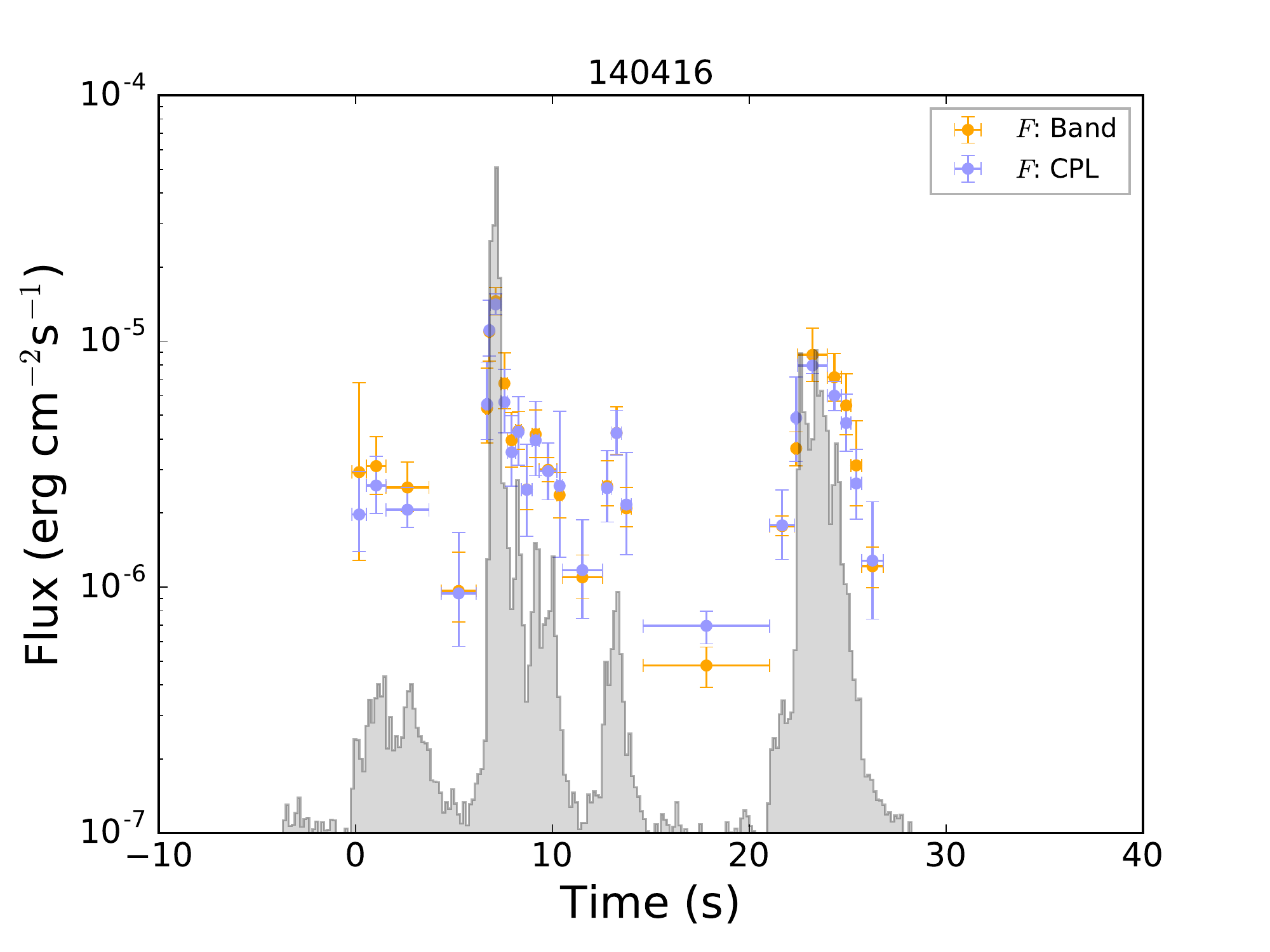}
\includegraphics[angle=0,scale=0.3]{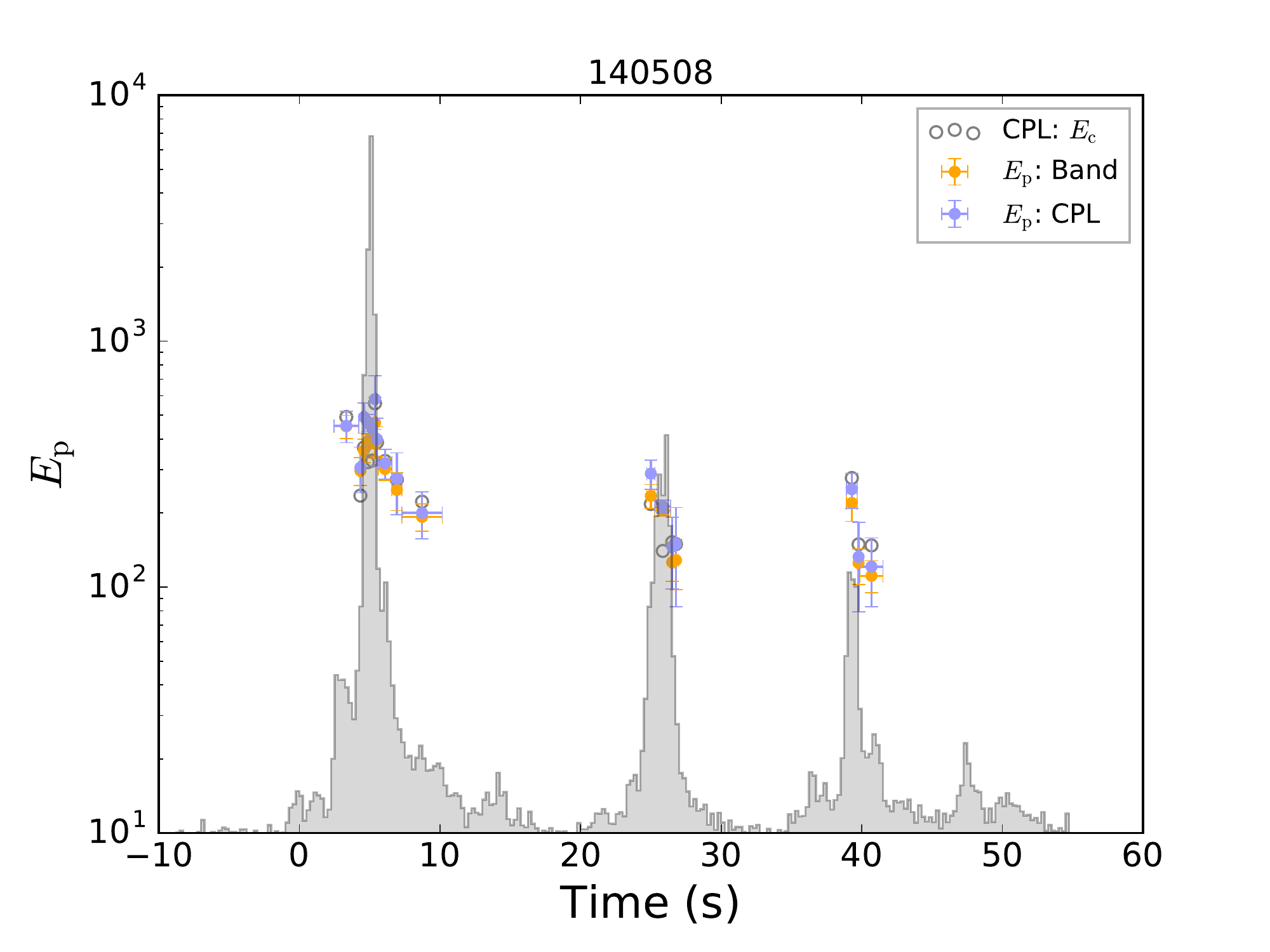}
\includegraphics[angle=0,scale=0.3]{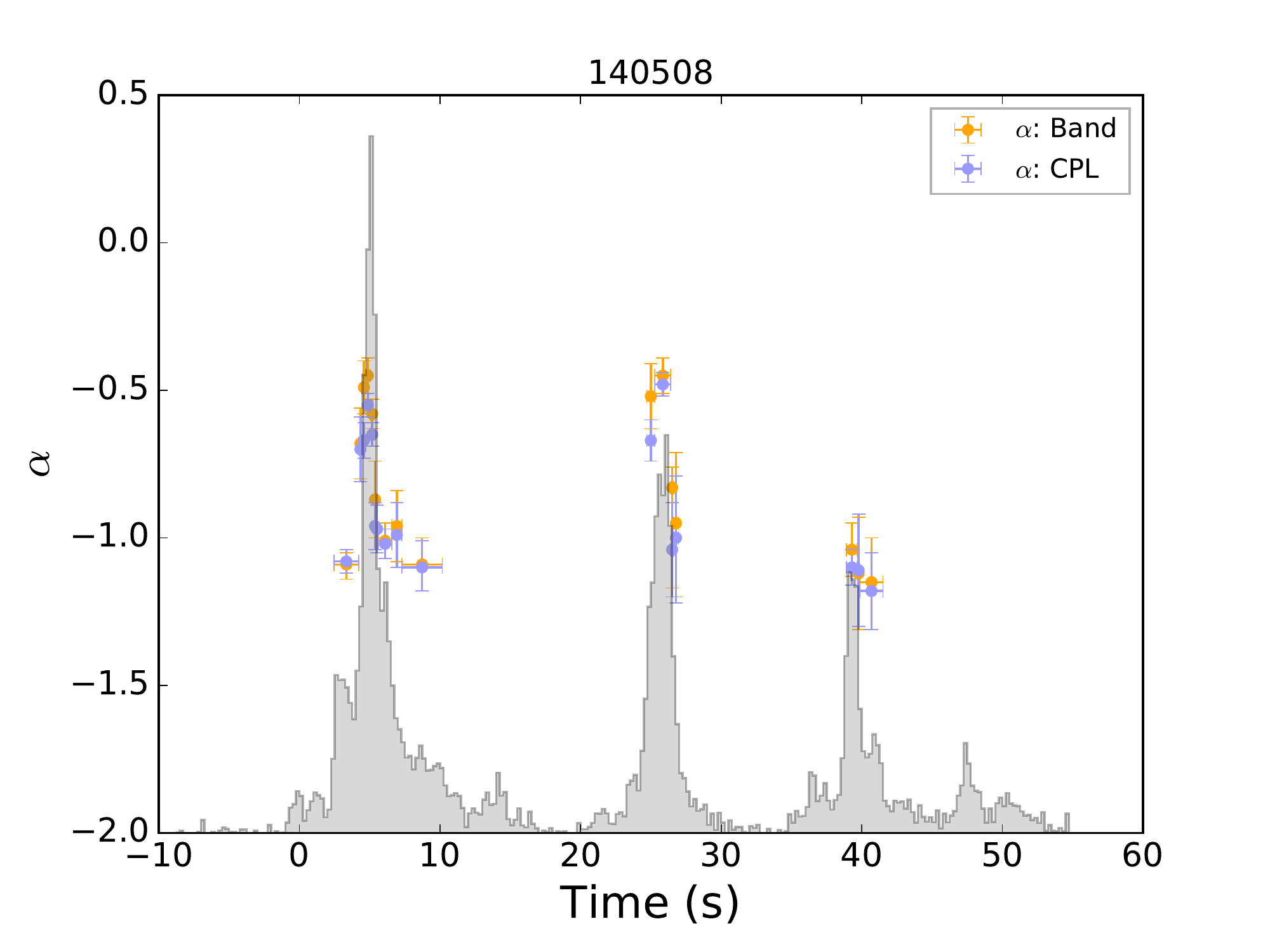}
\includegraphics[angle=0,scale=0.3]{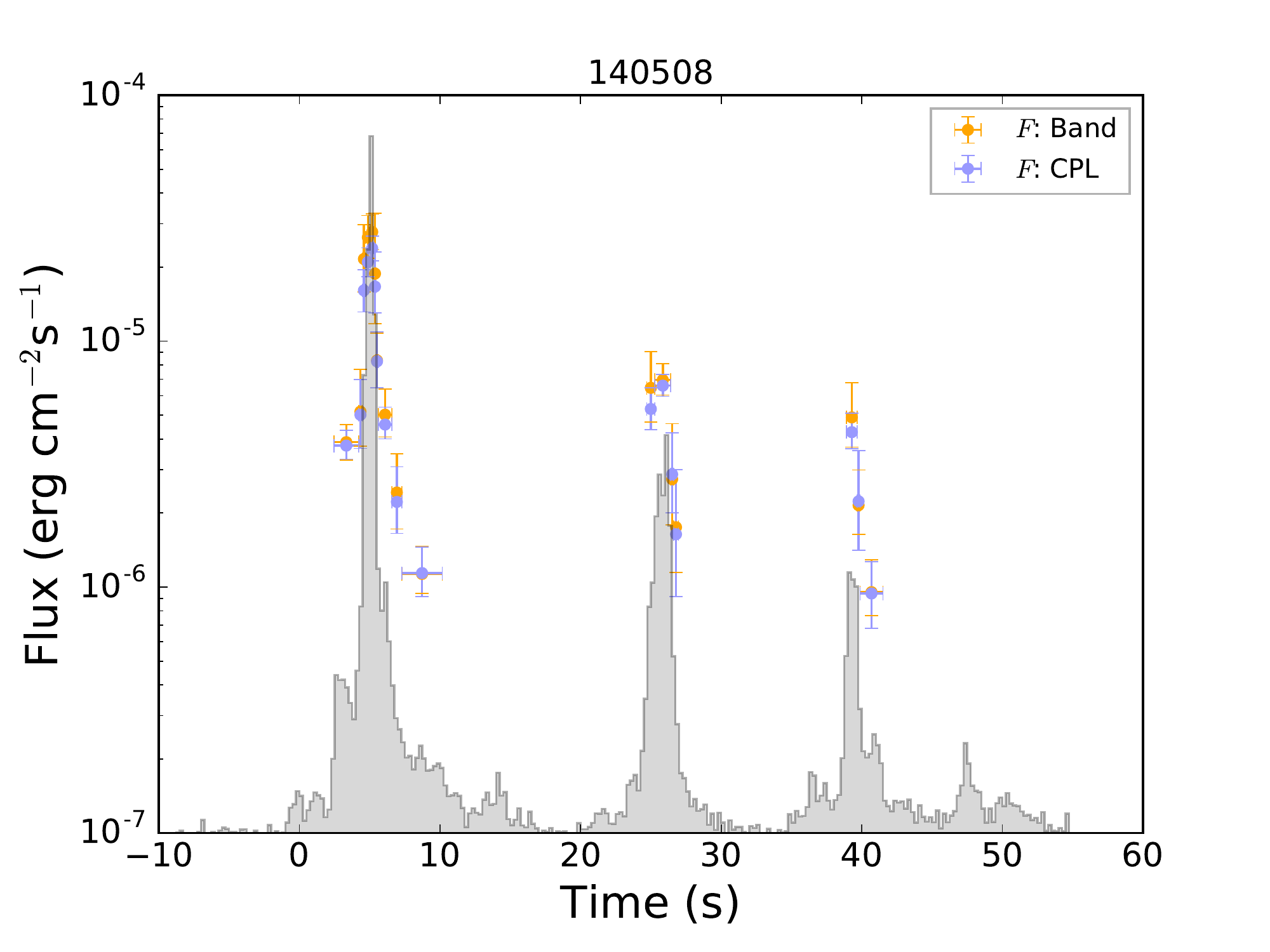}
\includegraphics[angle=0,scale=0.3]{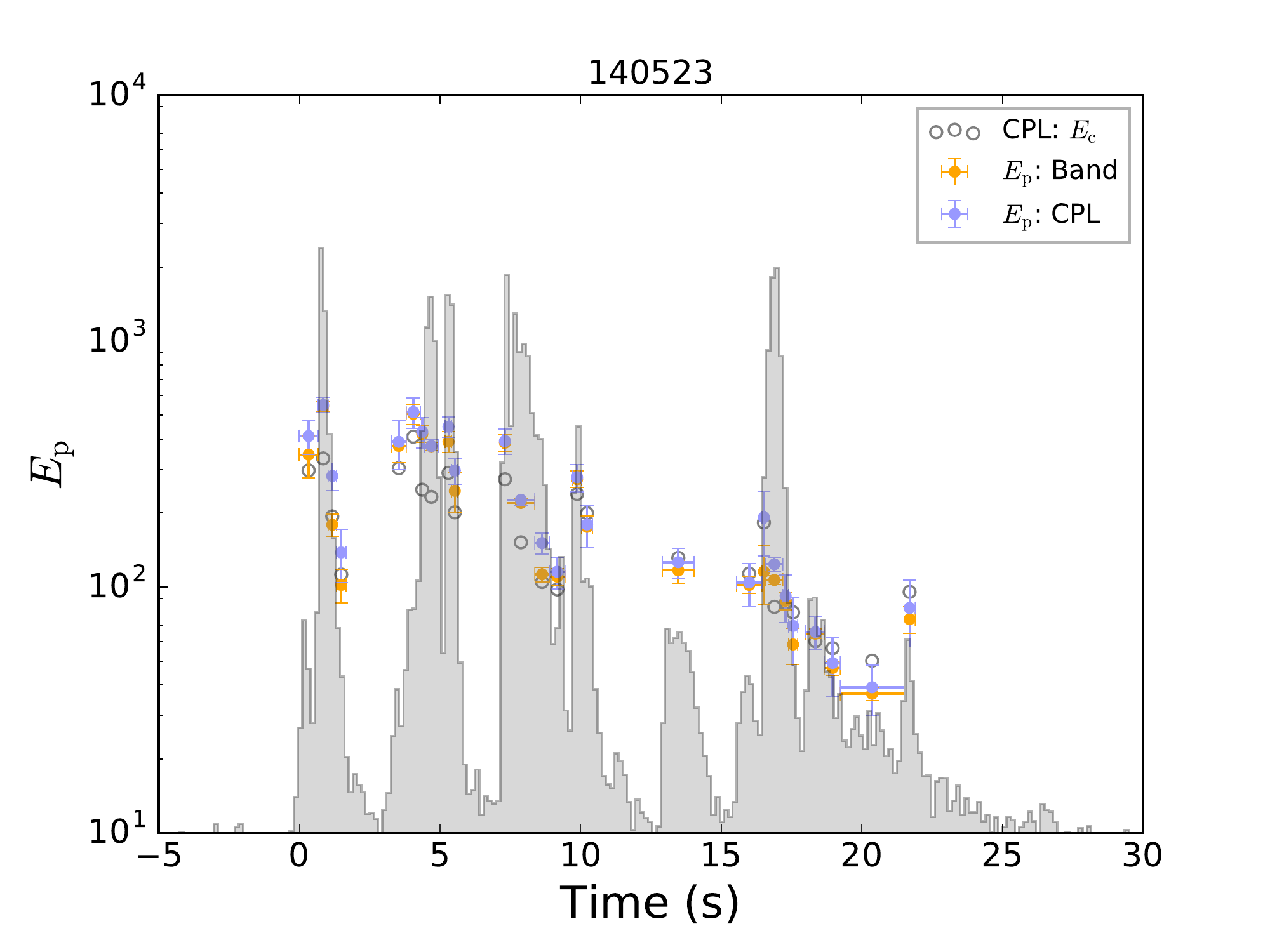}
\includegraphics[angle=0,scale=0.3]{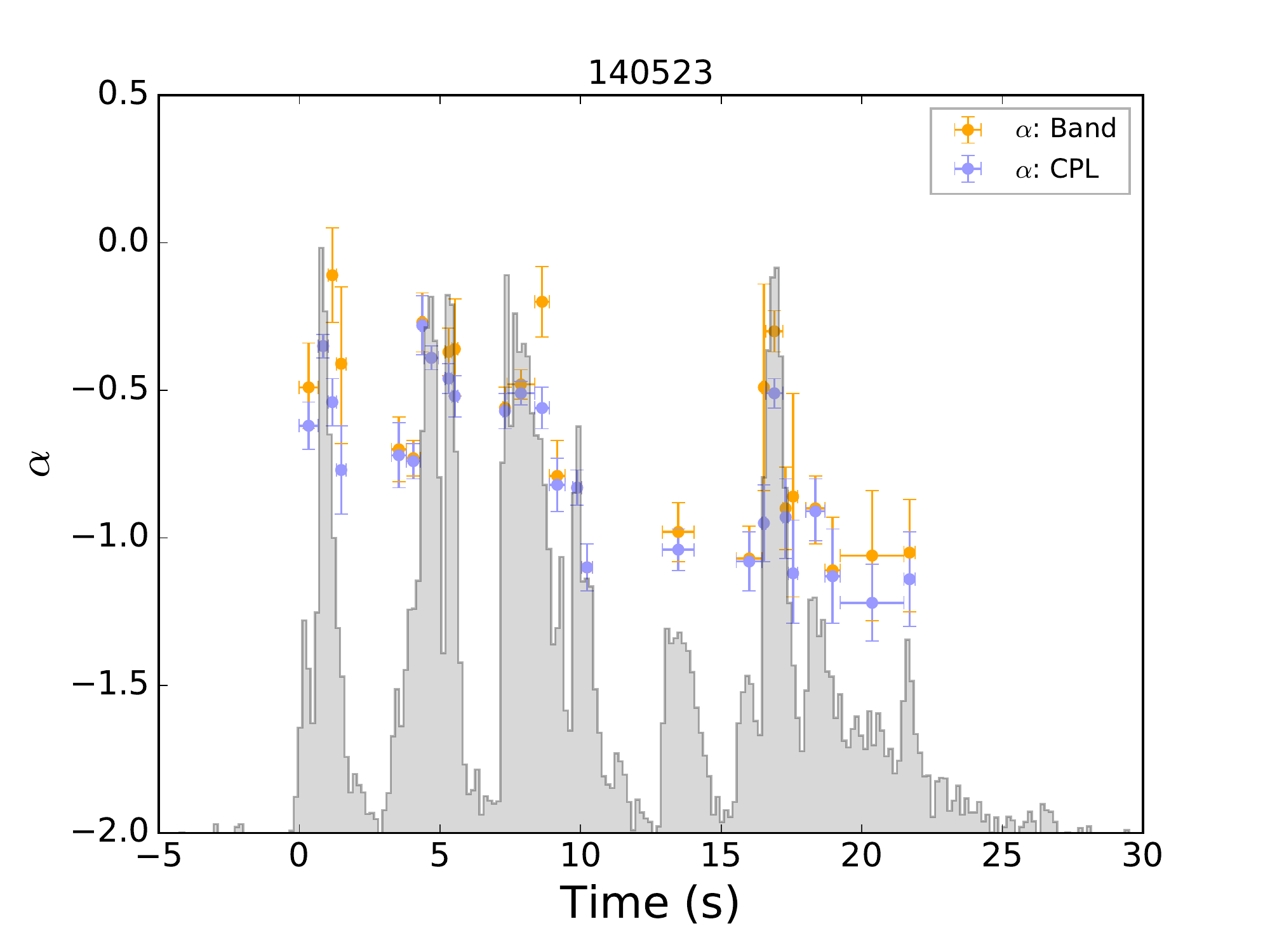}
\includegraphics[angle=0,scale=0.3]{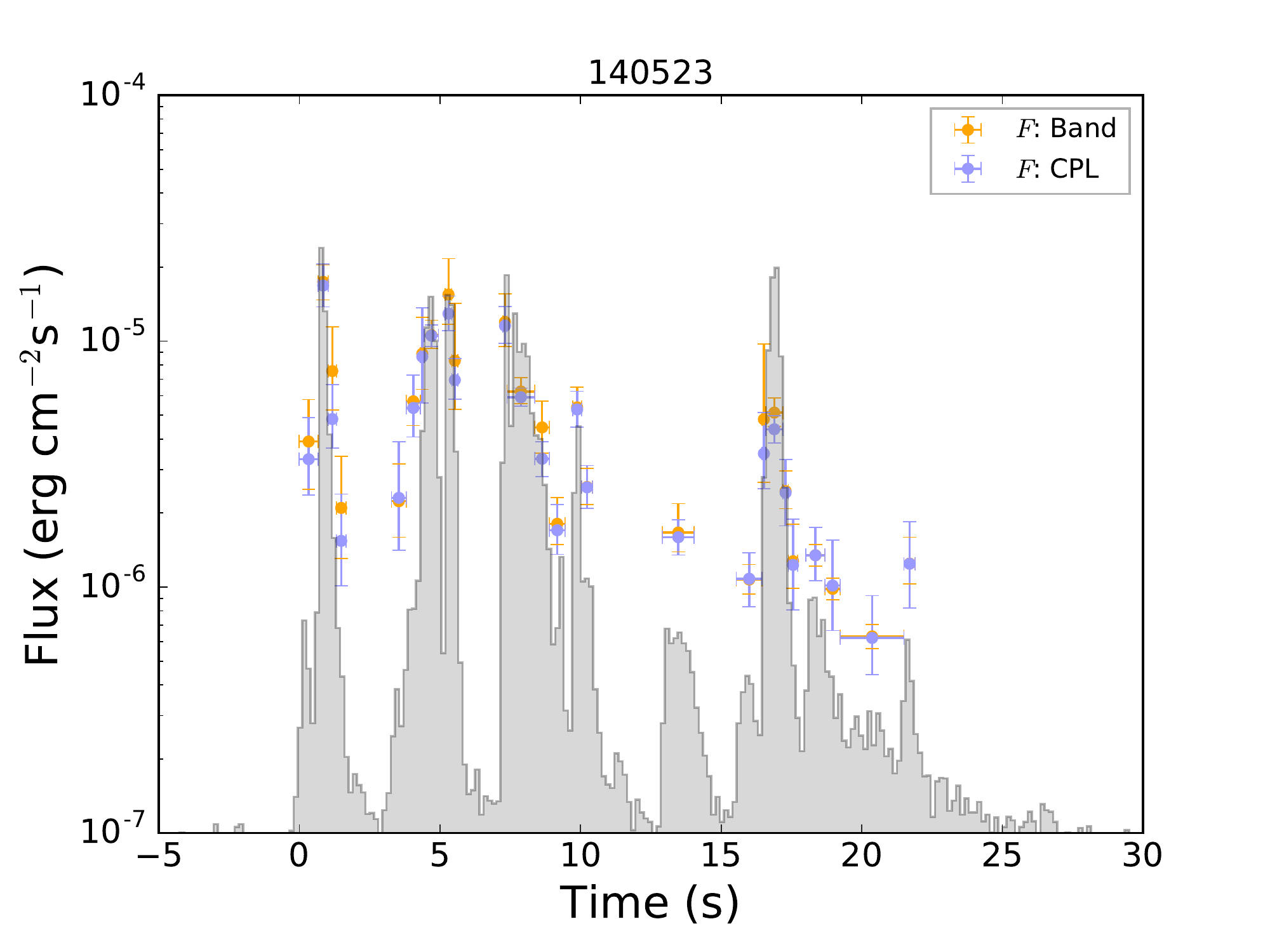}
\includegraphics[angle=0,scale=0.3]{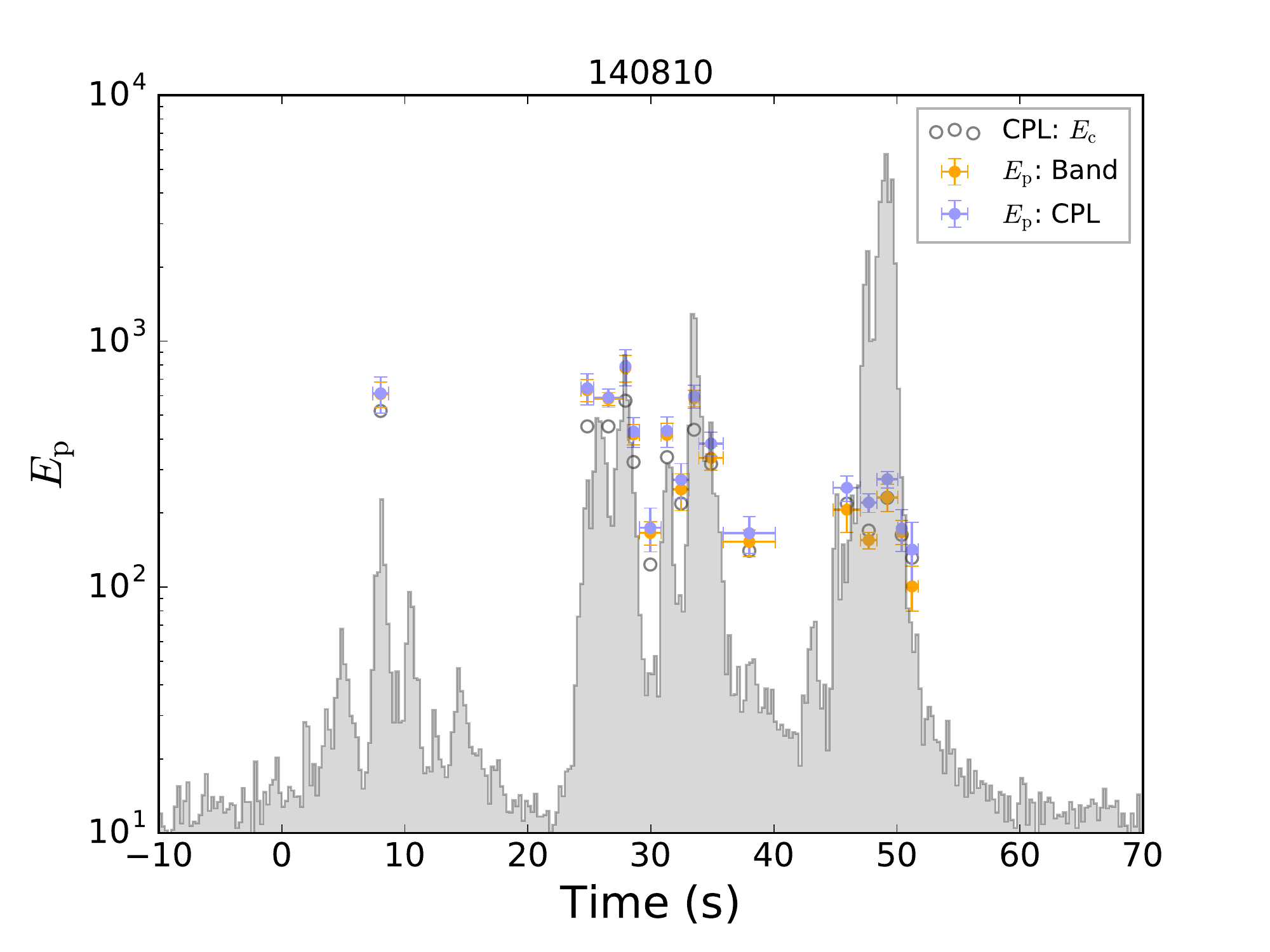}
\includegraphics[angle=0,scale=0.3]{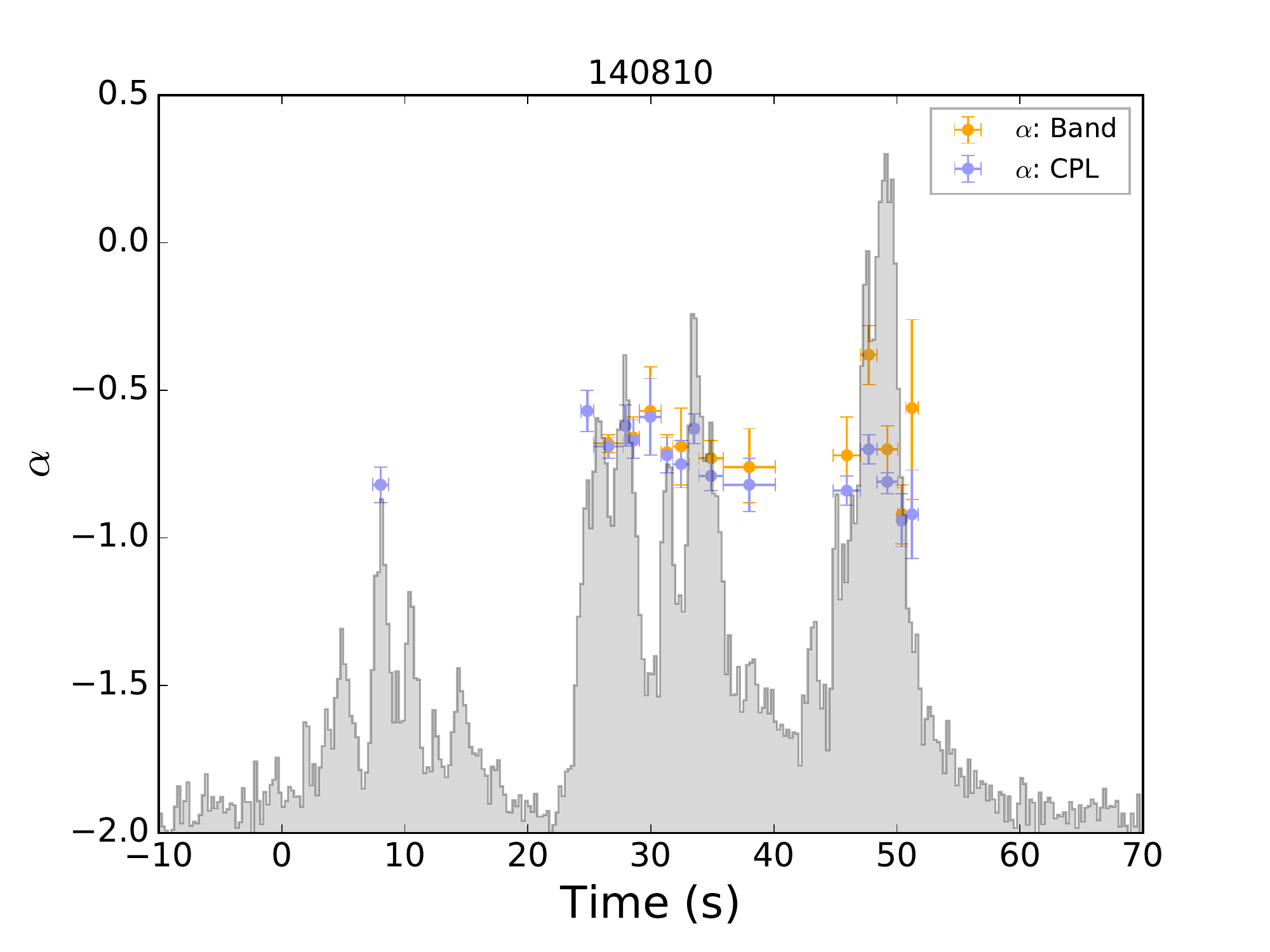}
\includegraphics[angle=0,scale=0.3]{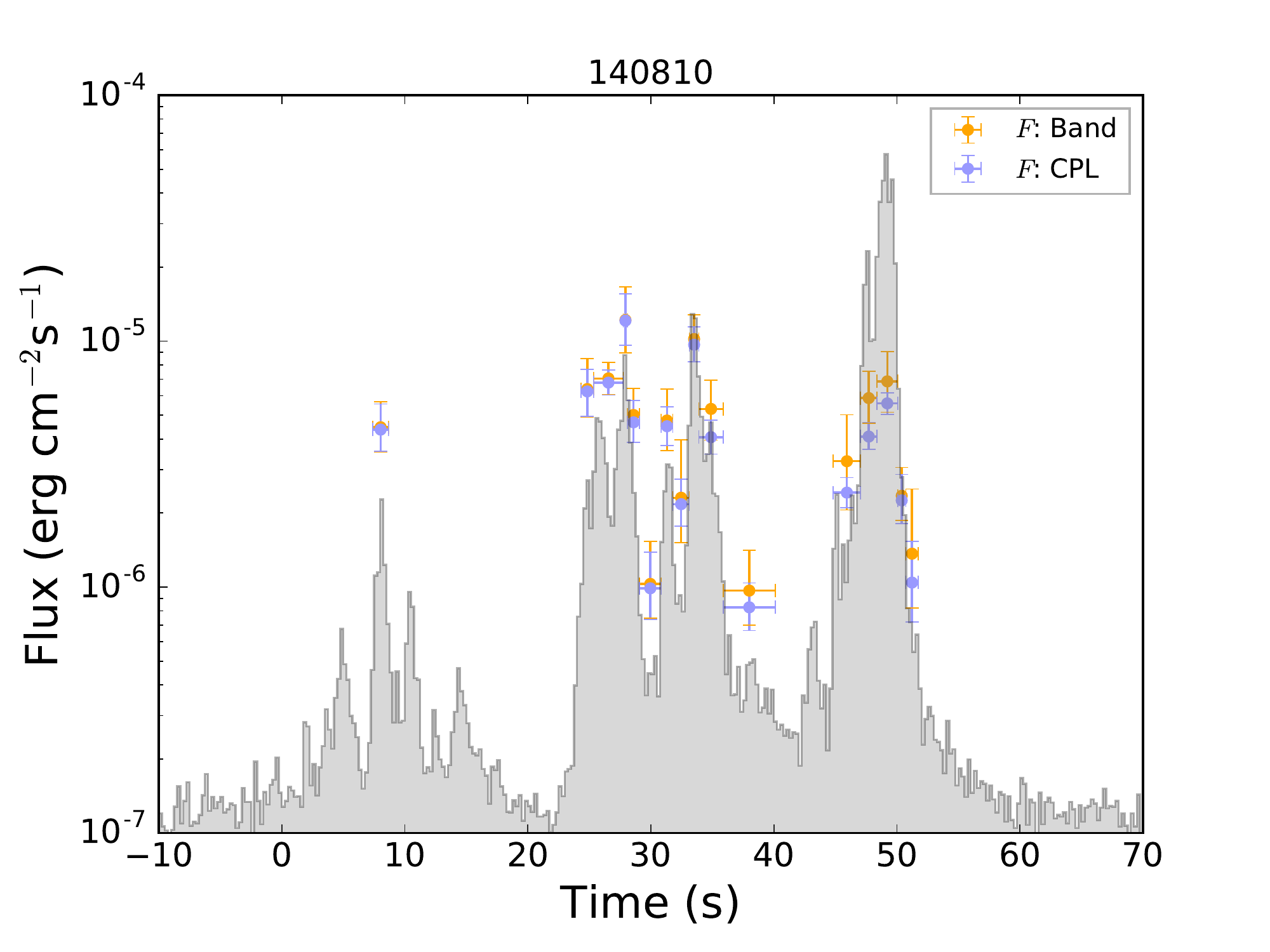}
\center{Fig. \ref{fig:evolution}--- Continued}
\end{figure*}
\begin{figure*}
\includegraphics[angle=0,scale=0.3]{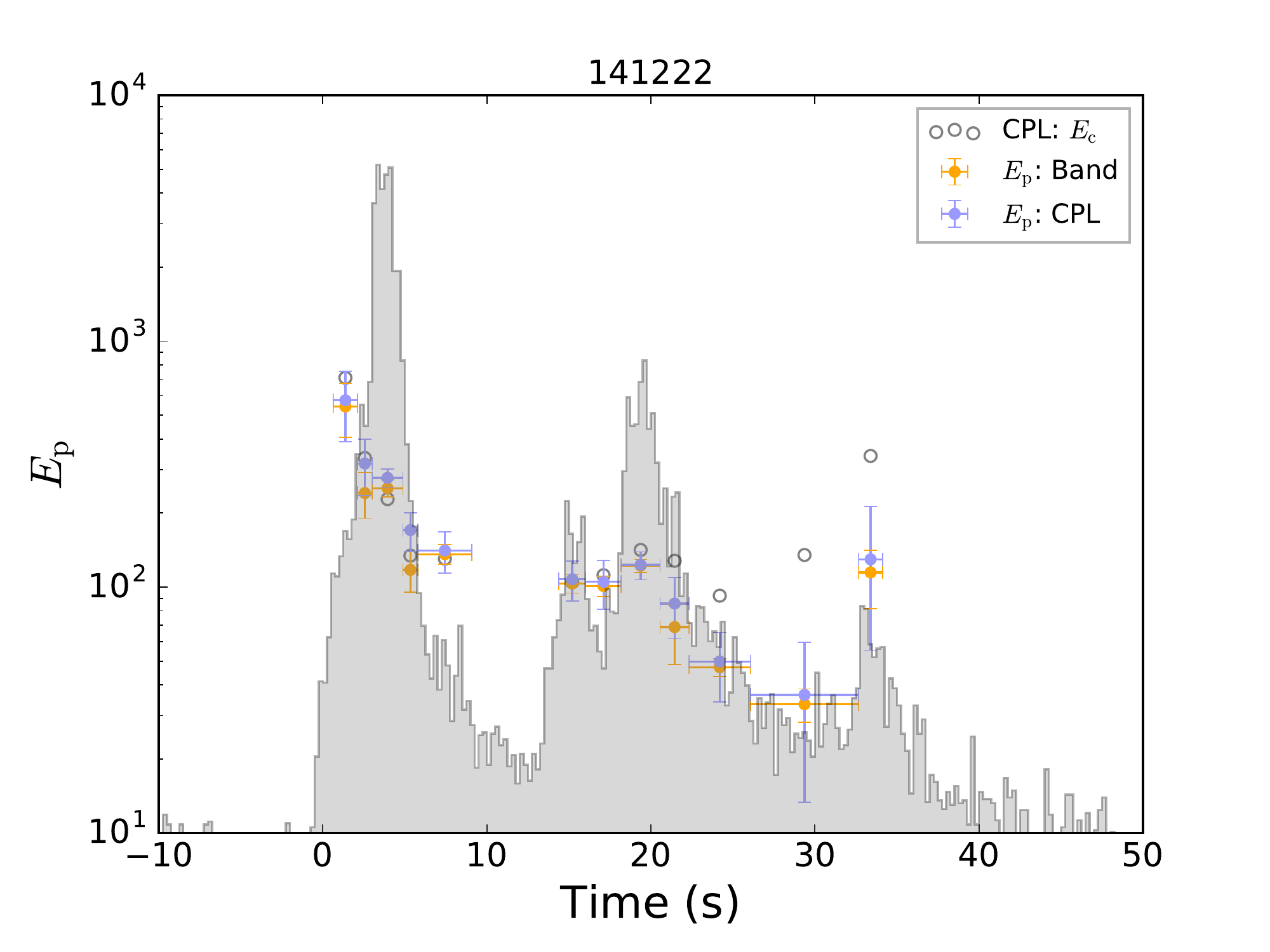}
\includegraphics[angle=0,scale=0.3]{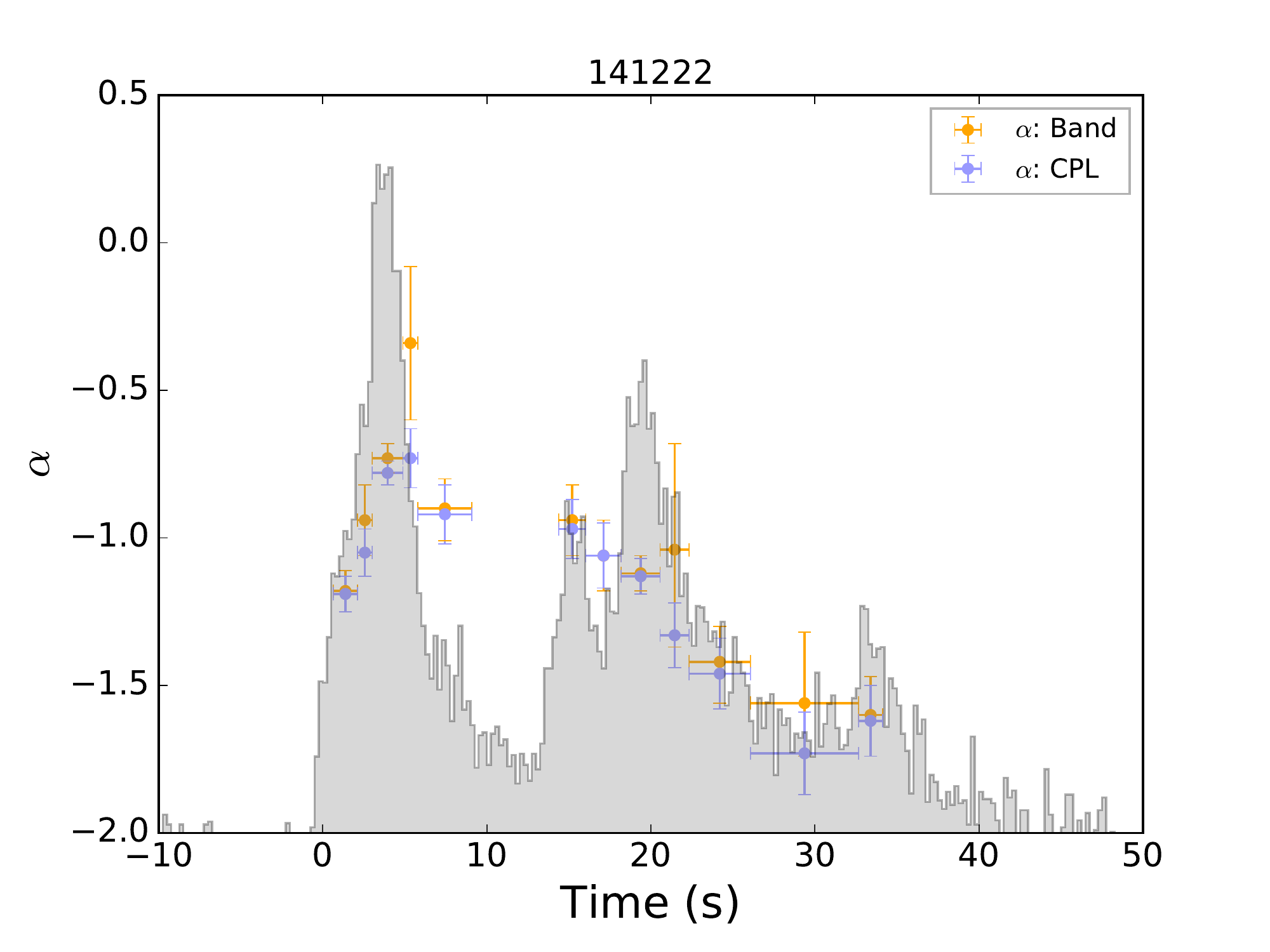}
\includegraphics[angle=0,scale=0.3]{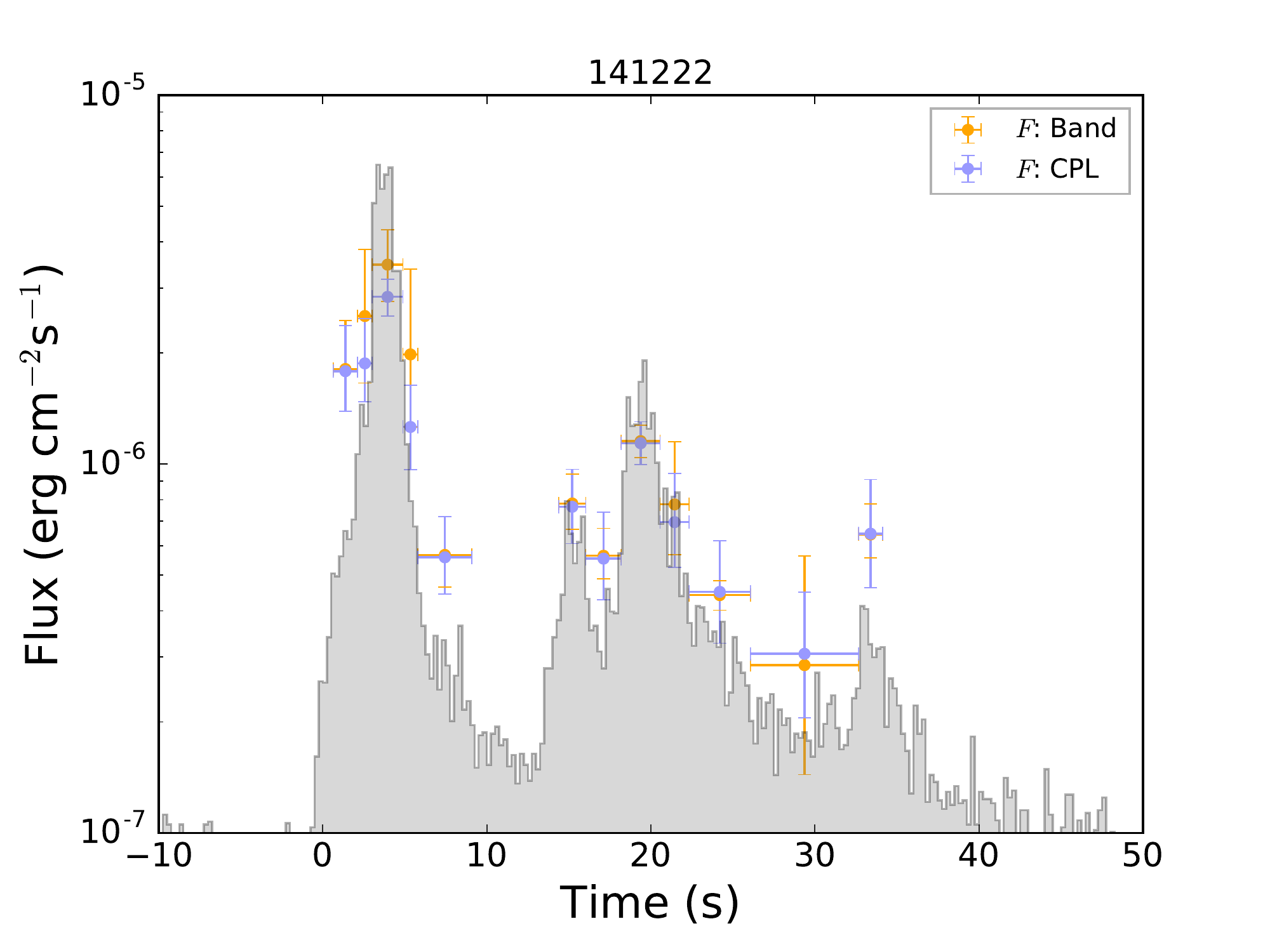}
\includegraphics[angle=0,scale=0.3]{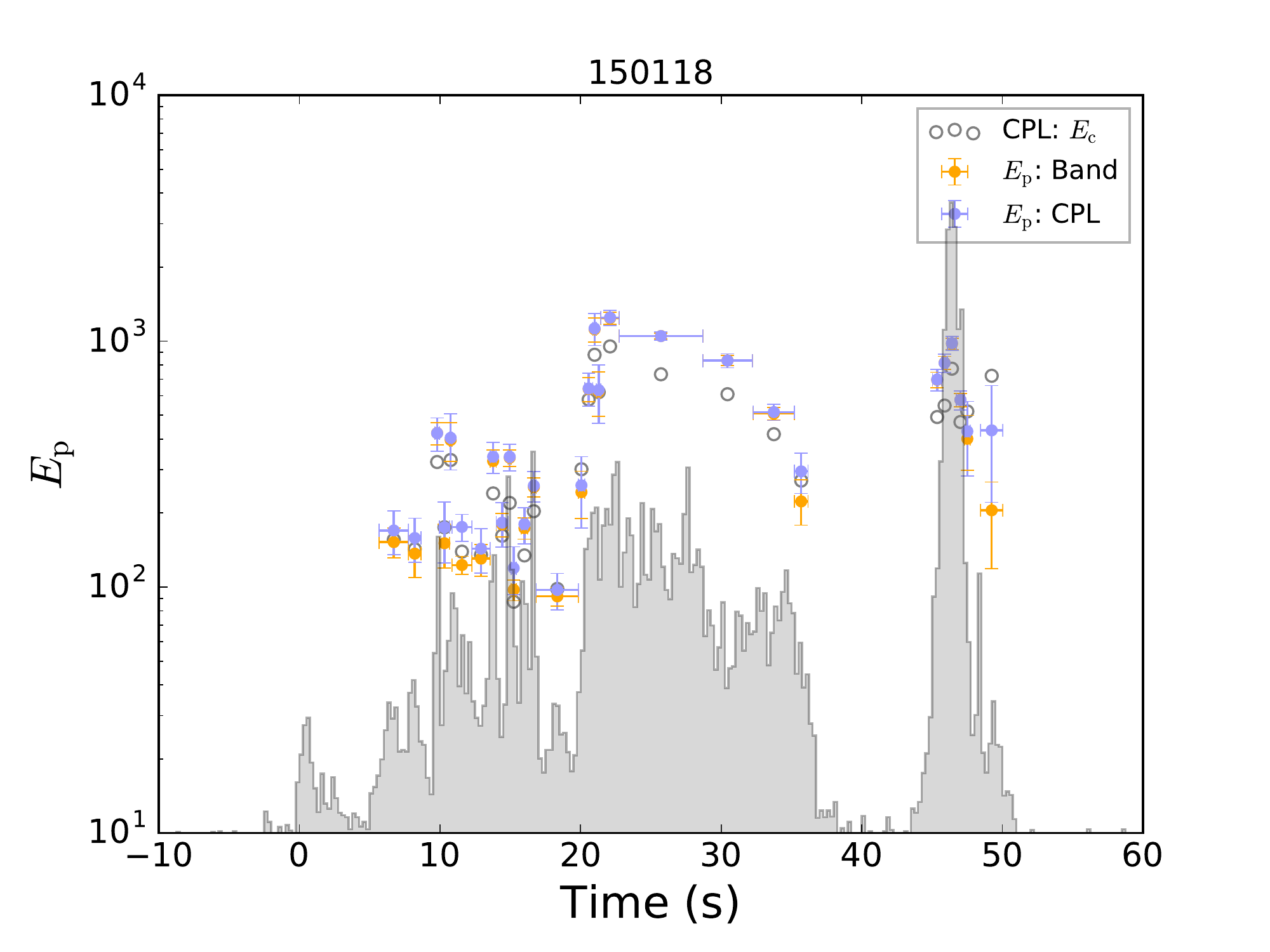}
\includegraphics[angle=0,scale=0.3]{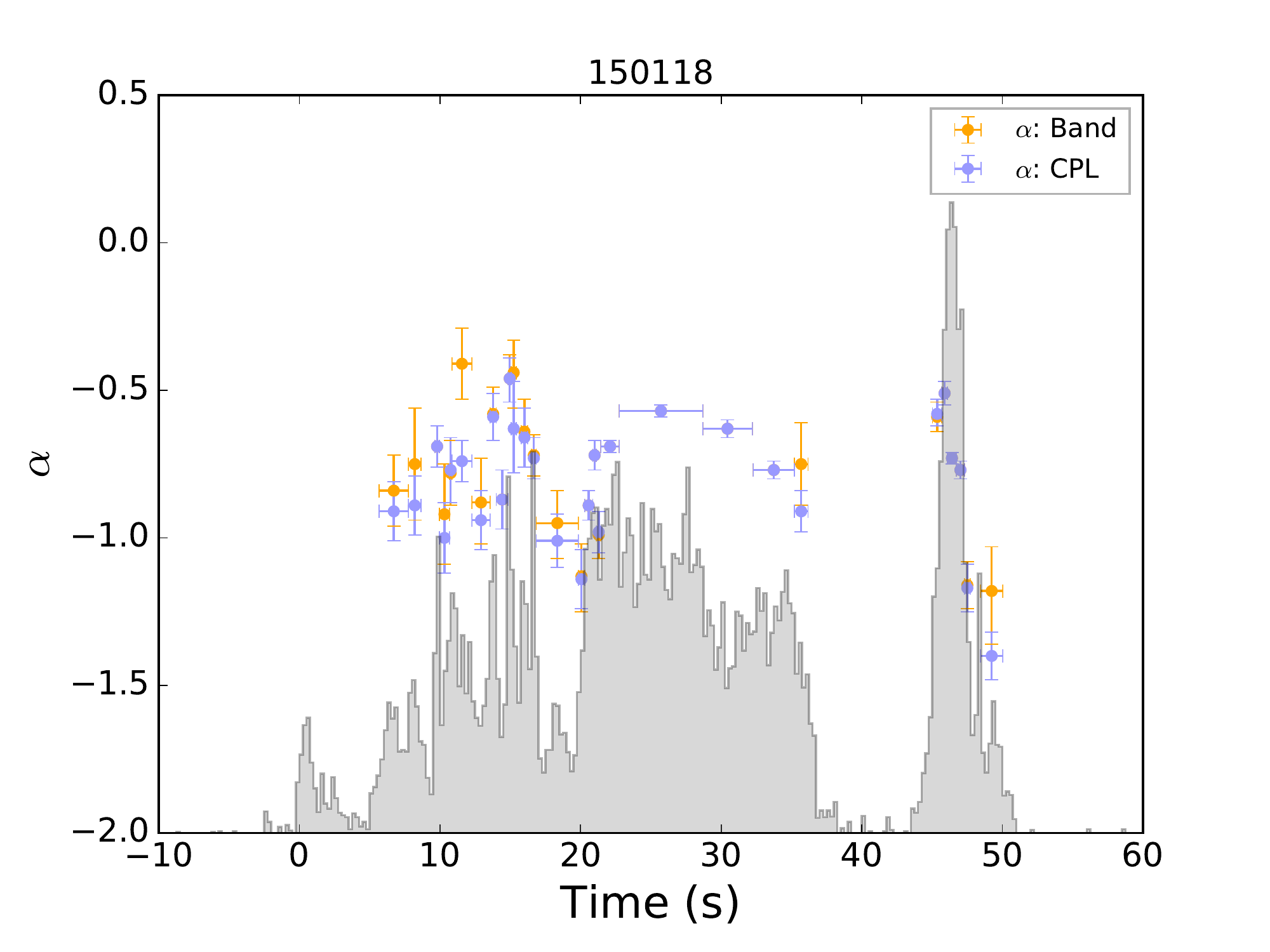}
\includegraphics[angle=0,scale=0.3]{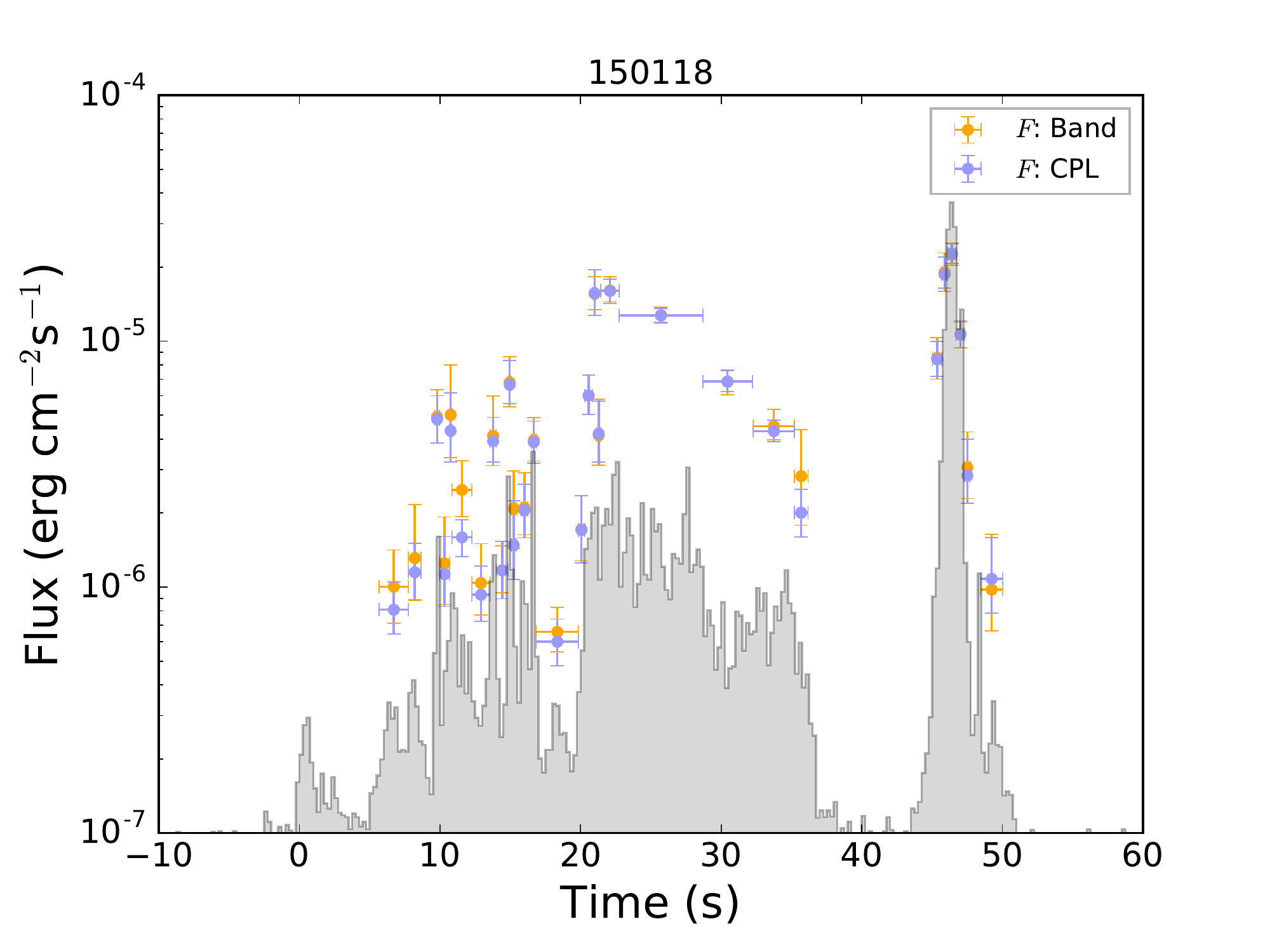}
\includegraphics[angle=0,scale=0.3]{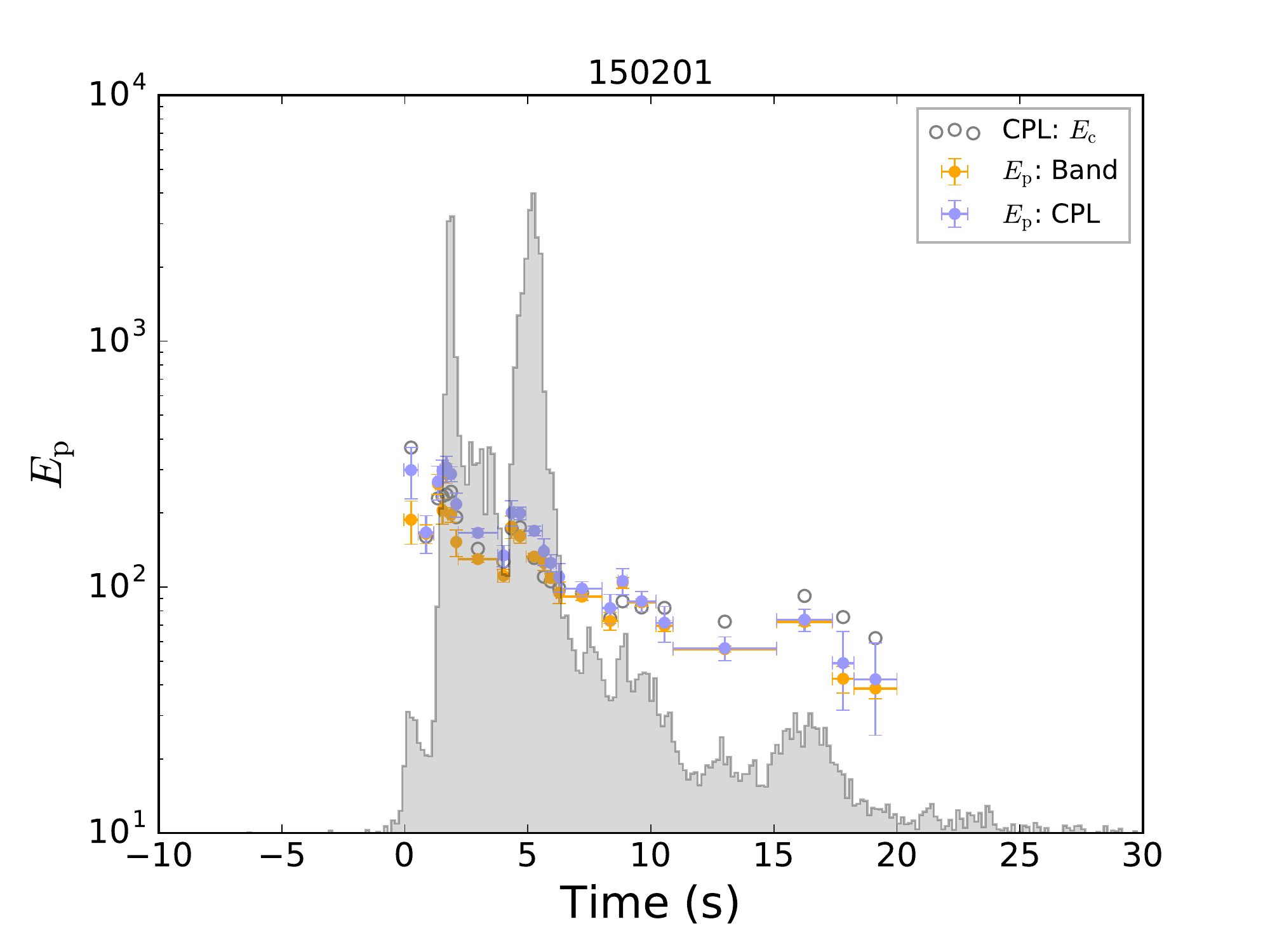}
\includegraphics[angle=0,scale=0.3]{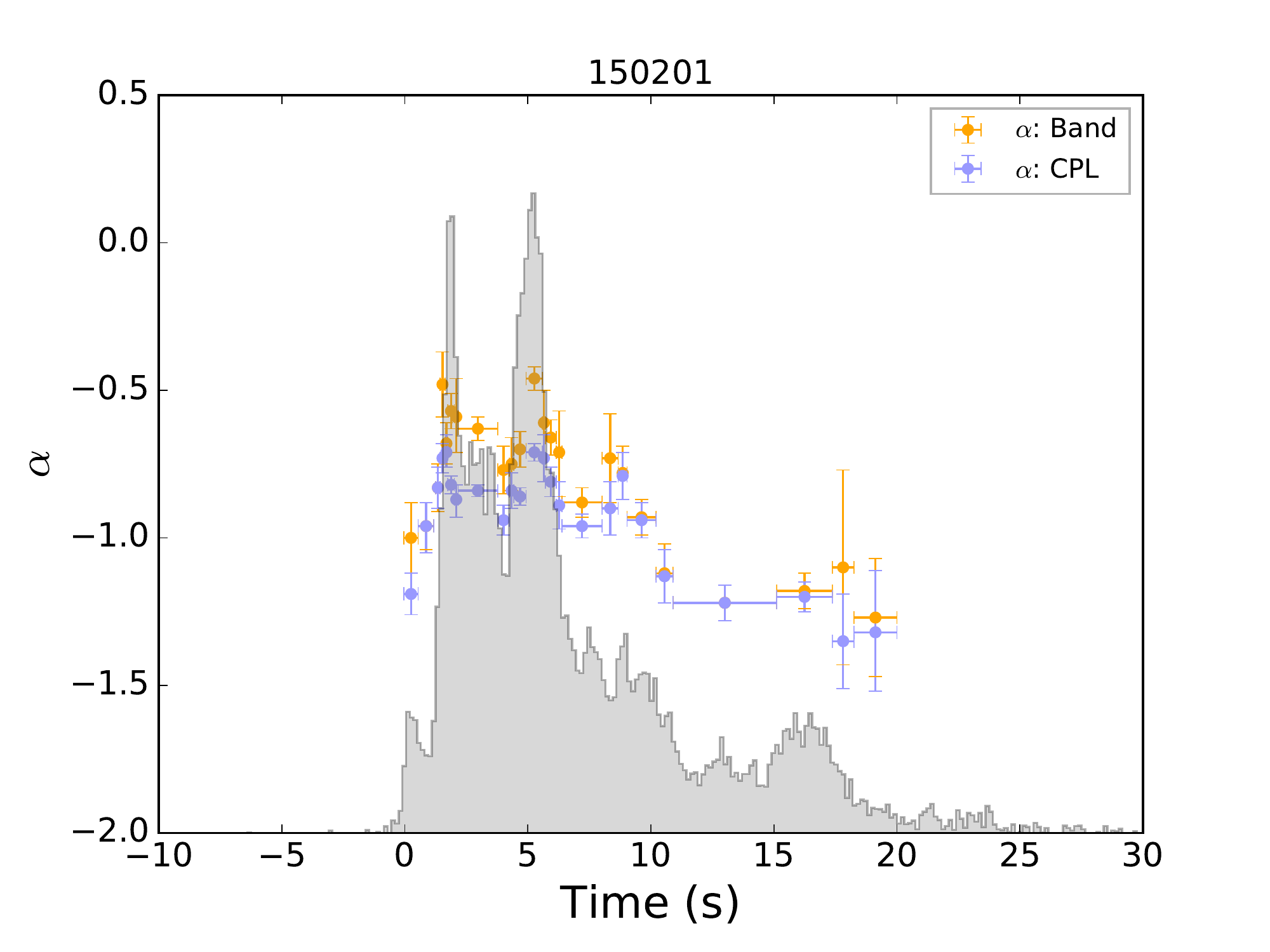}
\includegraphics[angle=0,scale=0.3]{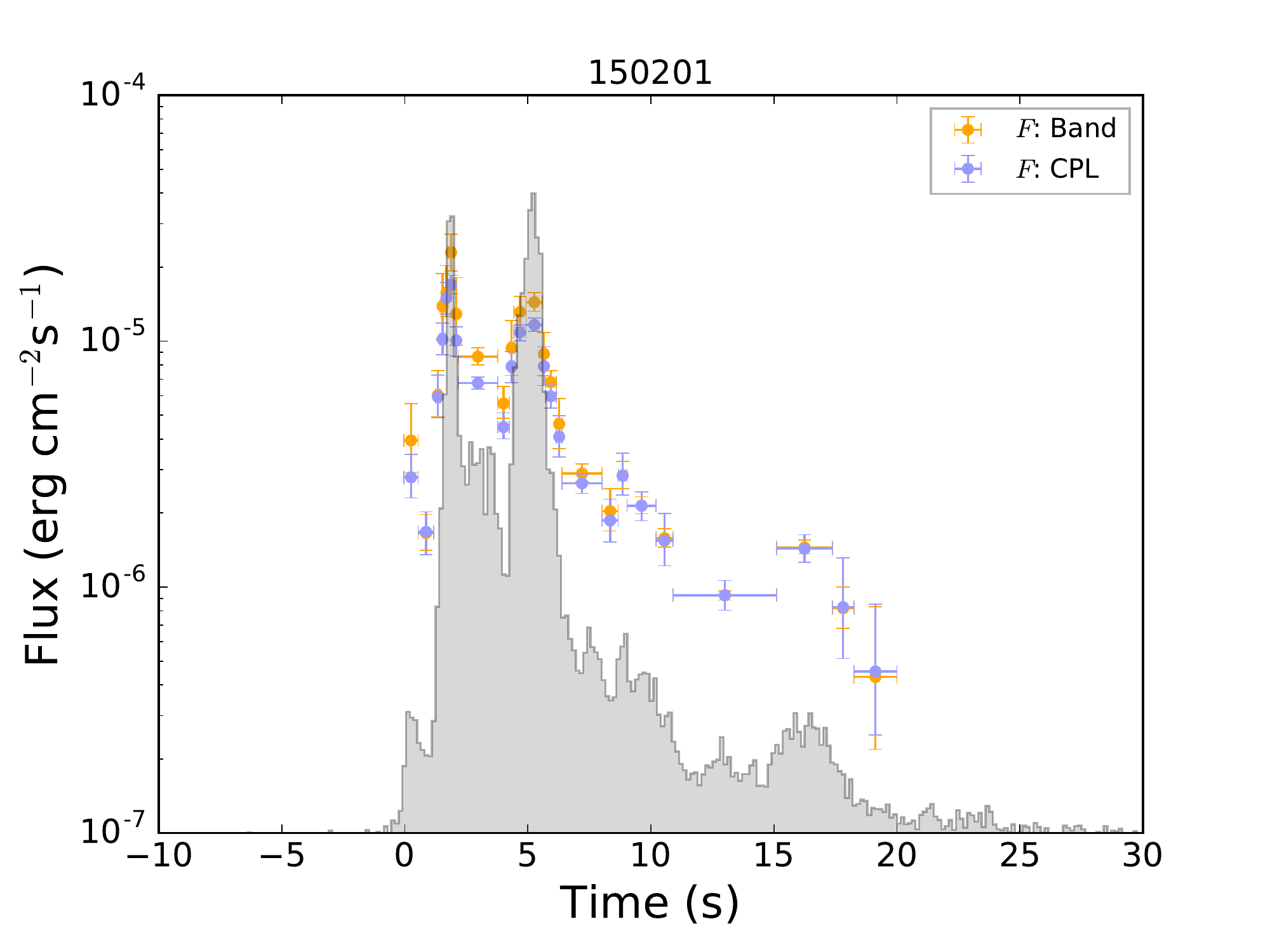}
\includegraphics[angle=0,scale=0.3]{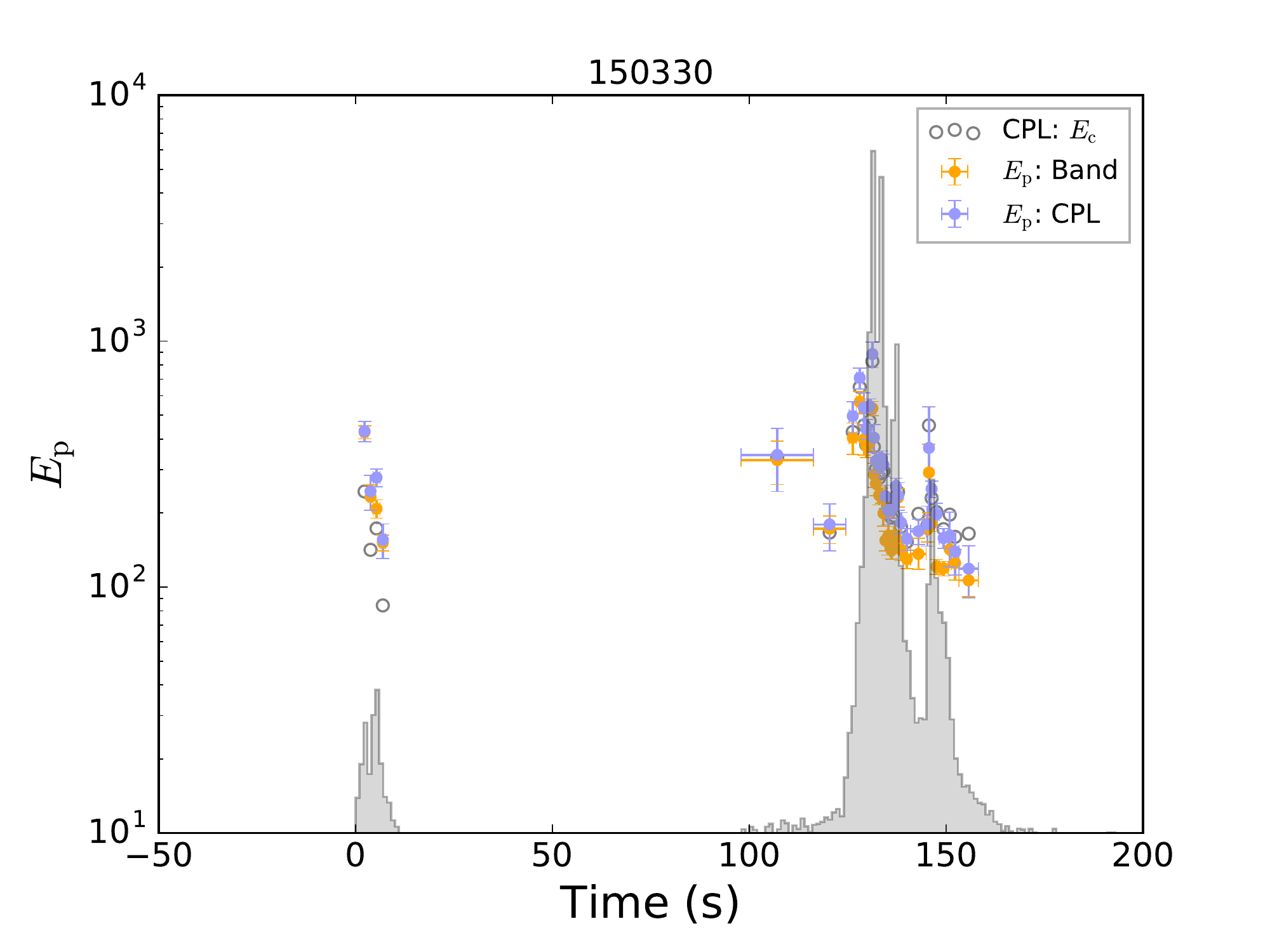}
\includegraphics[angle=0,scale=0.3]{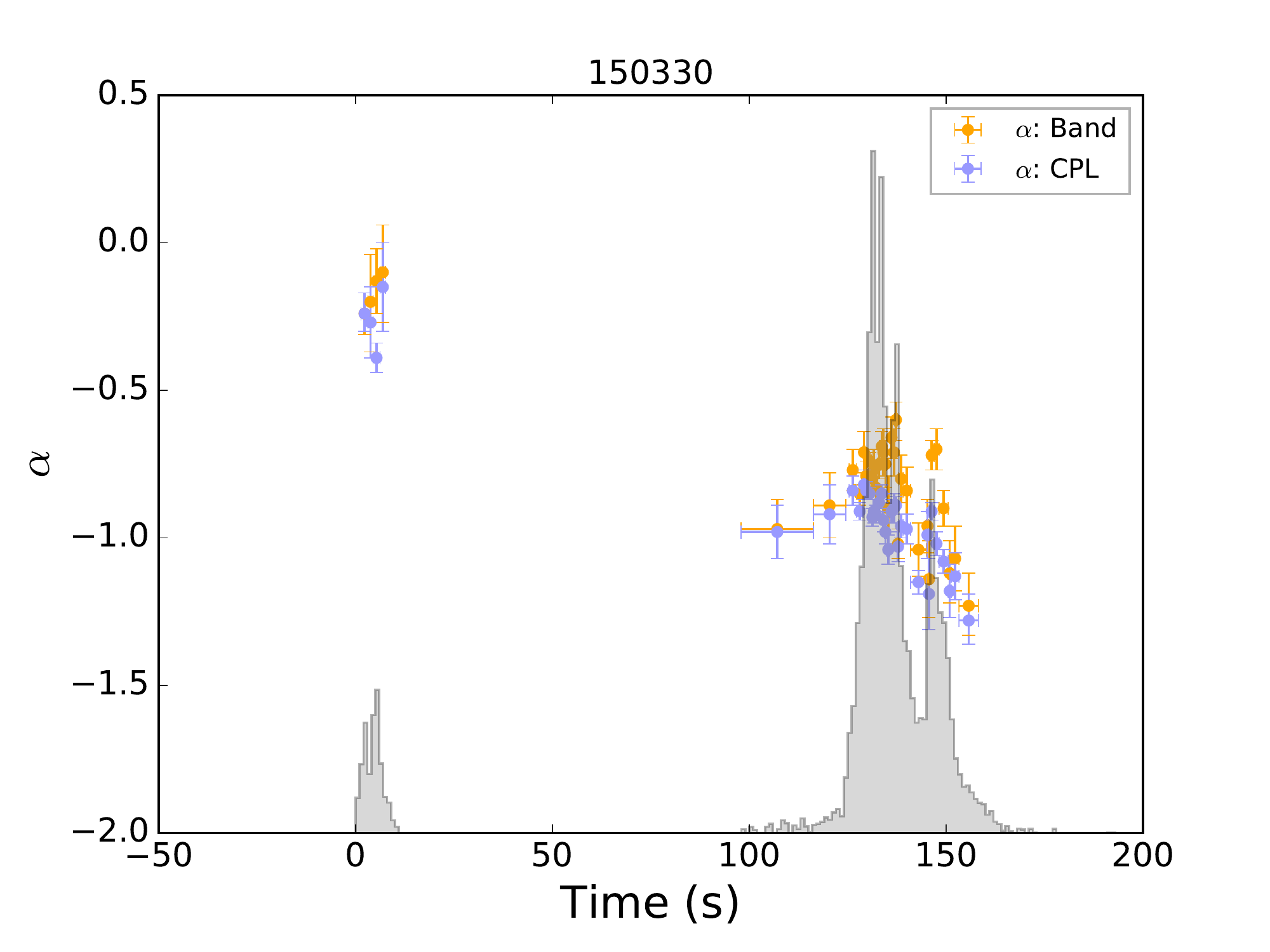}
\includegraphics[angle=0,scale=0.3]{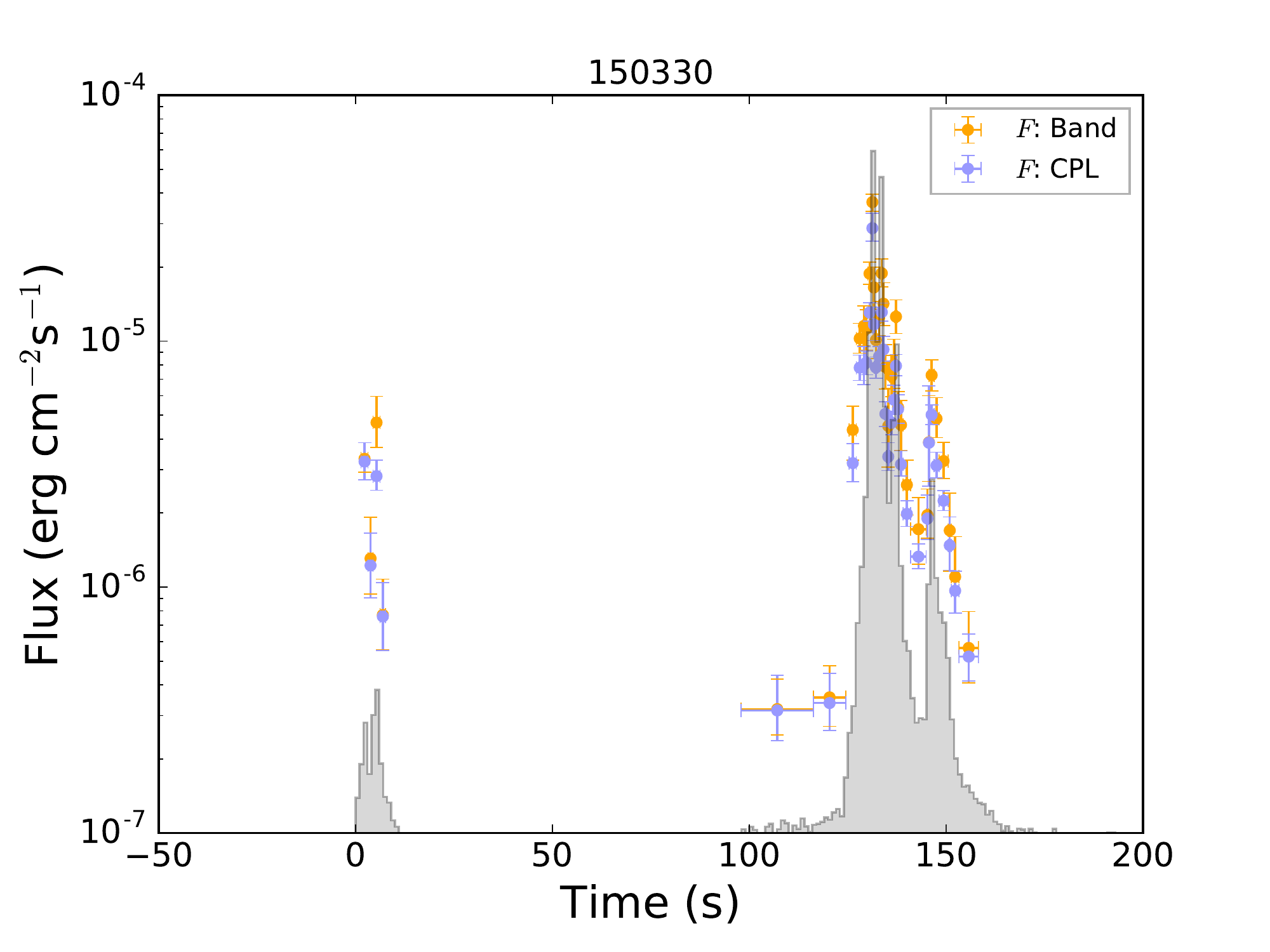}
\includegraphics[angle=0,scale=0.3]{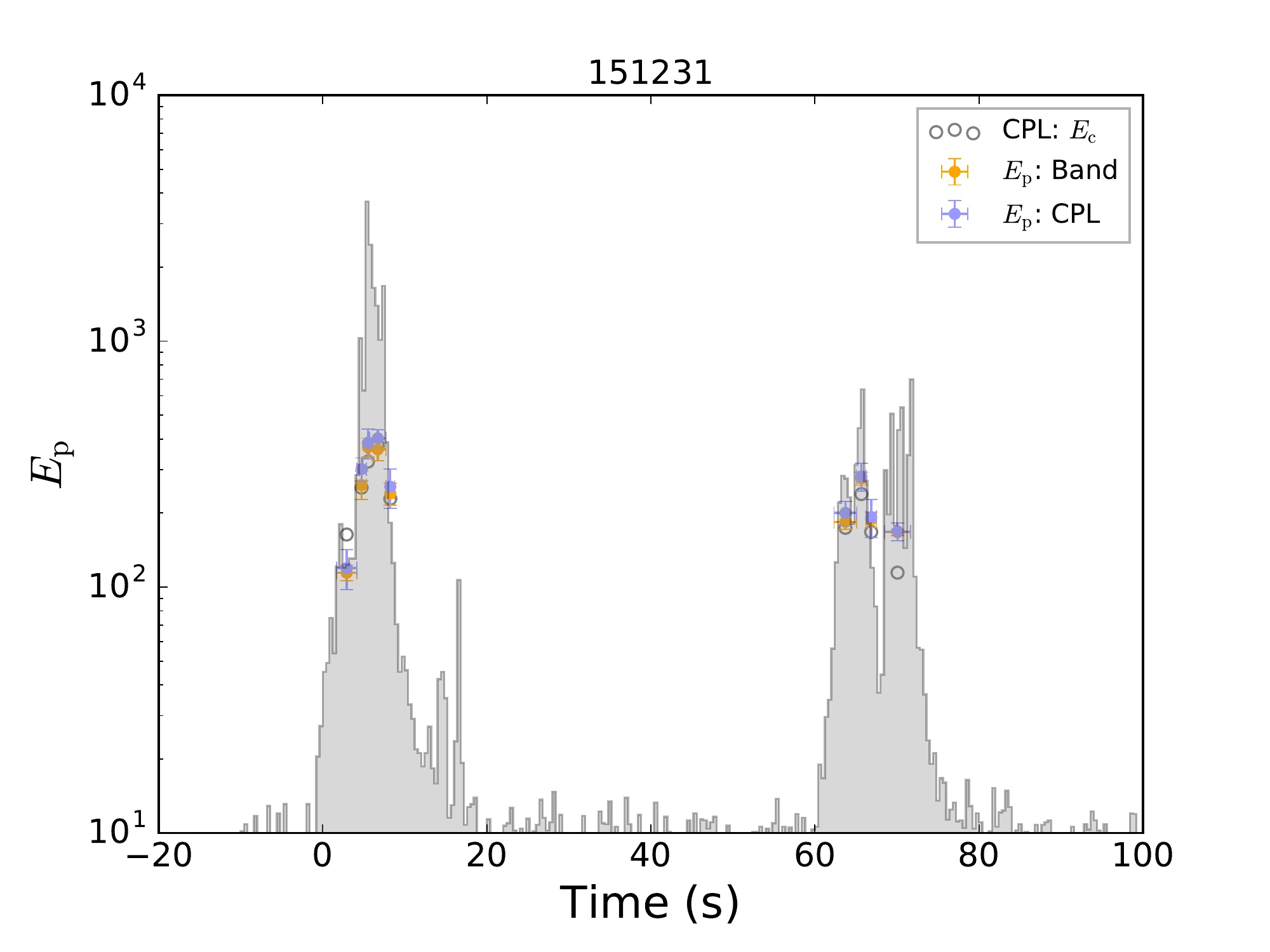}
\includegraphics[angle=0,scale=0.3]{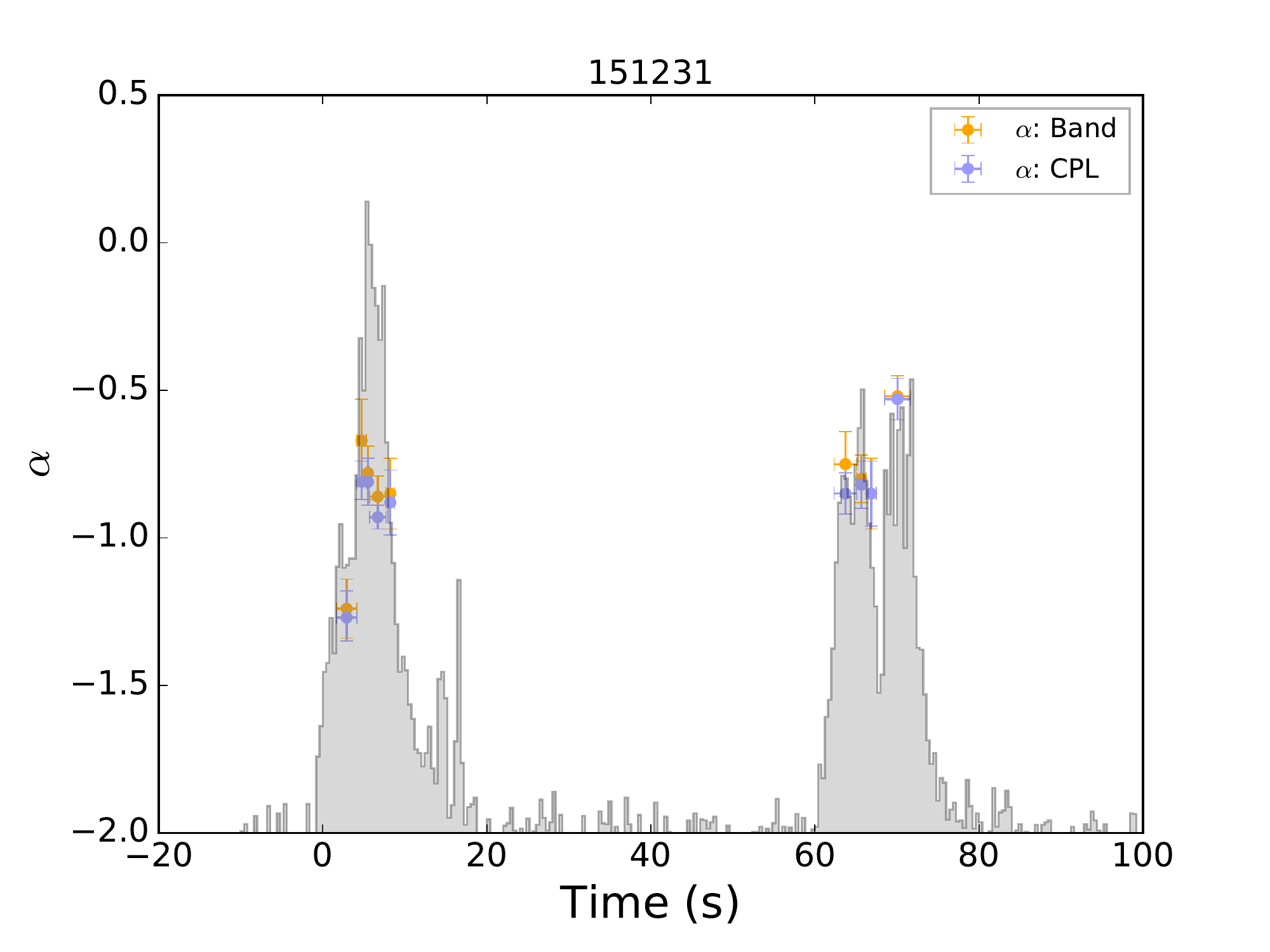}
\includegraphics[angle=0,scale=0.3]{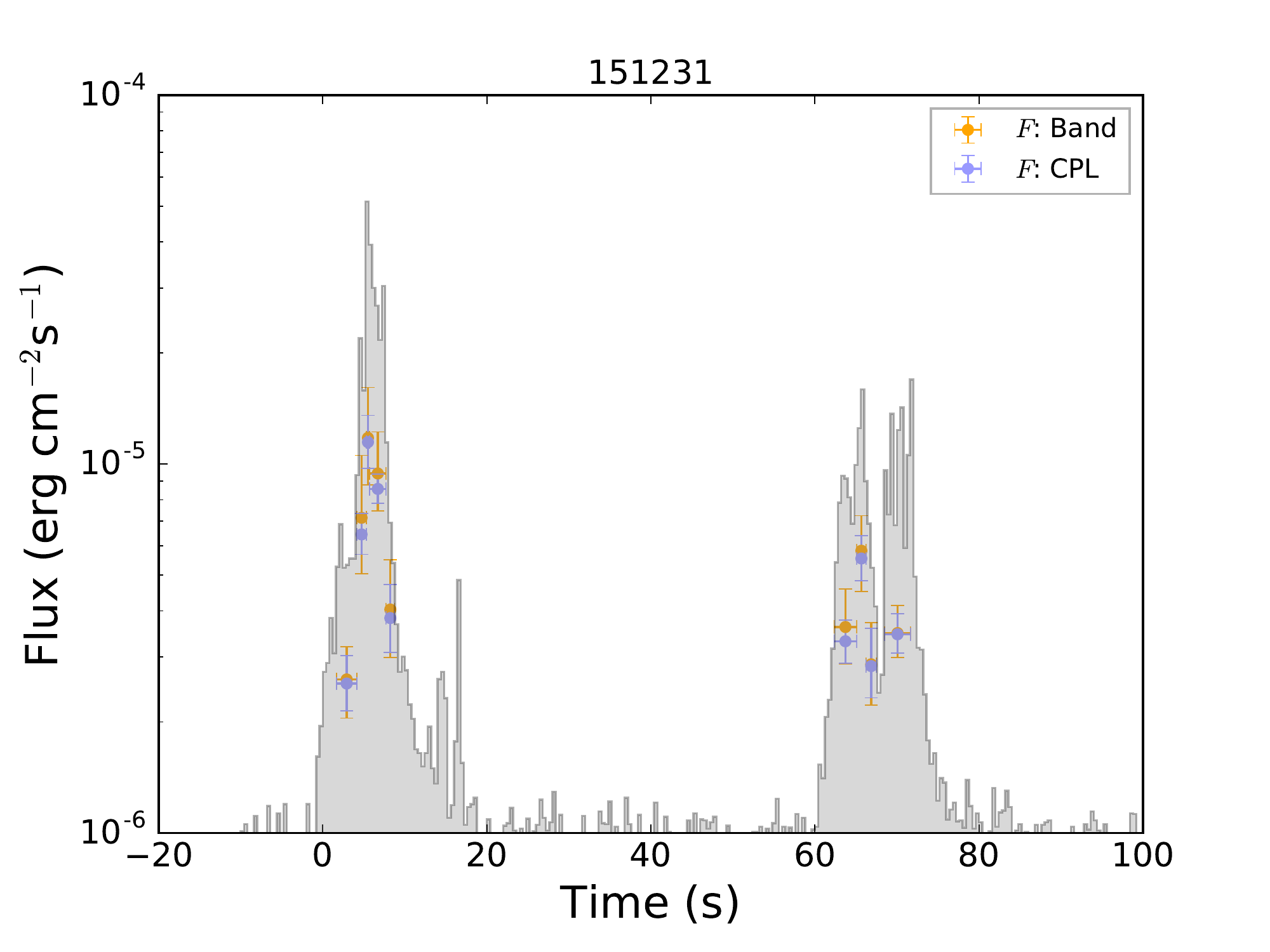}
\center{Fig. \ref{fig:evolution}--- Continued}
\end{figure*}
\begin{figure*}
\includegraphics[angle=0,scale=0.3]{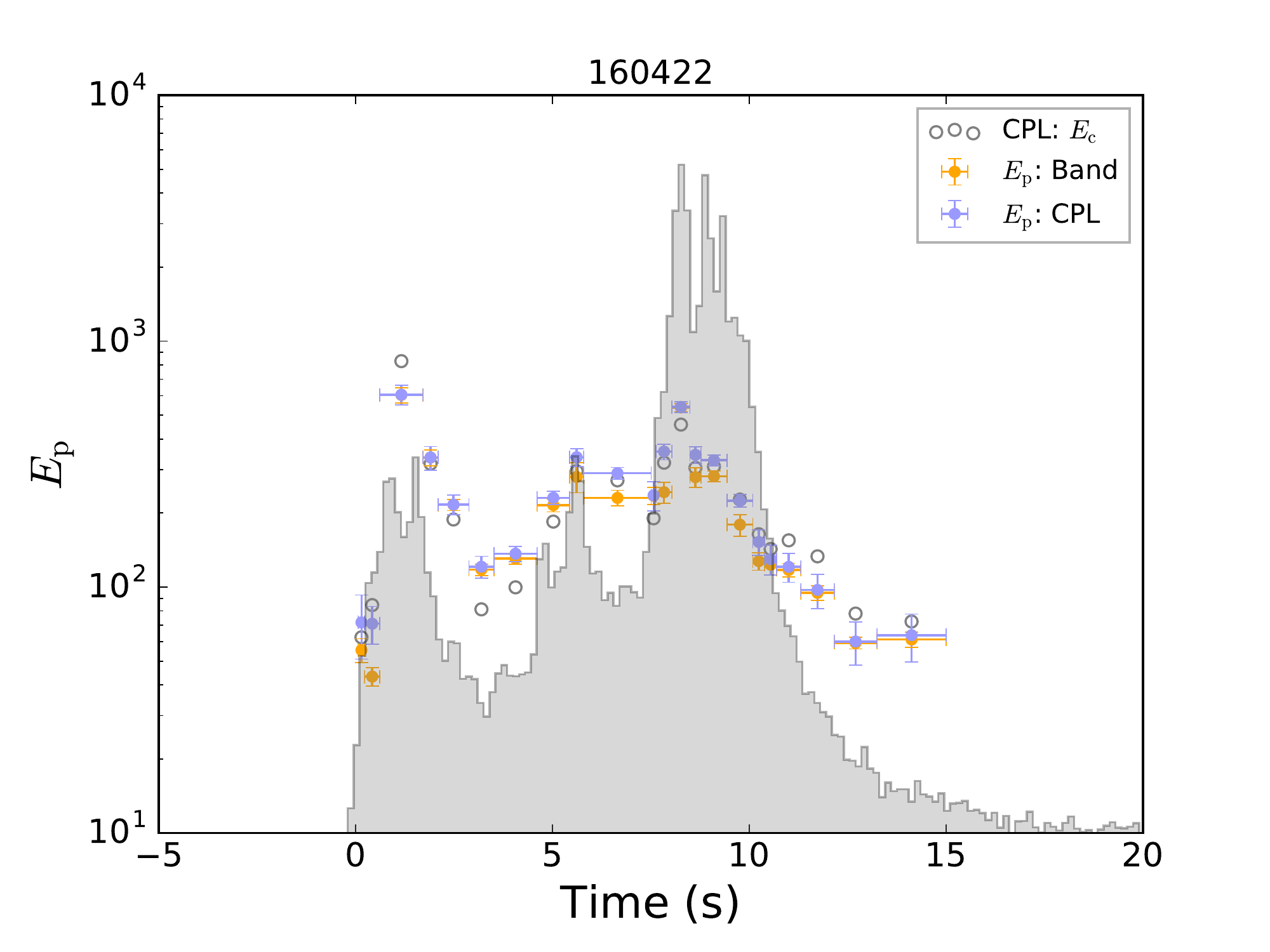}
\includegraphics[angle=0,scale=0.3]{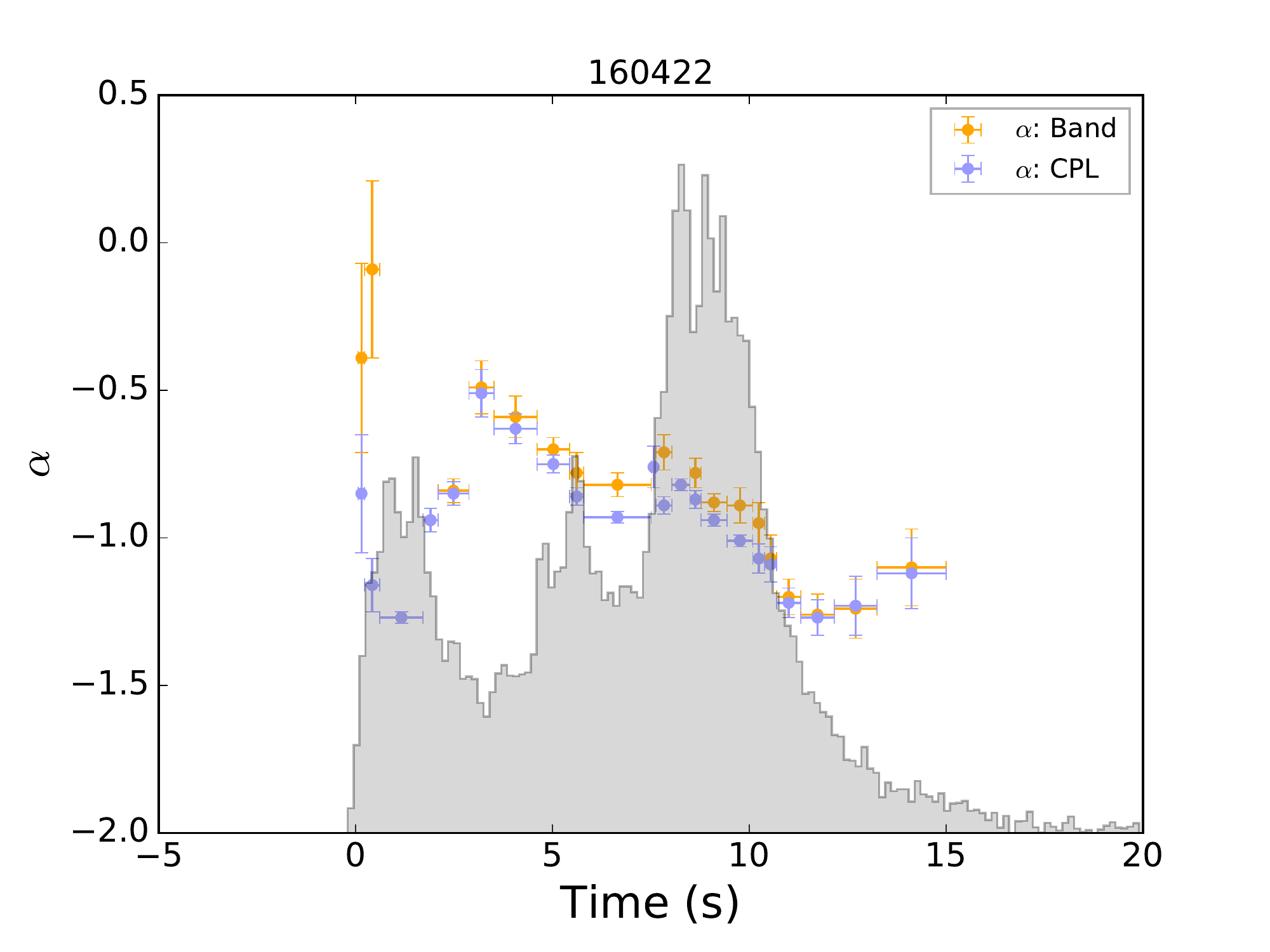}
\includegraphics[angle=0,scale=0.3]{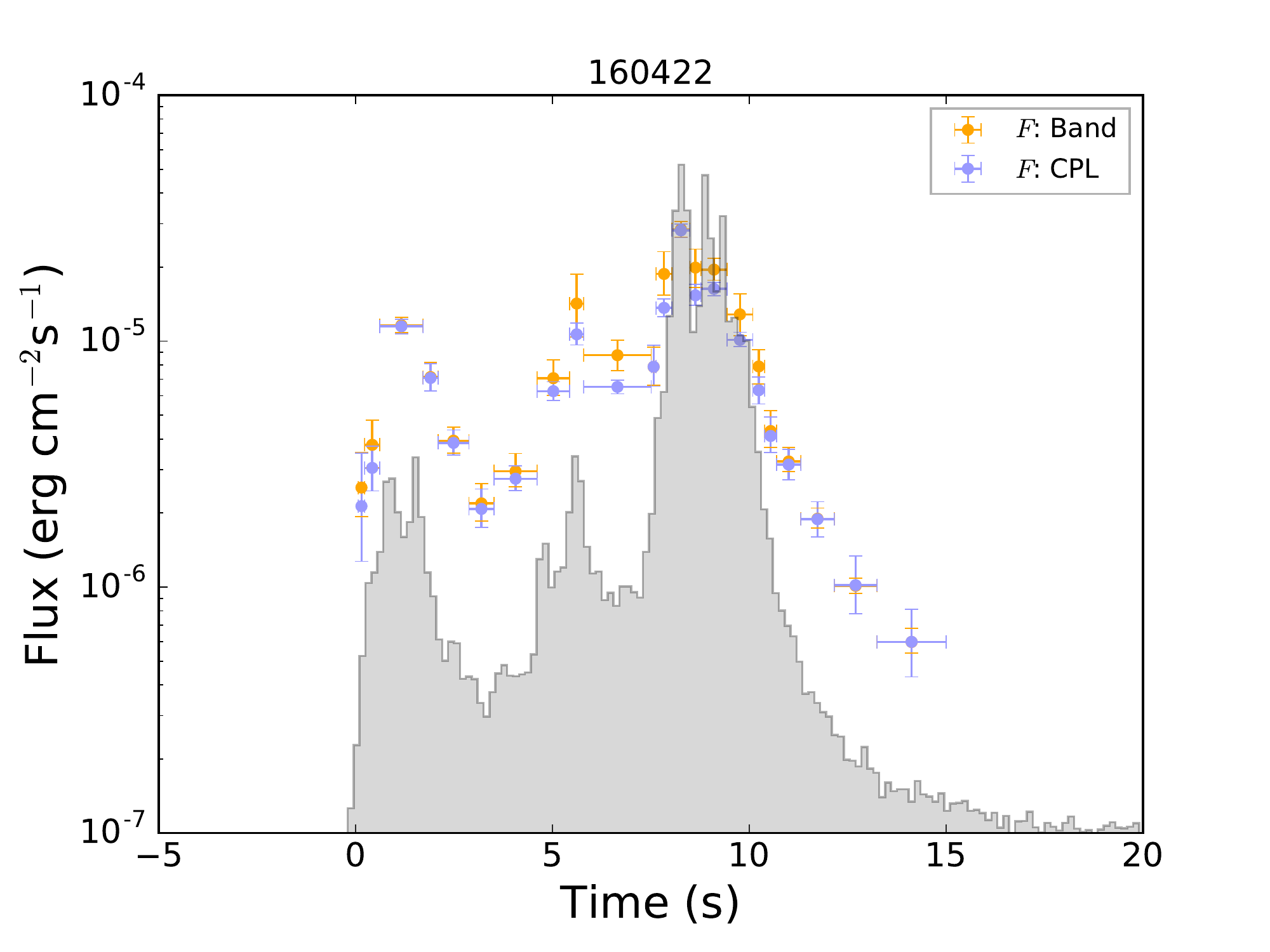}
\includegraphics[angle=0,scale=0.3]{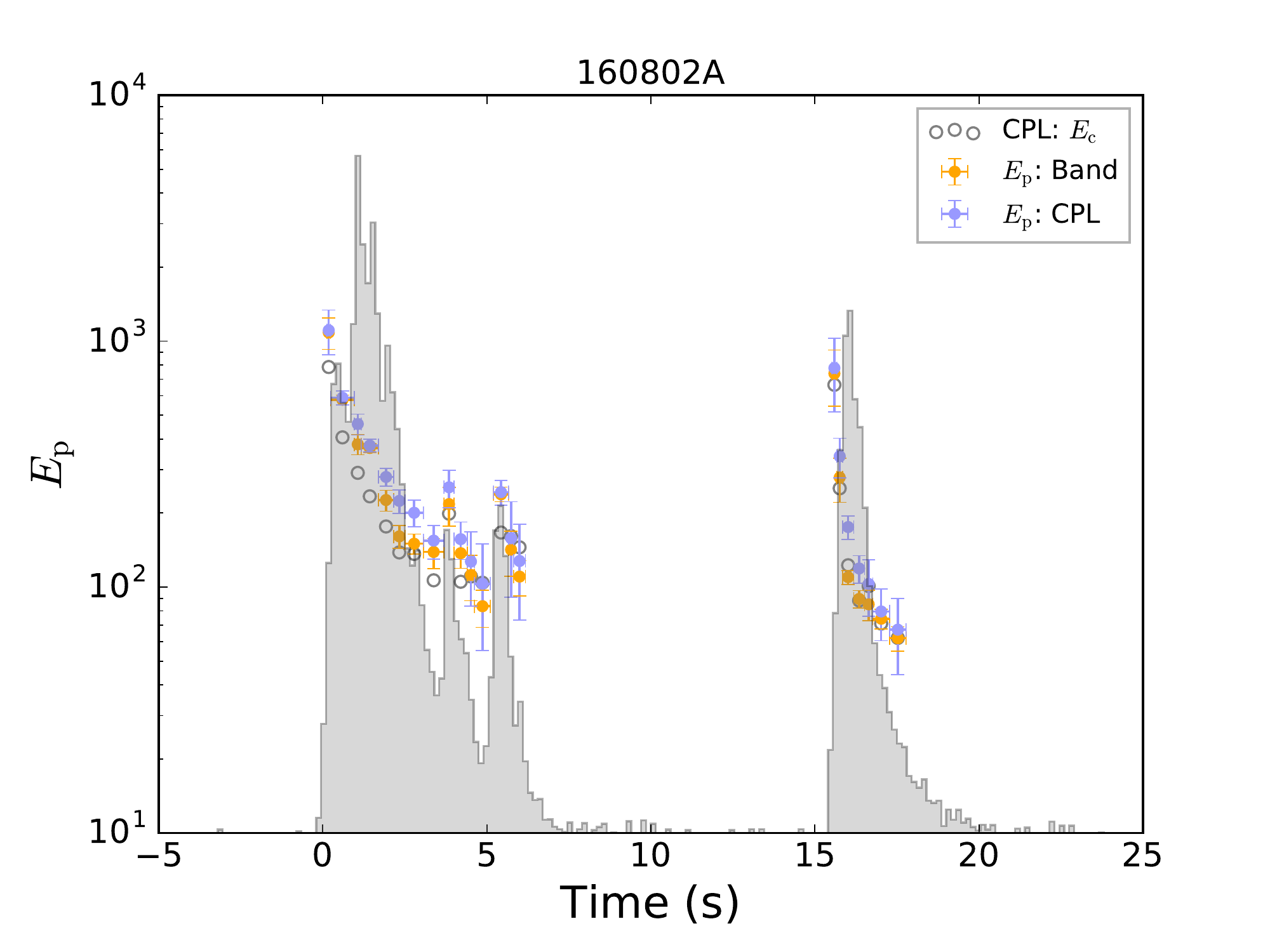}
\includegraphics[angle=0,scale=0.3]{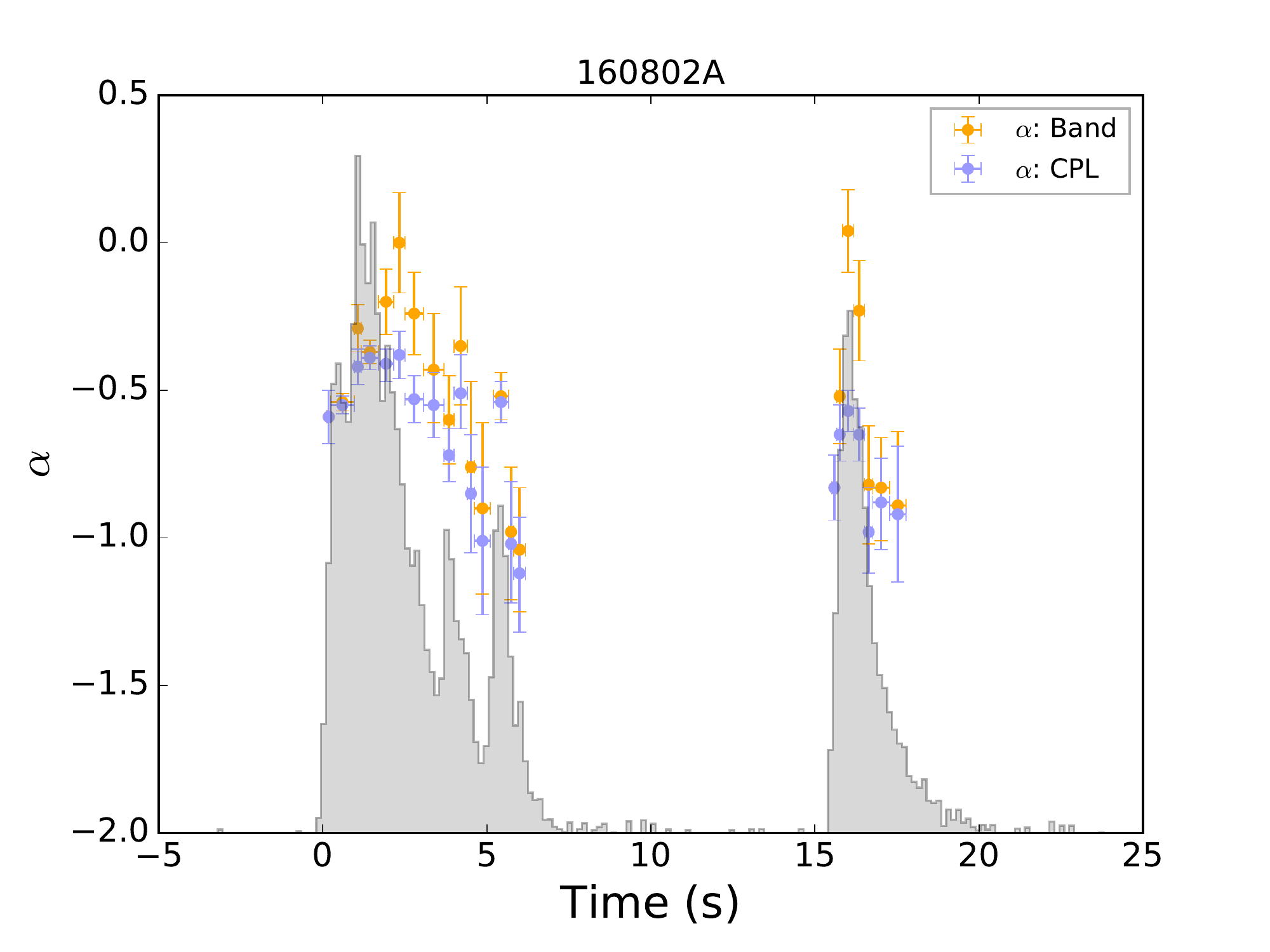}
\includegraphics[angle=0,scale=0.3]{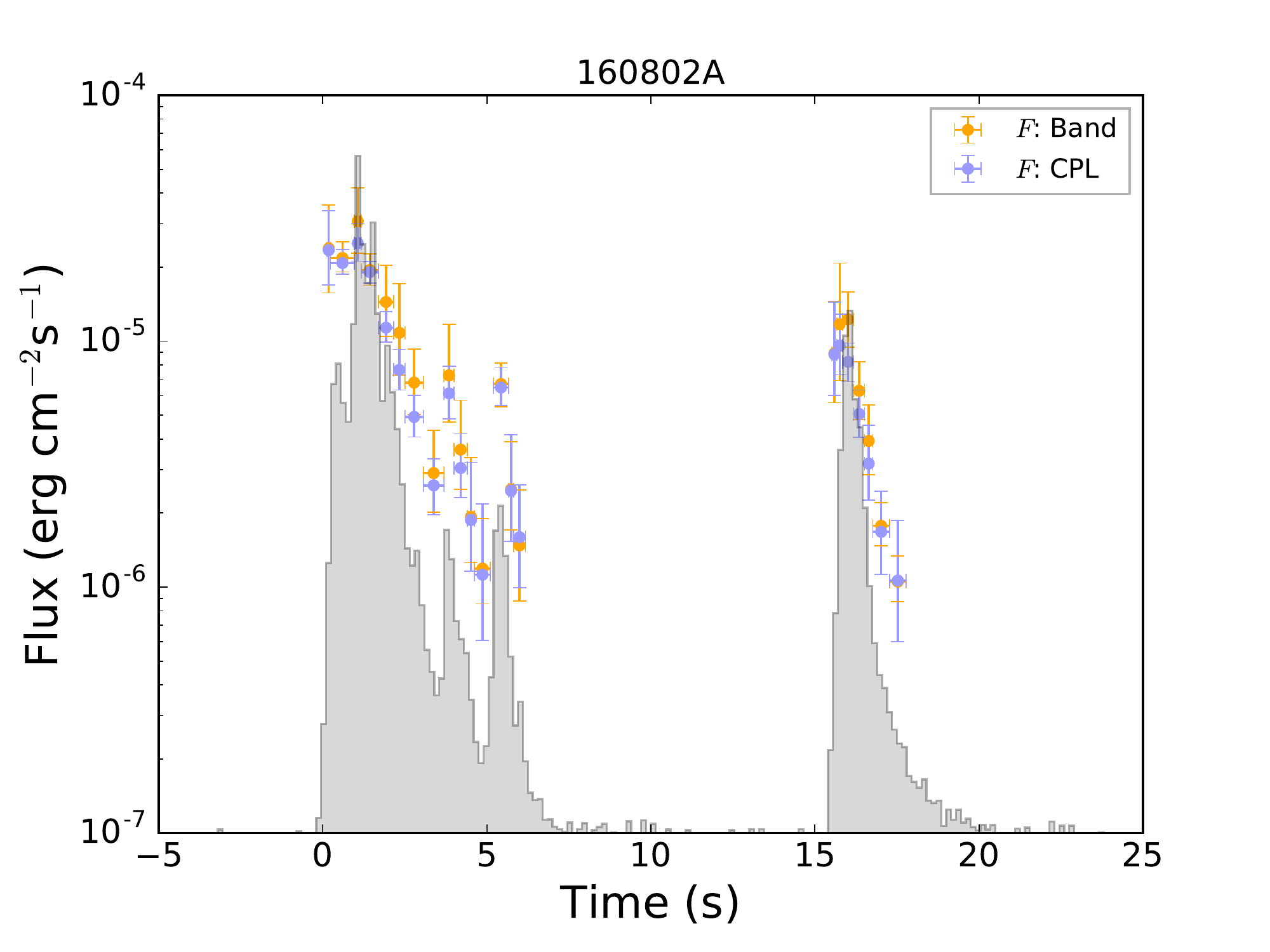}
\includegraphics[angle=0,scale=0.3]{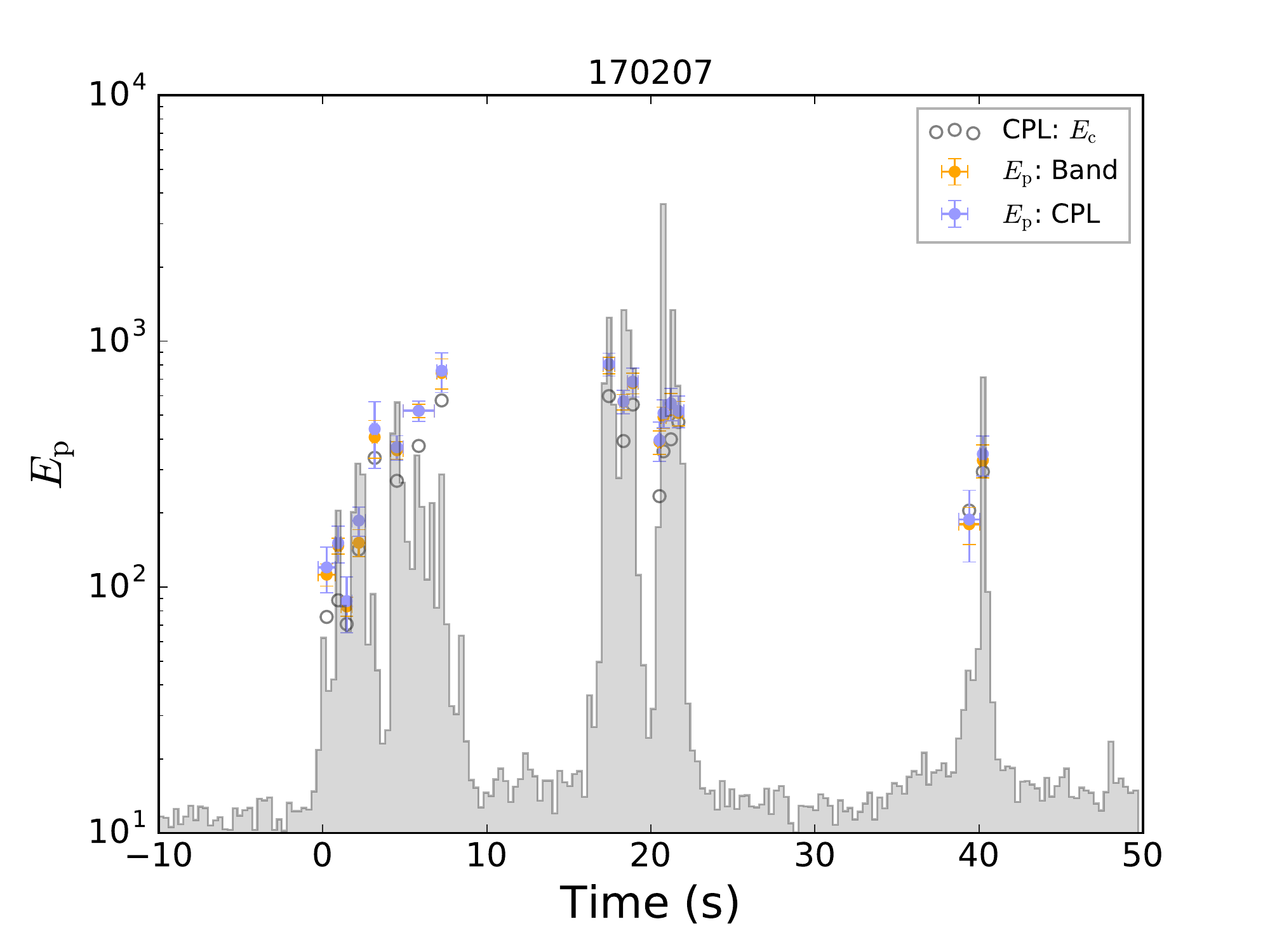}
\includegraphics[angle=0,scale=0.3]{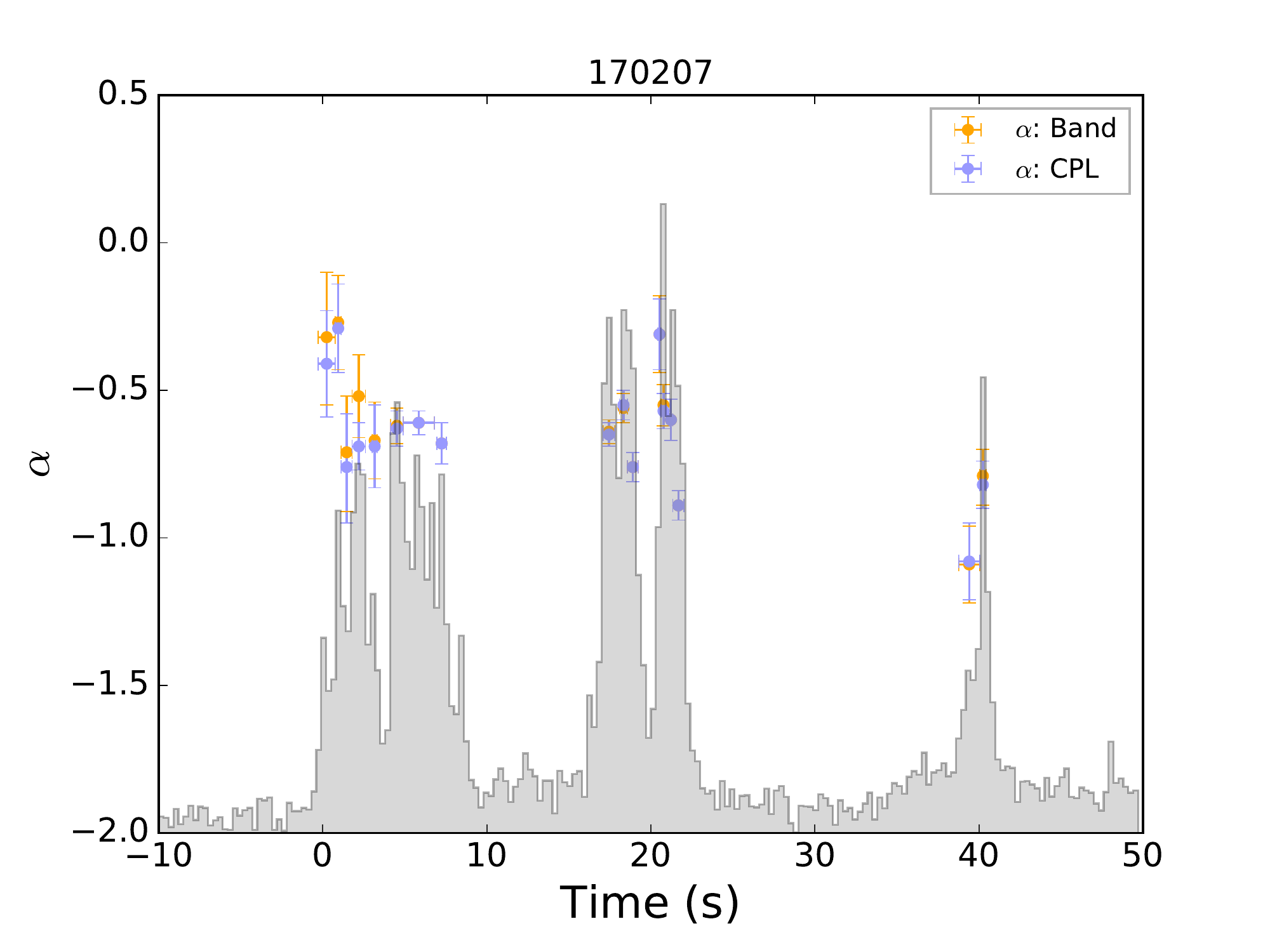}
\includegraphics[angle=0,scale=0.3]{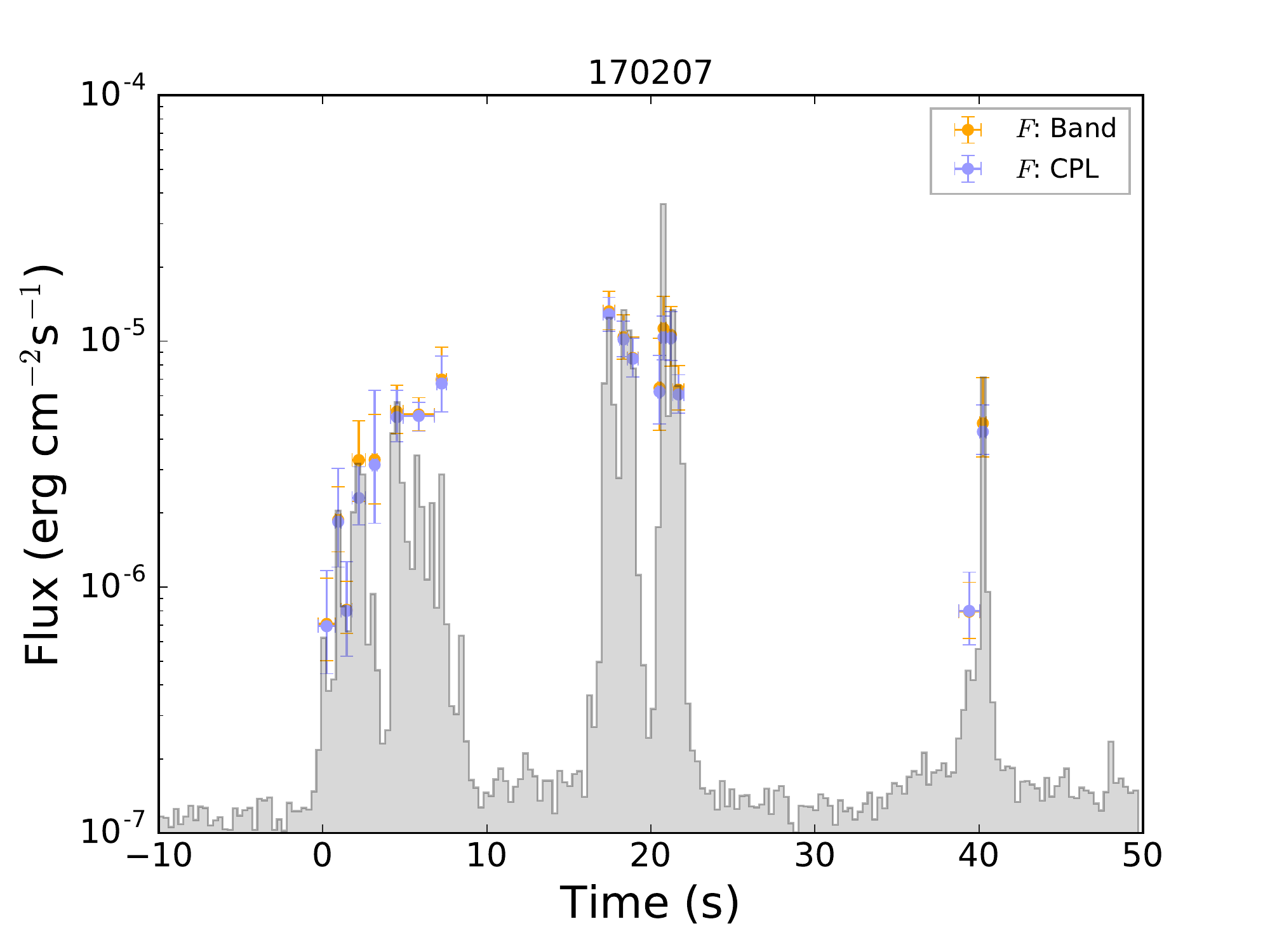}
\includegraphics[angle=0,scale=0.3]{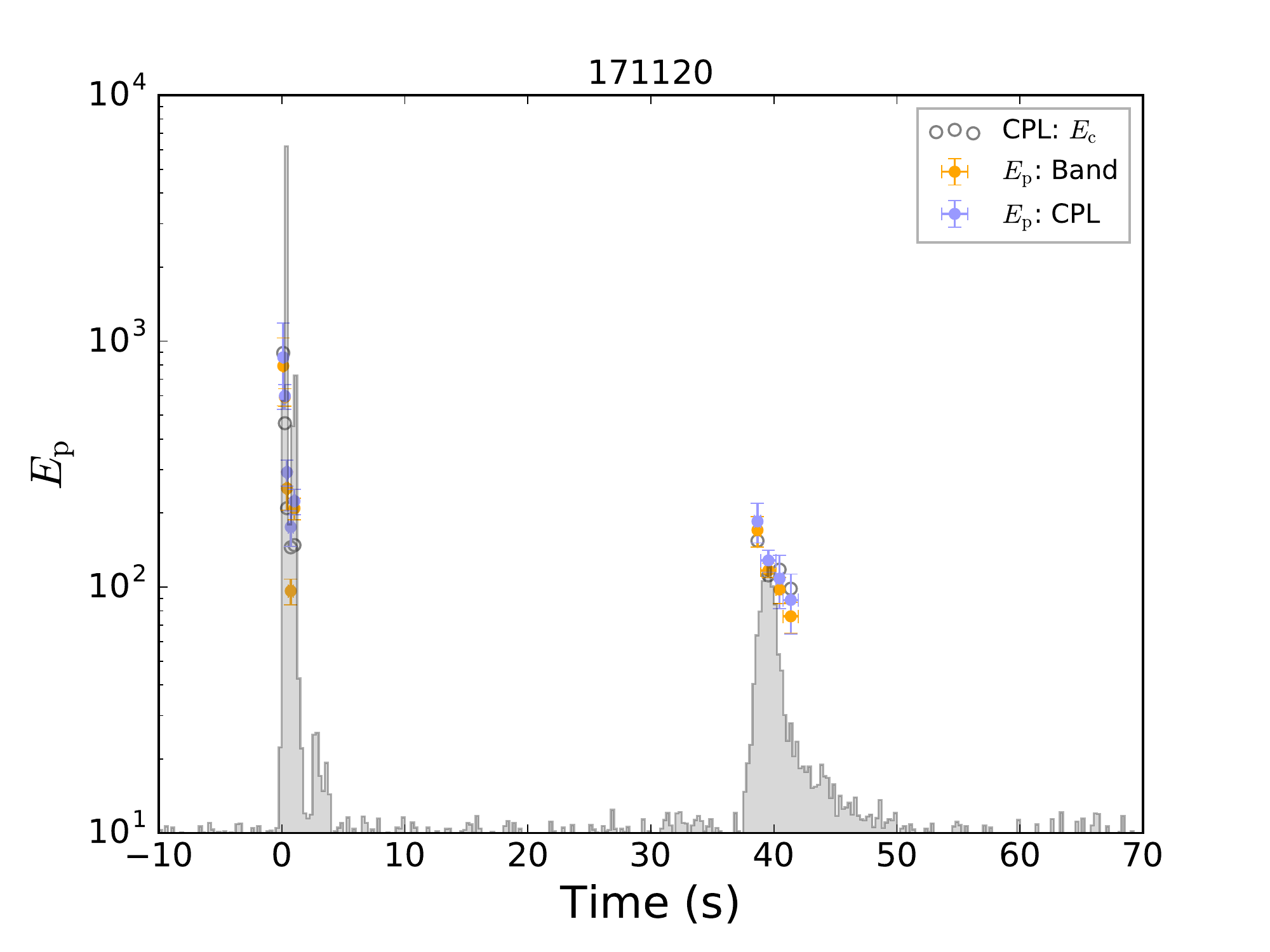}
\includegraphics[angle=0,scale=0.3]{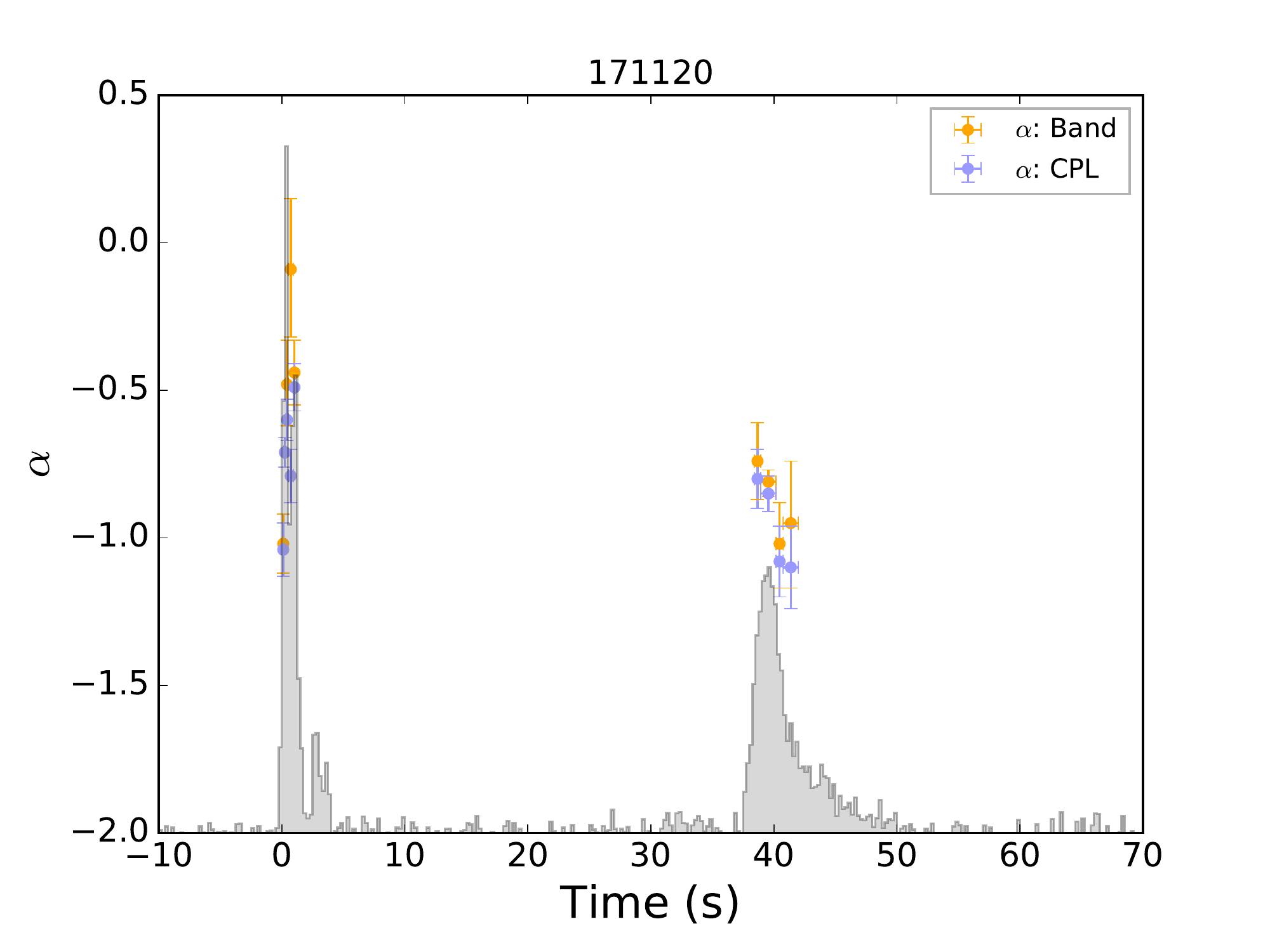}
\includegraphics[angle=0,scale=0.3]{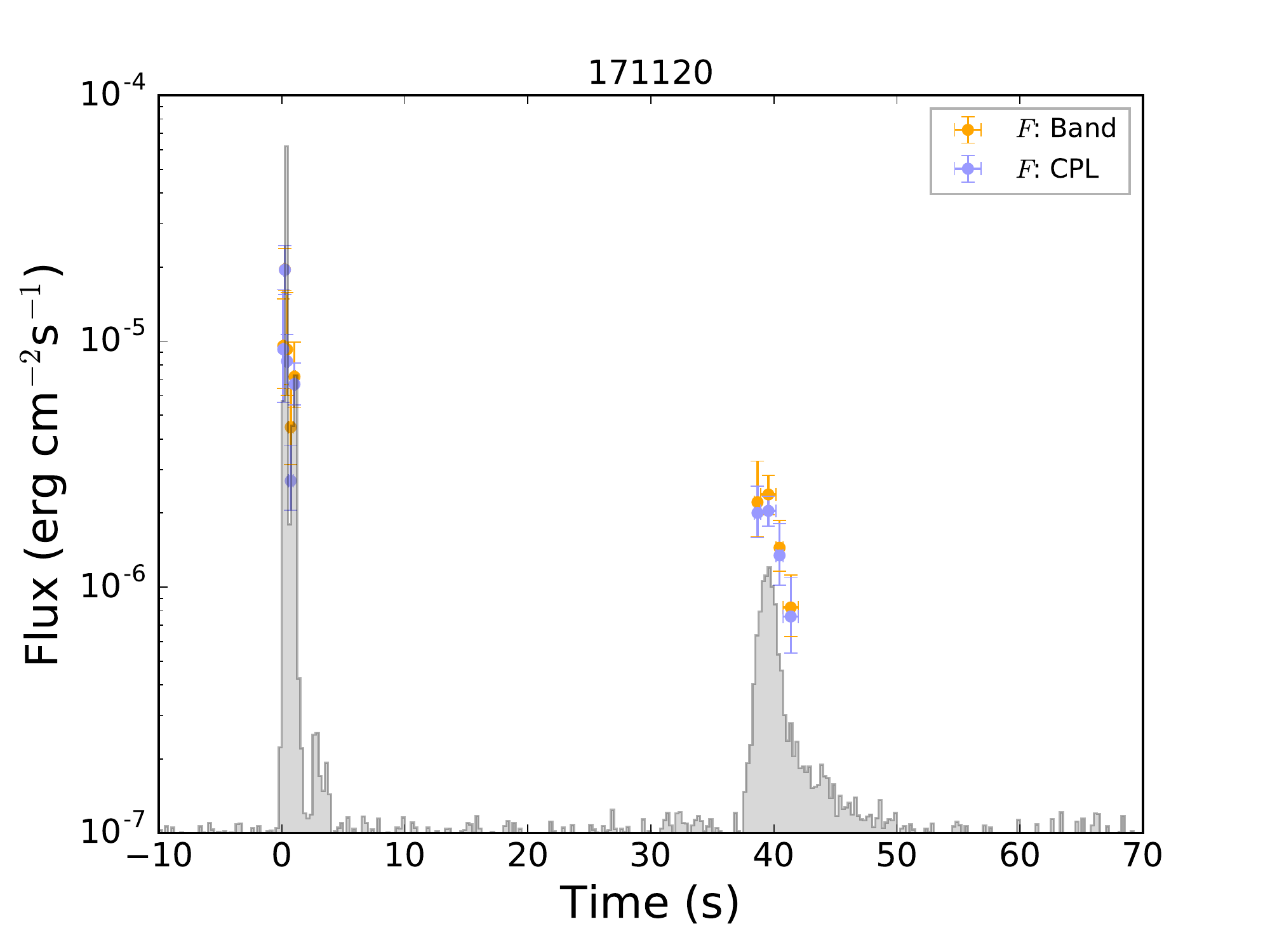}
\includegraphics[angle=0,scale=0.3]{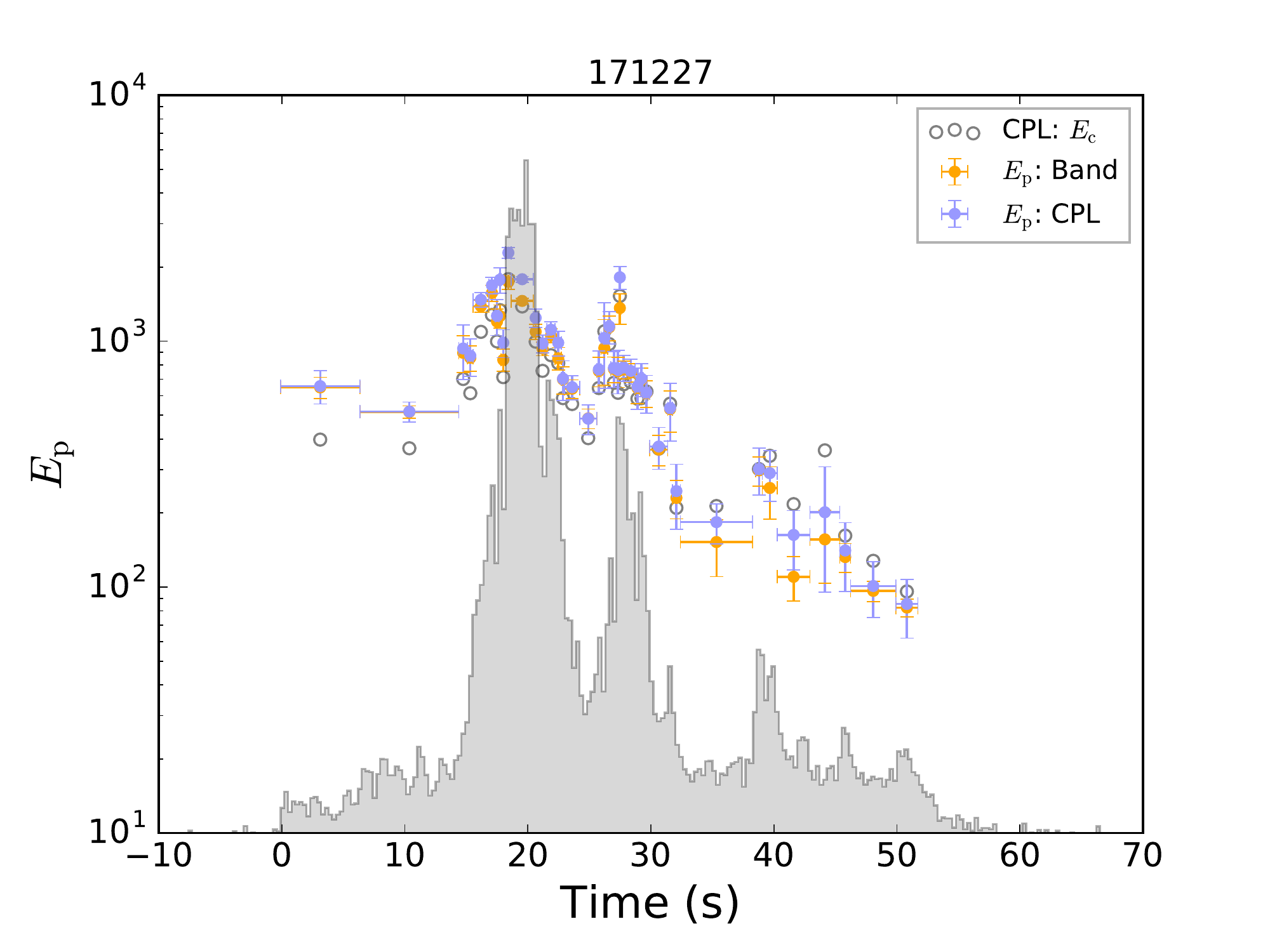}
\includegraphics[angle=0,scale=0.3]{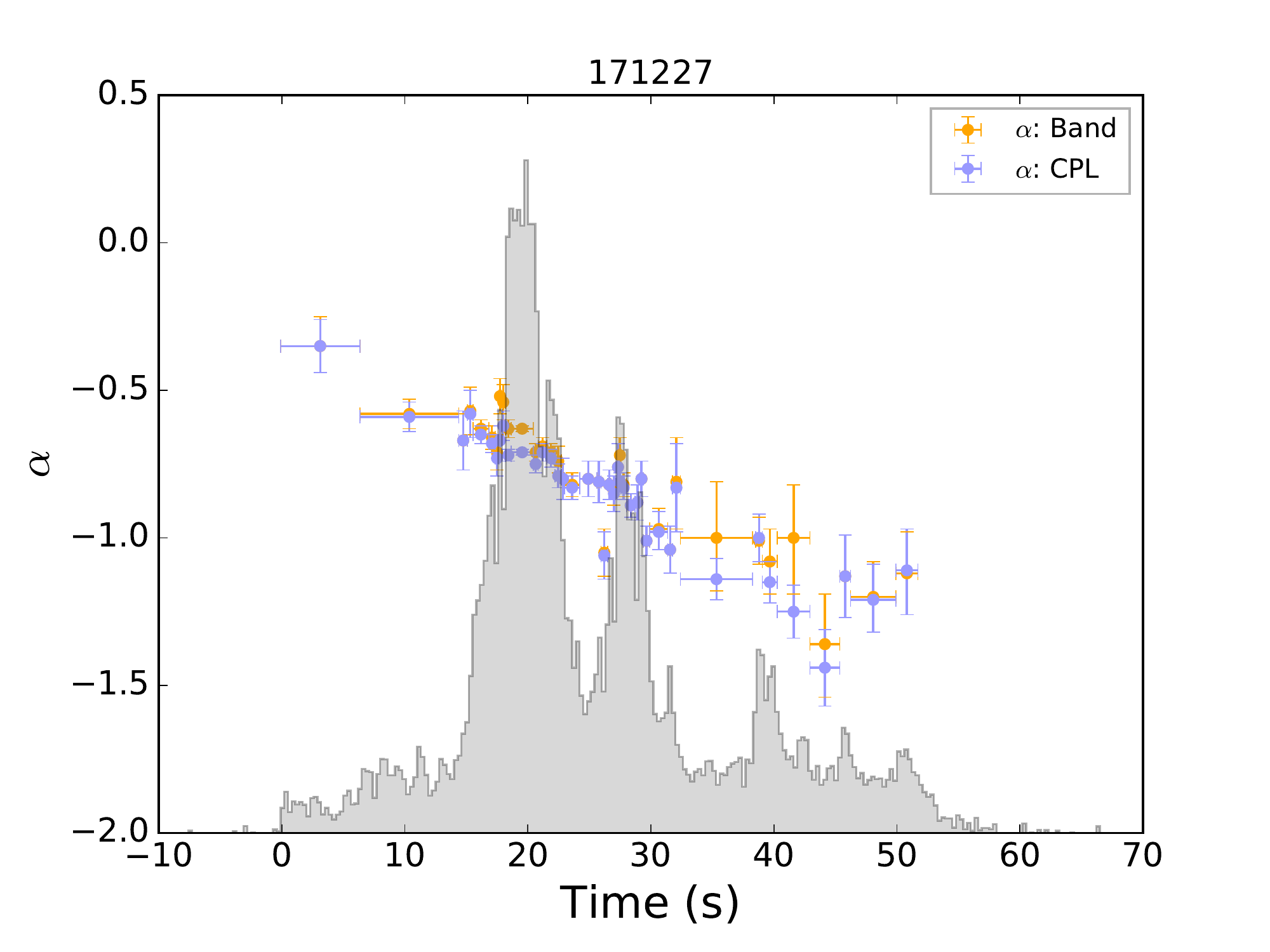}
\includegraphics[angle=0,scale=0.3]{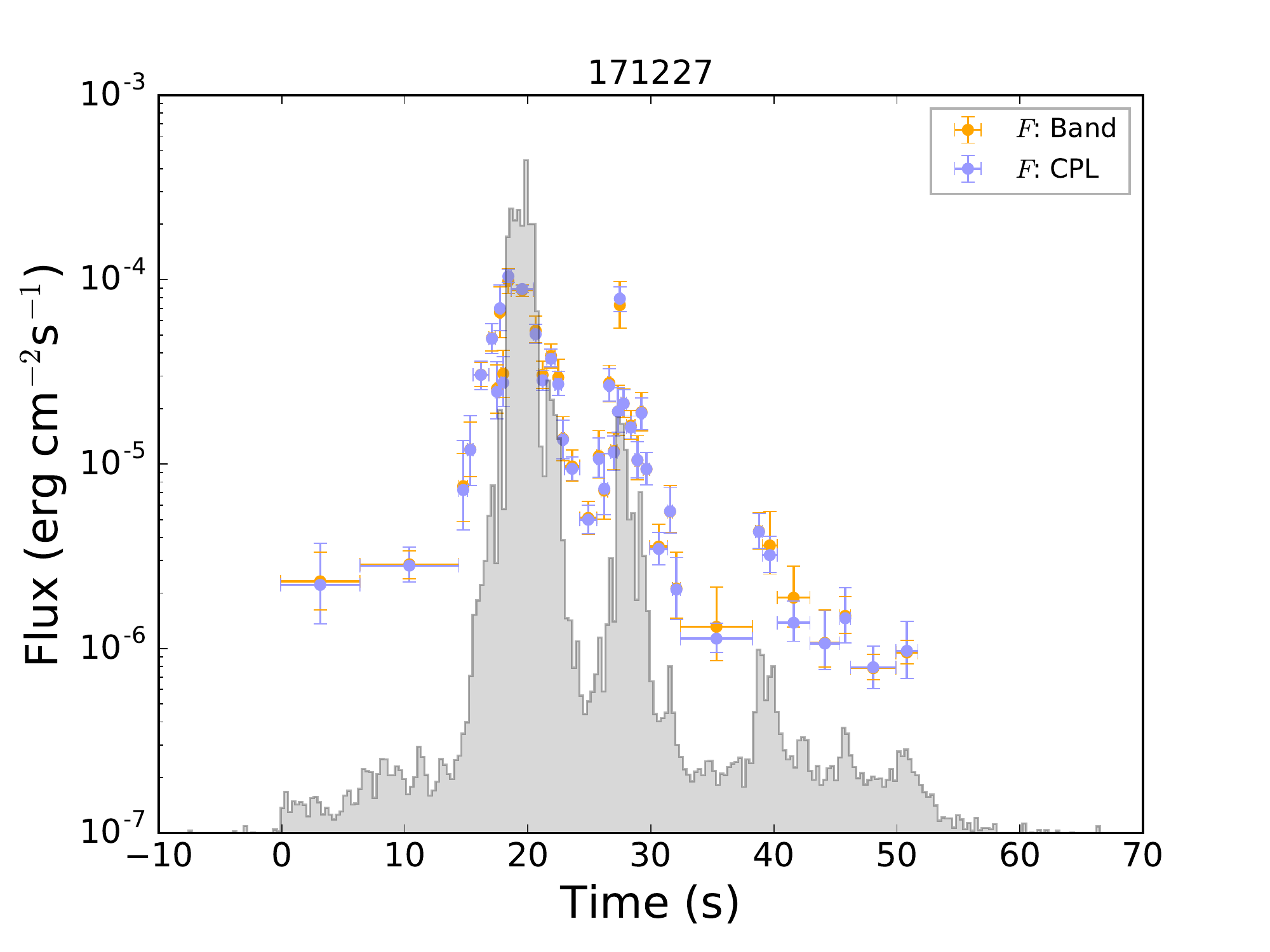}
\center{Fig. \ref{fig:evolution}--- Continued}
\end{figure*}
\begin{figure*}
\includegraphics[angle=0,scale=0.3]{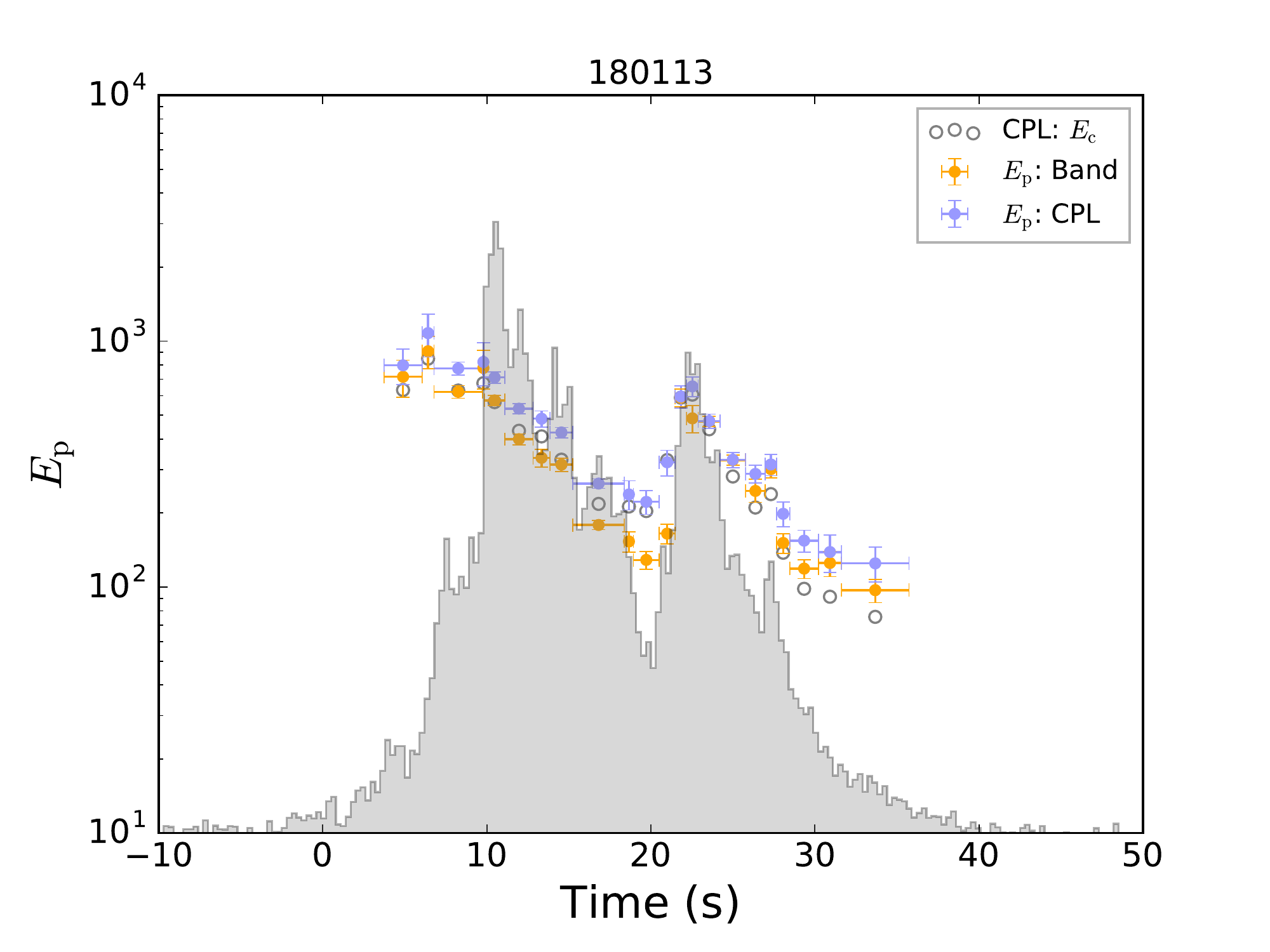}
\includegraphics[angle=0,scale=0.3]{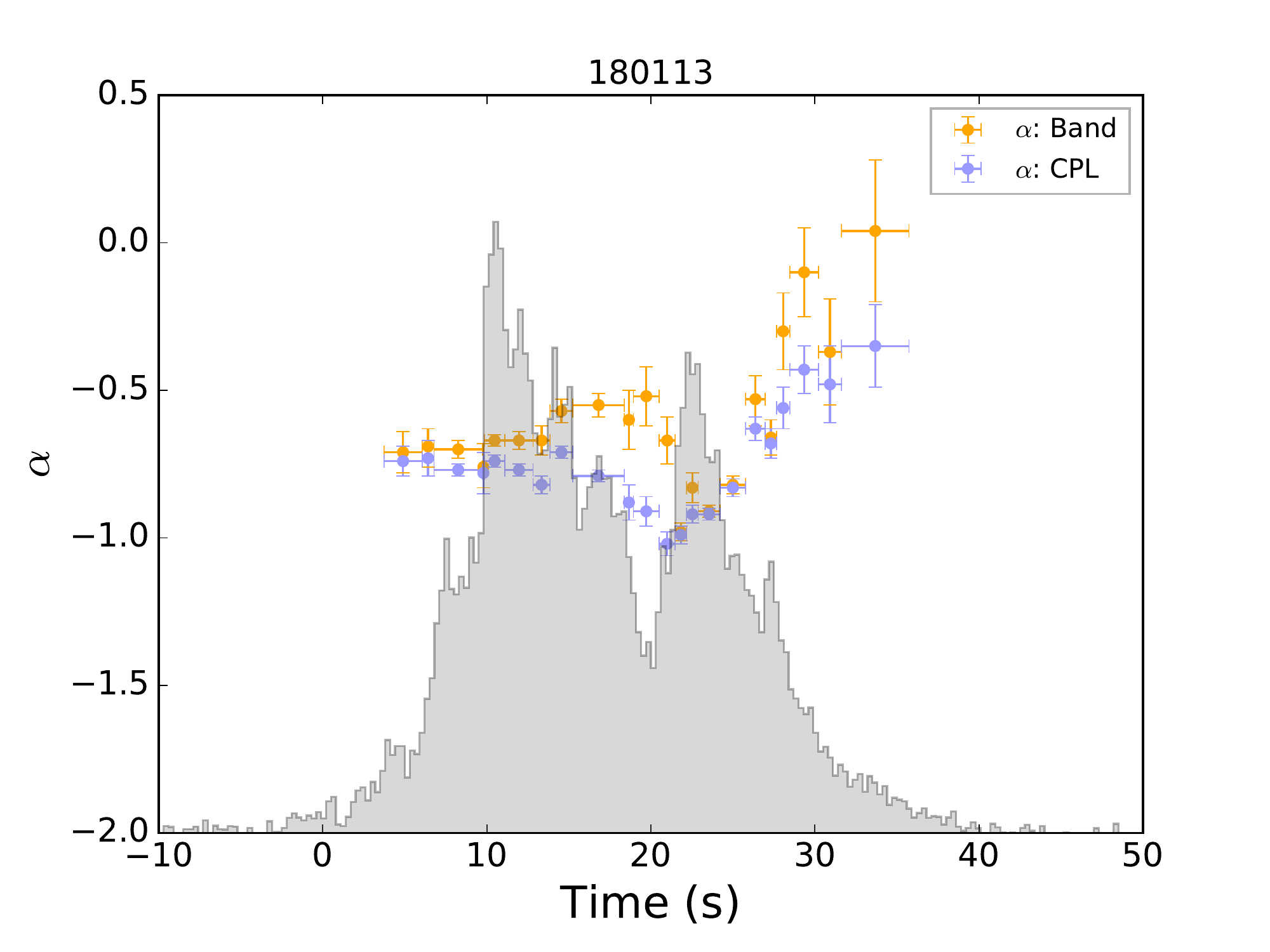}
\includegraphics[angle=0,scale=0.3]{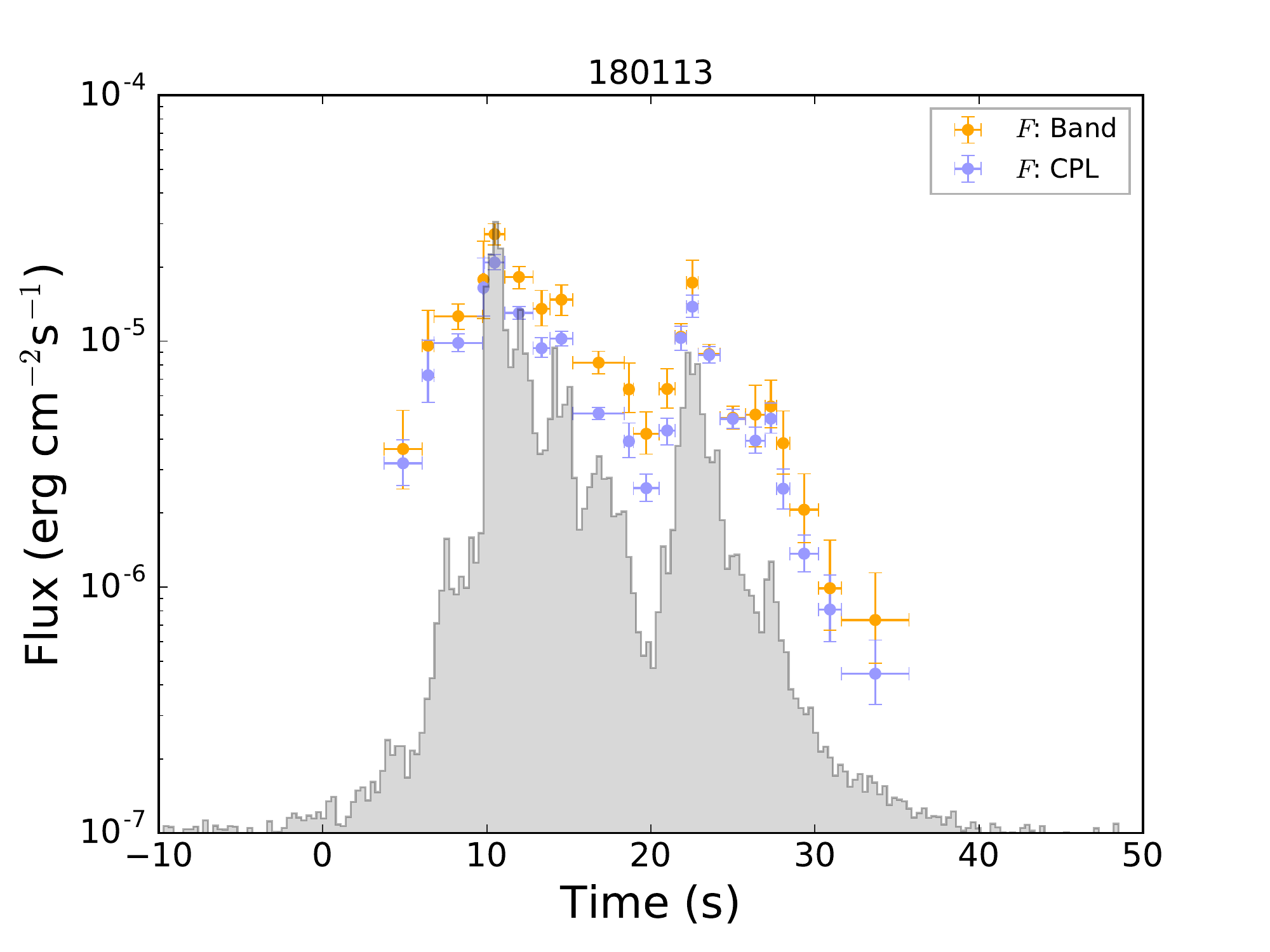}
\includegraphics[angle=0,scale=0.3]{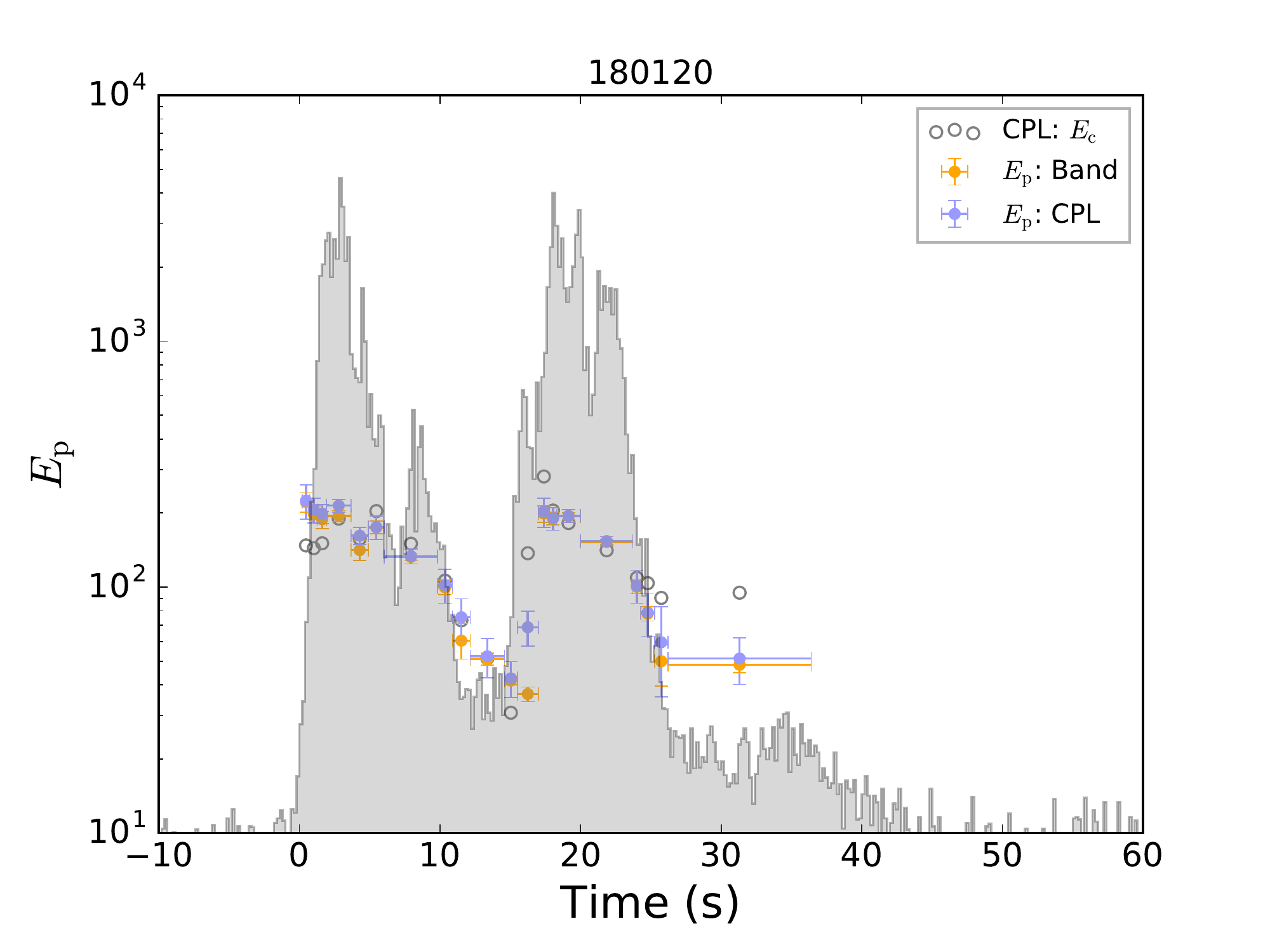}
\includegraphics[angle=0,scale=0.3]{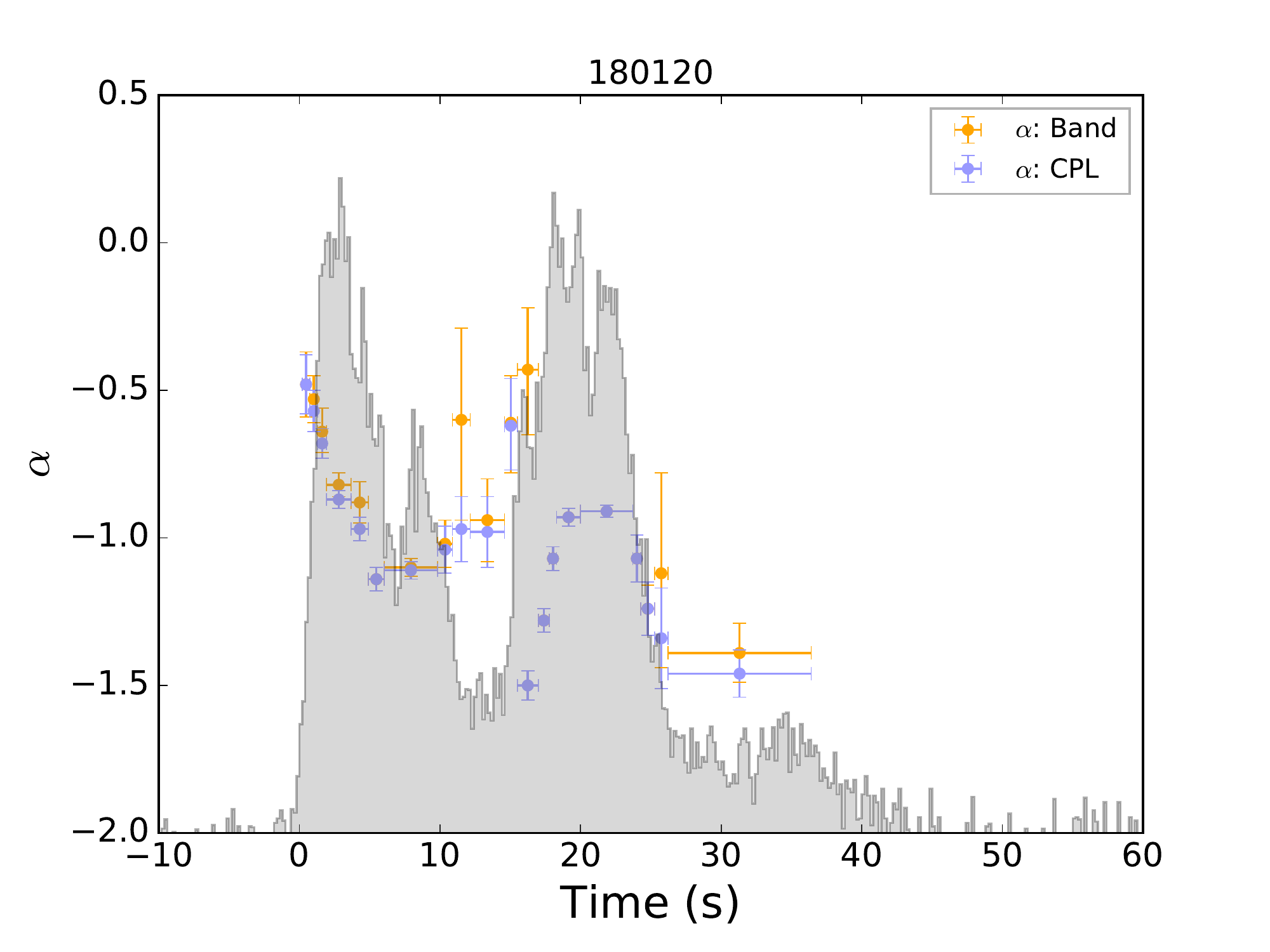}
\includegraphics[angle=0,scale=0.3]{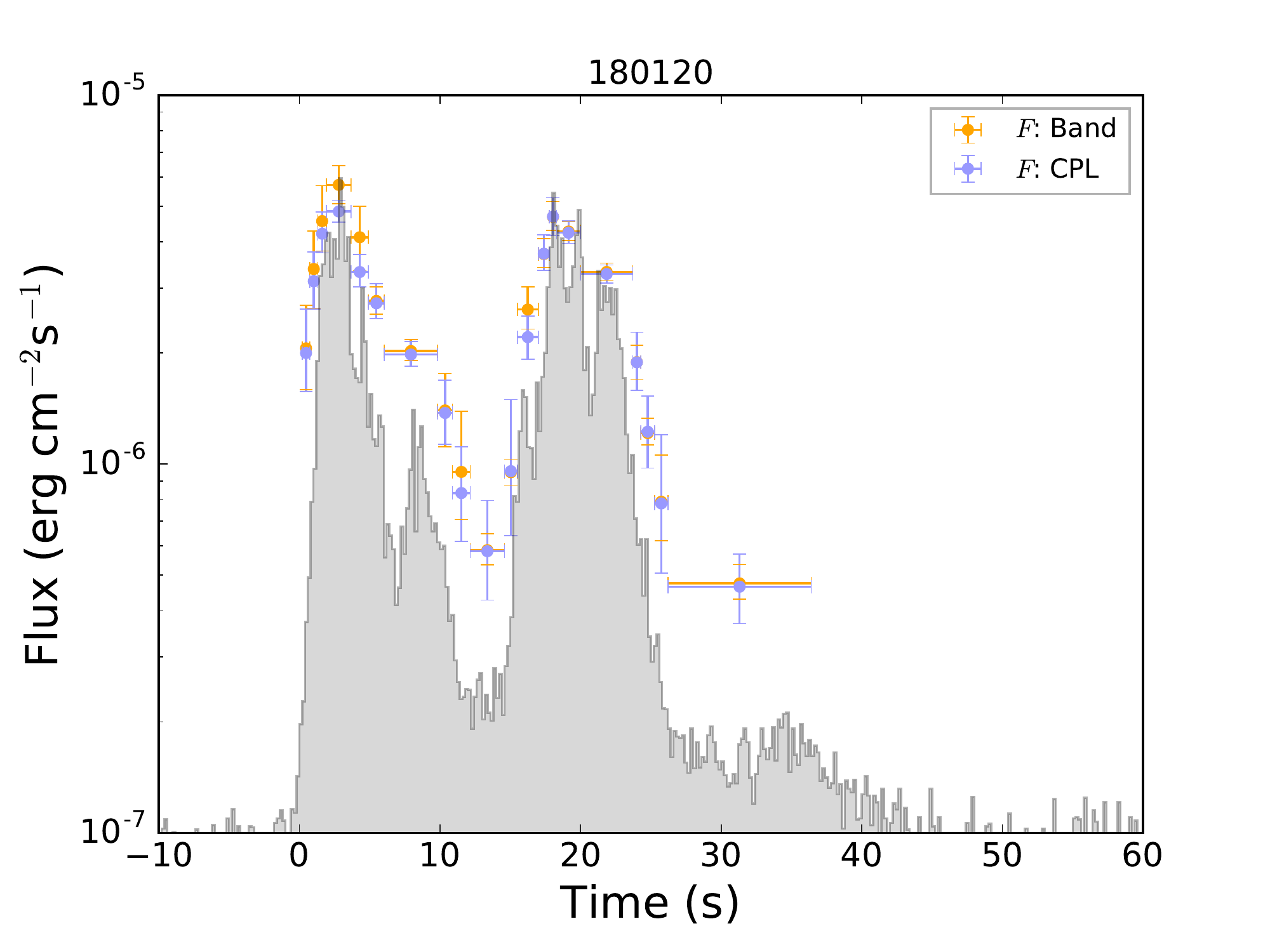}
\includegraphics[angle=0,scale=0.3]{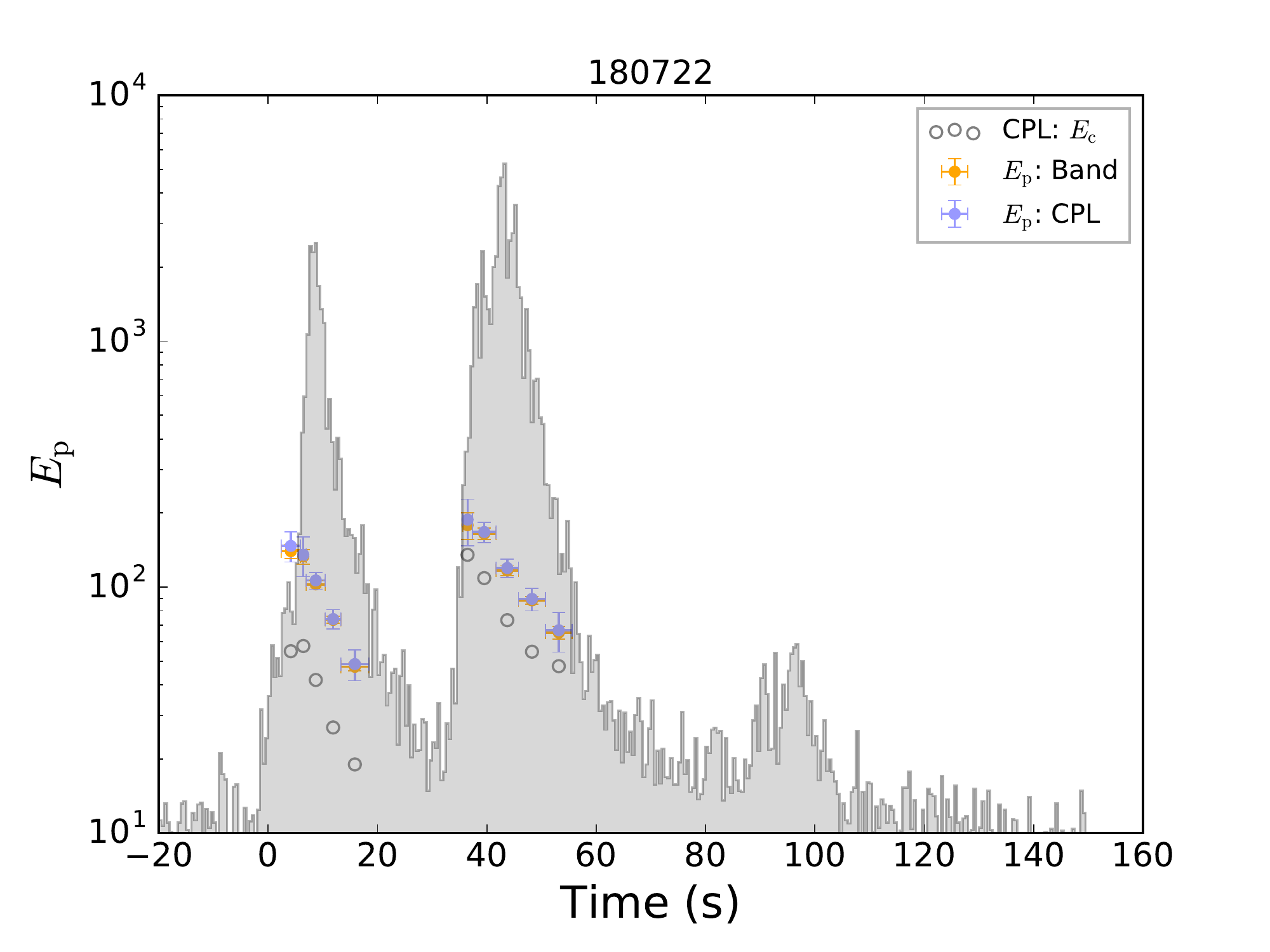}
\includegraphics[angle=0,scale=0.3]{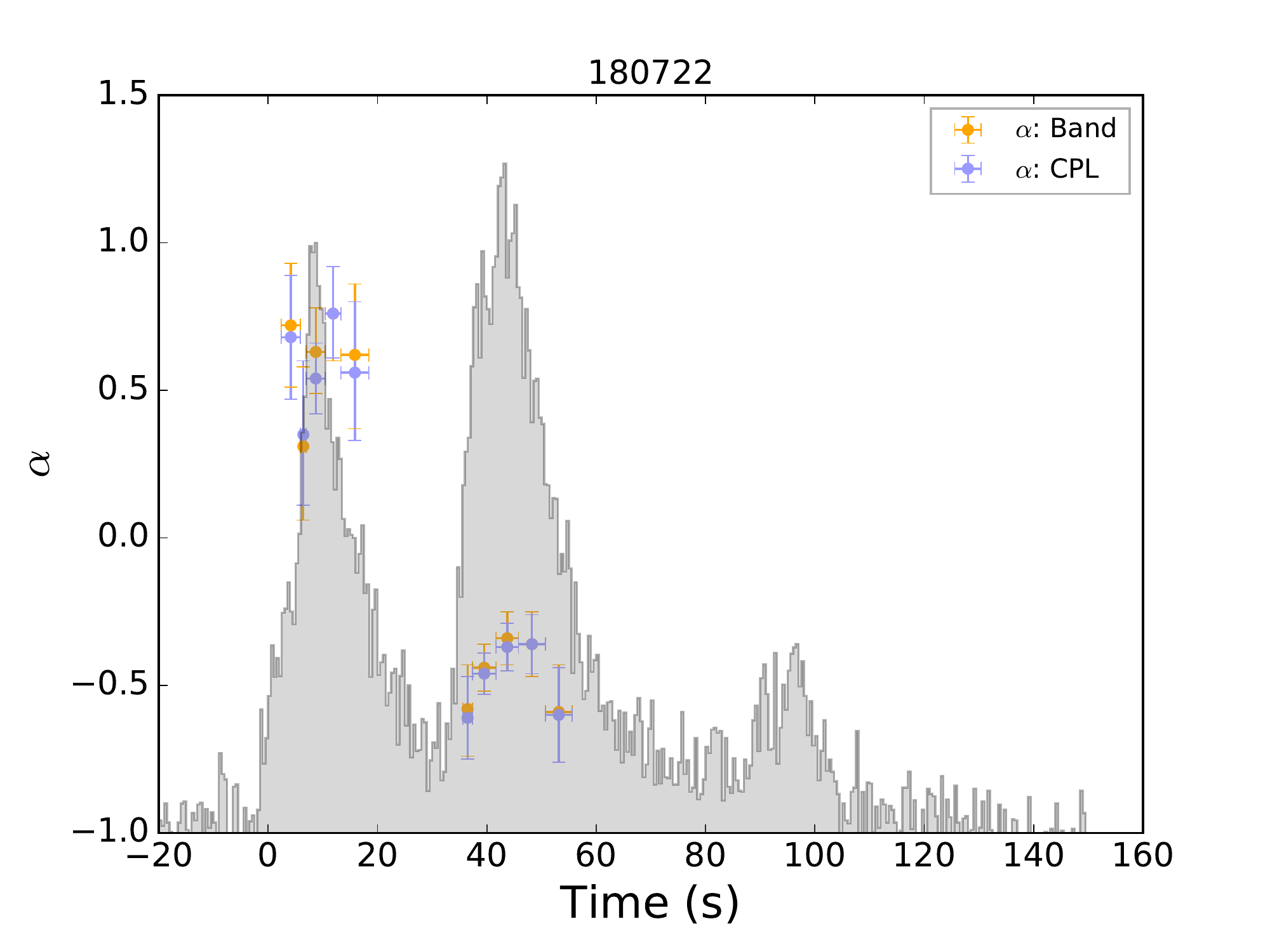}
\includegraphics[angle=0,scale=0.3]{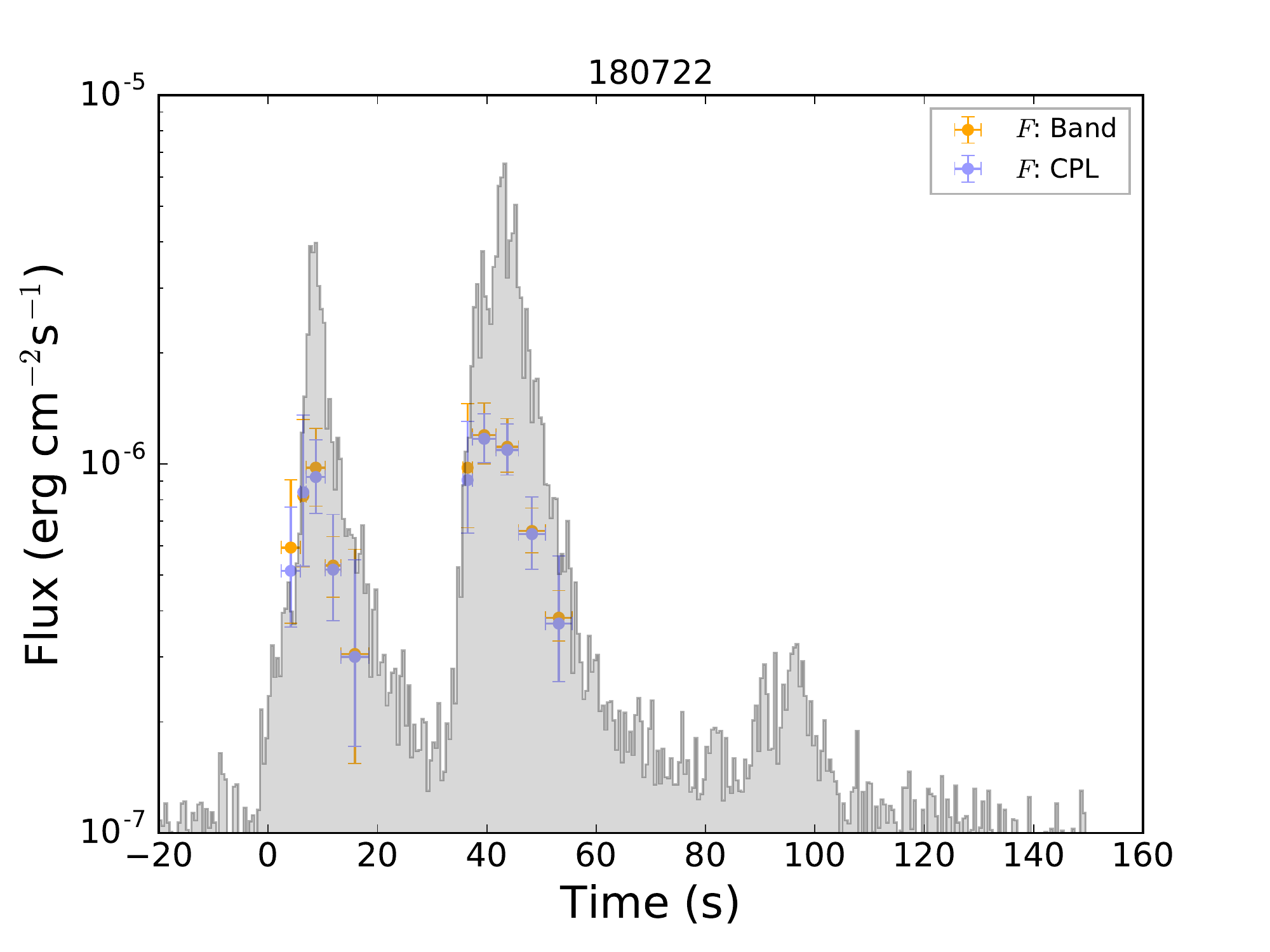}
\includegraphics[angle=0,scale=0.3]{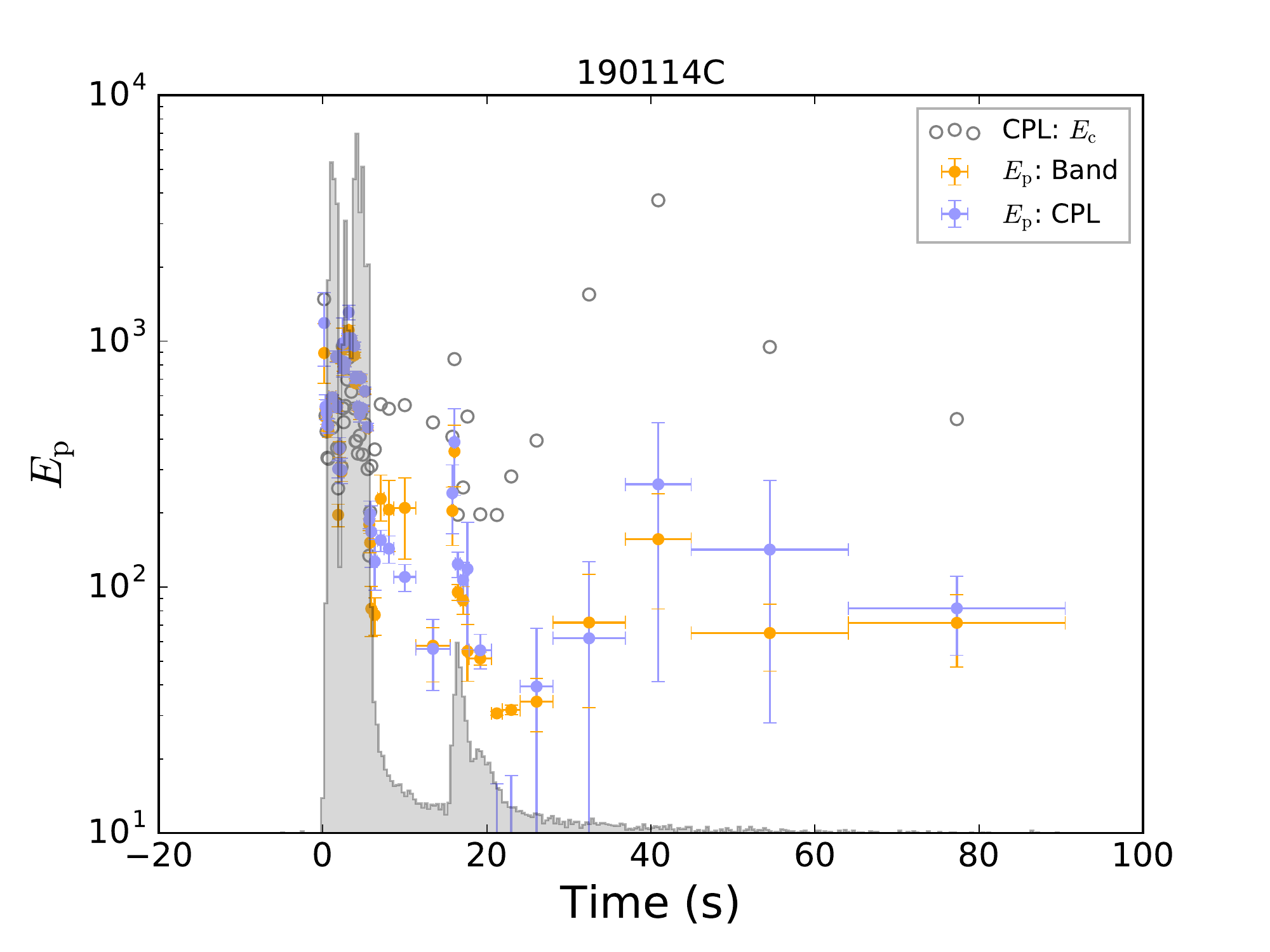}
\includegraphics[angle=0,scale=0.3]{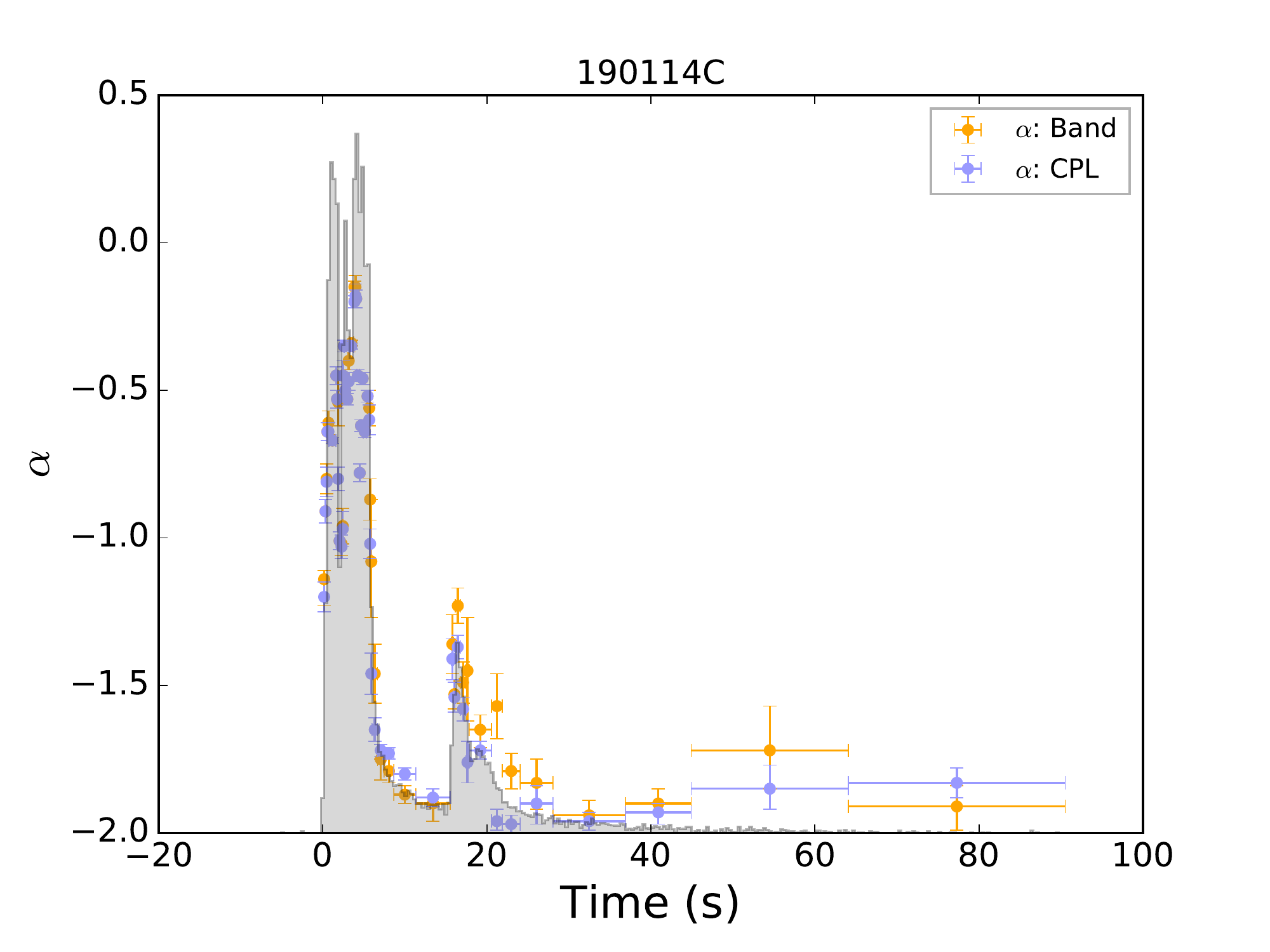}
\includegraphics[angle=0,scale=0.3]{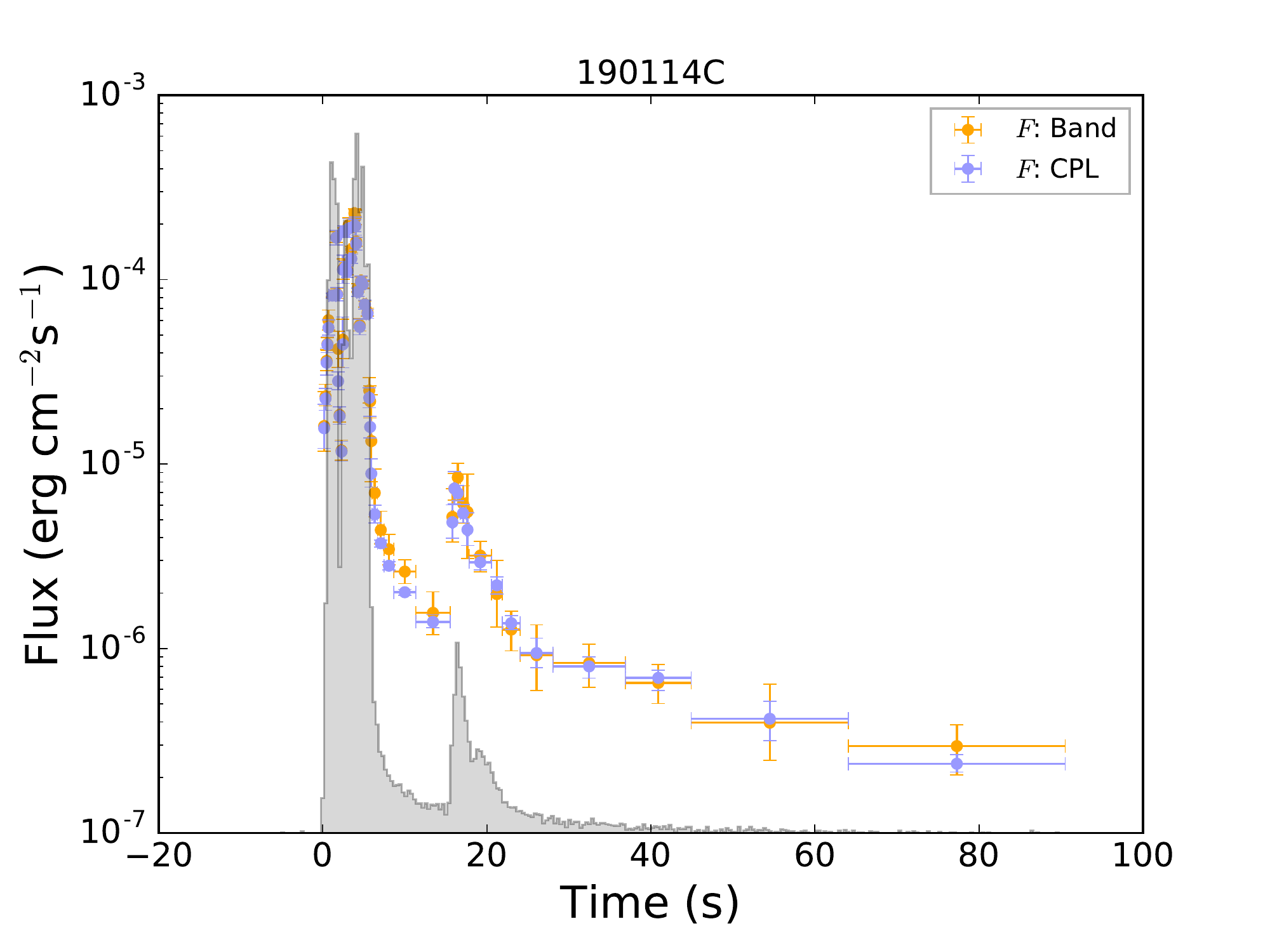}
\center{Fig. \ref{fig:evolution}--- Continued}
\end{figure*}

\clearpage
\begin{figure*}
\includegraphics[angle=0,scale=0.3]{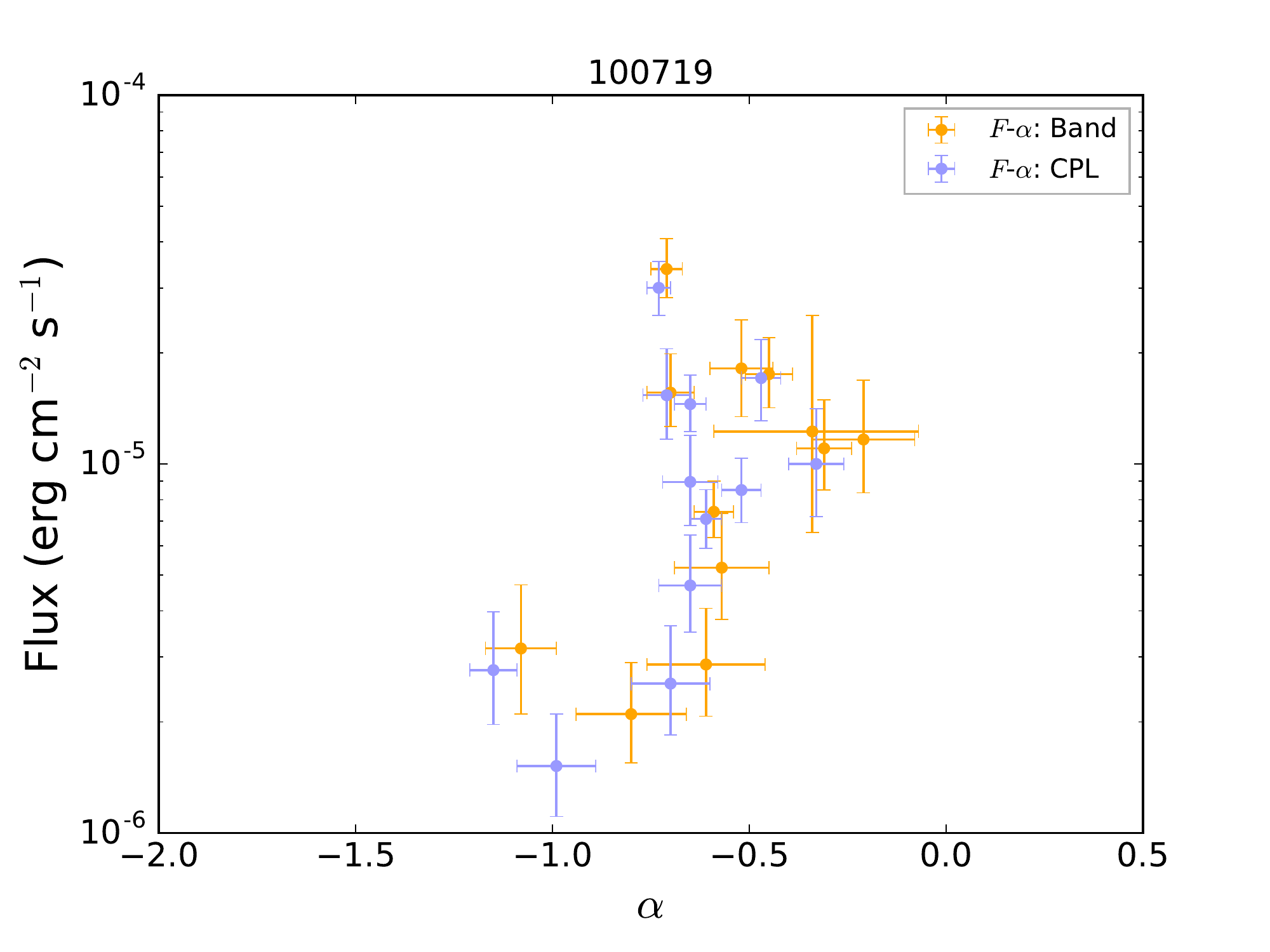}
\includegraphics[angle=0,scale=0.3]{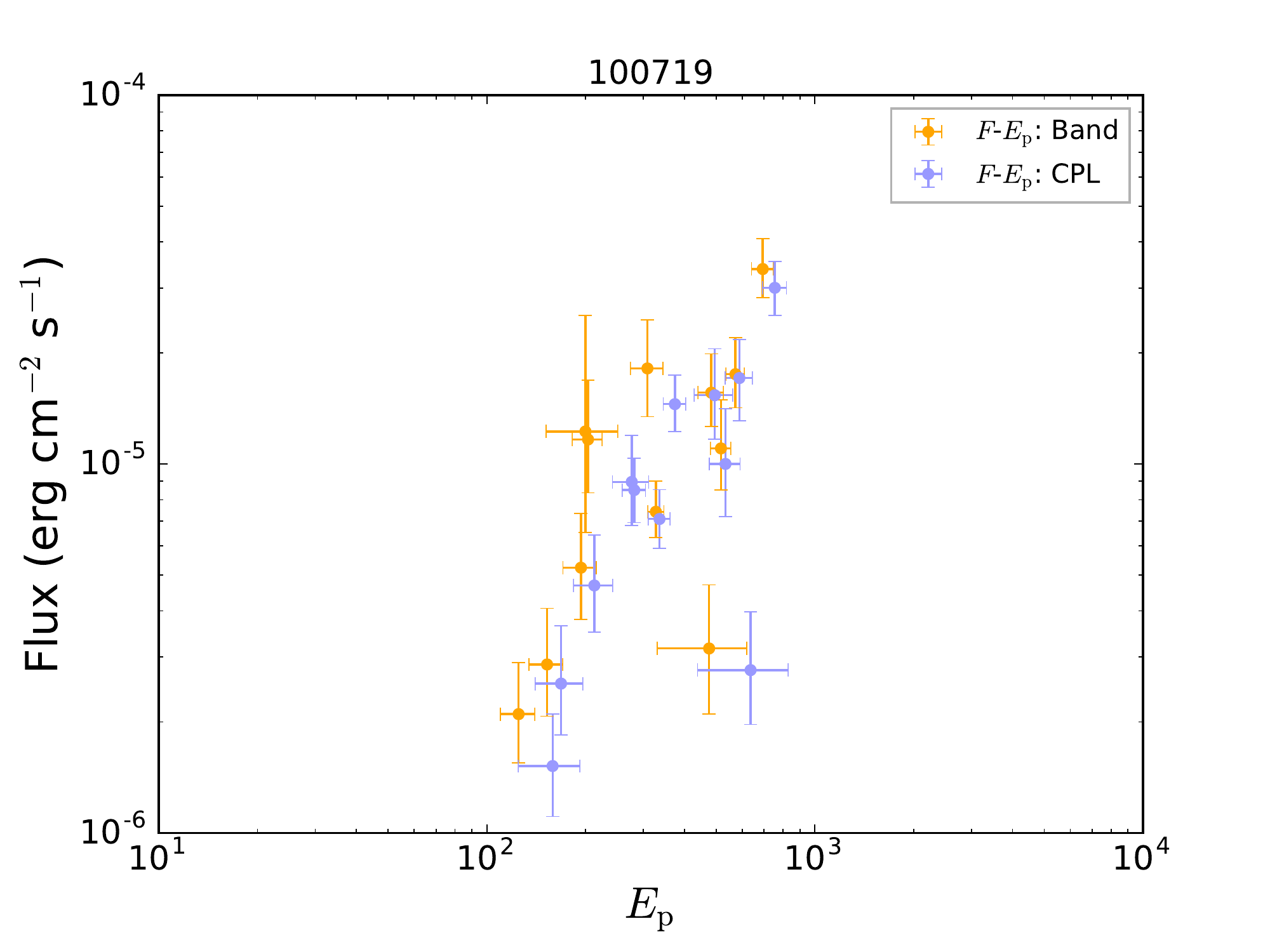}
\includegraphics[angle=0,scale=0.3]{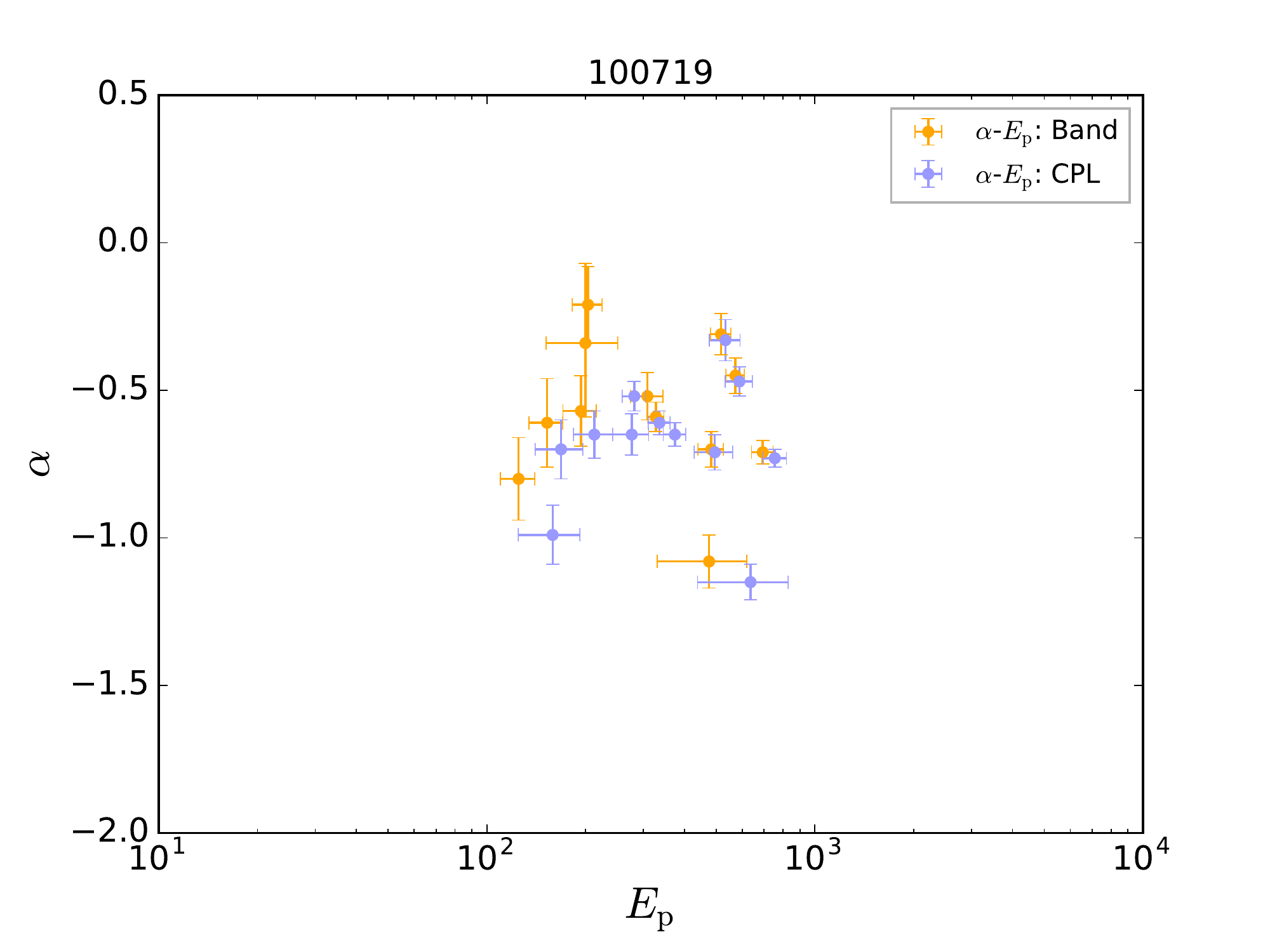}
\includegraphics[angle=0,scale=0.3]{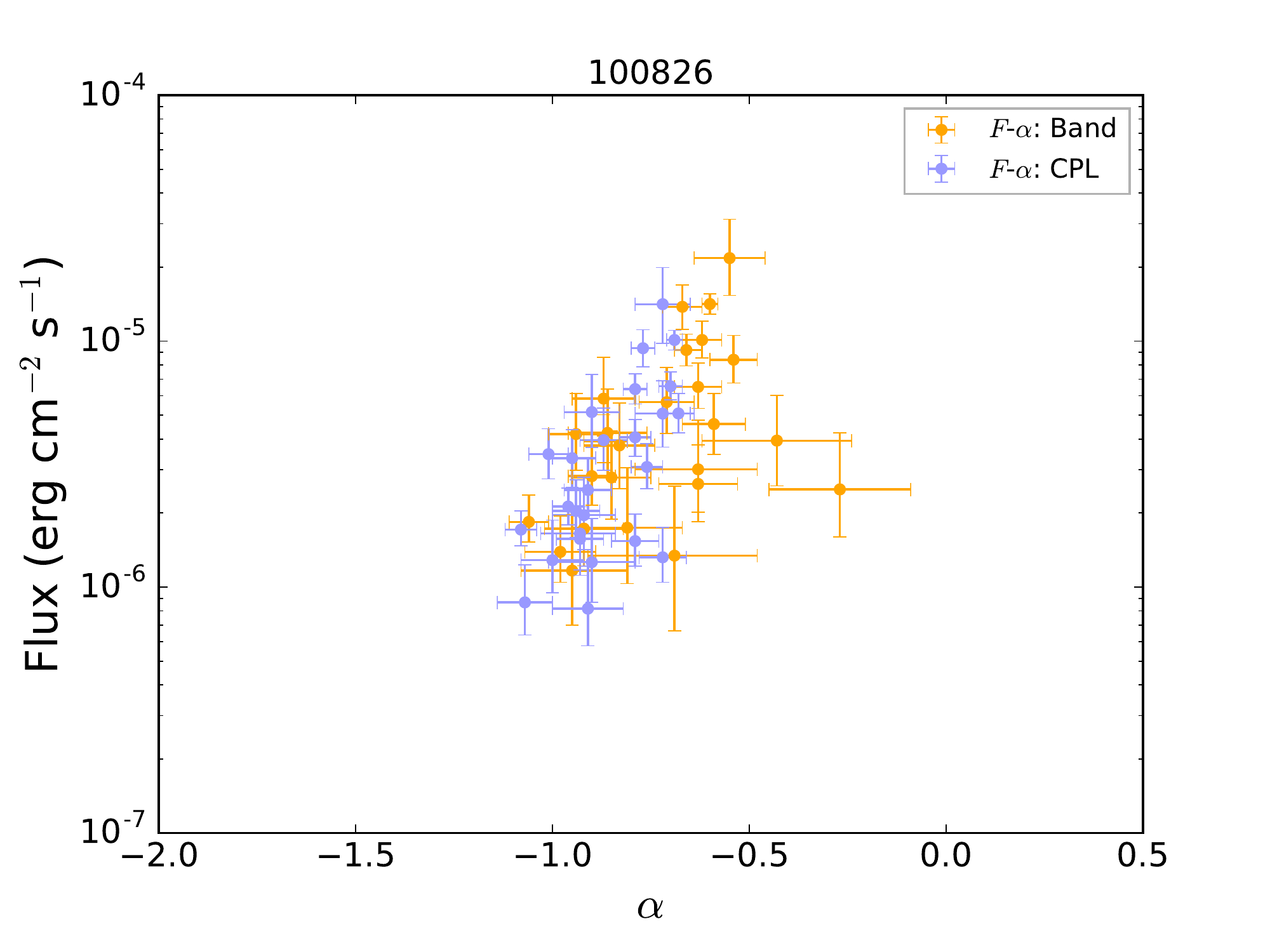}
\includegraphics[angle=0,scale=0.3]{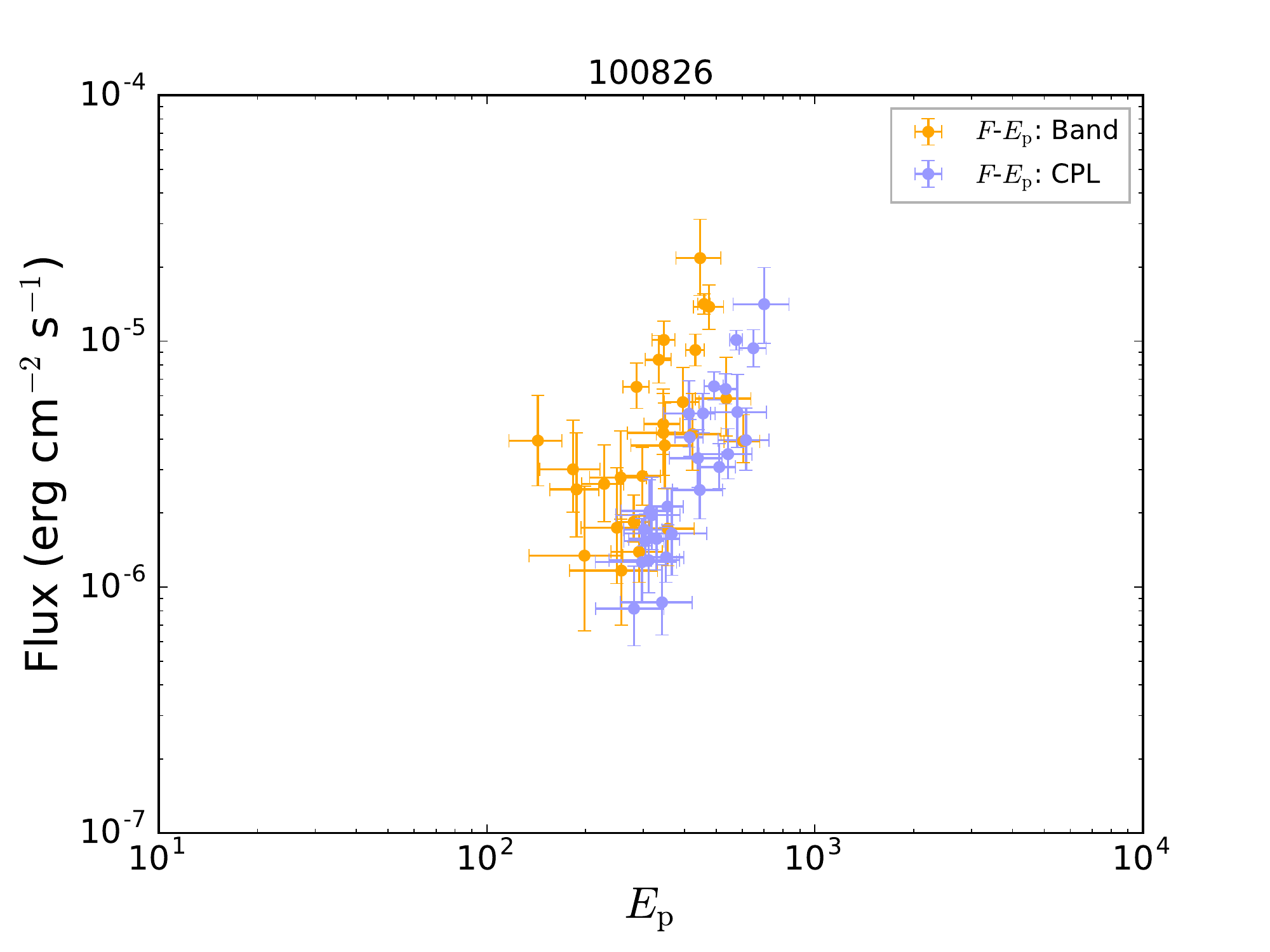}
\includegraphics[angle=0,scale=0.3]{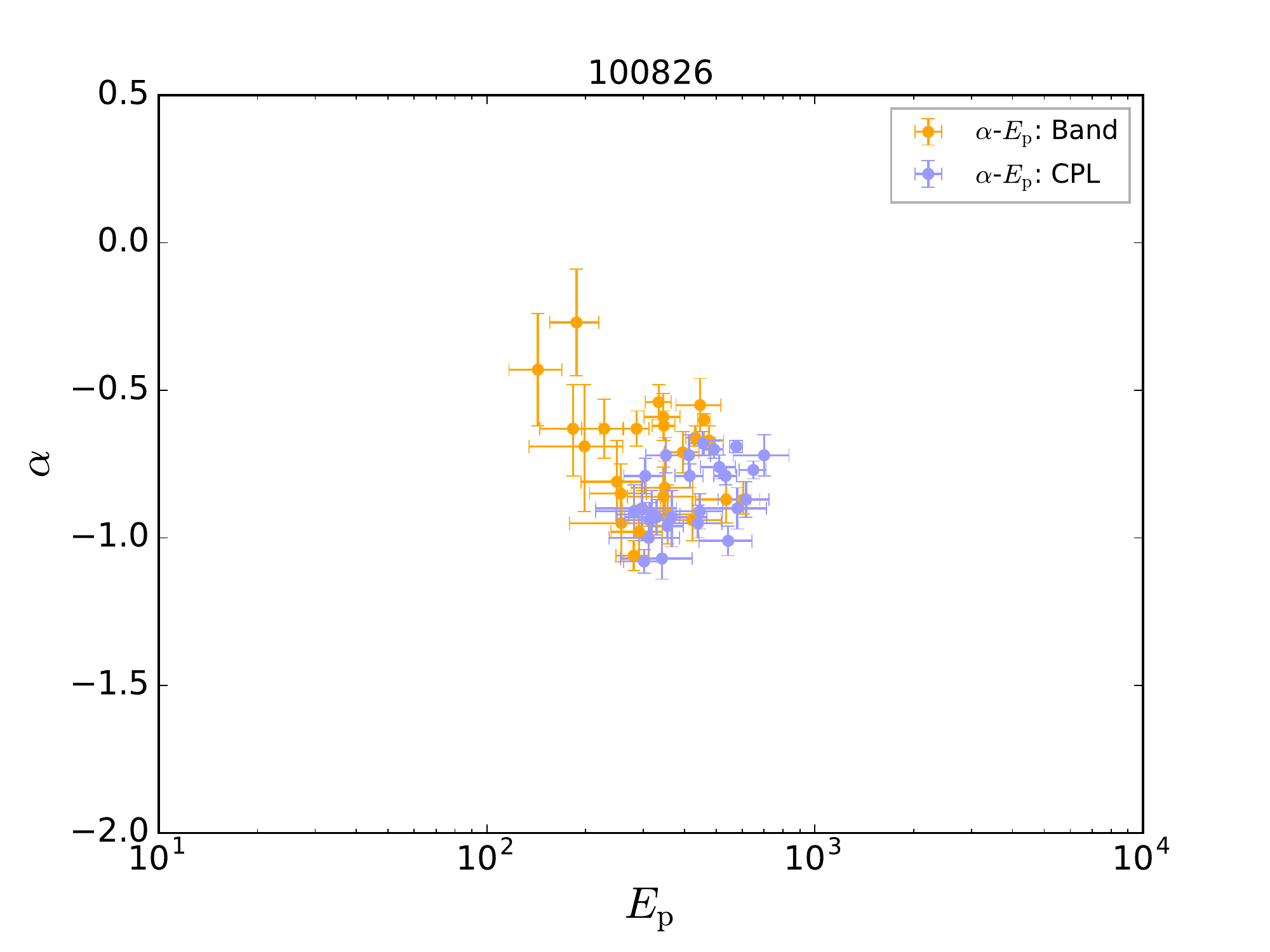}
\includegraphics[angle=0,scale=0.3]{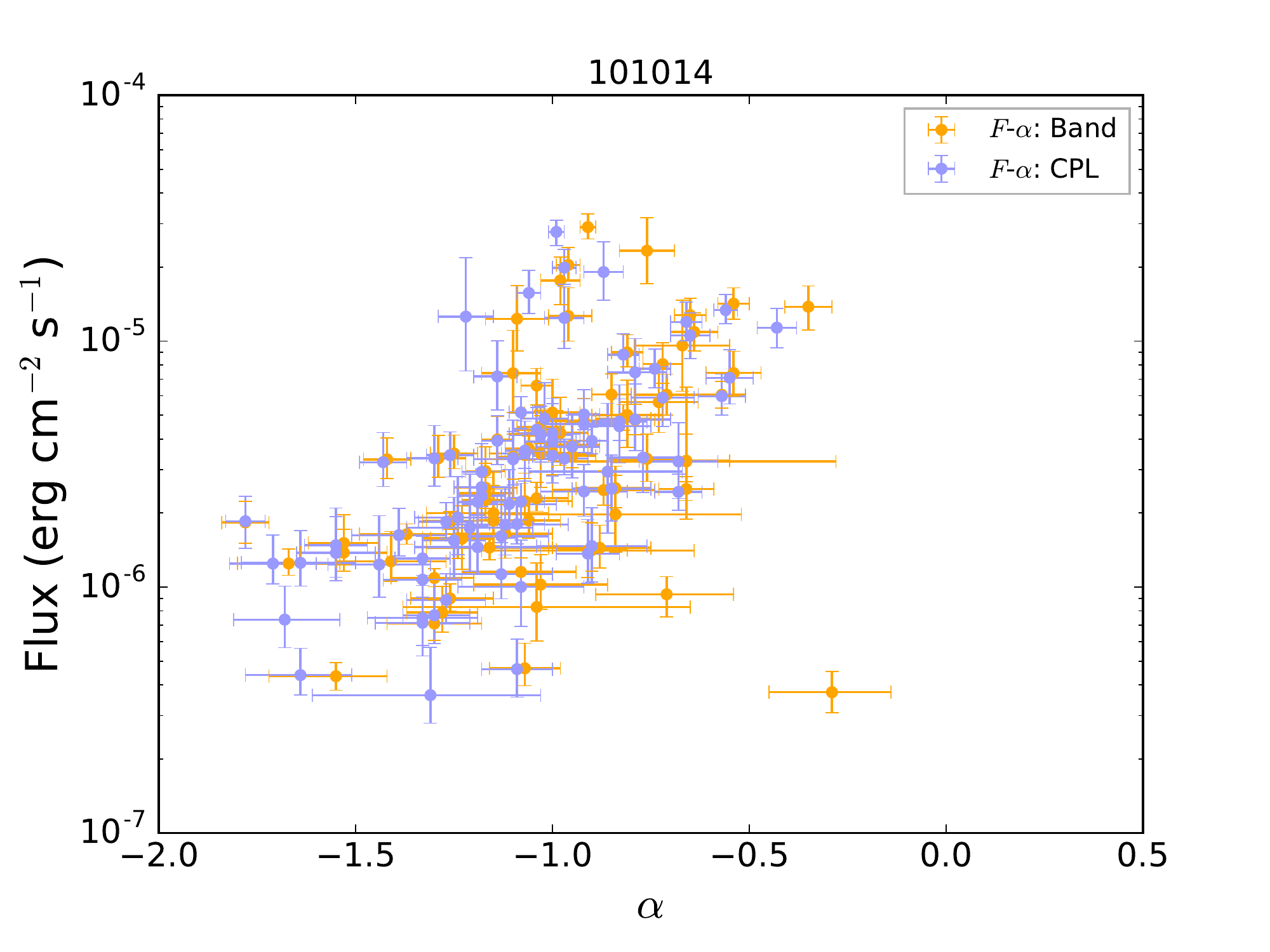}
\includegraphics[angle=0,scale=0.3]{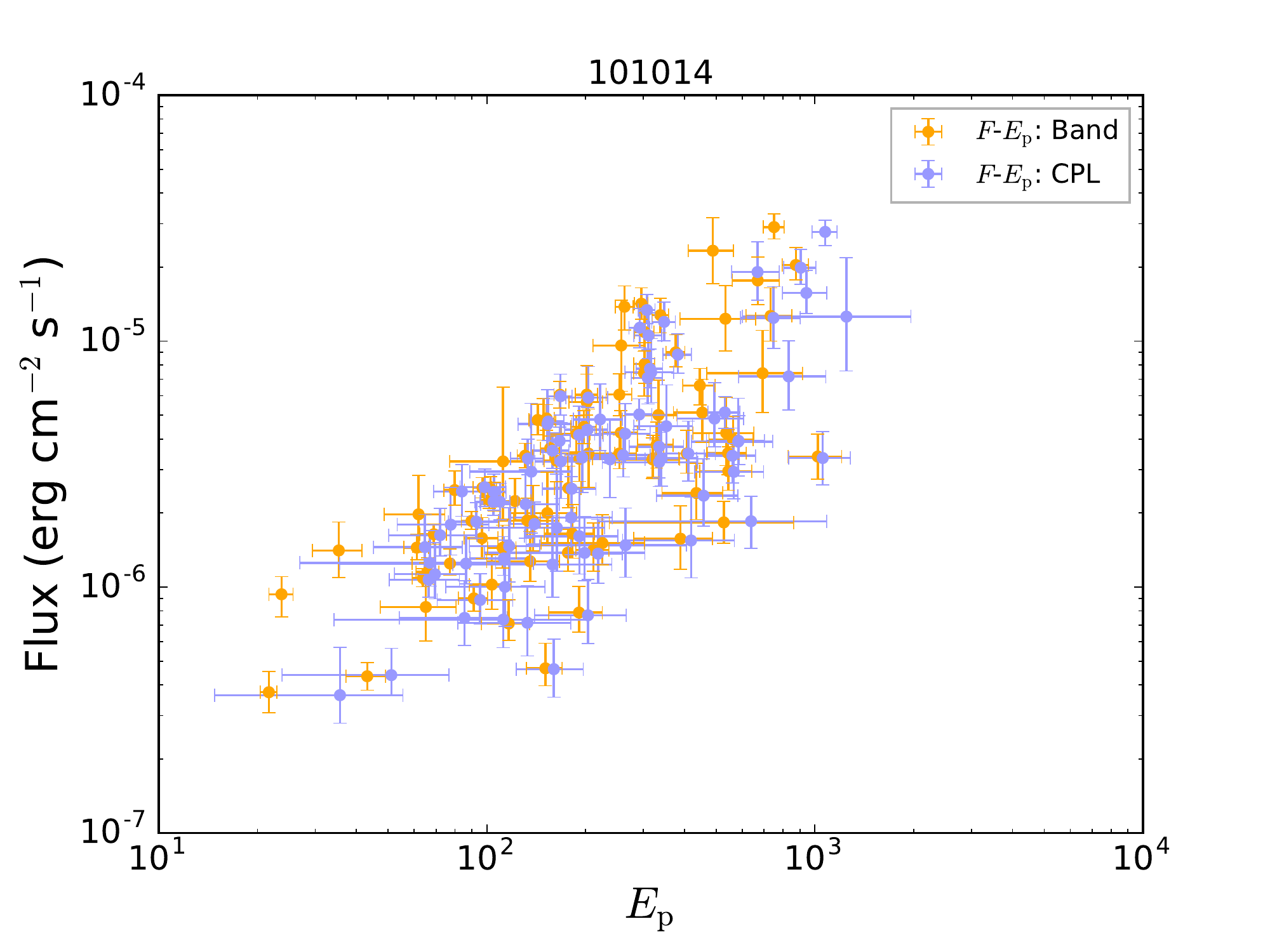}
\includegraphics[angle=0,scale=0.3]{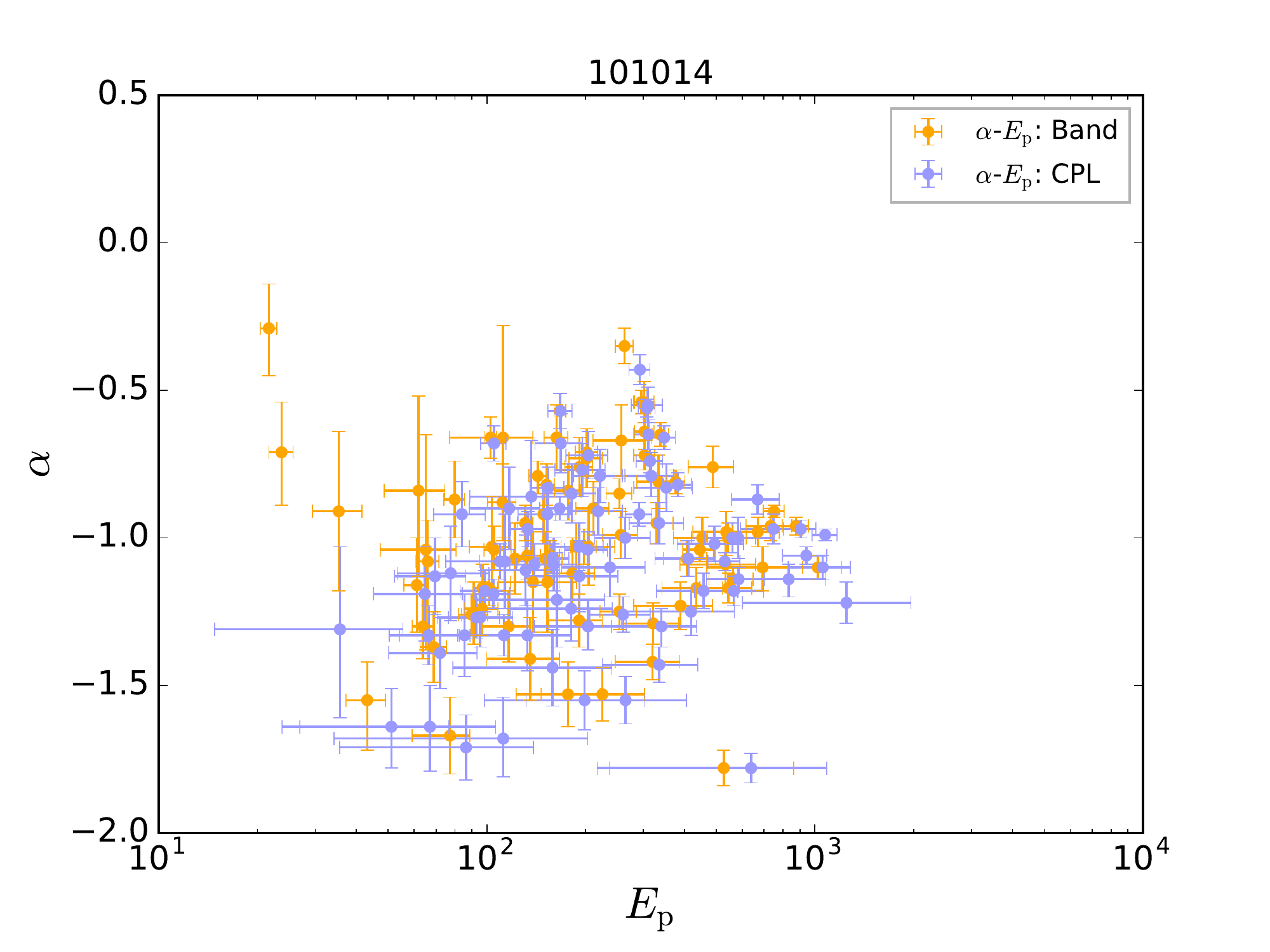}
\includegraphics[angle=0,scale=0.3]{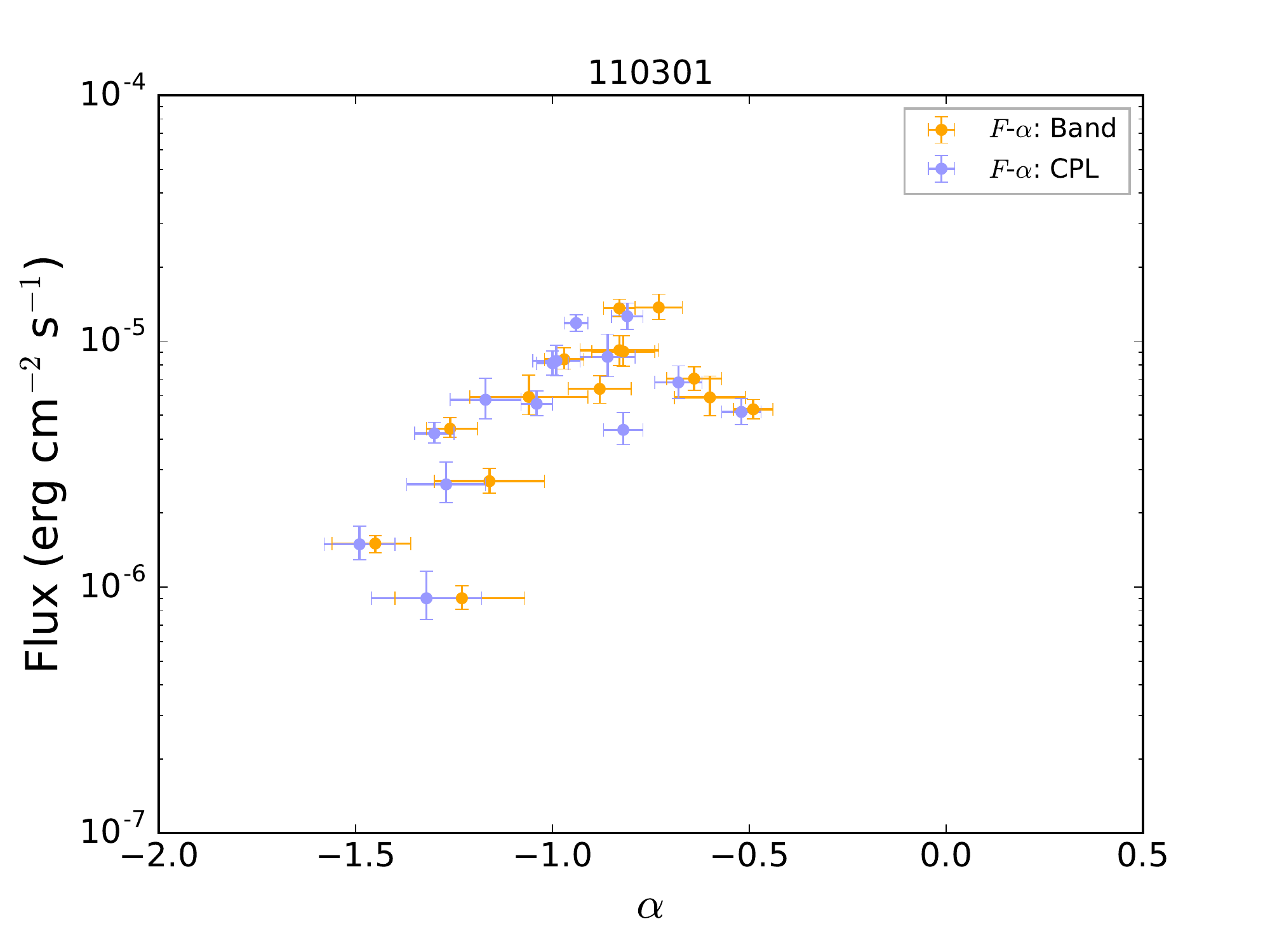}
\includegraphics[angle=0,scale=0.3]{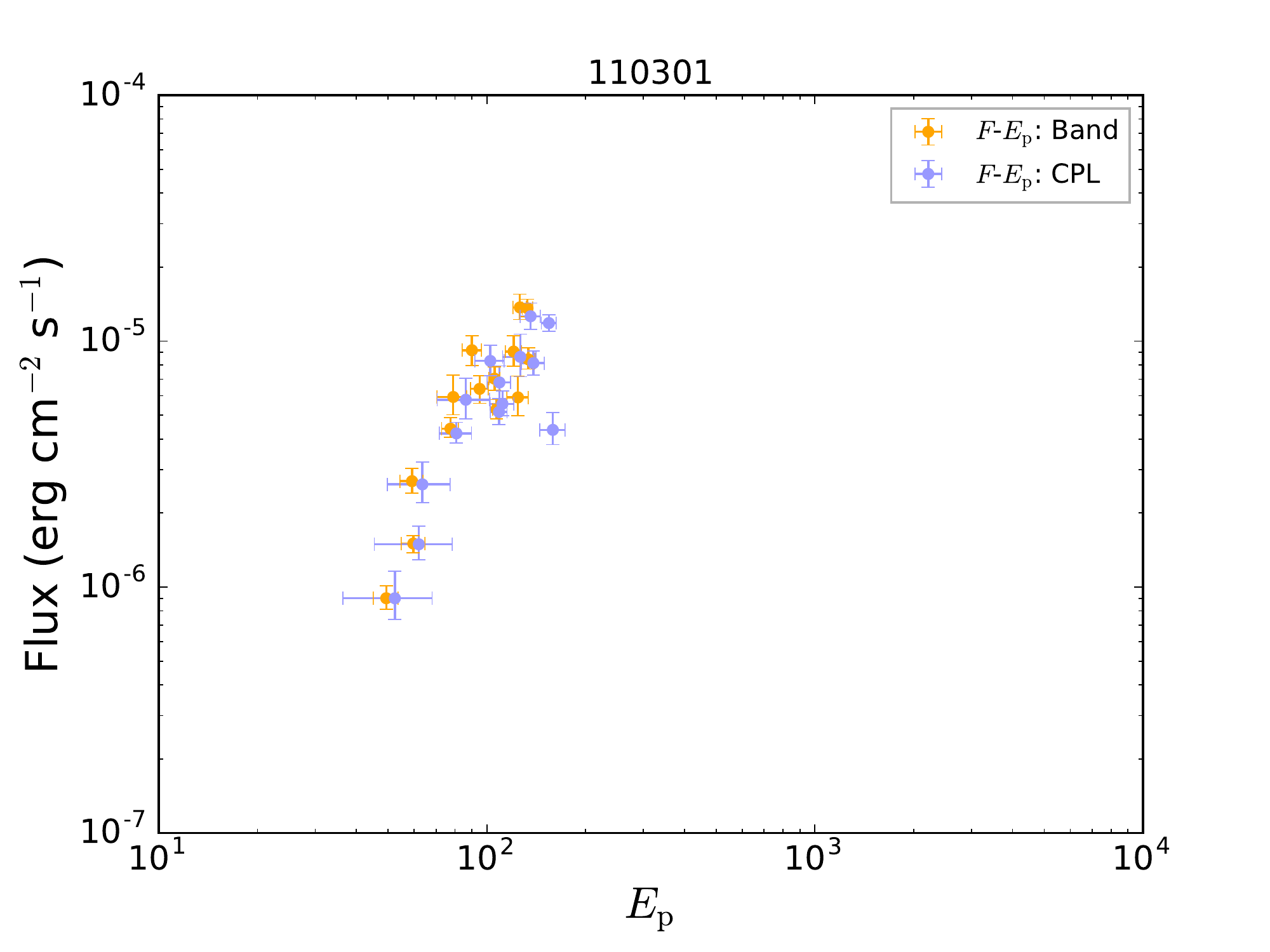}
\includegraphics[angle=0,scale=0.3]{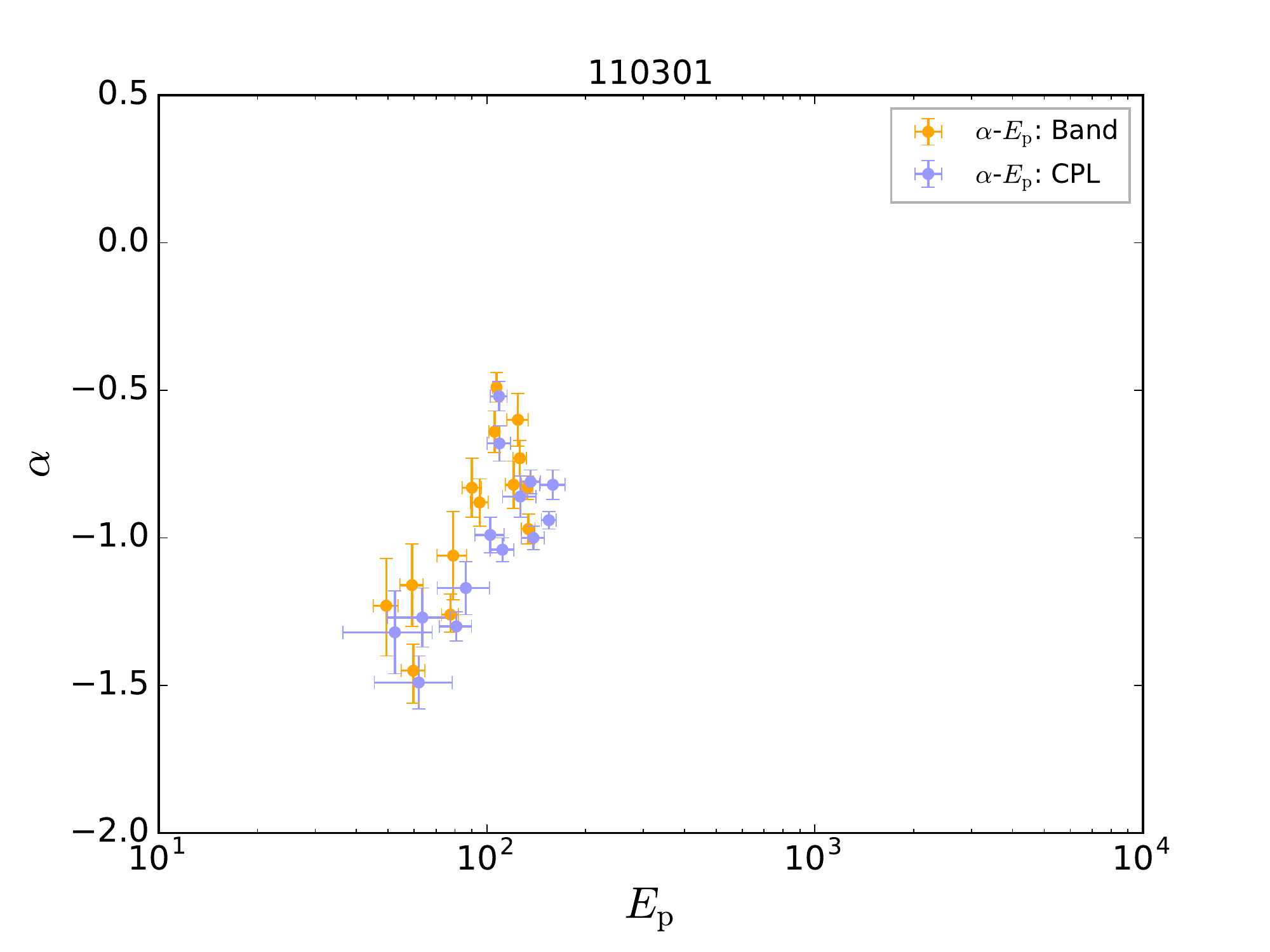}
\includegraphics[angle=0,scale=0.3]{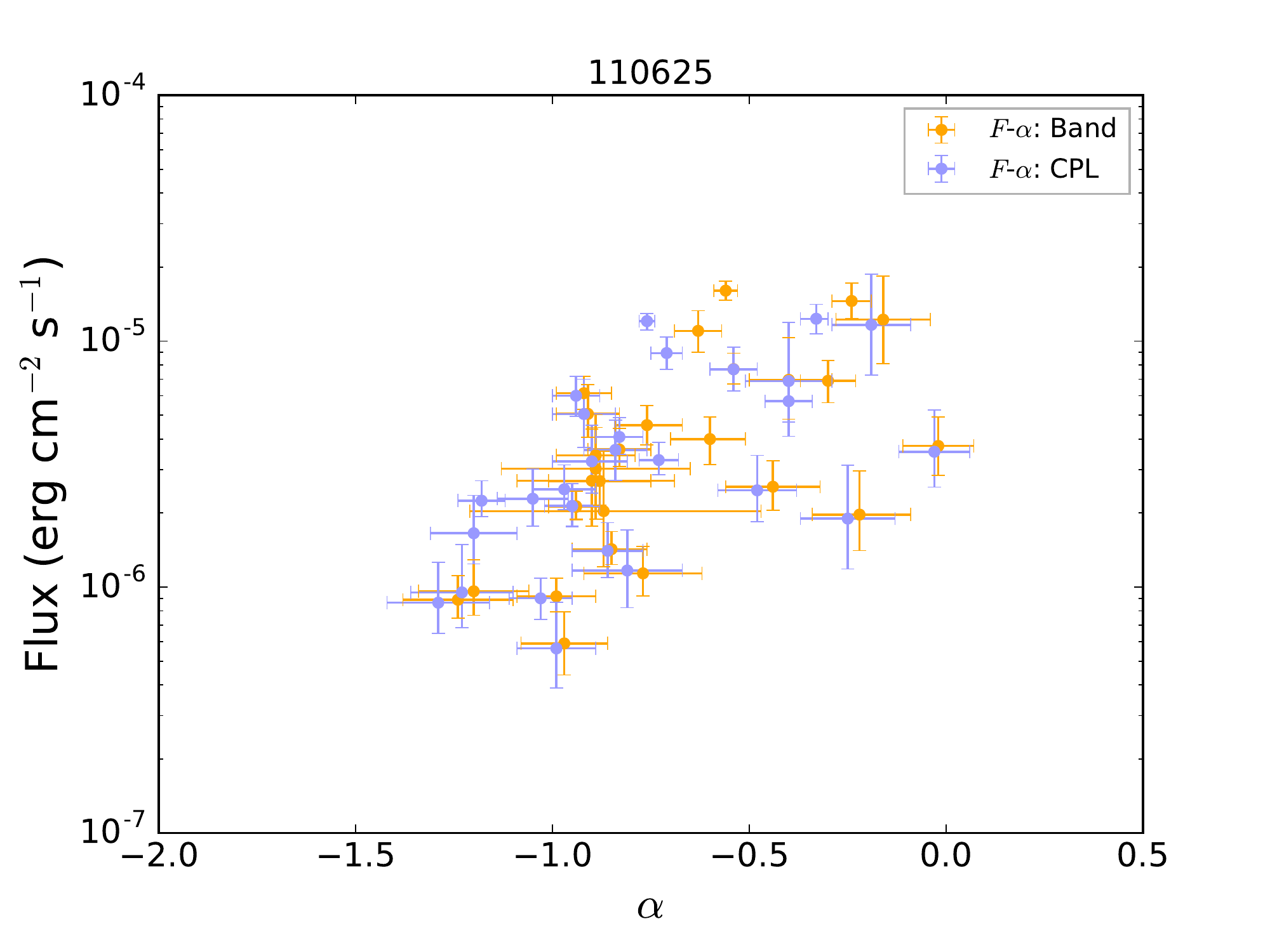}
\includegraphics[angle=0,scale=0.3]{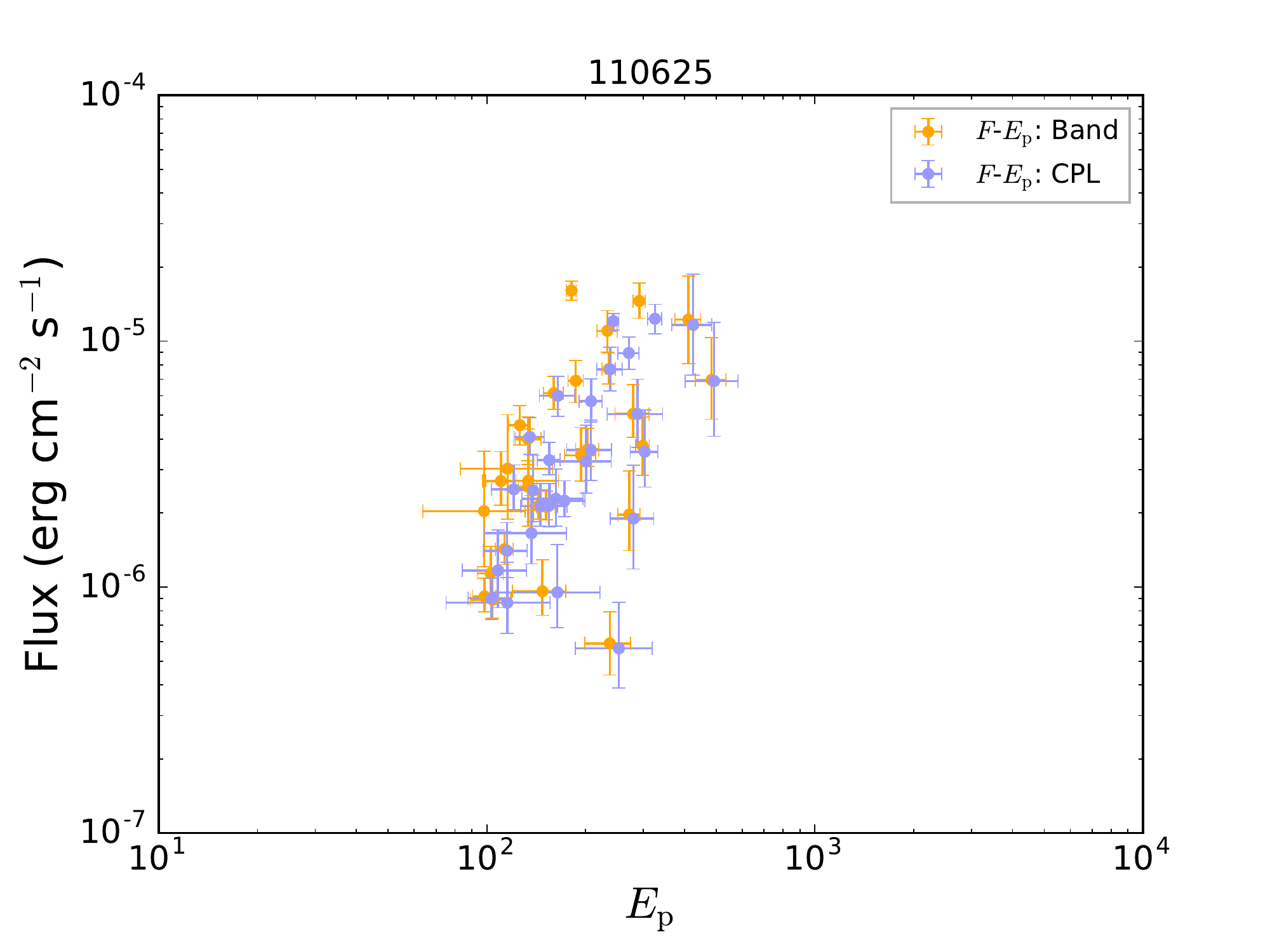}
\includegraphics[angle=0,scale=0.3]{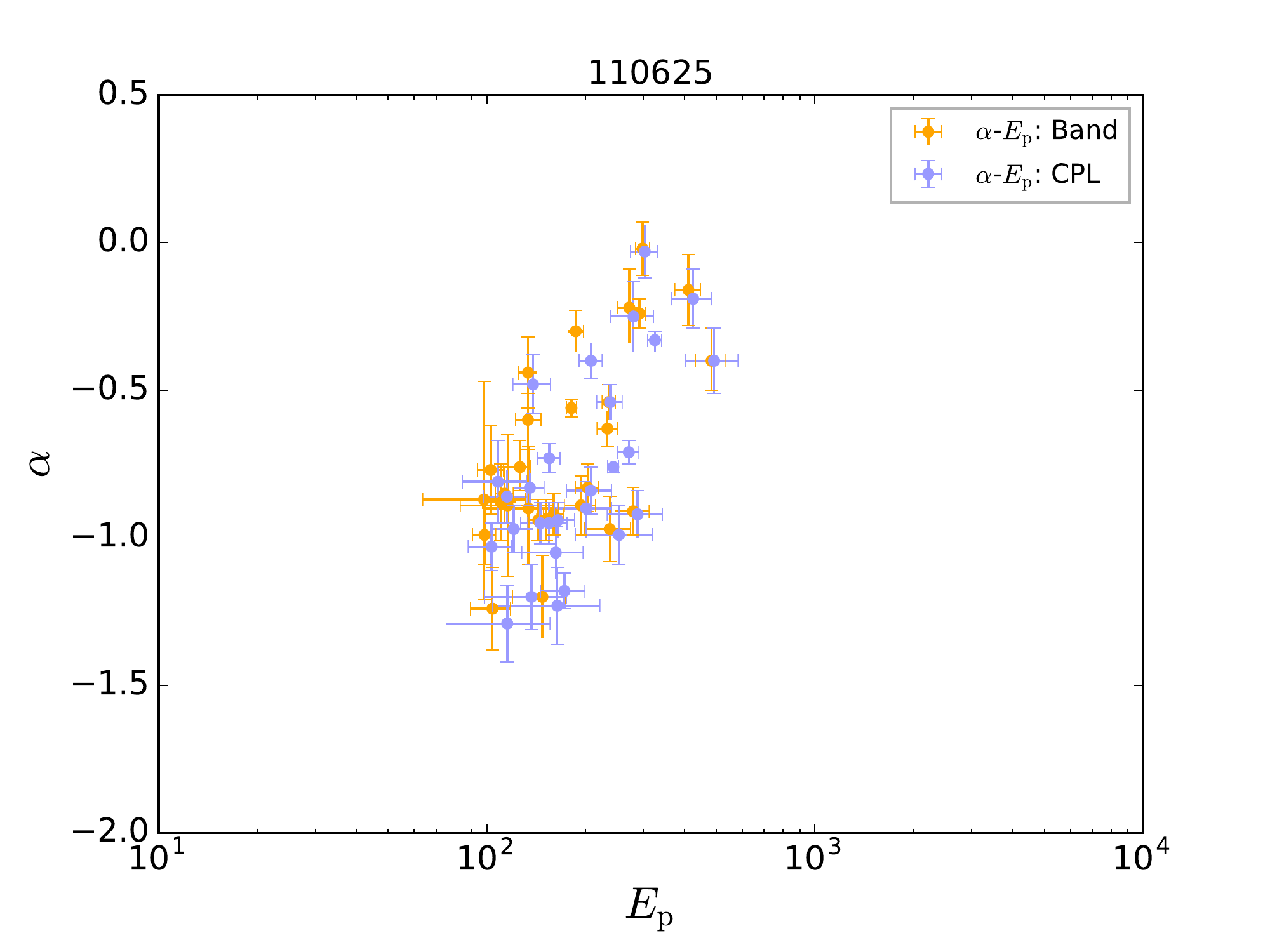}
\center{Fig. \ref{fig:relation3}--- Continued}
\end{figure*}
\begin{figure*}
\includegraphics[angle=0,scale=0.3]{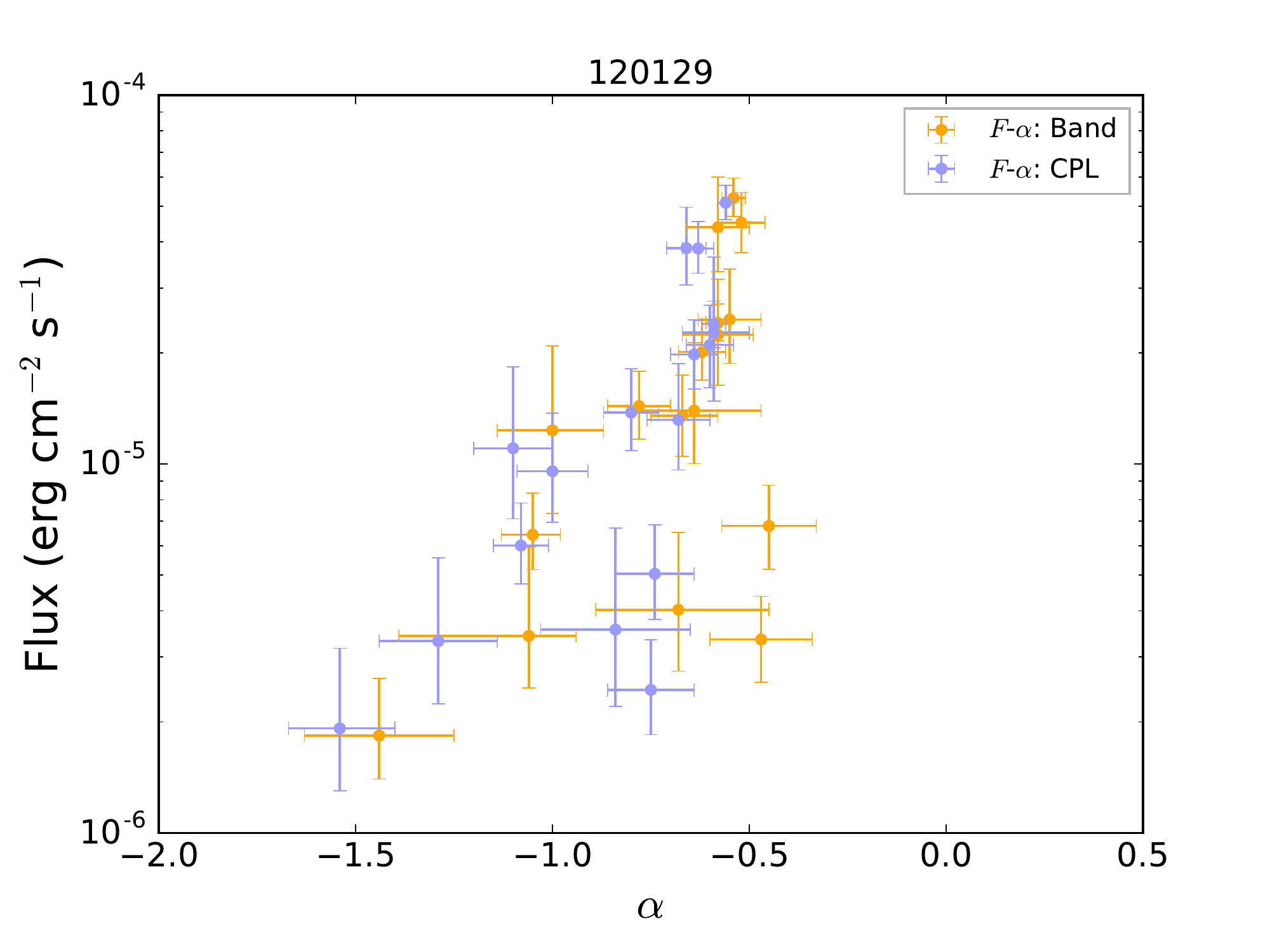}
\includegraphics[angle=0,scale=0.3]{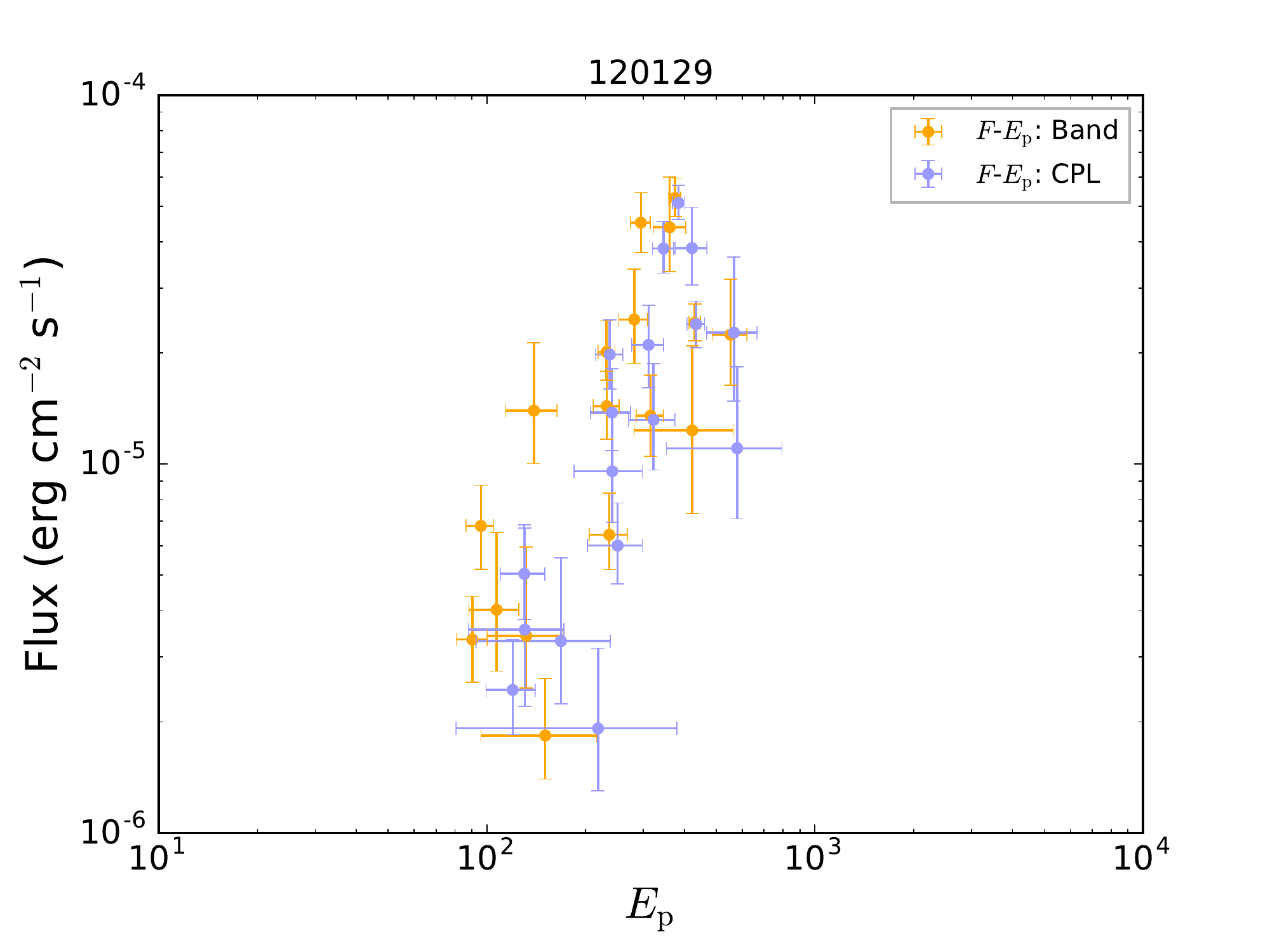}
\includegraphics[angle=0,scale=0.3]{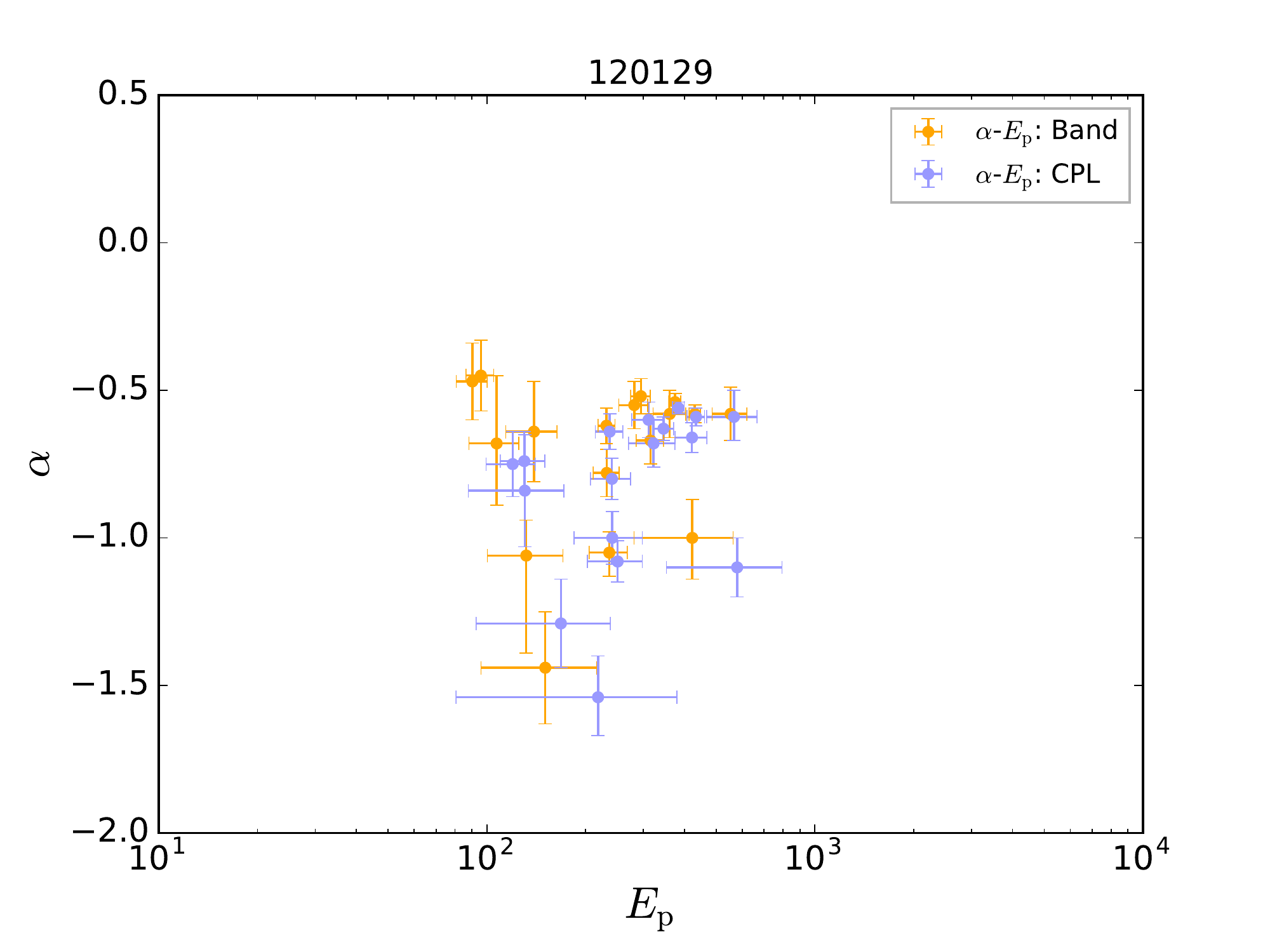}
\includegraphics[angle=0,scale=0.3]{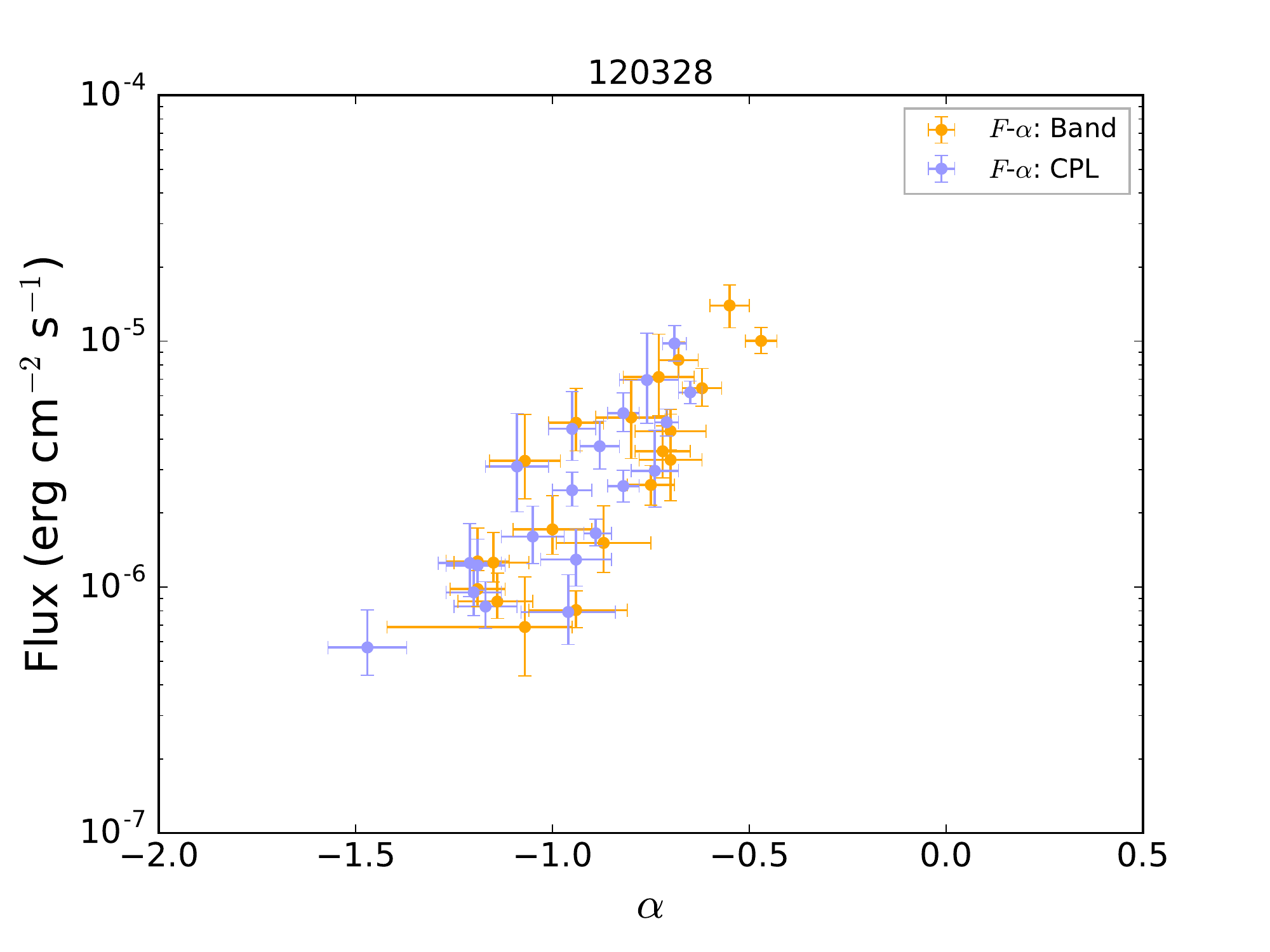}
\includegraphics[angle=0,scale=0.3]{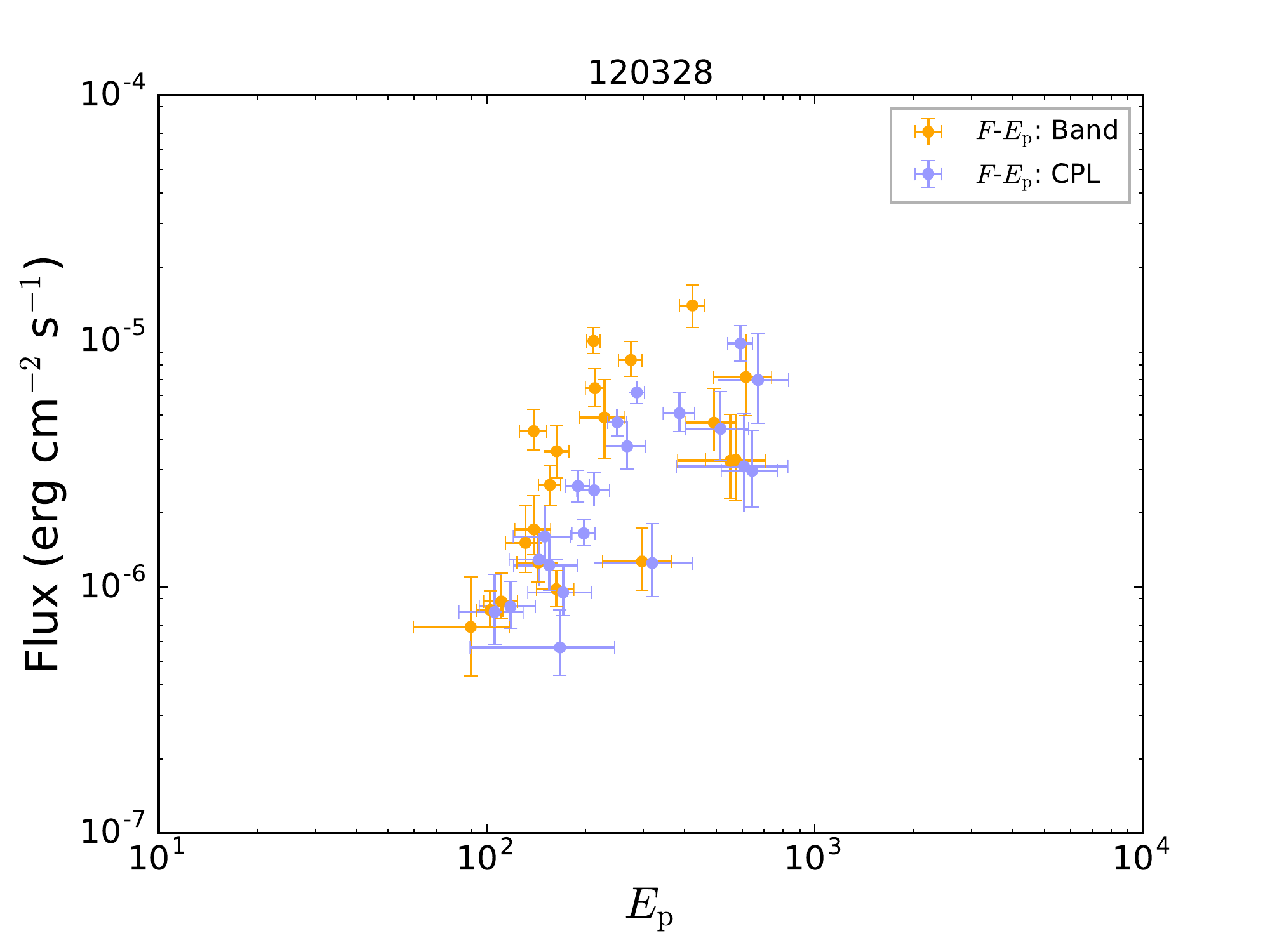}
\includegraphics[angle=0,scale=0.3]{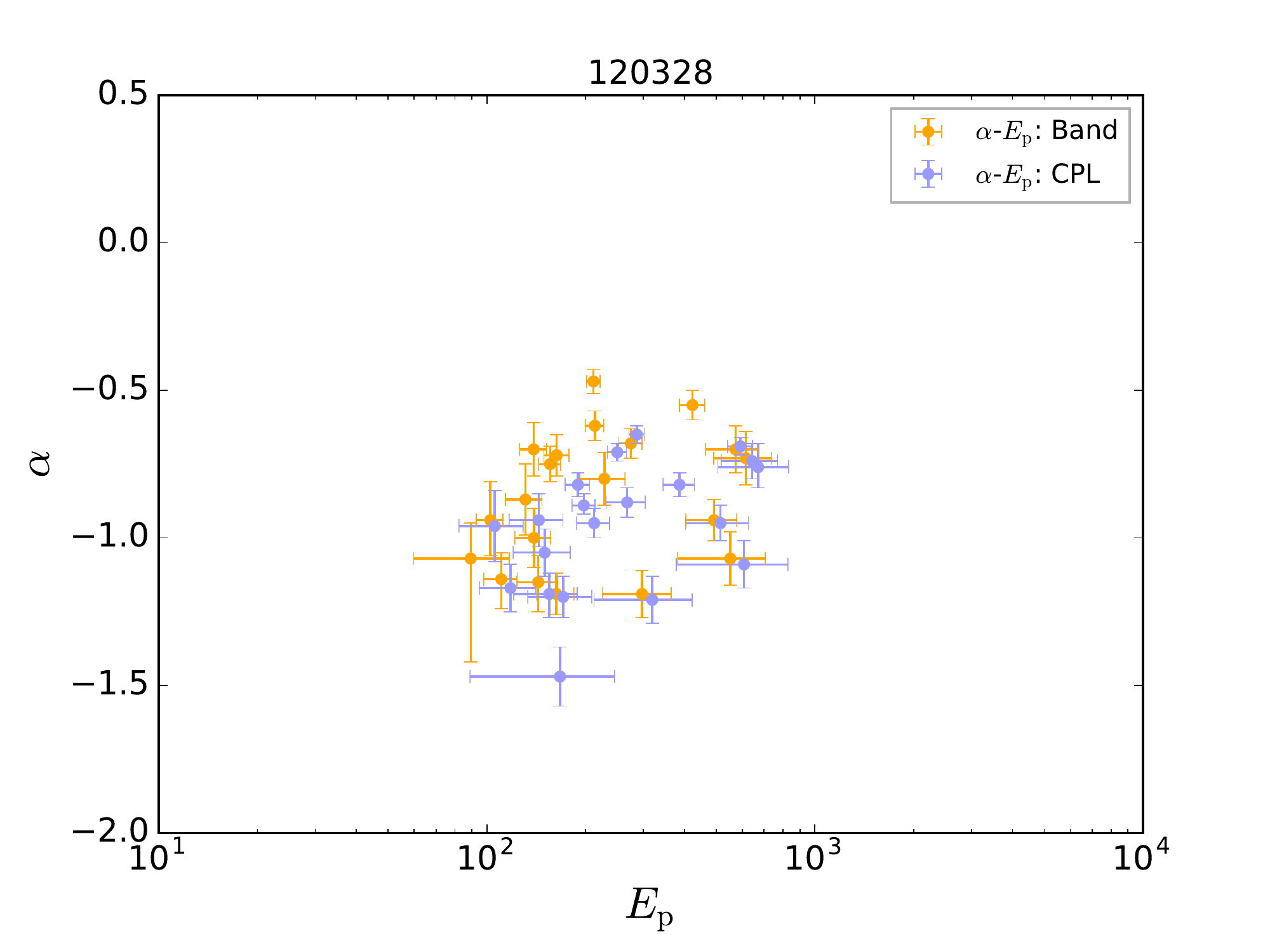}
\includegraphics[angle=0,scale=0.3]{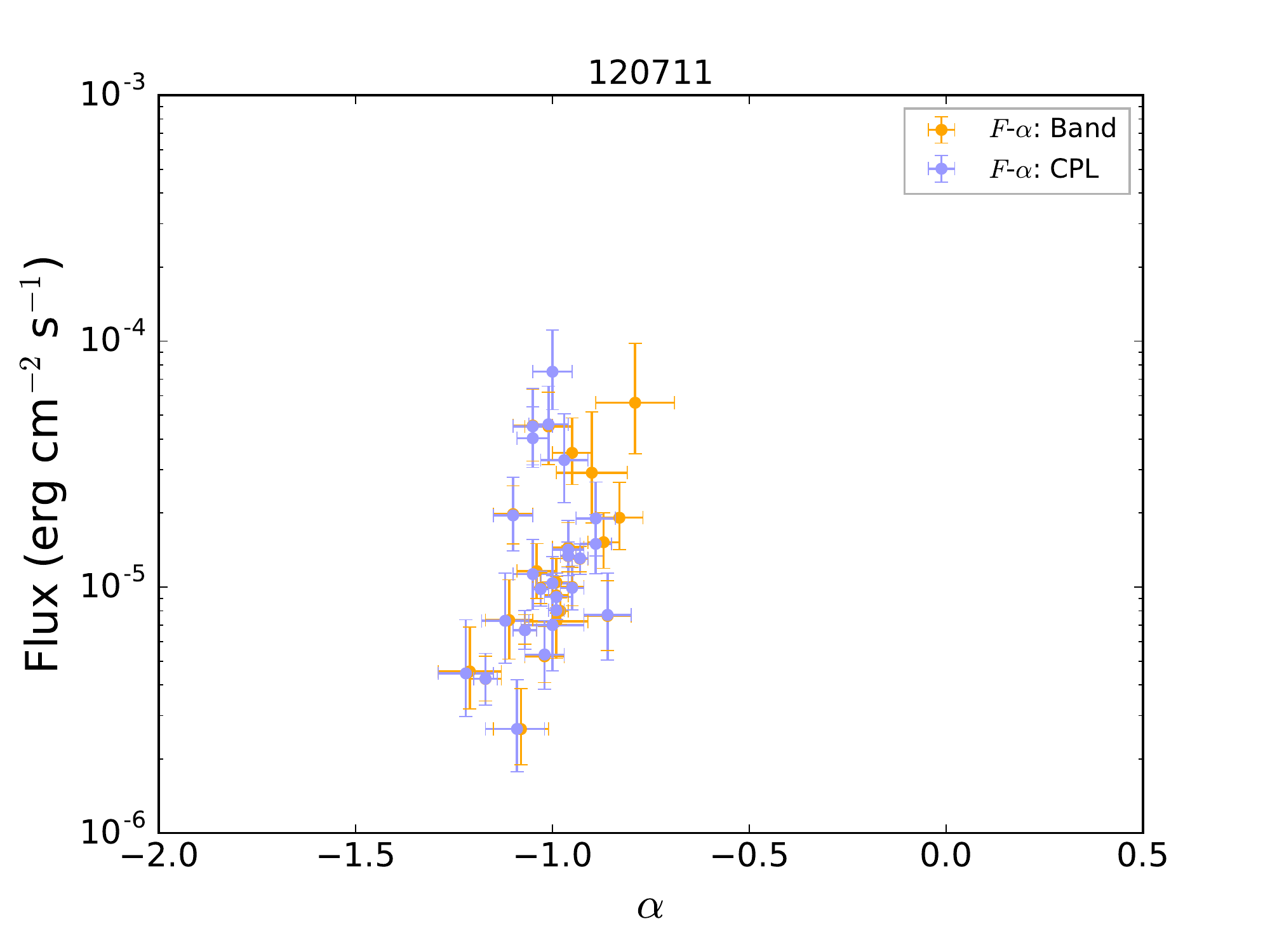}
\includegraphics[angle=0,scale=0.3]{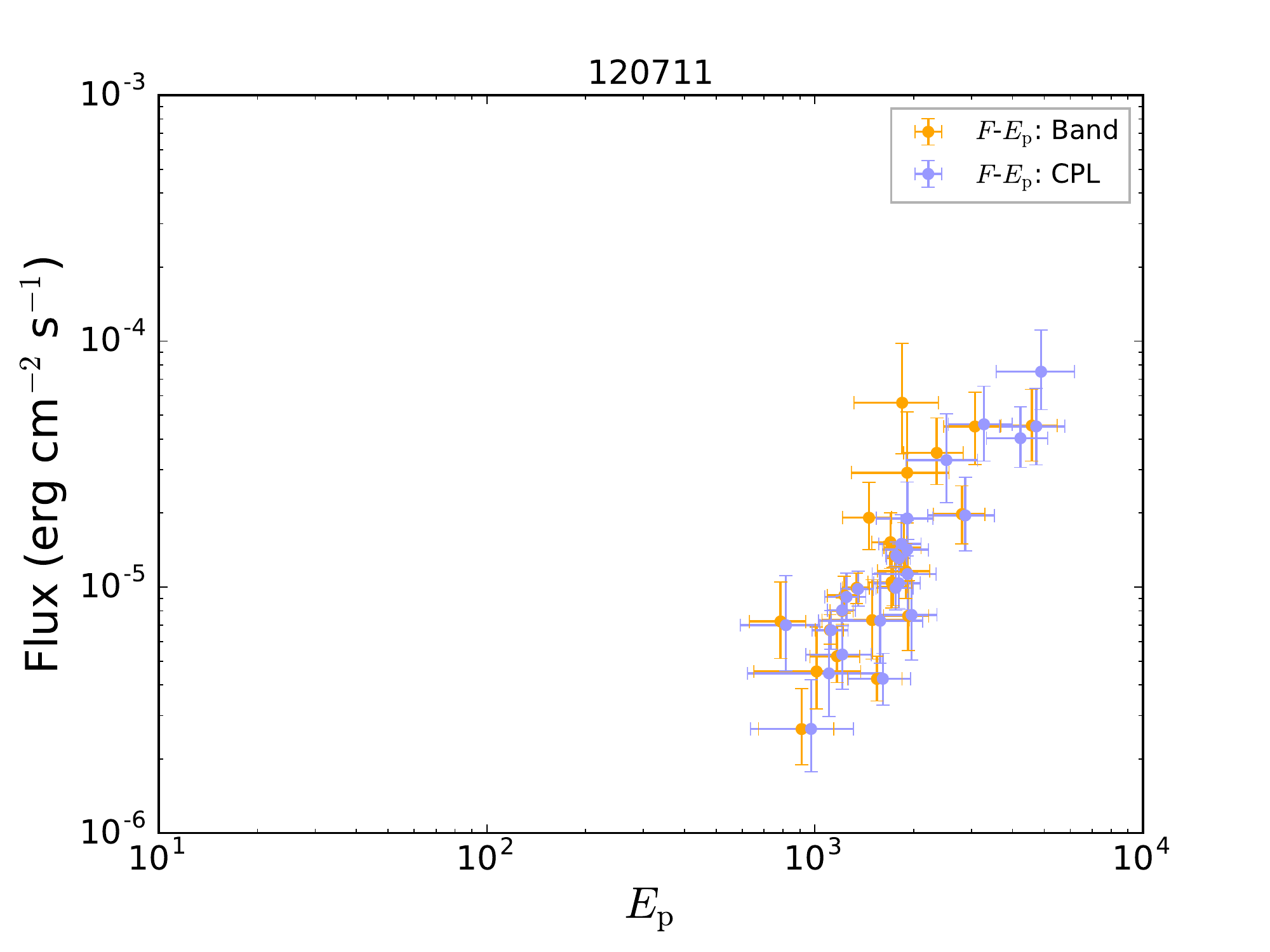}
\includegraphics[angle=0,scale=0.3]{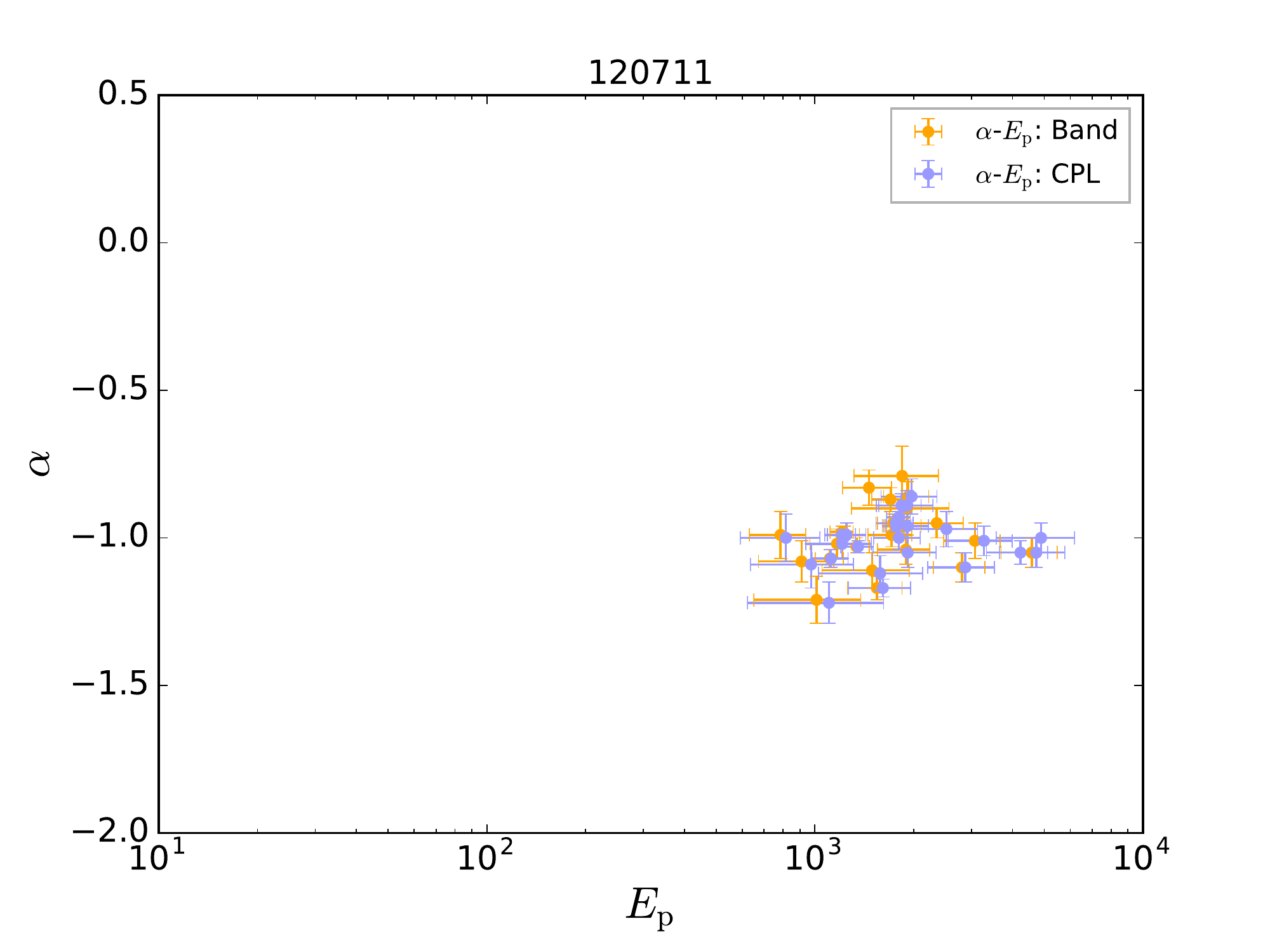}
\includegraphics[angle=0,scale=0.3]{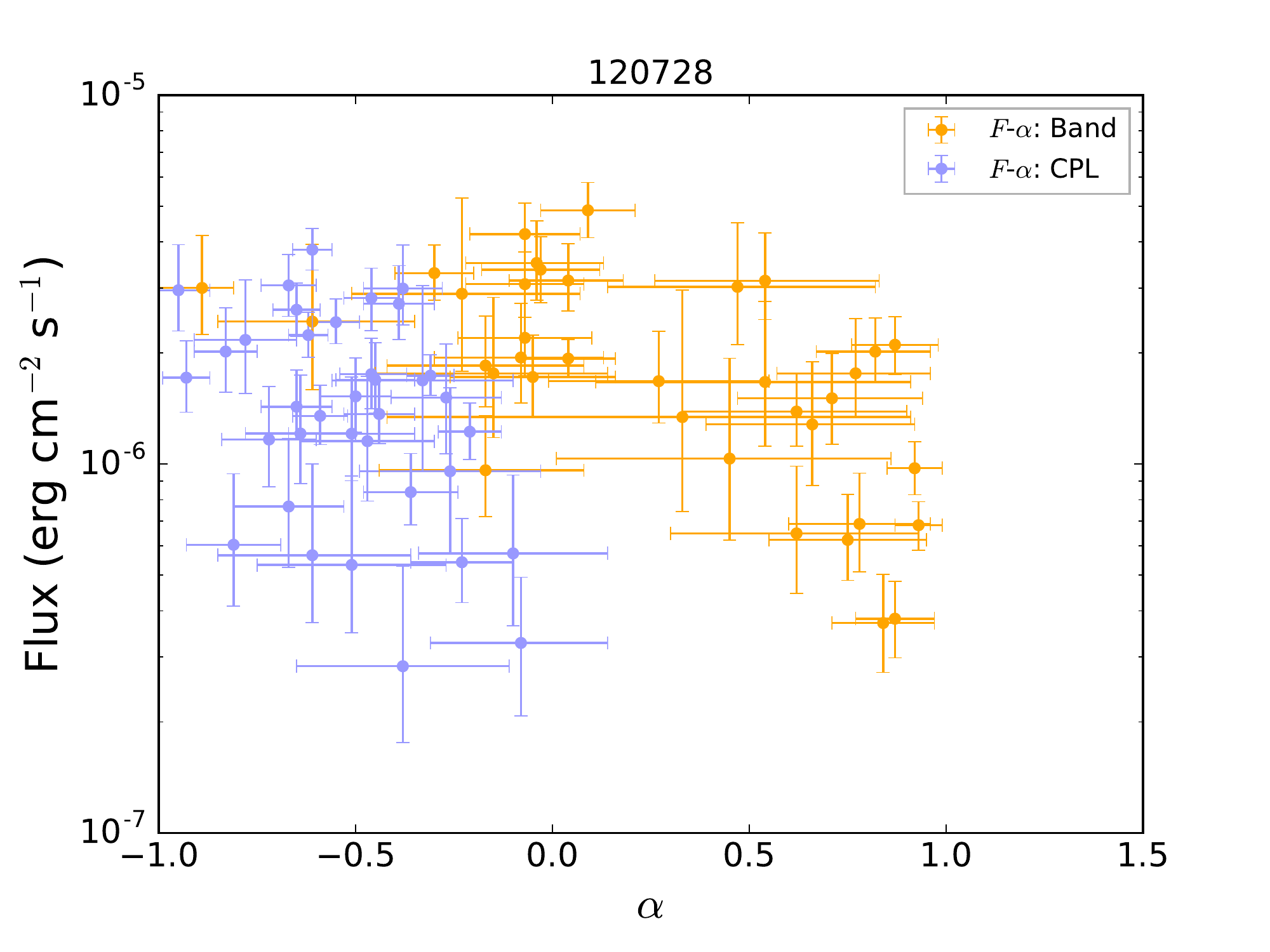}
\includegraphics[angle=0,scale=0.3]{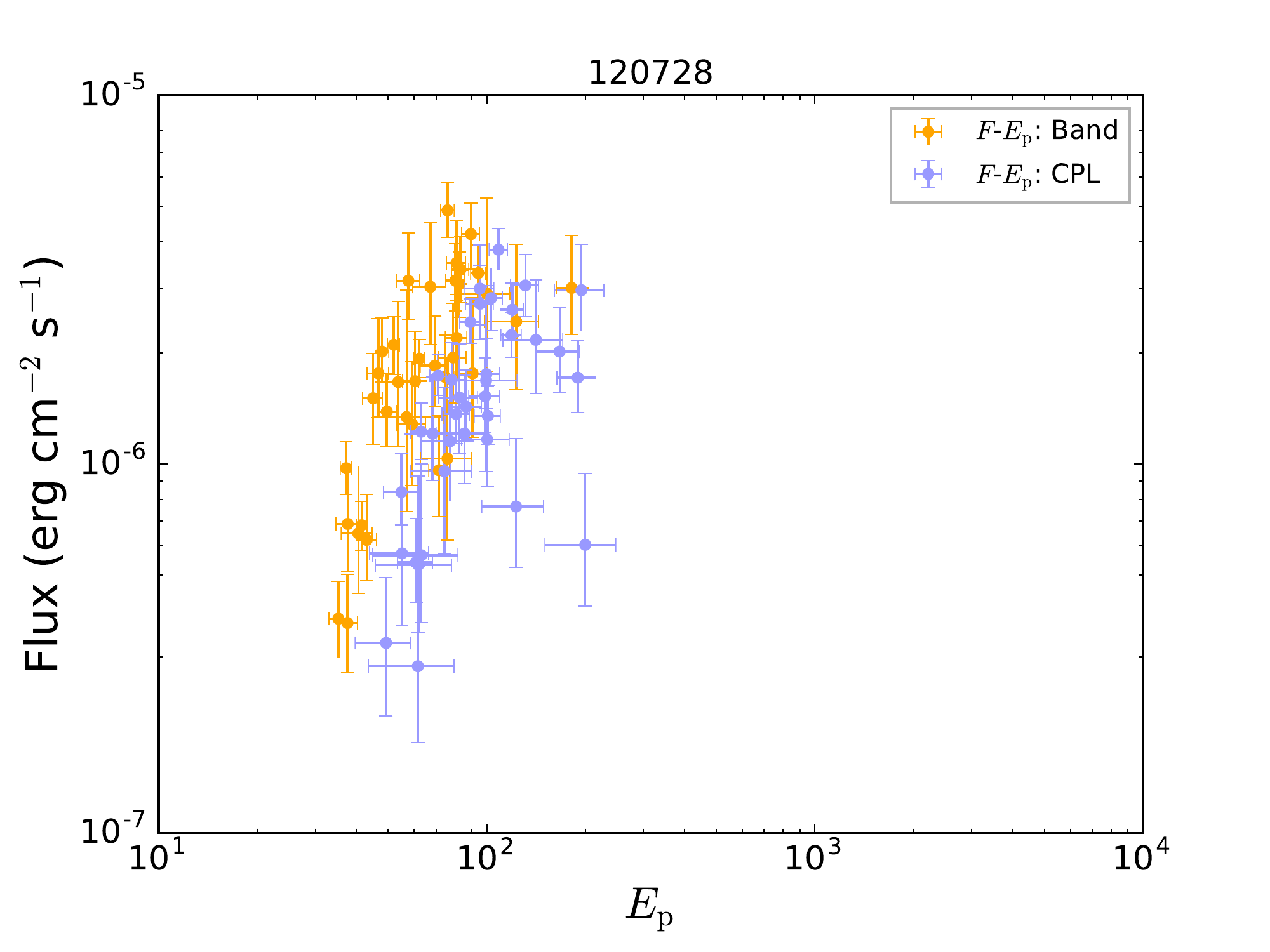}
\includegraphics[angle=0,scale=0.3]{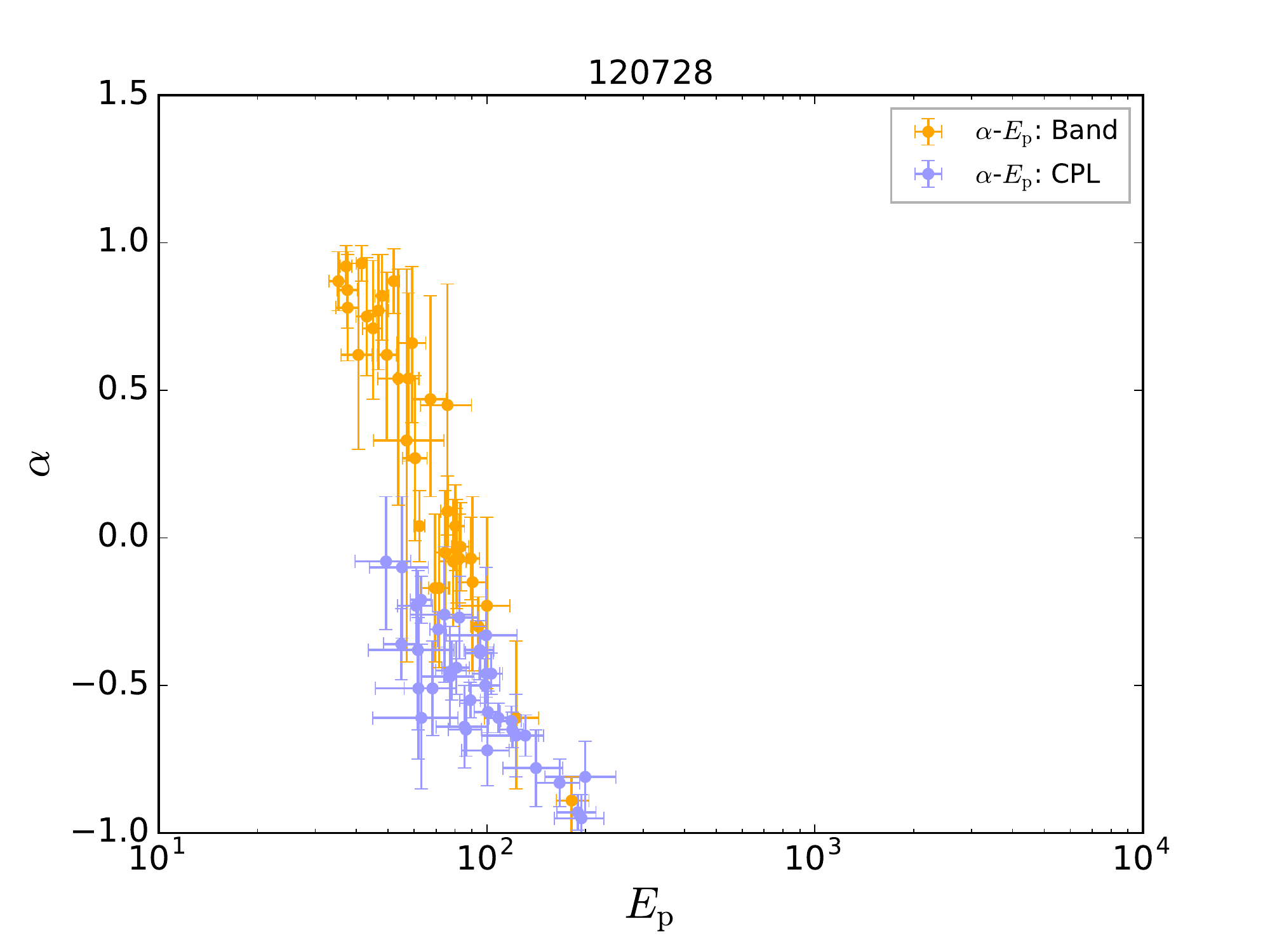}
\includegraphics[angle=0,scale=0.3]{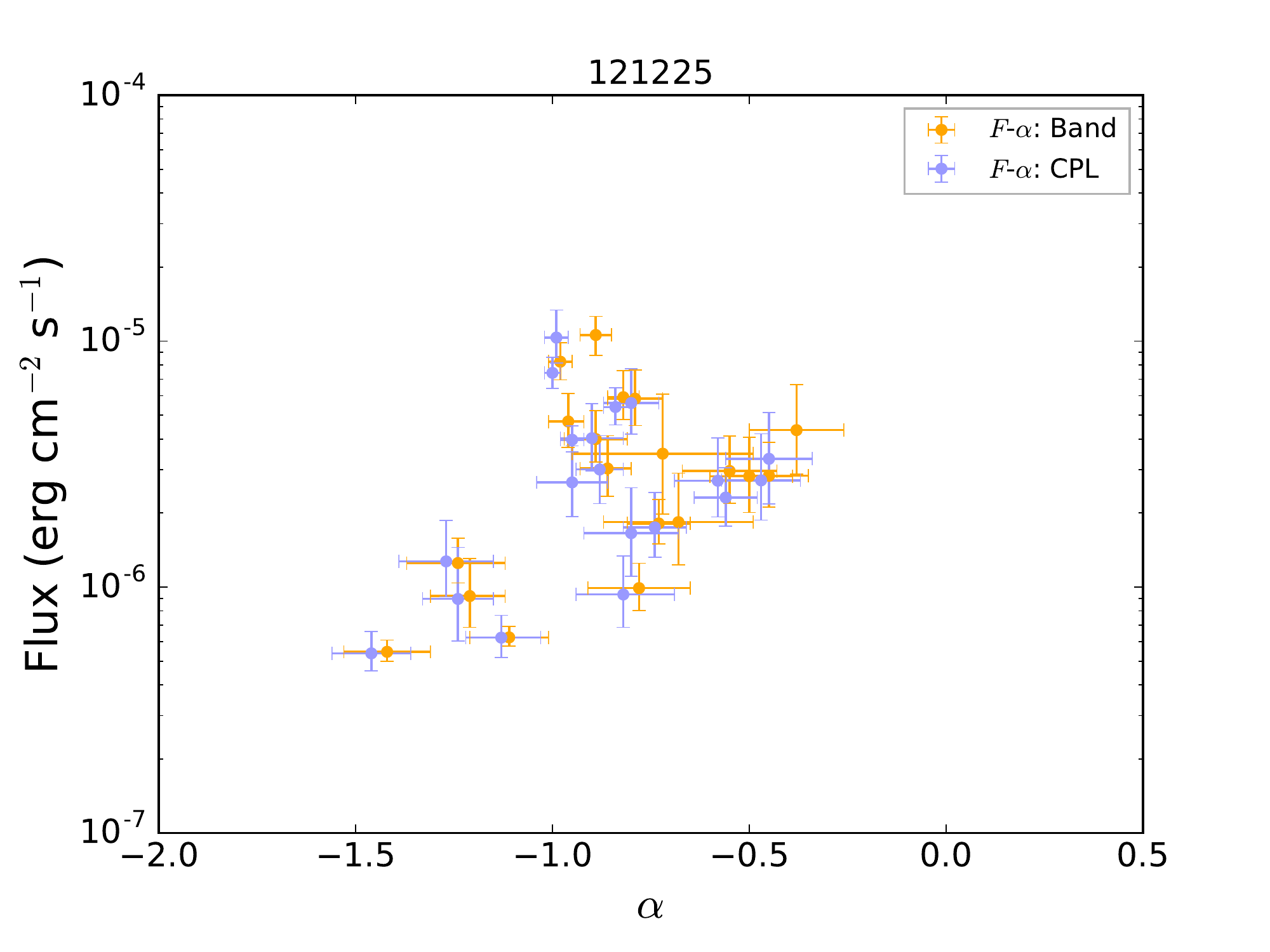}
\includegraphics[angle=0,scale=0.3]{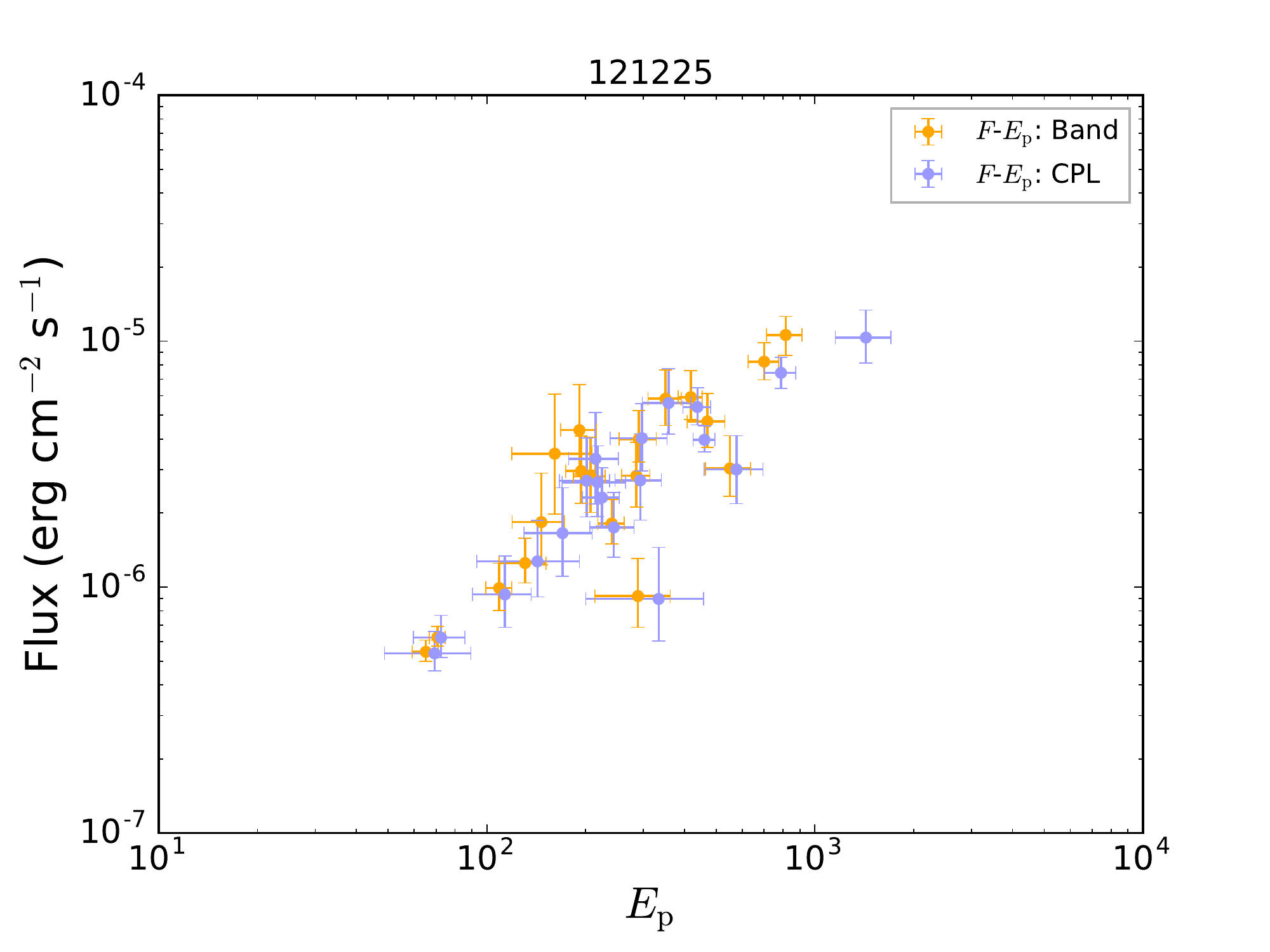}
\includegraphics[angle=0,scale=0.3]{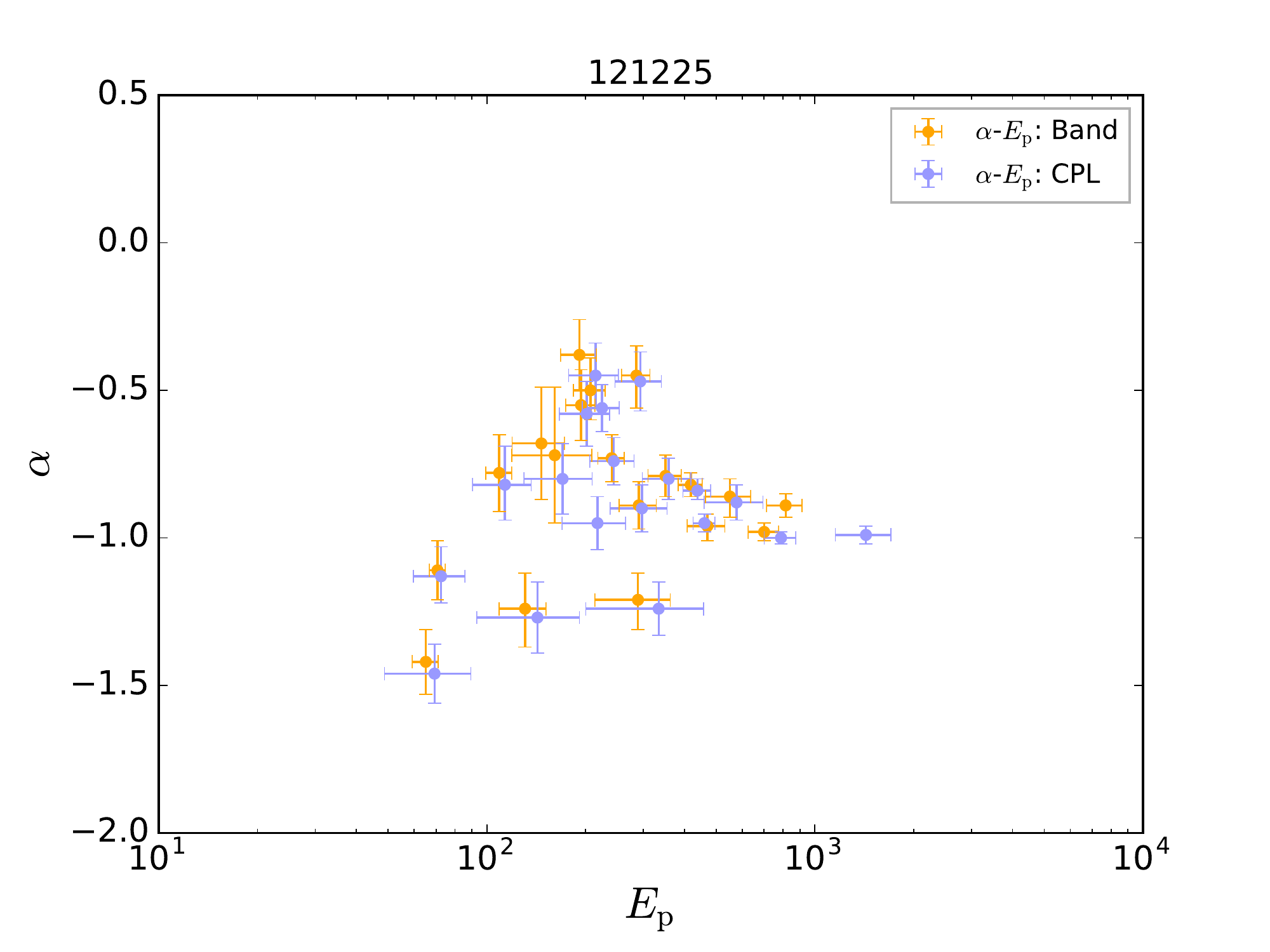}
\center{Fig. \ref{fig:relation3}--- Continued}
\end{figure*}
\begin{figure*}
\includegraphics[angle=0,scale=0.3]{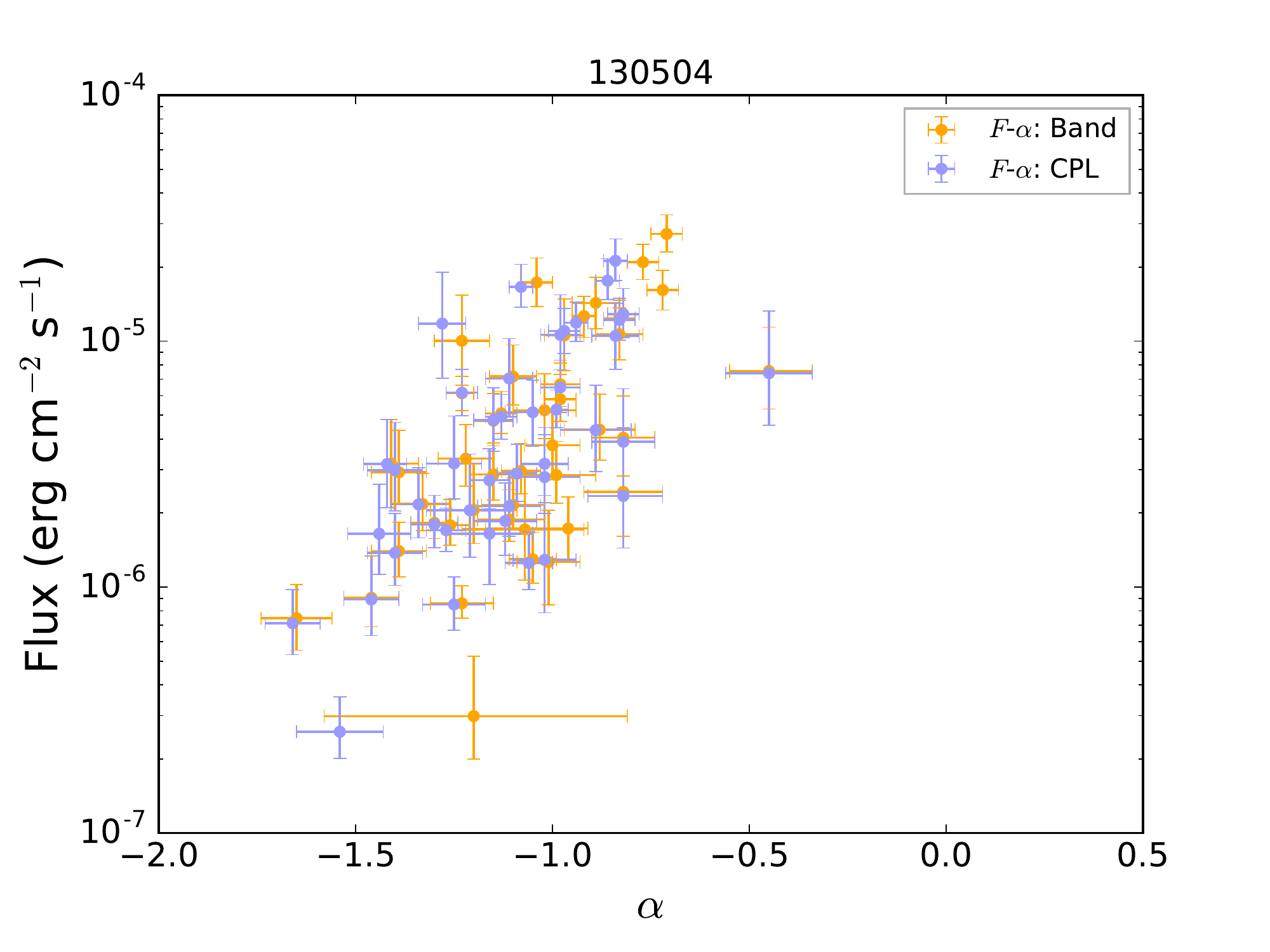}
\includegraphics[angle=0,scale=0.3]{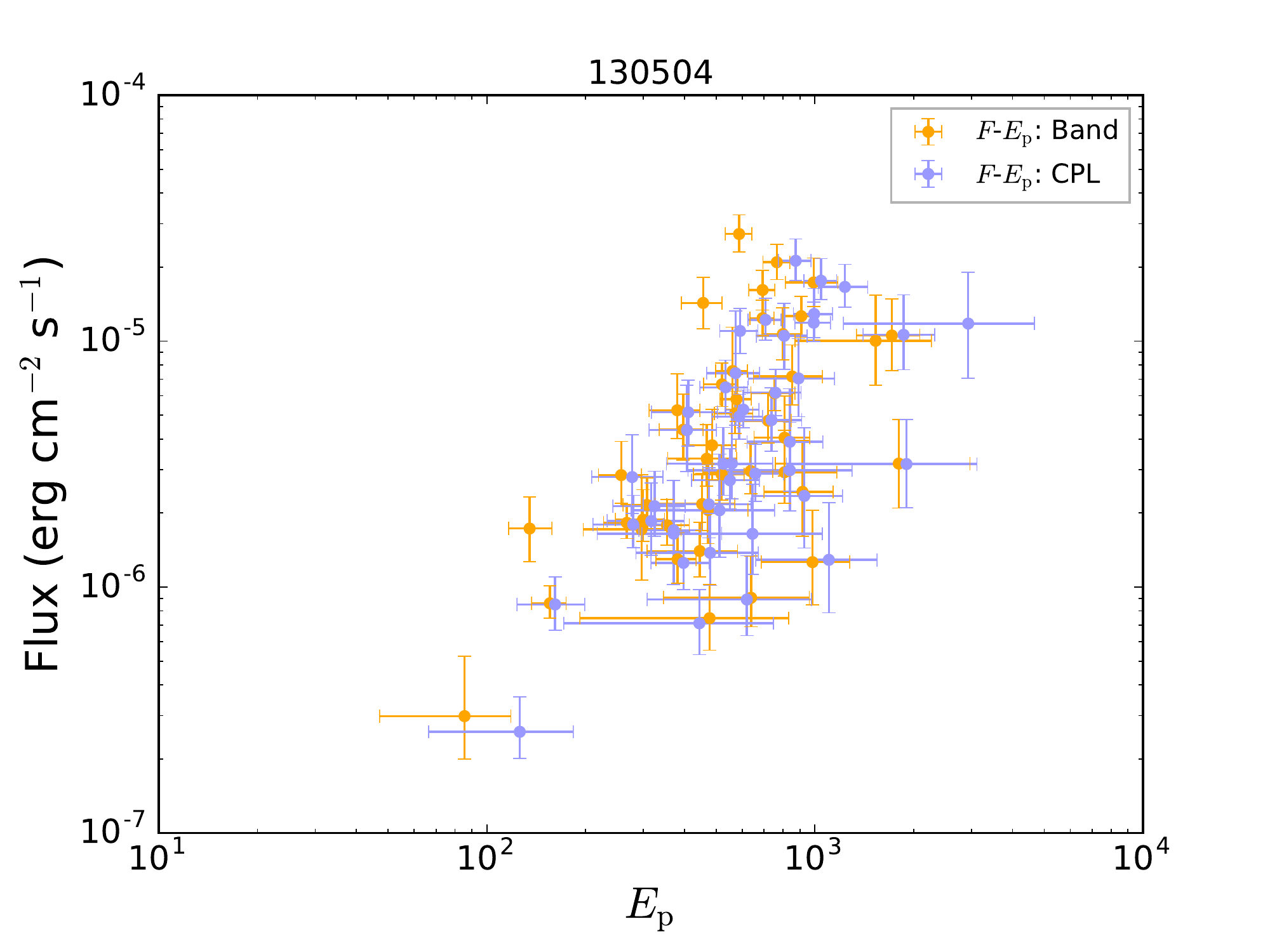}
\includegraphics[angle=0,scale=0.3]{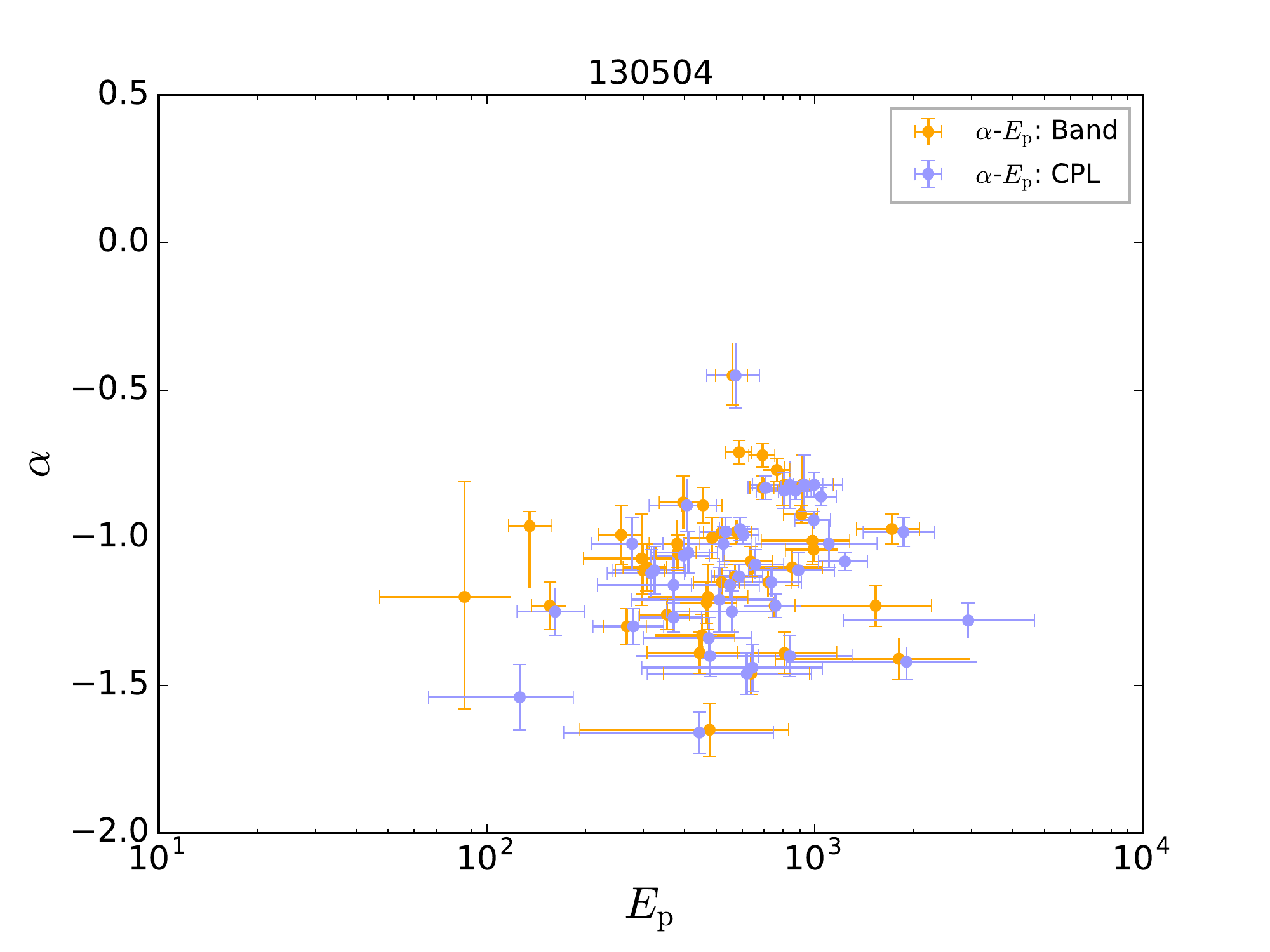}
\includegraphics[angle=0,scale=0.3]{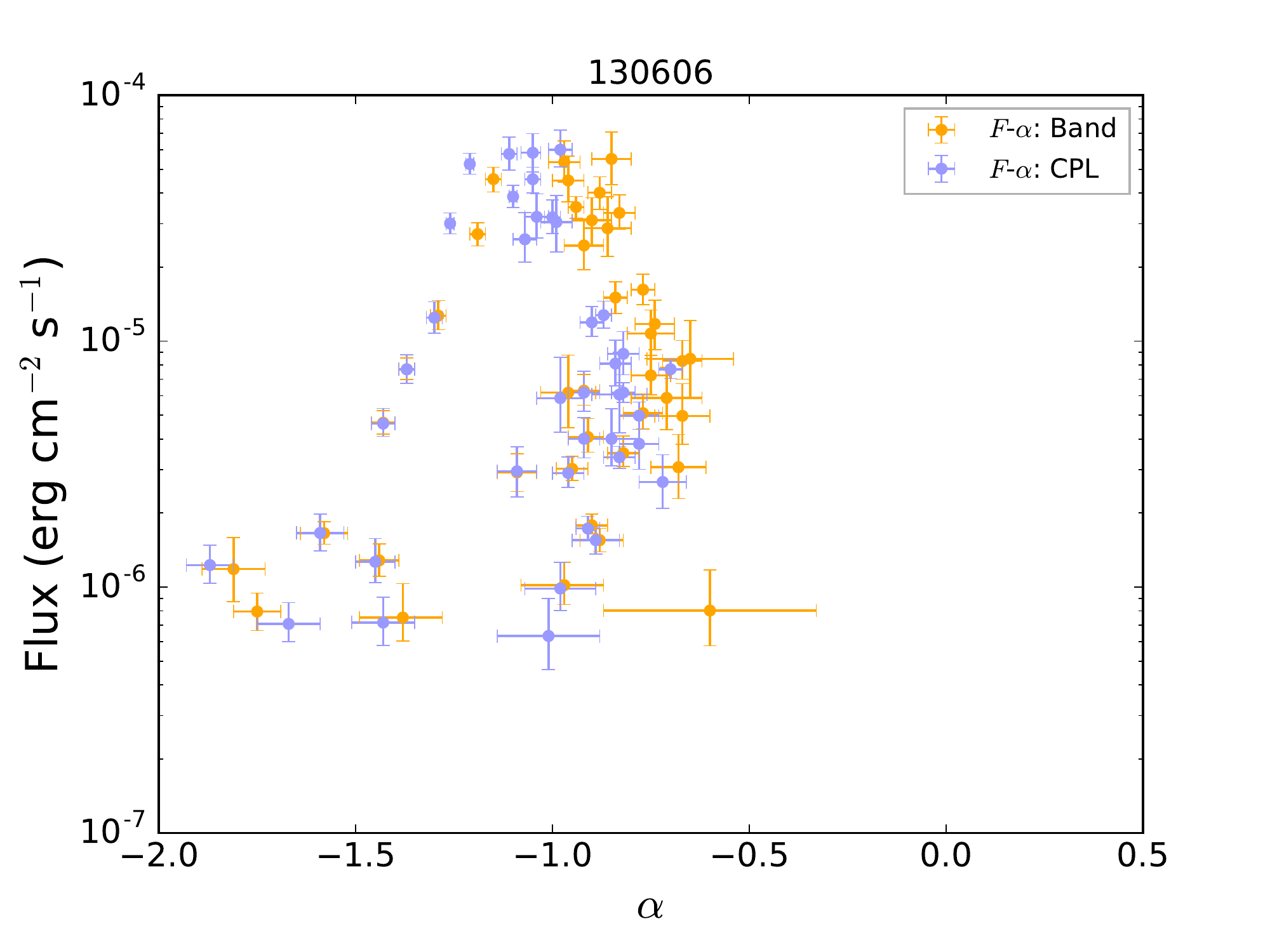}
\includegraphics[angle=0,scale=0.3]{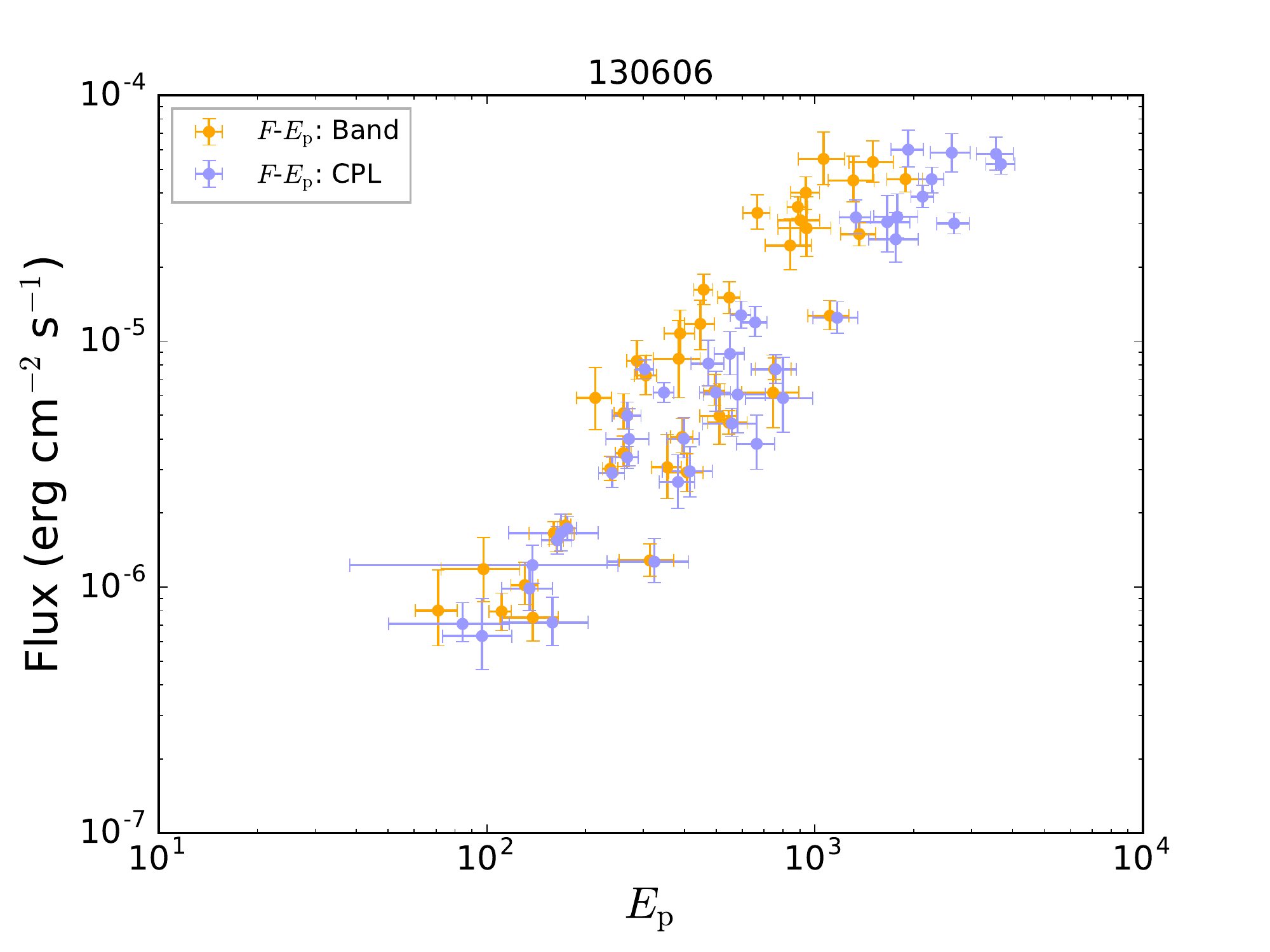}
\includegraphics[angle=0,scale=0.3]{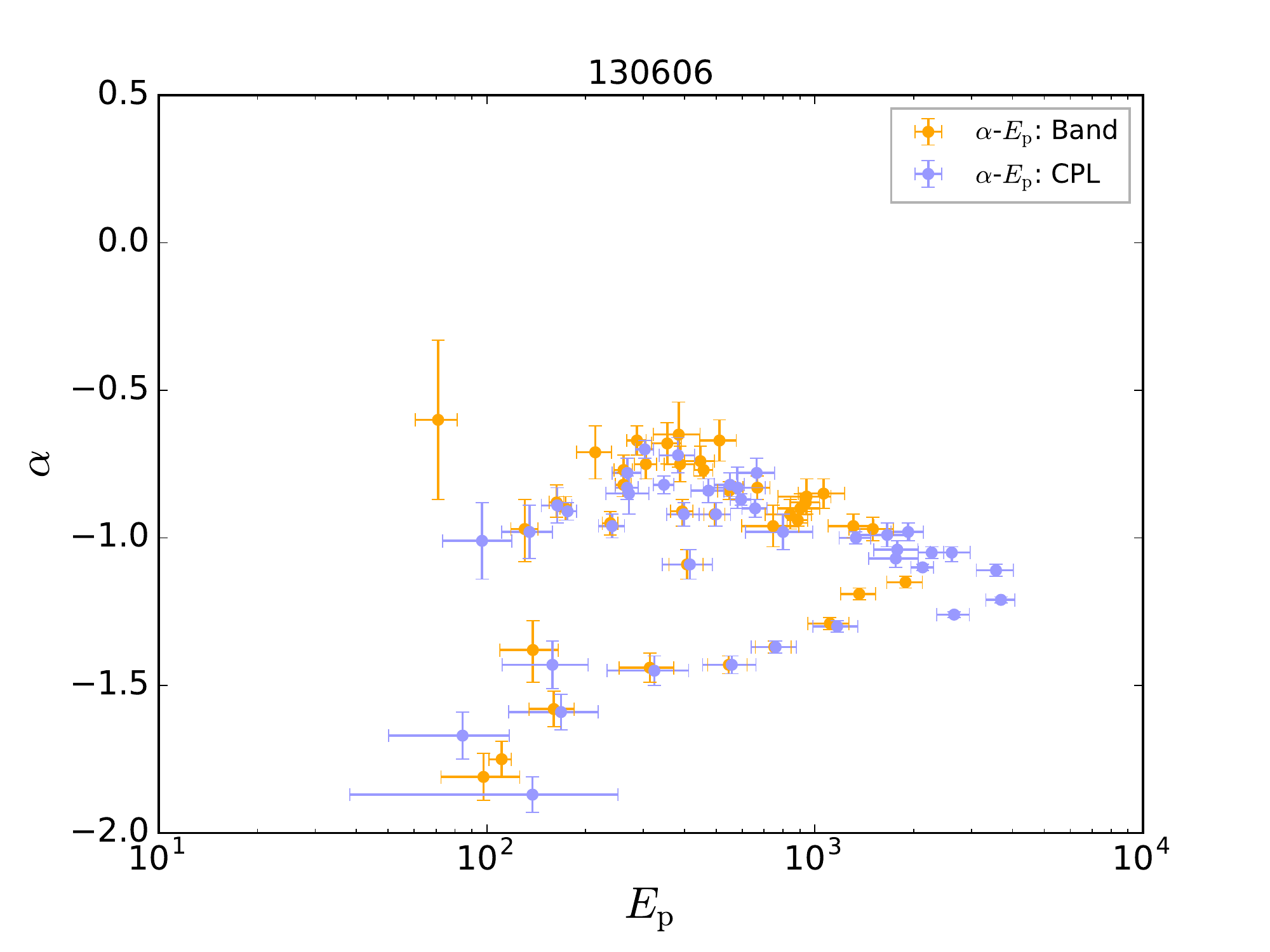}
\includegraphics[angle=0,scale=0.3]{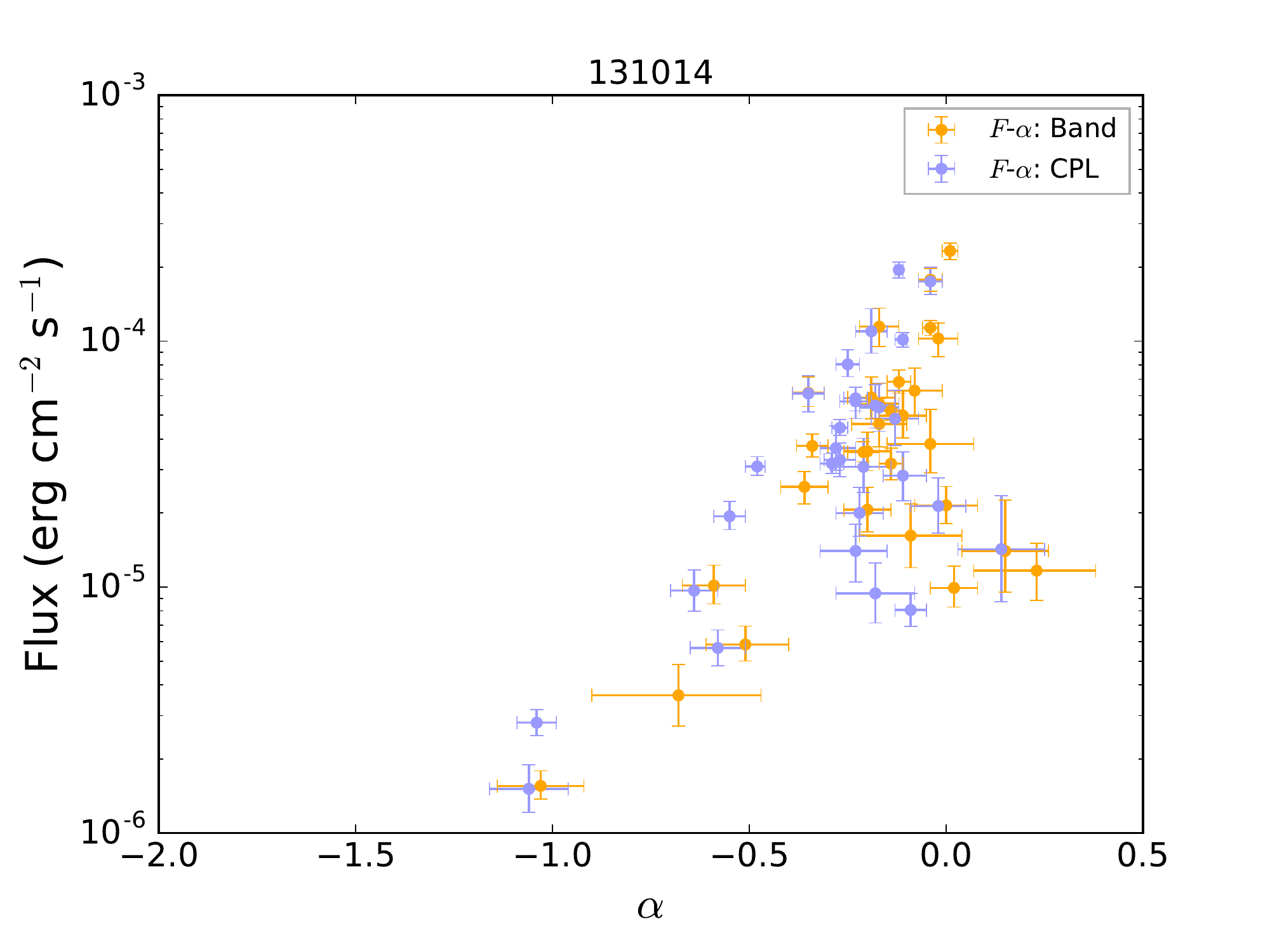}
\includegraphics[angle=0,scale=0.3]{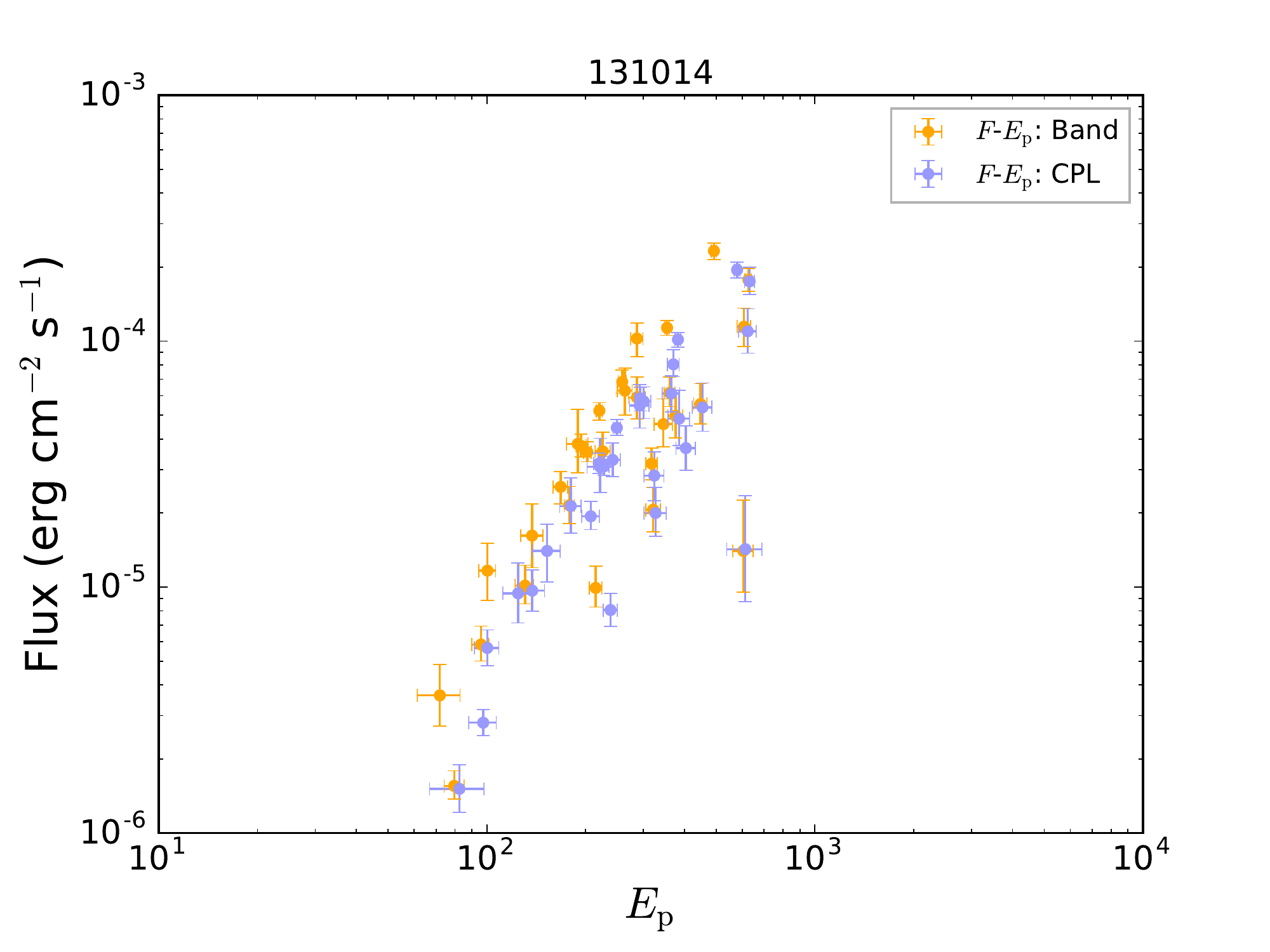}
\includegraphics[angle=0,scale=0.3]{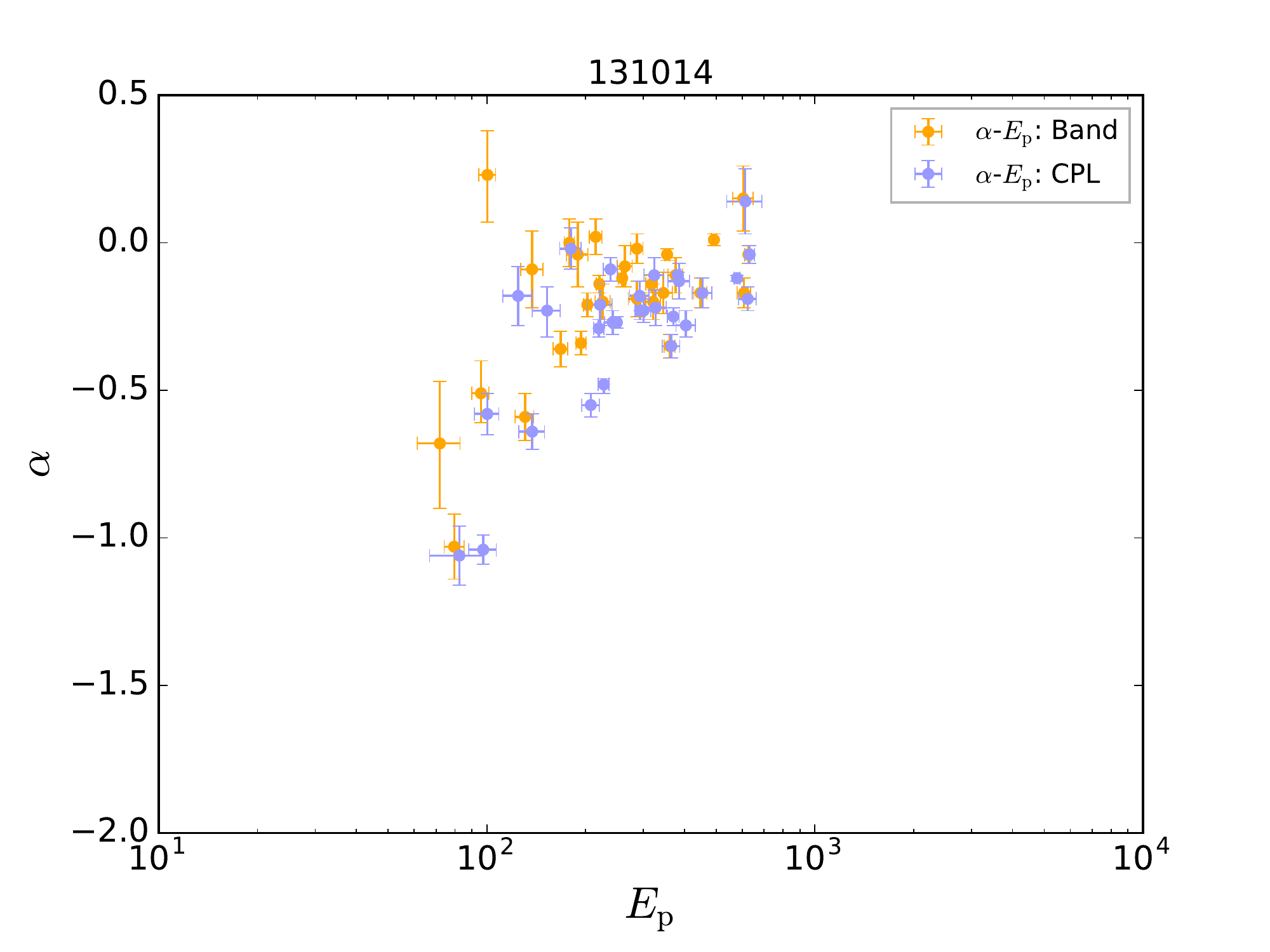}
\includegraphics[angle=0,scale=0.3]{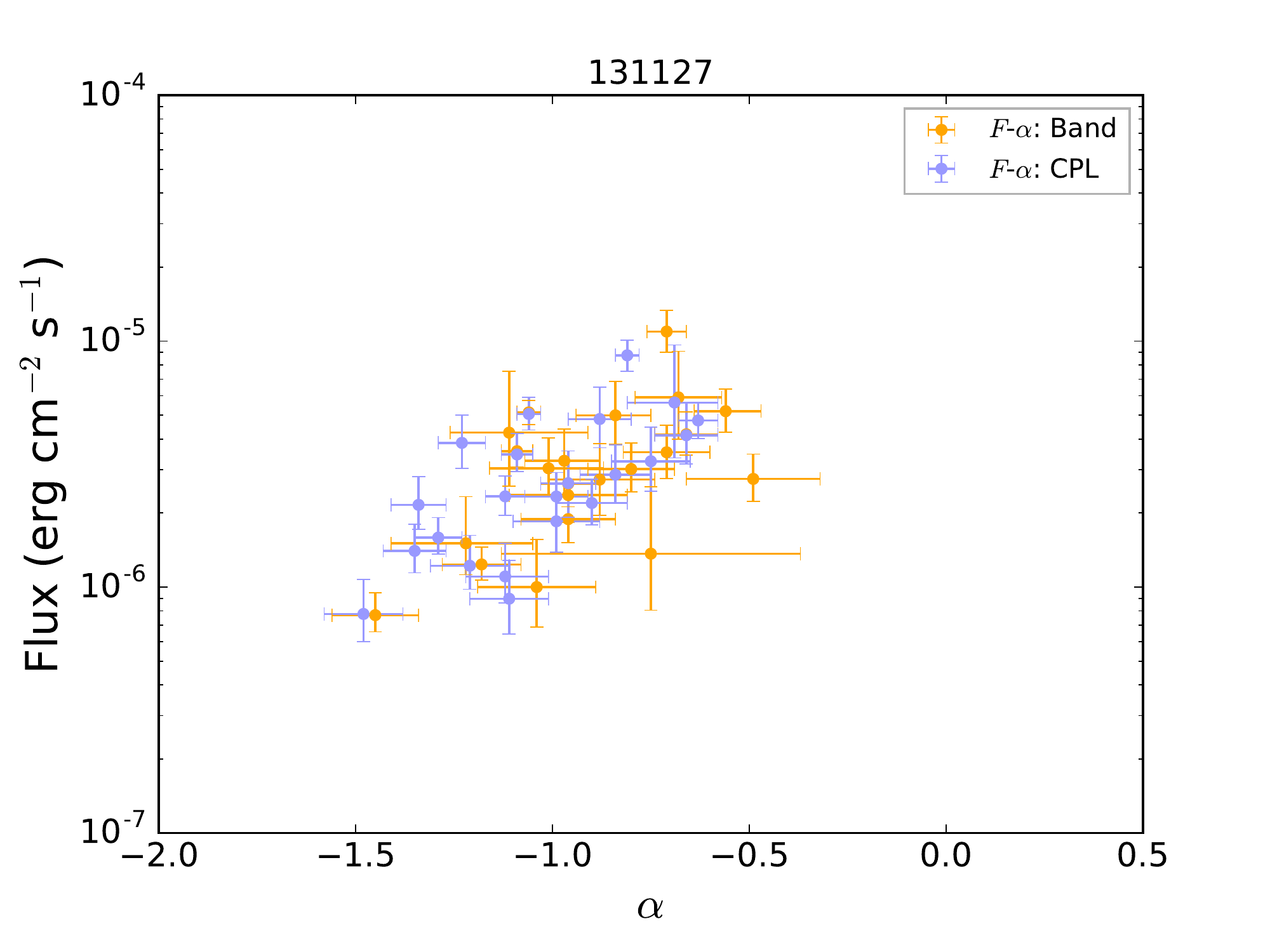}
\includegraphics[angle=0,scale=0.3]{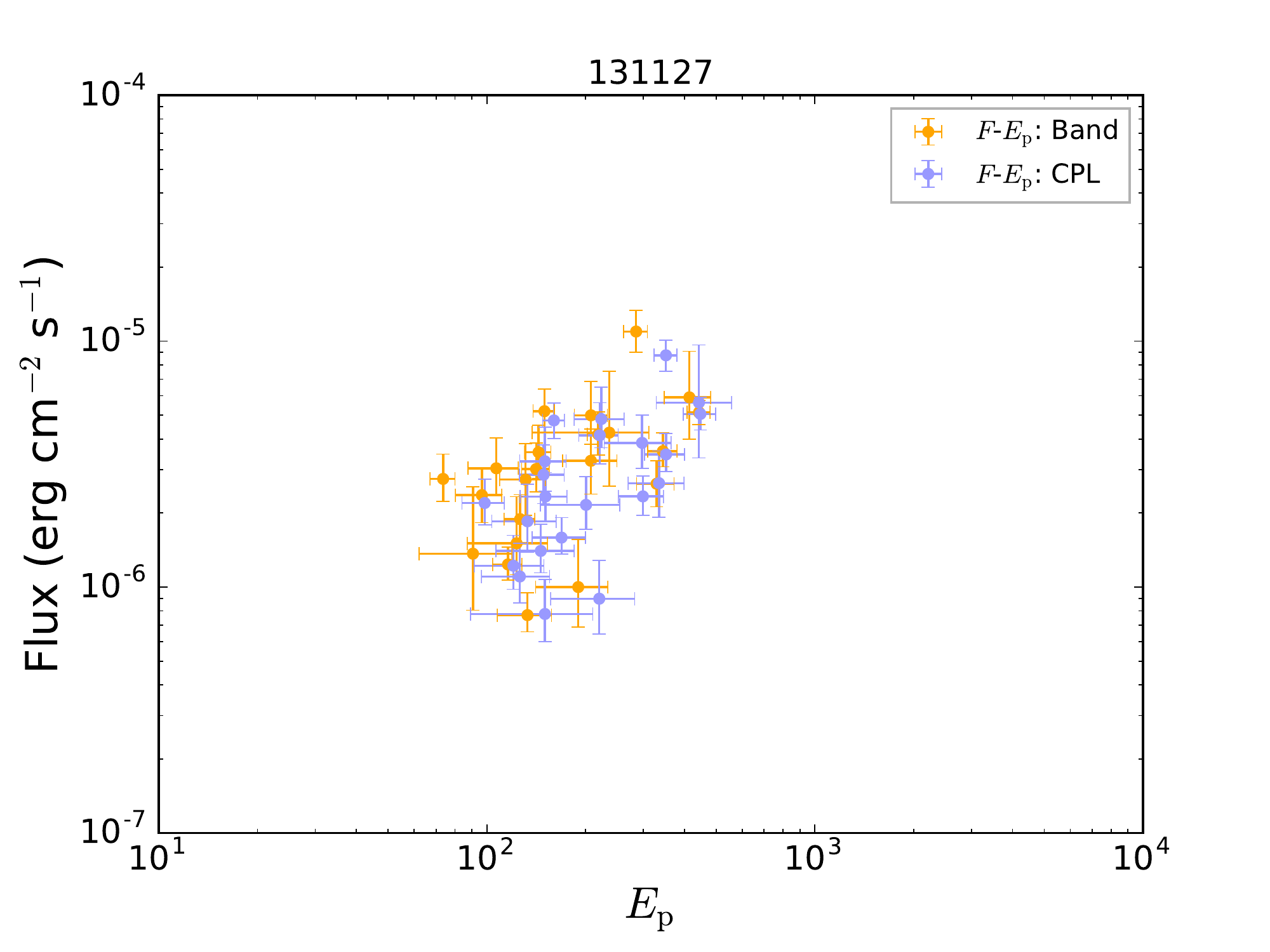}
\includegraphics[angle=0,scale=0.3]{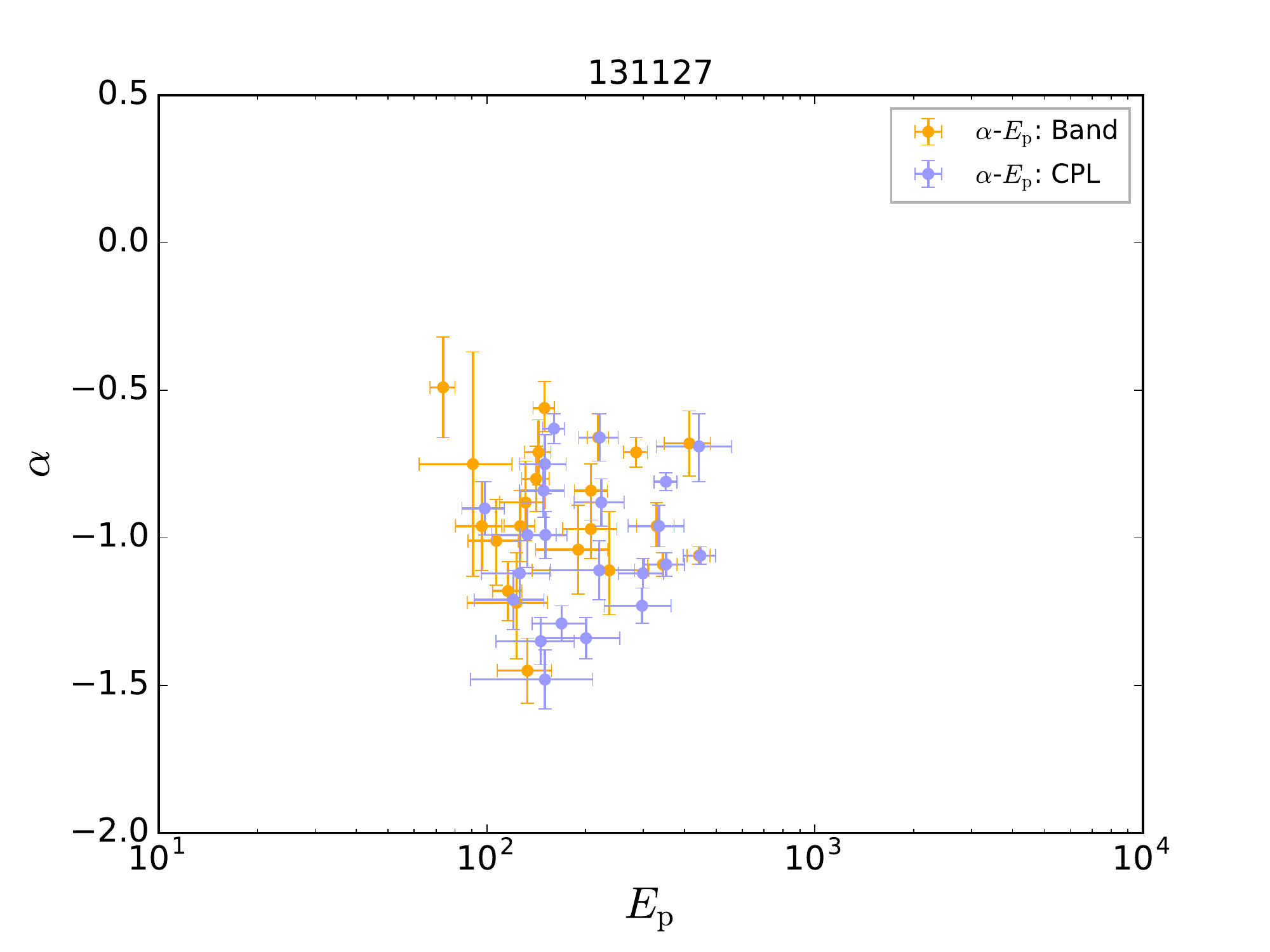}
\includegraphics[angle=0,scale=0.3]{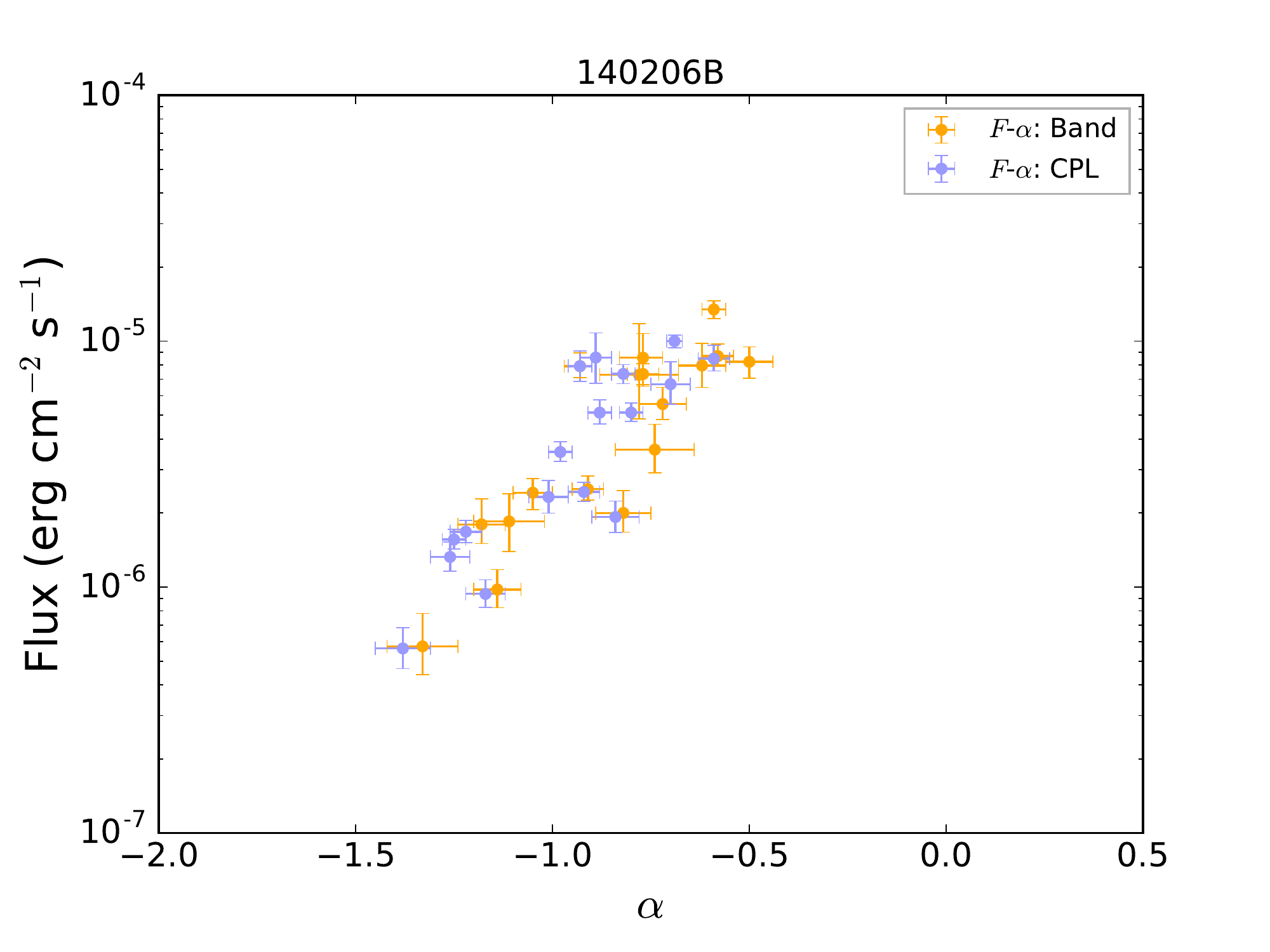}
\includegraphics[angle=0,scale=0.3]{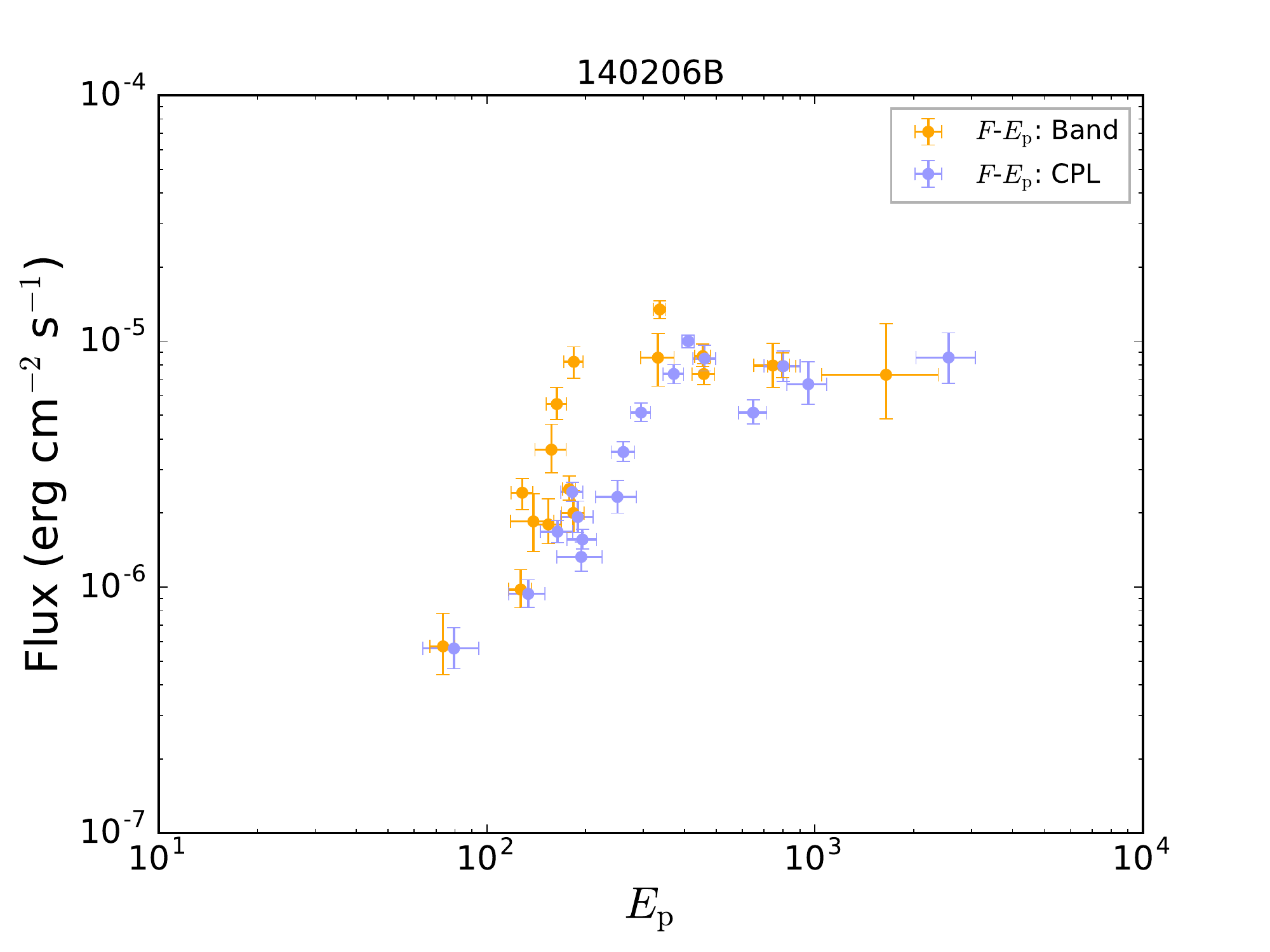}
\includegraphics[angle=0,scale=0.3]{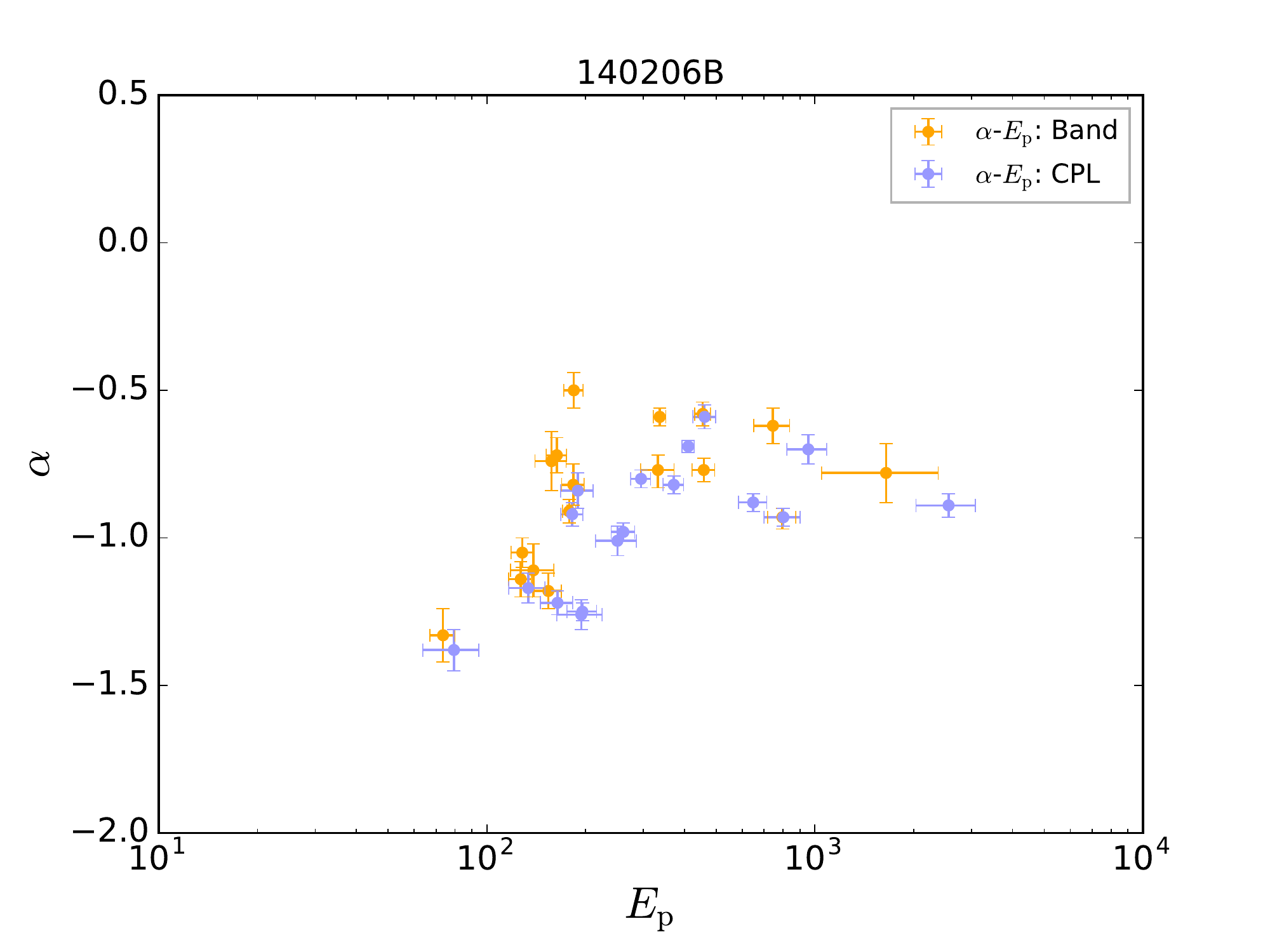}
\center{Fig. \ref{fig:relation3}--- Continued}
\end{figure*}
\begin{figure*}
\includegraphics[angle=0,scale=0.3]{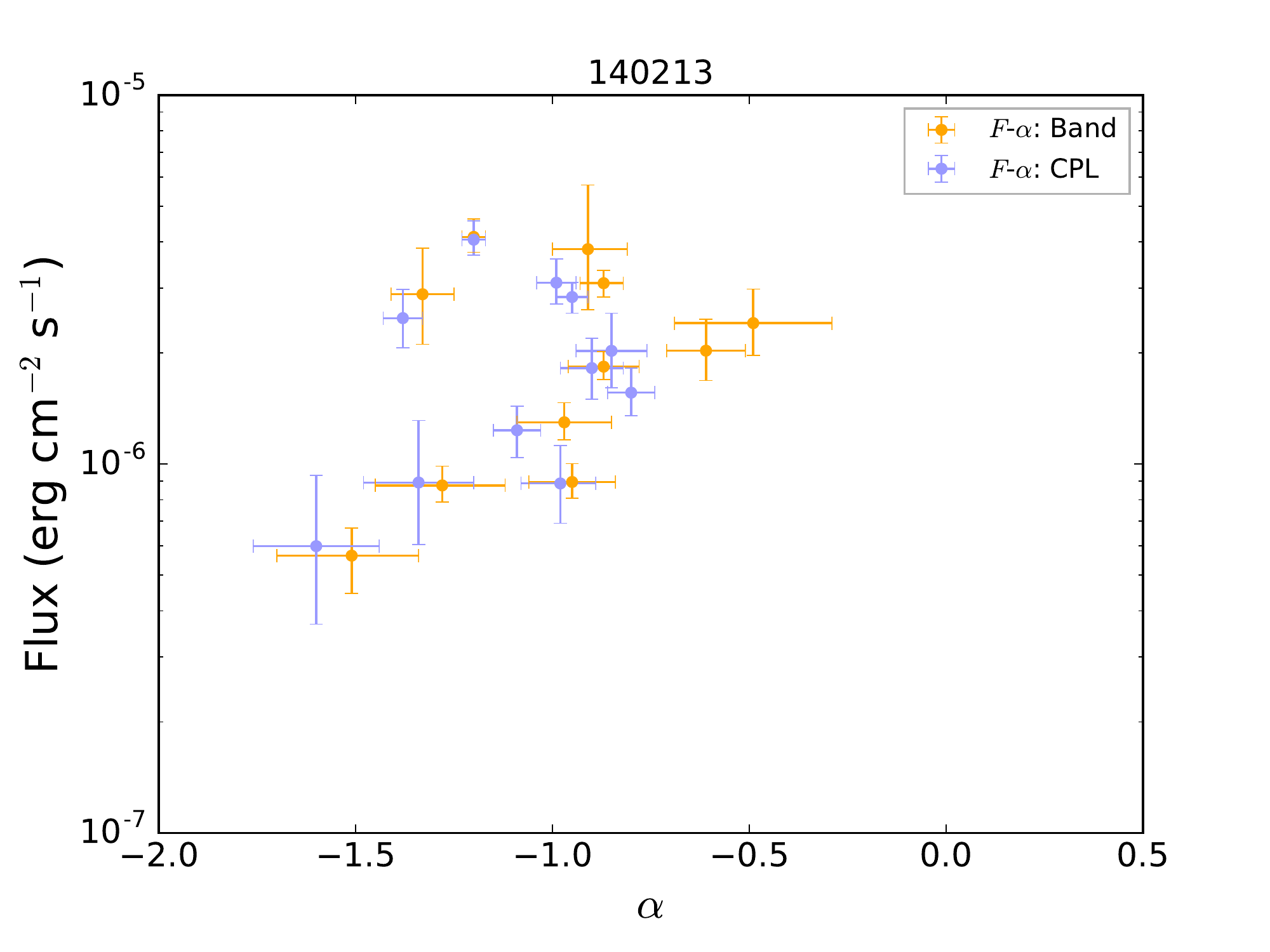}
\includegraphics[angle=0,scale=0.3]{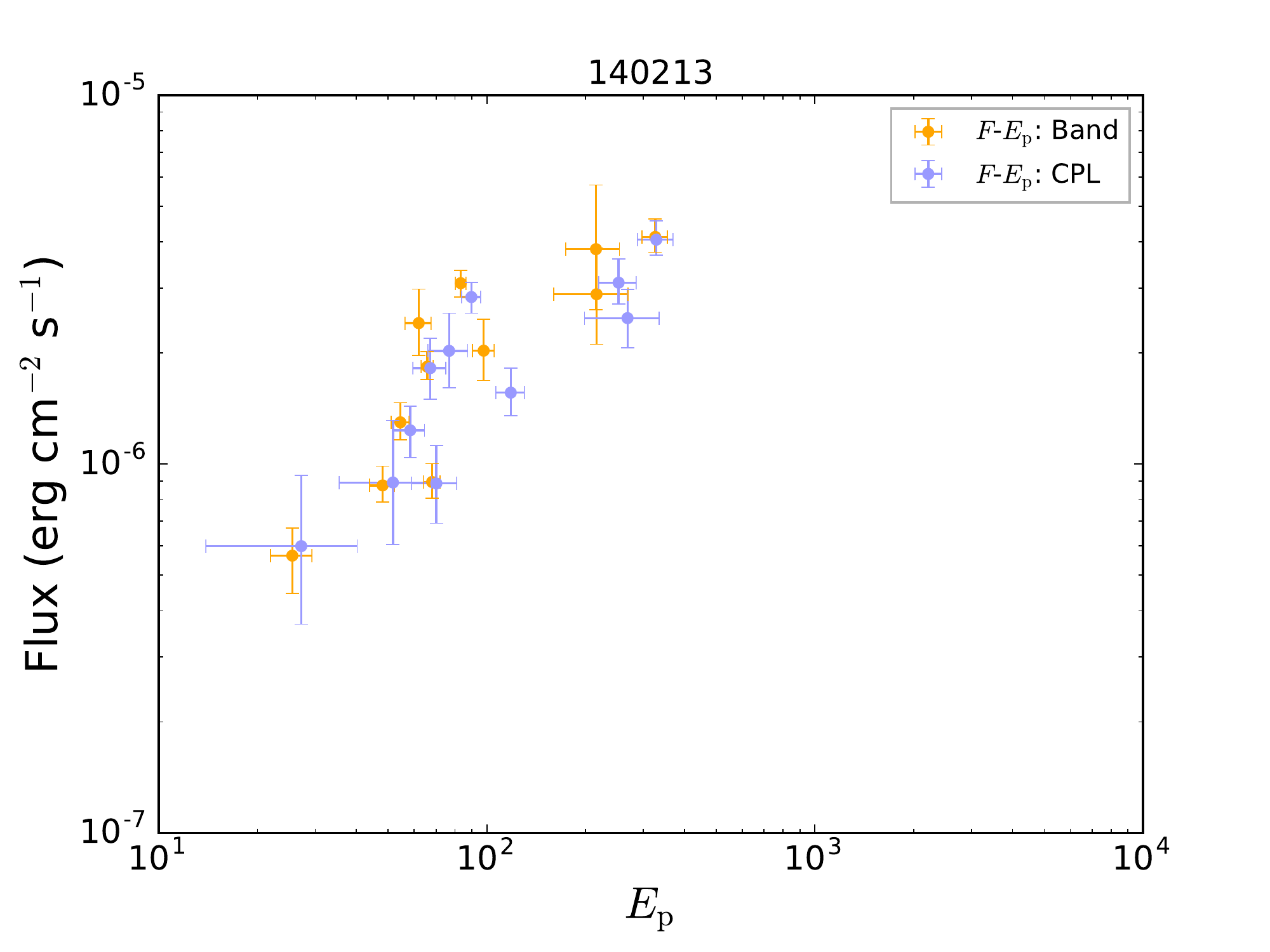}
\includegraphics[angle=0,scale=0.3]{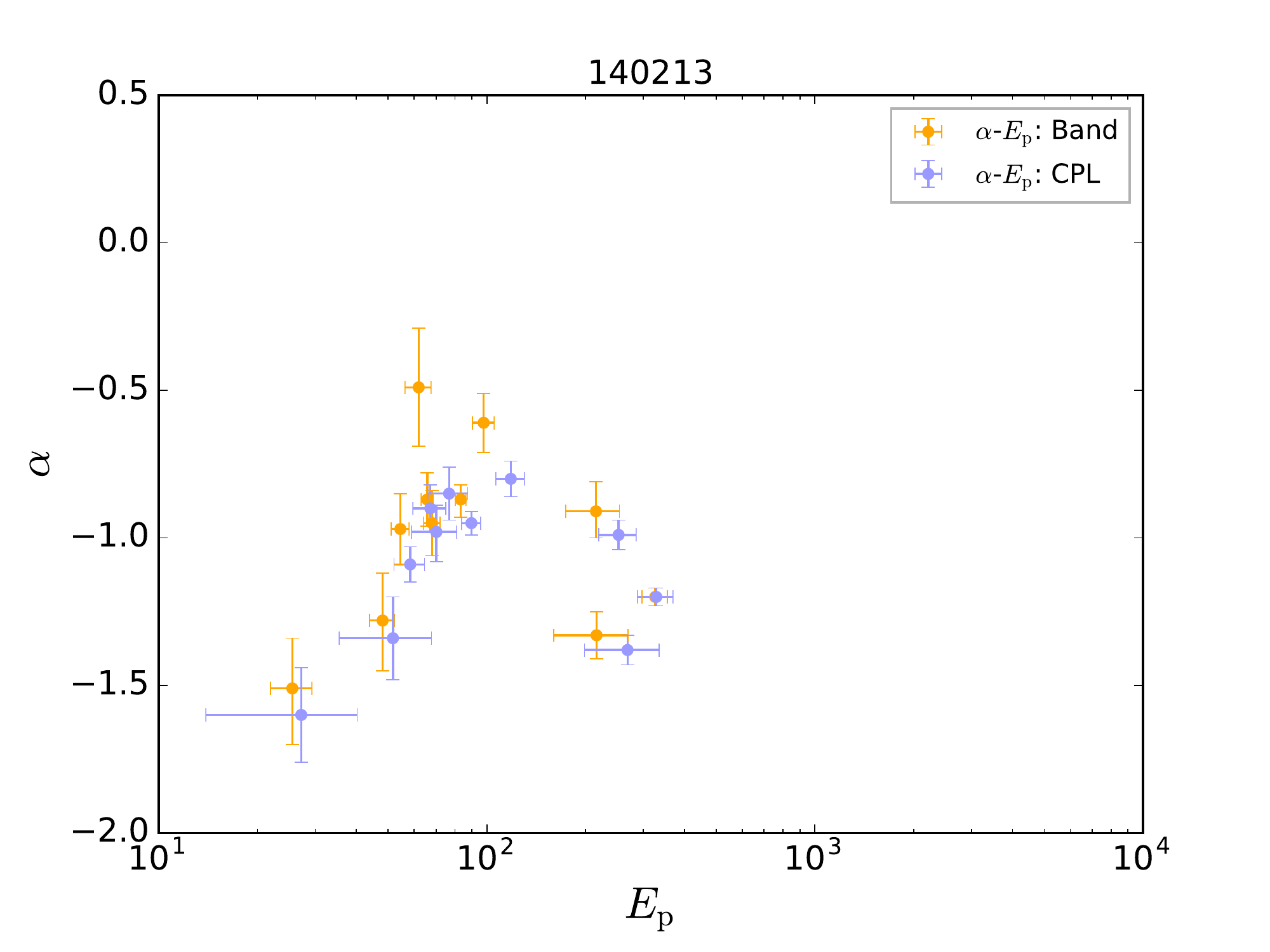}
\includegraphics[angle=0,scale=0.3]{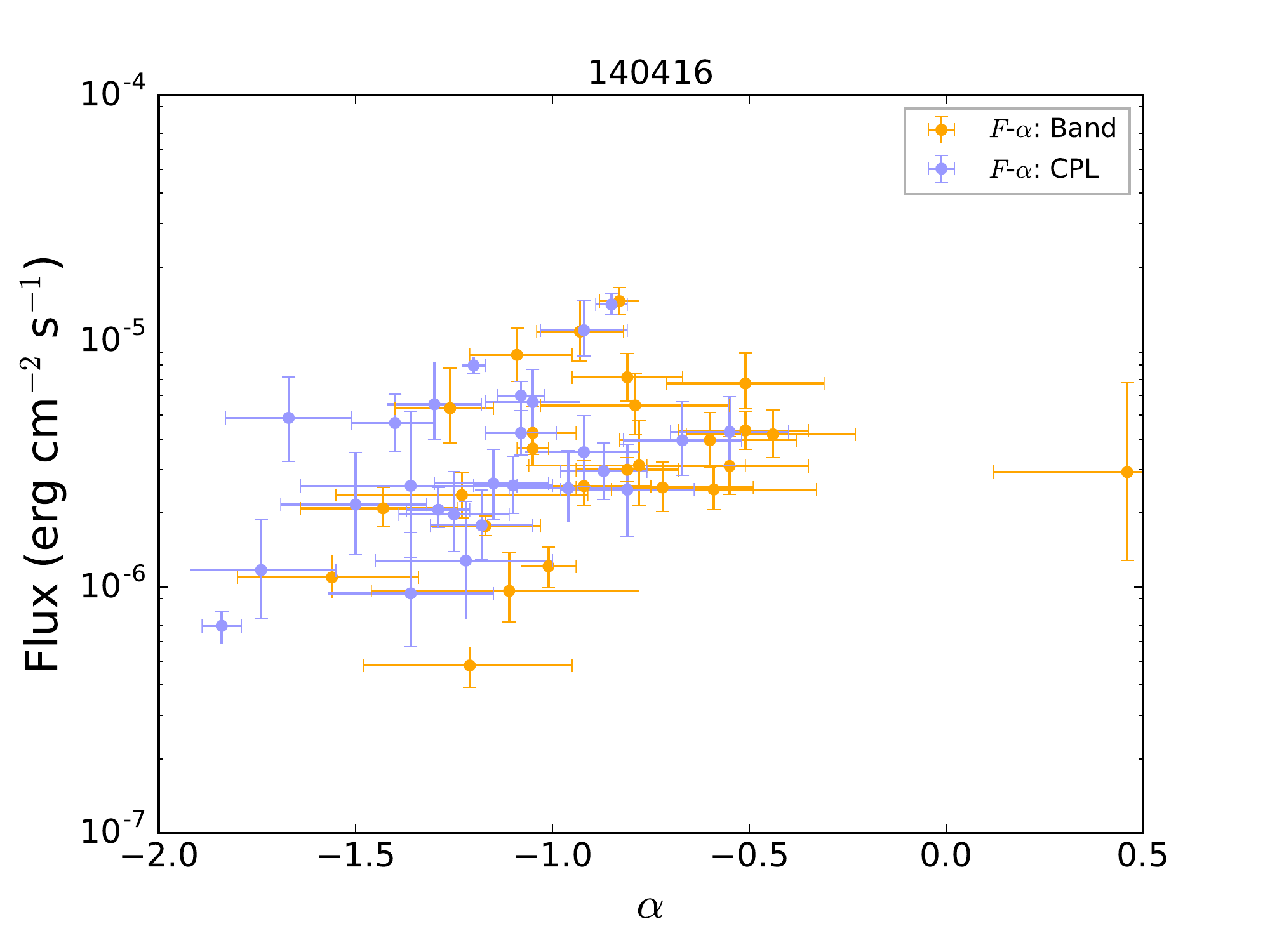}
\includegraphics[angle=0,scale=0.3]{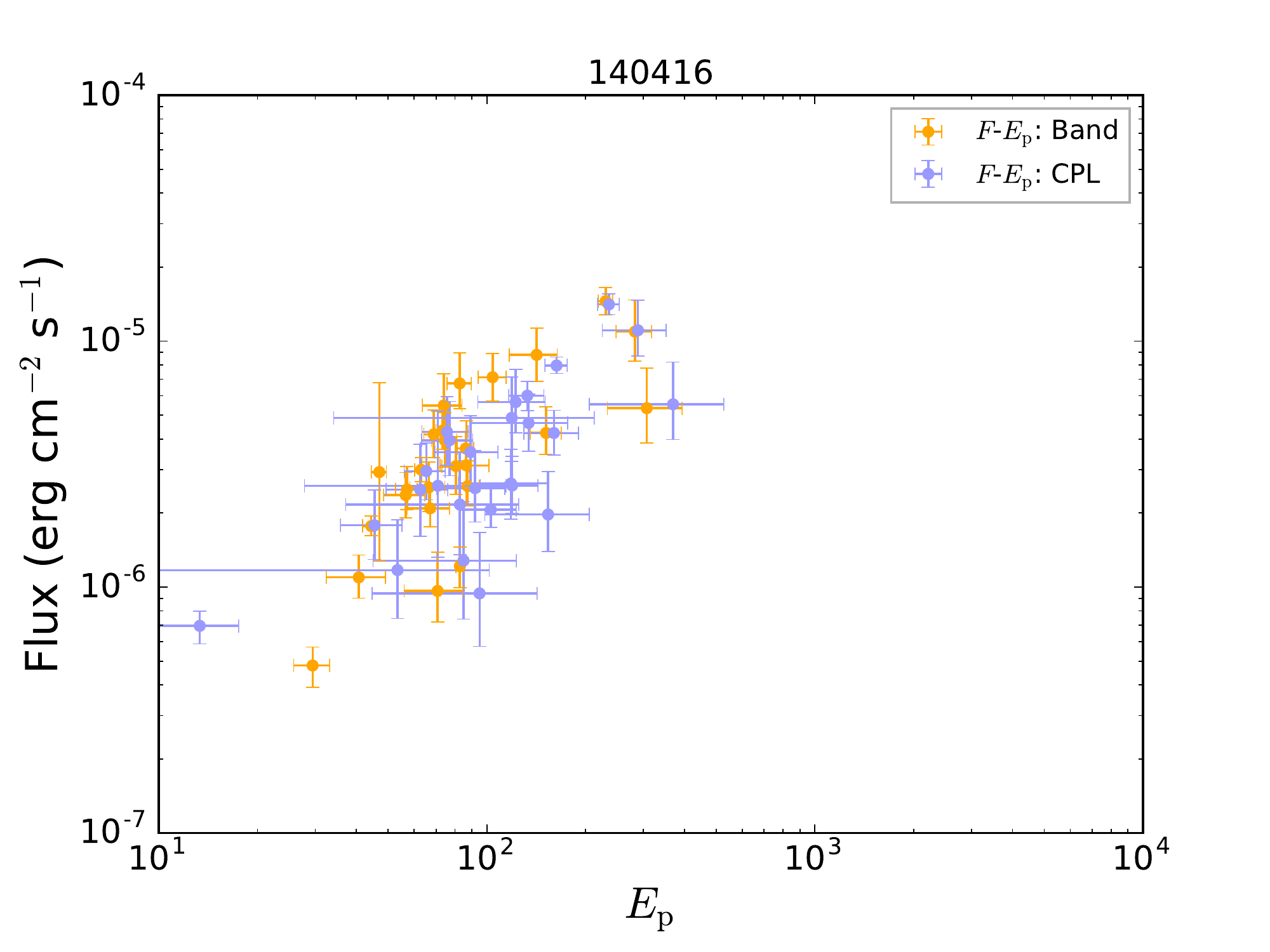}
\includegraphics[angle=0,scale=0.3]{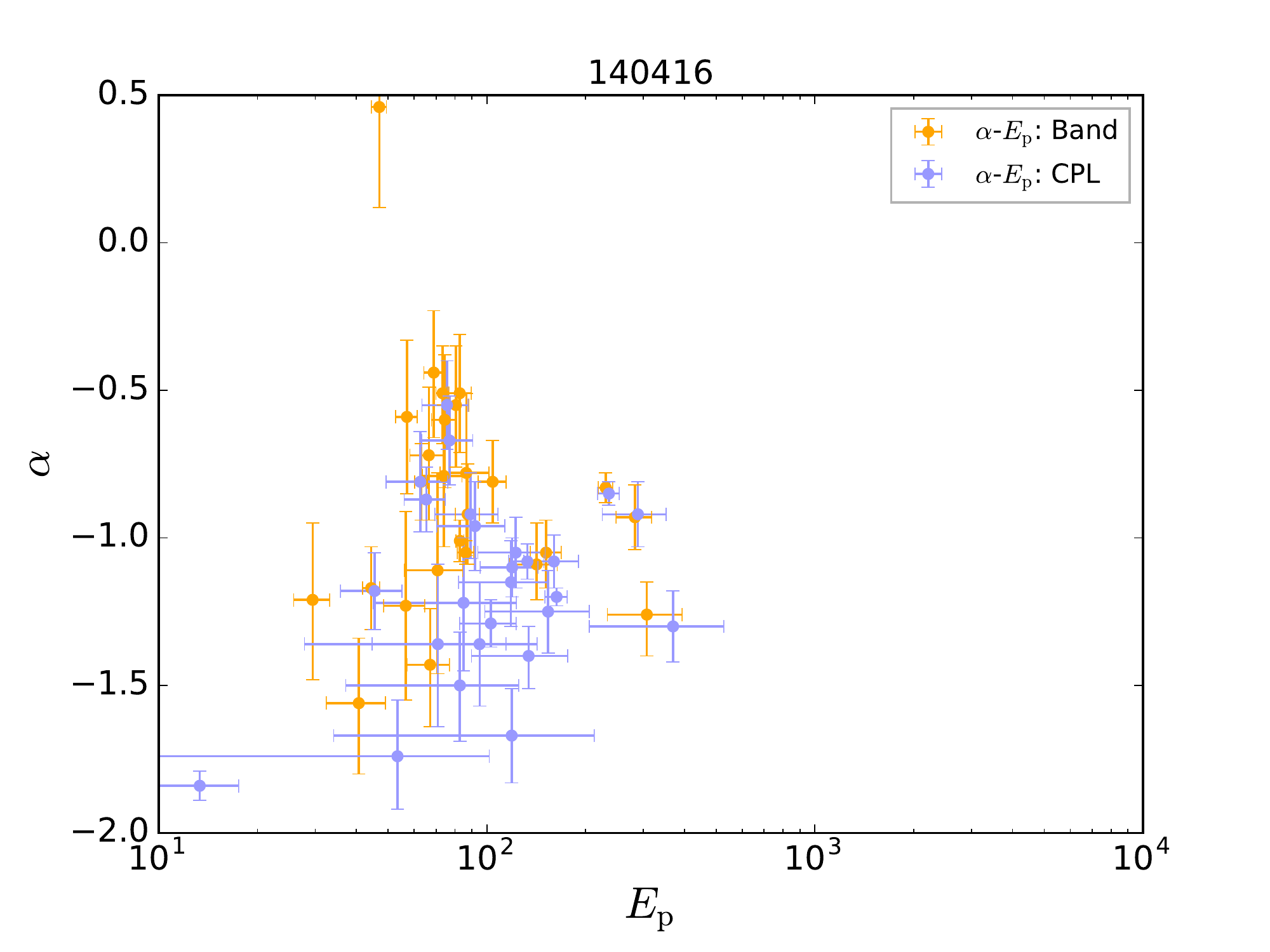}
\includegraphics[angle=0,scale=0.3]{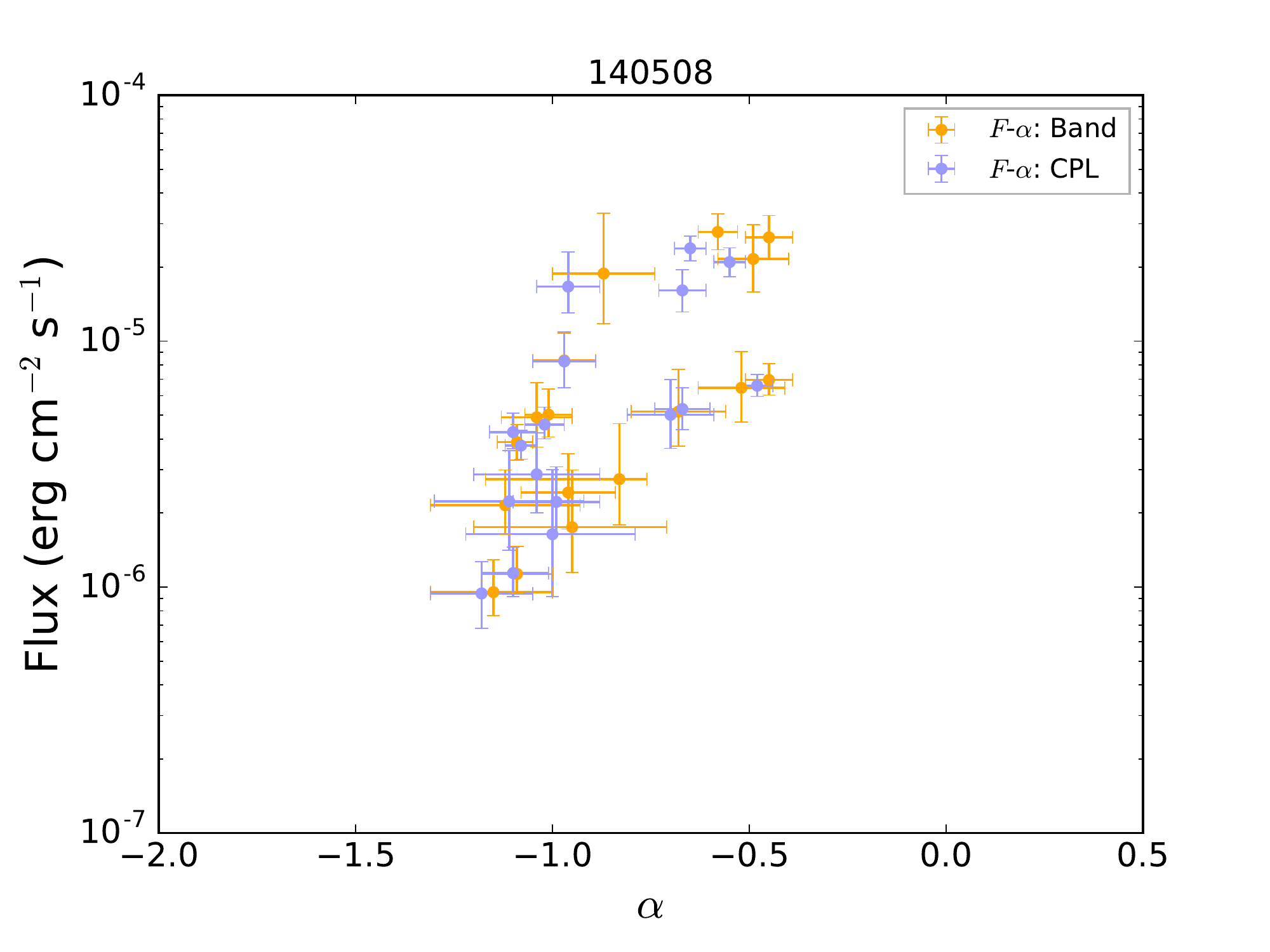}
\includegraphics[angle=0,scale=0.3]{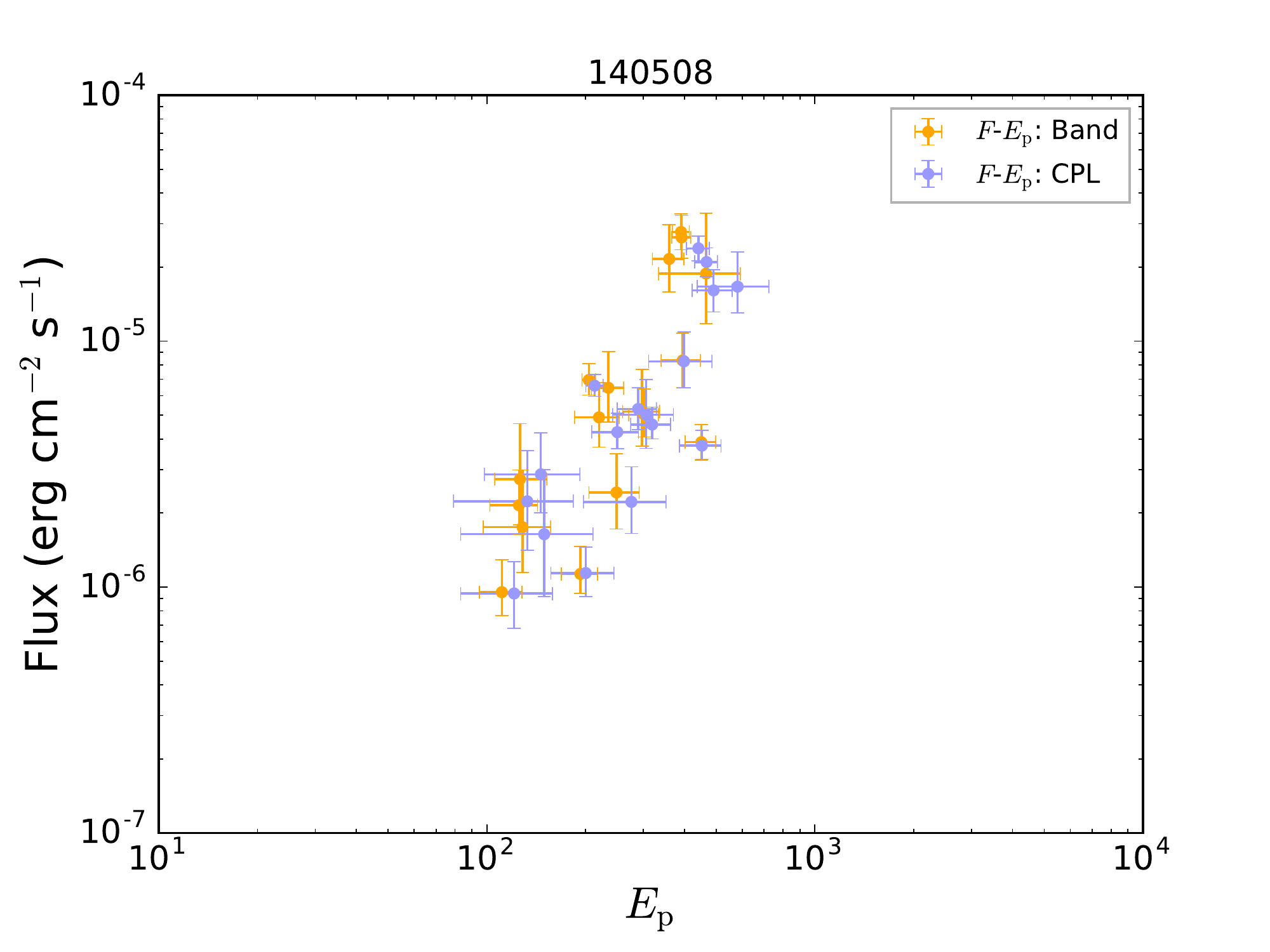}
\includegraphics[angle=0,scale=0.3]{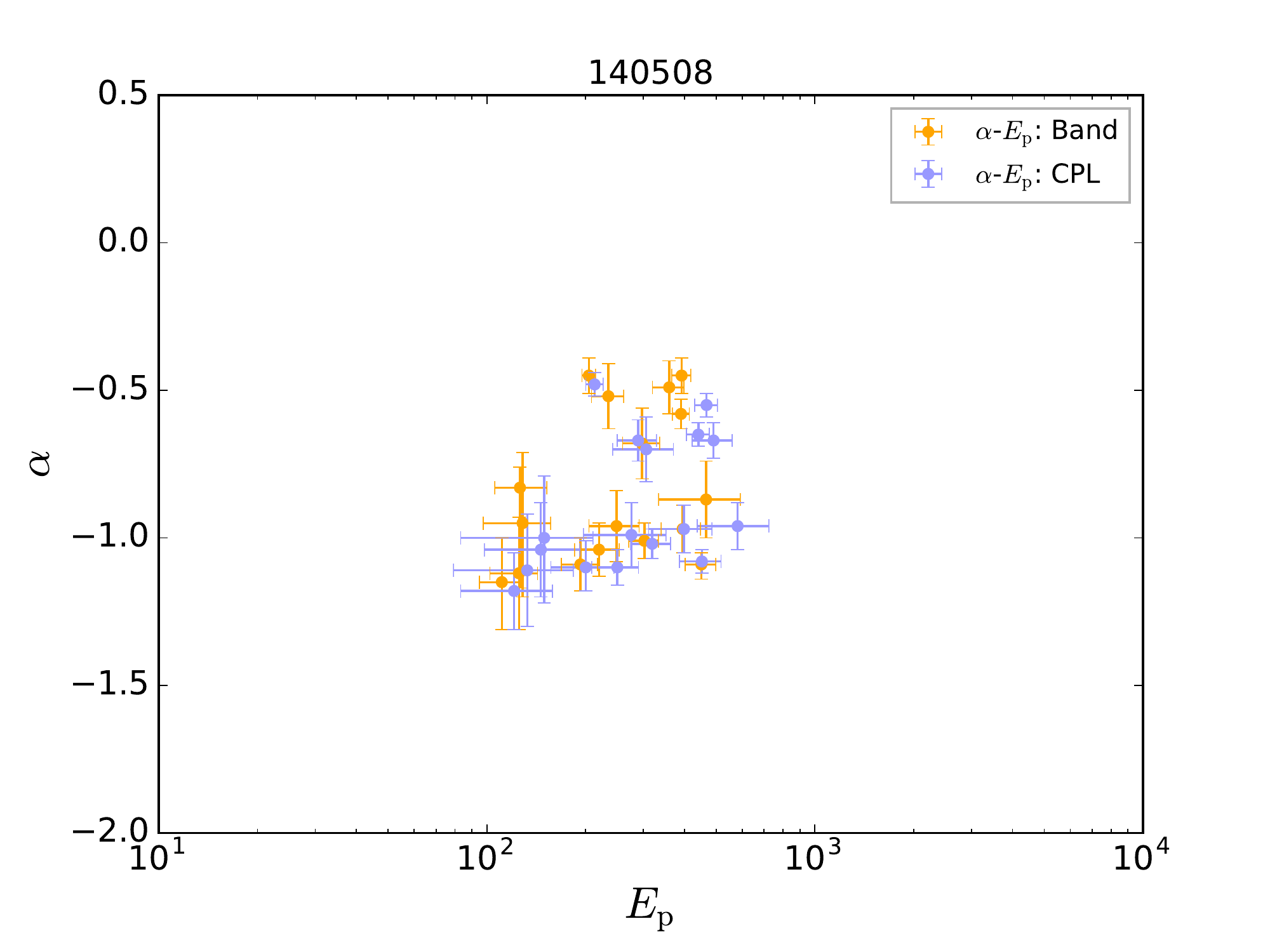}
\includegraphics[angle=0,scale=0.3]{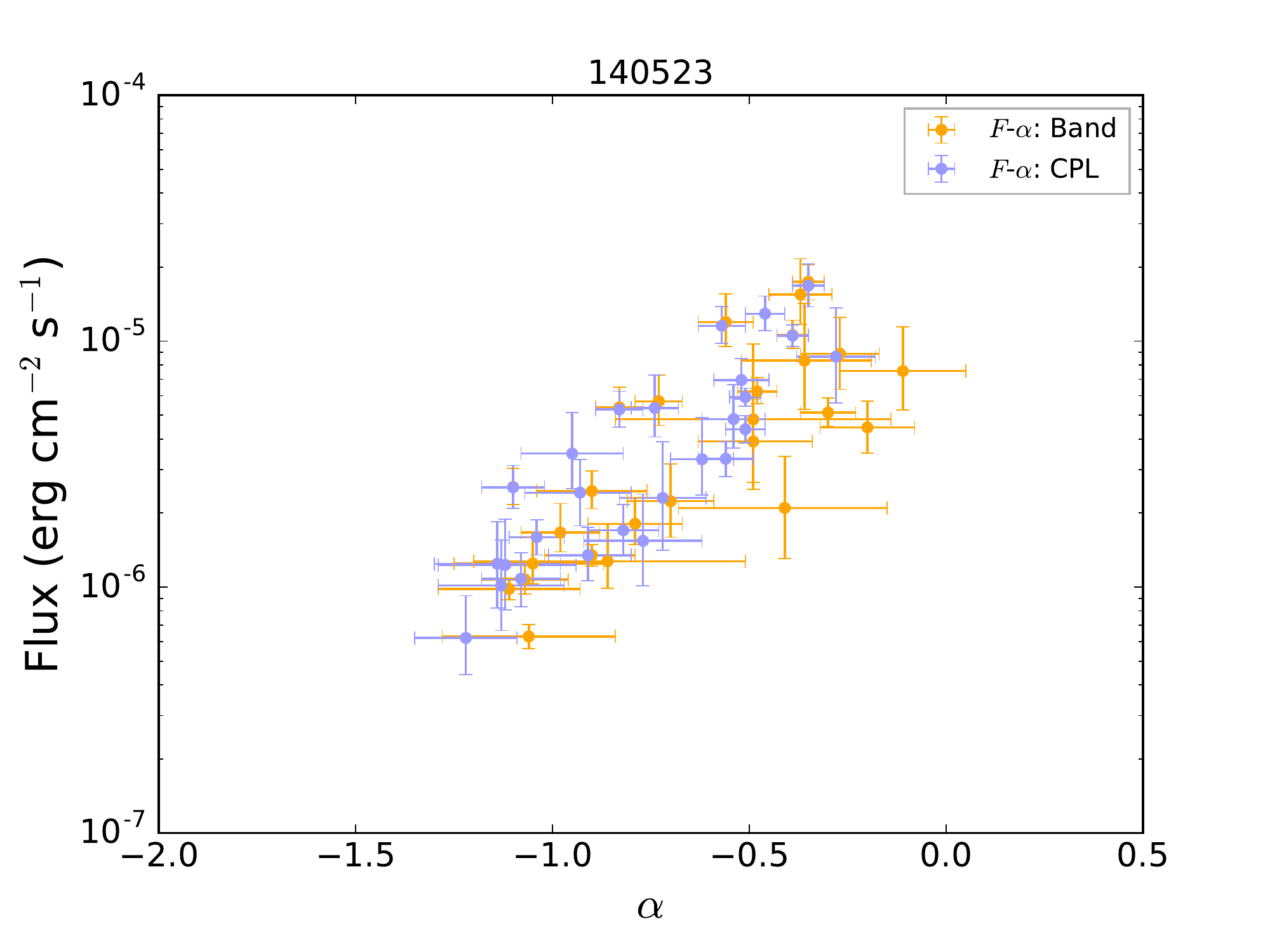}
\includegraphics[angle=0,scale=0.3]{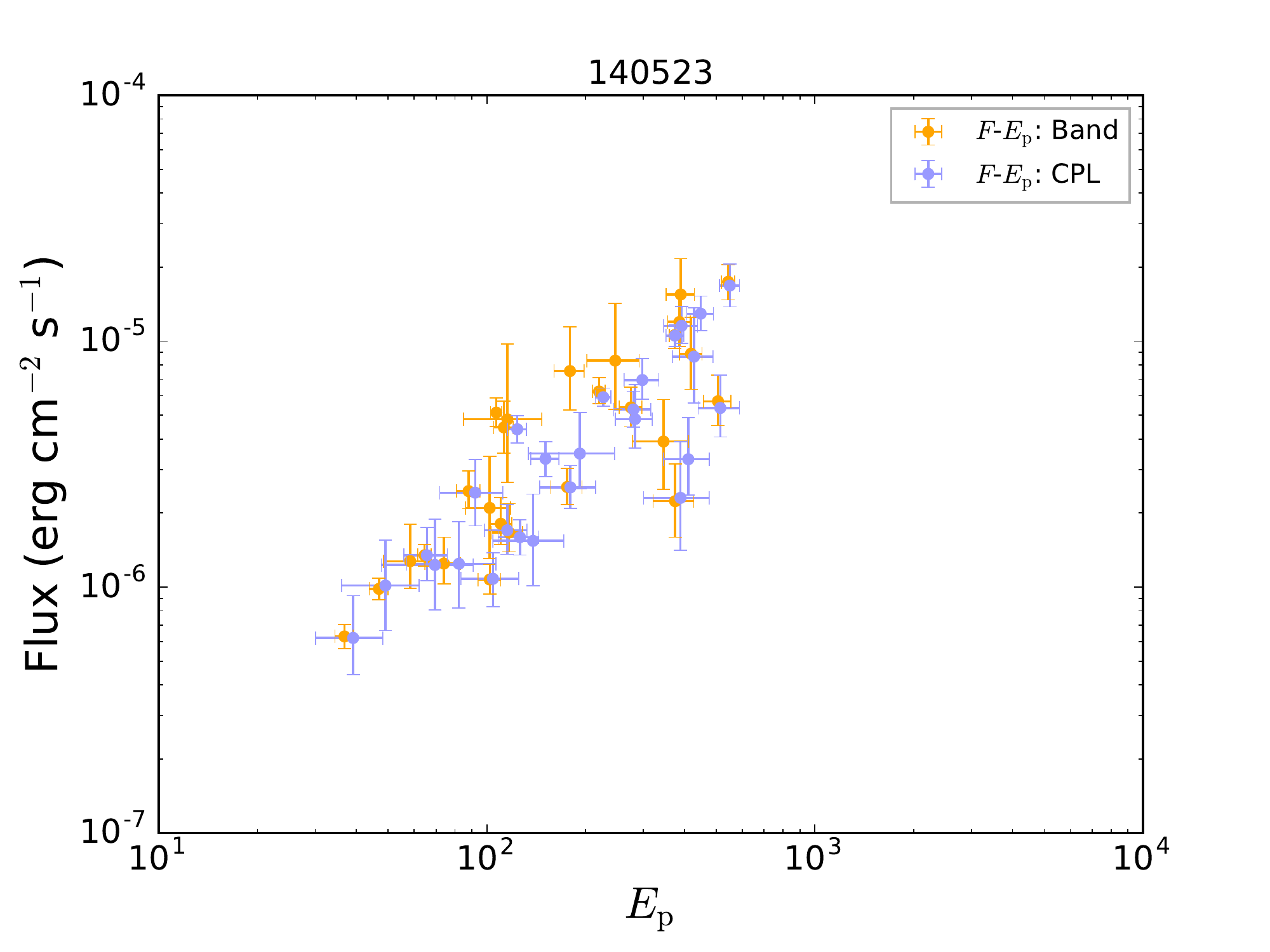}
\includegraphics[angle=0,scale=0.3]{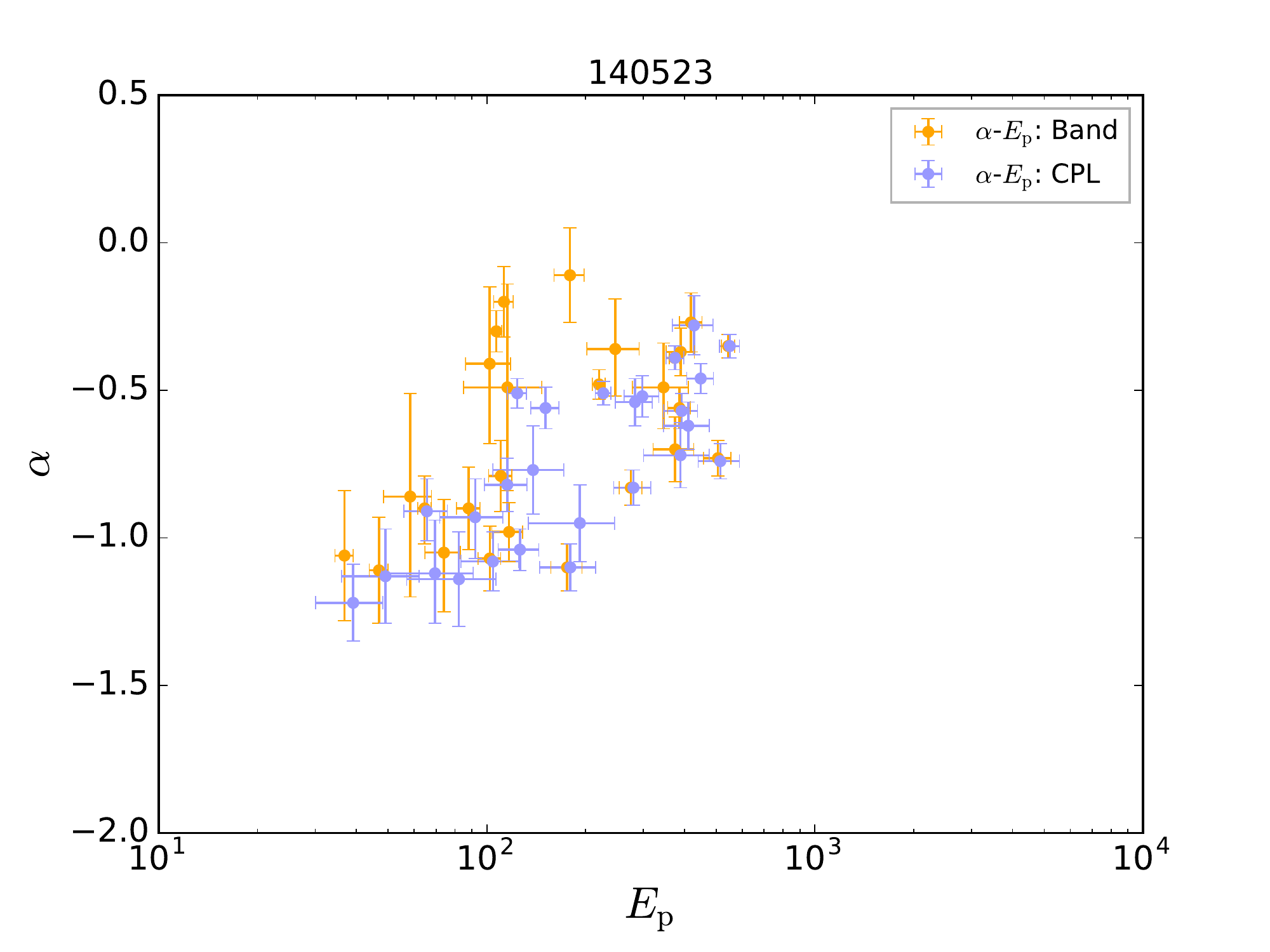}
\includegraphics[angle=0,scale=0.3]{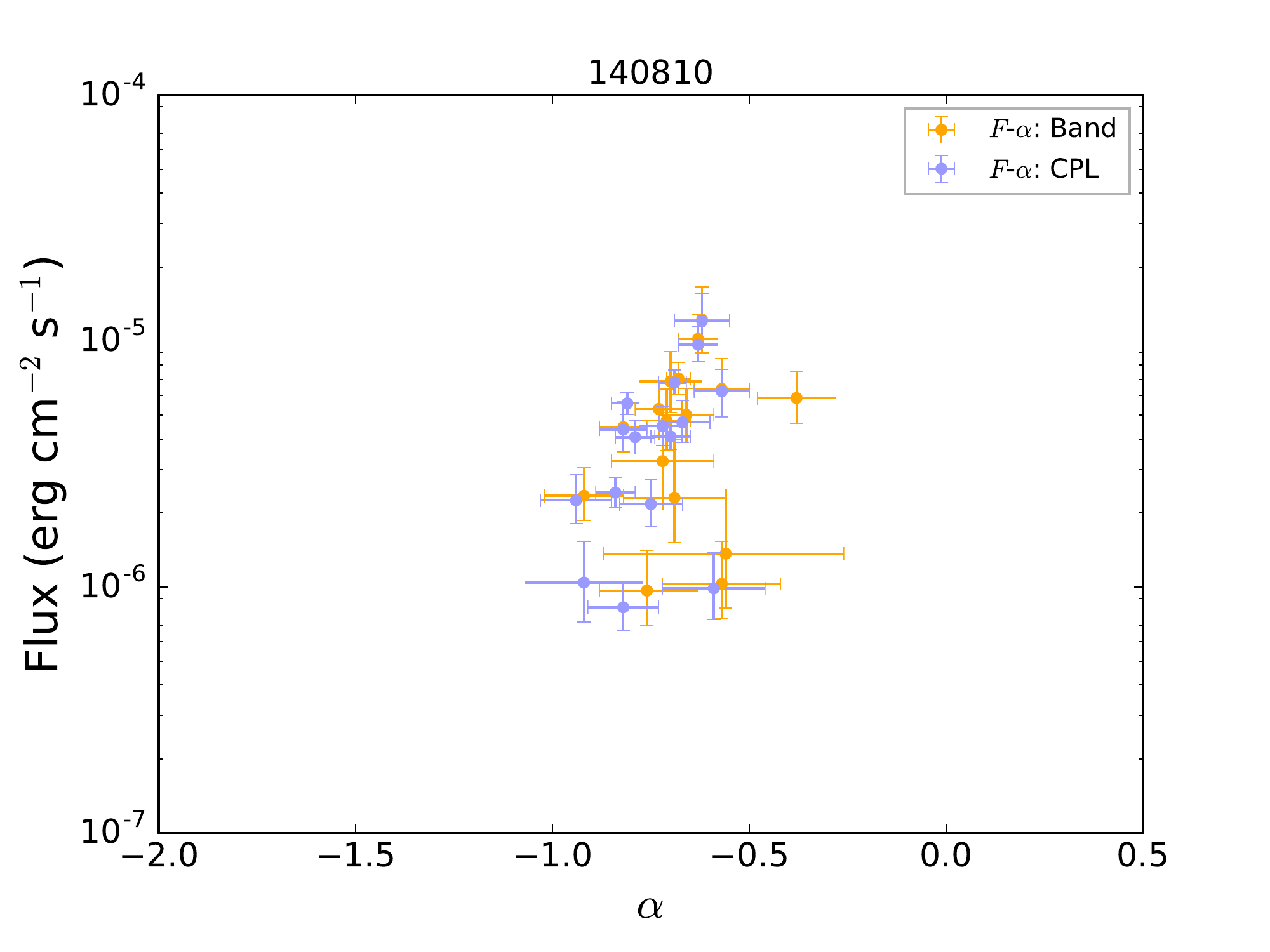}
\includegraphics[angle=0,scale=0.3]{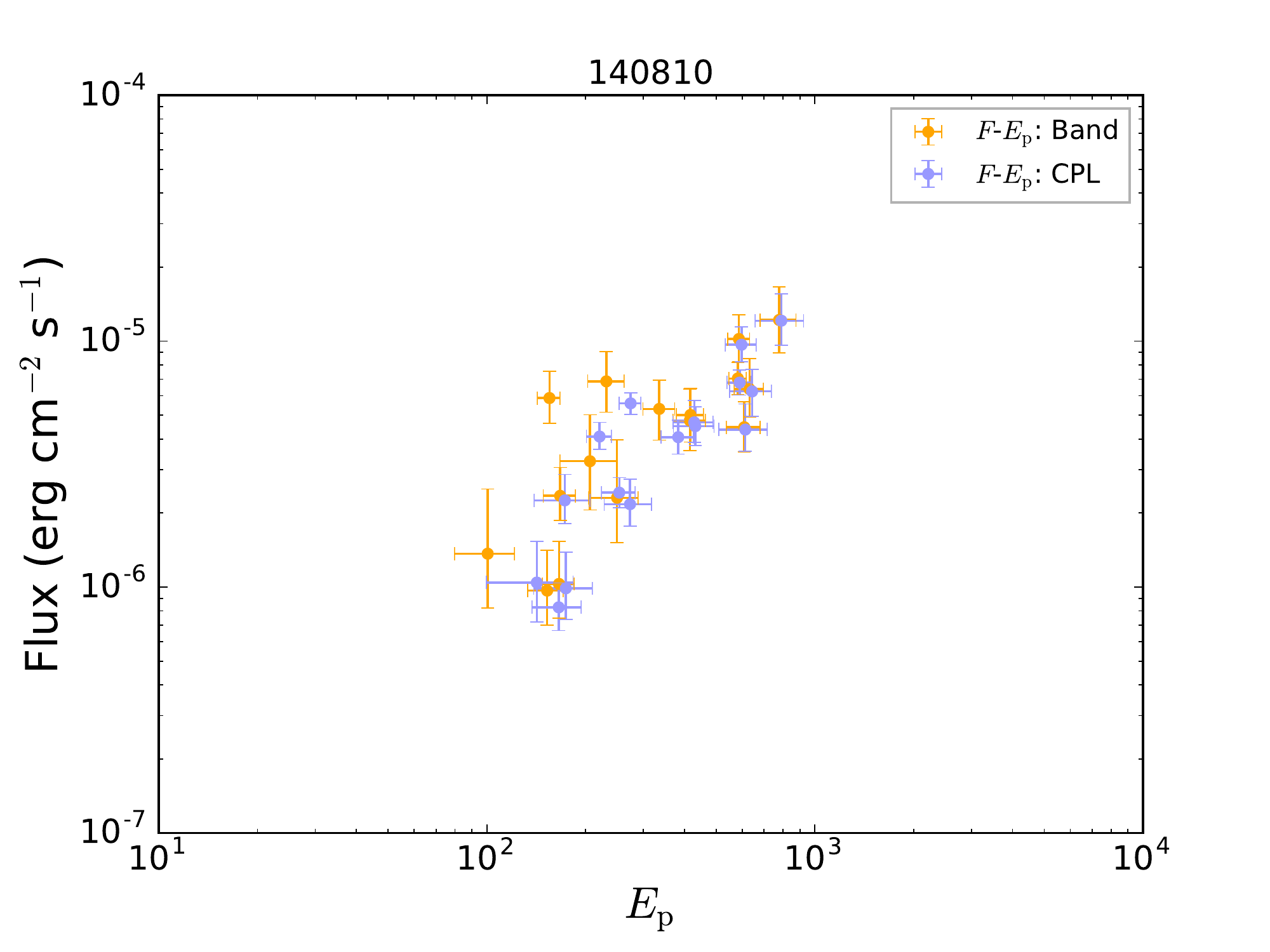}
\includegraphics[angle=0,scale=0.3]{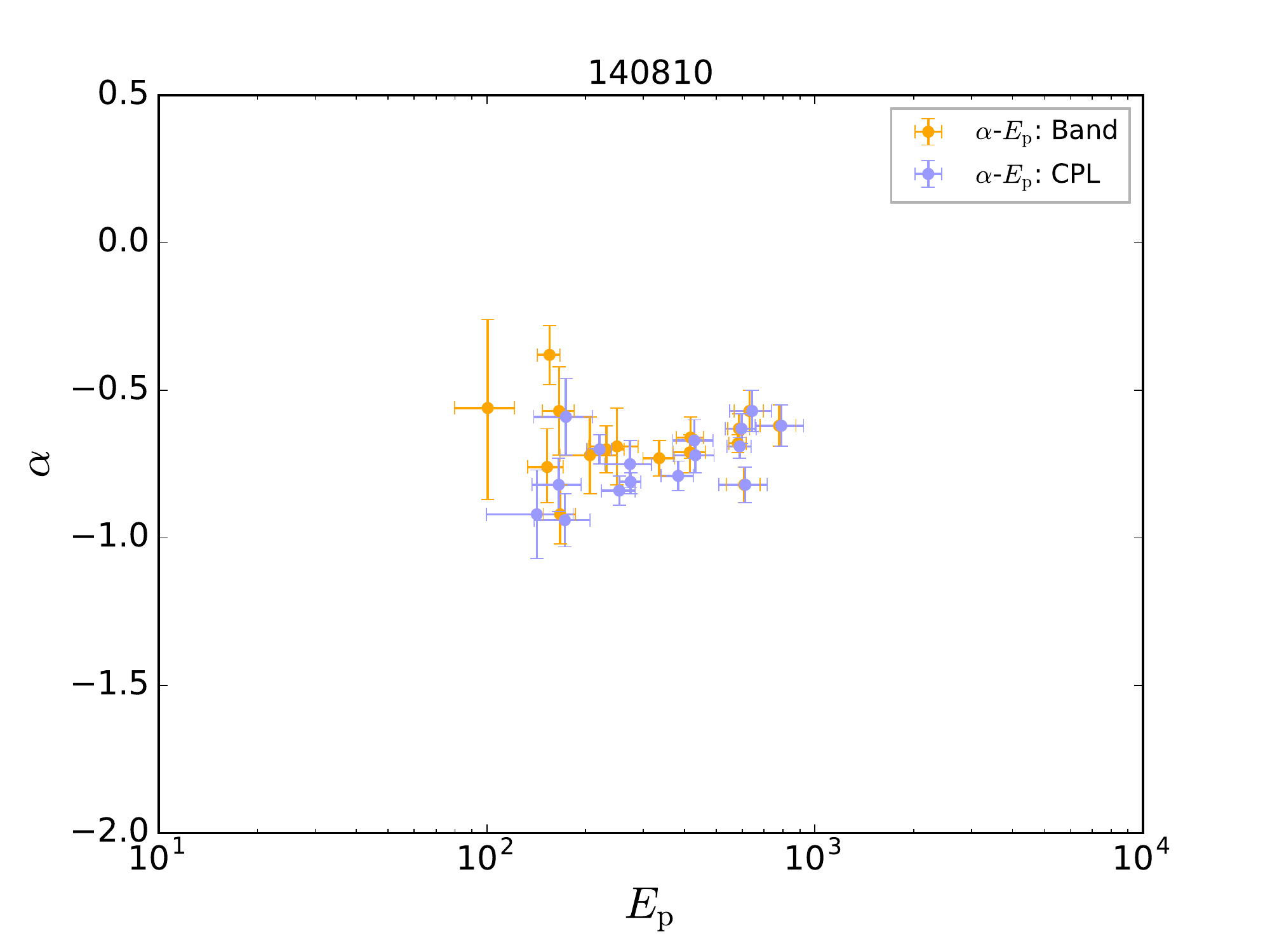}
\center{Fig. \ref{fig:relation3}--- Continued}
\end{figure*}
\begin{figure*}
\includegraphics[angle=0,scale=0.3]{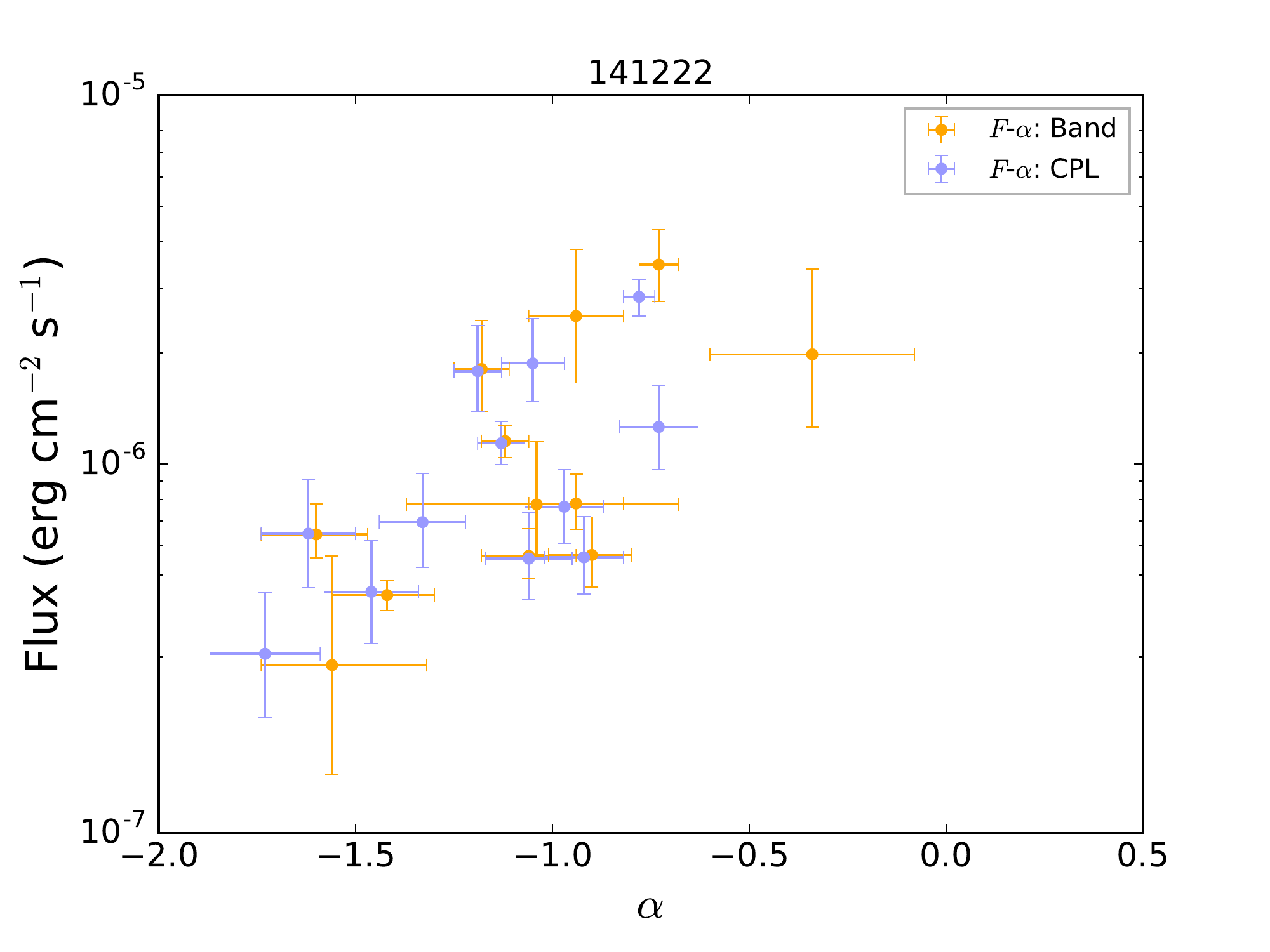}
\includegraphics[angle=0,scale=0.3]{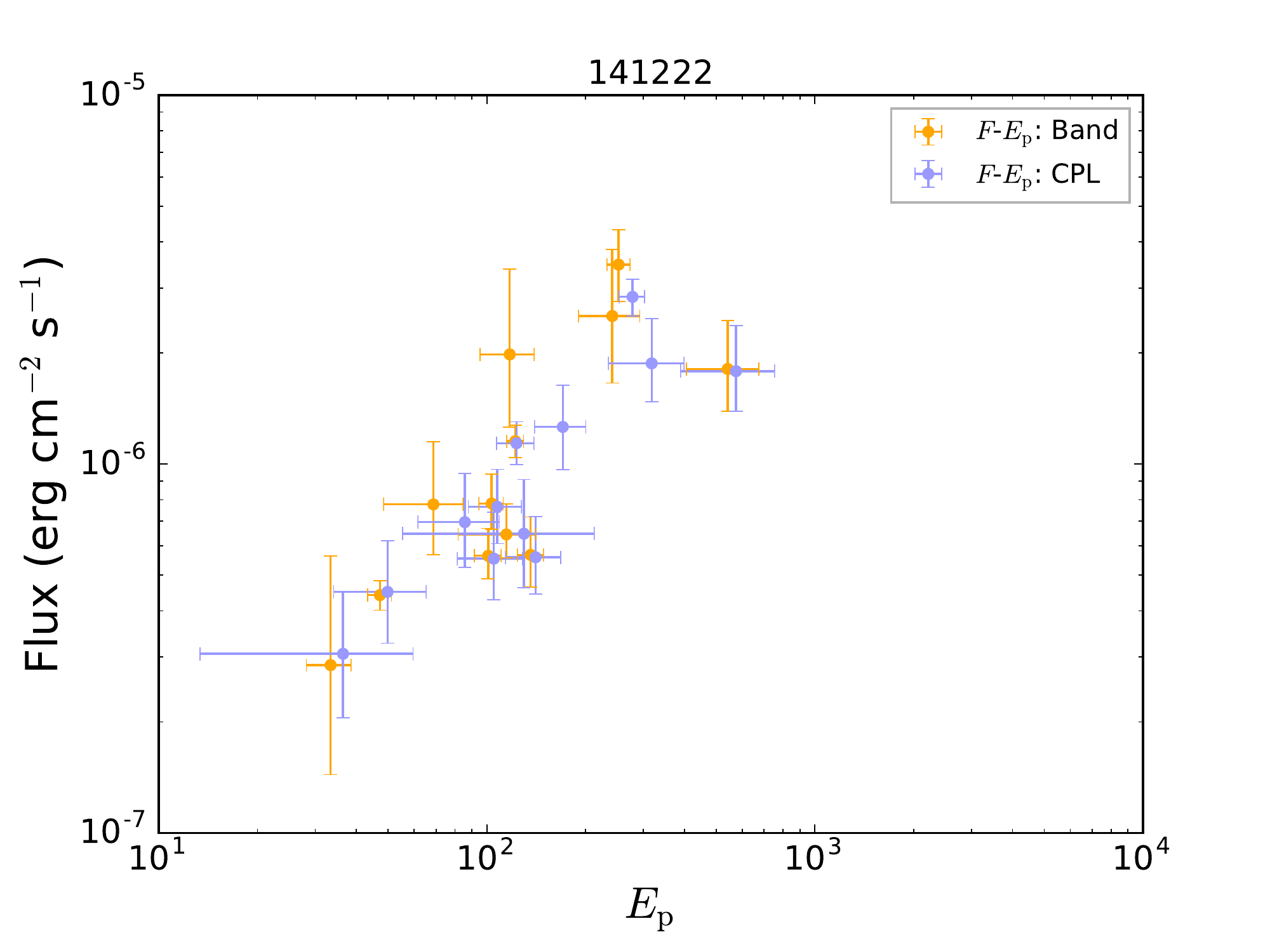}
\includegraphics[angle=0,scale=0.3]{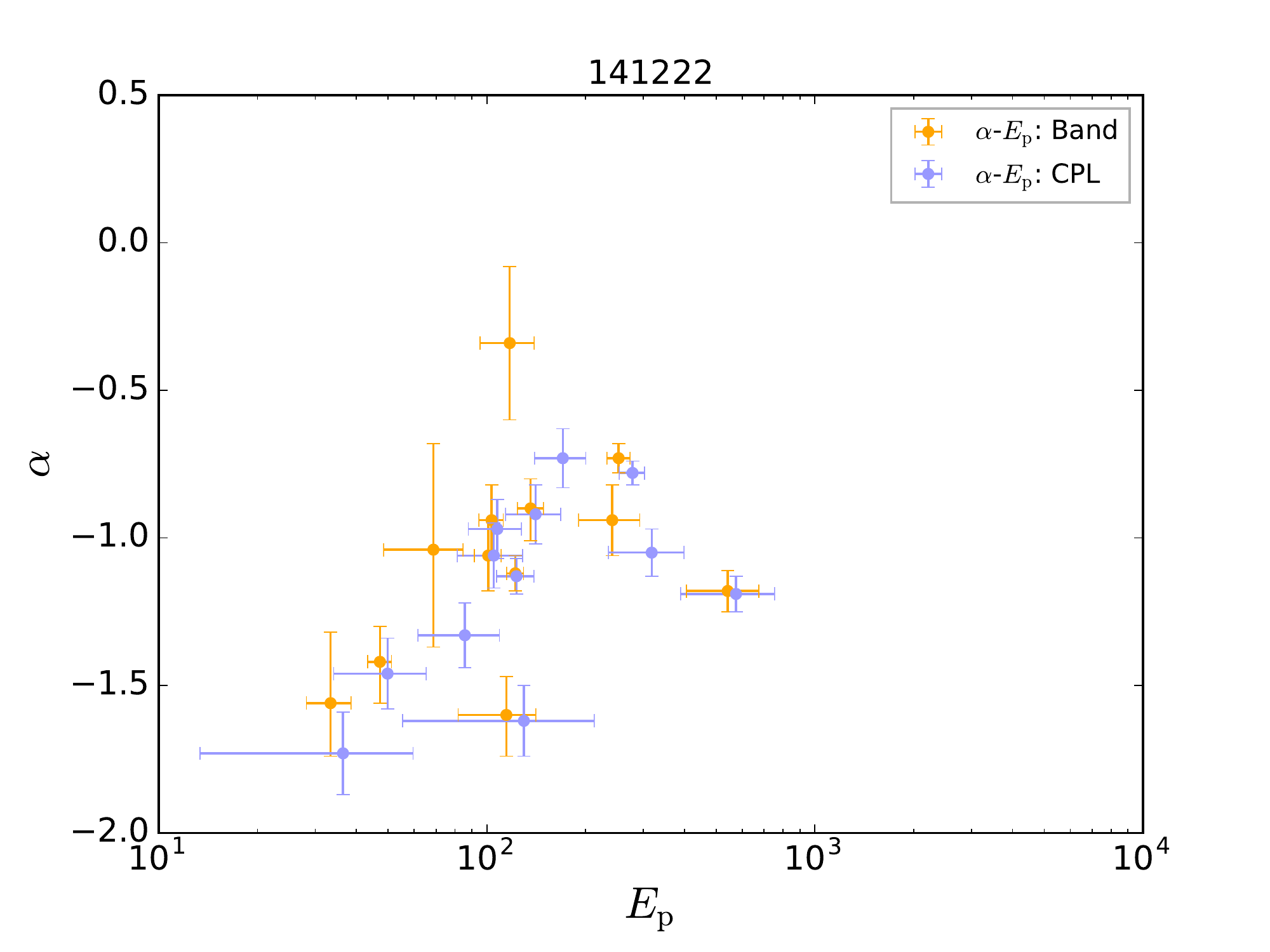}
\includegraphics[angle=0,scale=0.3]{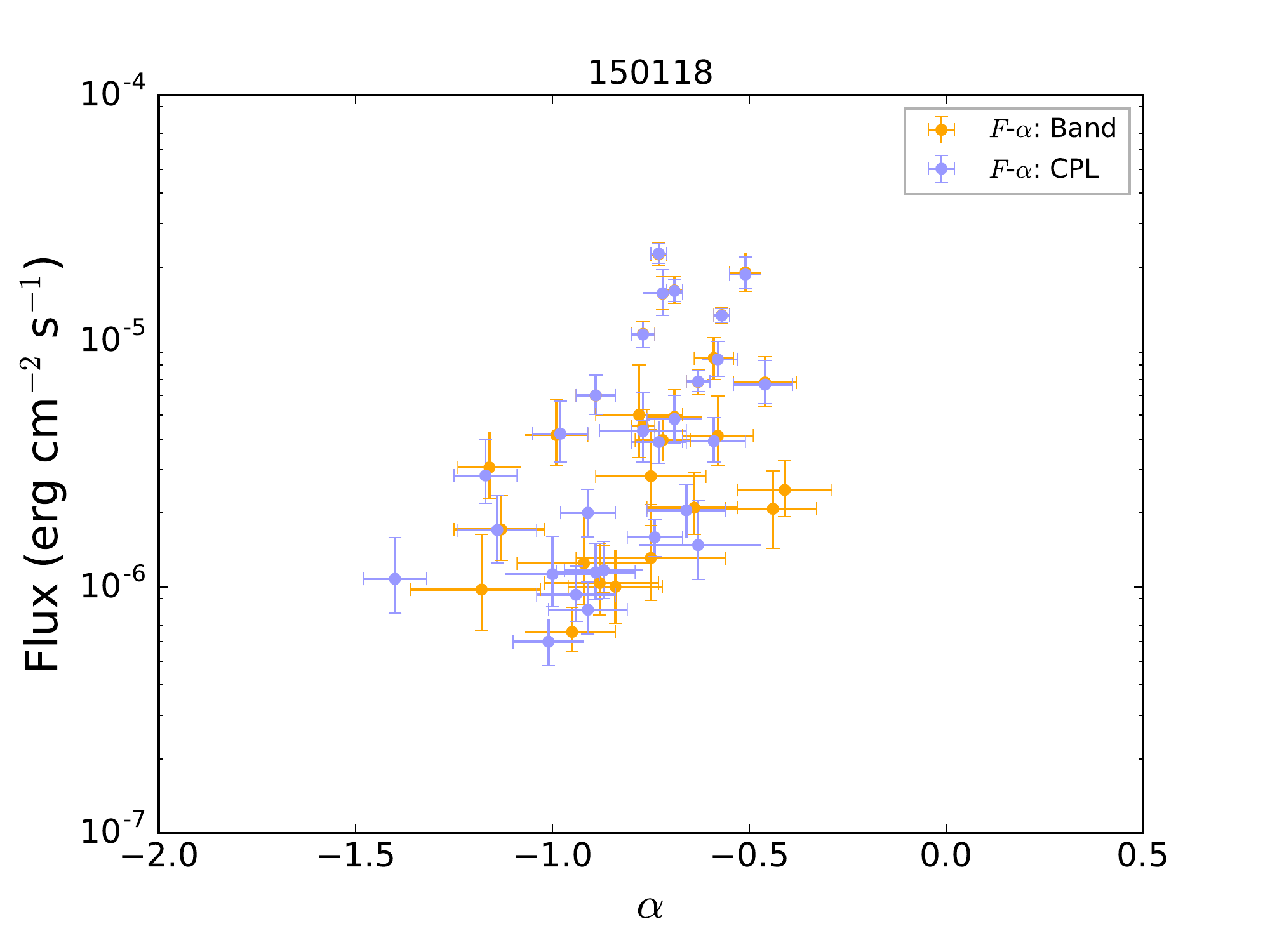}
\includegraphics[angle=0,scale=0.3]{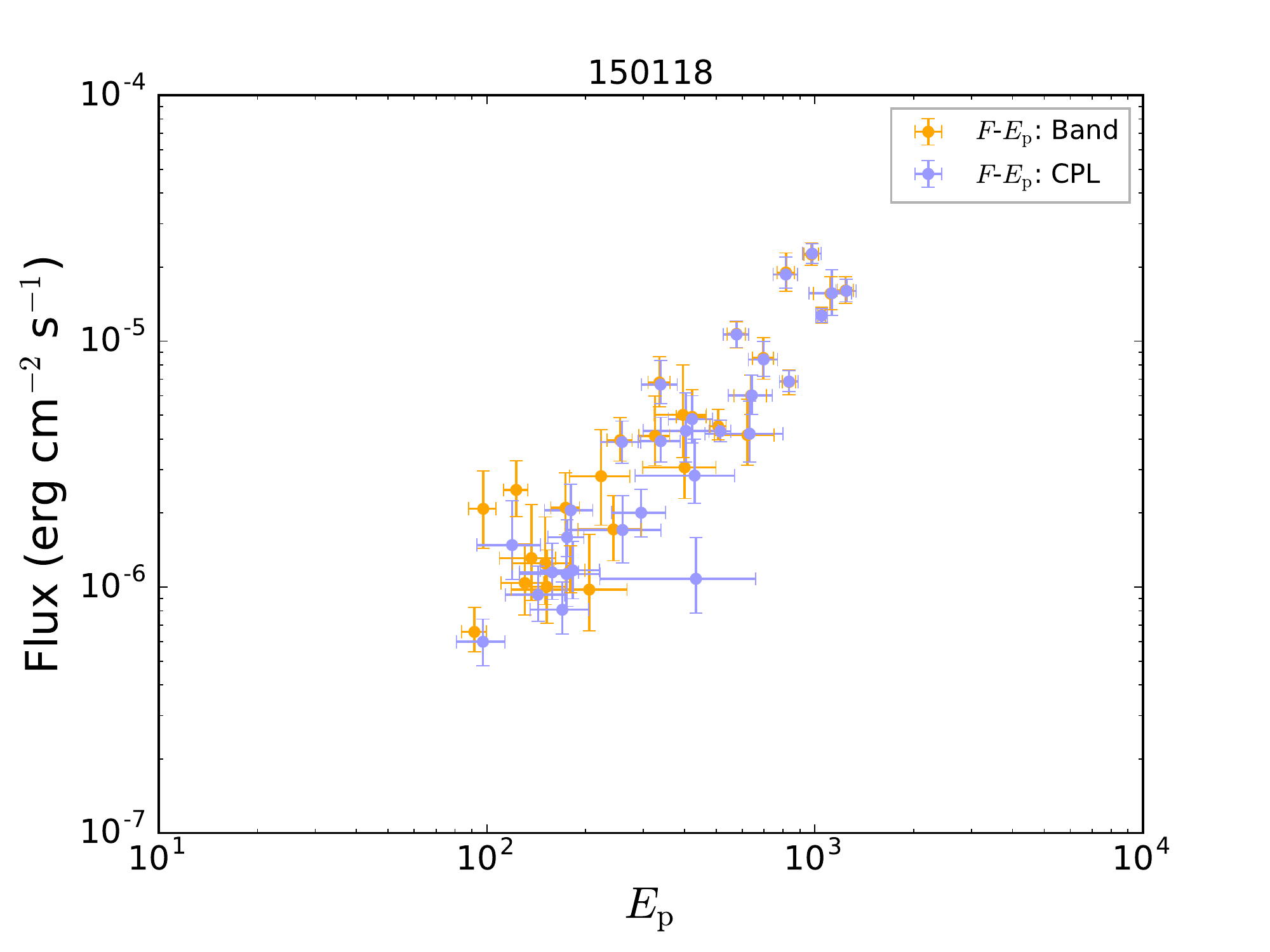}
\includegraphics[angle=0,scale=0.3]{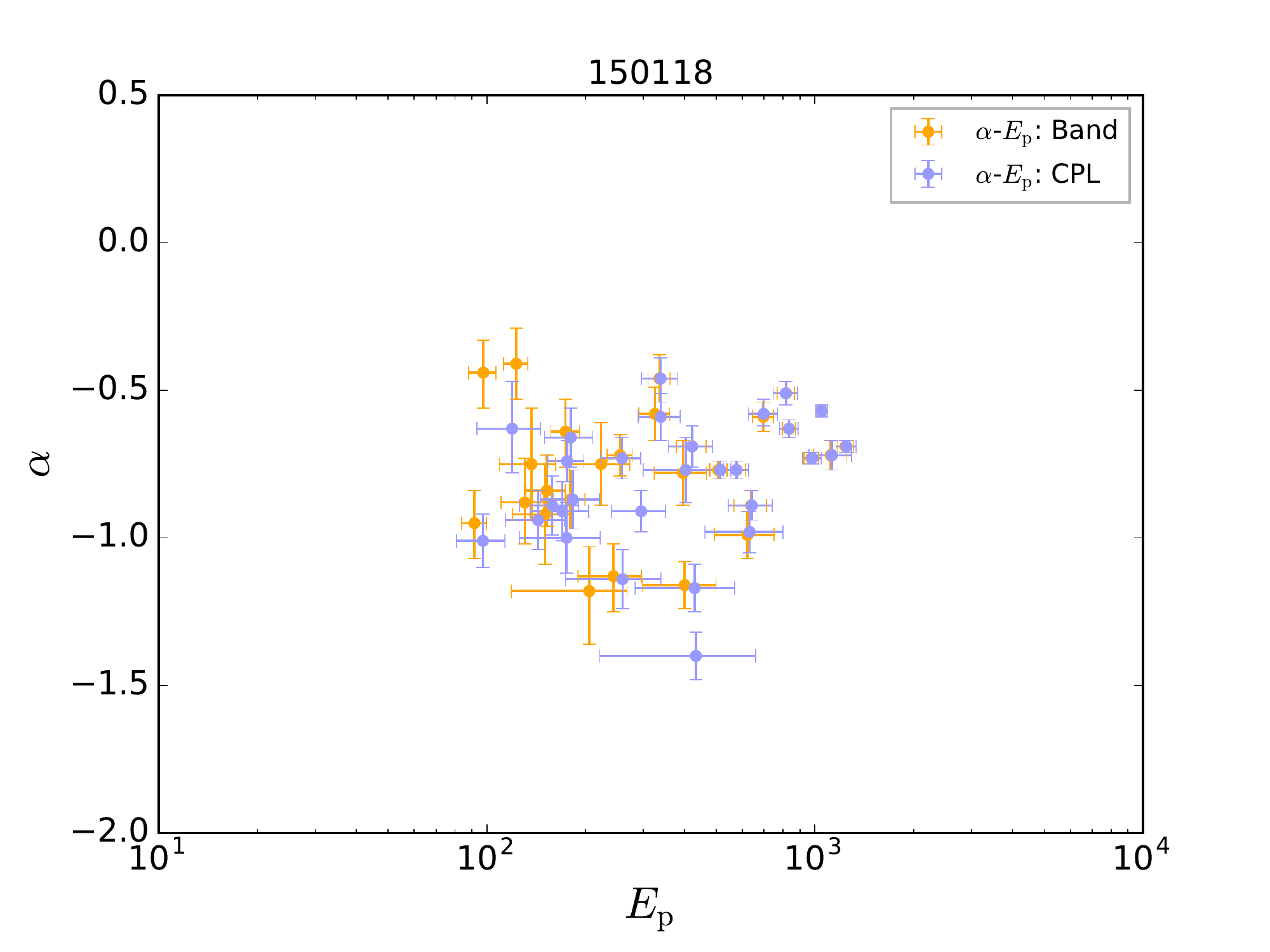}
\includegraphics[angle=0,scale=0.3]{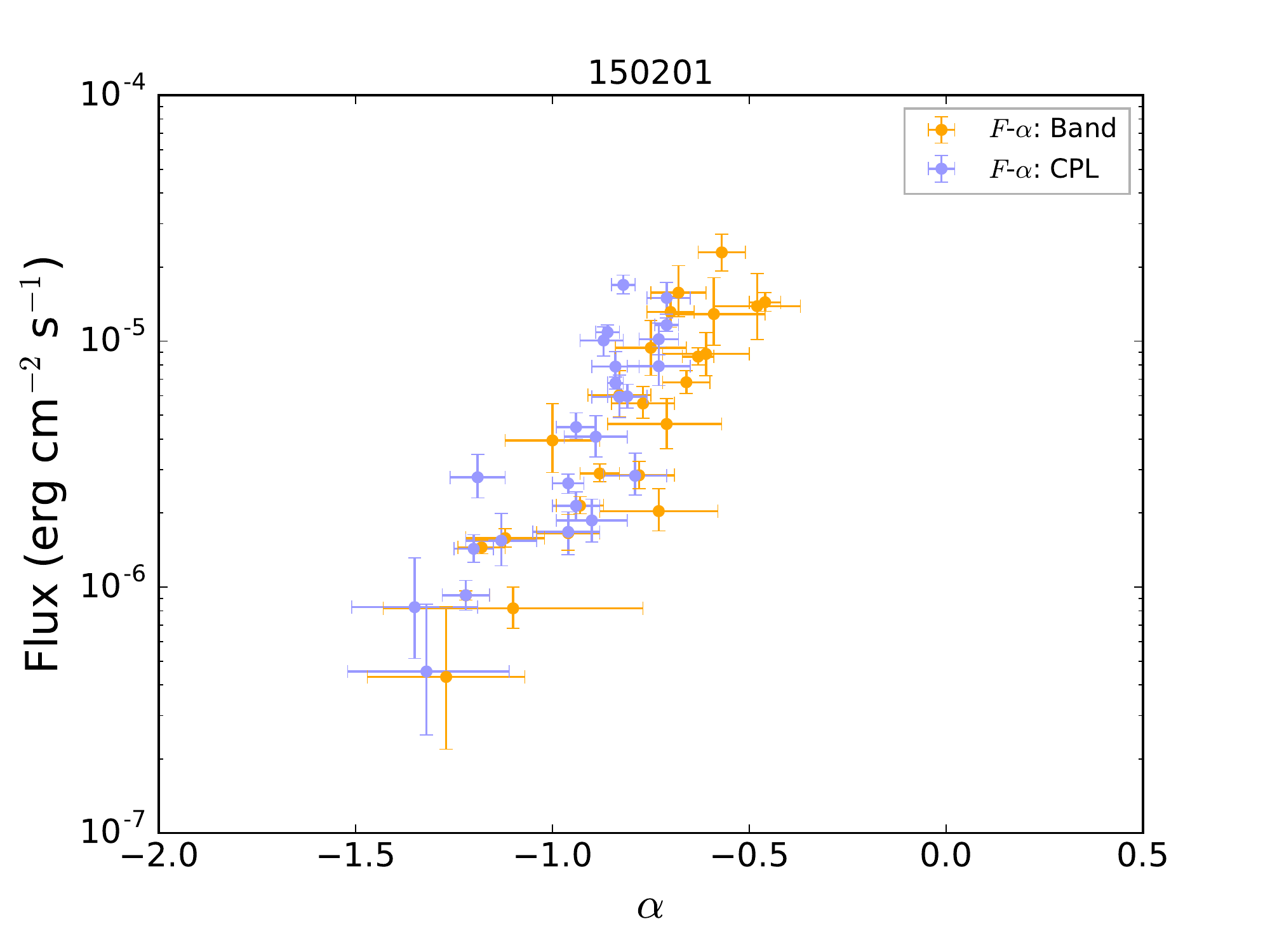}
\includegraphics[angle=0,scale=0.3]{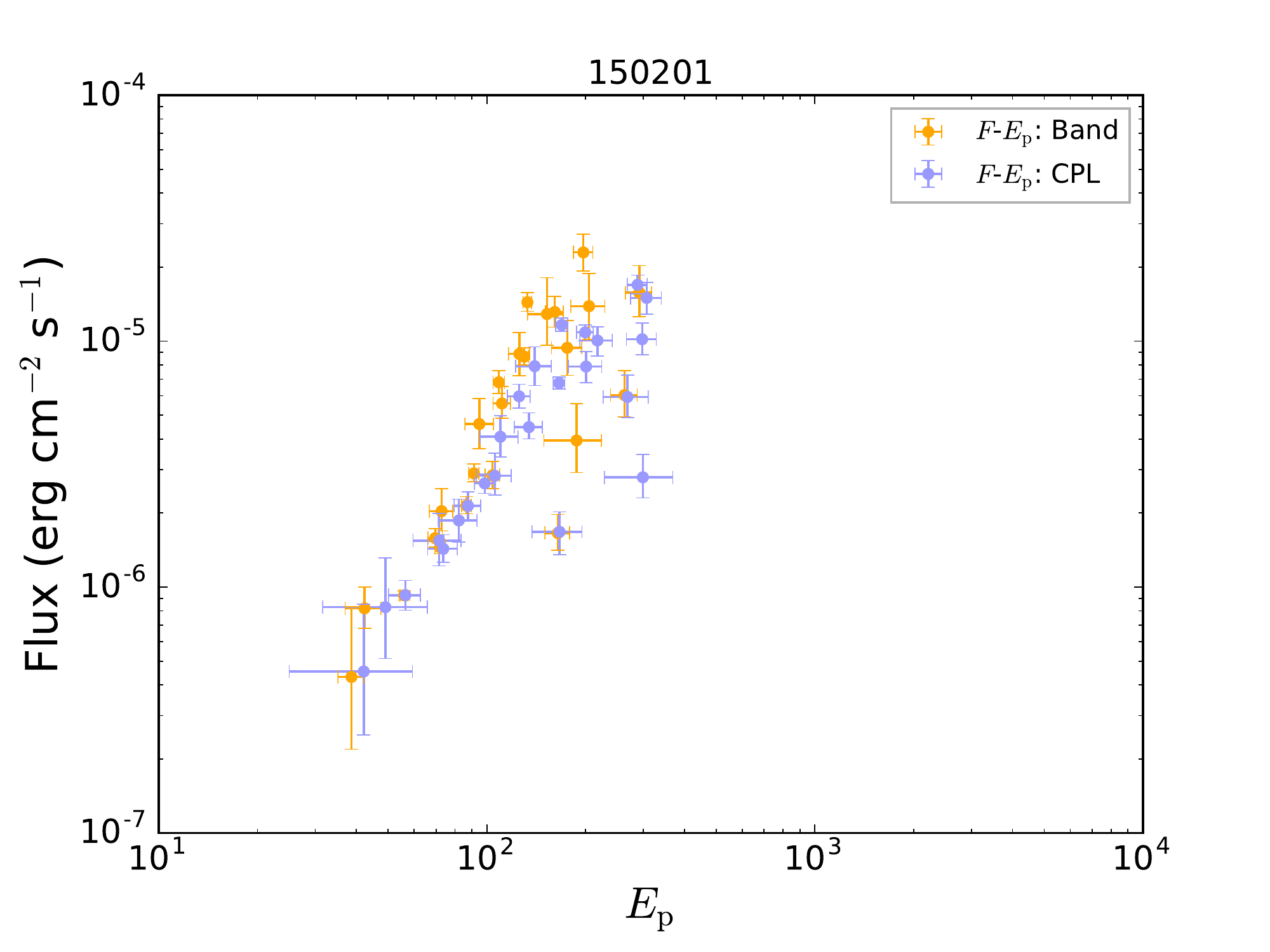}
\includegraphics[angle=0,scale=0.3]{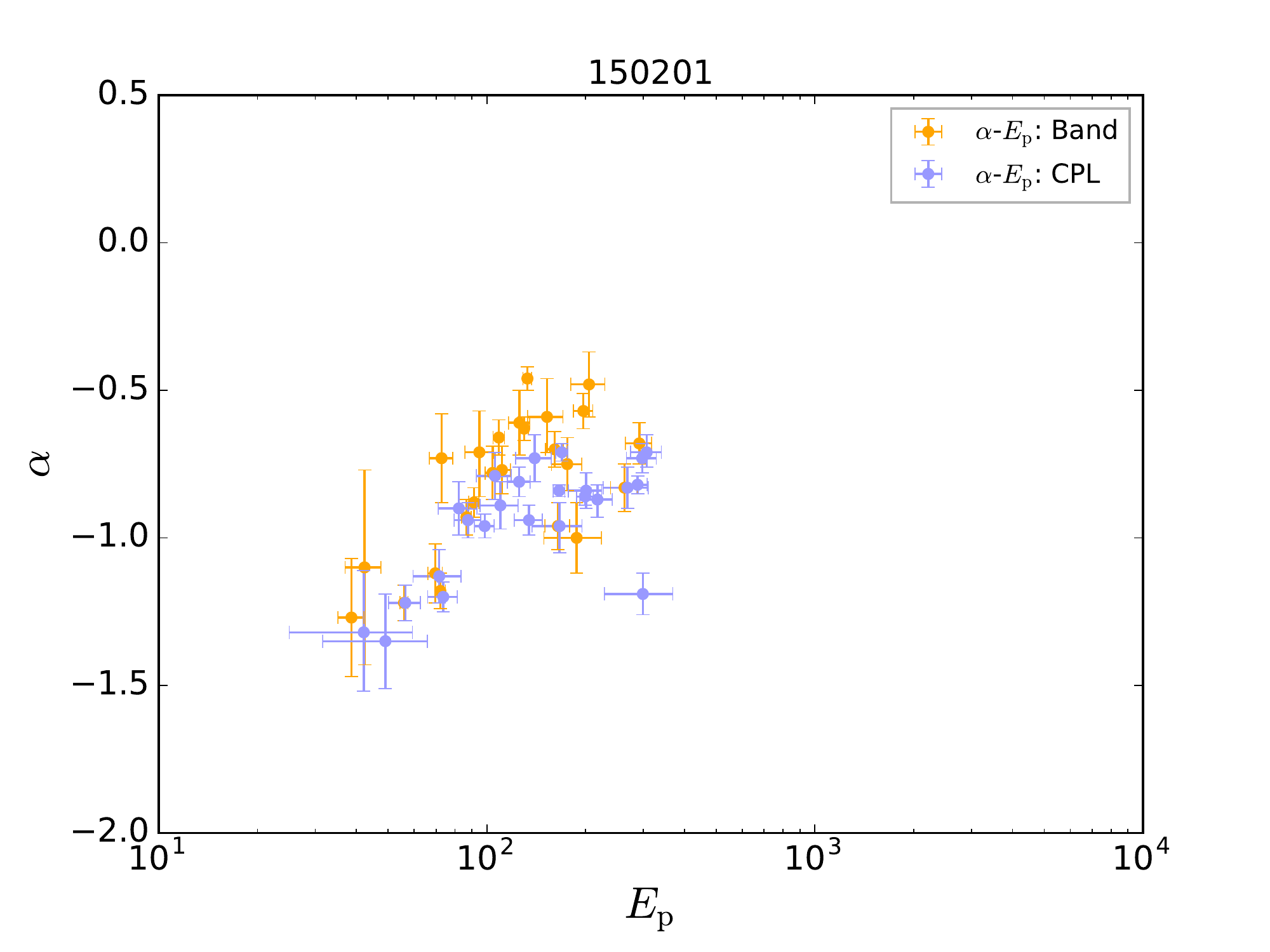}
\includegraphics[angle=0,scale=0.3]{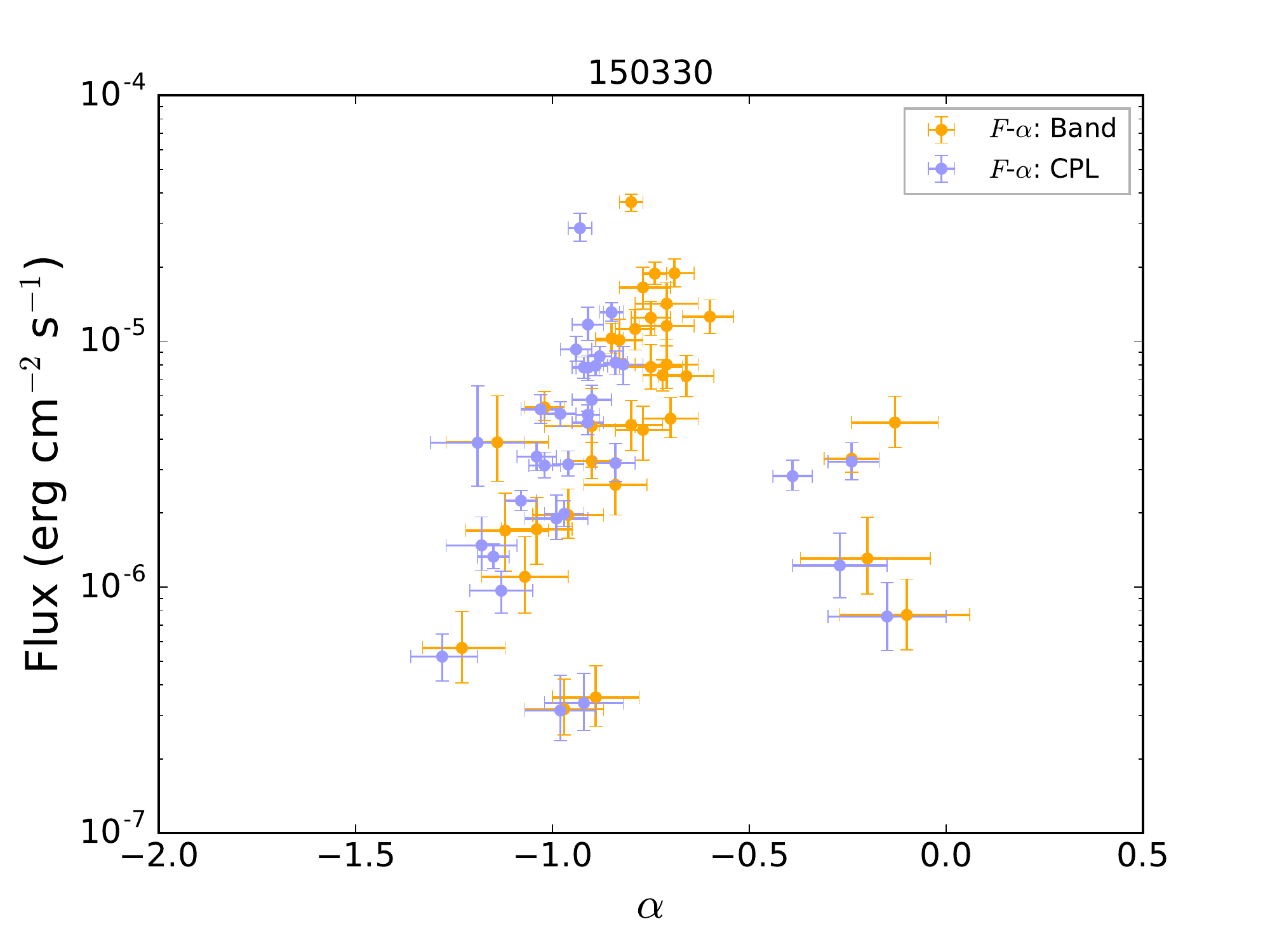}
\includegraphics[angle=0,scale=0.3]{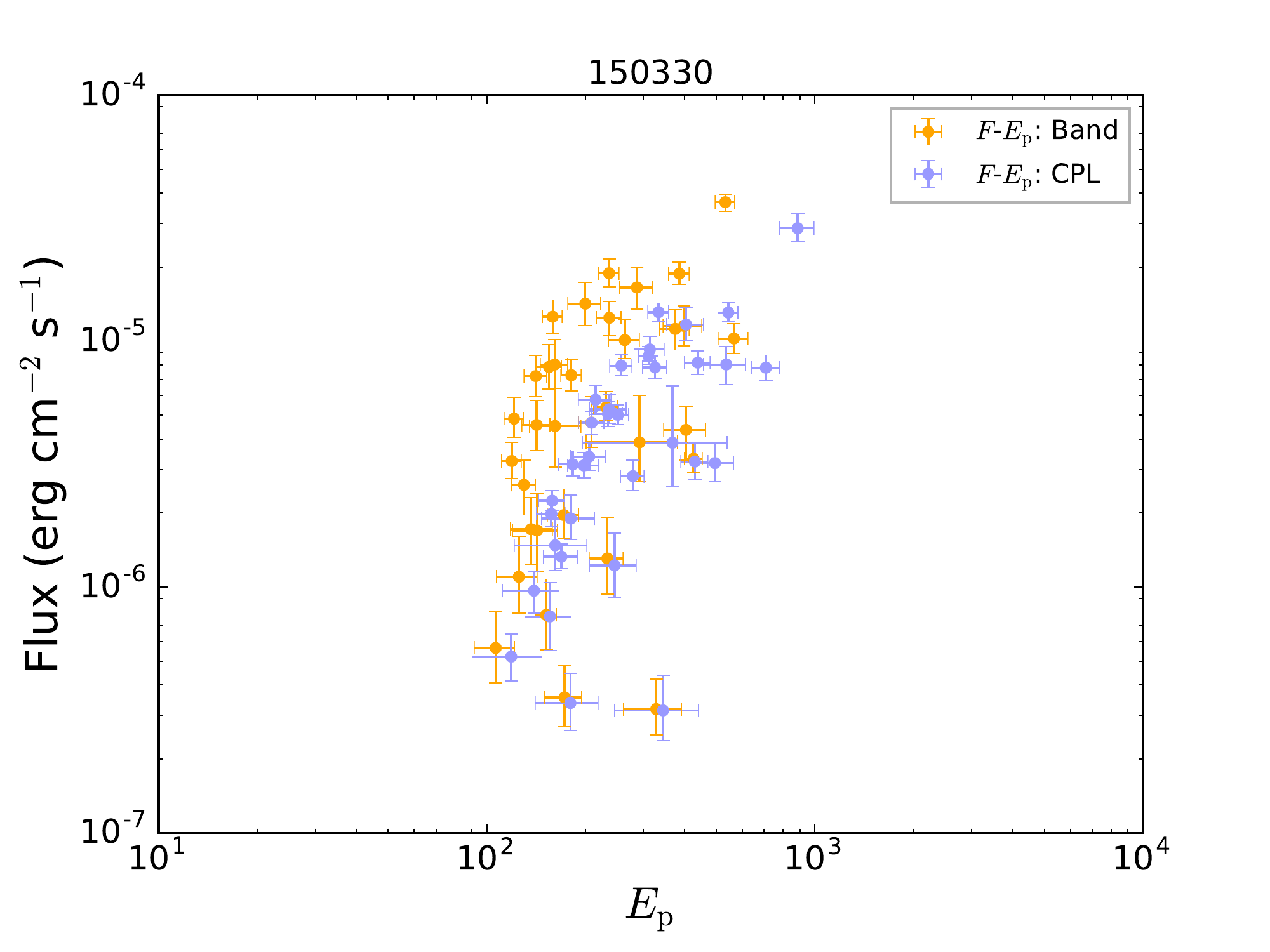}
\includegraphics[angle=0,scale=0.3]{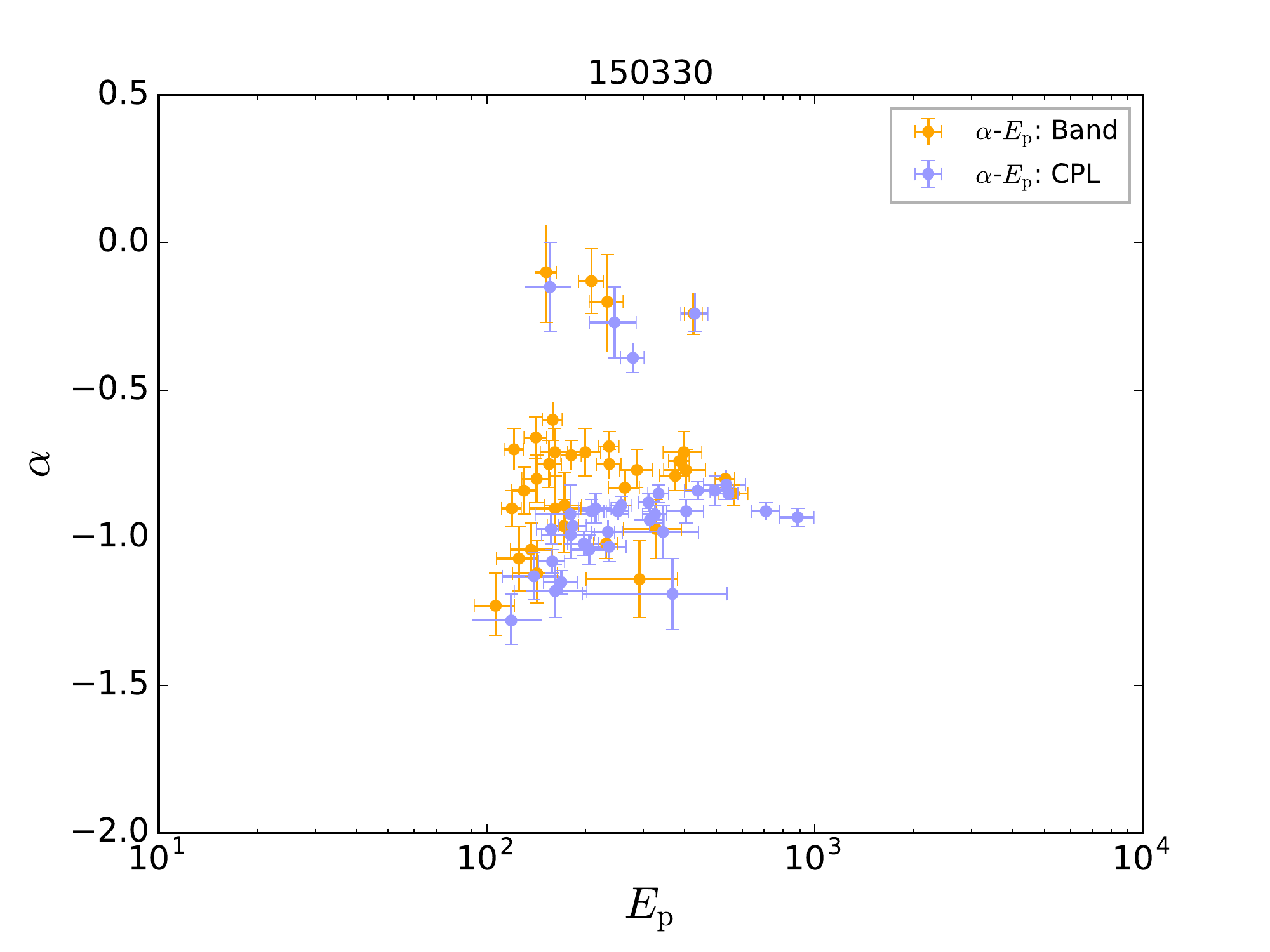}
\includegraphics[angle=0,scale=0.3]{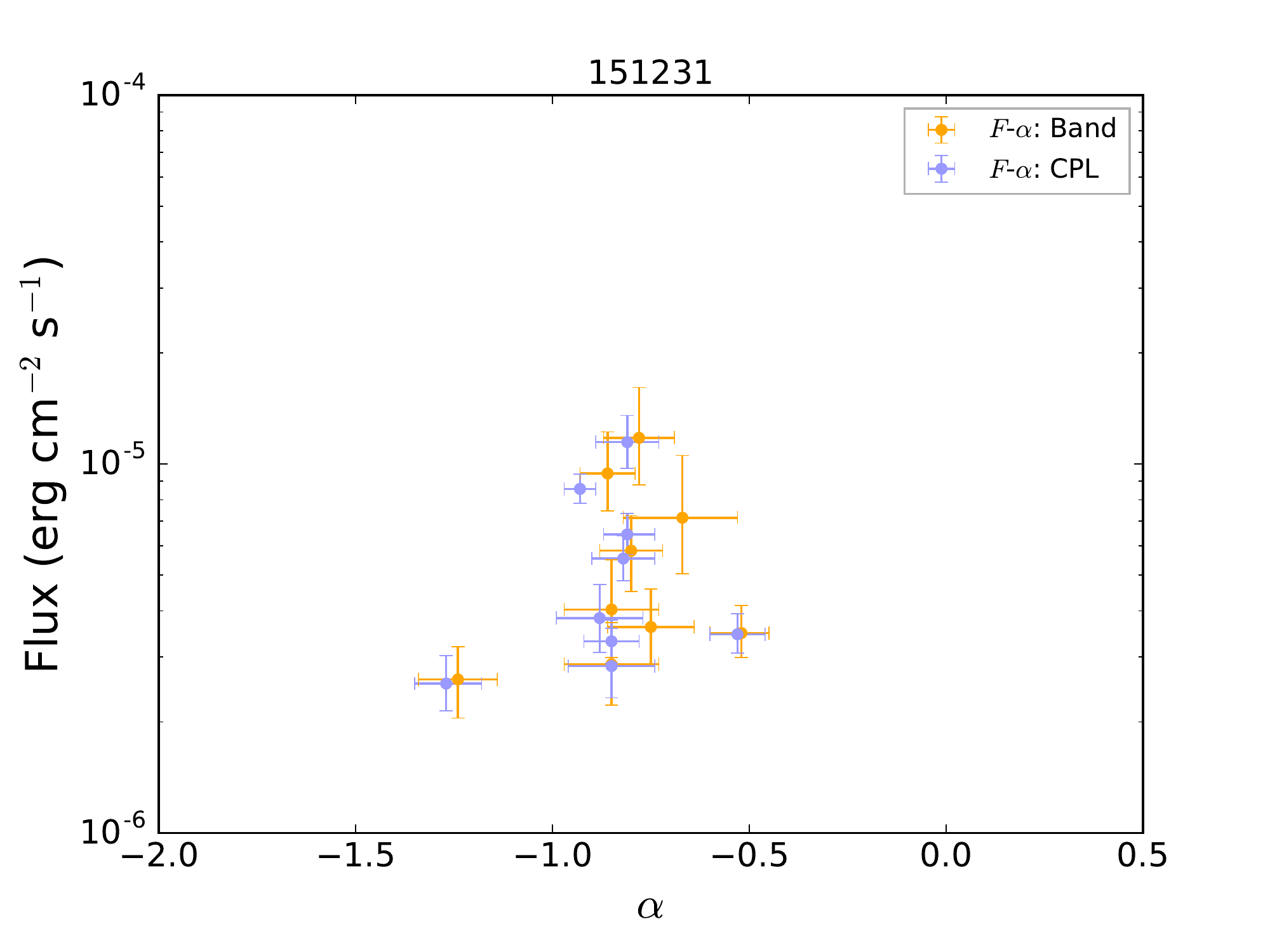}
\includegraphics[angle=0,scale=0.3]{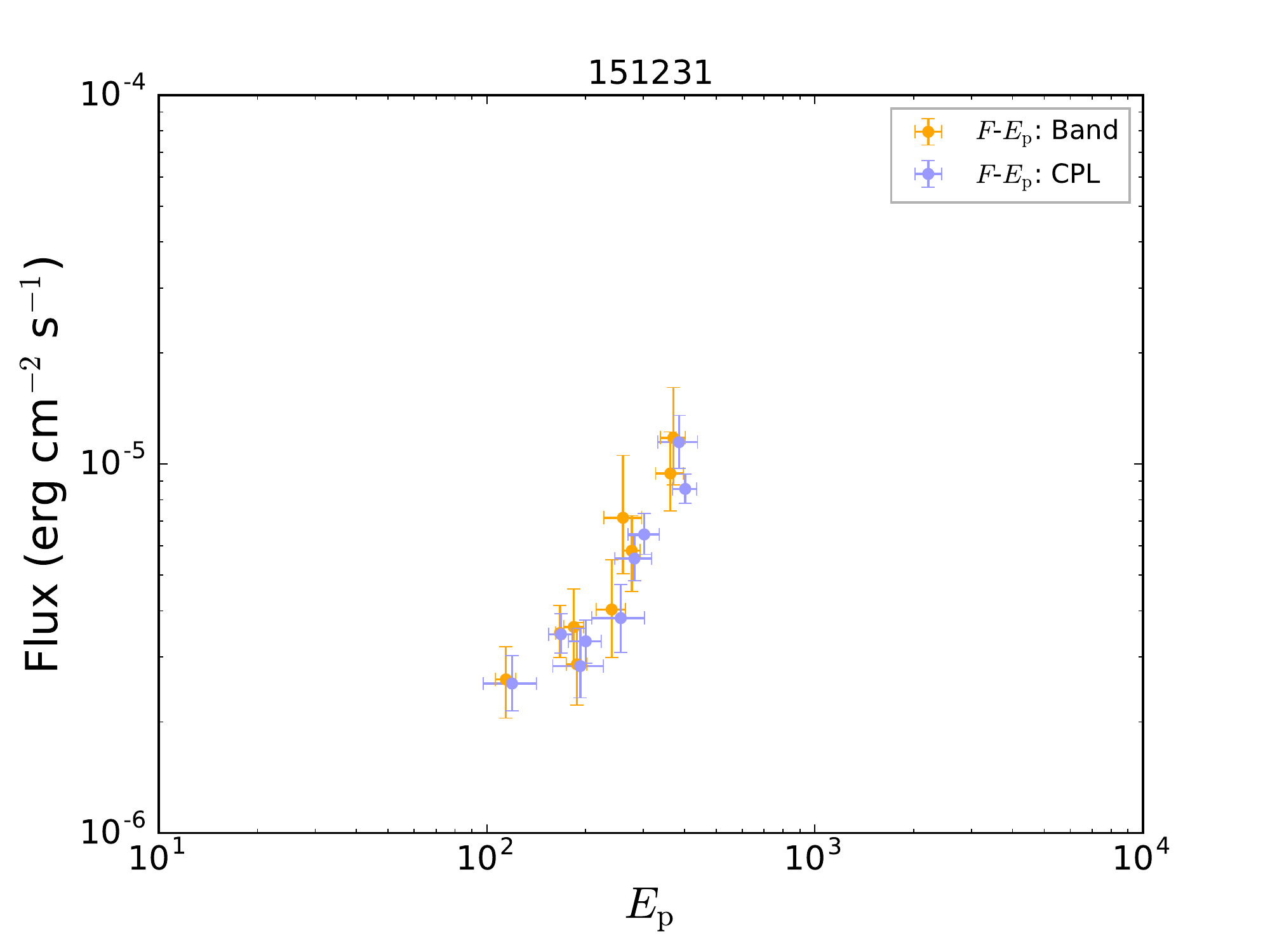}
\includegraphics[angle=0,scale=0.3]{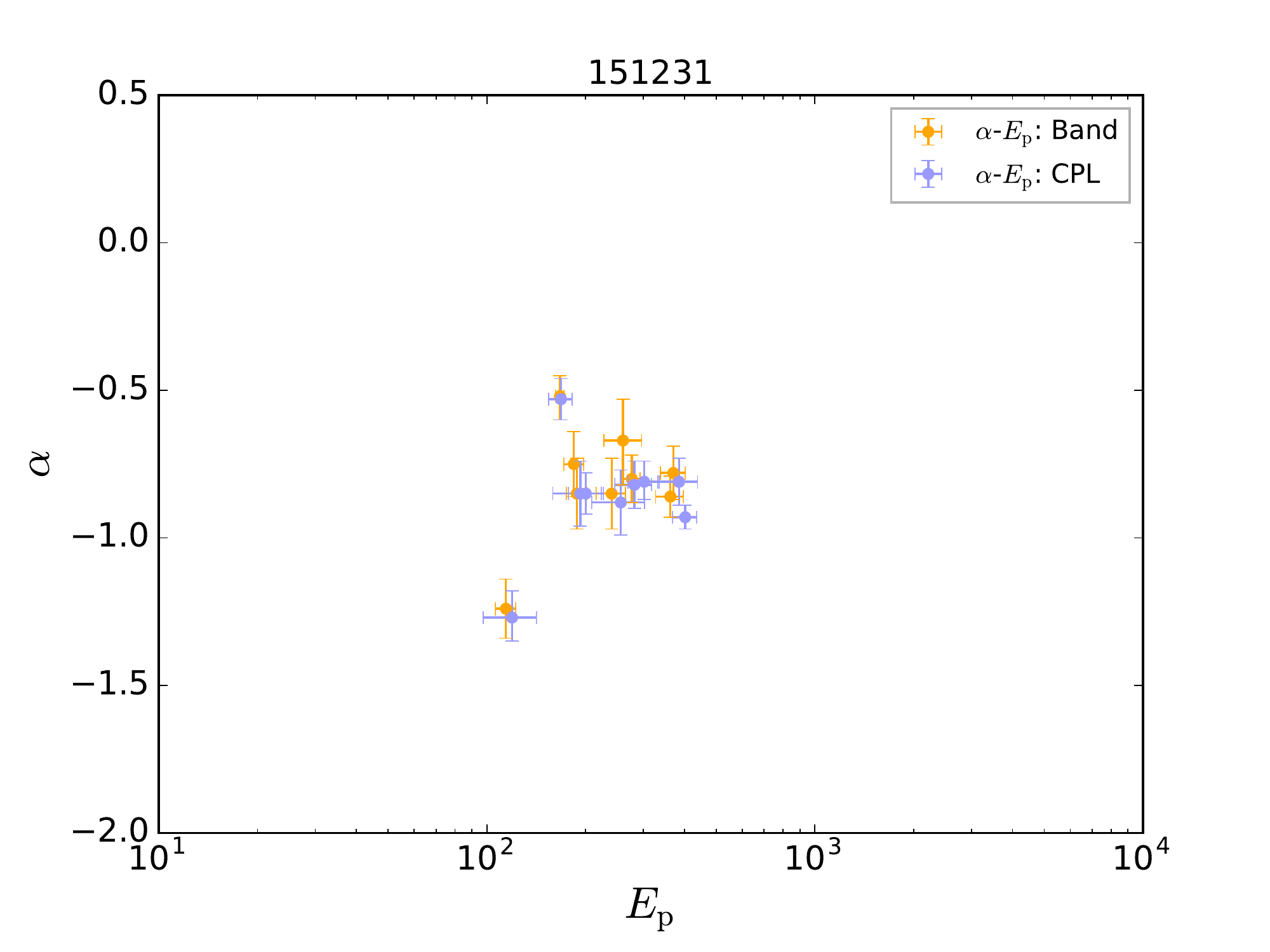}
\center{Fig. \ref{fig:relation3}--- Continued}
\end{figure*}
\begin{figure*}
\includegraphics[angle=0,scale=0.3]{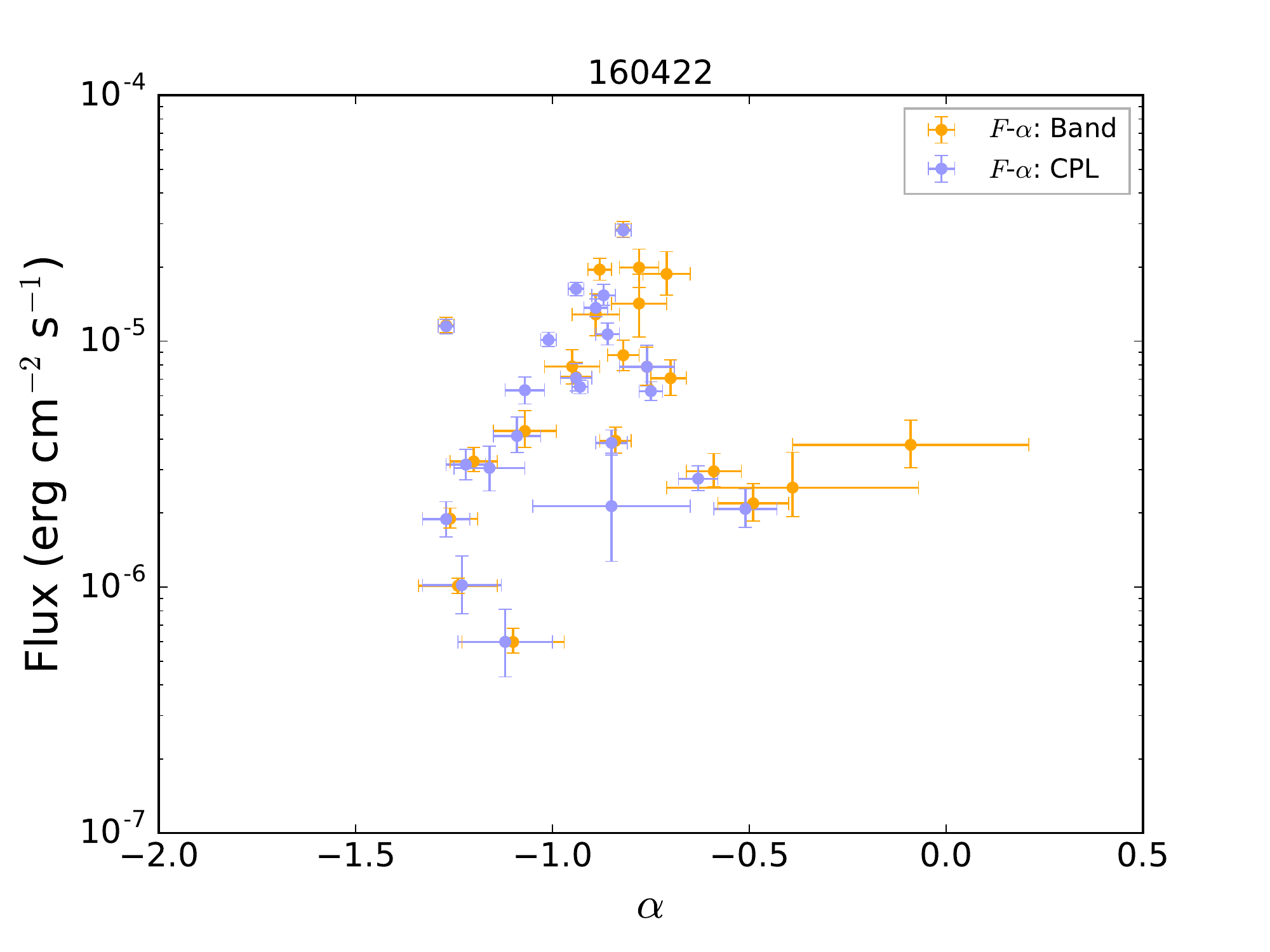}
\includegraphics[angle=0,scale=0.3]{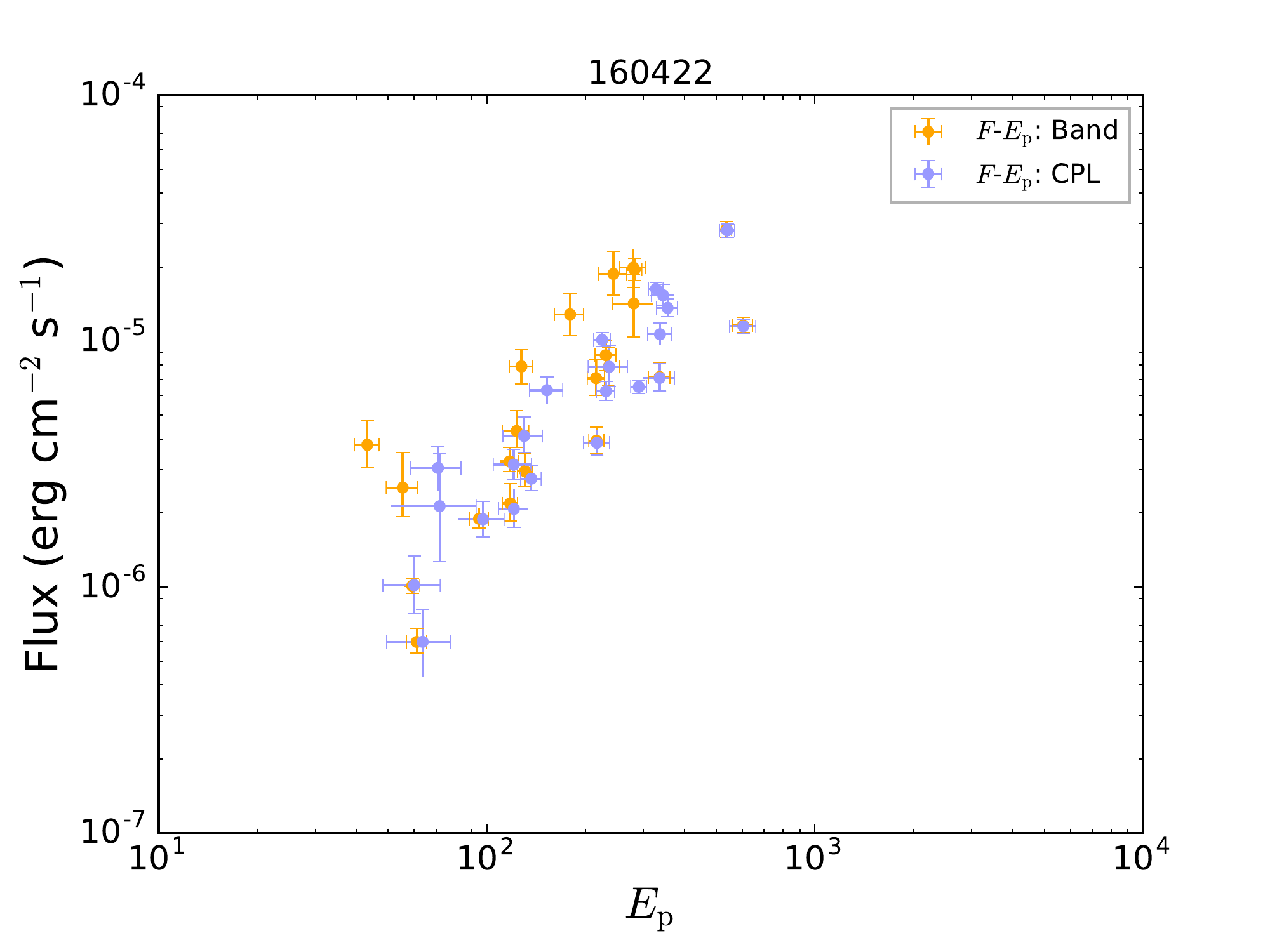}
\includegraphics[angle=0,scale=0.3]{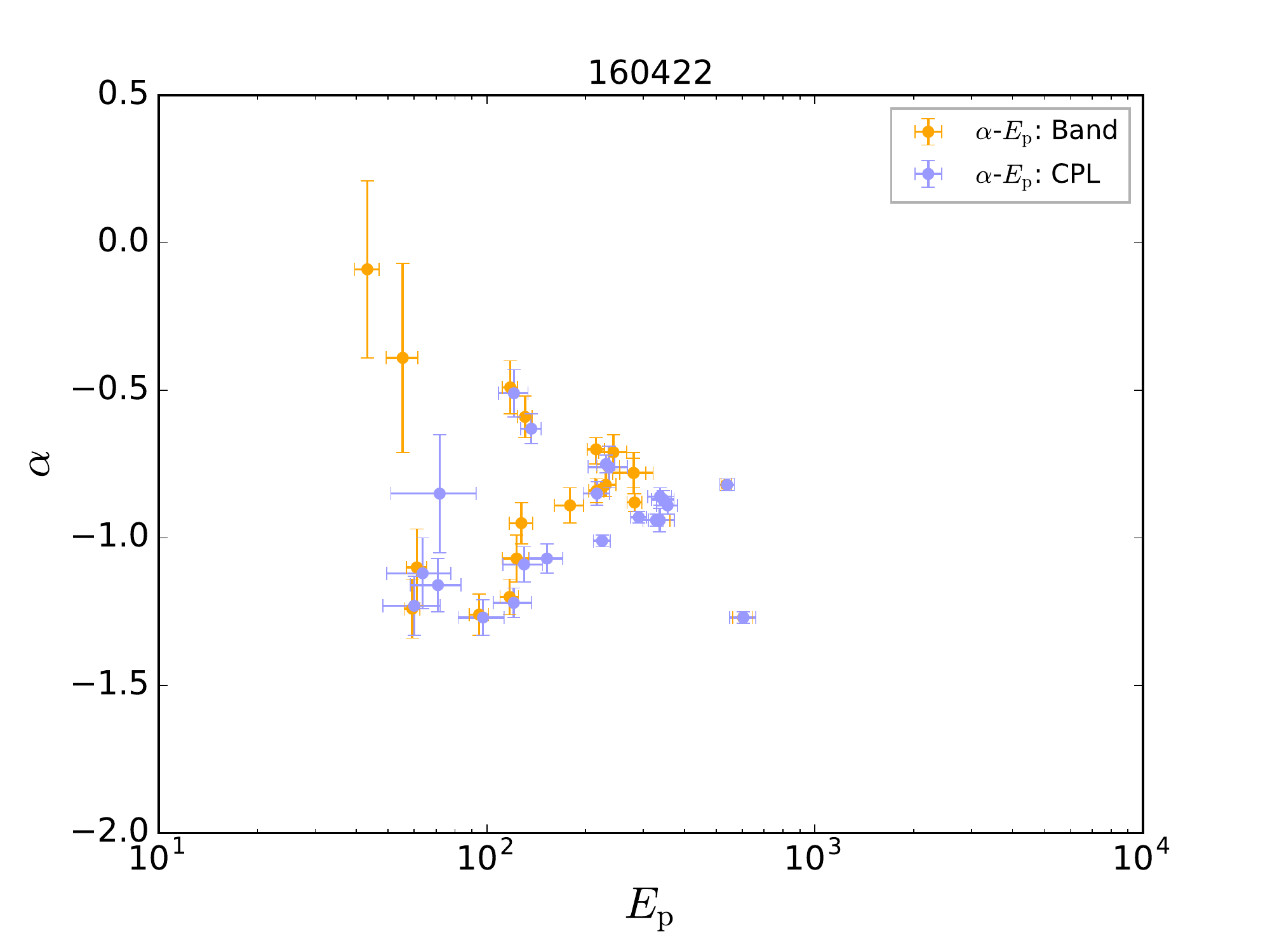}
\includegraphics[angle=0,scale=0.3]{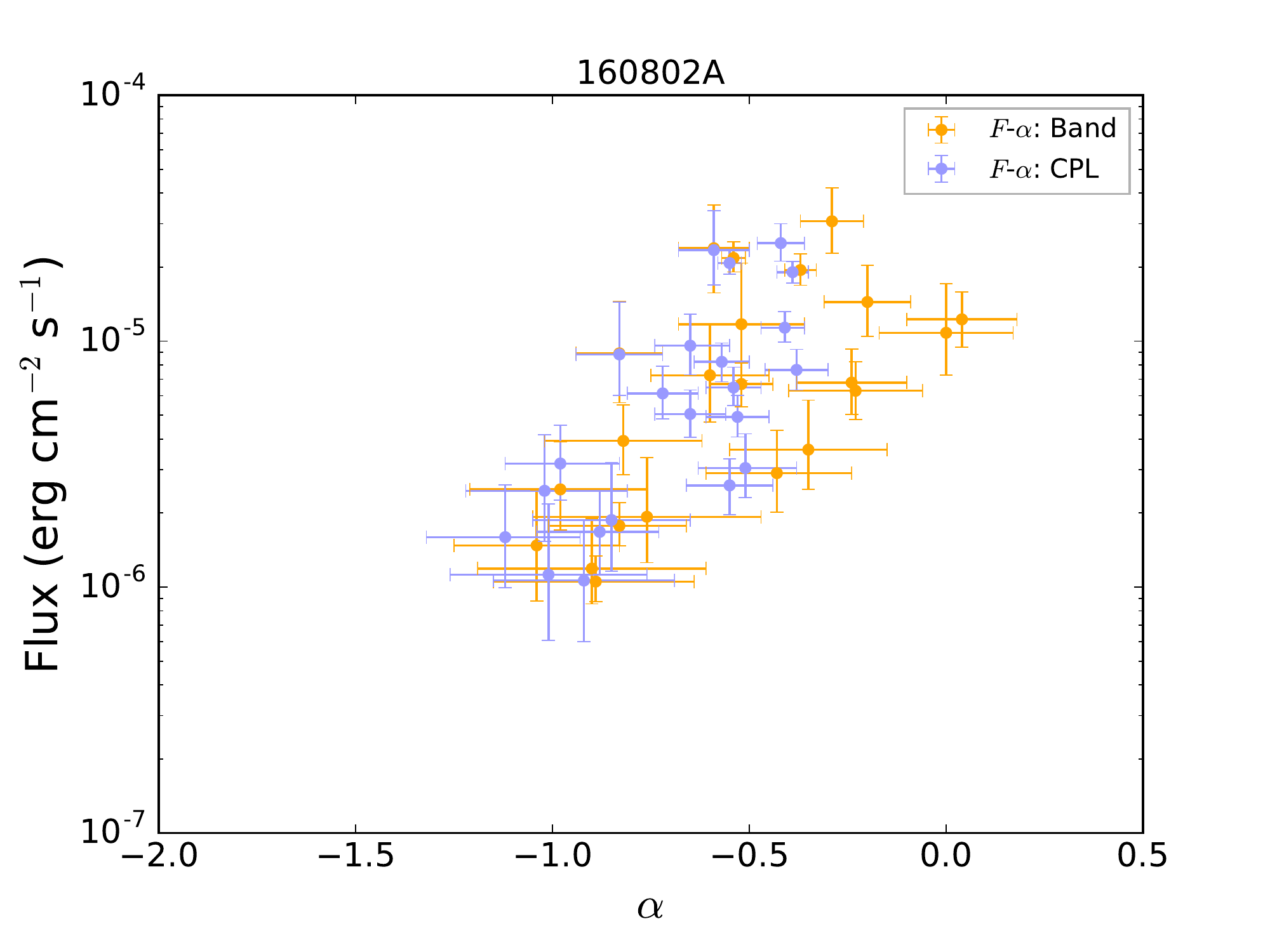}
\includegraphics[angle=0,scale=0.3]{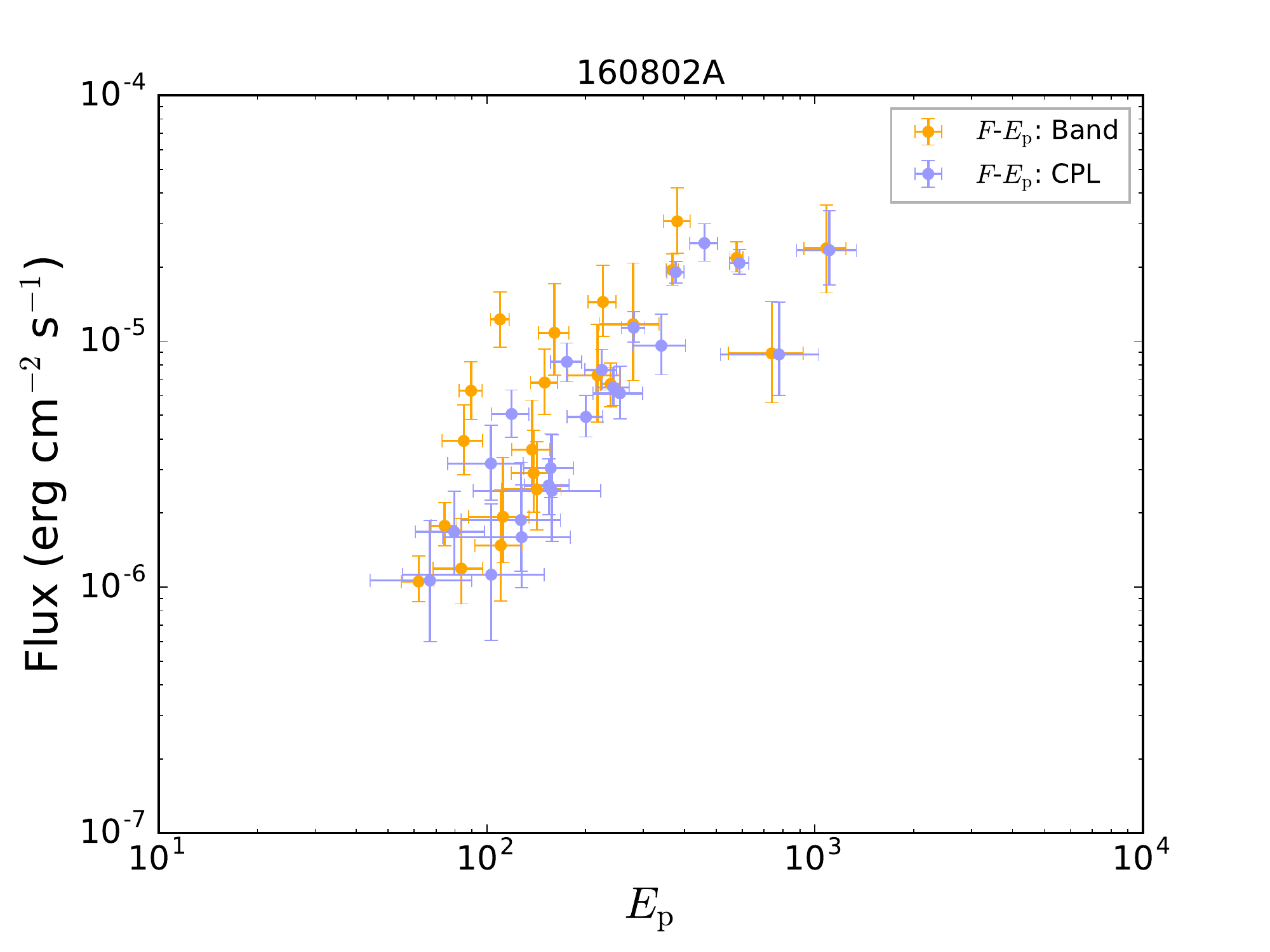}
\includegraphics[angle=0,scale=0.3]{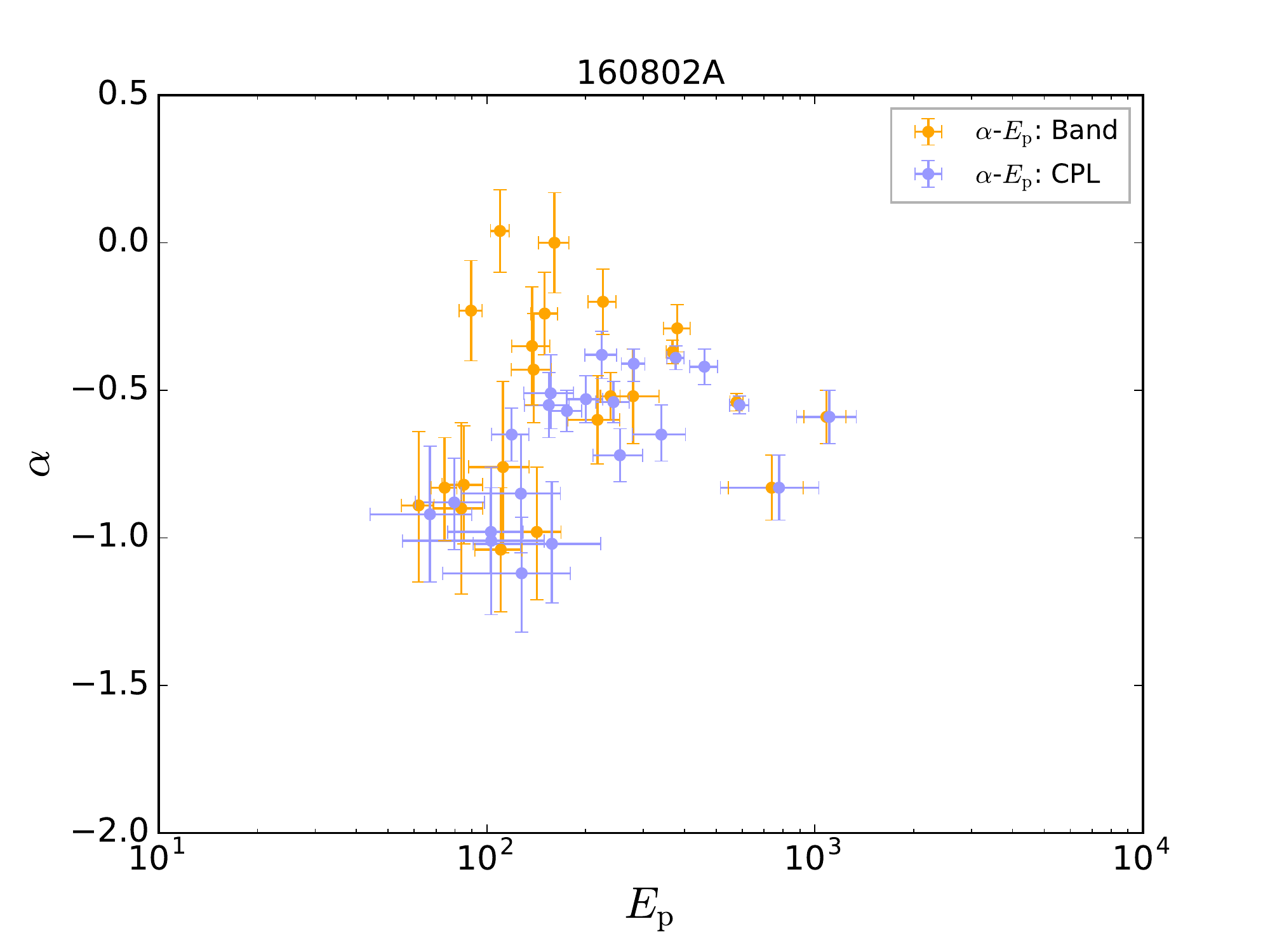}
\includegraphics[angle=0,scale=0.3]{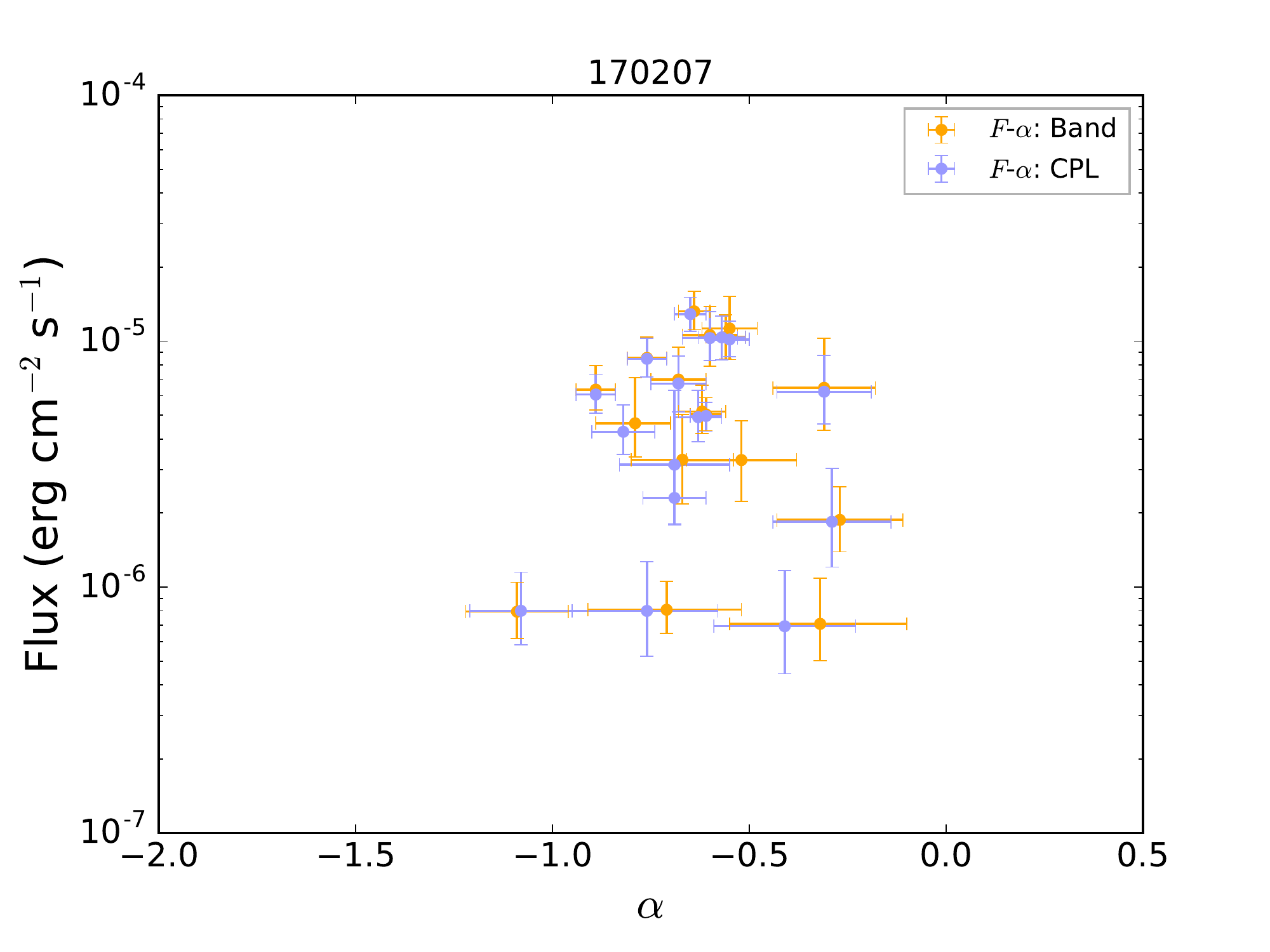}
\includegraphics[angle=0,scale=0.3]{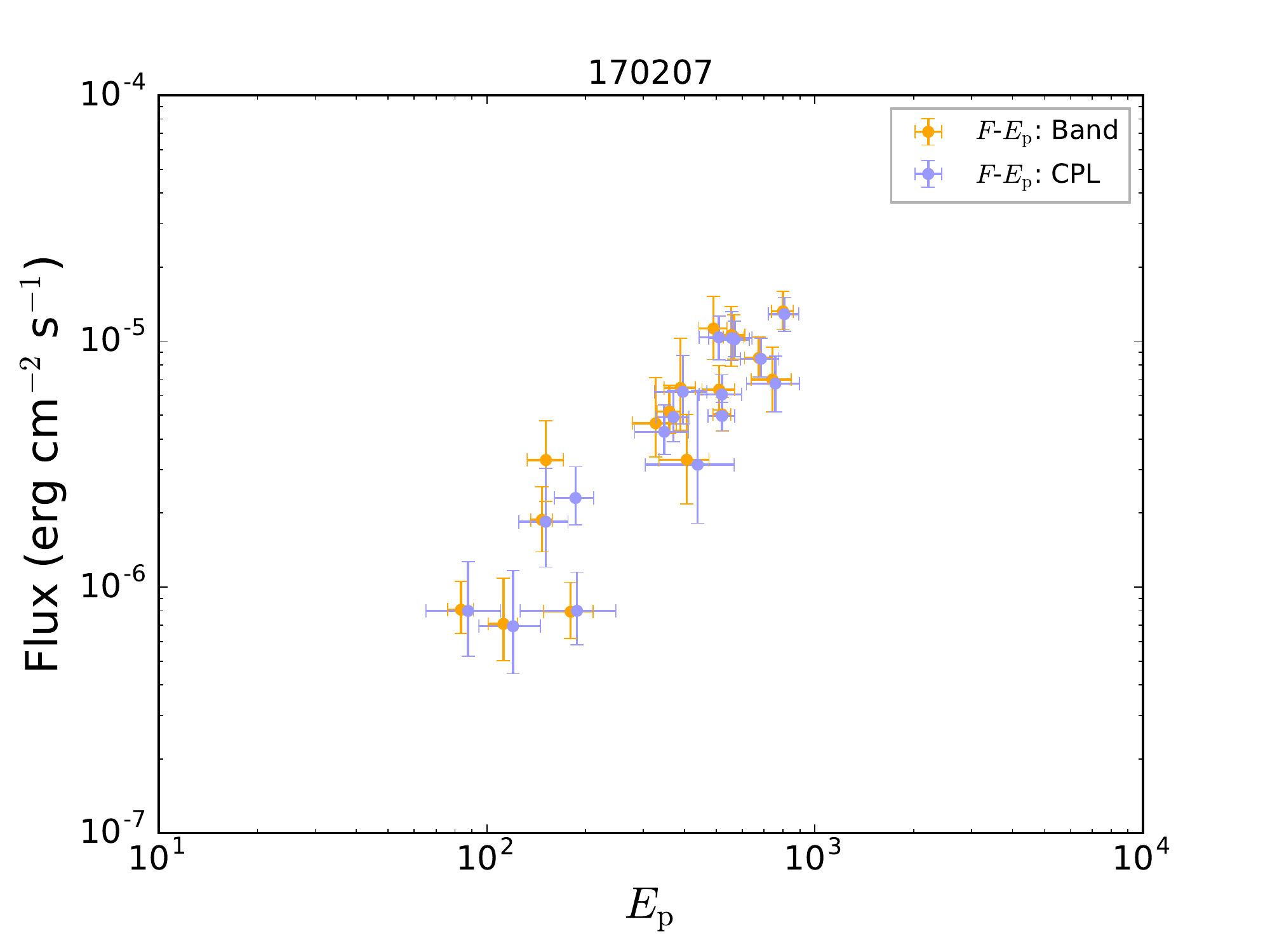}
\includegraphics[angle=0,scale=0.3]{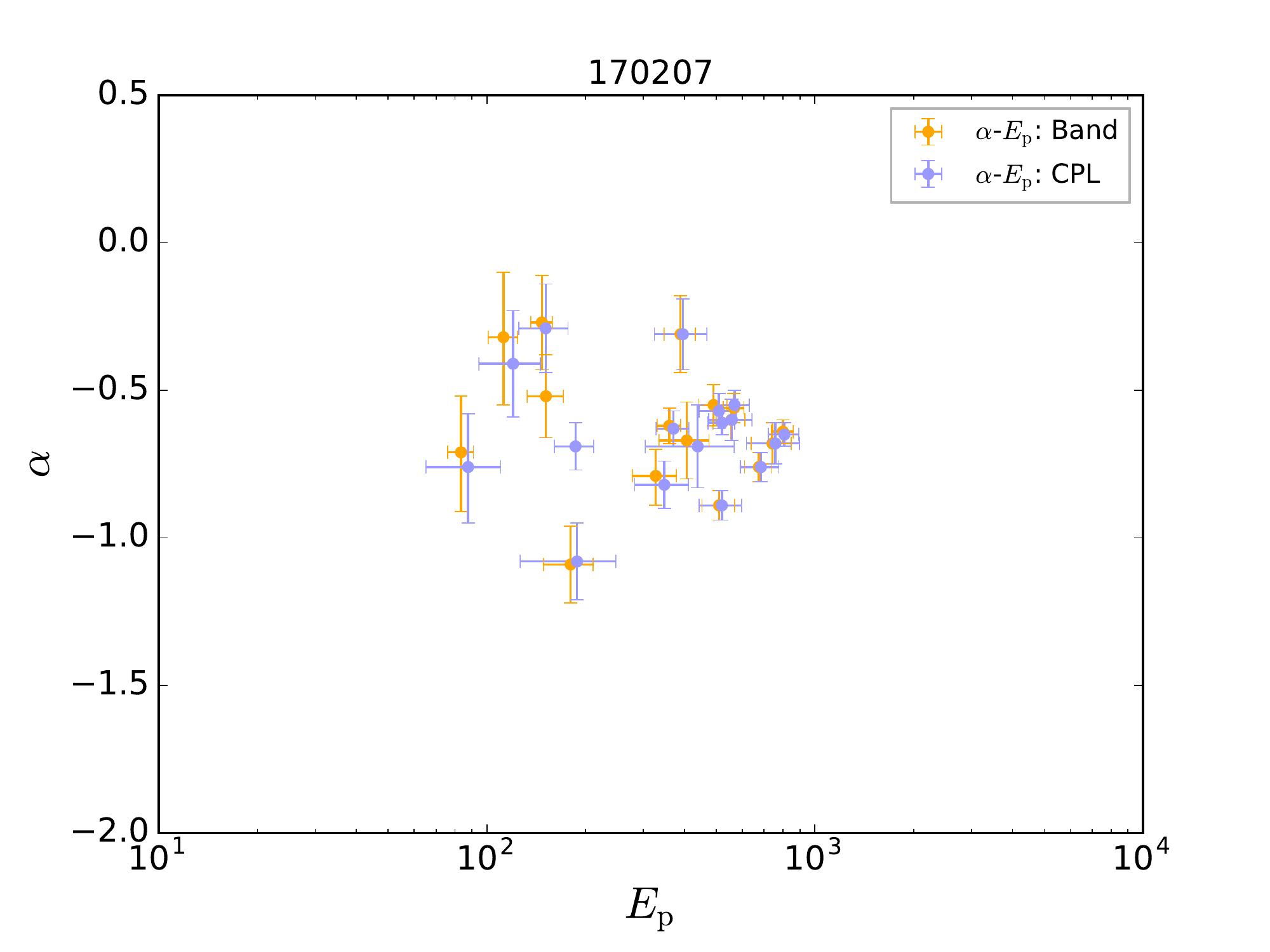}
\includegraphics[angle=0,scale=0.3]{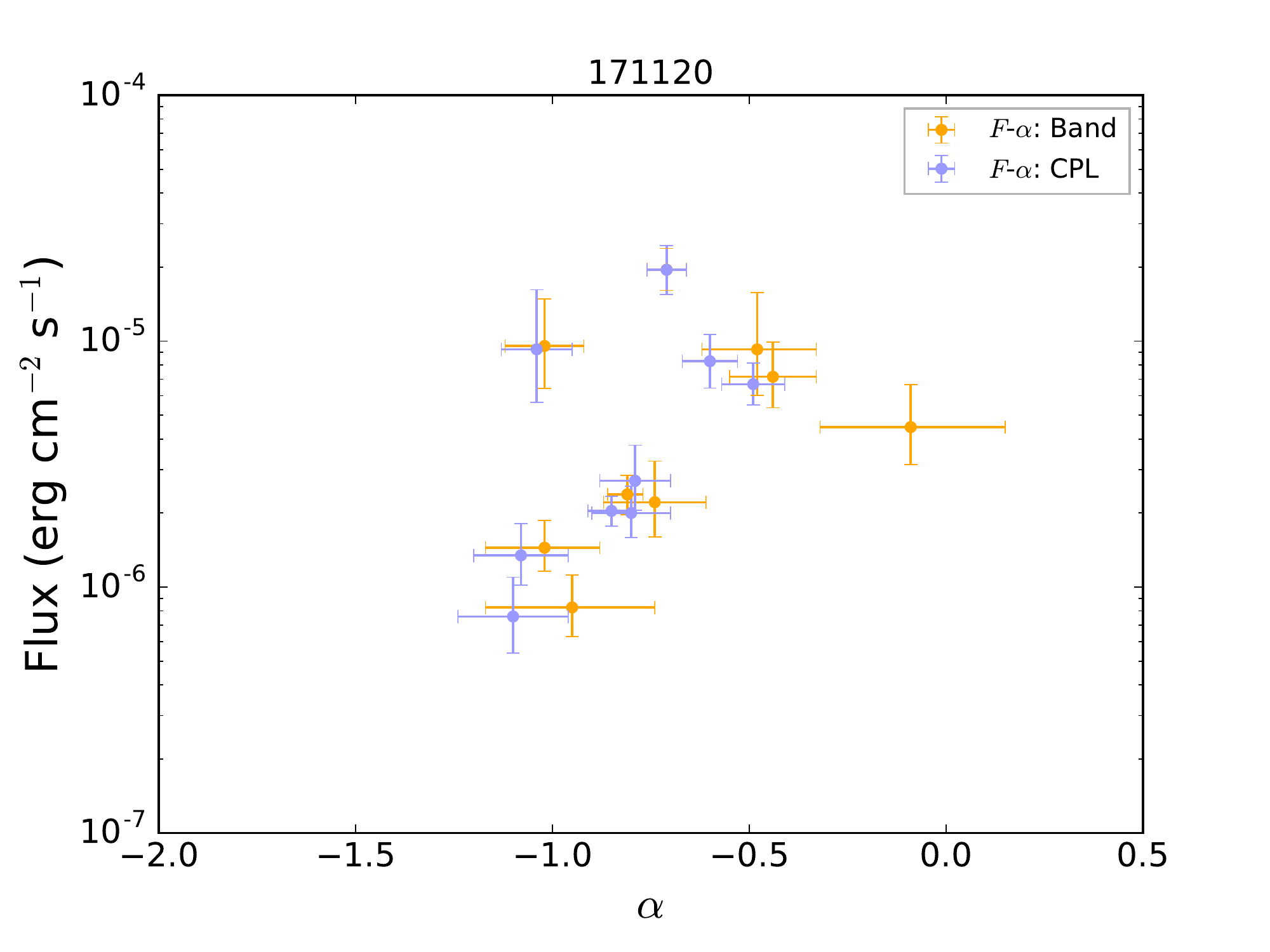}
\includegraphics[angle=0,scale=0.3]{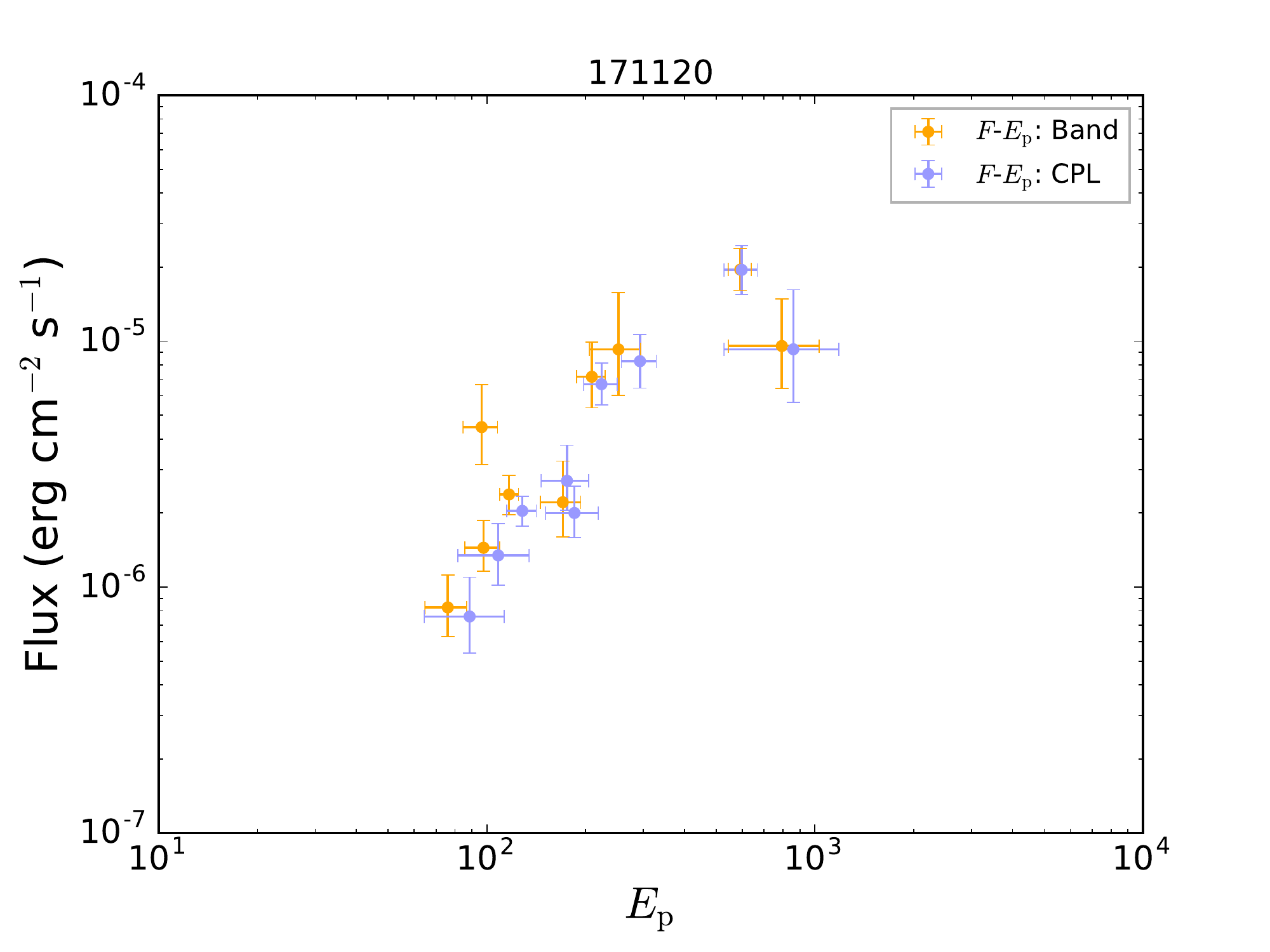}
\includegraphics[angle=0,scale=0.3]{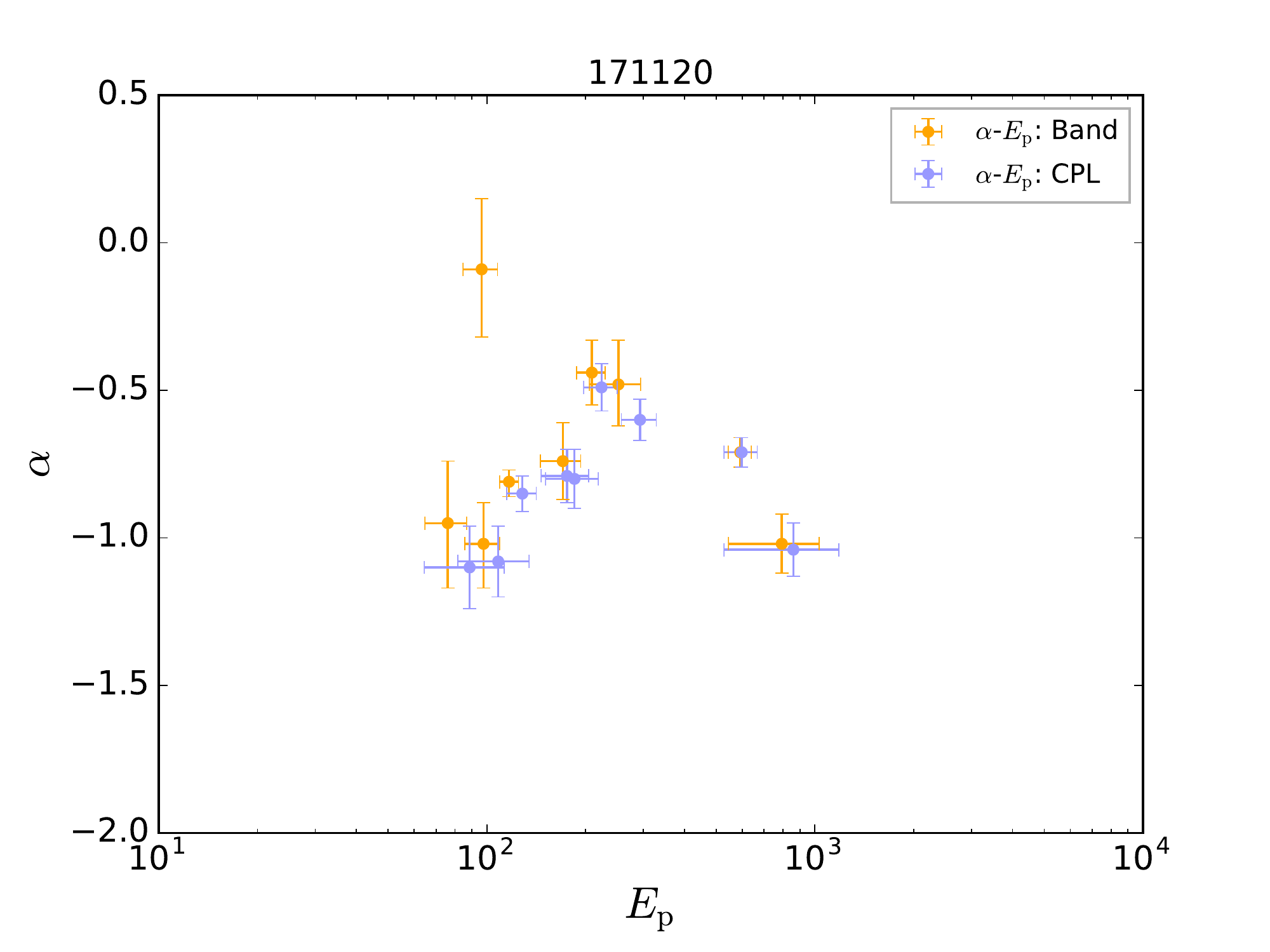}
\includegraphics[angle=0,scale=0.3]{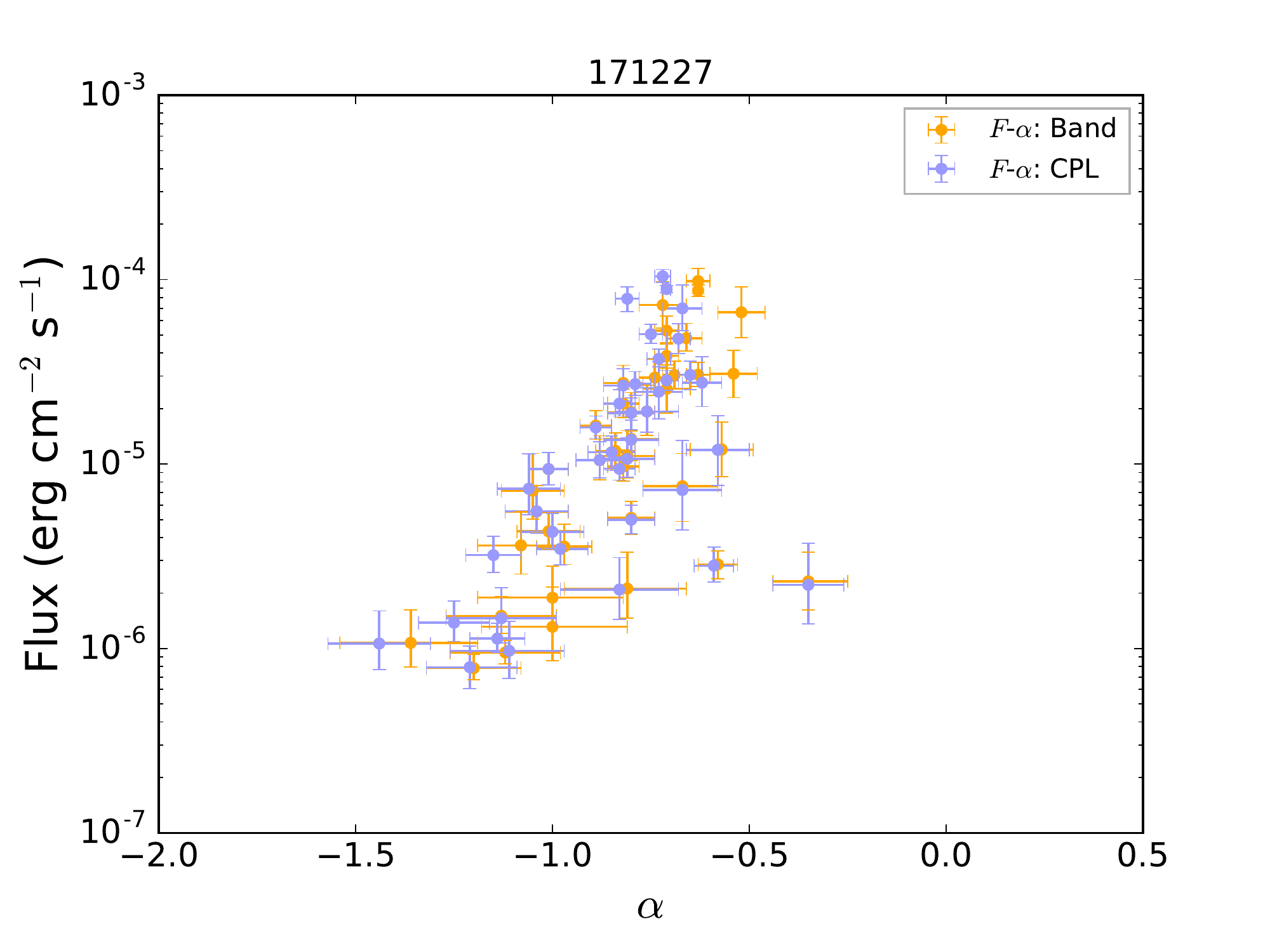}
\includegraphics[angle=0,scale=0.3]{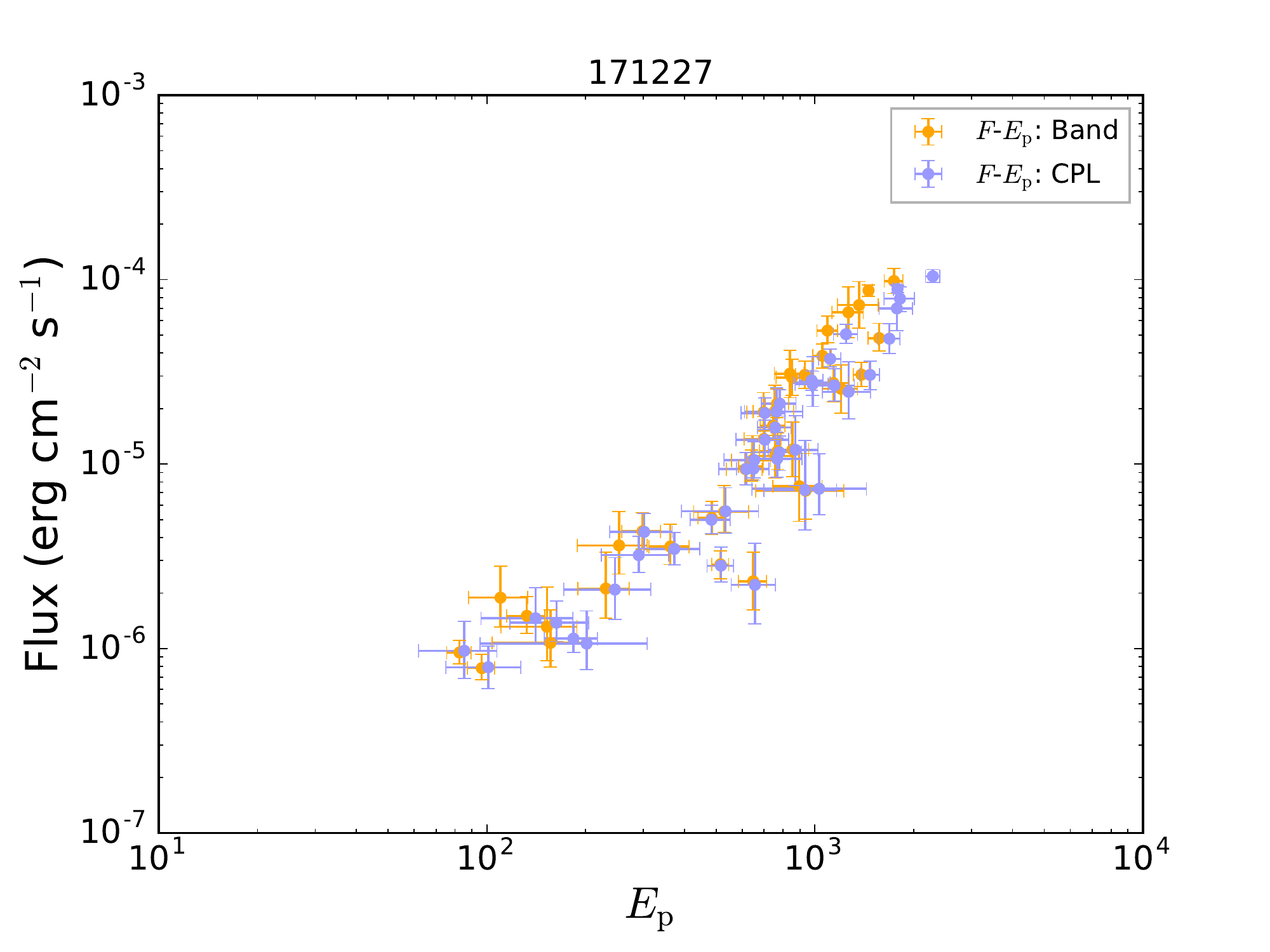}
\includegraphics[angle=0,scale=0.3]{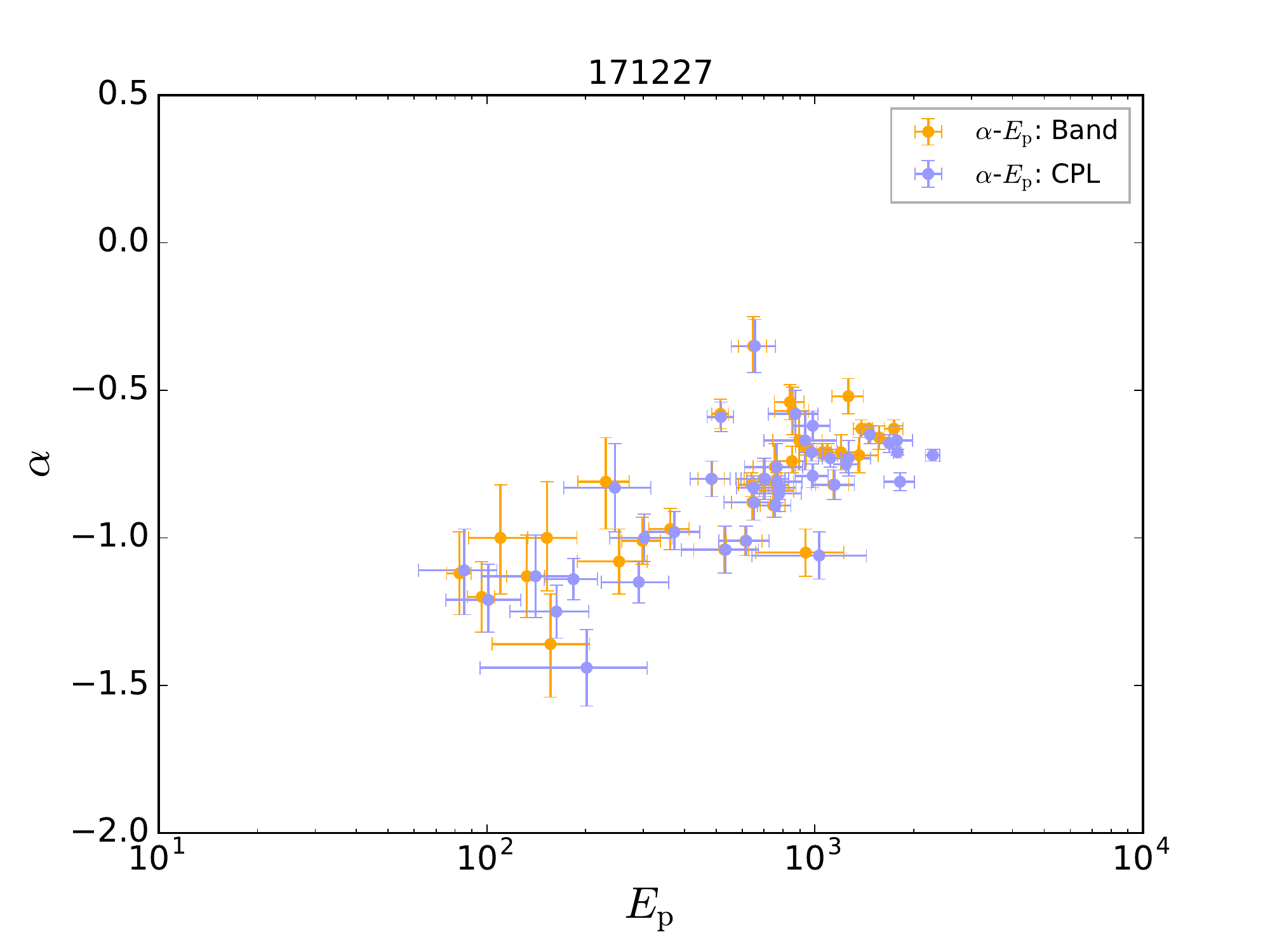}
\center{Fig. \ref{fig:relation3}--- Continued}
\end{figure*}
\begin{figure*}
\includegraphics[angle=0,scale=0.3]{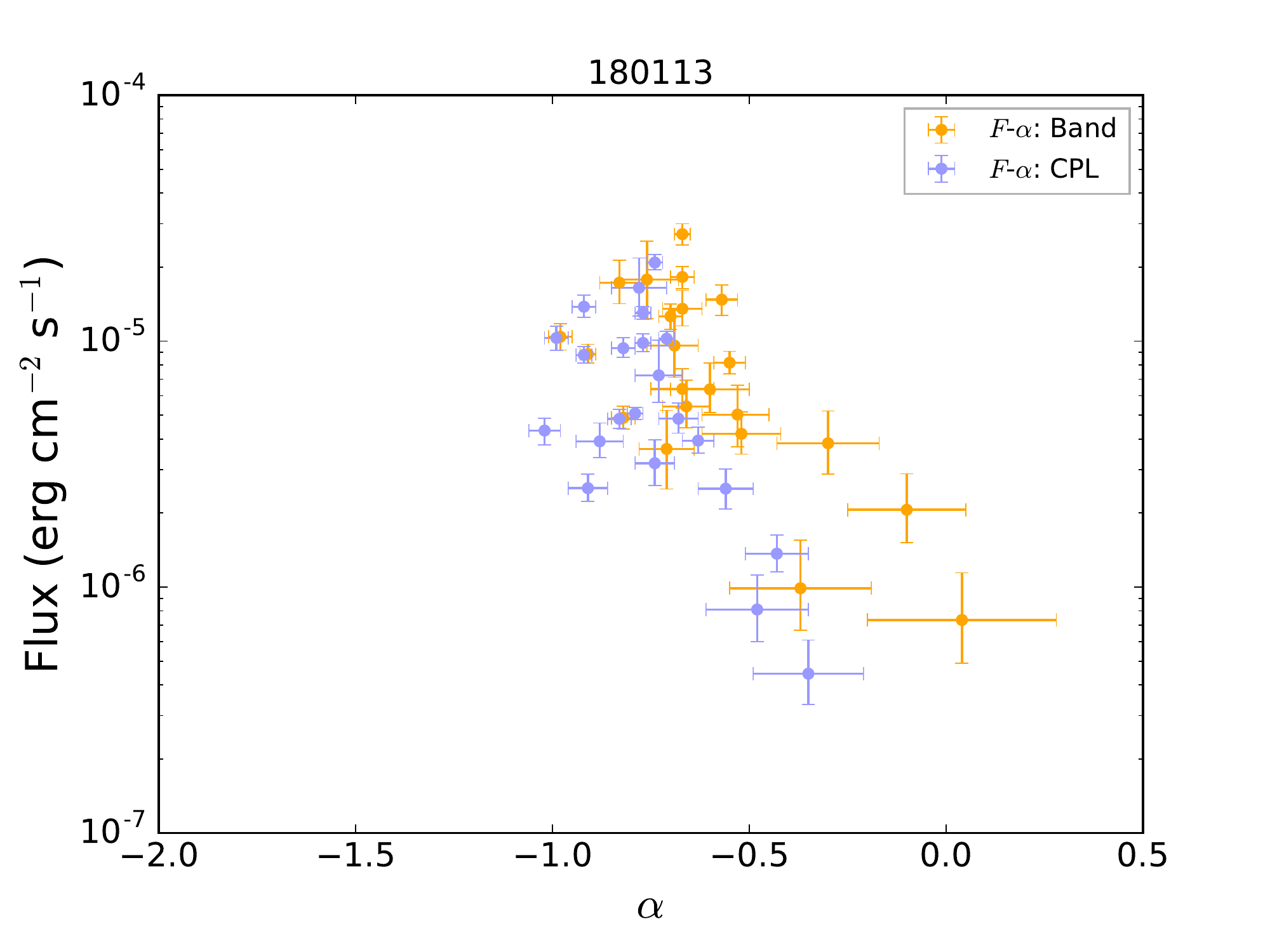}
\includegraphics[angle=0,scale=0.3]{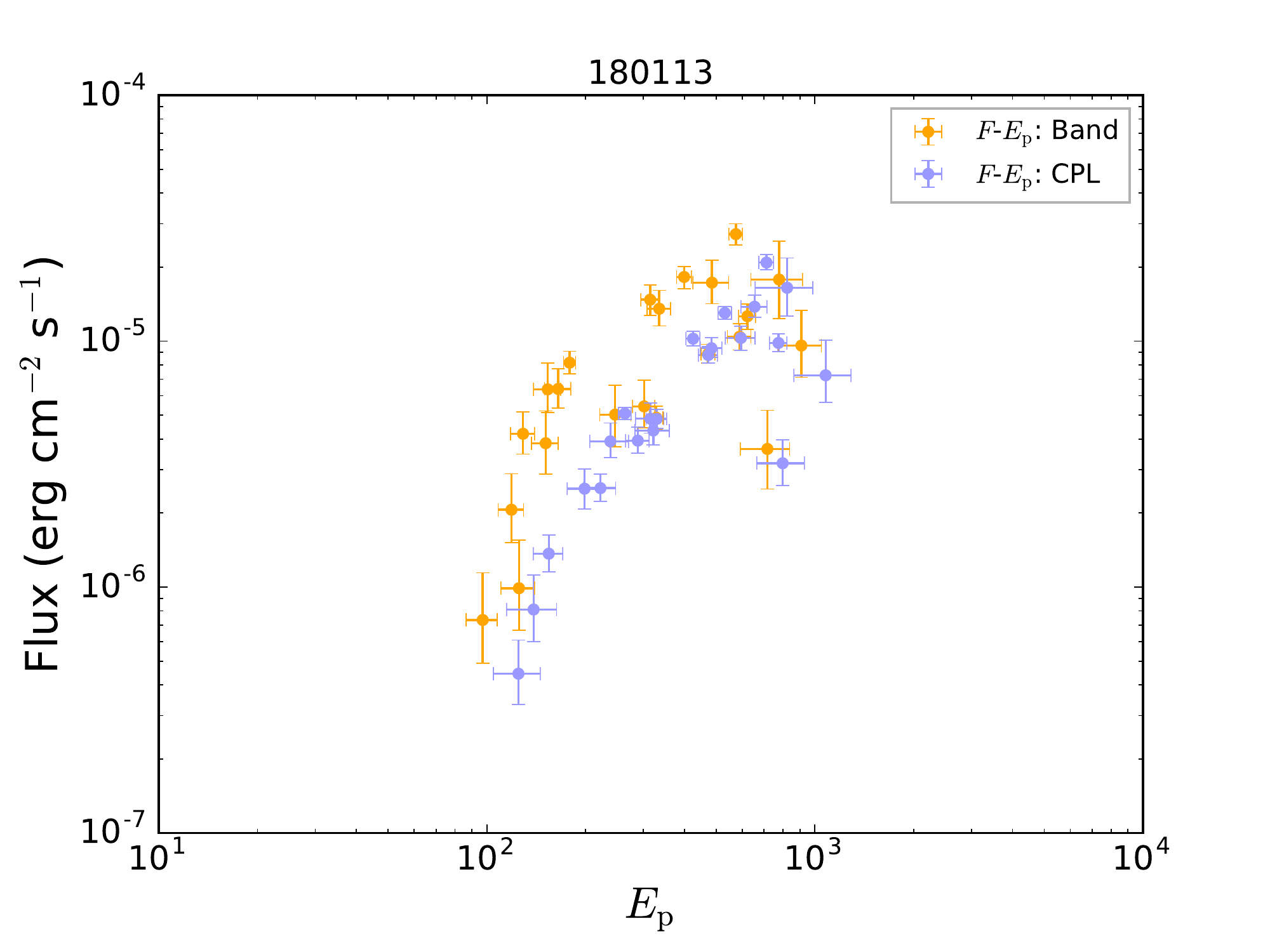}
\includegraphics[angle=0,scale=0.3]{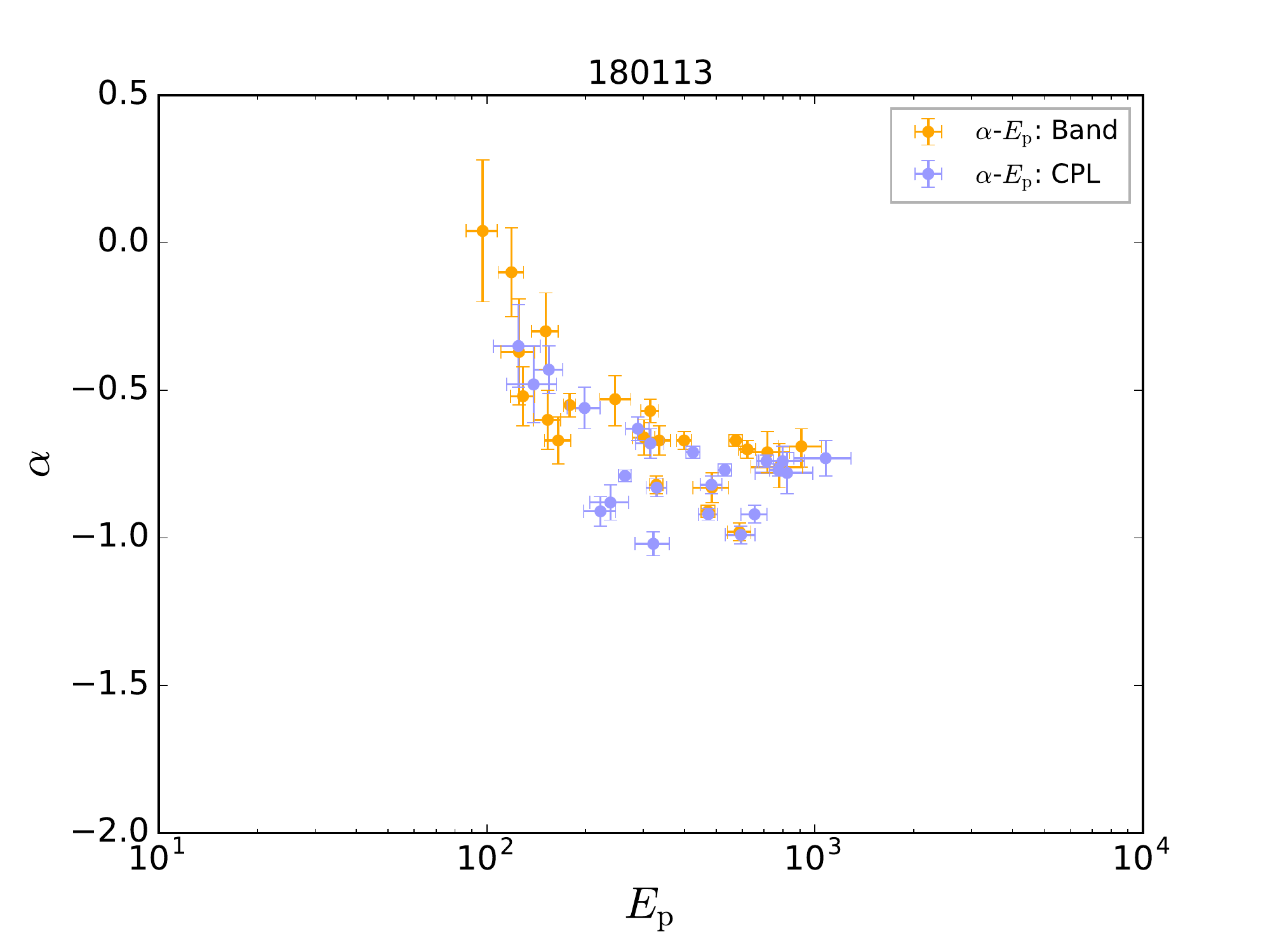}
\includegraphics[angle=0,scale=0.3]{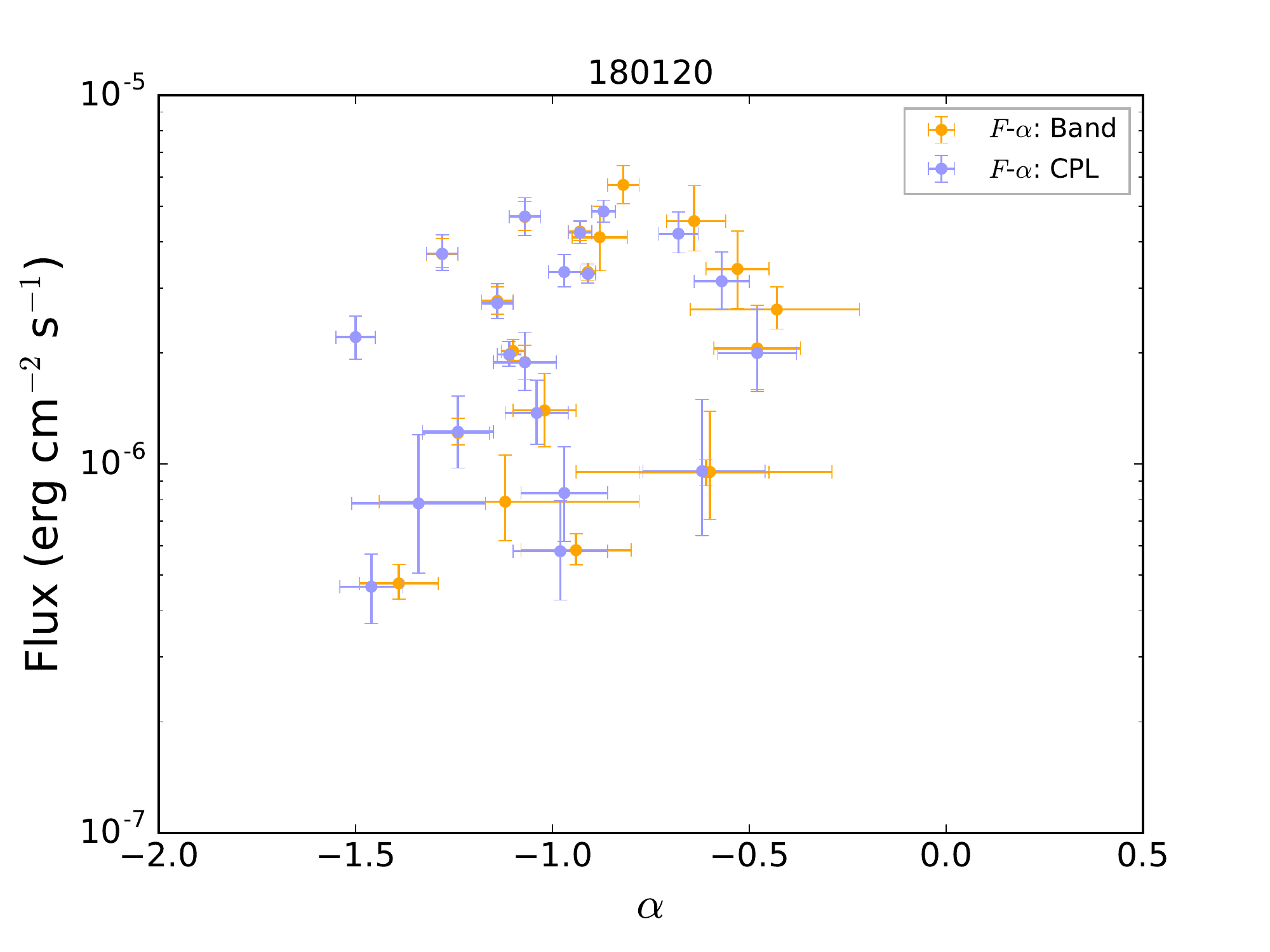}
\includegraphics[angle=0,scale=0.3]{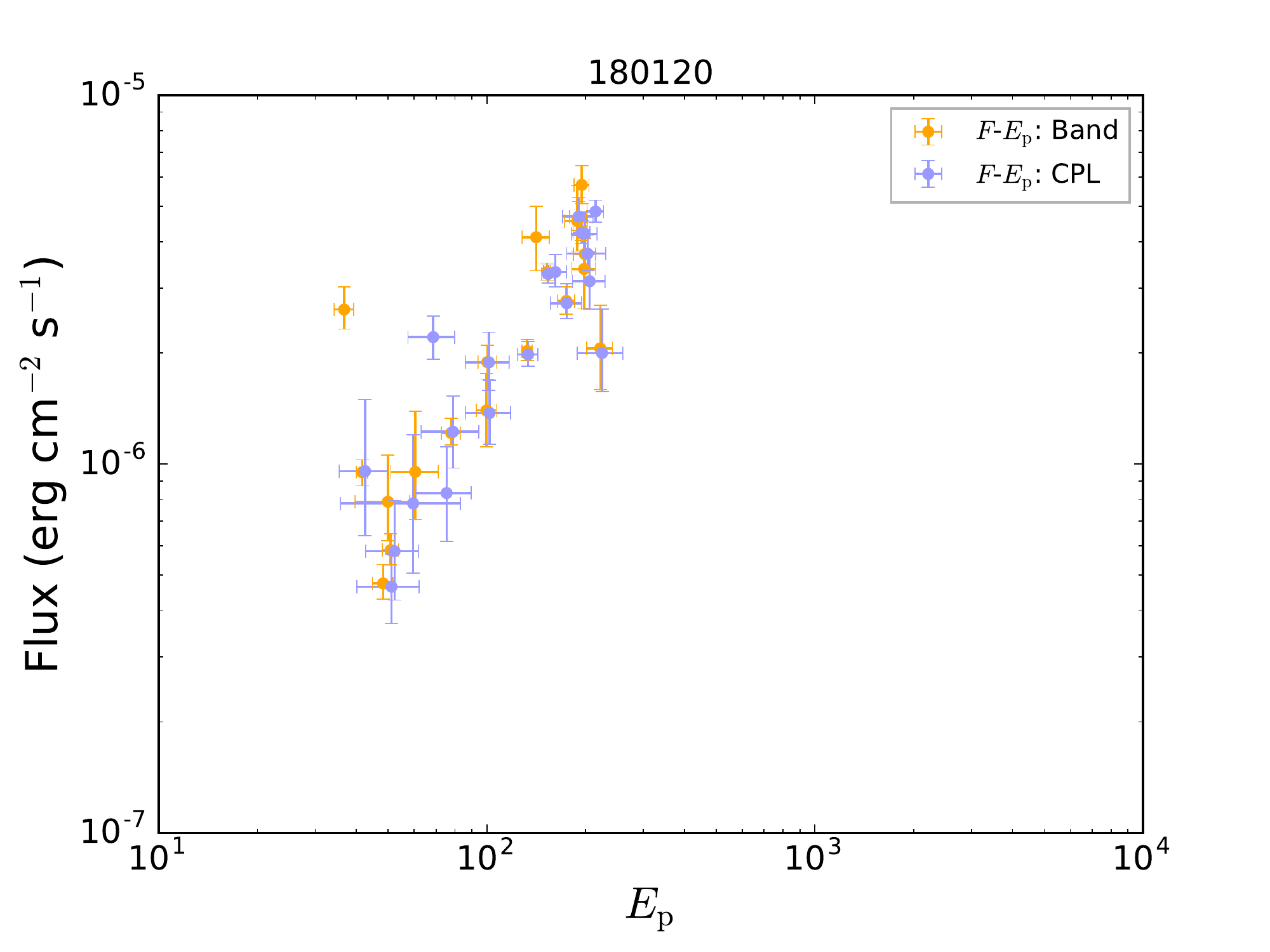}
\includegraphics[angle=0,scale=0.3]{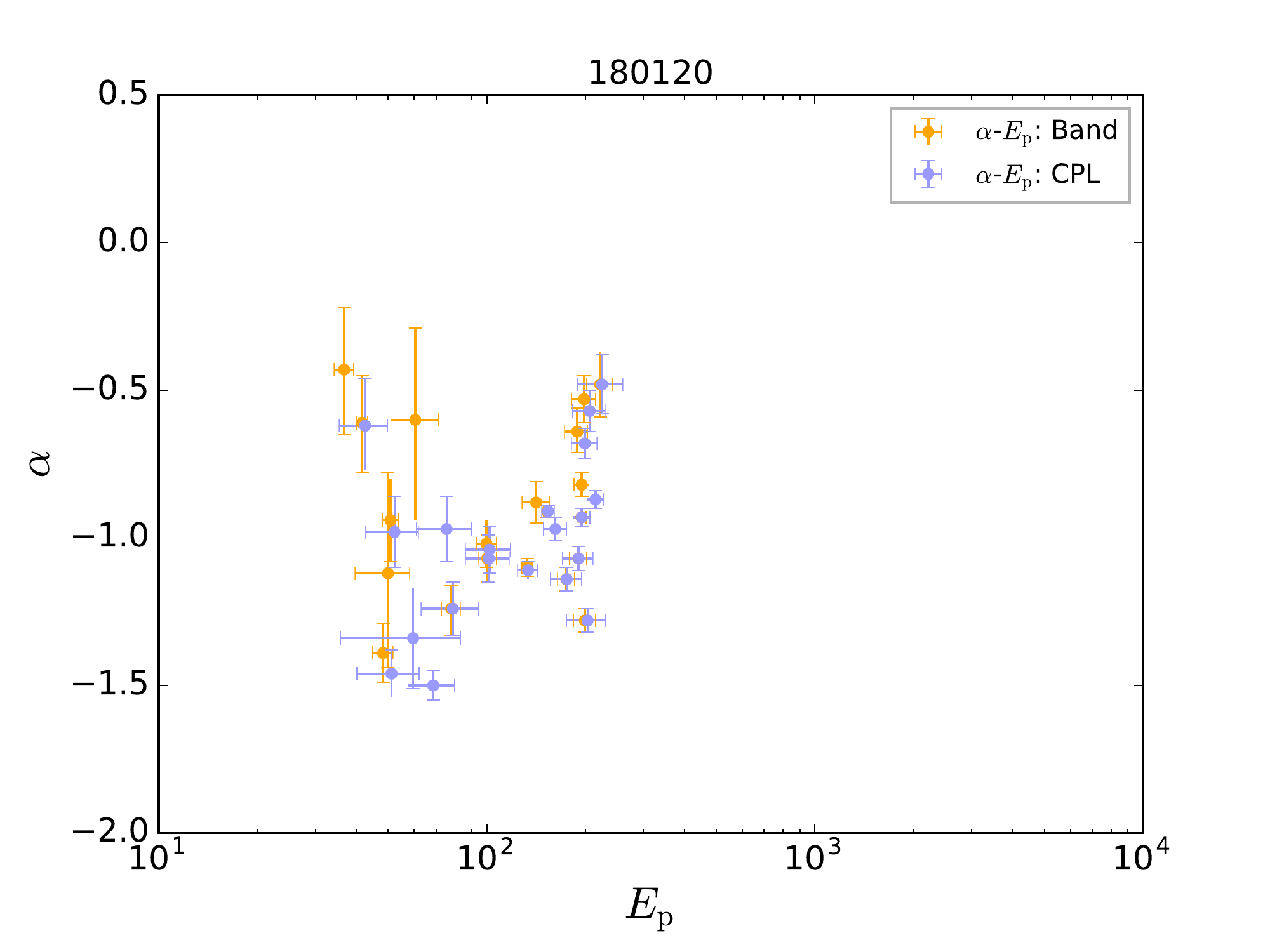}
\includegraphics[angle=0,scale=0.3]{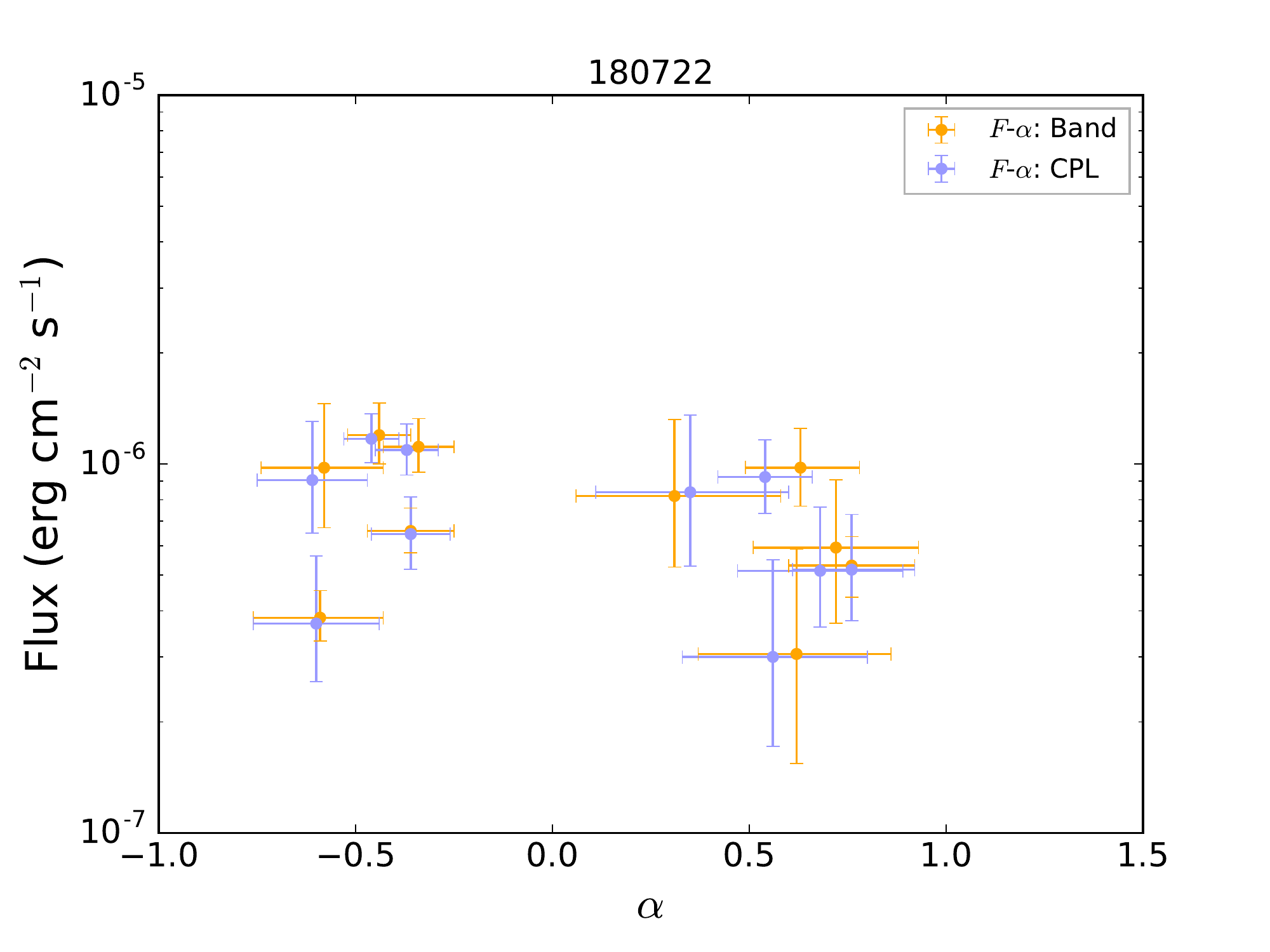}
\includegraphics[angle=0,scale=0.3]{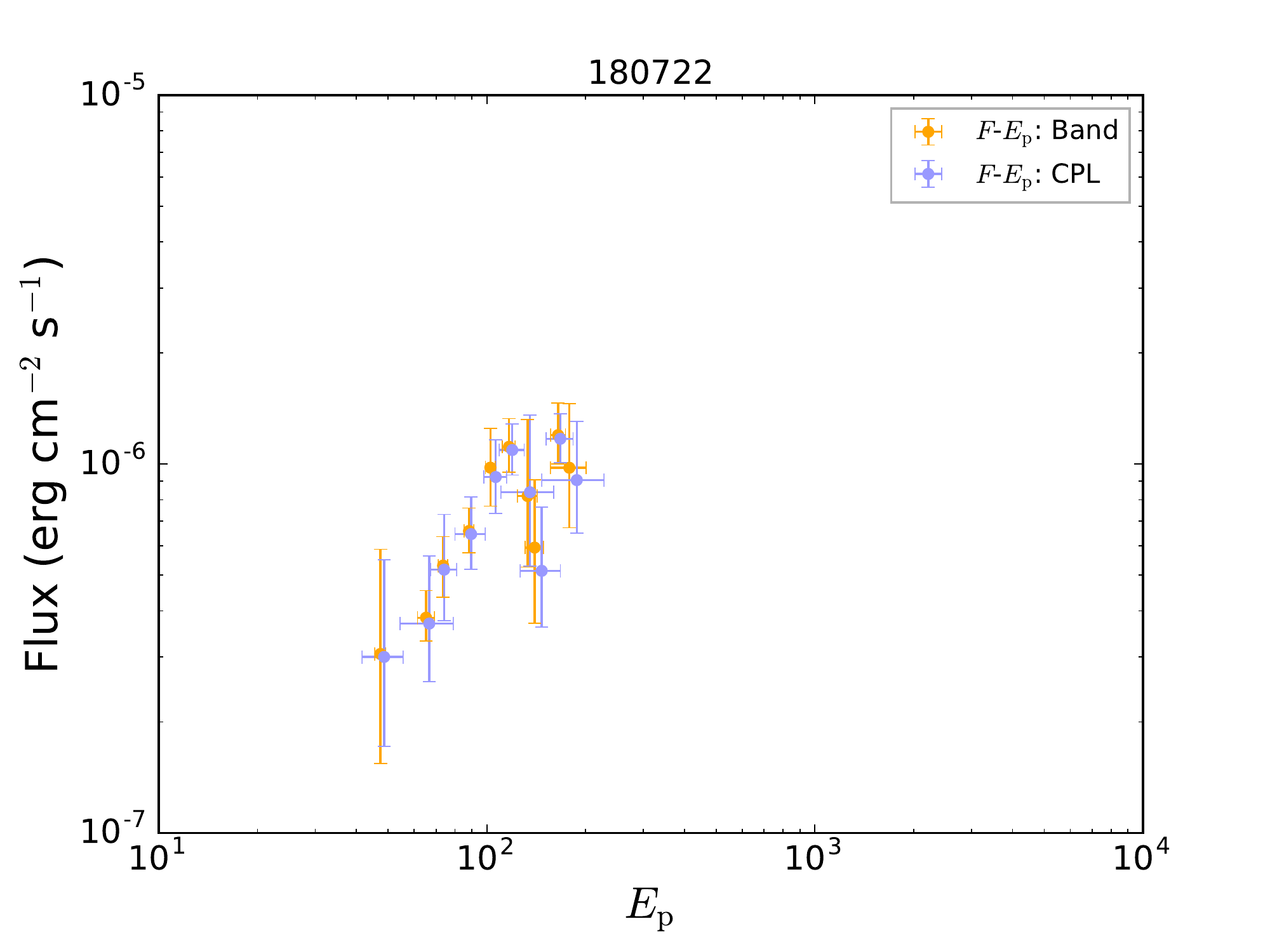}
\includegraphics[angle=0,scale=0.3]{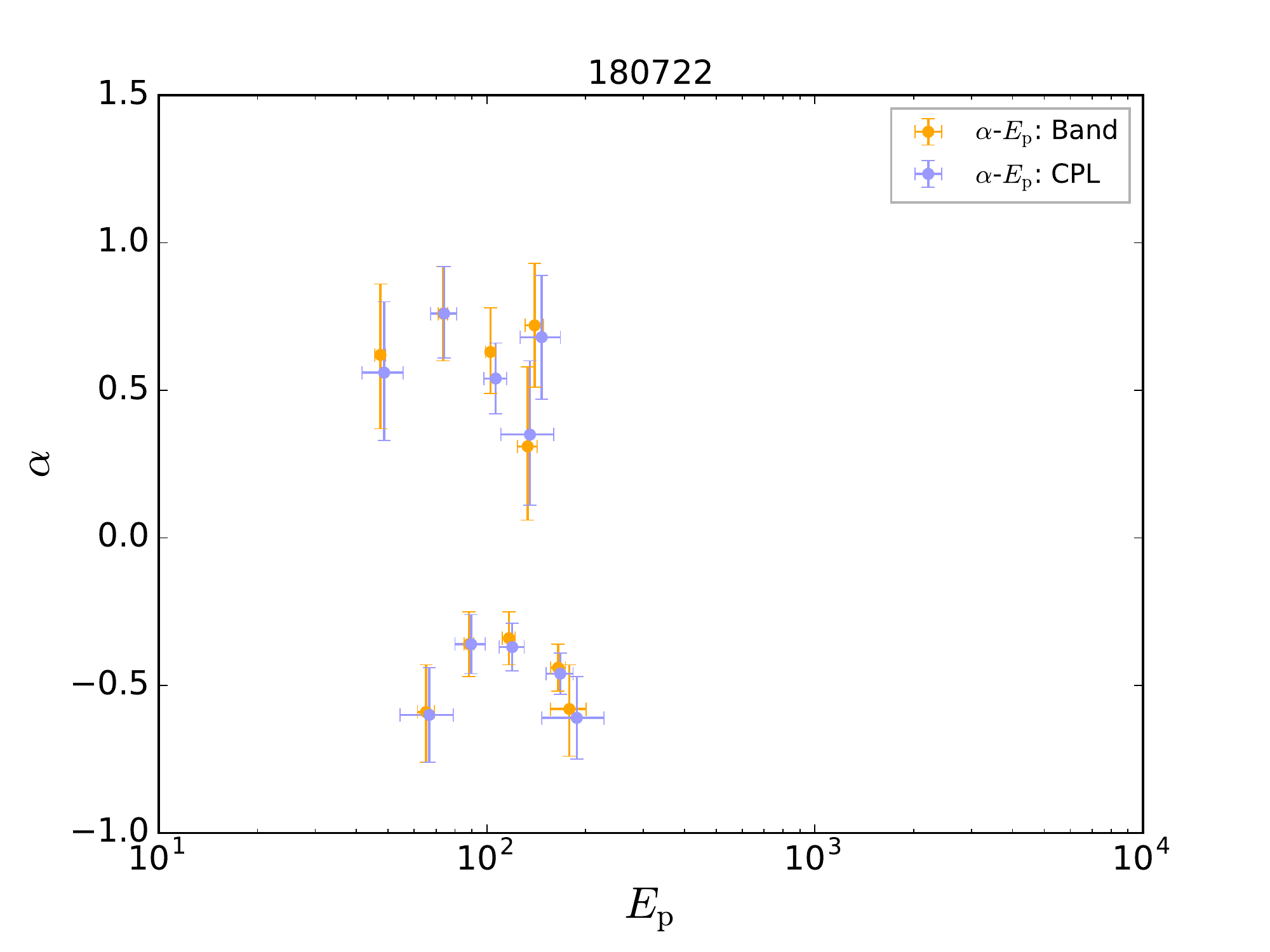}
\includegraphics[angle=0,scale=0.3]{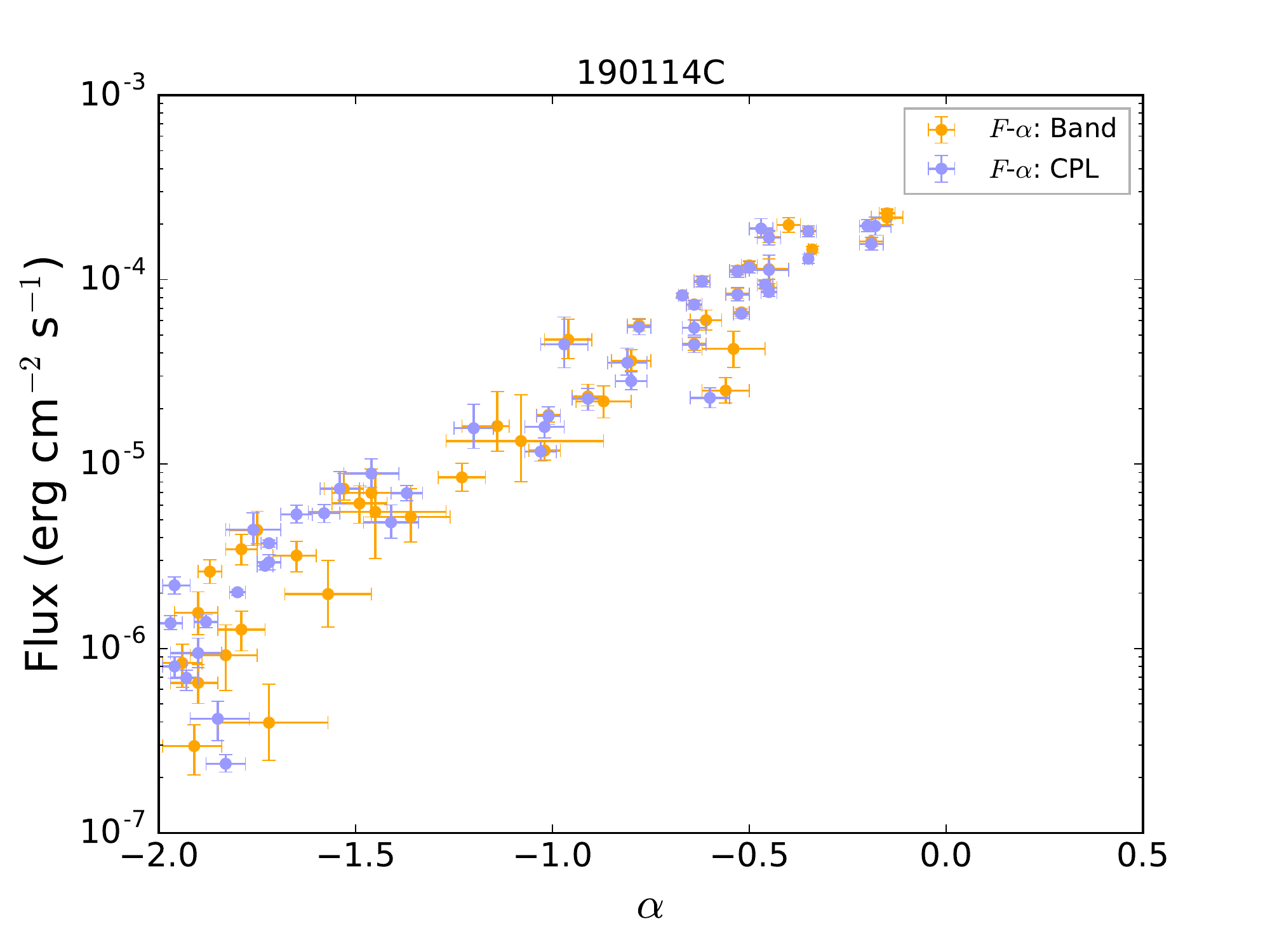}
\includegraphics[angle=0,scale=0.3]{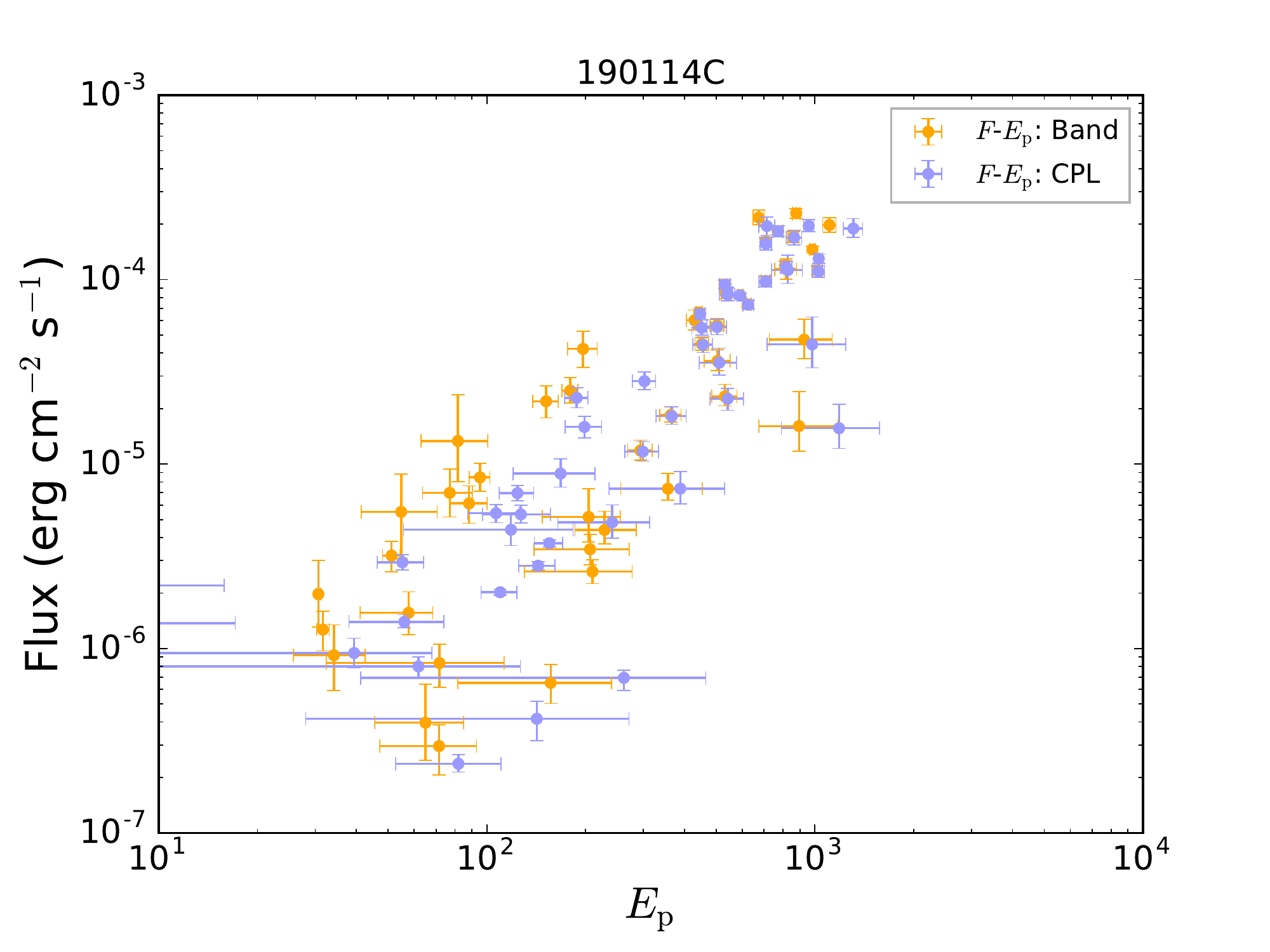}
\includegraphics[angle=0,scale=0.3]{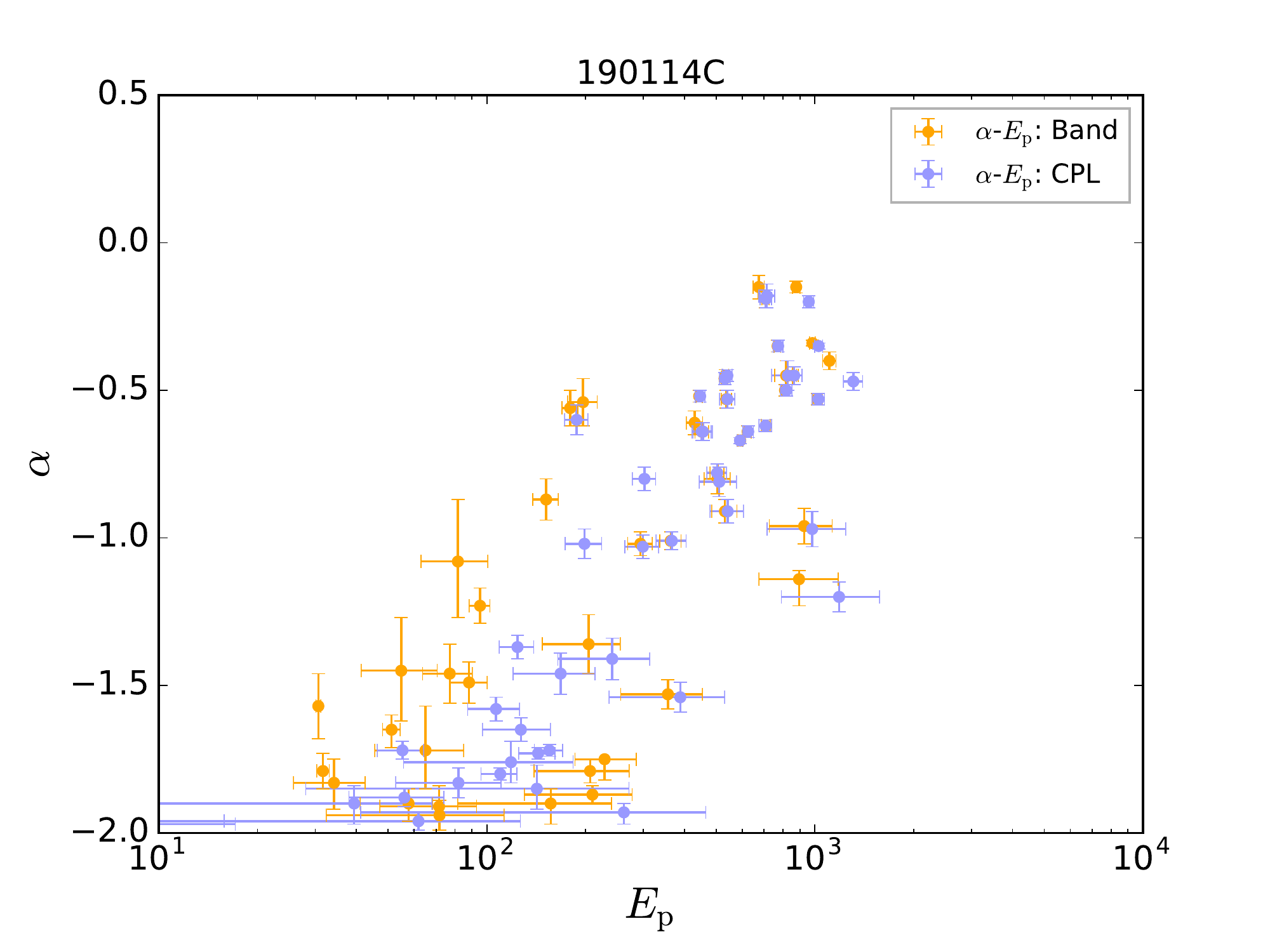}
\center{Fig. \ref{fig:relation3}--- Continued}
\end{figure*}

\end{document}